\documentclass{article}
\usepackage[utf8]{inputenc}
\usepackage{amsmath,amsfonts,mathtools}
\usepackage{subfig}
\usepackage{amssymb}
\usepackage{epsfig}
\usepackage{graphicx}
\usepackage{color}
\usepackage{hyperref}
\usepackage{comment}
\usepackage{cancel}
\usepackage{bbold}
\usepackage{lscape}
\usepackage{booktabs}
\usepackage[normalem]{ulem}
\usepackage[dvipsnames]{xcolor}
\usepackage{pbox}
\usepackage{multirow,bigdelim} % for braces in tables
\usepackage[dvipsnames]{xcolor}
\usepackage{float}
\usepackage[a4paper, margin=2.54cm]{geometry}
\usepackage{placeins}
\usepackage[english]{babel}
\usepackage{lineno}
\usepackage{pdflscape}
\usepackage{authblk}
\makeatletter
\renewcommand\maketitle{
{
\begin{center}
{\huge \@title \par}
\vspace{1cm}
{\large \textbf{The ECFA Early-Career Researchers Panel: Career Prospects and Diversity in Physics Programmes Working Groups}}
\\[0.8cm]
{\large \@date}
\\[0.8cm]
\begin{minipage}{0.82\textwidth}
\normalsize This document presents the outcomes of a comprehensive survey conducted among early career researchers (ECRs) in academic particle physics. Running from September 24, 2022, to March 3, 2023, the survey gathered responses from 759 ECRs employed in 39 countries. The study aimed to gain insights into the career prospects and experiences of ECRs while also delving into diversity and sociological aspects within particle physics research. The survey results are presented in a manner consistent with the survey choices. The document offers insights for the particle physics community, and provides a set of recommendations for enhancing career prospects, fostering diversity, and addressing sociological dimensions within this field.
\end{minipage}
\end{center}
\vspace{0.8cm}
\begin{flushleft}
{The ECFA Early-Career Researchers Panel (ECR) Panel: \href{mailto:ecfa-ecr-organisers@cern.ch}{ecfa-ecr-organisers@cern.ch}\\[0.5cm]\@author}
\end{flushleft}}}
\makeatother

\RequirePackage[style=numeric,backend=biber,sorting=none]{biblatex}
\begin{document}

\title{Results of the 2022 ECFA Early-Career Researchers Panel survey on career prospects and diversity}
\date{\today}

% TODO: check for accents on names and institutes!
%New:
\author[*,1]{Julia~Allen}
\author[2]{Kamil~Augsten}
\author[3]{Giovanni~Benato}
\author[4]{Neven~Blaskovic~Kraljevic}
\author[5]{Francesco~Brizioli}
\author[6]{Eleonora~Diociaiuti}
\author[7]{Viktoria~Hinger}
\author[8]{Armin~Ilg}
\author[*,9]{Kateřina~Jarkovská}
\author[10]{Katarína~Křížková~Gajdošová} % add accents
\author[11]{Magdalena~Kuich}
\author[*,12]{Aleksandra~Lelek}
\author[13]{Louis~Moureaux}
\author[*,14]{Holly~Pacey}
\author[*,15]{Guillaume~Pietrzyk}
\author[*,16]{G\'{e}raldine~R\"{a}uber}
\author[17]{Giulia~Ripellino}
\author[18]{Steven~Schramm}
\author[19]{Mariana~Shopova}
\author[20]{Pawel~Sznajder}
\author[21]{Abby~Waldron}

\affil[*]{Editor}
\affil[1]{University of Edinburgh, Edinburgh; United Kingdom}
\affil[2]{Faculty of Nuclear Sciences and Physical Engineering, Czech Technical University in Prague, Prague; Czech Republic}
\affil[3]{INFN Laboratori Nazionali del Gran Sasso, L'Aquila; Italy}
\affil[4]{MAX IV Laboratory, Lund University, Lund; Sweden}
\affil[5]{INFN Sezione di Perugia, Perugia; Italy}
\affil[6]{National Laboratory of Frascati and INFN, Frascati; Italy}
\affil[7]{Paul Scherrer Institut, Villigen; Switzerland}
\affil[8]{Physik-Institut, University of Zürich, Zürich; Switzerland}
\affil[9]{Faculty of Mathematics and Physics, Charles University, Prague; Czech Republic}
\affil[10]{Faculty of Nuclear Sciences and Physical Engineering, Czech Technical University in Prague, Prague; Czech Republic}
\affil[11]{University of Warsaw, Warsaw; Poland}
\affil[12]{Universiteit Antwerpen, Antwerpen; Belgium}
\affil[13]{Universit\"{a}t Hamburg, Hamburg; Germany}
\affil[14]{University of Oxford, Oxford; United Kingdom}
\affil[15]{IJCLab, CNRS/IN2P3, Université Paris-Saclay, Orsay; France}
\affil[16]{Institute of High Energy Physics, Austrian Academy of Sciences, Vienna; Austria}
\affil[17]{Department of Physics, Royal Institute of Technology, Stockholm; Sweden}
\affil[18]{D\'{e}partement de Physique Nucl\'{e}aire et Corpusculaire, Universit\'{e} de Gen\`{e}ve, Gen\`{e}ve; Switzerland}
\affil[19]{Institute for Nuclear Research and Nuclear Energy, Bulgarian Academy of Sciences, Sofia; Bulgaria}
\affil[20]{National Centre for Nuclear Research (NCBJ), Warsaw; Poland}
\affil[21]{Blackett Laboratory, Imperial College London, London; United Kingdom}

\maketitle

\tableofcontents

\FloatBarrier
\newpage
\section{Introduction}

This survey was created in order to comprehensively study the opinions of self-identifying early career researchers (ECRs) who work in academic particle physics on their career prospects and experiences within academia, in addition to understanding more about diversity and sociological aspects of particle physics research.
We also wanted to better understand what things ECRs would like to be changed or improved within academia.
A full list of the questions asked in the survey is provided in Appendix~\ref{app:questions}.

The survey was implemented in Google Forms and was distributed widely through:
\begin{itemize}
    \item ECFA national contacts;
    \item mailing lists for several collaborations (ALICE, ATLAS, CERN, CMS, Compass+Amber, EIC, FCC, LHCb, Mu2e, NA61/SHINE, NA62);
    \item national mailing lists of several panellist countries (Belgium, Czech Republic, Netherlands, Switzerland).
\end{itemize}
It collected 759 responses between the 24th September 2022 and the 3rd March 2023.

The survey was formatted primarily through multiple choice questions.
Where an `Other' category is presented, this was given as a free box for text.
Unless otherwise stated, the categories used in the presentation of the results correspond exactly to the choices presented in the survey.
For appropriate questions, respondents were allowed to select more than one choice, and this is indicated in the text.
Most of the questions were mandatory to answer, those that were voluntary are specified in the text. 

In Section~\ref{sec:answers} of the survey analysis, we present the responses to the questions asked in the survey, either as pie charts, histograms, or text.
Where appropriate, responses were grouped into broader categories to have a better sample size from which to draw conclusions.
In Section~\ref{sec:correlations} we study correlations between the answers given to different questions, in order to more comprehensively probe ECR views.
Finally, in Section~\ref{sec:conclusion}, we draw conclusions and suggest possible actions or events that the ECFA ECR panel, and the wider HEP community, could support in the future.
We kindly encourage readers with limited time to focus on Section~\ref{sec:conclusion} which can be easily followed without detailed study of the rest of the document.

This project was completed by the `Career Prospects' and `Diversity in Physics Programmes' working groups within the ECFA ECR Panel.
More information on the panel and its activities can be found in Refs.~\cite{ECFAECRPanel} and \cite{report2021-2022}.

%%%%%%%%%%%%%%%%%%%%%%%%%%%%%%%%%%%%%%%%%%%%%%%%%%%%%%%%%%%%%%%%%%%%%%%%%%%%%%%%%%%%%%%%%%%%%%%%
\section{Responses to questions}
\label{sec:answers}

\subsection{Demographics of respondents}

\begin{figure}[h!]
    \centering
    \subfloat[]{\label{fig:part1:Q1}\includegraphics[width=0.49\textwidth]{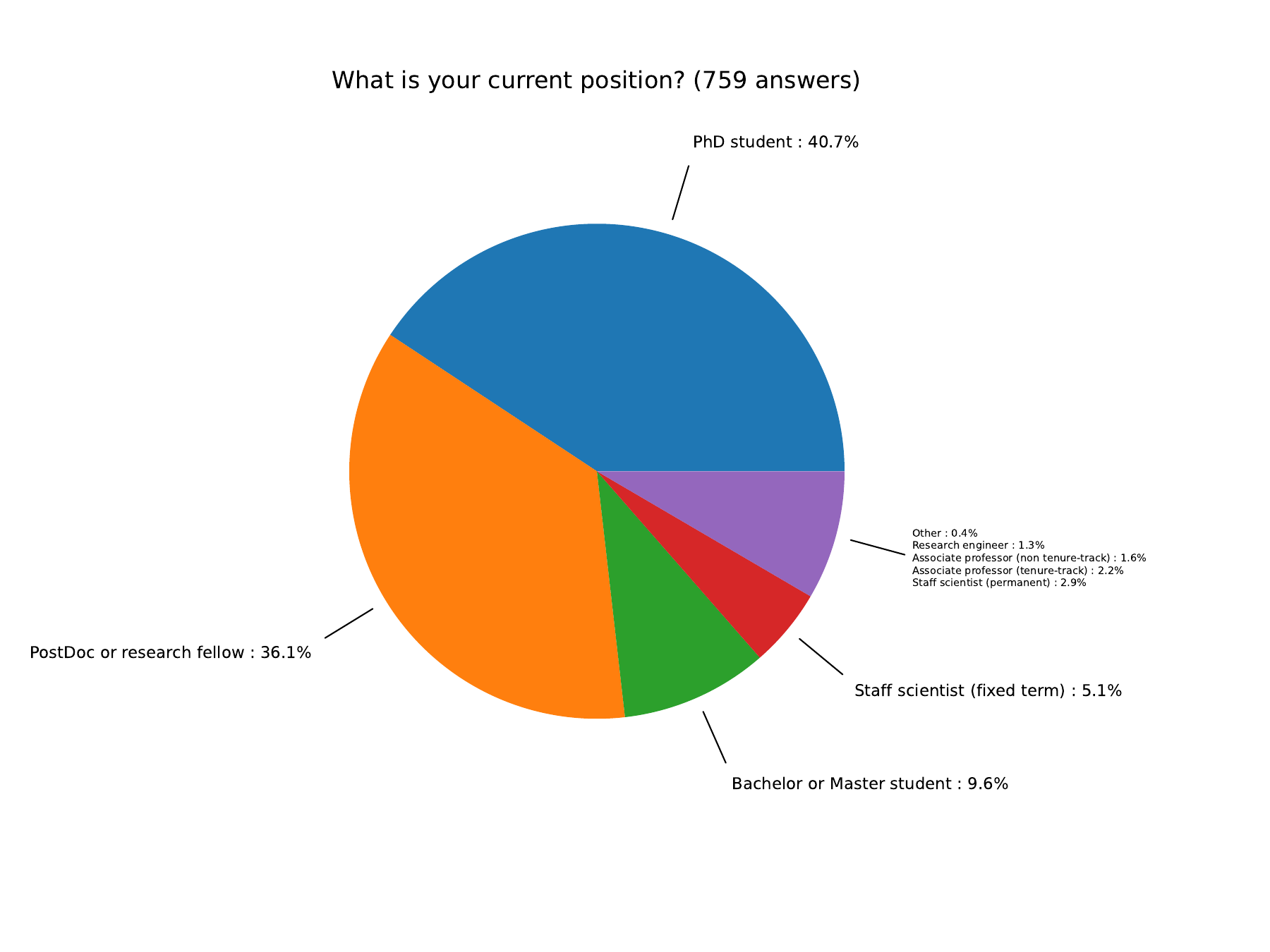}}
    \subfloat[]{\label{fig:part1:Q2}\includegraphics[width=0.49\textwidth]{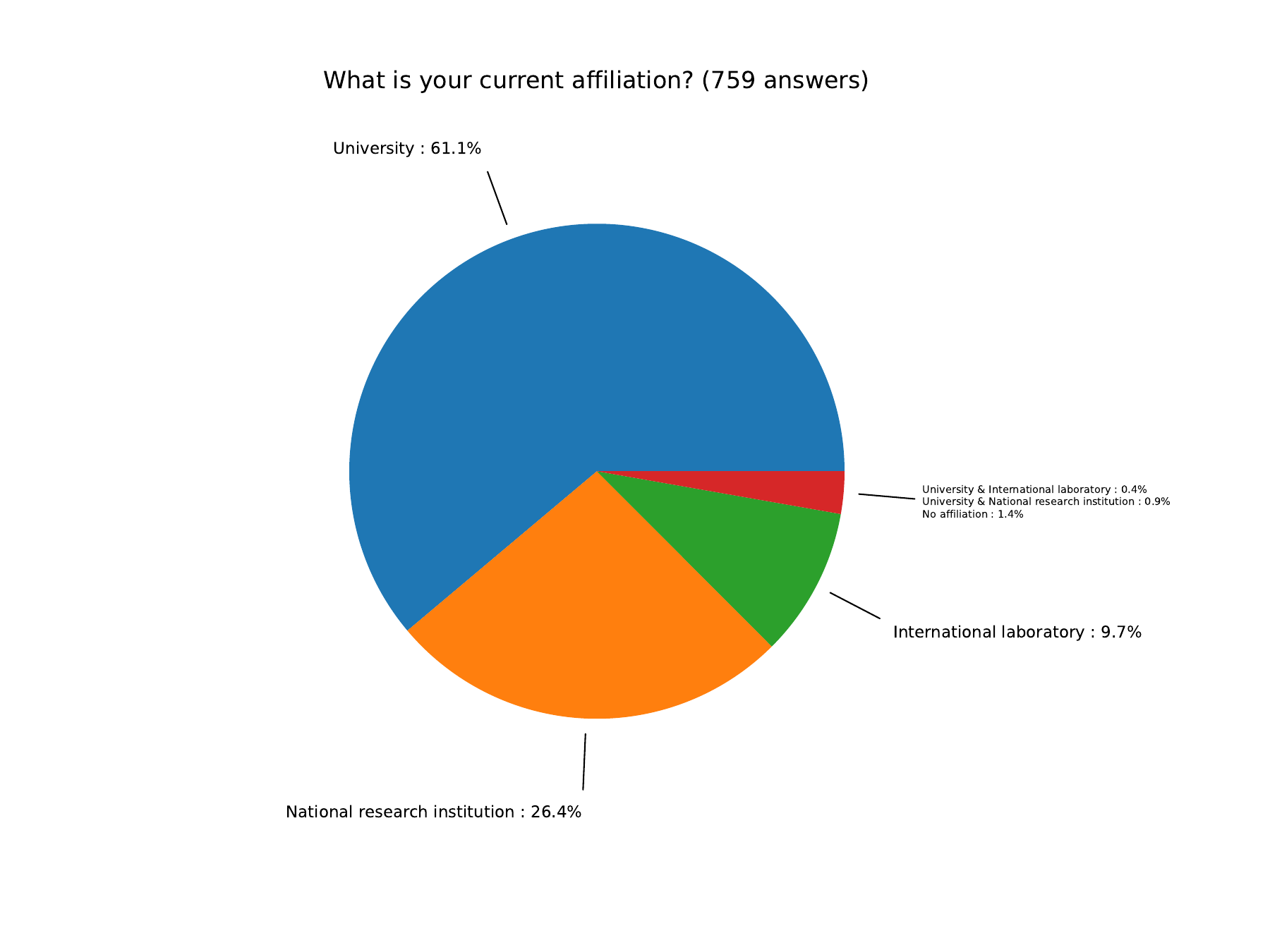}}\\
    \subfloat[]{\label{fig:part1:Q3}\includegraphics[width=0.49\textwidth]{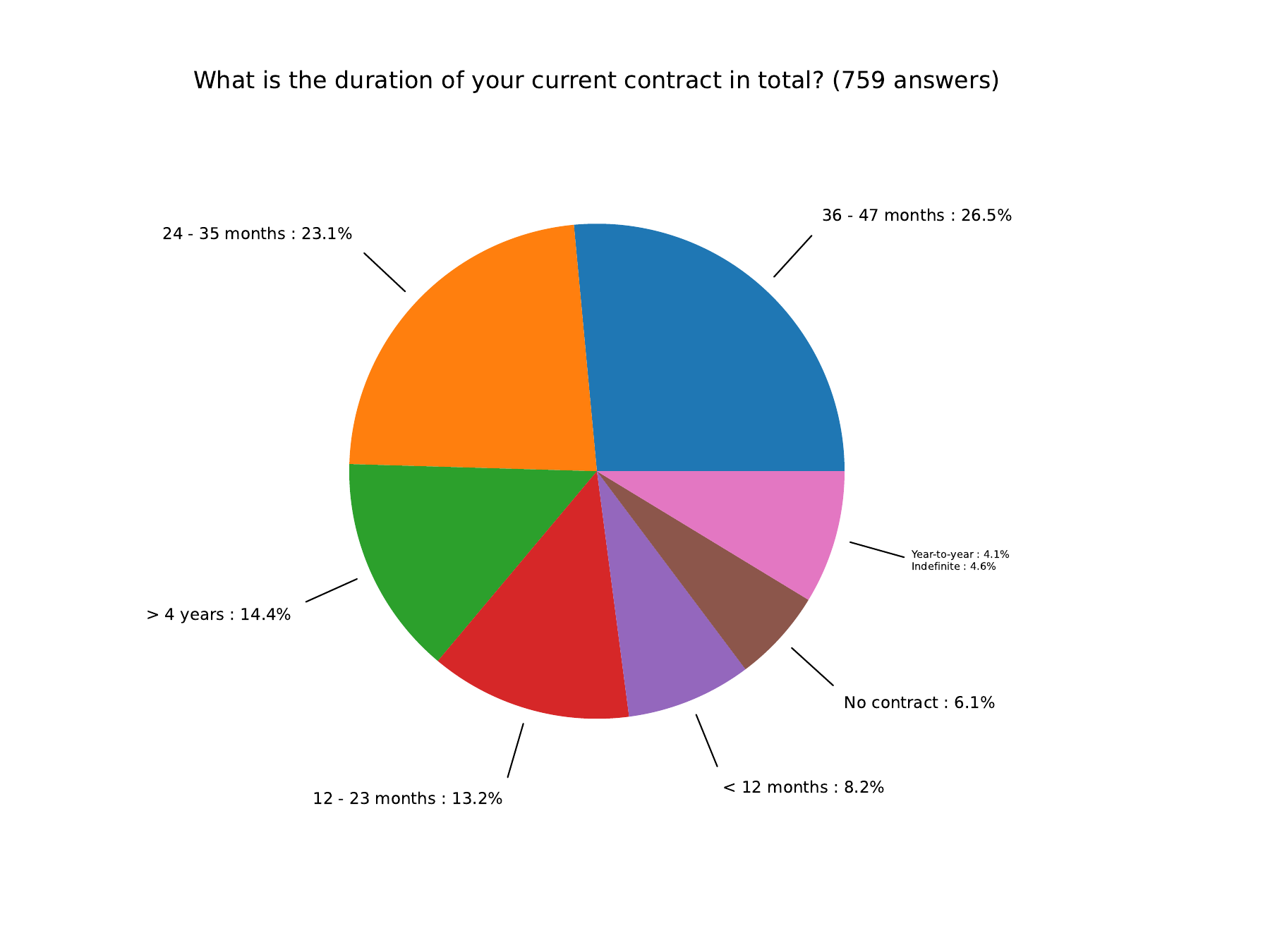}}
    \subfloat[]{\label{fig:part1:Q4}\includegraphics[width=0.49\textwidth]{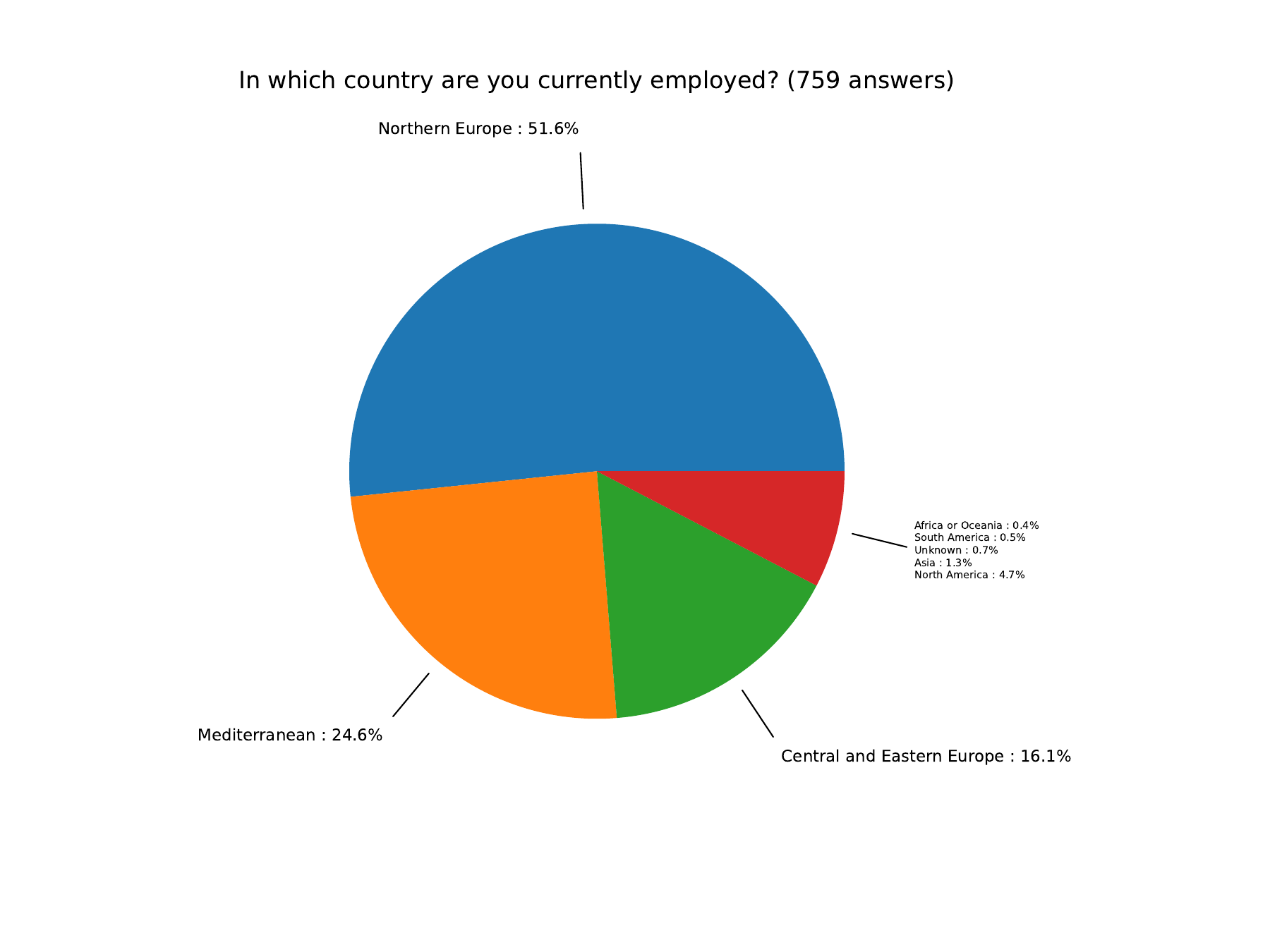}}\\
    \caption{(Q1--4) Pie charts of some respondent demographics.}
    \label{fig:part1:Q1Q4}
\end{figure}

In Figure~\ref{fig:part1:Q1}, the current position of the respondents is plotted, showing that the majority of the respondents are PhD students and slightly more than a third are PostDocs or research fellows.
Figure~\ref{fig:part1:Q2} presents the current affiliations of the respondents.
It indicates that the majority of respondents work at a university, just under half as many work in a national research institution, and just under half as many again work in an international laboratory. 
The duration of respondents' current contracts are shown in Figure~\ref{fig:part1:Q3}. The dominant durations are 36--47 months and 24--35 months.

\begin{figure}[h!]
    \centering
    \subfloat[]{\label{fig:part1:Q5}\includegraphics[width=0.49\textwidth]{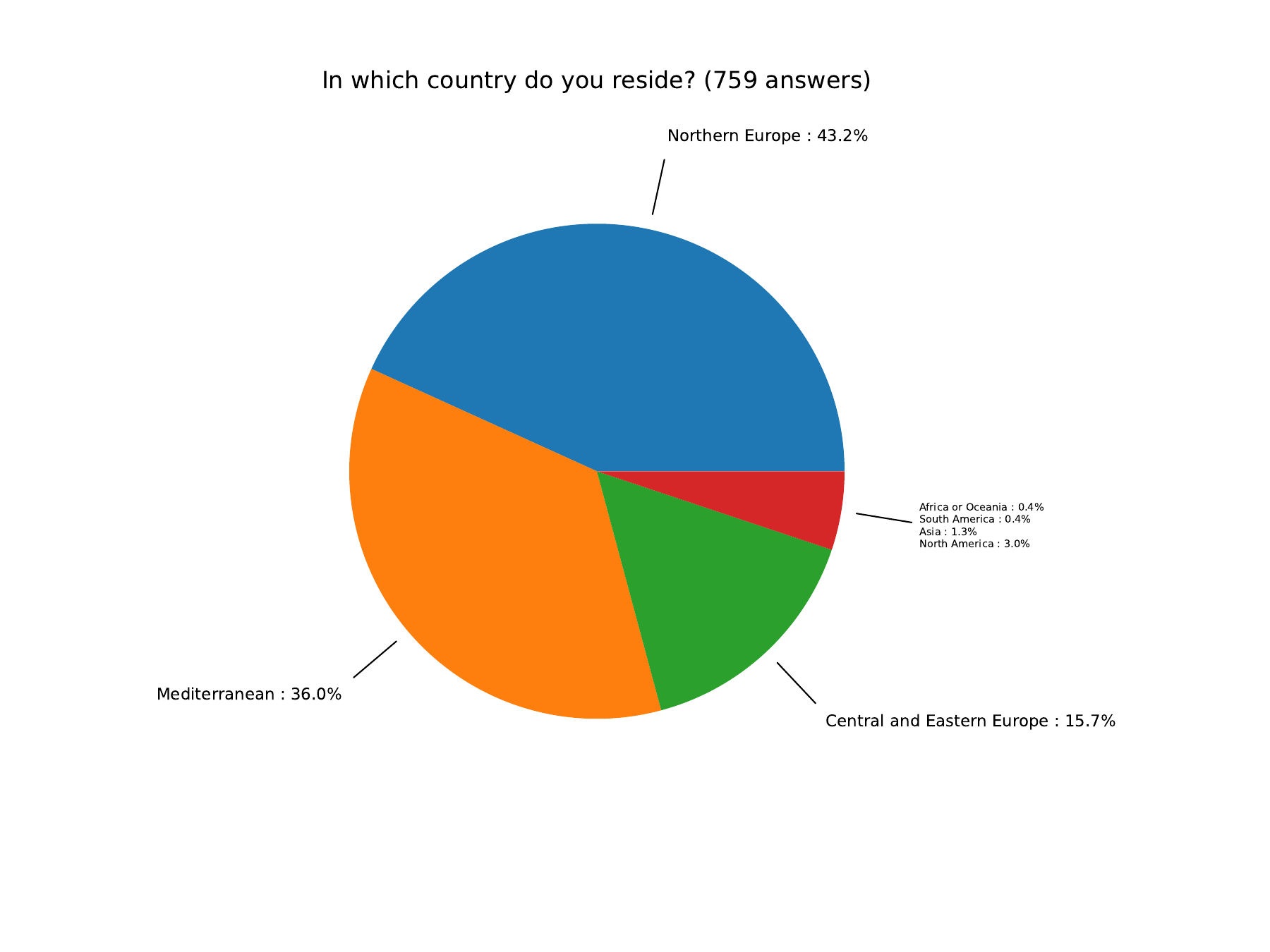}}
    \subfloat[]{\label{fig:part1:Q6}\includegraphics[width=0.49\textwidth]{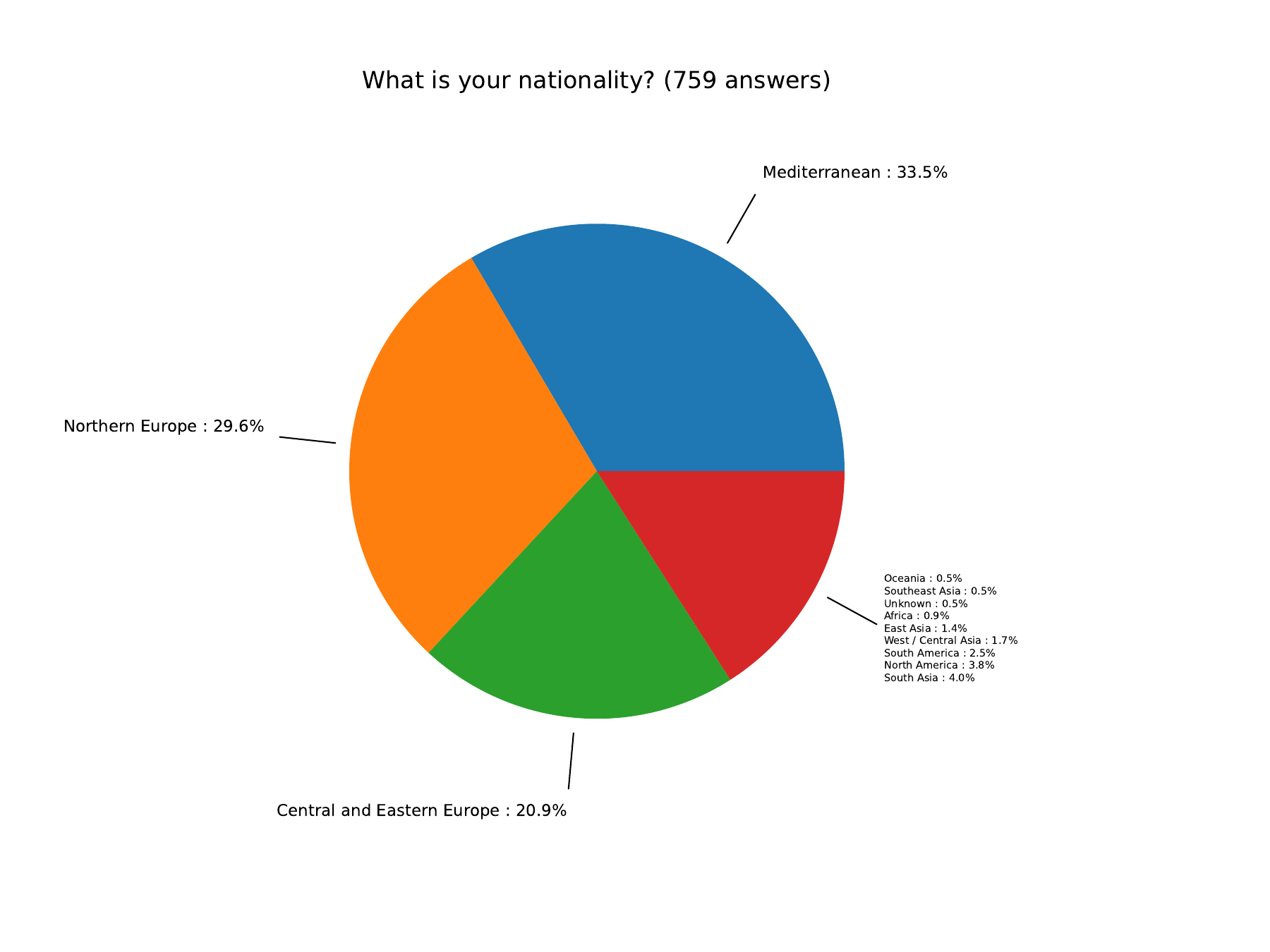}}\\
    \subfloat[]{\label{fig:part1:Q7}\includegraphics[width=0.49\textwidth]{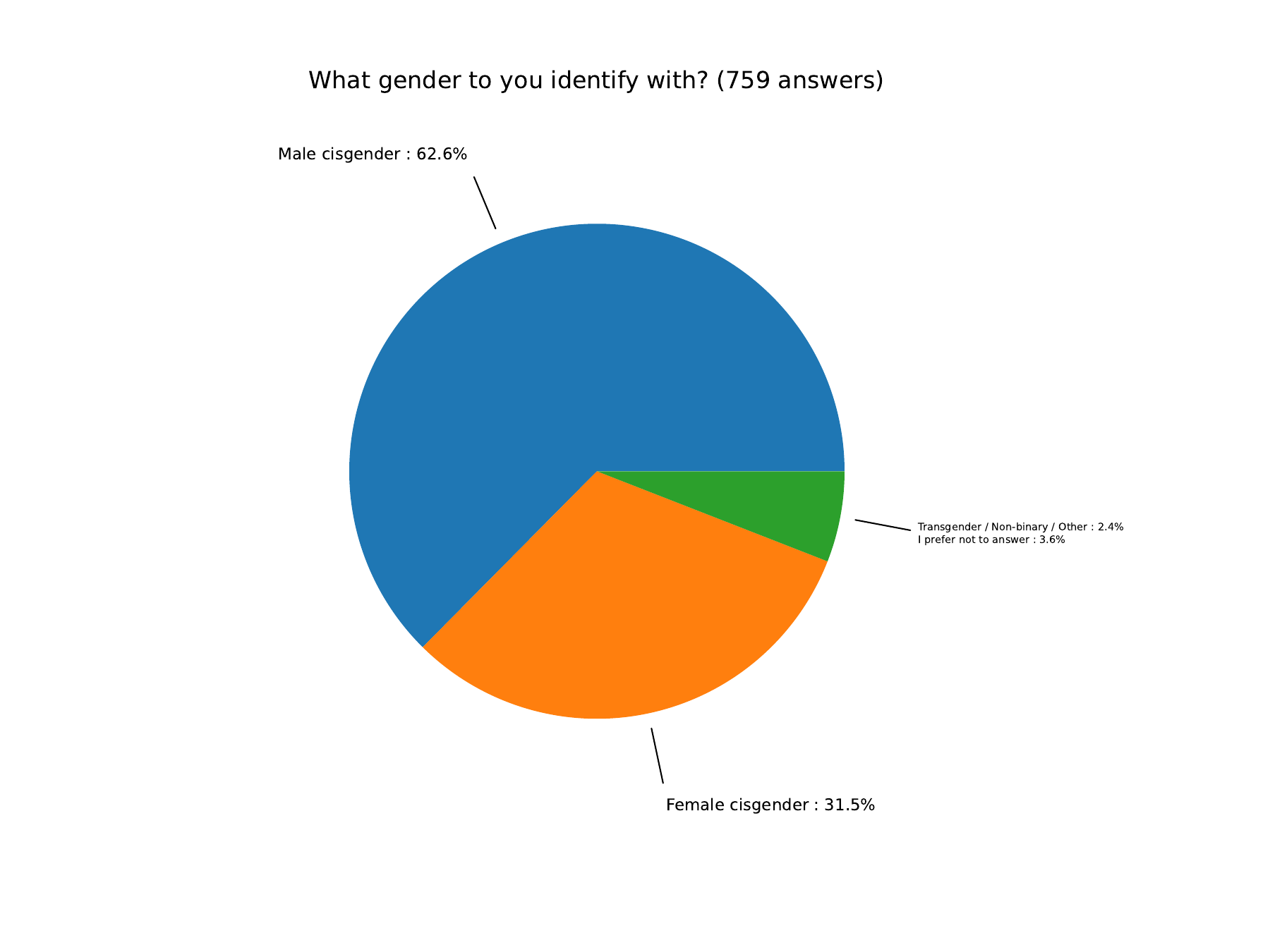}}
    \subfloat[]{\label{fig:part1:Q8}\includegraphics[width=0.49\textwidth]{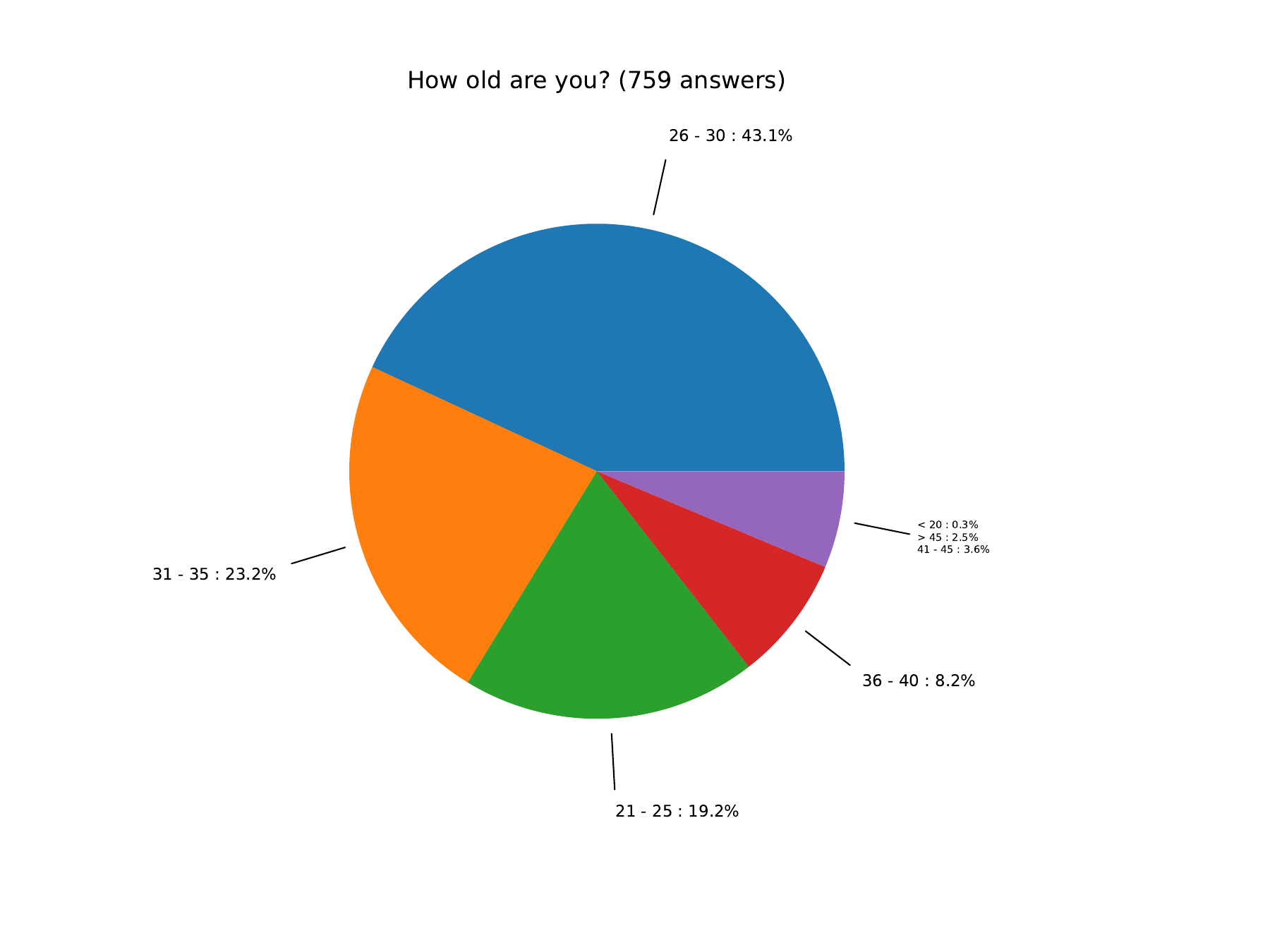}}
    \caption{(Q5--8) Pie charts of some more respondent demographics.}
    \label{fig:part1:Q5Q8}
\end{figure}

The participants represent 63 nationalities, employed in 39 countries and residing in 36 countries. 
The current countries of employment and of residence, grouped into geographical regions (see Appendix \ref{app:AppNationalityGroups} for how these groups were defined), are presented in Figures~\ref{fig:part1:Q4}--\ref{fig:part1:Q5}.
From both figures, we see that Northern Europe constitutes the majority of the answers, followed by Mediterranean and Central and Eastern Europe.  
The nationalities of the respondents are grouped into geographical regions in Figure~\ref{fig:part1:Q6}.
Around one third of the respondents are Mediterranean, and around another third are from Northern Europe.
A fifth are from Central and Eastern Europe, and remaining regions each represent a very small fraction of respondents. 

In Figure~\ref{fig:part1:Q7}, the gender the respondents identify with is illustrated.
More than a half of the respondents identify as a cisgender male, almost a third as a cisgender female and a much smaller group identify as a transgender, non-binary or another gender.
The age of the respondents is shown in Figure~\ref{fig:part1:Q8}.
Almost half of the respondents are aged between 26 and 30. Respondents aged between 31 and 35 or between 21 and 25 years old each correspond to slightly more than a fifth of the total.

\begin{figure}[h!]
    \centering
        \subfloat[]{\label{fig:part1:Q10}\includegraphics[width=0.49\textwidth]{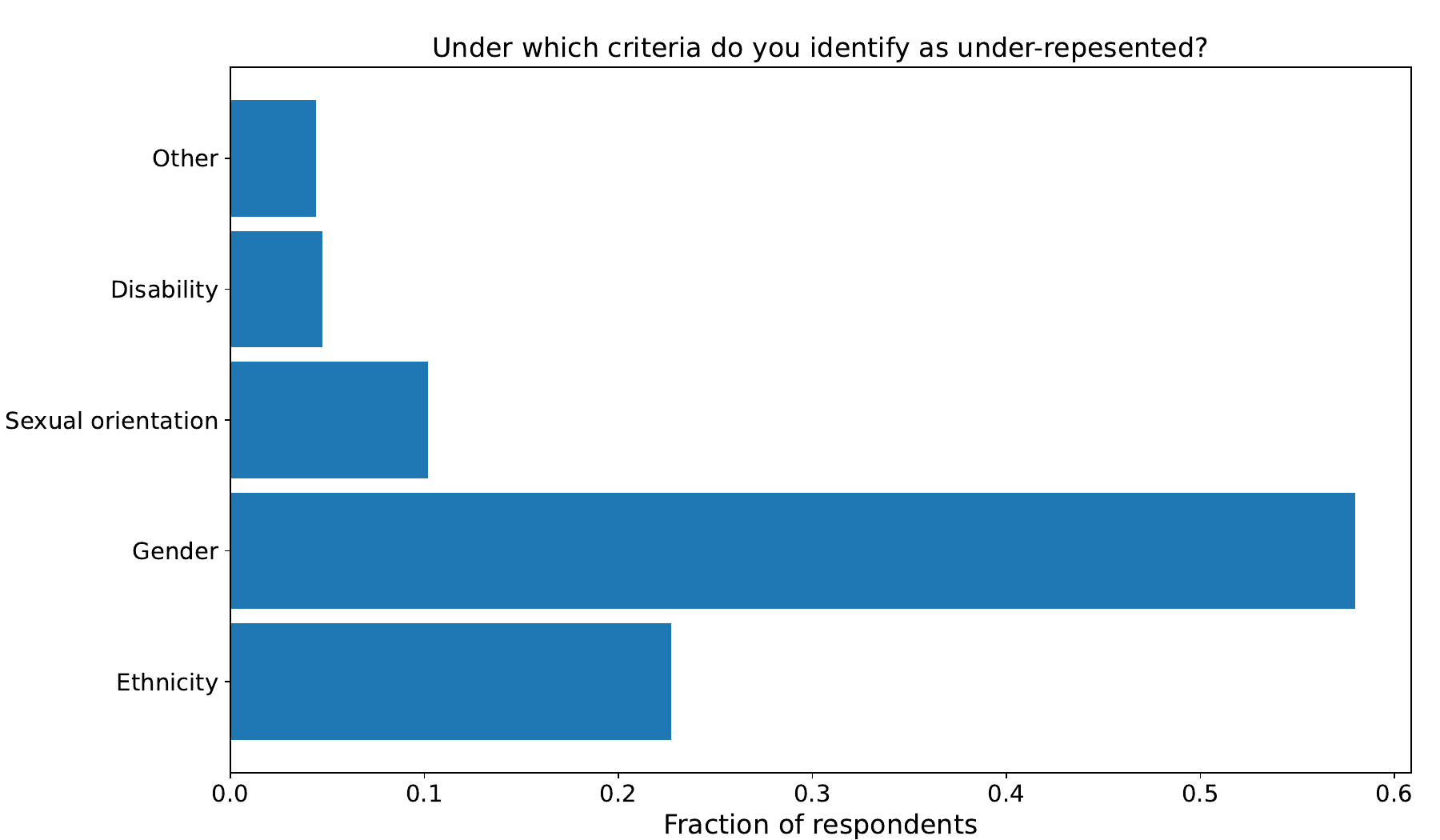}}
        \subfloat[]{\label{fig:part1:Q11}\includegraphics[width=0.49\textwidth]{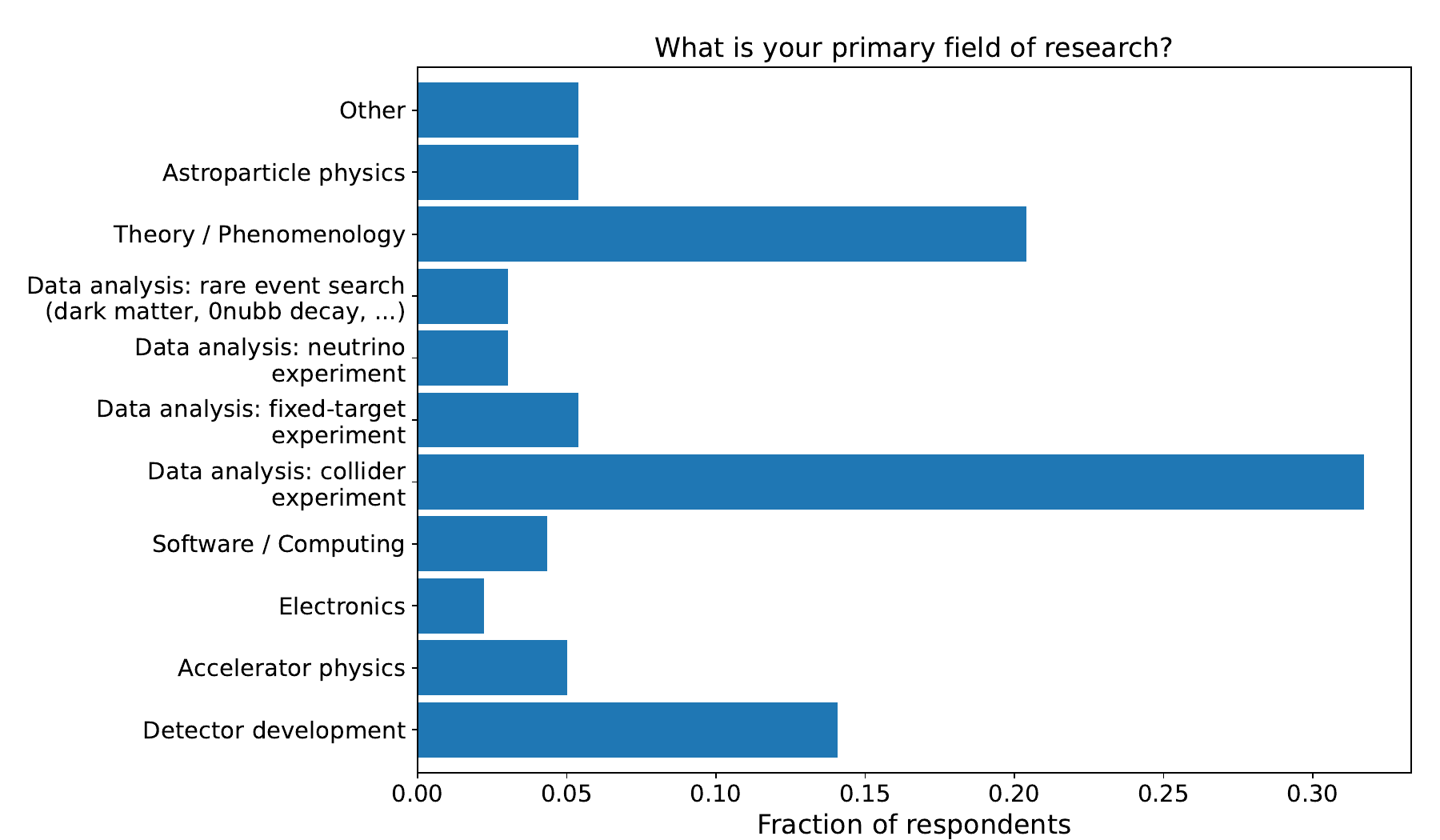}}\\
        \subfloat[]{\label{fig:part1:Q12}\includegraphics[width=0.49\textwidth]{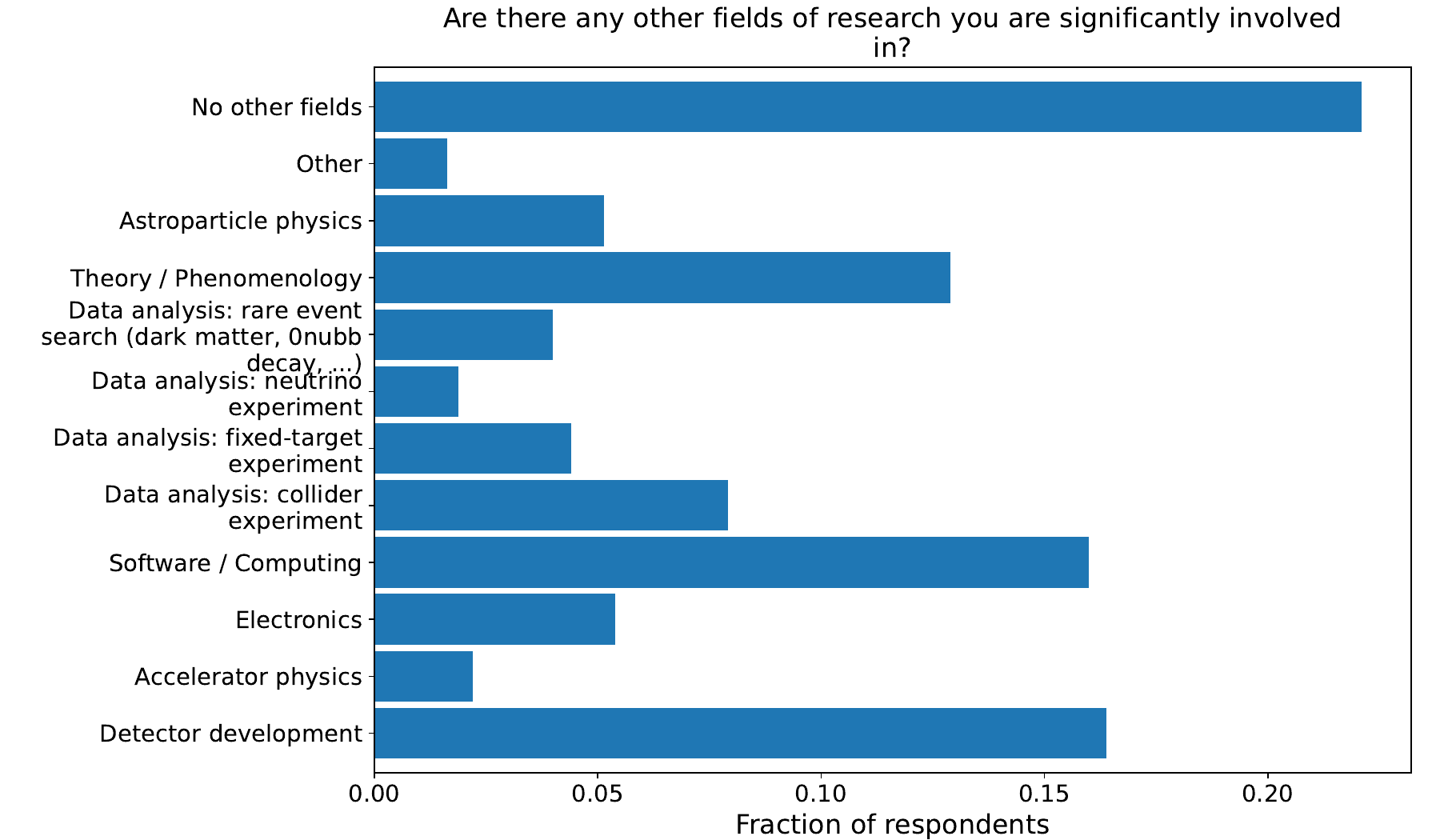}}
        \subfloat[]{\label{fig:part1:Q13}\includegraphics[width=0.49\textwidth]{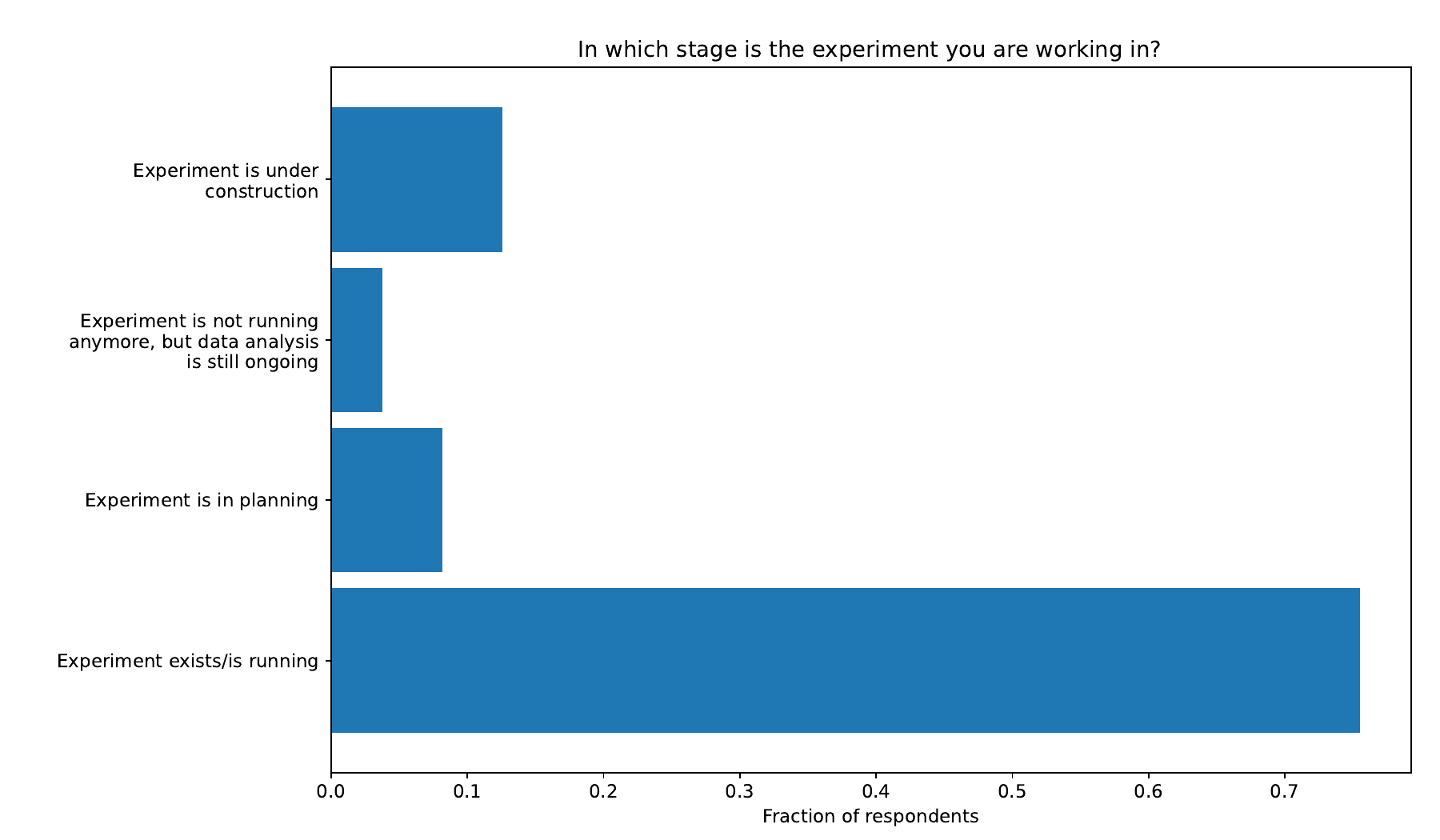}}
  \caption{(Q10--13) (a) The criterion under which respondents identify themselves as under-represented. (b) Respondents' primary field of research. (c) Other fields of research that respondents are significantly involved in. (d) The status of the experiments respondents are working on, with fractions given out of those who responded to the question. For (a) and (c) multiple answers were allowed per respondent. Questions (a) and (d) were voluntary.}
    \label{fig:part1:Q11Q12}
\end{figure}

We found that 26\% of participants identified as being part of an under-represented group within the physics community.
Figure~\ref{fig:part1:Q10} presents the criterion through which these respondents identify as under-represented.
More than one category could be selected and the question was not mandatory, though all respondents who identified as under-represented provided an answer.
Over 60\% state this is due to gender.
The next two largest criteria are ethnicity and sexual orientation. 
Within the `Other' category, 54\% are under-represented due to socio-economic background, and the remainder due to religion or political views.
% Note we checked that the same set of respondents who IDed as under-rep in Q9 gave a category in Q10.

Finally, we present the responses to questions concerning field of work.
The primary field of research of respondents is illustrated in Figure~\ref{fig:part1:Q11}.
Almost half of the respondents have a primary field of research related to data analysis (43.2\%).
% 43.16 = 31.71 + 5.39 + 3.03 + 3.03
Theory and phenomenology represent together approximately a fifth of the respondents, which is similar to the proportion who work in detector or accelerator physics (19.1\%).
% 19.1 = 5.0 + 14.1
Out of the `other' category, 31\% work with astrophysics and cosmology, 21\% work in nuclear or atomic physics, 15\% in Optics, 13\% in other experimental physics, 13\% in medical physics or biophysics, 5\% in engineering and 3\% in IT.

The majority of participants are not involved in an additional field of research, as shown in Figure~~\ref{fig:part1:Q12} (where multiple answers were possible).
For those that are, dominant additional fields are software/computing, detector development or theory/phenomenology.

We found that 75\% of respondents work on one experiment, and 1.4\% indicated that they work on more than one.
For respondents working on an experiment, the status of the experiment (for those who chose to answer this voluntary question) is shown in Figure \ref{fig:part1:Q13}.

%%%%%%%%%%%%%%%%%%%%%%%%%%%%%%%%%%%%%%%%%%%%%%%%%%%%%%%%%%%%%%%%%%%%%%%%%%%%%%%%%%%%%%%%%%%%%%%%%%%%%%%%%%%%%%%
%\FloatBarrier
%\pagebreak
\subsection{Work within a research group or collaboration}

In this section, we present the results of the survey concerning work as part of a research group or a collaboration.
For this survey, ``research group'' is defined as a group of researchers that work together on a daily basis, share the lab and/or office space, and have the same Principal Investigator(s) (PI(s)) and affiliation. 
Furthermore, ``collaboration'' is defined as a set of over 2 groups from different institutions in different cities, countries or continents that work together towards a common scientific goal, which could be a new measurement, the development of a new detector, or a new theory.

\begin{figure}[h!]
    \centering
        \includegraphics[width=0.6\textwidth]{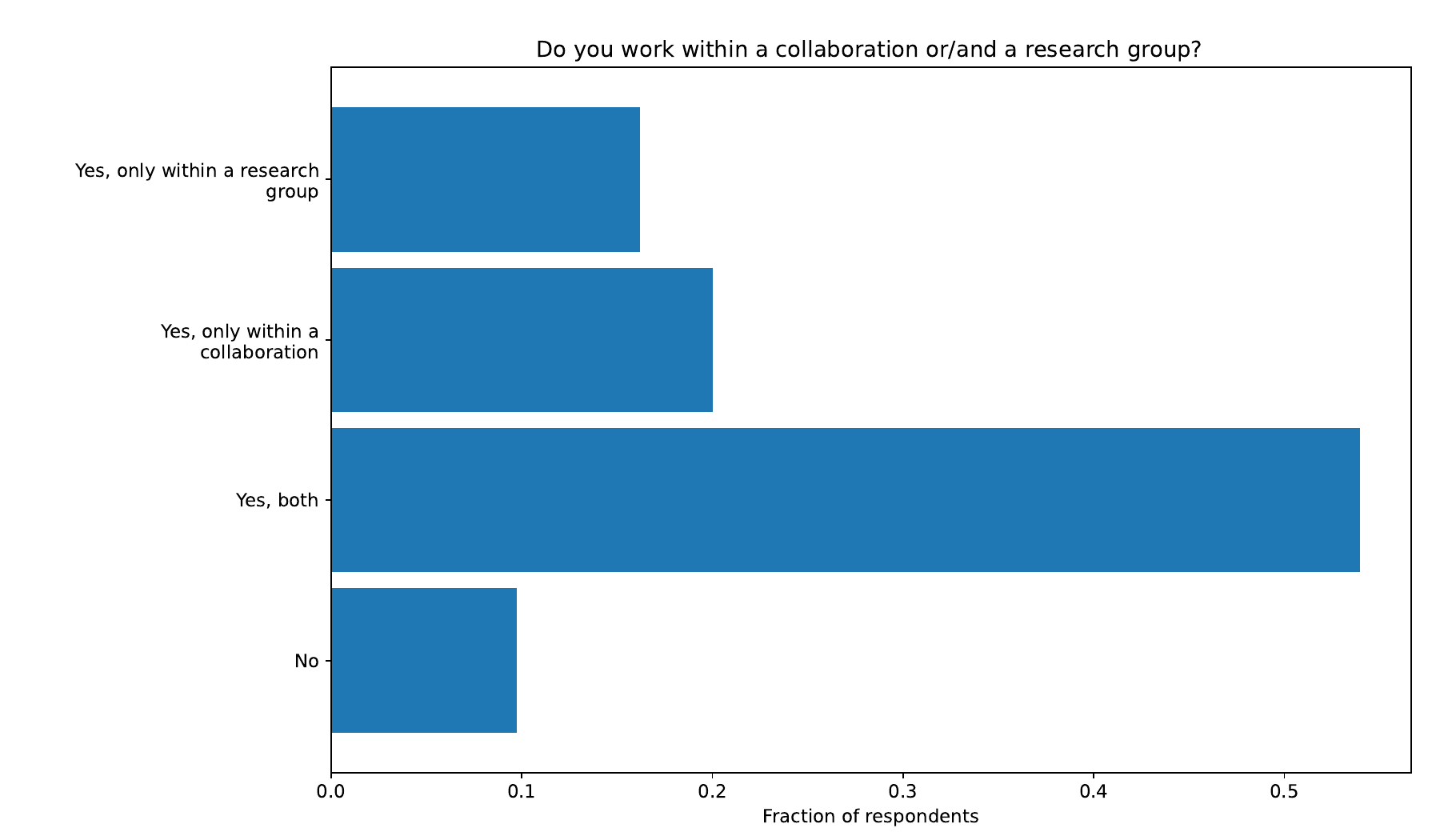}
    \caption{(Q14) Status of respondents' work within a collaboration or/and a research group.}
    \label{fig:part1:Q14}
\end{figure}

Respondents were first asked if they belong to a research group and/or a collaboration.
In Figure~\ref{fig:part1:Q14} we see that the majority of participants belong to both with over 50\% of responses.
Approximately 10\% of participants do not belong to either. 

\FloatBarrier
\pagebreak
%----------------------------------------------------------------------------------------
\subsubsection{Research groups}

\begin{figure}[h!]
    \centering
        \subfloat[]{\label{fig:part1:Q15}\includegraphics[width=0.49\textwidth]{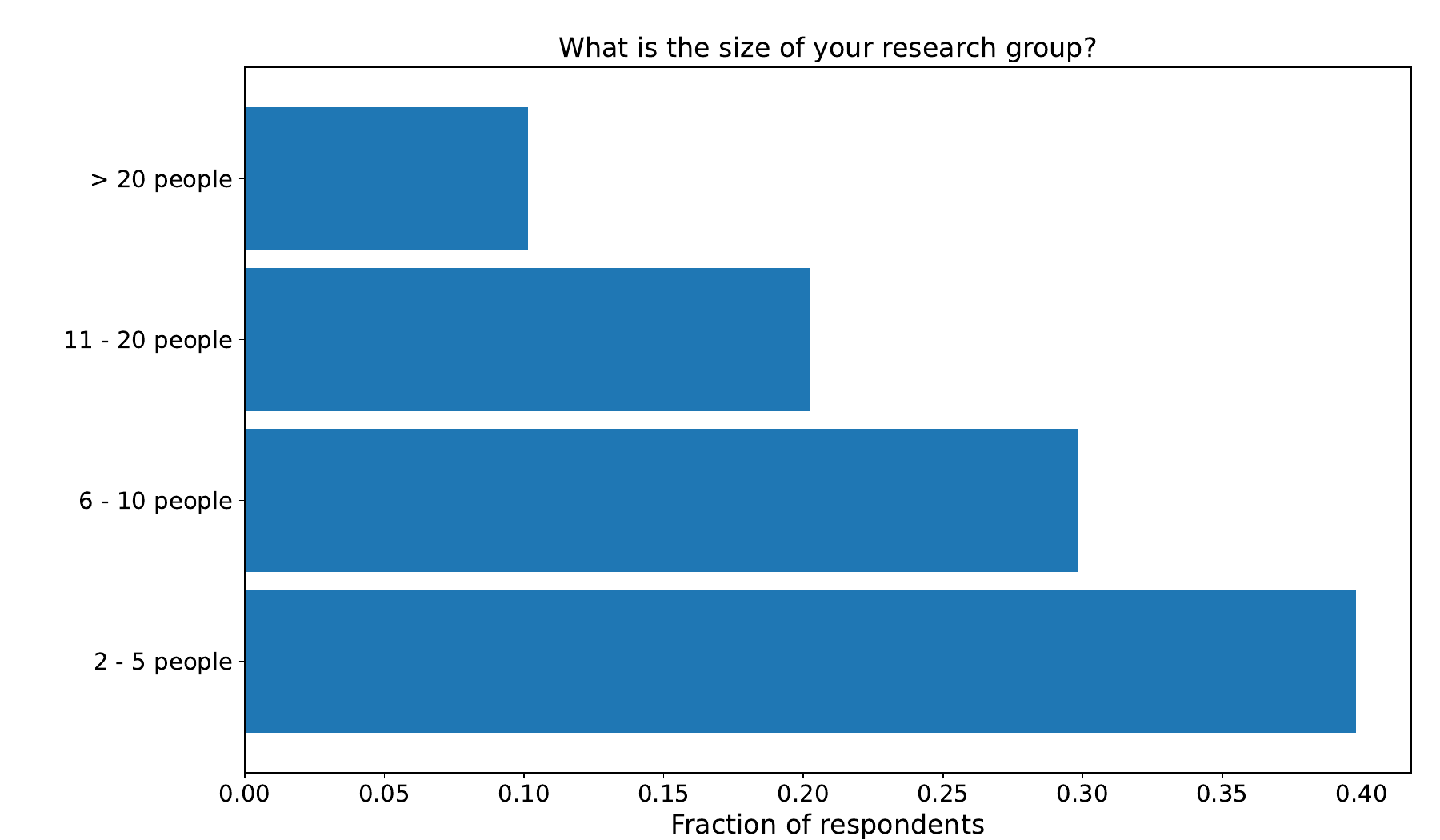}}
        \subfloat[]{\label{fig:part1:Q16}\includegraphics[width=0.49\textwidth]{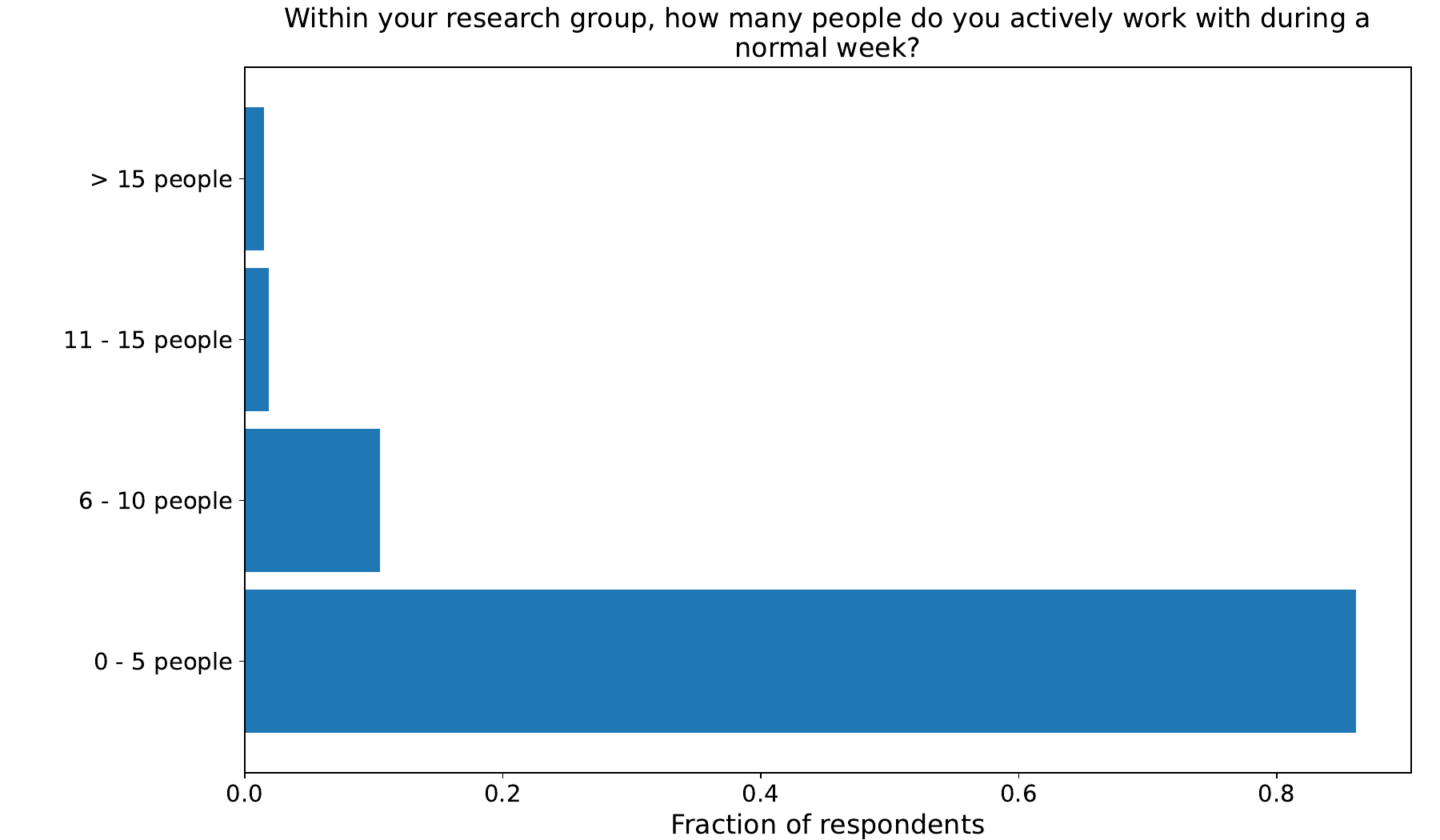}}
    \caption{(Q15--16) (a) Respondents' research group size. (b) The number of people in their research group with whom respondents actively work, during a normal week. For both plots, fractions are given out of all respondents who are part of a research group.}
    \label{fig:part1:rgroup_you_work_with}
\end{figure}

The questions about research group were answered only by respondents who are members of one.
The size of their research group is illustrated in Figure~\ref{fig:part1:Q15}.
We observe that the smaller the group, the greater the response fraction. 
The biggest fraction of respondents, which amounts to almost 40\%, are part of a small research group, composed of 2--5 people.
Approximately 10\% work in a research group of more than 20 people.

The number of people with whom the respondent actively work during a normal week is given in Figure~\ref{fig:part1:Q16}.
The majority of respondents in a research group (over 80\%) work with 0--5 people.
The second largest fraction of the respondents, roughly 10\%, actively work with 6--10 people.

\begin{figure}[h!]
    \centering
        \includegraphics[width=0.6\textwidth]{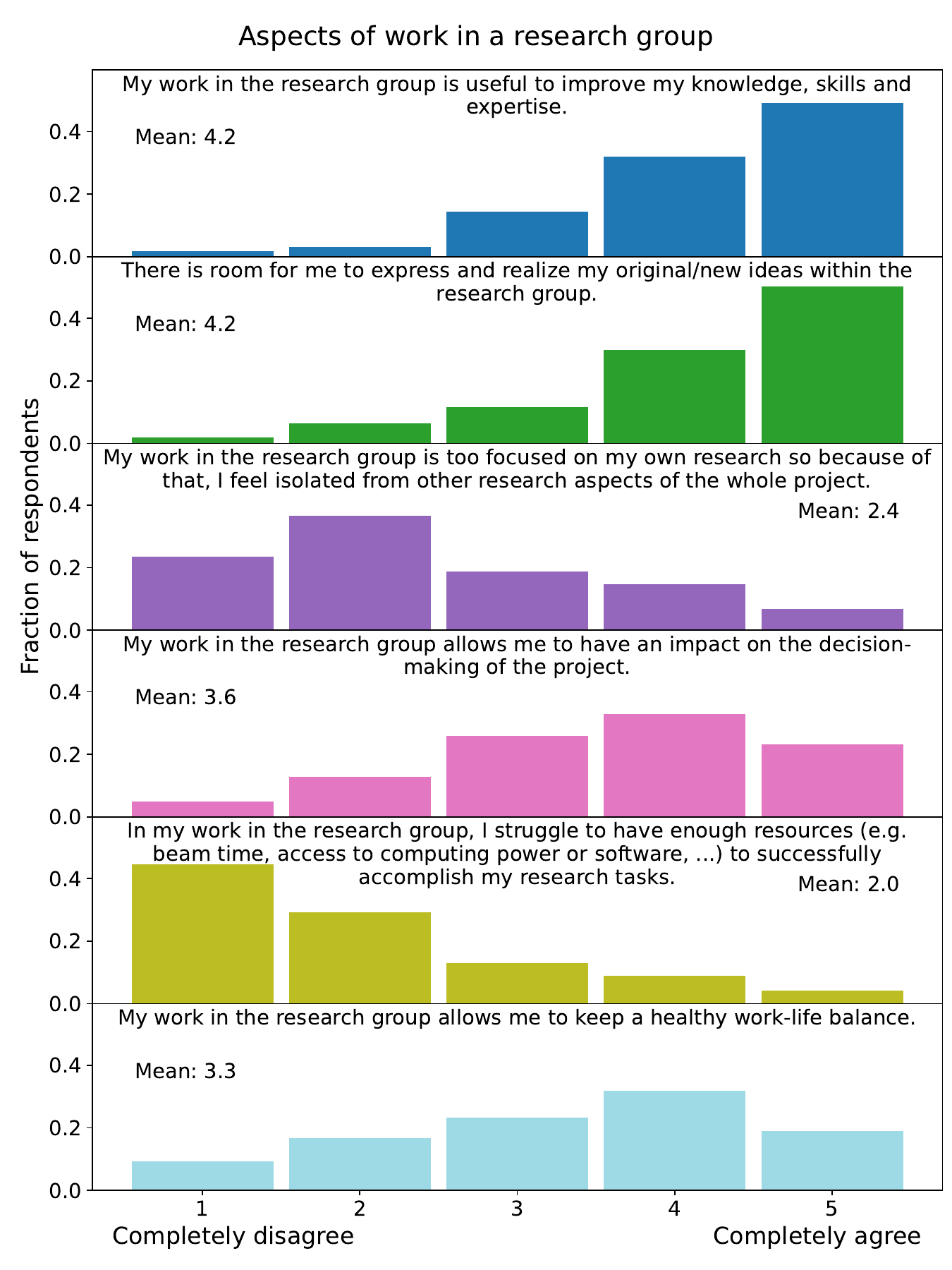}
    \caption{(Q17--22) Aspects of respondents' work in their research group. Fractions are given out of all respondents who are part of a research group and answered the questions.}
    \label{fig:part1:rgroup_aspects_of_work}
\end{figure}

Respondents views on aspects of work in a research group are presented in Figure~\ref{fig:part1:rgroup_aspects_of_work}, which presents a set of voluntary questions.
Respondents agree quite strongly that their work in the group is useful to improve their knowledge, skills and expertise, and are positive about their ability to express original/new ideas.
Respondents are generally positive about their ability to impact their research group's decision-making, and slightly less so about their work-life balance. 
We also observe that respondents do not generally struggle to obtain resources for work within their group, or feel isolated from group project aspects outside of their own contribution.

\begin{figure}[h!]
    \centering
        \includegraphics[width=0.6\textwidth]{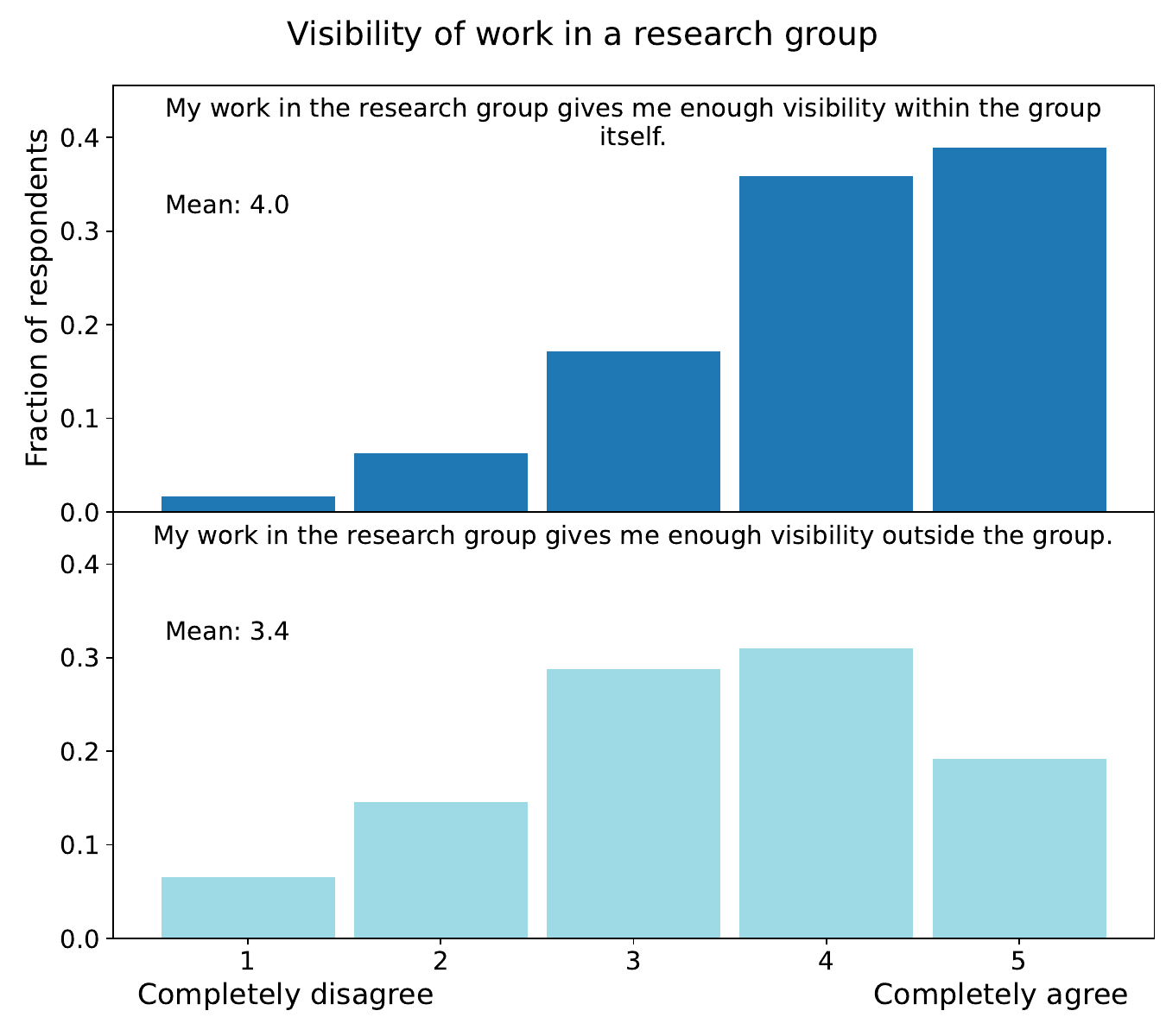}
    \caption{(Q23--24) Visibility of respondents due to work within their research group. Fractions are given out of all respondents who are part of a research group and answered the questions.}
    \label{fig:part1:rgroup_visibility}
\end{figure}

The visibility by working in a research group is treated in Figure~\ref{fig:part1:rgroup_visibility}, through a set of voluntary questions.
Respondents are generally satisfied with their visibility within their research group.
They feel less strongly about visibility outside their group, but the response remains positive.

\begin{figure}[h!]
    \centering
        \includegraphics[width=0.6\textwidth]{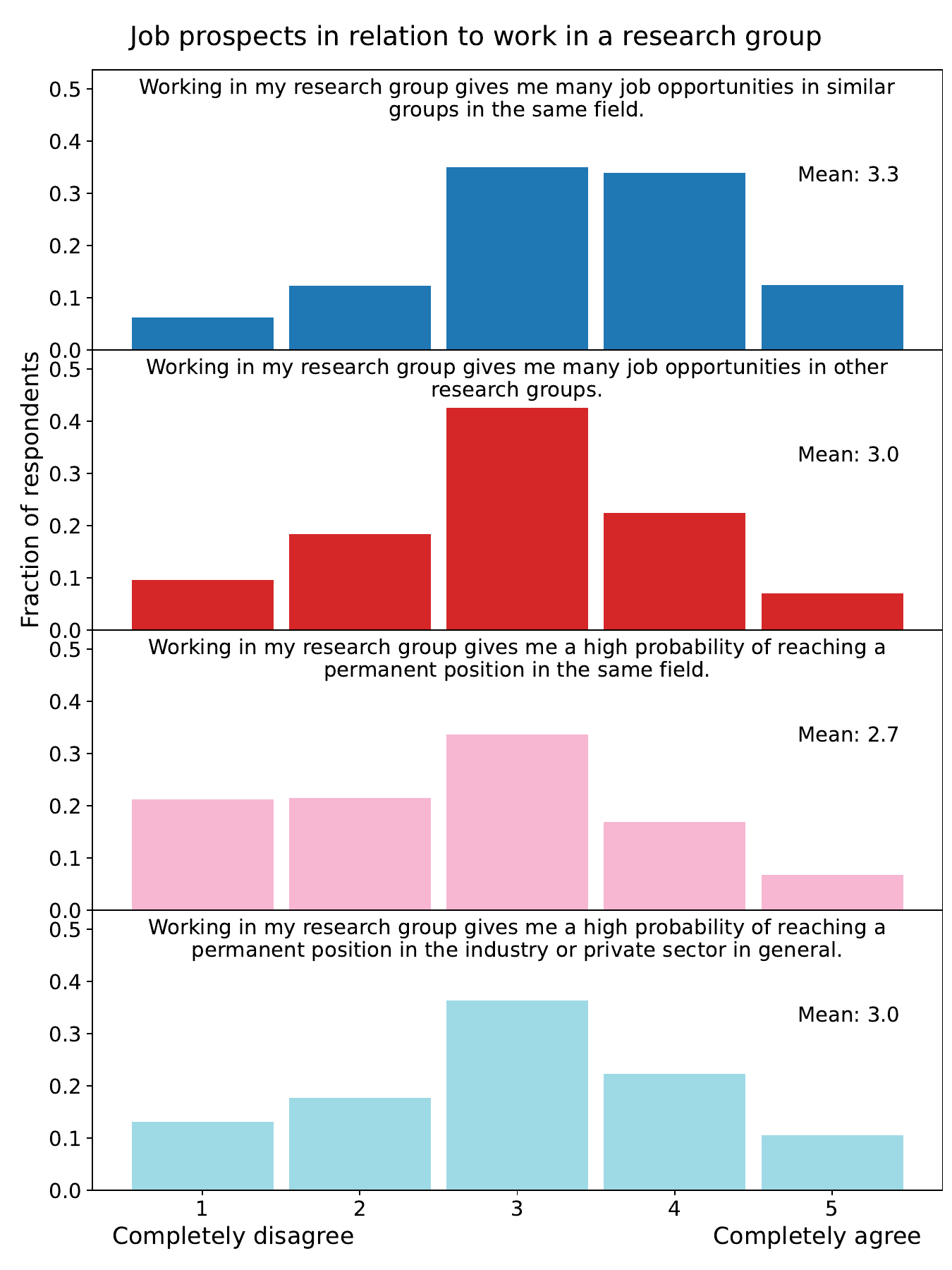}
    \caption{(Q25--28) Respondents' views on how work in their research group affects their job prospects. Fractions are given out of all respondents who are part of a research group and answered the questions.}
    \label{fig:part1:rgroup_job}
\end{figure}

Views on how working in a research group affects respondents' job prospects are presented in Figure~\ref{fig:part1:rgroup_job}, which considers a set of voluntary questions.
We see that respondents are somewhat positive about their ability to get job opportunities in similar groups, but neutral about their job opportunities in other groups.
Respondents are weakly negative about their ability to reach a permanent position within academia but neutral about their prospects outside of it.

\begin{figure}[h!]
    \centering
        \includegraphics[width=0.7\textwidth]{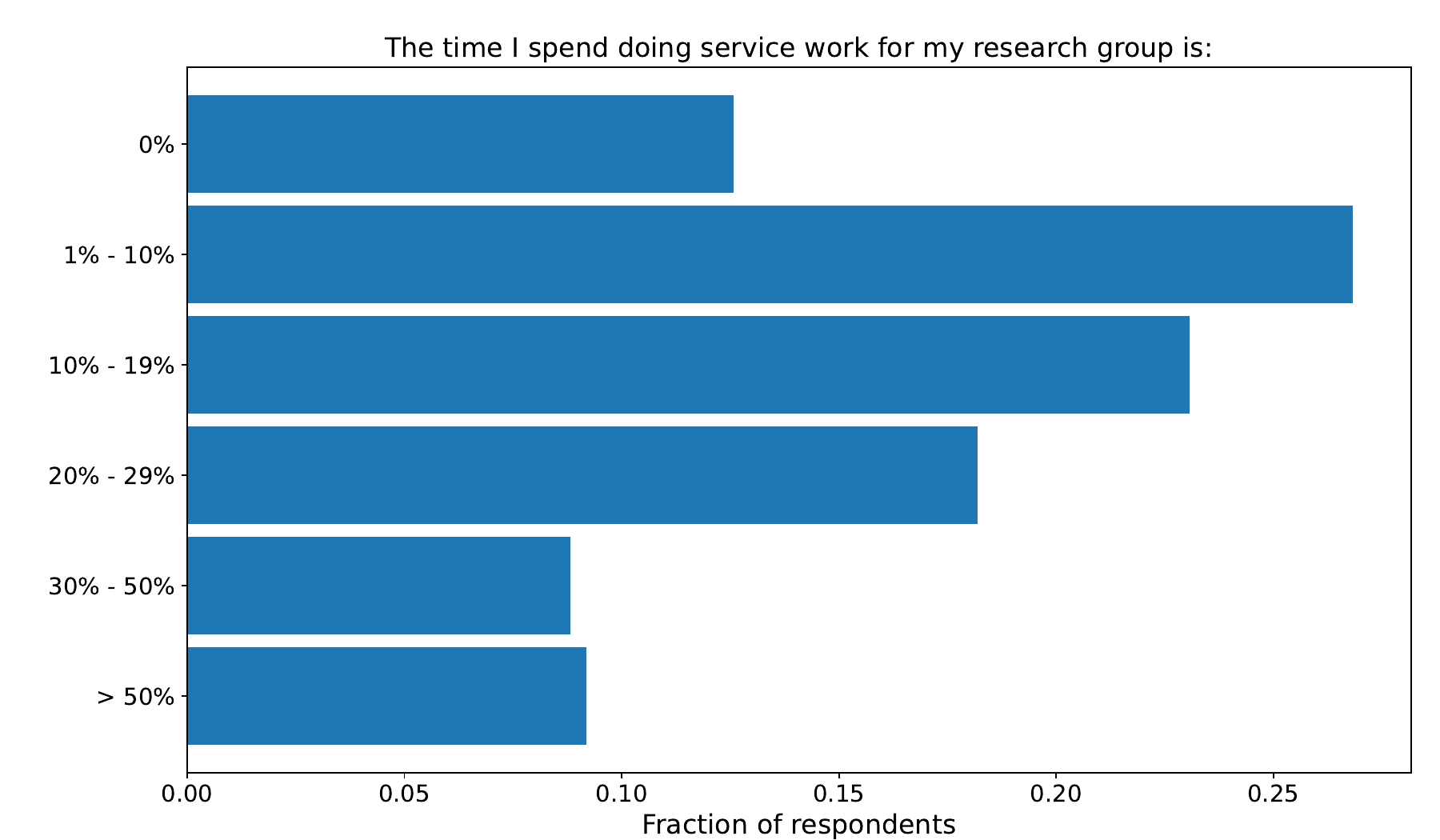}
    \caption{(Q29) The time respondents spend doing service work for their research group. Fractions are given out of all respondents who are part of a research group and answered the questions.}
    \label{fig:part1:rgroup_amount_of_service_work}
\end{figure}

\begin{figure}[h!]
    \centering
        \includegraphics[width=0.6\textwidth]{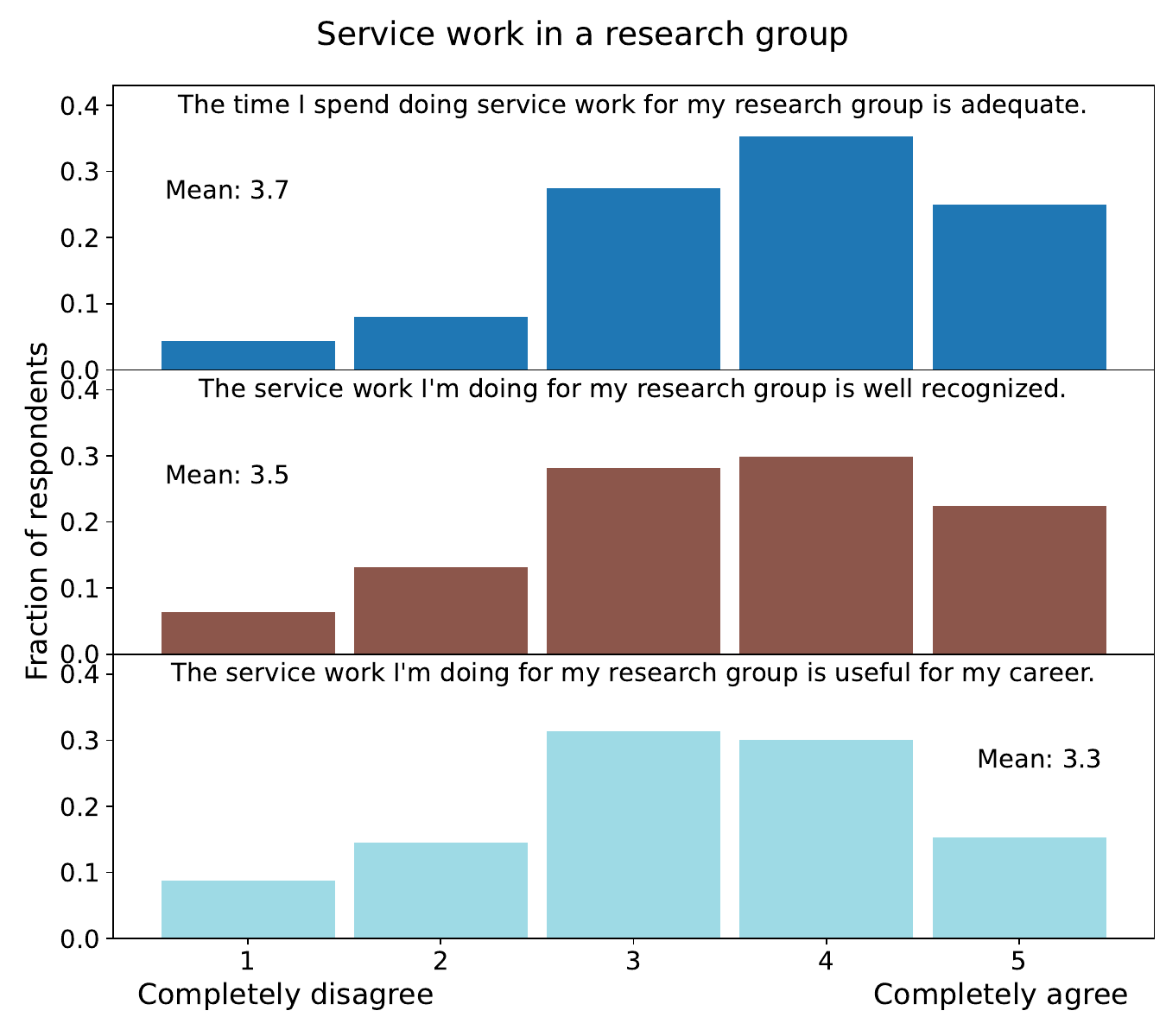}
    \caption{(Q30--32) Respondents' views on the service work they do for their research group. Fractions are given out of all respondents who are part of a research group and answered the questions..}
    \label{fig:part1:rgroup_service_work}
\end{figure}

Answers to questions relating to the service work done for the research group are illustrated in Figures~\ref{fig:part1:rgroup_amount_of_service_work} and~\ref{fig:part1:rgroup_service_work}, which present voluntary questions.
Almost one third of the respondents who are in a research group spend 1--10\% of their time doing service work for the group.
More respondents do no service work at all than do it for more than half of their time.

The respondents generally agree that the time spent doing service work for the research group is adequate.
They weakly agree that their service work is well-recognised, but even more weakly agree that it is useful for their careers.

\FloatBarrier
\pagebreak

\FloatBarrier
\subsubsection{Collaborations}

\begin{figure}[h!]
    \centering
        \subfloat[]{\label{fig:part1:Q33}\includegraphics[width=0.49\textwidth]{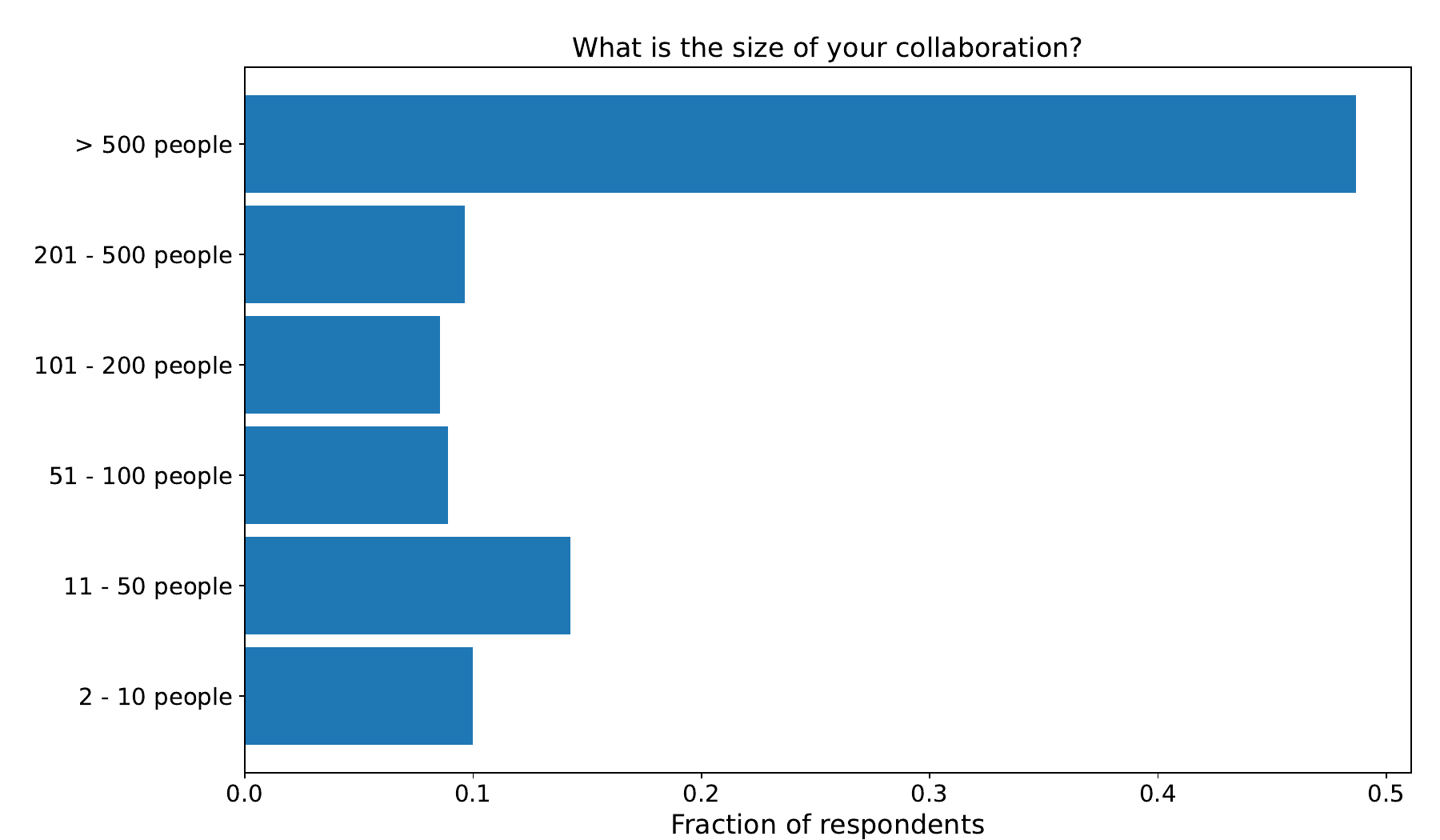}}
        \subfloat[]{\label{fig:part1:Q34}\includegraphics[width=0.49\textwidth]{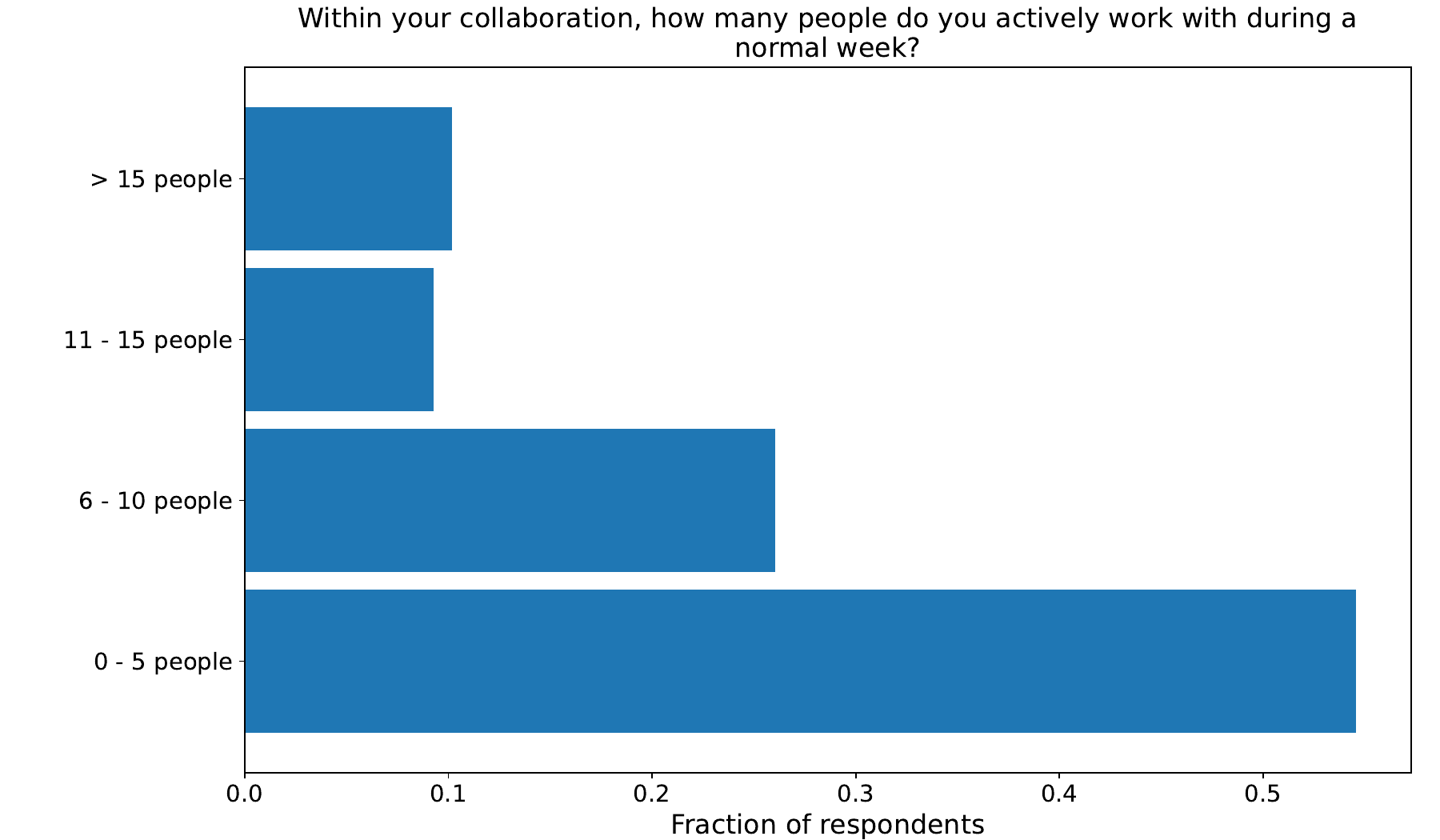}}
    \caption{(Q33--34) (a) Respondents' collaboration size. (b) The number of people in their collaboration with whom respondents actively work, during a normal week. For both plots, fractions are given out of all respondents who are part of a collaboration.}
    \label{fig:part1:collab-you-work-with}
\end{figure}

\begin{figure}[h!]
    \centering
        \includegraphics[width=0.6\textwidth]{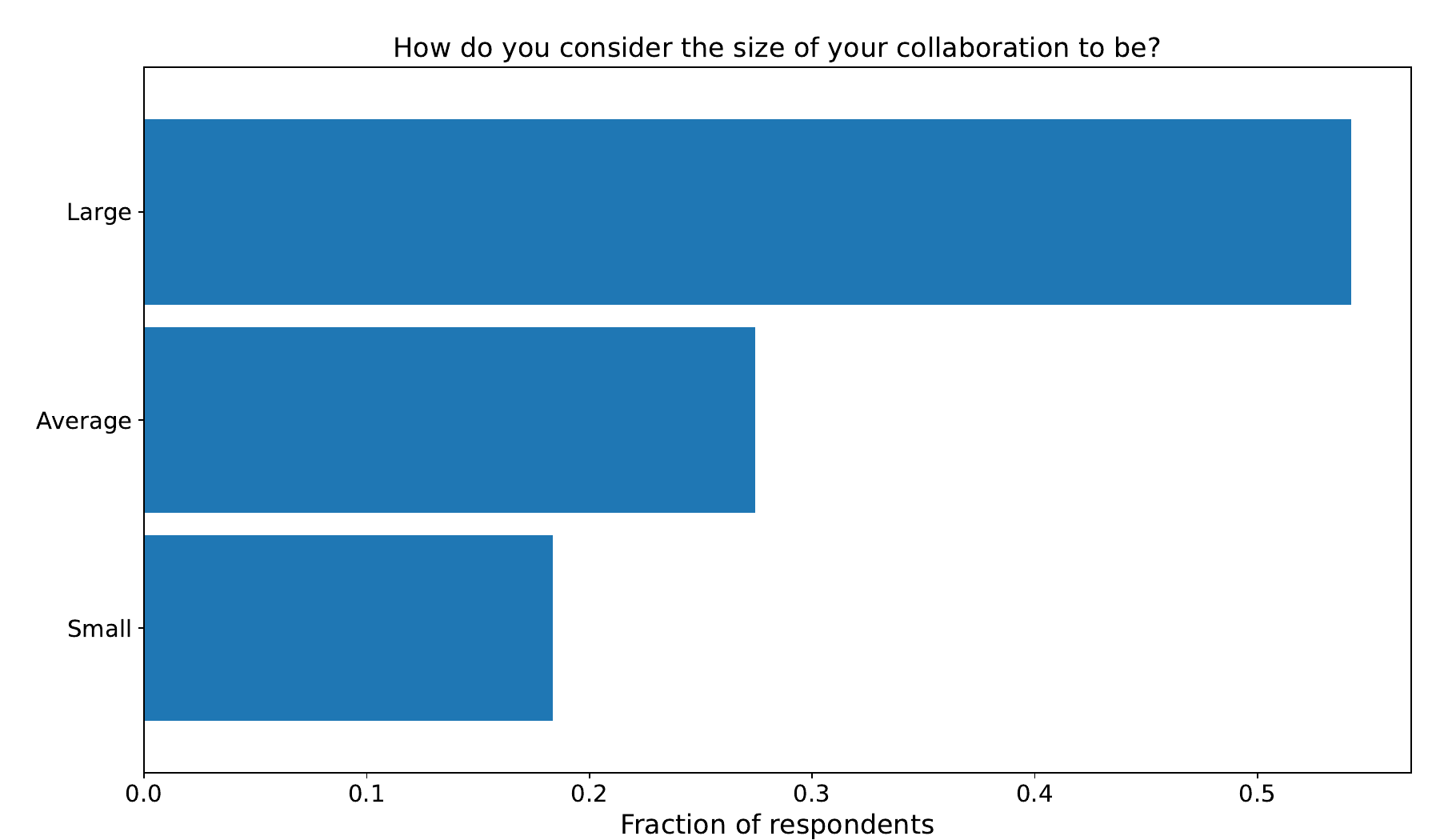}
    \caption{(Q35) Respondents' opinions on the size of their collaboration. Fractions are given out of all respondents who are part of a collaboration.}
    \label{fig:part1:consider-collab-size}
\end{figure}

Questions about collaborations were only answered by respondents who are members of a collaboration.
The largest fraction of respondents who are in a collaboration are part of large collaborations with over 500 people, as shown in Figure~\ref{fig:part1:Q33}.
The distribution of respondents in collaborations of fewer than 500 people are fairly uniform.
Although a significant portion of the respondents reported that they are part of a large collaboration, we see in Figure~\ref{fig:part1:Q34} that most respondents actively work with $0-5$ people during normal week within their collaboration.
Most of the respondents consider their collaboration to be large, as Figure~\ref{fig:part1:consider-collab-size} shows, in line with the previous question's answers.

\begin{figure}[h!]
    \centering
        \includegraphics[width=0.6\textwidth]{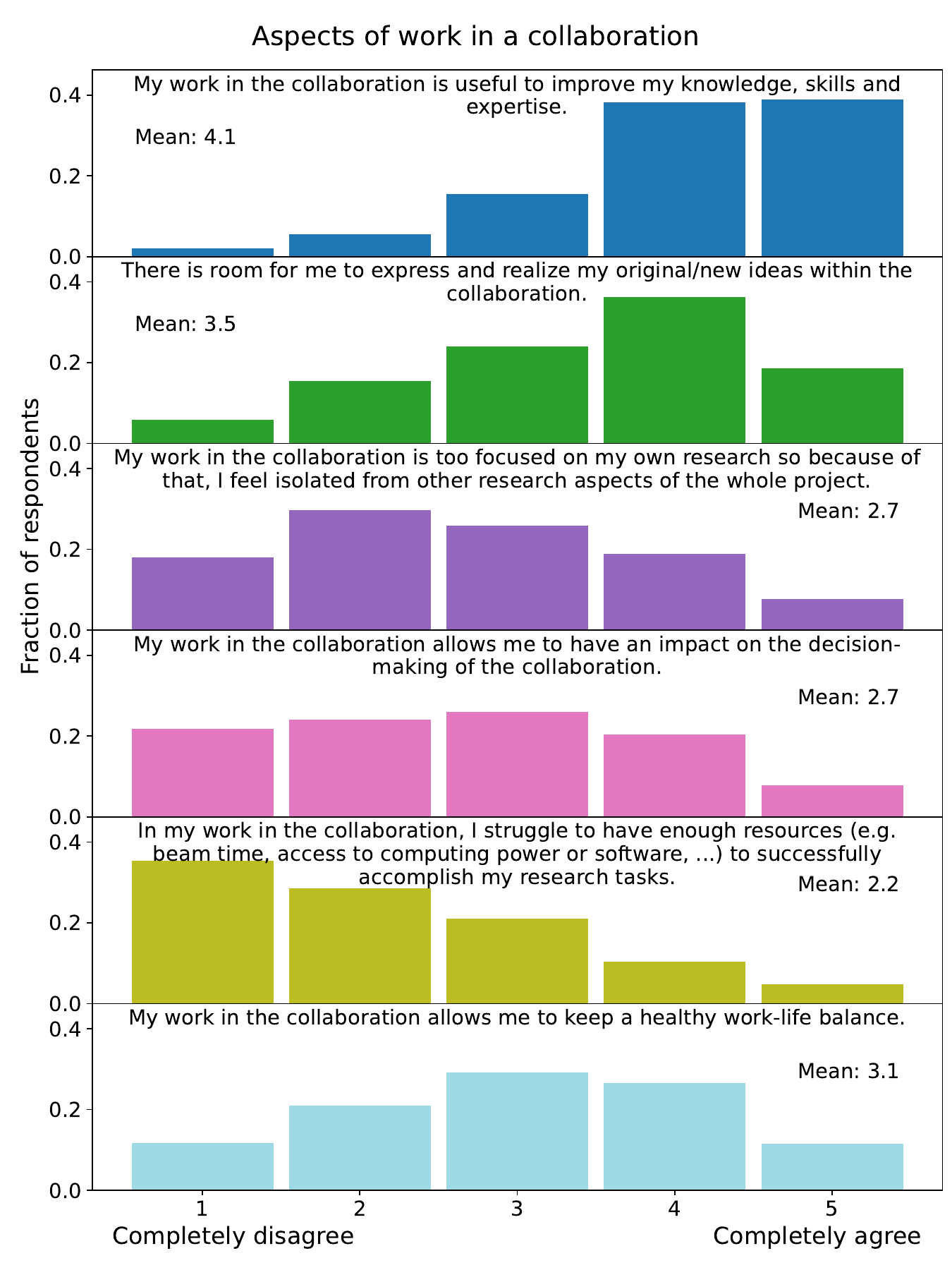}
    \caption{(Q36-41) Aspects of respondents' work in their collaboration. Fractions are given out of all respondents who are part of a collaboration and answered the questions.}
    \label{fig:part1:aspects_of_work_in_collab}
\end{figure}

Several aspects of the work respondents do in collaborations are summarised in Figure~\ref{fig:part1:aspects_of_work_in_collab}, which presents a set of voluntary questions.
The work is generally viewed as useful for improving one's knowledge, skills and expertise.
Furthermore, respondents are positive about their ability to express and realise their original/new ideas within a collaboration but less so than in a research group.
Respondents reported that their work in the collaboration is not too focused on their own research.
On the other hand, they are more negative about their ability to impact decision-making in the collaboration, more so than in a research group.
Respondents generally do not have to struggle to have enough resources to successfully accomplish their research task within the collaboration.
Finally, there was no strong opinion about the collaboration allowing respondents to maintain a healthy work-life balance.

\begin{figure}[h!]
    \centering
        \includegraphics[width=0.6\textwidth]{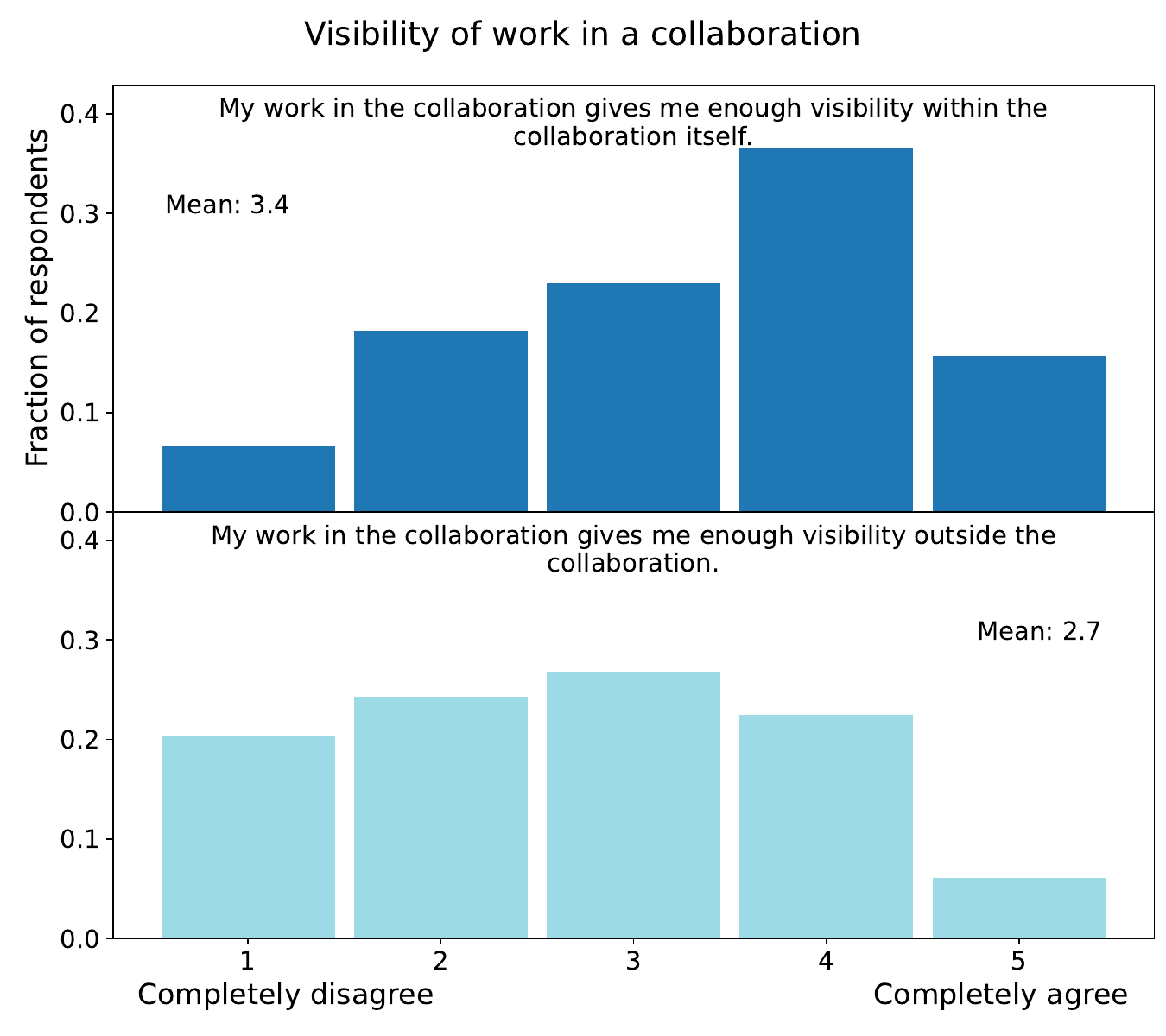}
    \caption{(Q42--43) Visibility of respondents due to work within their collaboration. Fractions are given out of all respondents who are part of a collaboration and answered the questions.}
    \label{fig:part1:visibility_collaboration}
\end{figure}

Visibility achieved by working in a collaboration is presented in Figure~\ref{fig:part1:visibility_collaboration}, which considers a set of voluntary questions.
Respondents are generally satisfied with their visibility within the collaboration itself, however they are generally dissatisfied with their visibility outside of their collaboration.
We observe more disagreement than in answers to analogous questions concerning visibility related to respondents' research groups.

\begin{figure}[h!]
    \centering
        \includegraphics[width=0.6\textwidth]{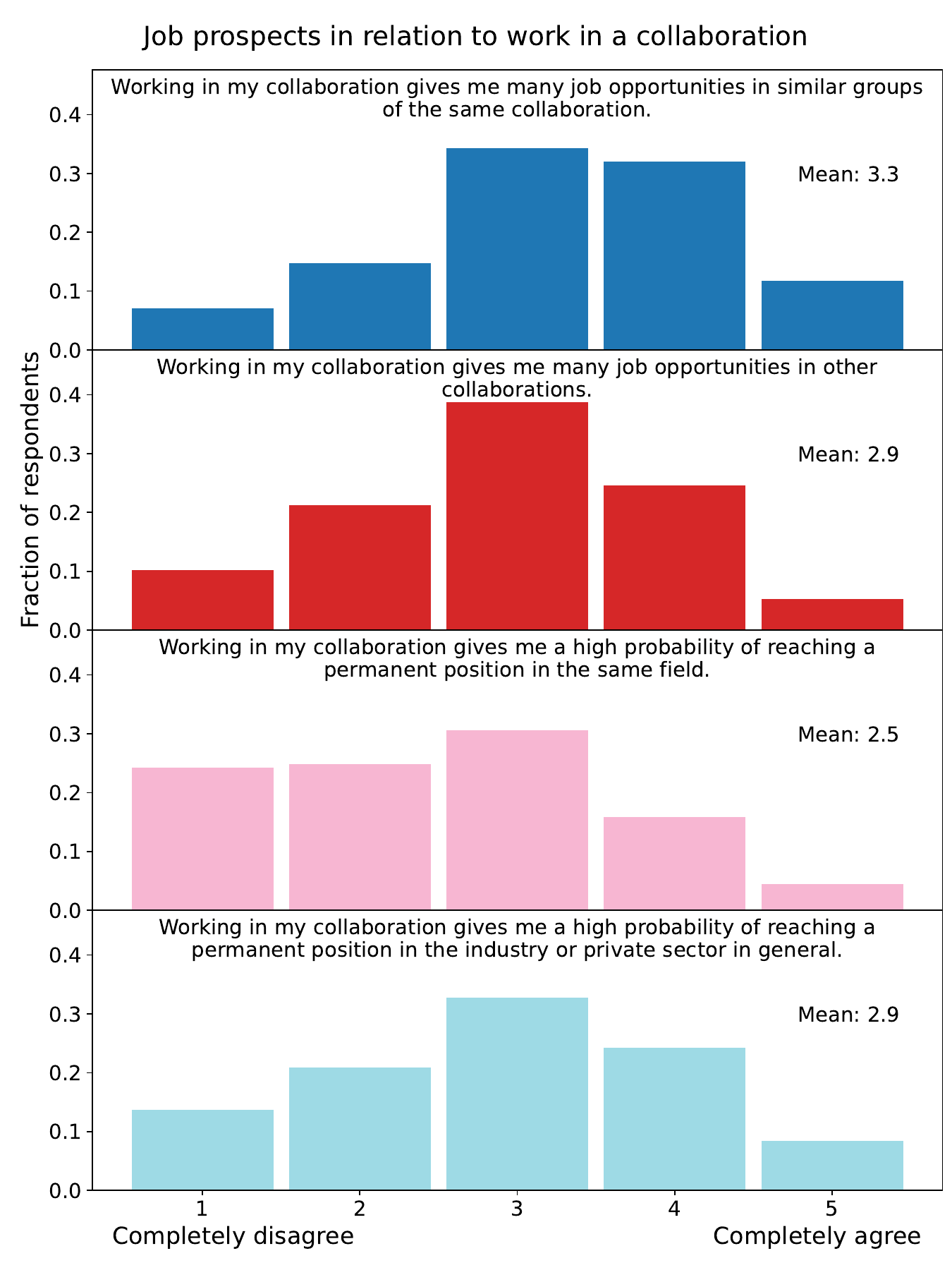}
    \caption{(Q44--47) Respondents' views on how work in their collaboration affects their job prospects. Fractions are given out of all respondents who are part of a collaboration and answered the questions.}
    \label{fig:part1:job_collaboration}
\end{figure}

Responses to questions about job prospects in relation to collaboration work are summarised in Figure~\ref{fig:part1:job_collaboration} (which presents a set of voluntary questions) and show similar behaviour to the responses about the research groups.
Respondents are weakly positive about their ability to get job opportunities in similar collaborations, but neutral about their job opportunities in other collaborations.
Additionally, we see that respondents are weakly negative about their ability to reach a permanent position within academia but neutral about their prospects outside it.

\begin{figure}[h!]
    \centering
        \includegraphics[width=0.7\textwidth]{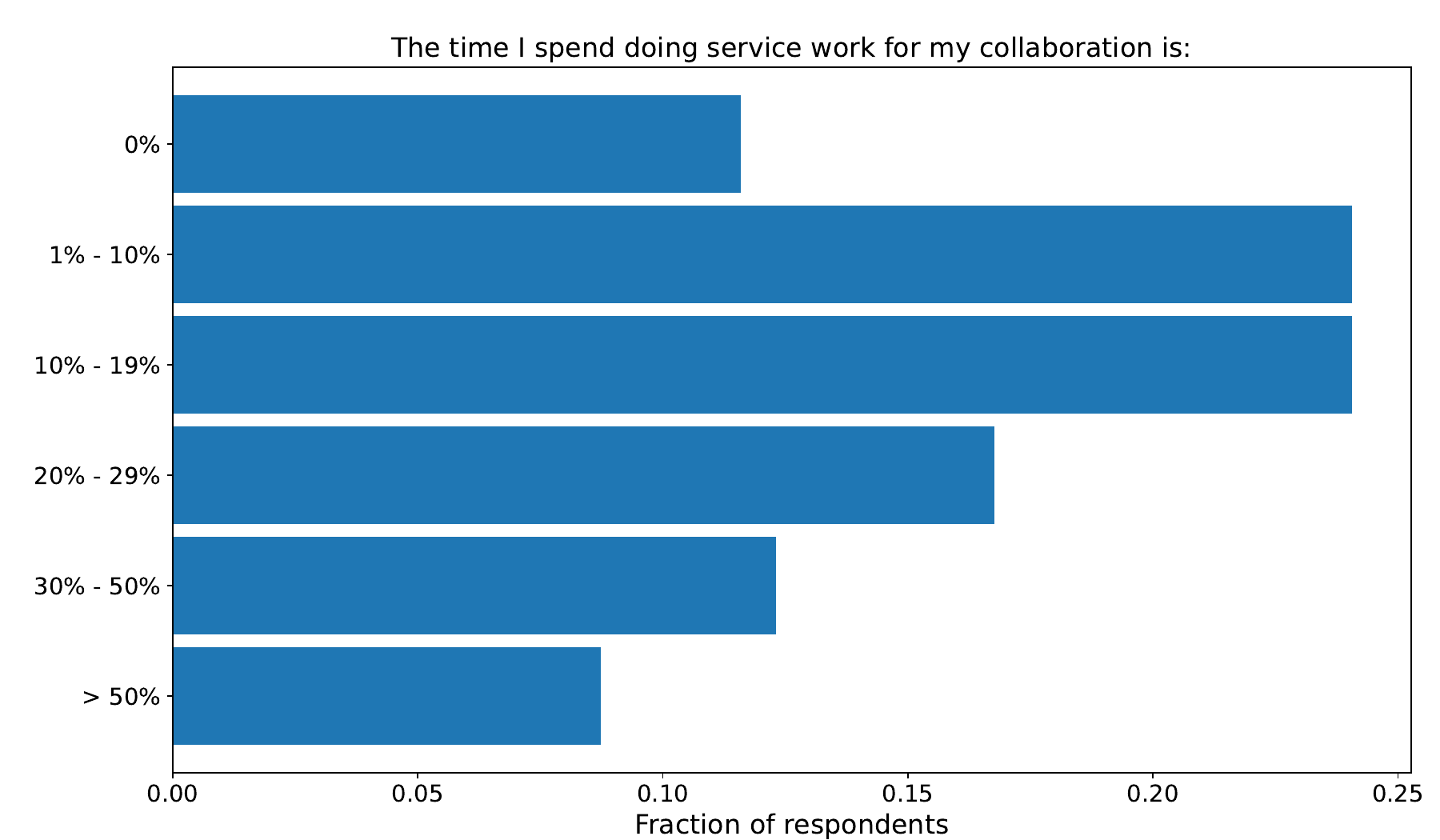}
    \caption{(Q48) The time respondents spend doing service work for their collaboration. Fractions are given out of all respondents who are part of a collaboration and answered the question.}
    \label{fig:part1:amount-of-service-work}
\end{figure}

\begin{figure}[h!]
    \centering
        \includegraphics[width=0.6\textwidth]{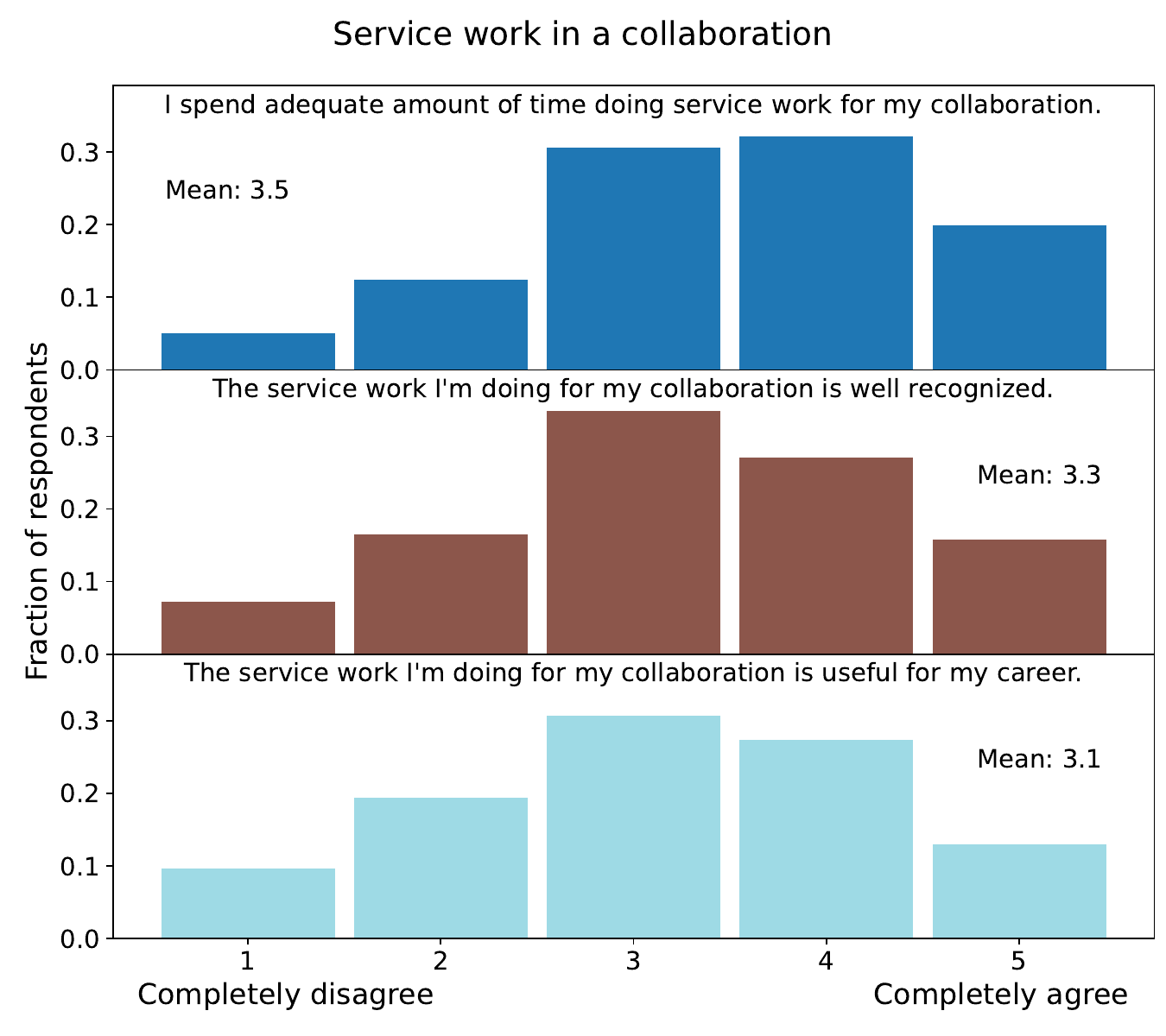}
    \caption{(Q49--51) Respondents' views on the service work done for their collaboration. Fractions are given out of all respondents who are part of a collaboration and answered the questions.}
    \label{fig:part1:service_work_collaboration}
\end{figure}

The majority of respondents who are part of a collaboration spend $\leq 20\%$ time doing service work for the collaboration, as we show in Figure~\ref{fig:part1:amount-of-service-work}.
Only $\sim 10\%$ of respondents do no service work for a collaboration at all and, on the other hand, only $\sim 10\%$ spend more than $50\%$ of their time doing service work.
Other aspects of service work are summarised in Figure~\ref{fig:part1:service_work_collaboration}.
Overall, respondents feel somewhat positive about the amount of time they spent doing service work in their collaboration, less so about how well-recognised it is, and quite neutral about how useful it is for their careers.
There is more disagreement than in the analogous questions relating to research group.

%%%%%%%%%%%%%%%%%%%%%%%%%%%%%%%%%%%%%%%%%%%%%%%%%%%%%%%%%%%%%%%%%%%%%%%%%%%%%%%%%%%%%%%%%%%%%%%%%%%%%%%%%%%%%%%
\FloatBarrier
\subsection{The diversity of Physics programs}

\begin{figure}[htb!]
    \centering
        \includegraphics[width=0.7\textwidth]{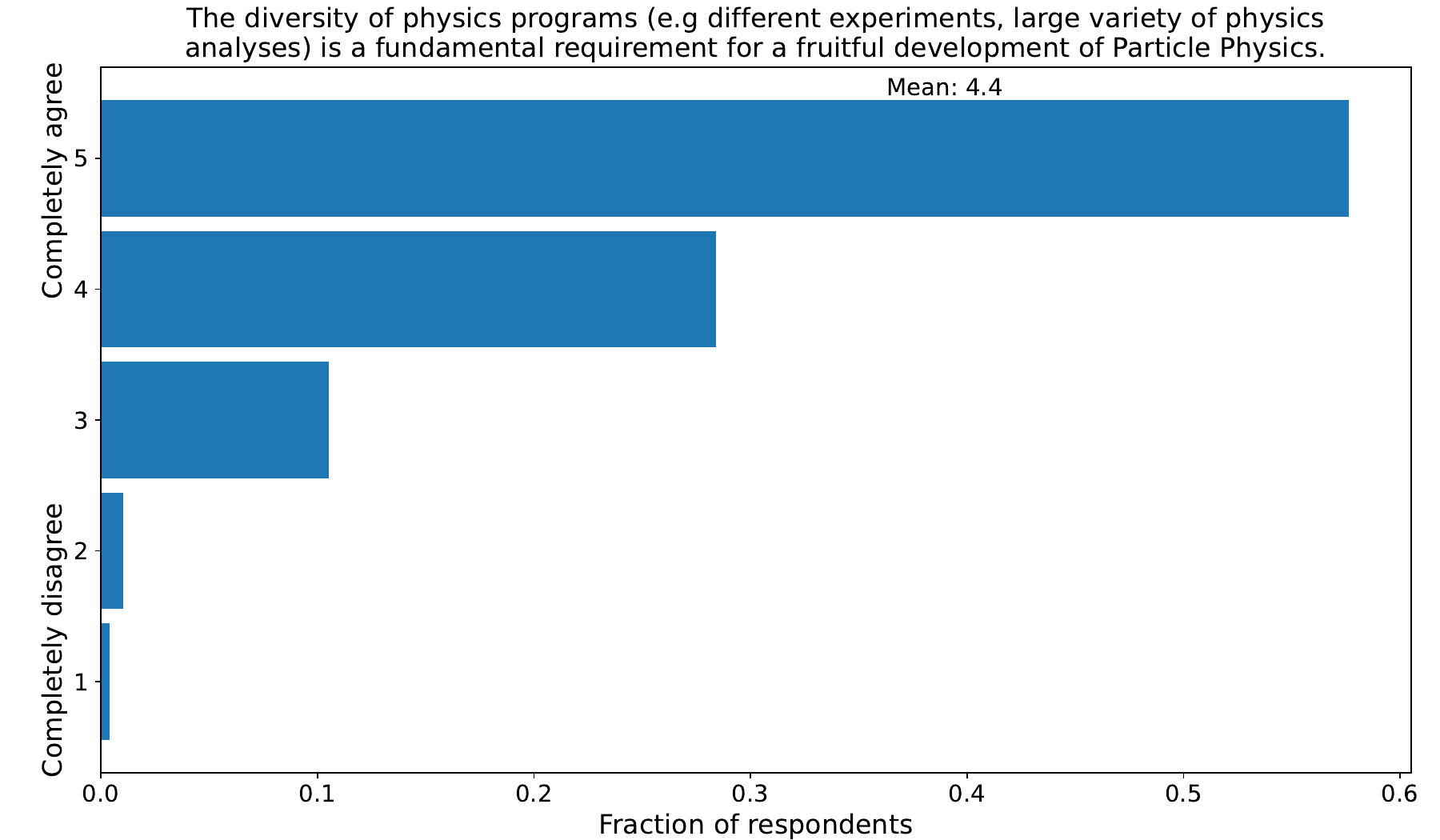}
    \caption{(Q52) The level of respondents' agreement with the the diversity of physics programs (e.g different experiments, large variety of physics analyses) being a fundamental requirement for a fruitful development of Particle Physics. This question was voluntary.}
    \label{fig:part1:Q52}
\end{figure}

We now move to considering questions related to respondents agreement on the importance of a diverse physics programme.
Firstly, we show respondents' level of agreement with the diversity of physics programs (e.g different experiments, large variety of physics analyses) being a fundamental requirement for a fruitful development of particle physics, in Figure~\ref{fig:part1:Q52}.
The vast majority of the respondents agree with this statement. 

\begin{figure}[h]
    \centering
        \subfloat[]{\label{fig:part1:Q53}\includegraphics[width=0.49\linewidth]{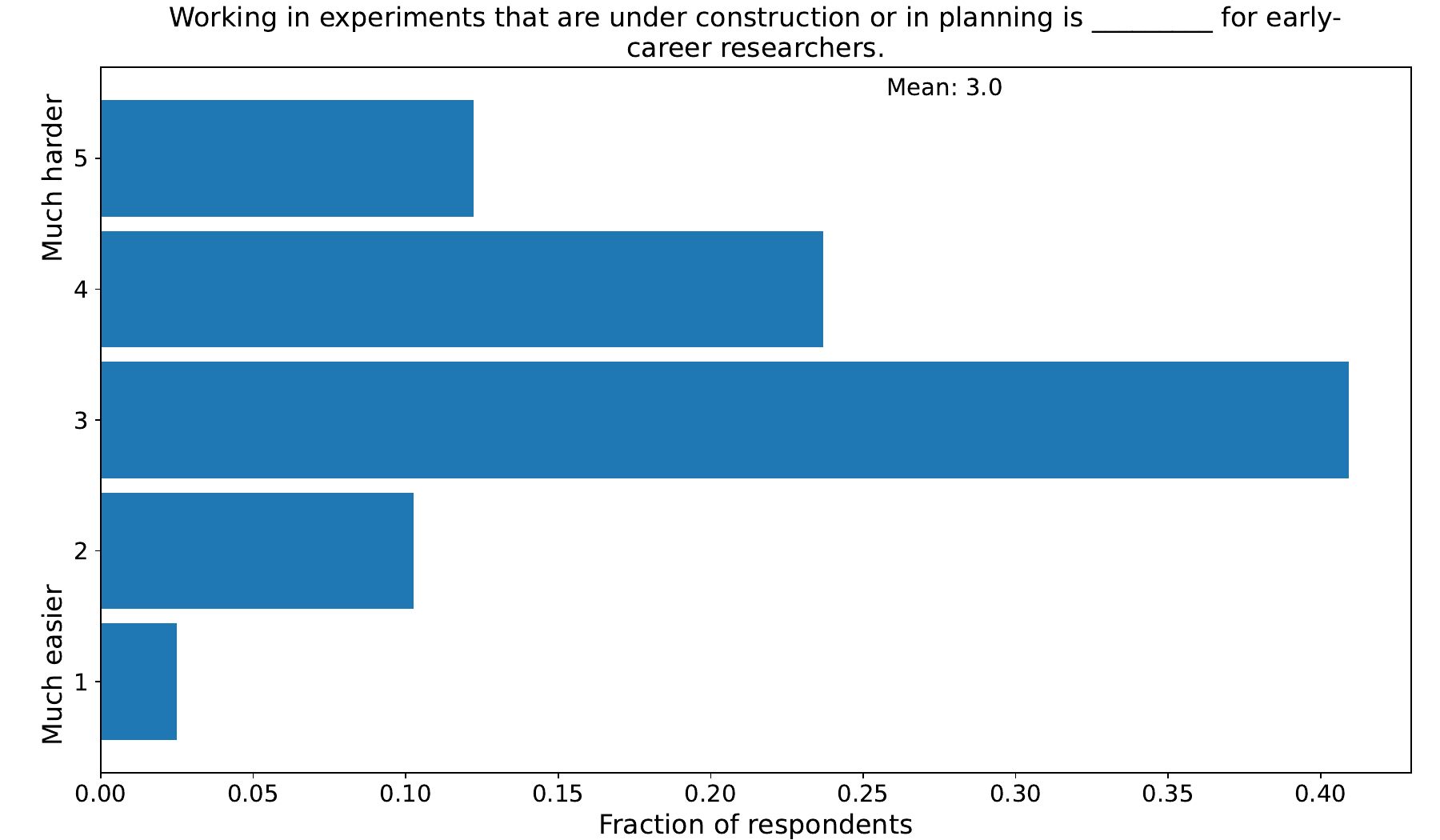}}
        \subfloat[]{\label{fig:part1:Q54}\includegraphics[width=0.49\linewidth]{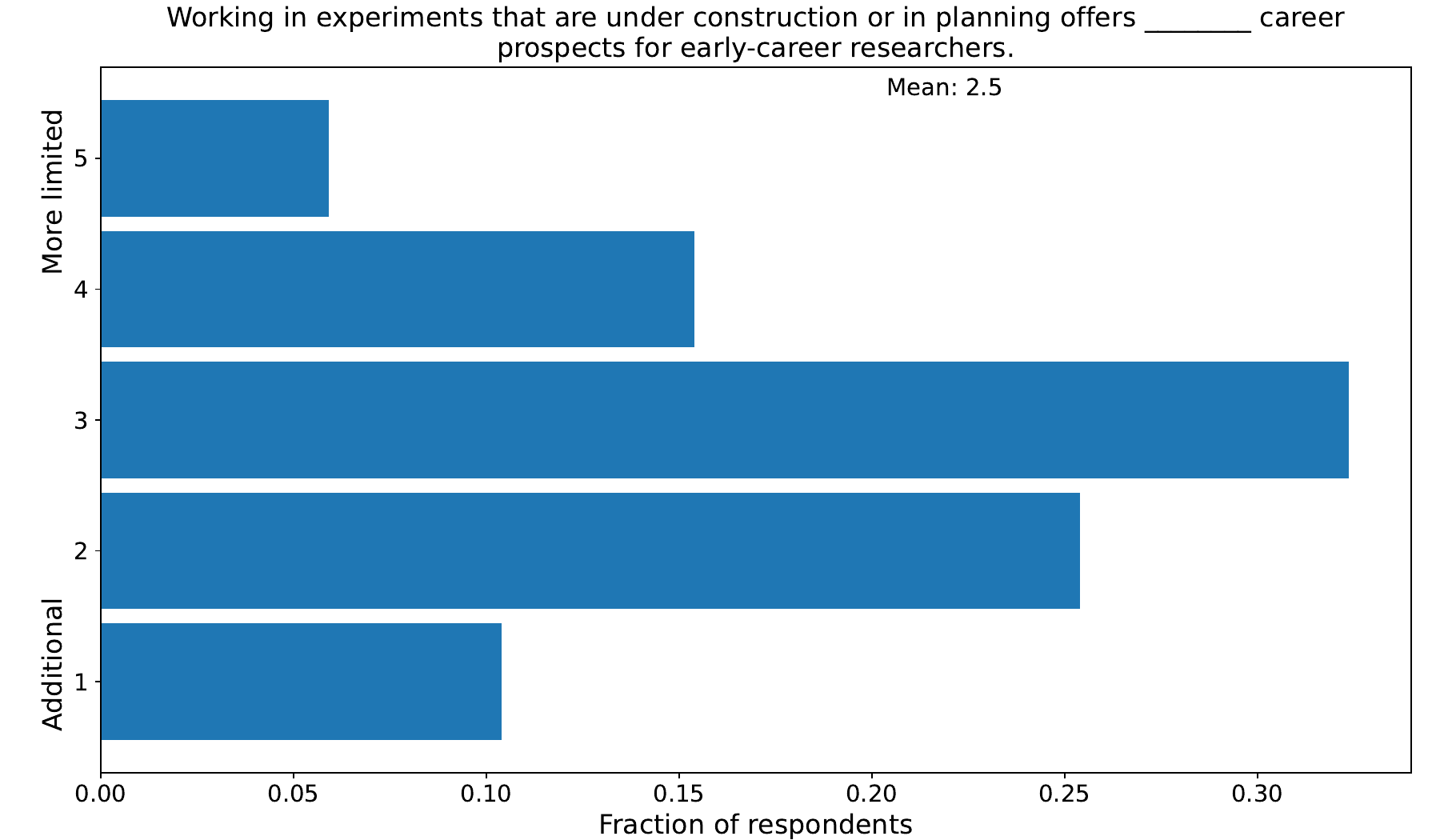}}
    \caption{(Q53--54) Respondents' views on the difficulty of working in experiments that are under construction and how this work affects career prospects. These questions were voluntary.}
    \label{fig:part1:Q53Q54}
\end{figure}

Secondly, we probe views on working on experiments that are under construction. 
Two questions were asked, shown in Figure~\ref{fig:part1:Q53Q54}.
We can see that respondents on average believe that working in the experiments under construction is harder for ECRs but offers additional career prospects. 

%%%%%%%%%%%%%%%%%%%%%%%%%%%%%%%%%%%%%%%%%%%%%%%%%%%%%%%%%%%%%%%%%%%%%%%%%%%%%%%%%%%%%%%%%%%%%%%%%%%%%%%%%%%%%%%
\FloatBarrier
\subsection{Career perspectives and information sharing}

%----------------------------------------------------------------------------------------
\subsubsection{Information on career prospects}

\begin{figure}[h!]
    \centering
        \includegraphics[width=0.55\textwidth]{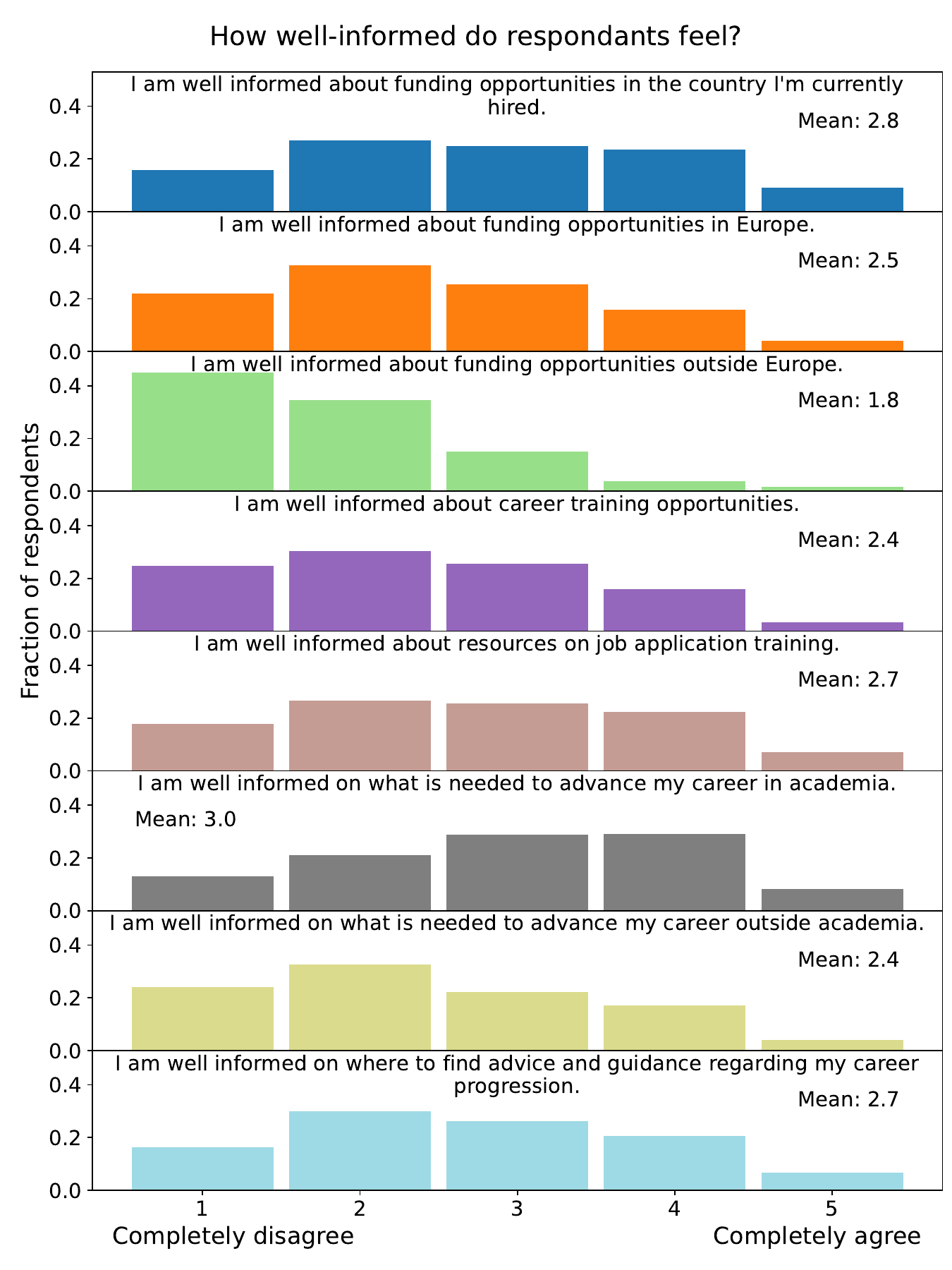}
    \caption{(Q55--62) How well-informed respondents feel about aspects of future career planning.}
    \label{fig:part1:FeelInformed}
\end{figure}

In Figure~\ref{fig:part1:FeelInformed}, respondents are asked how well-informed they are about funding and training opportunities, resources, and what is needed to advance their careers.
The majority of respondents do not feel well-informed about function opportunities in their current country of employment.
This majority grows when considering other countries in Europe, and is substantially larger when considering funding opportunities outside of Europe.
The majority of respondents do not agree that they are well-informed about opportunities for career training, resources on training for job applications, or where to find advice and guidance regarding career progression.\footnote{We would like to point out that for training in Instrumentation, ECFA hosts a list of schools relevant for ECRs in Ref.~\cite{instSchools}.}
On the other hand, they are relatively well-informed about what is needed to advance their careers in academia.
Far fewer respondents agree that they are well-informed on what is needed to advance their careers outside of academia.

\begin{figure}[h!]
    \centering
        \includegraphics[width=0.6\textwidth]{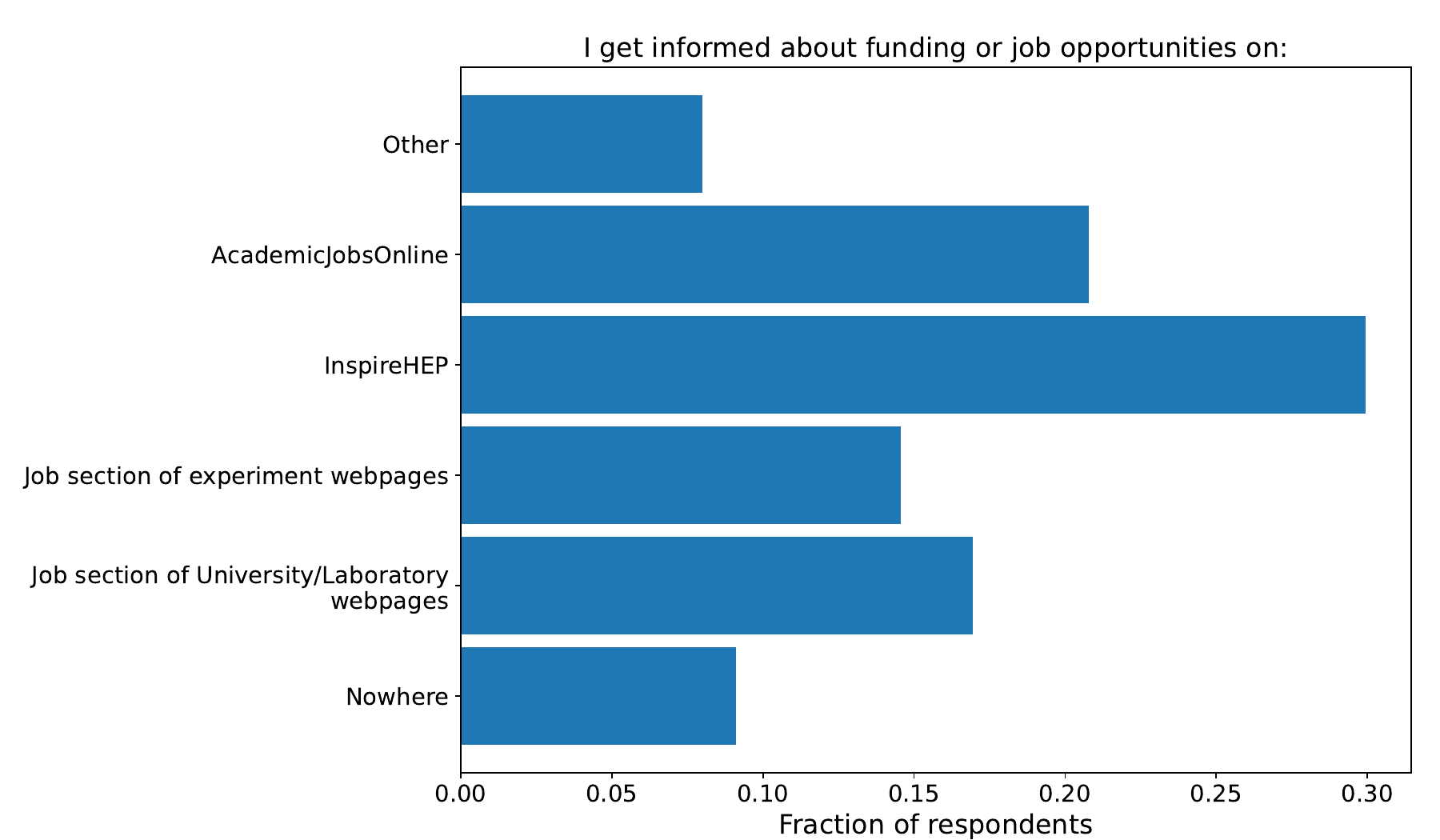}
    \caption{(Q63) Resources respondents use to learn about academic job vacancies. Multiple answers were allowed per respondent. Fractions are given out of all respondents and empty responses are not shown.}
    \label{fig:part1:Q63}
\end{figure}

The resources respondents use to find information on job vacancies are shown in Figure~\ref{fig:part1:Q63}.
Multiple answers could be entered by each respondent, and the majority of respondents use InspireHEP~\cite{inspireHEP}, followed by AcademicJobsOnline~\cite{AcademicJobsOnline}.
Of responses entering the `Other' category: 32\% indicated email; 15\% word of mouth; 4.2\% from discussion with their supervisor; 18\% LinkedIn~\cite{LinkedIn}; 18\% other websites; 3.5\% national or funding agency resources; 3.5\% social media; 2.1\% institute resources other than web pages; and 2.8\% industry websites.
Some examples of websites/companies reported by respondents include:
Researchgate~\cite{Researchgate}, Indeed~\cite{Indeed}, EURAXESS~\cite{EURAXESS}, Stepstone~\cite{Stepstone}, Xing~\cite{Xing}, jobs.ac.uk~\cite{JobsAcUK}, academics.de~\cite{AcademicsDE}, PolytechnicPositions~\cite{PolytechnicPositions}, findaphd~\cite{FindaPHD}, CERN Alumni~\cite{CERNAlumni}, AAS Job Register~\cite{AASJobRegister} and EuroScienceJobs~\cite{EuroScienceJobs}.

\begin{figure}[h!]
    \centering
        \includegraphics[width=0.55\textwidth]{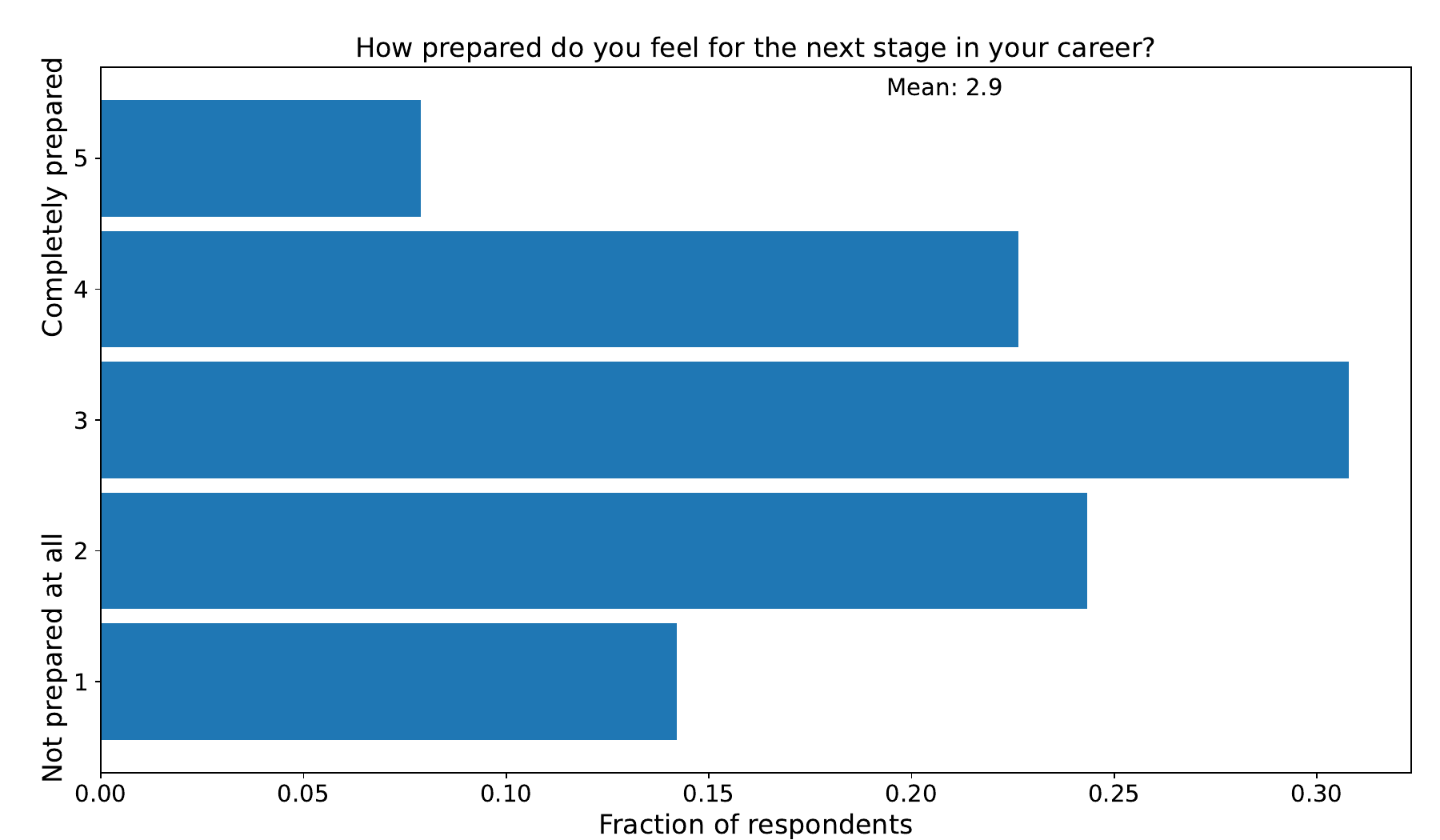}
    \caption{(Q64) How prepared respondents feel for the next stage in their career.}
    \label{fig:part1:Q64}
\end{figure}

Figure~\ref{fig:part1:Q64} shows how prepared respondents feel for the next stage in their career.
Slightly more respondents feel unprepared than prepared, but most don't have a strong view on this statement.

\begin{figure}[h!]
    \centering
        \includegraphics[width=0.6\textwidth]{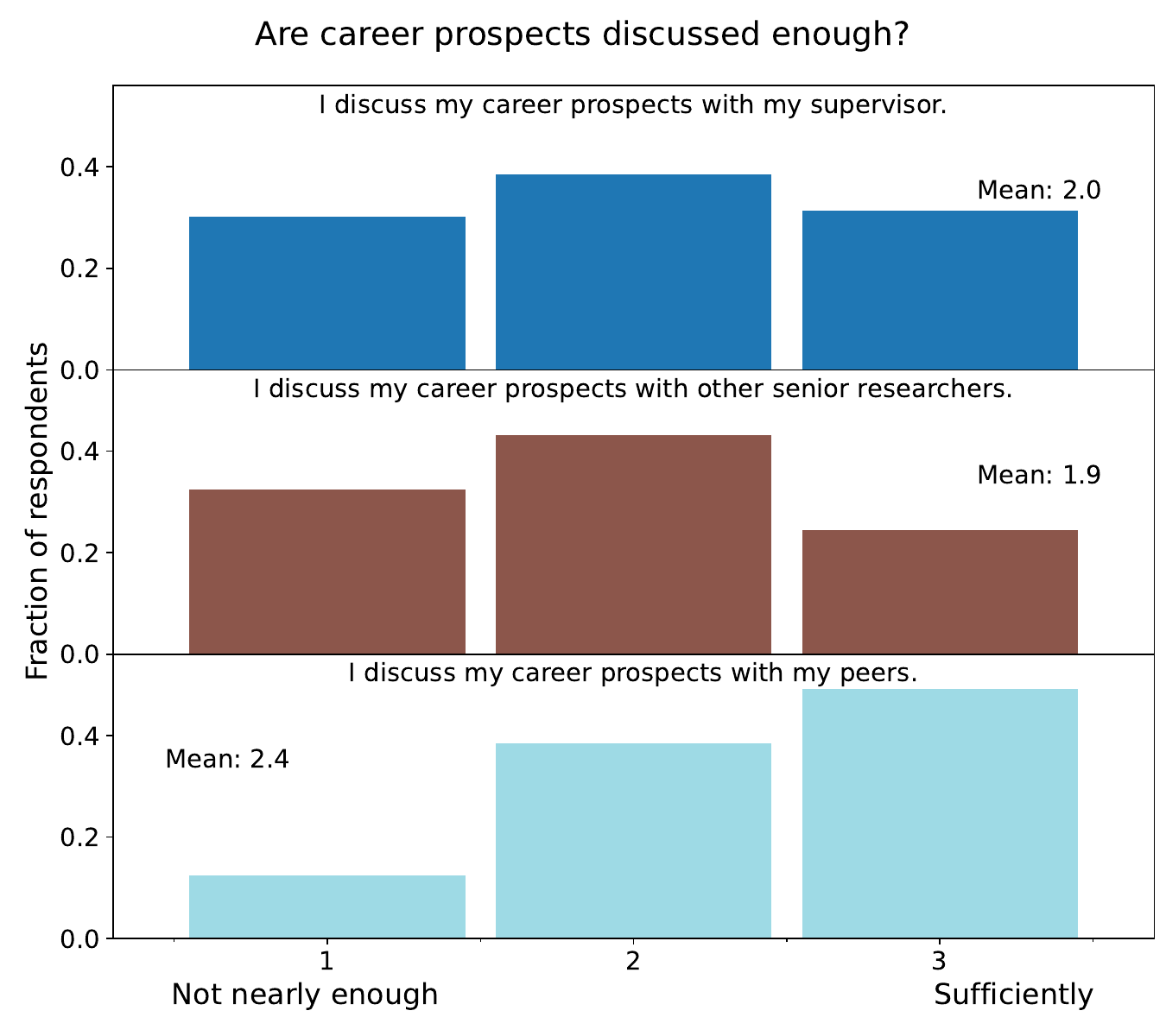}
    \caption{(Q65--67) How much respondents agree that they discuss their career prospects with various people.}
    \label{fig:part1:Discussion}
\end{figure}

Focusing further on discussion of career prospects, we show how well respondents agree that they discuss their career prospects enough with various people in Figure~\ref{fig:part1:Discussion}.
Around half of respondents discuss career prospects enough with their peers, whilst around $30\%$ of respondents each agree or disagree that they discuss this enough with their supervisor.
Less than $30\%$ of respondents feel that they discuss their career prospects enough with other senior researchers.

%----------------------------------------------------------------------------------------
\FloatBarrier
\subsubsection{Valuing research skills}

We now address a series of questions regarding possible attributes that a high-quality researcher should possess for a successful career in academia.

\begin{figure}[h!] % Ola
    \centering
        \includegraphics[width=0.7\textwidth]{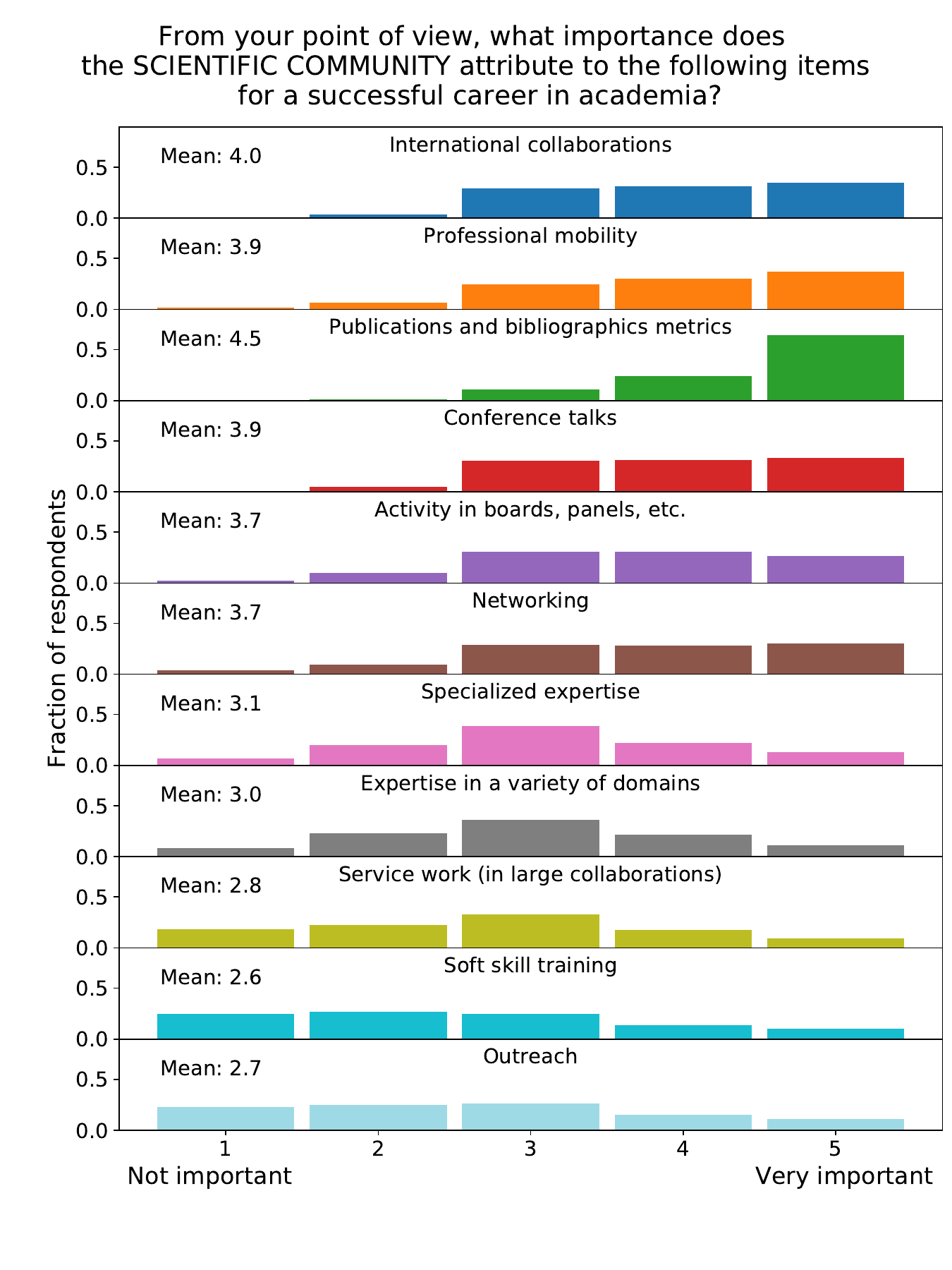}
    \caption{(Q69) Respondents' views on the importance of various aspects of research attributed to being a high-quality researcher and having successful career in academia according to the scientific community.}
    \label{fig:part1:Q69}
\end{figure}

\begin{figure}[h!] % Ola
    \centering
        \includegraphics[width=0.7\textwidth]{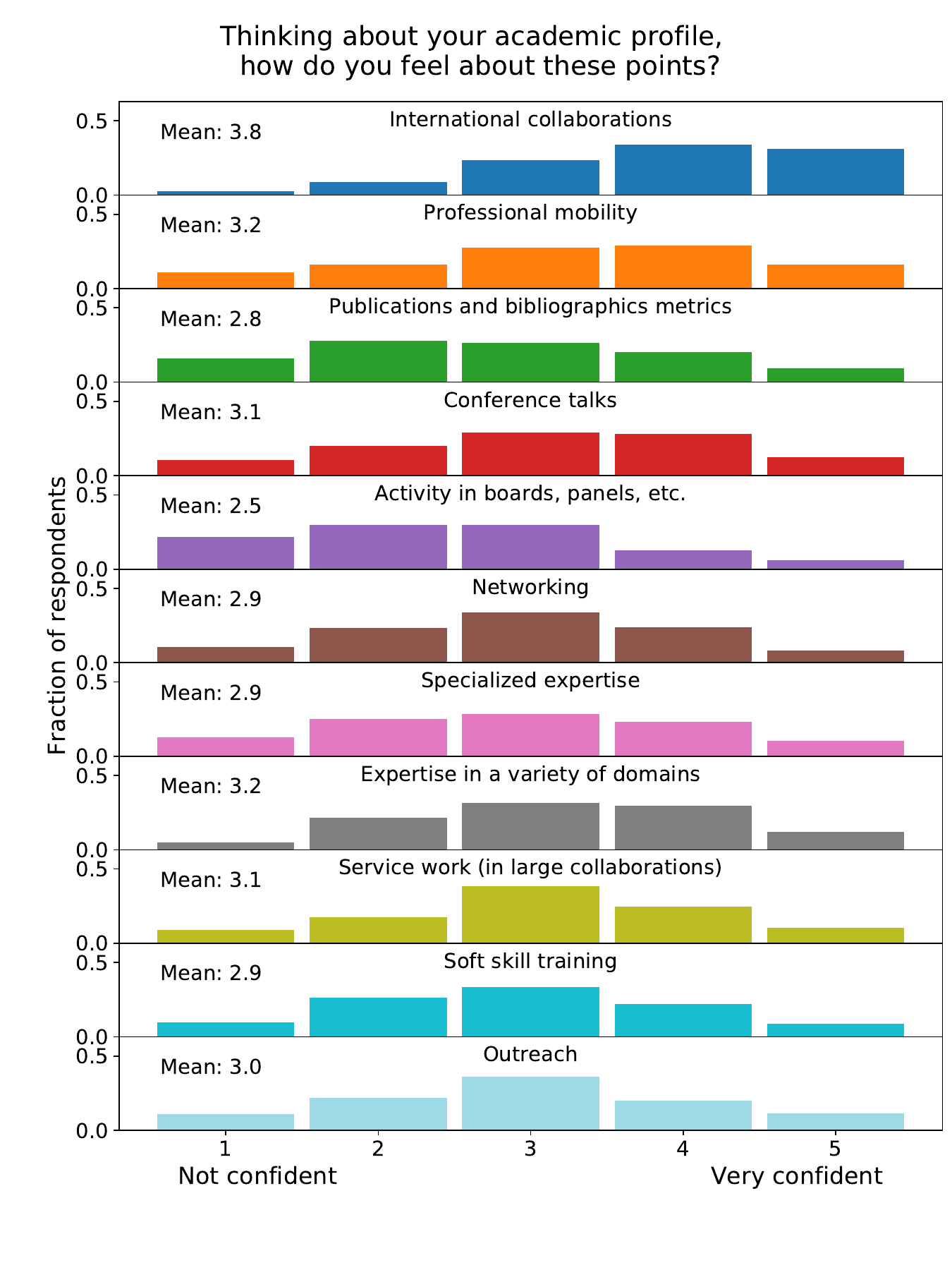}
    \caption{(Q70) Respondents' views on their fulfilment of various aspects attributed to being a high-quality researcher and having a successful career in academia.}.
    \label{fig:part1:Q70}
\end{figure}

Figure~\ref{fig:part1:Q68} shows what importance respondents personally attribute to a selection of metrics (listed in Appendix~\ref{app:questions} and in the figure) for a high-quality researcher to possess.
In Figure~\ref{fig:part1:Q69}, we instead show what importance respondents believe the scientific community attributes to the same set of items.
Respondents views on how the items are fulfilled for themselves are shown in Figure~\ref{fig:part1:Q70}.
We can conclude that respondents acknowledge all the attributes as important for a high-quality researcher with service work being the only attribute in the list with average personal importance below $3$.
For almost all attributes the level of personal importance is higher than the level of self-confidence, with service work as the only exception.

A particularly interesting observation is which of the attributes respondents consider to be underrated or overrated by the scientific community.
Respondents believe that the importance of professional mobility, publications and bibliographic metrics, conference talks, and activity in boards, panels, etc. are overrated by the scientific community.
On the other hand, they find specialised expertise, expertise in a variety of domains, soft skills, and outreach underrated.

\begin{figure}[h!] % Ola
    \centering
        \includegraphics[width=0.7\textwidth]{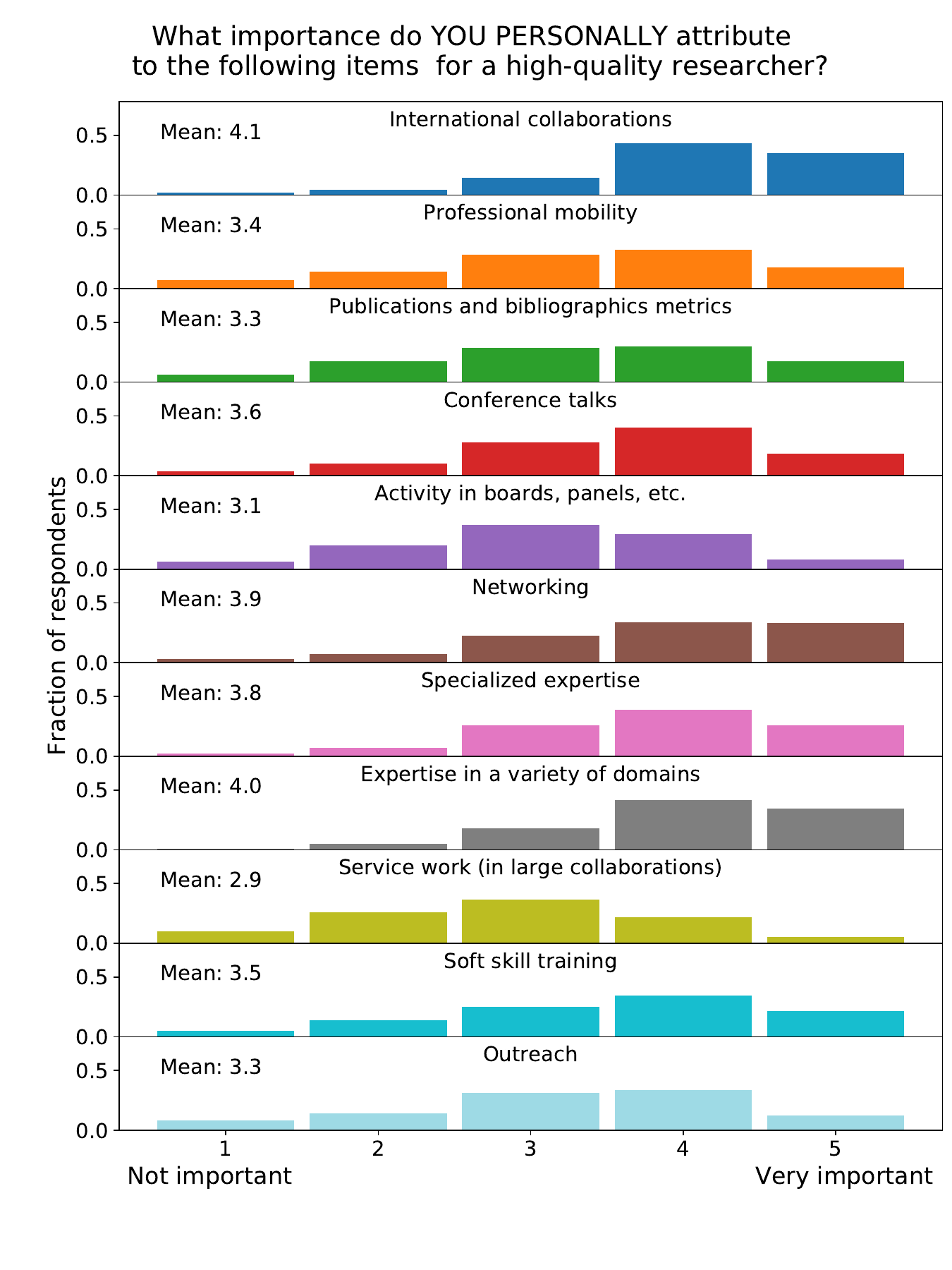}
    \caption{(Q68) Respondents' own views on the importance of various aspects of research to being a high-quality researcher and having a successful career in academia.}
    \label{fig:part1:Q68}
\end{figure}

%%%%%%%%%%%%%%%%%%%%%%%%%%%%%%%%%%%%%%%%%%%%%%%%%%%%%%%%%%%%%%%%%%%%%%%%%%%%%%%%%%%%%%%%%%%%%%%%%%%%%%%%%%%%%%%
\FloatBarrier
\subsection{Work-life balance, career planning and mobility}

%----------------------------------------------------------------------------------------
\subsubsection{Work-life balance}

\begin{figure}[h!] % Ola
    \centering
        \includegraphics[width=0.75\textwidth]{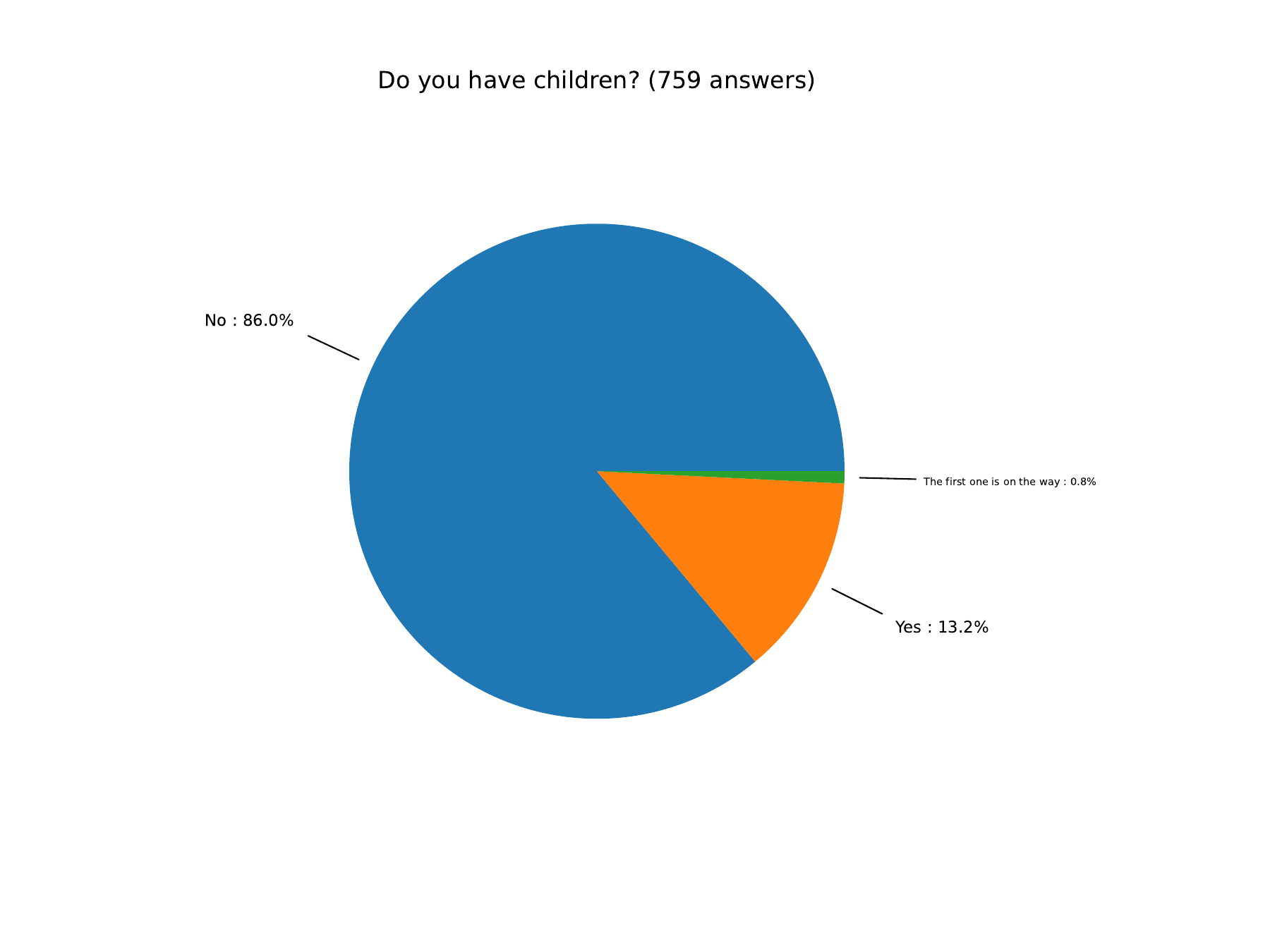}
    \caption{(Q71) Pie chart showing what fraction of respondents have children.}
    \label{fig:part1:Q71}
\end{figure}

\begin{figure}[h!] % Ola
    \centering
        \includegraphics[width=0.7\textwidth]{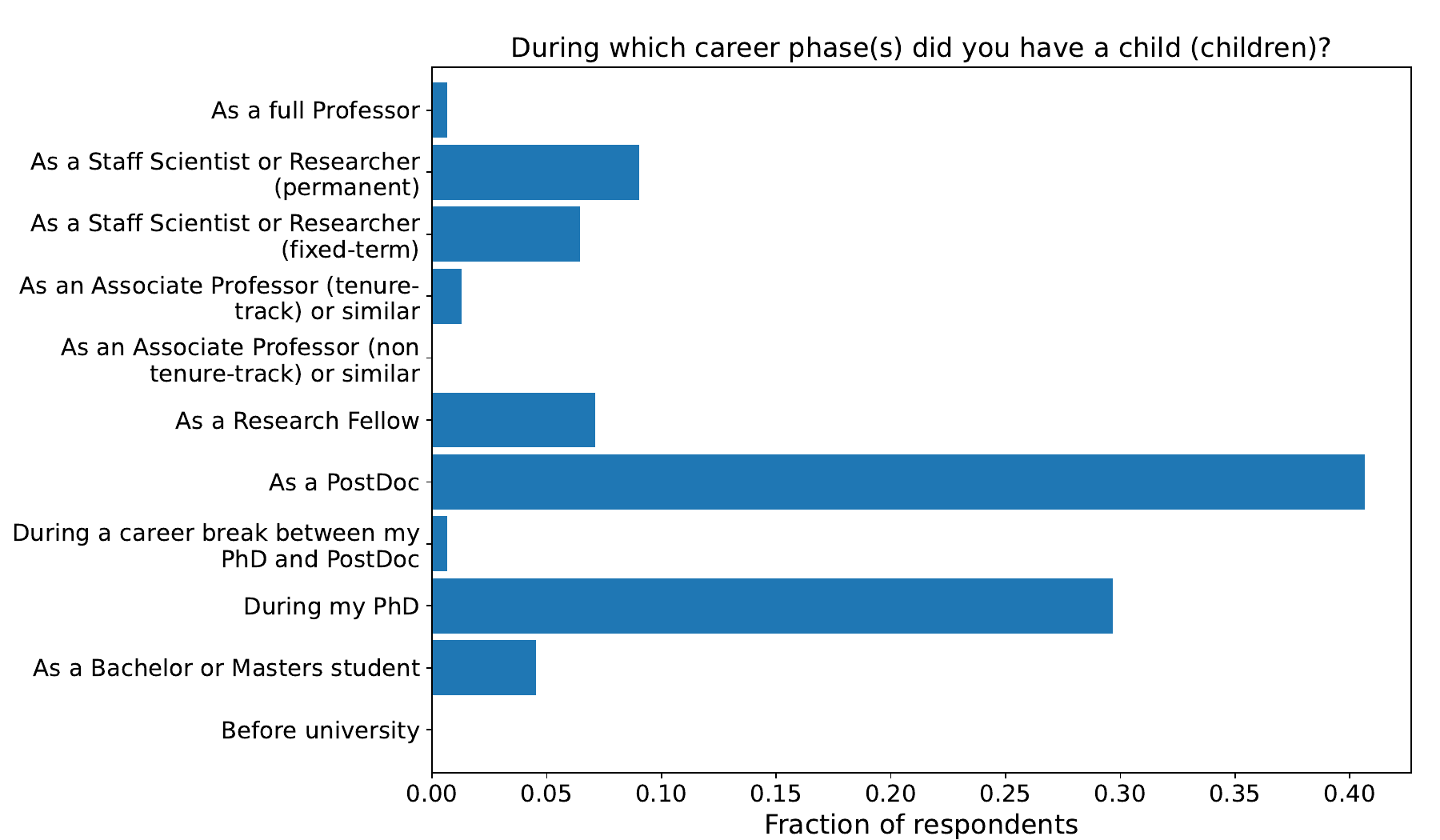}
    \caption{(Q72) The career phases during which respondents had children. Multiple answers per respondent were allowed, and responses are shown as a fraction of all responses from respondents who have had children.}
    \label{fig:part1:Q72}
\end{figure}

Figure~\ref{fig:part1:Q71} illustrates what portion of respondents have children.
We can see that $86\%$ of the respondents do not.
Figure~\ref{fig:part1:Q72} shows at which career phase(s) respondents had a child (children).
The majority of children were born during PostDoc and PhD phases, but we emphasise that the majority of survey respondents are PhD students and post-doctoral researchers.
The `Other' category related to a respondent having a child during a career break between PhD and PostDoc.

% Note we checked that the same set of respondents who said they have children in Q71 gave a career stage in Q72.

\begin{figure}[h!] % Ola
    \centering
        \includegraphics[width=0.6\textwidth]{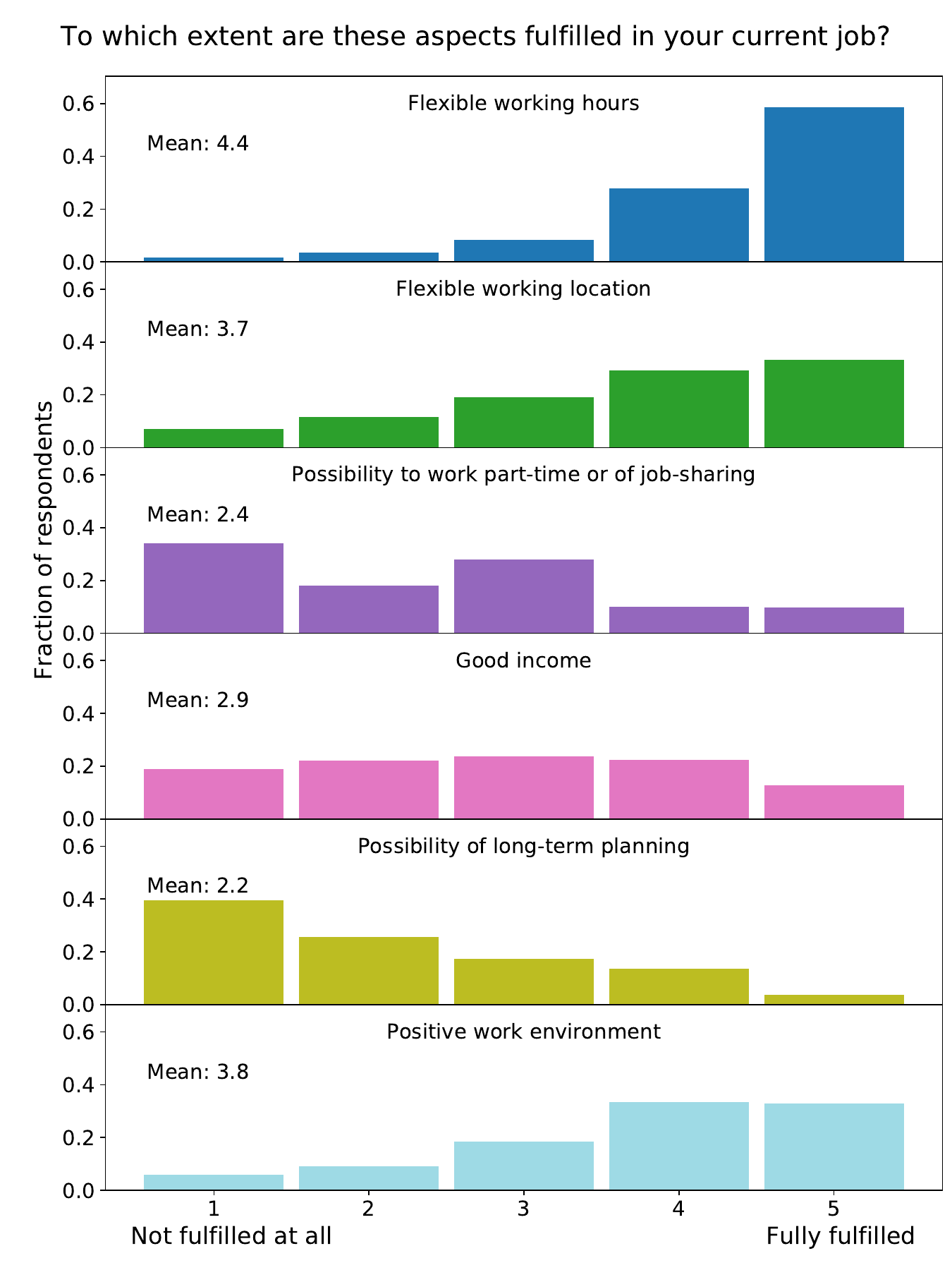}
    \caption{(Q74) Respondent's views on how different aspects of a good work-life balance are fulfilled in their current job. Fractions are given out of all respondents.}
    \label{fig:part1:Q74}
\end{figure}

\begin{figure}[h!] % Ola
    \centering
        \includegraphics[width=0.6\textwidth]{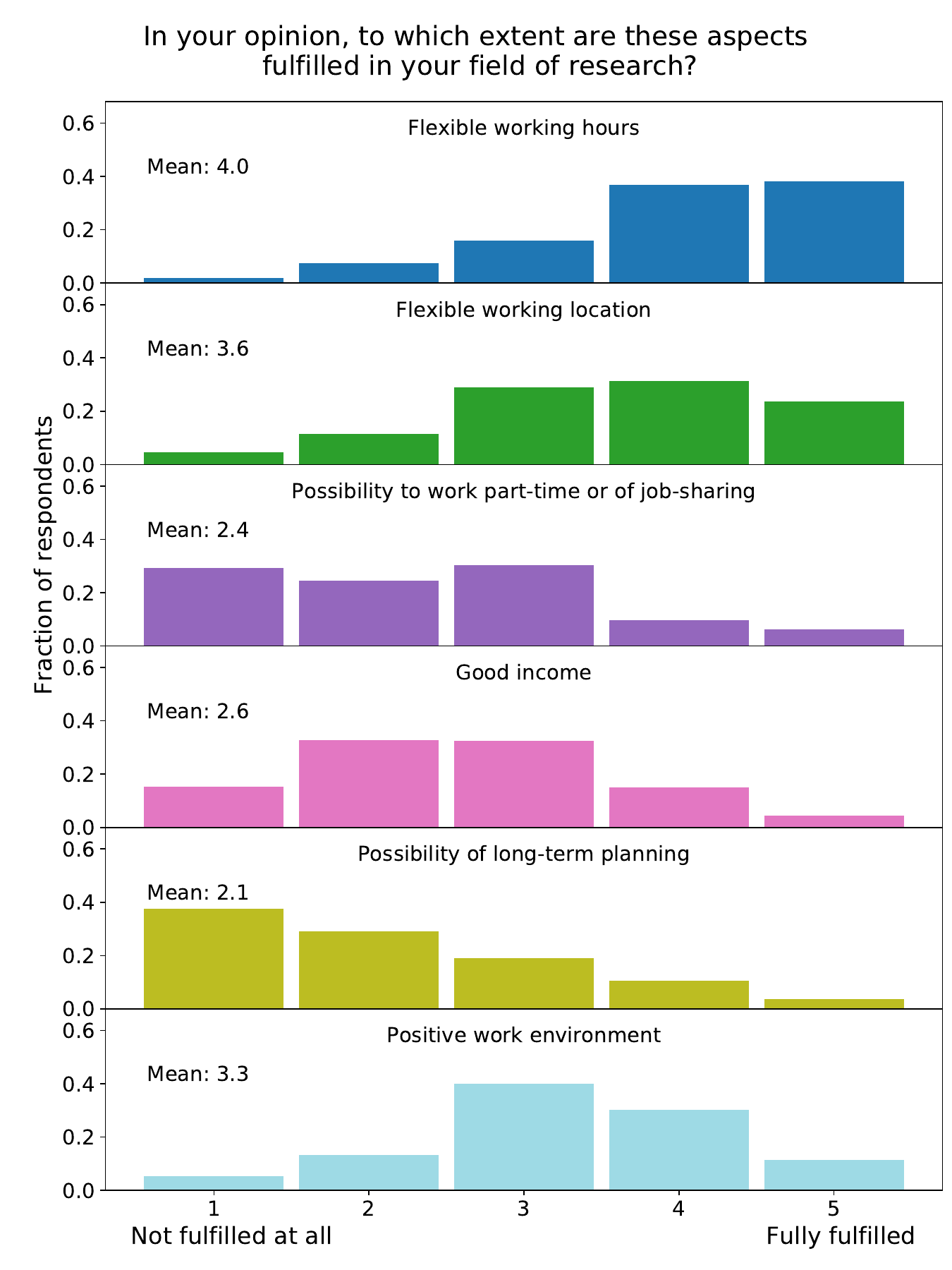}
    \caption{(Q75) Respondent's views on how different aspects of a good work-life balance are fulfilled in their field of research. Fractions are given out of all respondents.}
    \label{fig:part1:Q75}
\end{figure}

Figure~\ref{fig:part1:Q73} shows how important a set of items (listed in Appendix~\ref{app:questions} and in the figure) are to respondents in order to have a good work-life balance.
Figure~\ref{fig:part1:Q74} shows correspondingly how fulfilled each item is in respondents' current job, and Figure~\ref{fig:part1:Q75} shows how fulfilled they are believed to be in respondents' field of research as a whole.

We observe that respondents find all the plotted aspects to be important, with the highest importance assigned to a positive work environment and the lowest to the possibility of part-time work or job-sharing.
Flexible working hours is also one of the most important items, and the aspect most fulfilled, on average, in respondents' current jobs.
The biggest discrepancy between what respondents find important for a proper work-life balance and the fulfilment in reality is the possibility of long-term planning.
The average level of fulfilment in the current jobs of respondents is only 2.2, whilst the average importance score is 4.4.
For all aspects, respondents perceive that aspects of a good work-life balance are fulfilled at least as successfully in their current job as in general in their research field.

\begin{figure}[h!] % Ola
    \centering
        \includegraphics[width=0.6\textwidth]{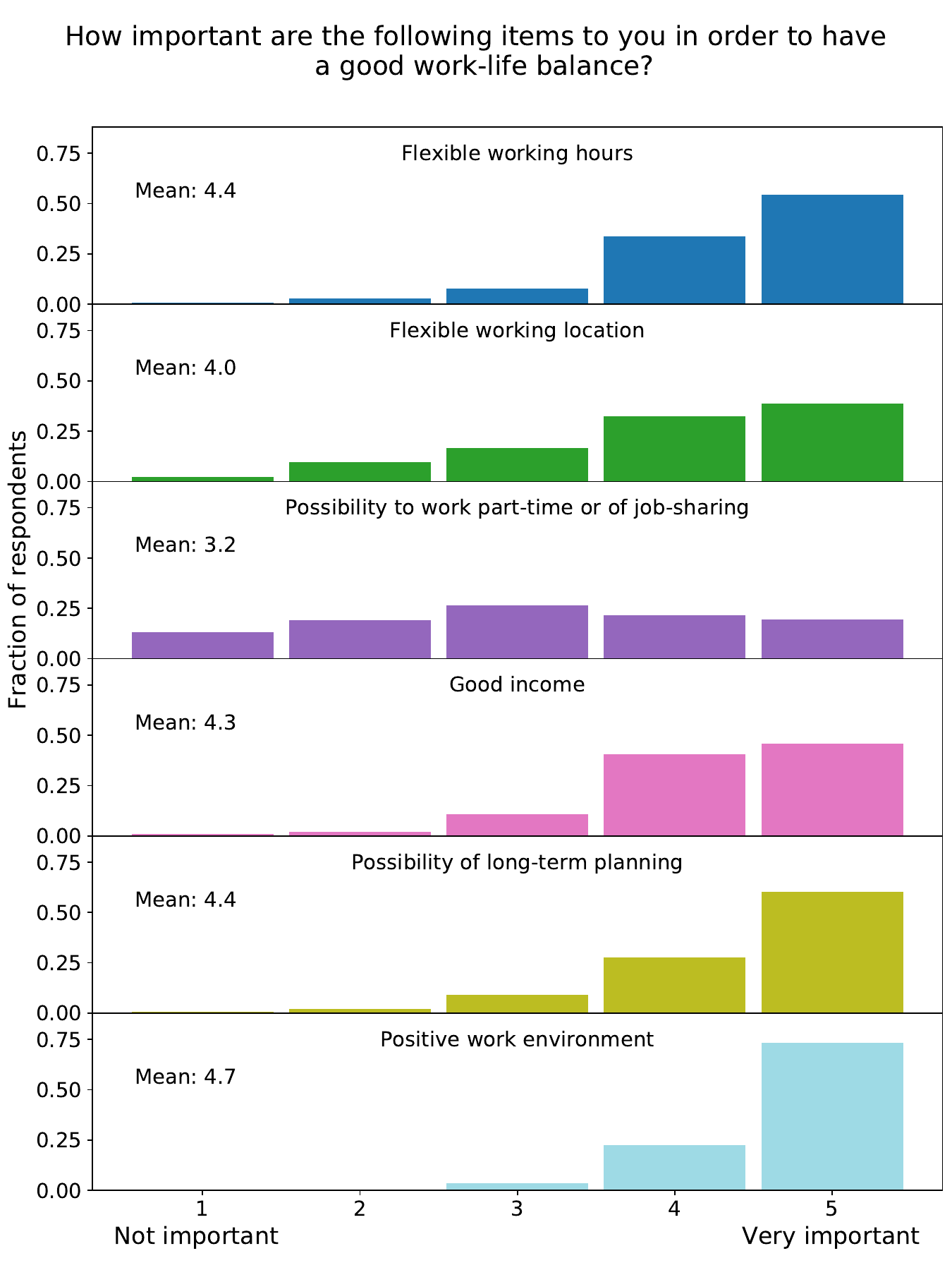}
    \caption{(Q73) Respondents' views on the importance of different aspects of a good work-life balance. Fractions are given out of all respondents.}
    \label{fig:part1:Q73}
\end{figure}

\begin{figure}[h!] % Ola
    \centering
        \includegraphics[width=0.6\textwidth]{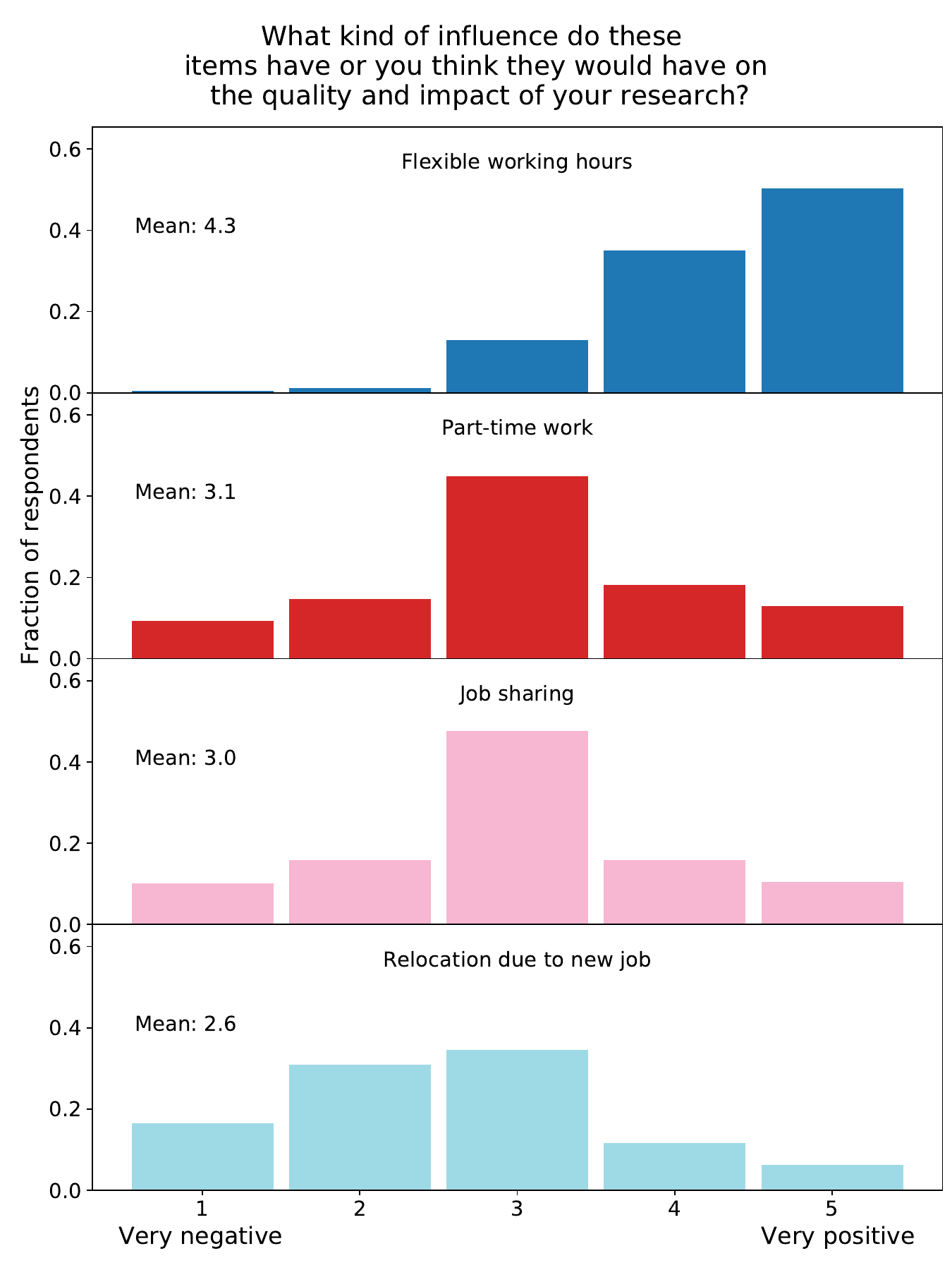}
    \caption{(Q76) Respondents' views on how positive or negative the impact on their research is, or could be, for different aspects of a good work-life balance. Fractions are given out of all respondents.}
    \label{fig:part1:Q76}
\end{figure}

Next, respondents' opinions on the impact (or potential impact) of various aspects of work-life balance on their research was studied.
The answers are summarised in Figure~\ref{fig:part1:Q76}.
According to the respondents, flexible working hours would have a strong positive impact on research, whilst having to relocate is viewed rather negatively.

\begin{figure}
    \centering
        \includegraphics[width=0.6\textwidth]{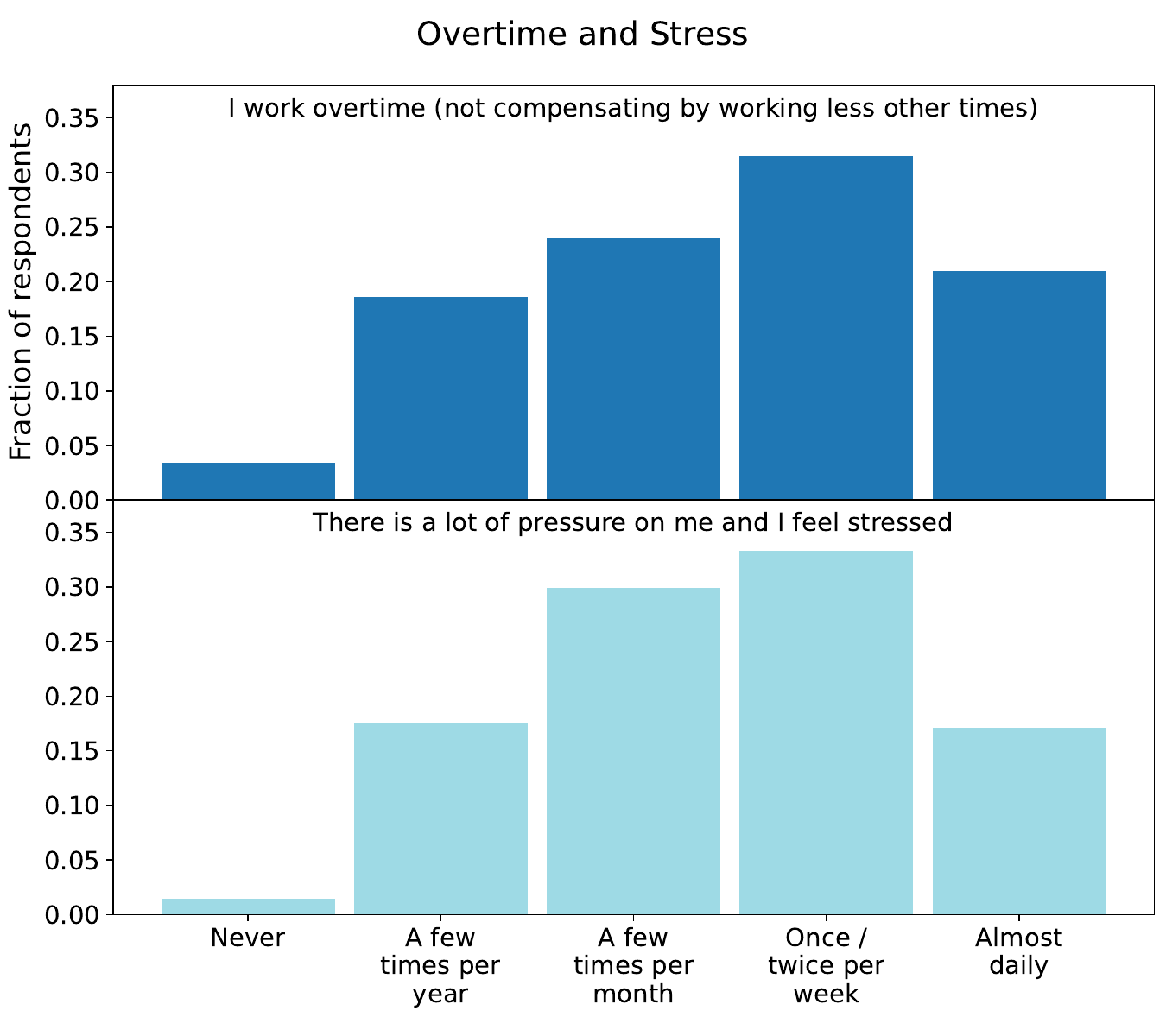}
    \caption{(Q79--80) How often respondents work overtime, or feel stressed and under pressure. Fractions given include the 1.6\% of respondents who didn't answer these questions, which is not shown in the plot.}
    \label{fig:part1:overtimeStress}
\end{figure}

Figure~\ref{fig:part1:overtimeStress} demonstrates that, unfortunately, most respondents to the survey feel stressed and work overtime very frequently.
We see that $21\%$ of respondents work overtime almost daily, whilst $50\%$ of respondents feel stressed and under a lot of pressure at least once a week.

%----------------------------------------------------------------------------------------
\FloatBarrier
\pagebreak
\subsubsection{Career mobility and leaving academia}

\begin{figure}[h!] % Ola
    \centering
        \includegraphics[width=0.7\textwidth]{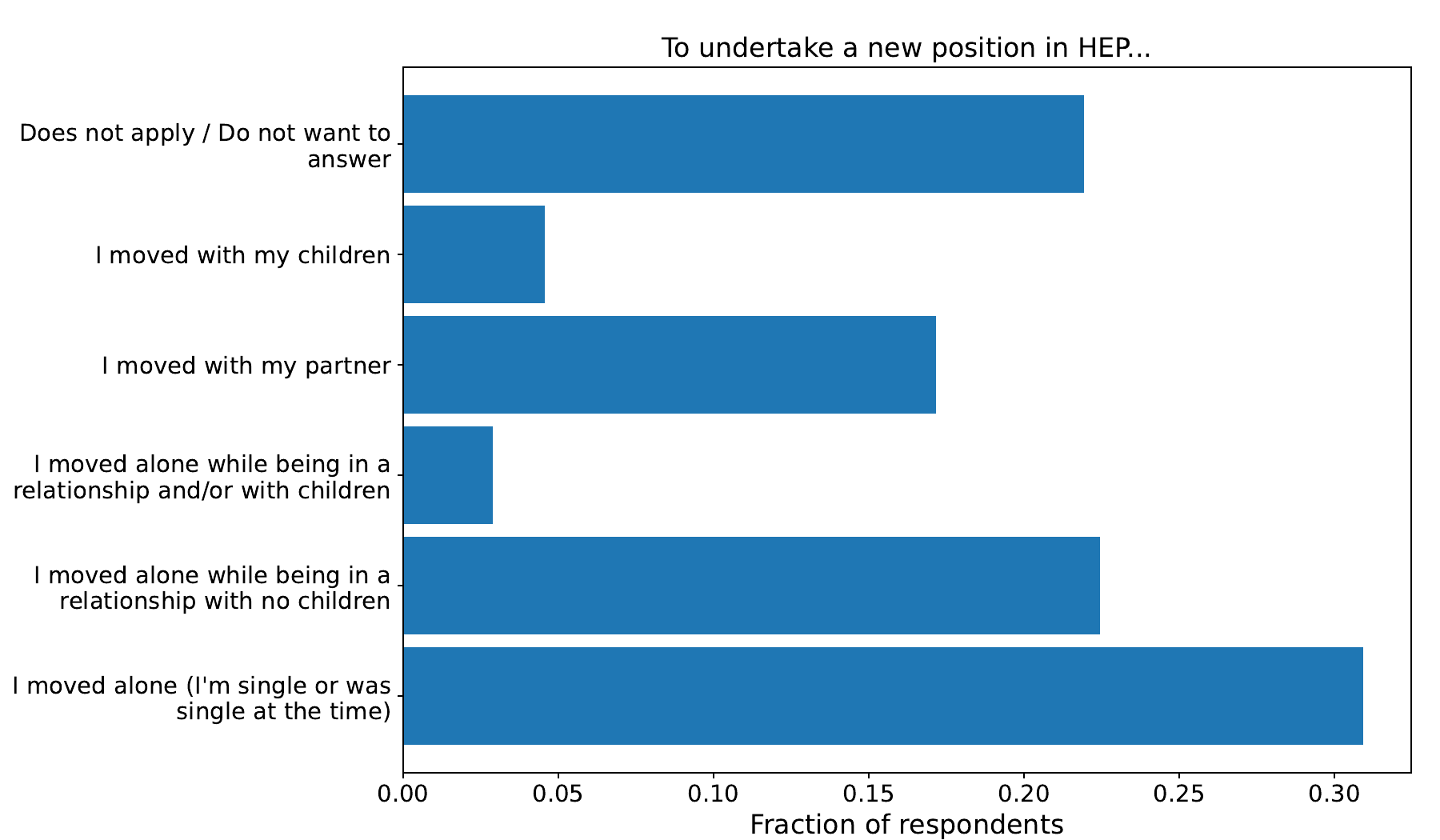}
    \caption{(Q77) Respondents' family situation while moving to undertake a new position in HEP. Multiple answers could be selected per respondent and fractions are given out of all respondents.}
    \label{fig:part1:Q77}
\end{figure}

Respondents' experiences connected with career mobility were studied next.
Figure~\ref{fig:part1:Q77} shows respondents' family situation, if they moved in order to undertake a new position in HEP.
Multiple answers could be selected per respondent.
The majority of the respondents moved alone since they were single at that time.
However, almost as many respondents moved alone despite being in a relationship and/or having children.

\begin{figure}[h!] % Ola
    \centering
        \includegraphics[width=0.7\textwidth]{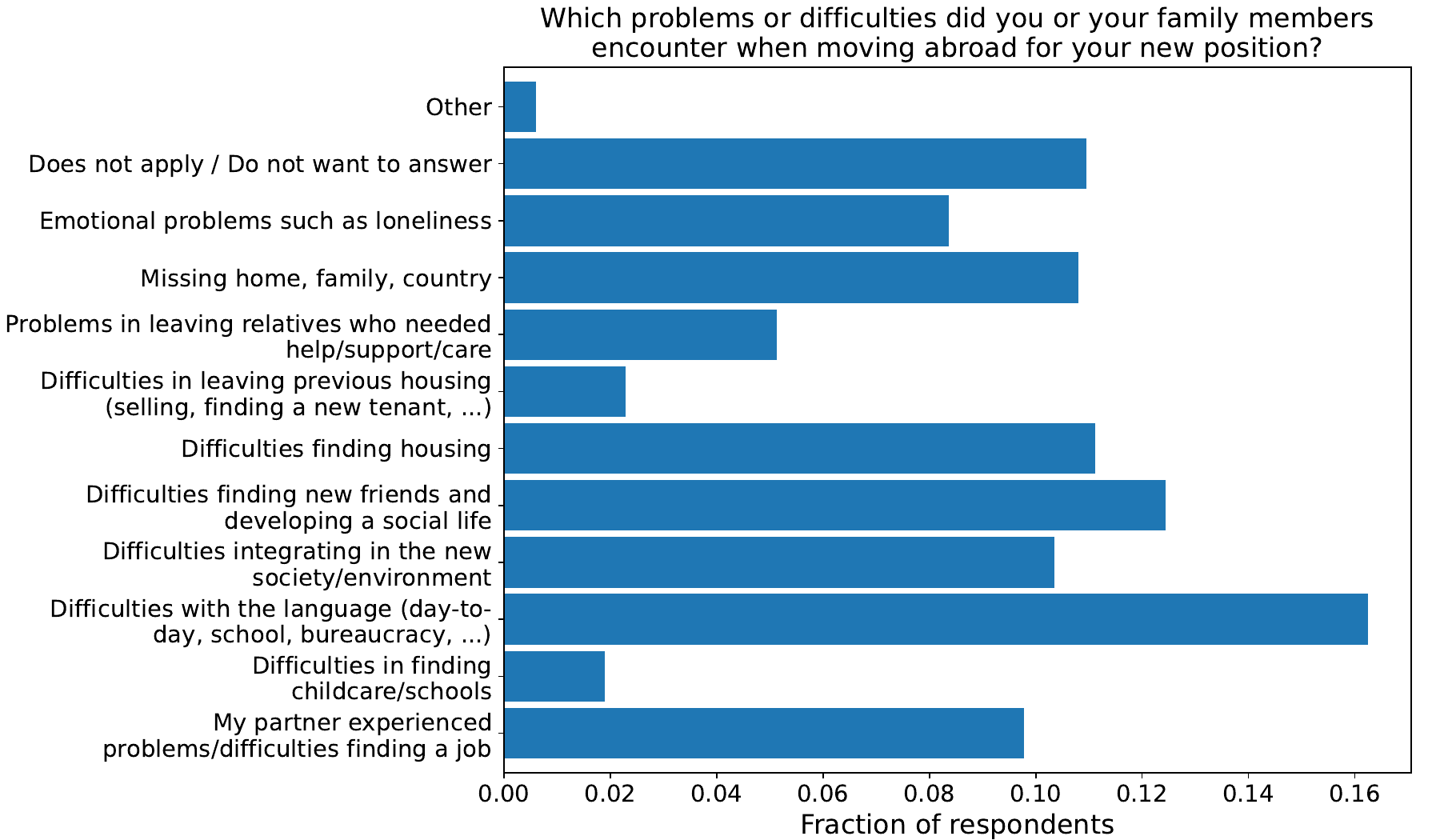}
    \caption{(Q78) The problems experienced by respondents and their family members while moving to undertake a new position in HEP. Fractions are shown out of all respondents who answered the question. Multiple answers could be selected by each respondent and question was not obligatory.}
    \label{fig:part1:Q78}
\end{figure}

Views on which problems or difficulties respondents or their family members encountered when moving abroad for their new position are presented in Figure~\ref{fig:part1:Q78}.
Multiple answers were allowed per respondent and the question was voluntary.
We observe that several problems are very common and concern: difficulties with language; finding new friends and developing a social life; difficulties in finding housing; and missing home, family, or country.
The least encountered problem is difficulties in finding childcare/schools, which is not unexpected as only approximately $13\%$ of respondents have children.
Within the `Other' category: 56\% pointed to relationship problems, 31\% cited difficulties with VISA/administration, and 13\% mentioned COVID-19 related problems.
The mobility questions triggered many additional comments in the survey.
This illustrates the importance of this aspect to the respondents and its emotional impact.
Some of the most common or serious comments shared by the respondents in the open `Other' box are quoted in Appendix~\ref{app:quotes}.

% Note that every response in 'other' clearly fit into an existing category to they were reassigned.

\begin{figure}
    \centering
        \includegraphics[width=0.7\textwidth]{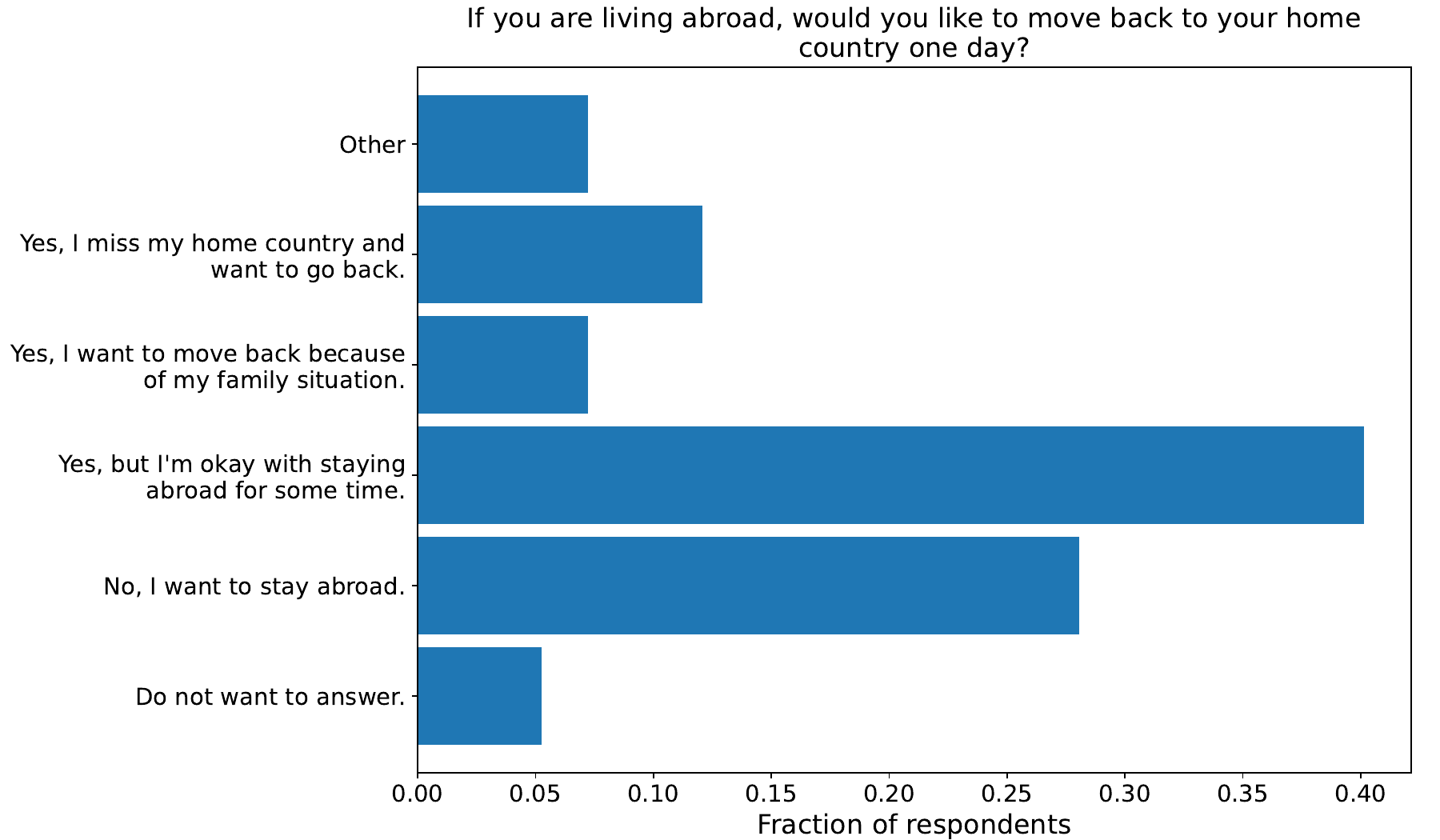}
    \caption{(Q81) The opinions of respondents who are living abroad, on whether they would like to move back to their home country. Fractions are given out of all respondents who are currently living abroad.}
    \label{fig:part1:Q81}
\end{figure}

We summarise the views of respondents currently living abroad (over $52\%$ of the total) on whether they would like to move back to their home country in the future in Figure~\ref{fig:part1:Q81}.
Overall, most respondents who are living abroad would like to go home eventually.
Out of respondents in the `Other' category: 48\% were unsure and said it would depend on the circumstances (such as their partner's location and the quality of life); 31\% didn't mind; and 21\% said they would like to stay abroad but cannot, for example due to a lack of job opportunities or the socio-political climate.

\begin{figure}
    \centering
        \includegraphics[width=0.7\textwidth]{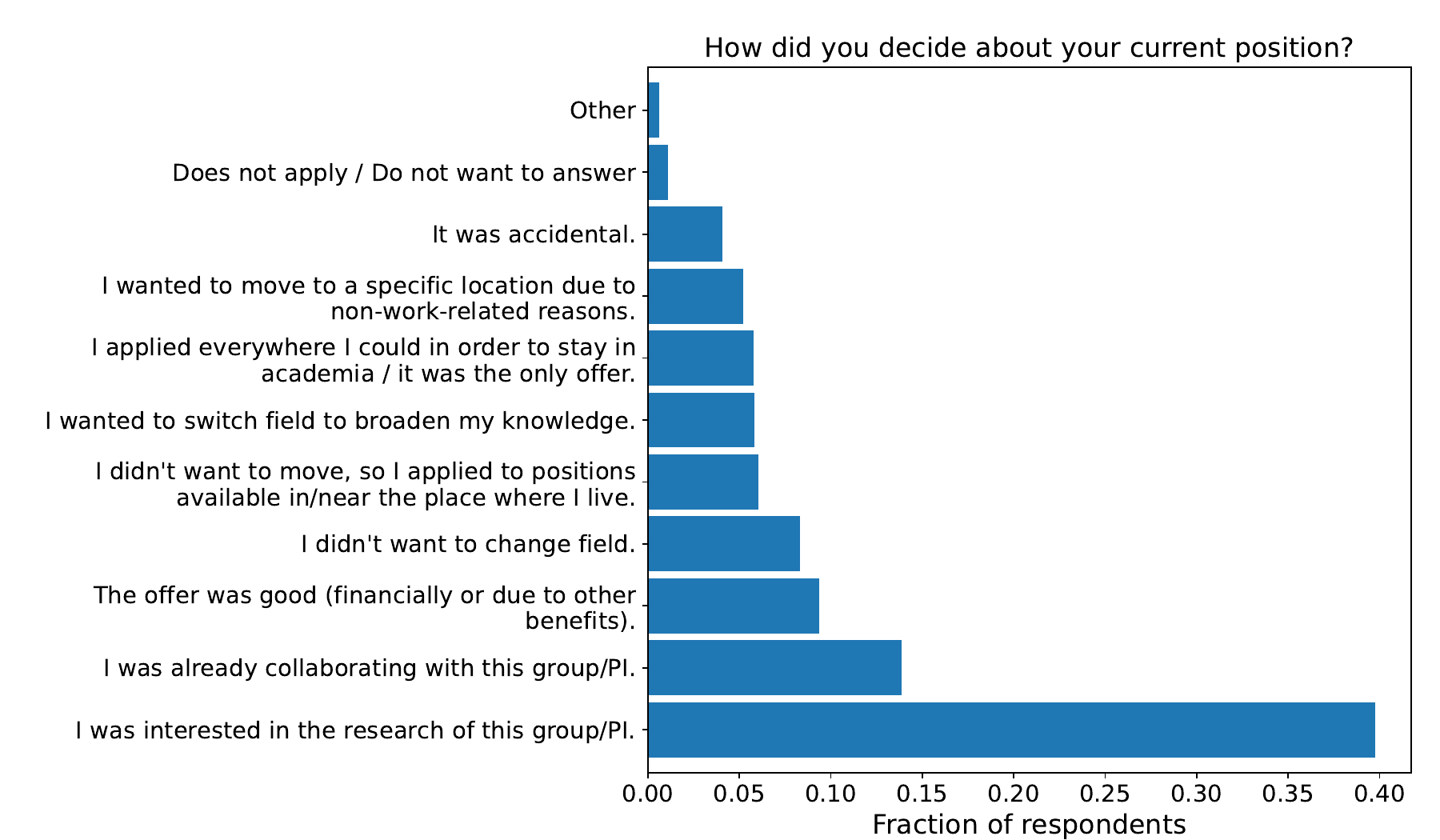}
    \caption{(Q82) The importance of various factors, to respondents, in choosing their current position. Fractions are shown out of all respondents. Multiple answers could be selected per respondent.}
    \label{fig:part1:Q82}
\end{figure}

Figure~\ref{fig:part1:Q82} shows what factors respondents used when choosing their current position (with multiple answers allowed per respondent).
By far the strongest factor is being interested in the work of the group/PI, followed by already collaborating with the group/PI, having good terms in the offer, and not wanting to change field.
Slightly more respondents prioritised applying everywhere possible to stay in academia compared to applying for positions with the aim of moving to a specific location.
73\% of the responses in the `Other' category related to wanting to work in the group because of its good reputation in research, work environment, etc., and 27\% to getting tenure or an increased chance of tenure.

\begin{figure}
    \centering
        \includegraphics[width=0.7\textwidth]{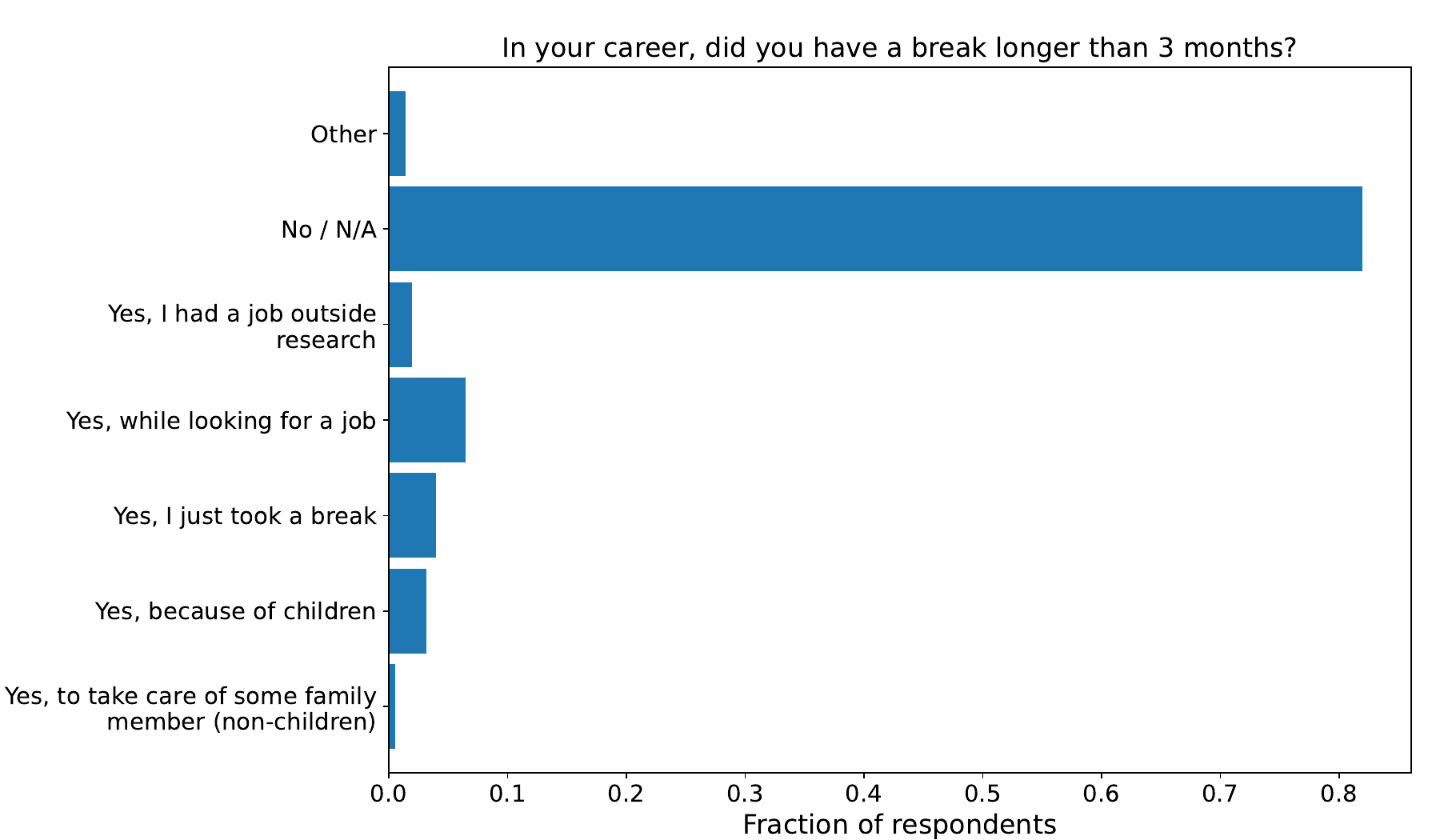}
    \caption{(Q83) Histogram showing Whether respondents have had a career break of more than three months or not, and --- if so --- why. Fractions are given out of all respondents.}
    \label{fig:part1:Q83}
\end{figure}

Answers to whether respondents have had a career break of longer than 3 months, and if so why, are shown in Figure~\ref{fig:part1:Q83}.
Only 18\% of respondents have, most while looking for a new job.
Within the `Other' category: 63\% took a break due to illness and 38\% due to the COVID-19 pandemic.
Following this, we asked respondents whether they have ever changed field within physics.
$30\%$ of respondents answered yes.

% note as the caption and text says, respondents NOT living abroad are excluded here (since it's not relevant to the question) and fractions calculated for the remainining categories/respondents.

\begin{figure}
    \centering
        \includegraphics[width=0.7\textwidth]{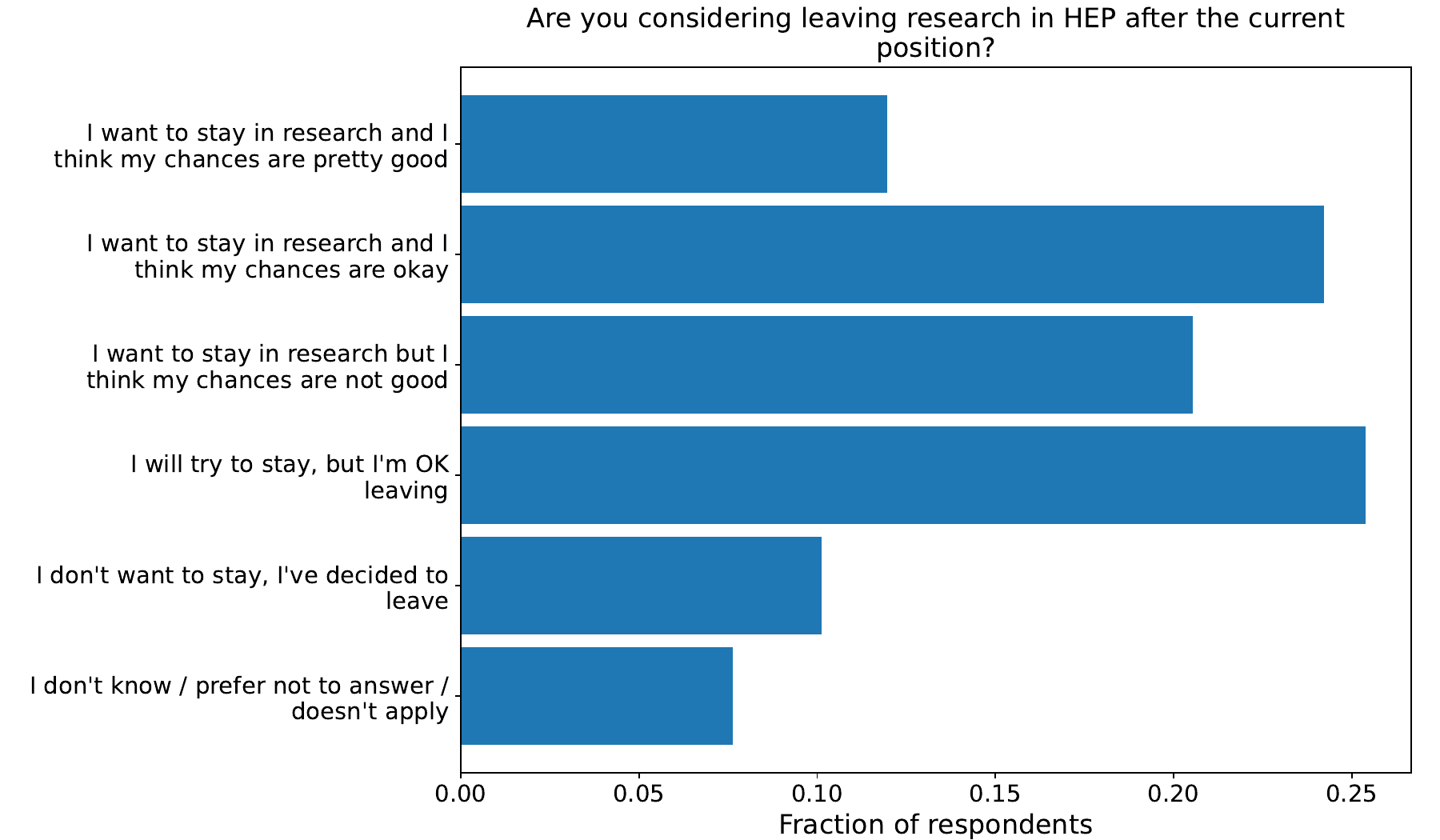}
    \caption{(Q85) Histogram showing whether respondents are considering leaving research in HEP after their current position or not. Fractions are given out of all respondents.}
    \label{fig:part1:Q85}
\end{figure}

In Figure~\ref{fig:part1:Q85}, we show answers to whether respondents want to leave research in HEP after their current position.
Whilst $57\%$ of respondents explicitly want to stay in HEP, only $12\%$ think that their chances are pretty good.
In contrast $10\%$ of respondents want to leave.

\begin{figure}[htb!] % Ola
    \centering
        \includegraphics[width=0.75\textwidth]{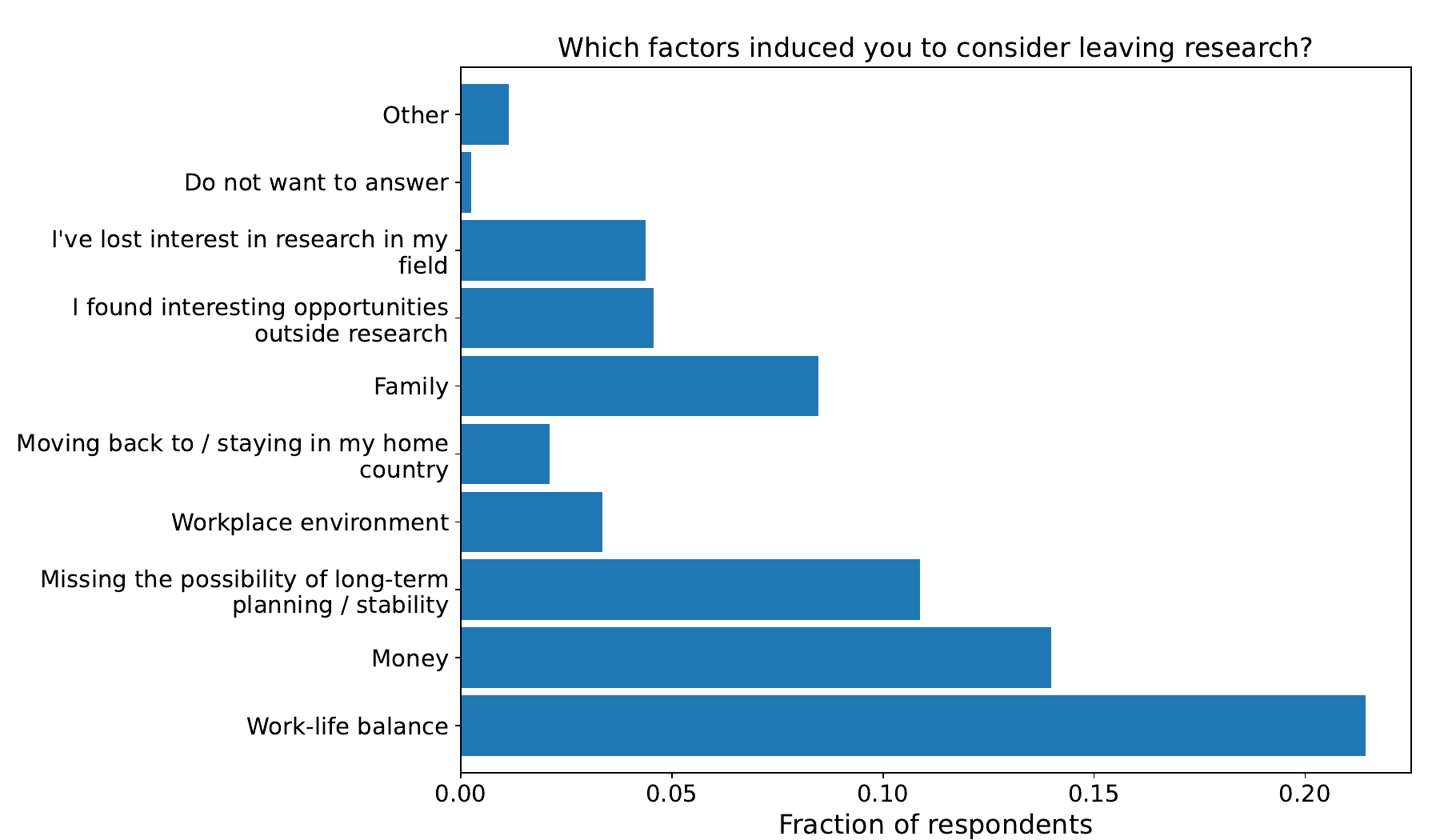}
    \caption{(Q86) Factors that induced respondents to consider leaving research. Fractions are given from all respondents, and empty answers are not shown. Multiple answers were allowed per-respondent.}
    \label{fig:part1:Q86}
\end{figure}

Factors which cause respondents to consider leaving research are summarised in Figure~\ref{fig:part1:Q86}, with multiple answers allowed per respondent.
We see that the three most common factors inducing respondents to consider leaving research are work-life balance, money and missing the possibility of long-term planning.
The least common option, from those presented, was ``moving back to my home country''.
Within the `Other' category, 32\% cited systemic problems in academia, 16\% lack of job opportunities, 21\% social/economic/political factors, and the remainder other issues.
Some of the most striking or common responses given in the open ``Other' box are quoted in Appendix~\ref{app:quotes}.
\clearpage
% of Other: 32\% systemic problems in academia, 16\% lack of job opportunities, 21\% social/economic/political factors, 32\% other-other

%%%%%%%%%%%%%%%%%%%%%%%%%%%%%%%%%%%%%%%%%%%%%%%%%%%%%%%%%%%%%%%%%%%%%%%%%%%%%%%%%%%%%%%%%%%%%%%%%%%%%%%%%%%%%%%

\subsection{Discriminatory or abusive treatment}

\begin{figure}[htb!]
    \centering
        \includegraphics[width=0.75\textwidth]{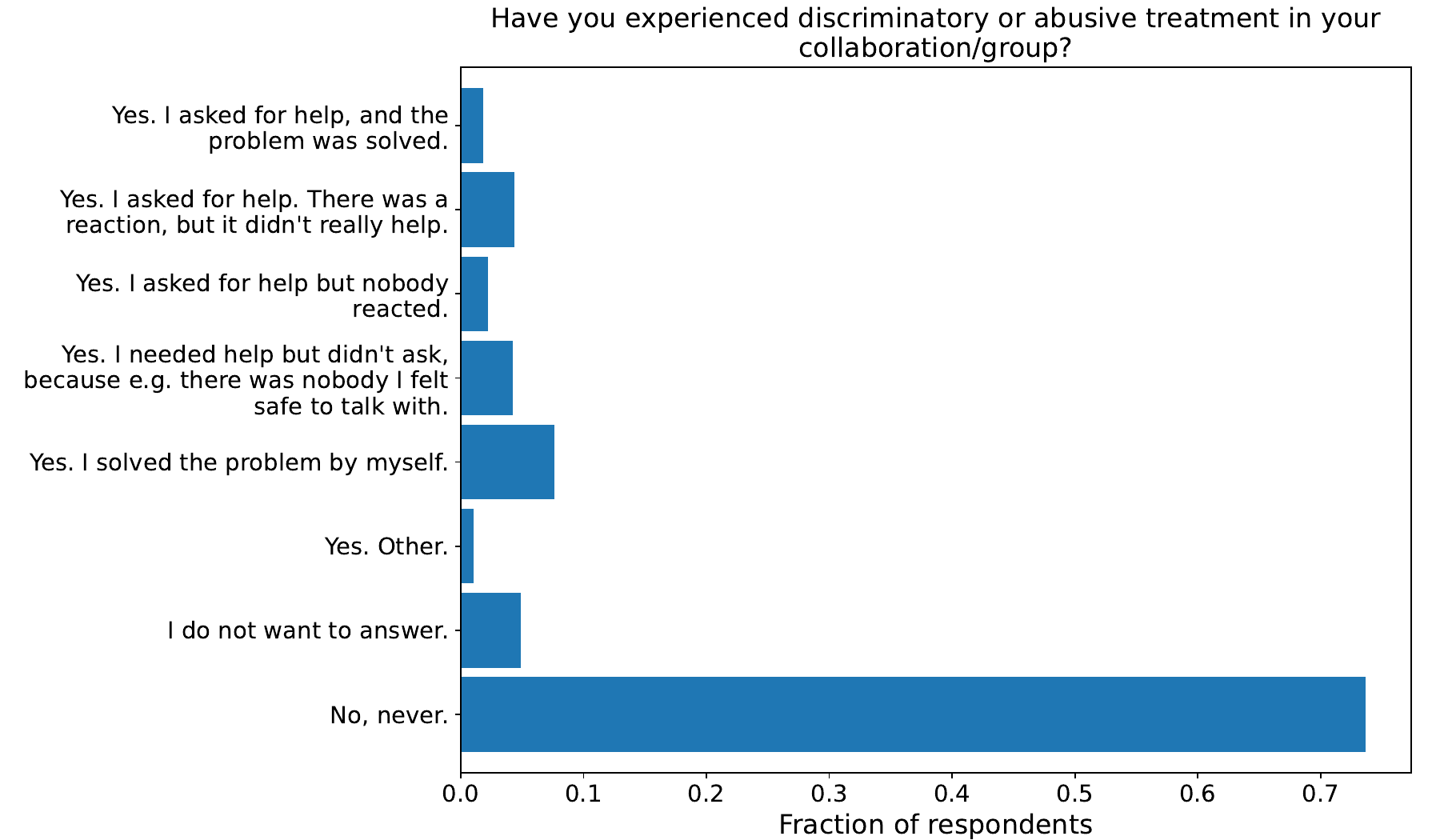}
    \caption{(Q87) Histogram showing whether respondents have experienced discriminatory or abusive treatment in their collaboration/group or not. Fractions are given out of all respondents.}
    \label{fig:part1:Q87}
\end{figure}

In Figure~\ref{fig:part1:Q87}, we address the question of whether respondents have experienced discriminatory or abusive treatment in their collaboration/group.
Whilst $74\%$ of respondents have never experienced this treatment, a non-negligible $21\%$ of respondents have.
Half of the respondents who have experienced this treatment either felt unable to seek help or asked for help but found it unhelpful.
Two respondents pointed out a general culture of sexism in their collaboration/group.

\begin{figure}[htb!]
    \centering
        \includegraphics[width=0.75\textwidth]{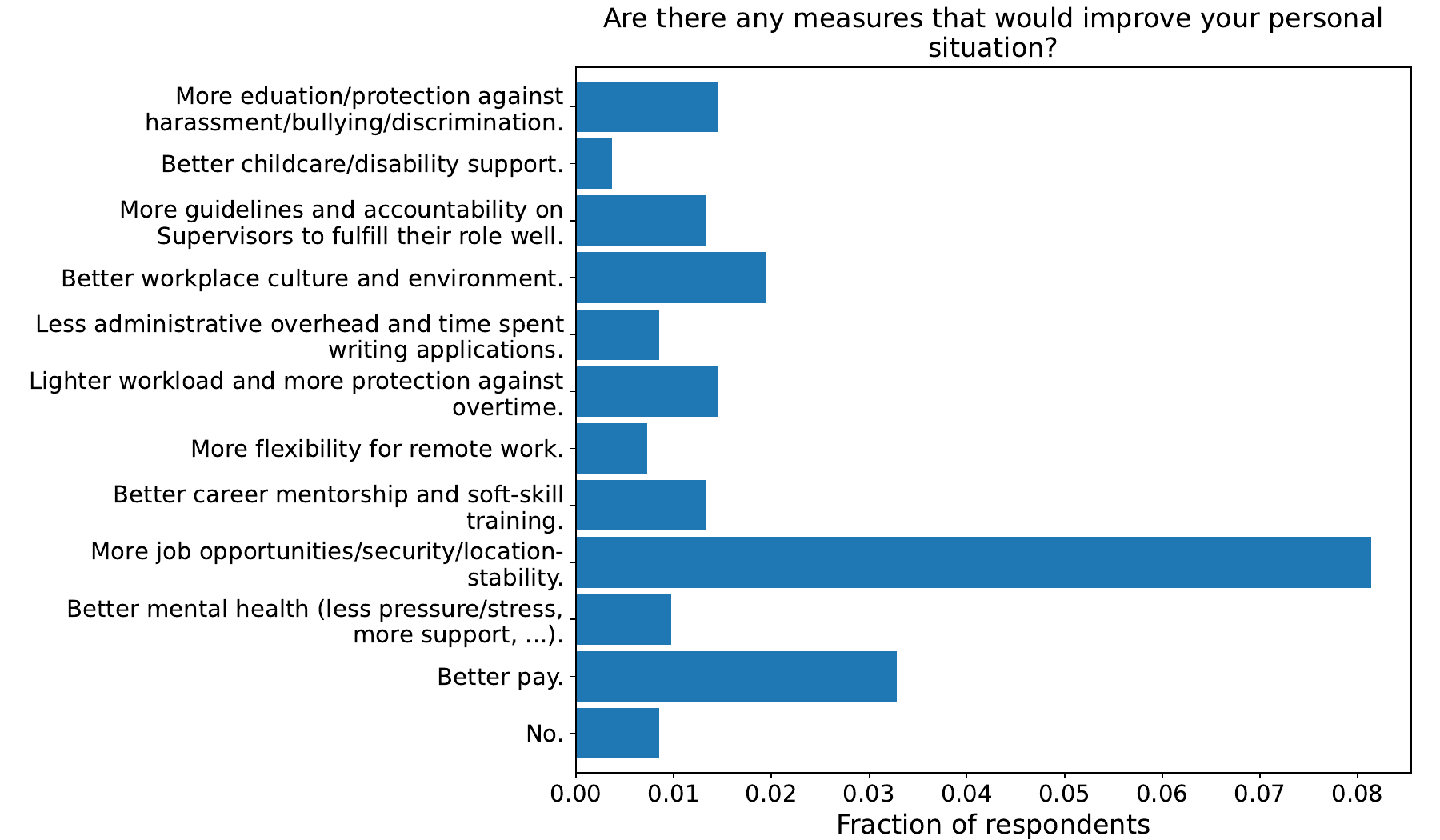}
    \caption{(Q88) The categories of measures that respondents said would improve their personal situation. The fraction given is out of all respondents, empty responses are not shown.}
    \label{fig:part1:Q88}
\end{figure}

After this, respondents were given a open box to answer whether there are any measures that would improve their personal situation.
Answers were provided by $19\%$ of respondents, and where multiple distinct points were made, all were considered separately, though a lot of the topics are connected.
Responses were grouped into categories for plotting purposes.
As Figure~\ref{fig:part1:Q88} shows, responses overwhelmingly relate to the academic job market.
Respondents want more job opportunities in general, and jobs with a longer or permanent contract in particular.
Several respondents stressed the problems with family and long term social connections that come with having to frequently move, which were also noted for previous questions.
One respondent pointed out that shorter contracts make it harder to make time for `high-risk' research, and relating to this, three focused on the amount of time sunk into writing job/grant applications as part of the administrative overhead category.
The joint-second most frequently mentioned categories related to better pay and better workplace culture/environment.
Examples of improvements to workplace culture/environment include: more open communication and knowledge transfer with all levels of seniority; more academic support and respect; better organisation and administrative support.
Related to this, several respondents called for more education and protection against harassment, bullying and discrimination.
Specific examples here include training to prevent gender bias and improved Ombudsman services.
Several respondents also desired better/more-equal childcare support, or better flexibility for remote work (largely for family reasons).
Similar numbers of respondents discussed two related categories: better training in soft-skills and more career mentorship; and for their supervisors to have more guidelines and accountability in order to fulfil their roles well.

As reflects the data in Figure~\ref{fig:part1:overtimeStress}, several respondents wished for a lighter workload: by having fewer tasks to deal with at once, by having more employment protection against unpaid overtime, or by creating a culture where overtime work is not encouraged.
Additionally, a few respondents wished for lower stress and pressure, or better support with their mental health and self-confidence.

%%%%%%%%%%%%%%%%%%%%%%%%%%%%%%%%%%%%%%%%%%%%%%%%%%%%%%%%%%%%%%%%%%%%%%%%%%%%%%%%%%%%%%%%%%%%%%%%%%%%%%%%%%%%%%%
\FloatBarrier
\subsection{Recognition and visibility}

\begin{figure}[h!]
    \centering
        \includegraphics[width=0.6\textwidth]{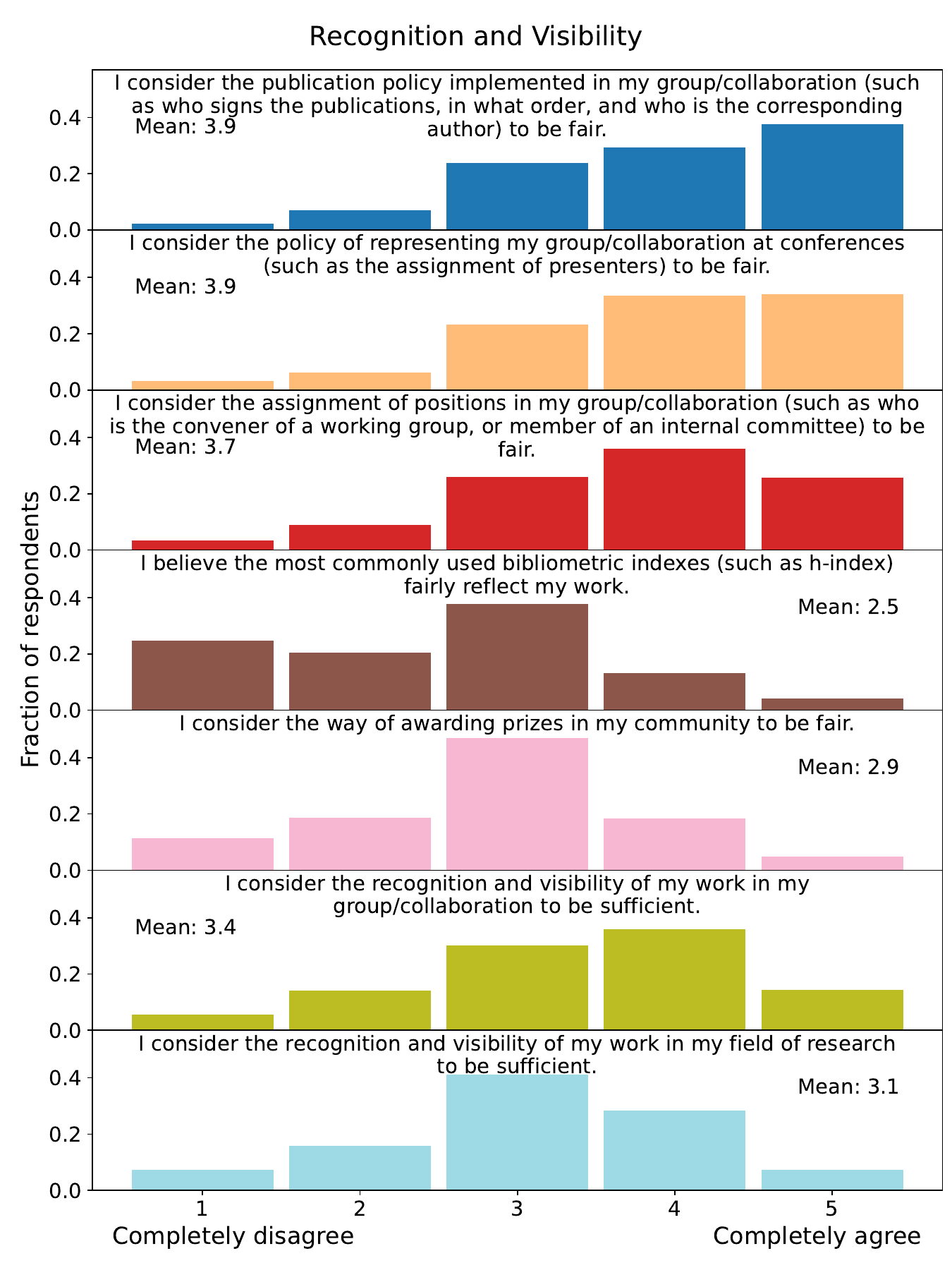}
    \caption{(Q89--95) Respondents' feelings about recognition and visibility of their work. Fractions are given out of all respondents who answered the questions.}
    \label{fig:part1:FairnessRecognition}
\end{figure}

In Figure~\ref{fig:part1:FairnessRecognition}, we asked respondents for their level of agreement with statements regarding the recognition and visibility they receive. Only the questions regarding bibliometric indexes and awarding prizes were mandatory. 
In the top two panels it is shown that a clear majority of respondents perceive their group/collaboration's policies on publications and conferences to be fair.
Respondents also agree, though less strongly, that the assignment of positions within their group/collaboration is fair.
The respondents' opinion on commonly used bibliometric indices (such as the h-index) is more bimodal, with most either strongly disagreeing or feeling unsure that these fairly reflect work done.
Respondents also don't have a strong overall opinion about whether the way prizes are awarded in their community is fair.
Finally, respondents generally agree (though not strongly) that recognition and visibility of their work at the research group/collaboration level is fair, but are more neutral at research field level.

%%%%%%%%%%%%%%%%%%%%%%%%%%%%%%%%%%%%%%%%%%%%%%%%%%%%%%%%%%%%%%%%%%%%%%%%%%%%%%%%%%%%%%%%%%%%%%%%%%%%%%%%%%%%%%%
\FloatBarrier
\subsection{Open questions}

In the last part of the survey, 3 open questions were asked. 
Two of these were used in conjunction with the rest of the analysis, to produce a set of recommendations for the Panel and the particle physics community, which are discussed in Section~\ref{sec:conclusion}.

\begin{figure}
    \centering
        \includegraphics[width=0.7\textwidth]{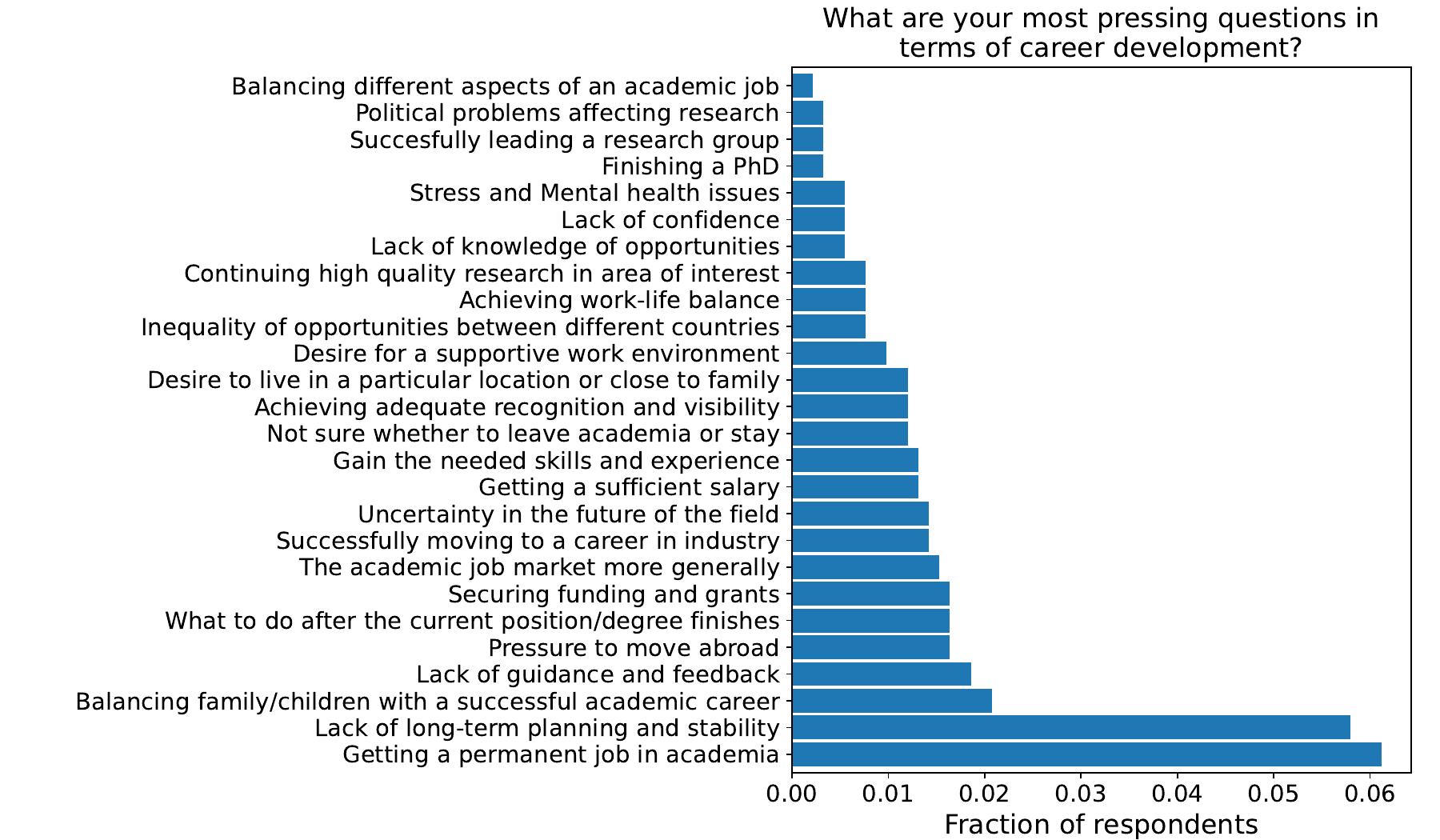}
    \caption{(Q97) The categories of pressing questions respondents have about their careers. The fraction given is out of all respondents, empty responses are not shown. Multiple answers could be selected per respondent.}
    \label{fig:part1:Q97}
\end{figure}

The third open-form question asked for respondents' more pressing questions about their careers.
The responses, allowing for multiple per persons, were grouped into categories and are presented in Figure~\ref{fig:part1:Q97}.
Consistent to what was seen in Figure~\ref{fig:part1:Q88}, the most pressing questions related to getting a permanent job in academia, and the lack of long-term planning and stability.

%%%%%%%%%%%%%%%%%%%%%%%%%%%%%%%%%%%%%%%%%%%%%%%%%%%%%%%%%%%%%%%%%%%%%%%%%%%%%%%%%%%%%%%%%%%%%%%%%%%%%%%%%%%%%%%
\FloatBarrier
\section{Correlations between questions}
\label{sec:correlations}

For this section we considered correlations between the responses given to different questions in the survey.
We considered both the demographics of respondents (questions 1-14 as defined in Appendix~\ref{app:questions}), and other questions.
While we acknowledge the potential existence of interesting correlations not shown here, we had to make choices in what it was feasible to study.
We present a subset of plots which we found to be interesting, or to yield a strong correlation.
We also remind the reader that despite a large number of responses to our survey overall, when considering certain categories the sample size is limited.
One should avoid drawing conclusions which could just be due to statistical fluctuations or a biased dataset from those who chose to respond to the survey.
Thus, we have also focused our observations on patterns observed in categories with a higher sample size.

%%%%%%%%%%%%%===================================================================================================
\FloatBarrier
\subsection{Correlations between respondent demographics}

\begin{figure}[h!]
    \centering
        \includegraphics[width=0.9\textwidth]{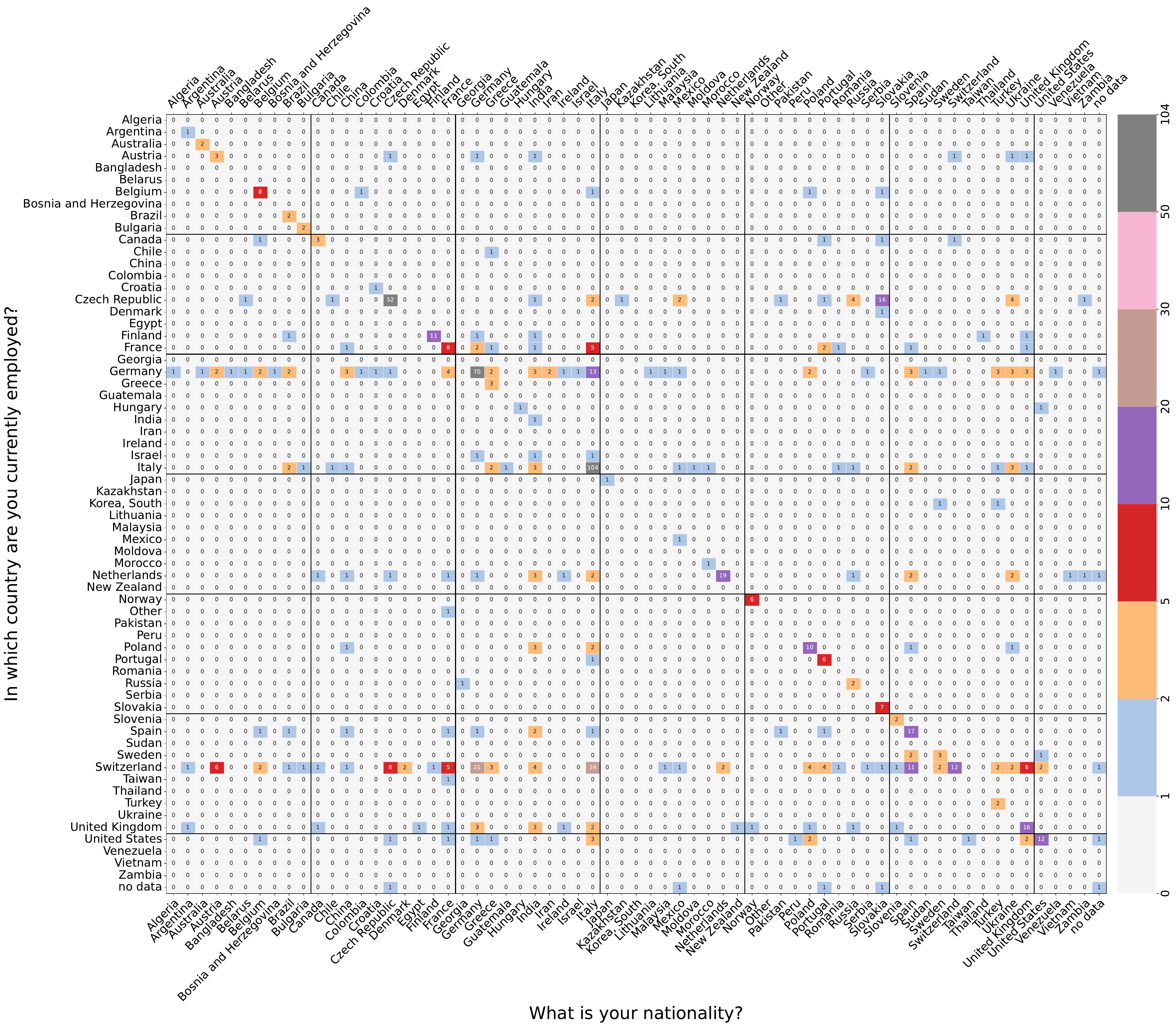}
    \caption{(Q4 v Q6) Correlations between respondents' current country of employment and nationality.}
    \label{fig:part2:Q4vQ6_full}
\end{figure}

\begin{figure}[h!]
    \centering
        \includegraphics[width=0.7\textwidth]{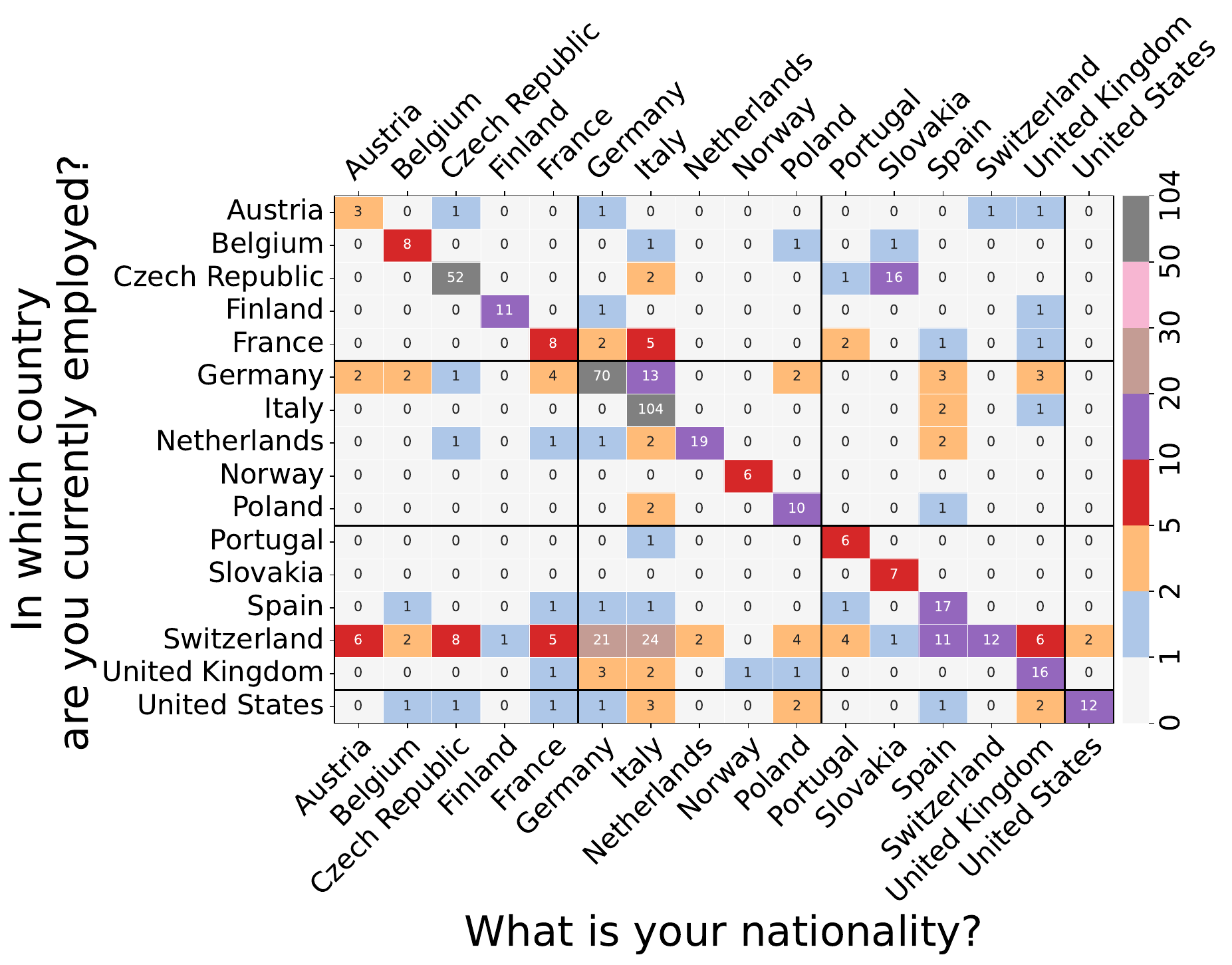}
    \caption{(Q4 v Q6) Correlations between respondents' current country of employment and nationality, for categories with at least 5 responses.}
    \label{fig:part2:Q4vQ6_5}
\end{figure}

To begin, correlations between respondents' nationality, country of residence, and country of employment were investigated in two-dimensional plots.
In Figure~\ref{fig:part2:Q4vQ6_full}, the place of work and the nationality are shown in full, and in Figure~\ref{fig:part2:Q4vQ6_5}, the rows and columns containing more than 5 answers are considered.
The plots highlights that the nationality of the respondent doesn't necessarily match the place of work.
For example, Switzerland and Germany show a more diverse set of nationalities than Italy and Spain. 
We expect that the trends observed in Switzerland and France in this section are largely due to respondents who are based at CERN.

\begin{figure}[h!]
    \centering
        \includegraphics[width=0.9\textwidth]{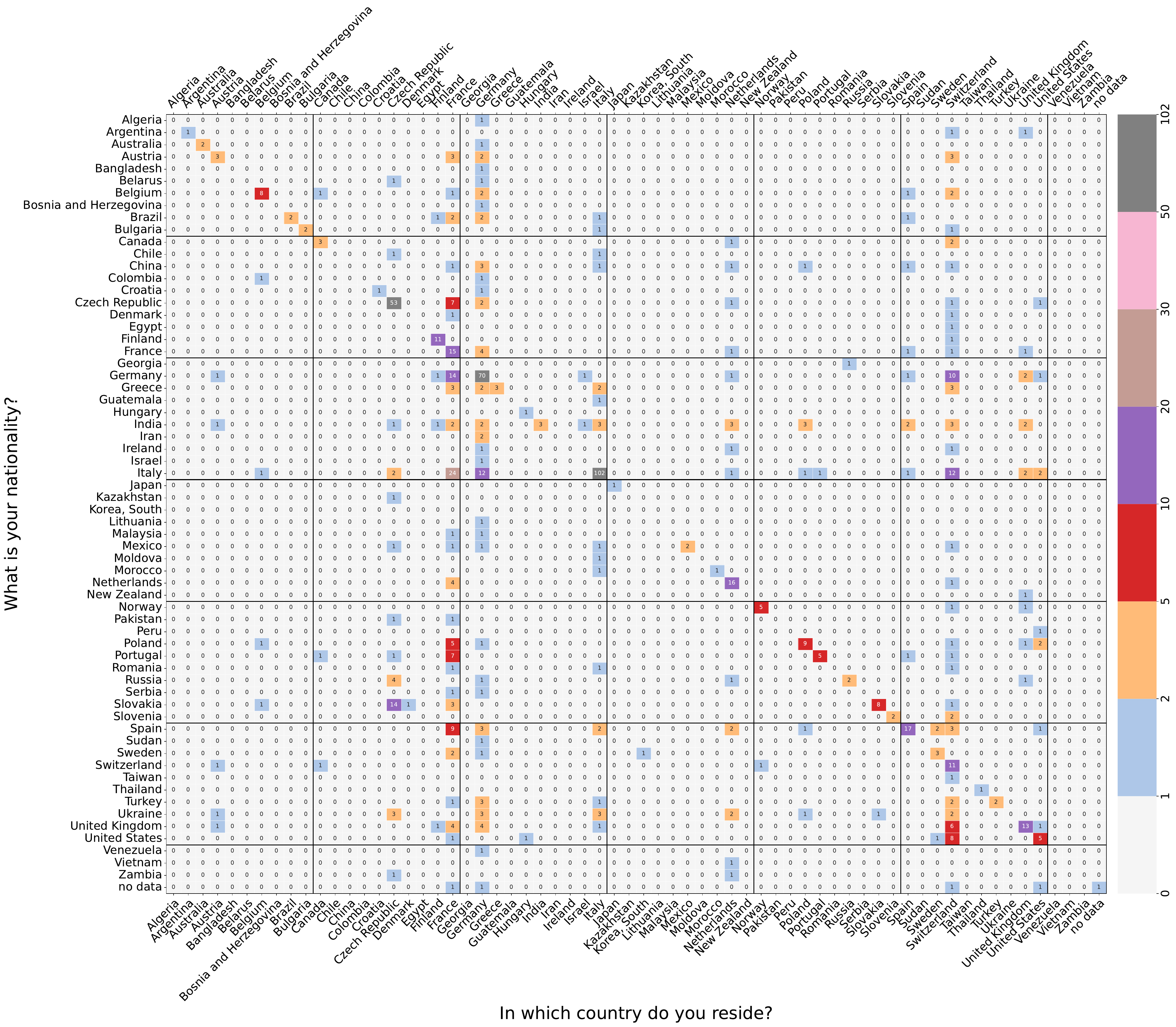}
    \caption{(Q6 v Q5) Correlations between respondents' current country of residence and nationality.}
    \label{fig:part2:Q6vQ5_full}
\end{figure}

\begin{figure}[h!]
    \centering
        \includegraphics[width=0.7\textwidth]{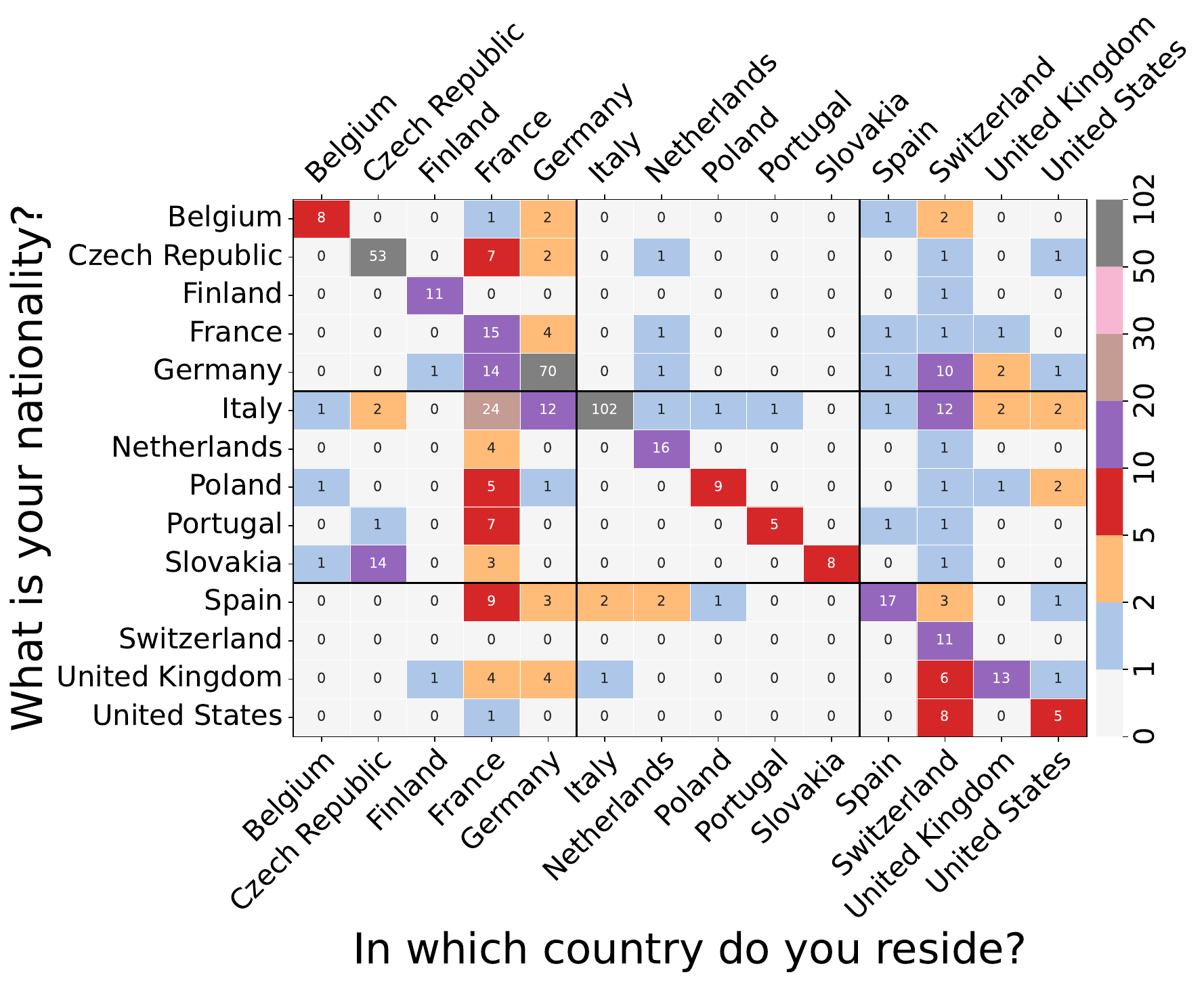}
    \caption{(Q6 v Q5) Correlations between respondents' current country of residence and nationality, for categories with at least 5 responses.}
    \label{fig:part2:Q6vQ5_5}
\end{figure}

In Figures~\ref{fig:part2:Q6vQ5_full}--\ref{fig:part2:Q6vQ5_5}, the place of residence and the nationality are studied.
We see the largest spread in nationalities for those residing in Switzerland, Germany, and France.
We also see the largest spread in countries of residence for Indian and Italian respondents.

\begin{figure}[h!]
    \centering
        \includegraphics[width=0.9\textwidth]{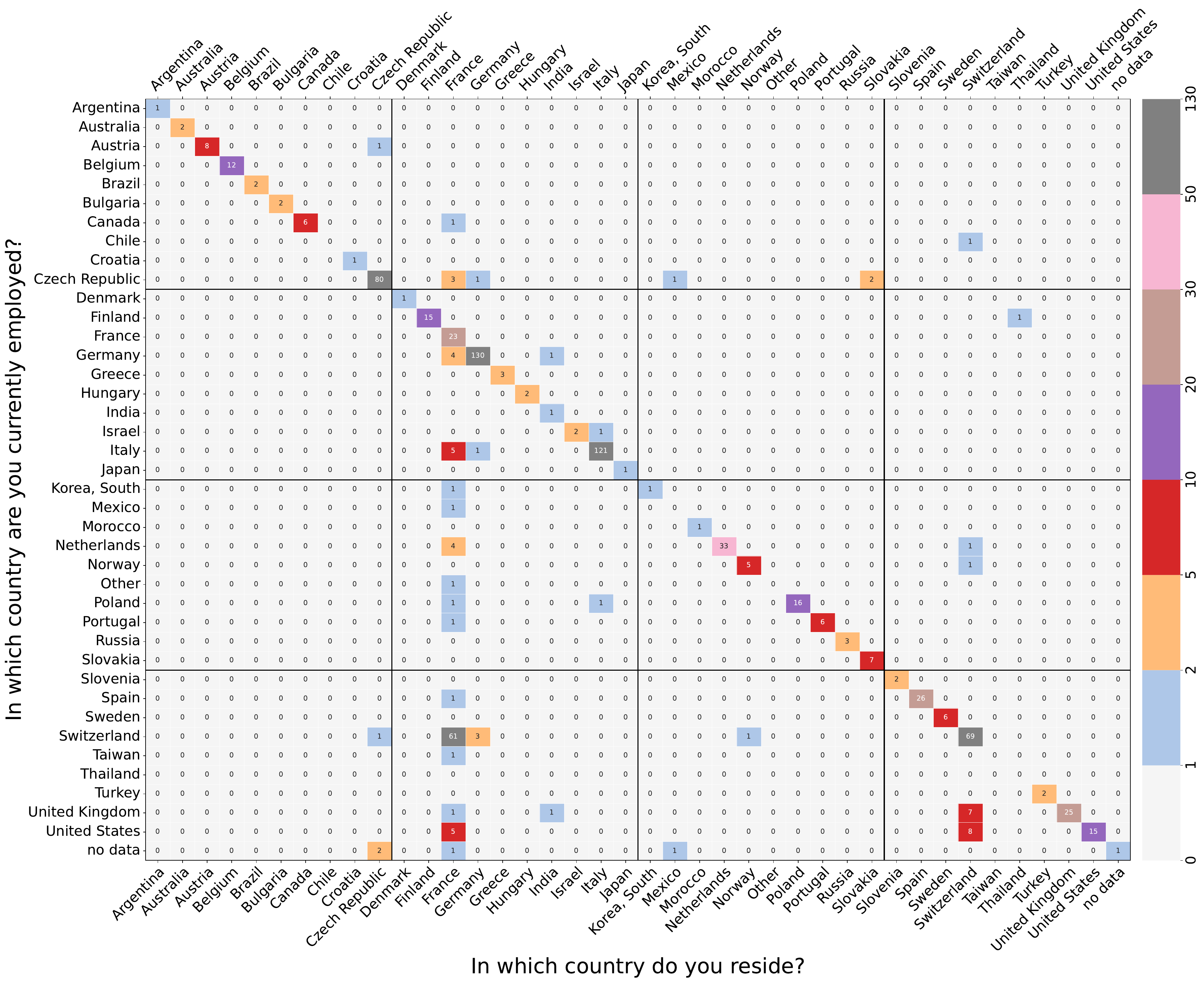}
    \caption{(Q4 v Q5) Correlations between respondents' current countries of employment and residence.}
    \label{fig:part2:Q4vQ5_full}
\end{figure}

\begin{figure}[h!]
    \centering
        \includegraphics[width=0.7\textwidth]{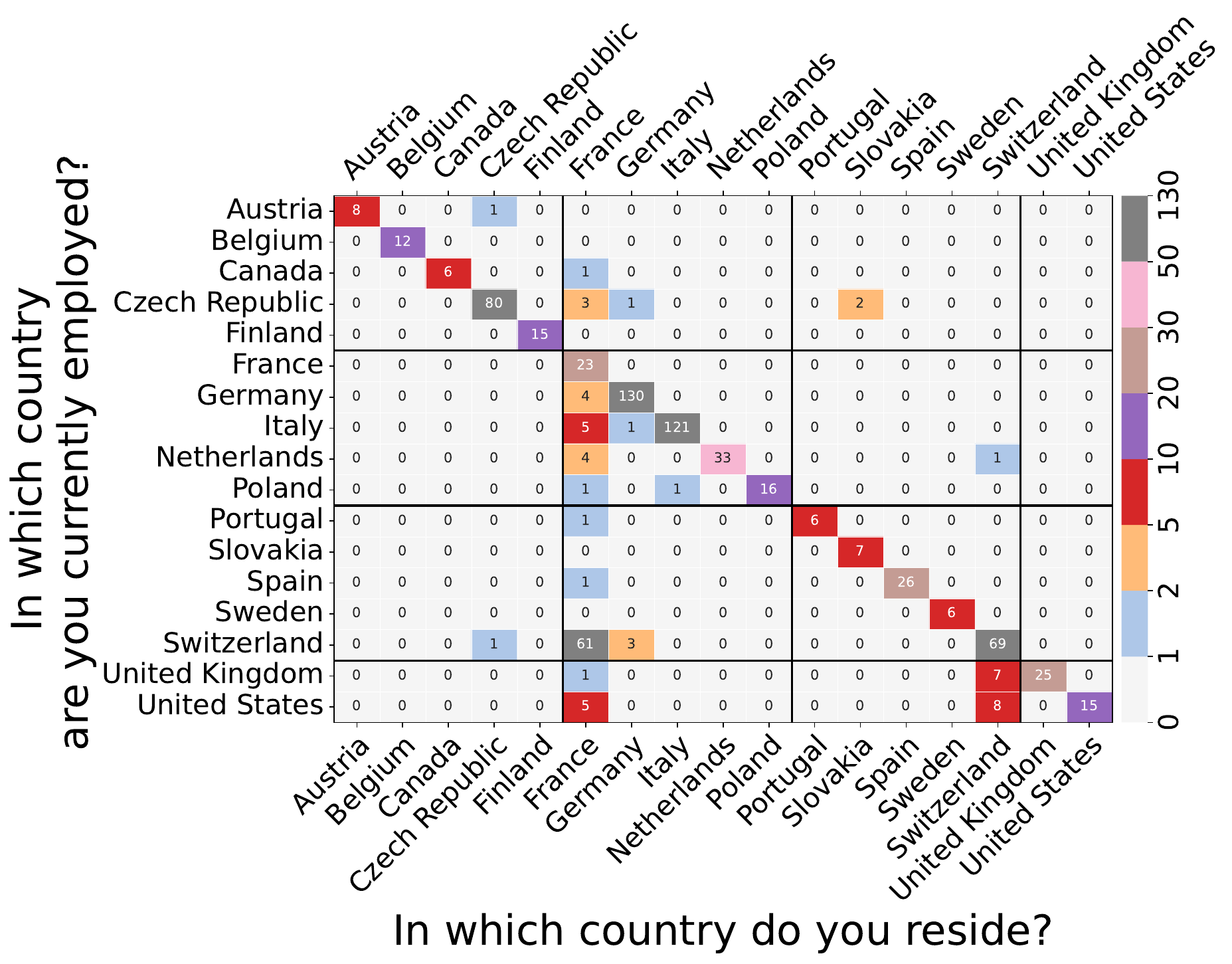}
    \caption{(Q4 v Q5) Correlations between respondents' current countries of employment and residence, for categories with at least 5 responses.}
    \label{fig:part2:Q4vQ5_5}
\end{figure}

In Figures~\ref{fig:part2:Q4vQ5_full}--\ref{fig:part2:Q4vQ5_5}, countries of residence and employment are studied.
This highlights the fact that most of the respondents reside in the country in which they are employed, except for a couple of countries.
61 of respondents residing in France are employed in Switzerland, whilst only 23 respondents are employed in France. 31 respondents reside in France but are employed in another country (apart from Switzerland).
For respondents employed by the US, a similar number reside in France or Switzerland as reside in the US. 

\begin{figure}[ht!]
    \centering
        \subfloat[]{\label{fig:part2:Q1vQ4}\includegraphics[width=0.49\textwidth]{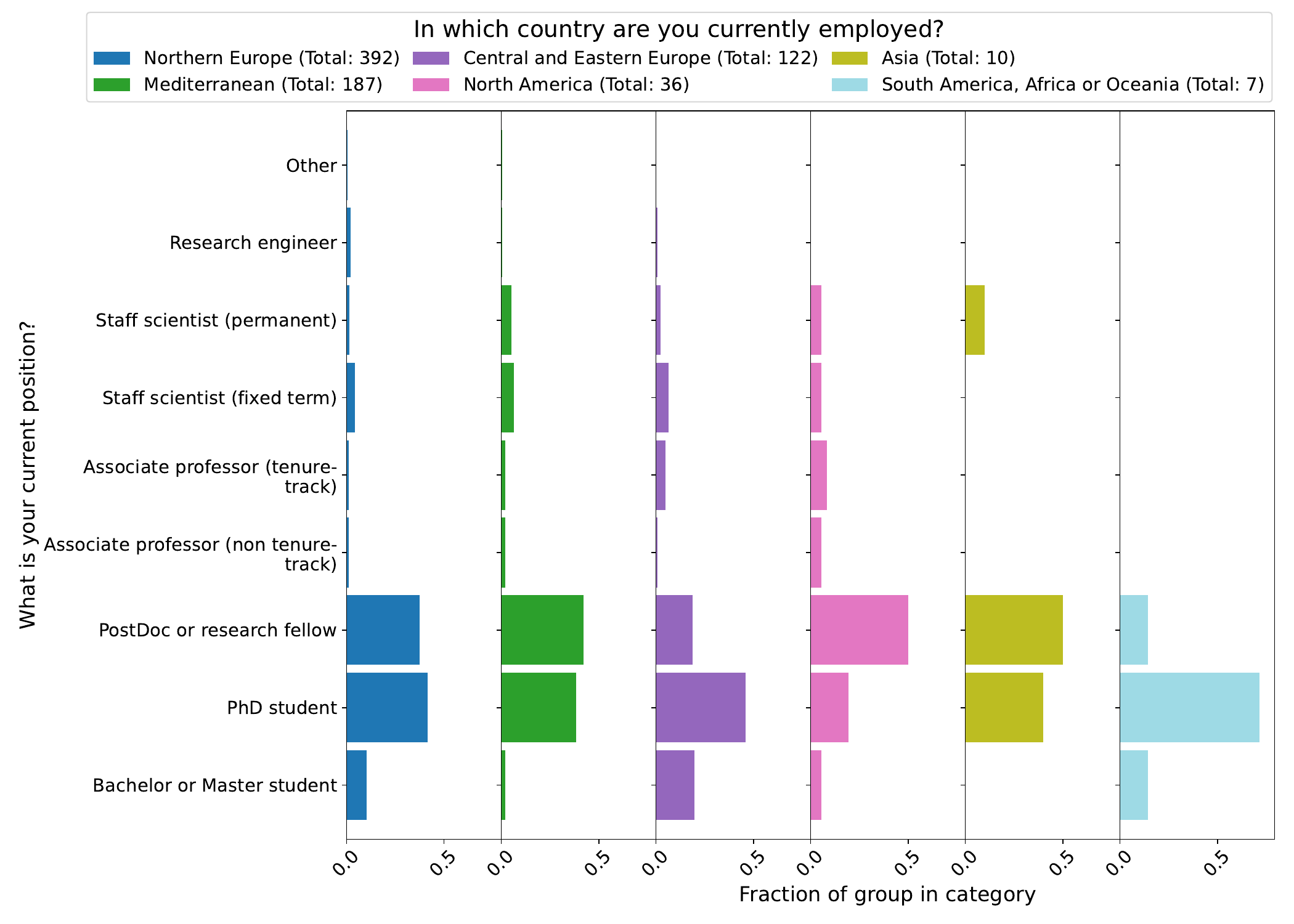}}
        \subfloat[]{\label{fig:part2:Q1vQ8}\includegraphics[width=0.49\textwidth]{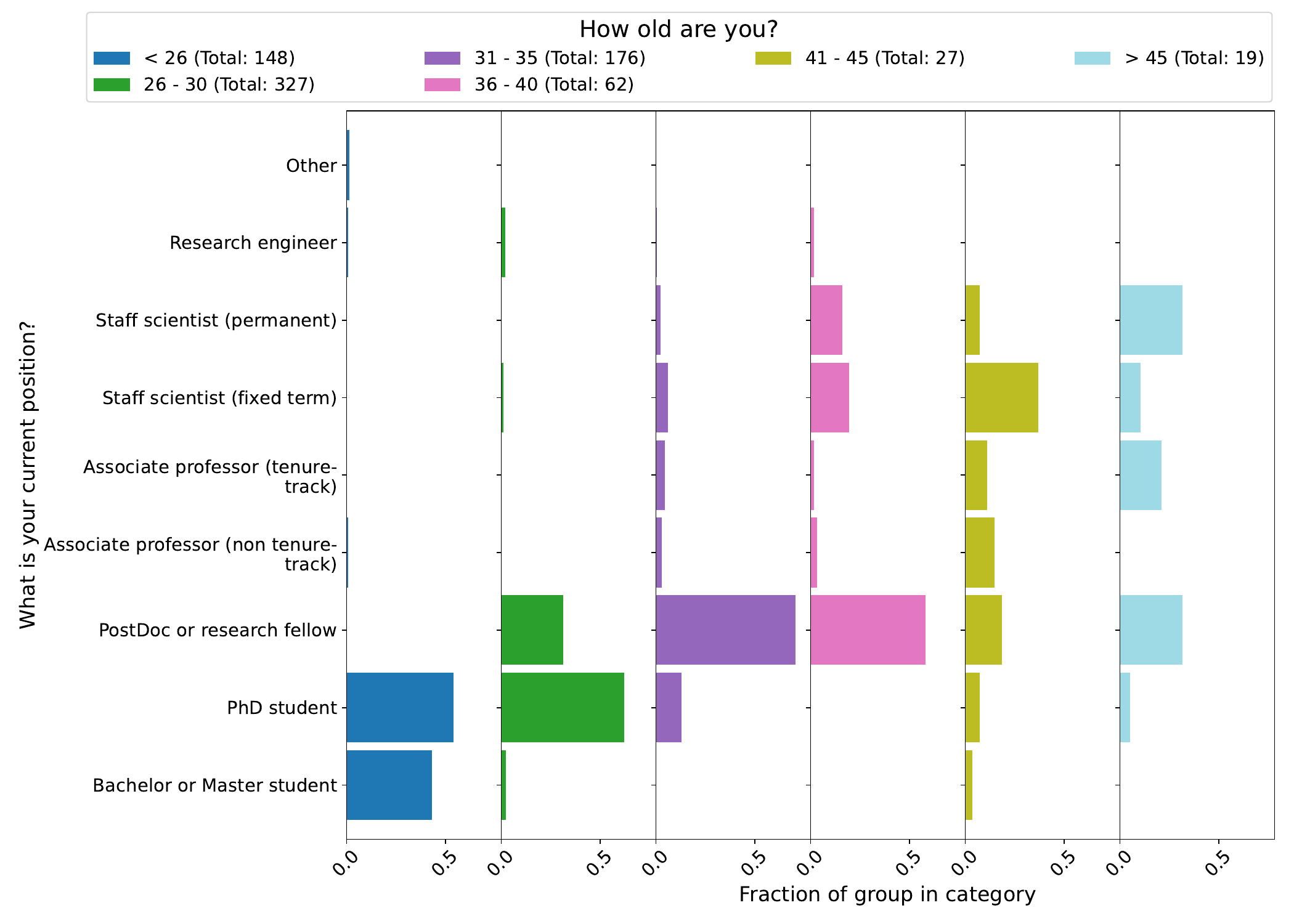}}
    \caption{(Q1 v Q4,8) Correlations between respondents' current position and age or country of employment. Fractions are given out of all respondents, or for all respondents who answer the question in (a).}
    \label{fig:part2:Q1vQ4Q8}
\end{figure}

\FloatBarrier
\pagebreak
Following this, we studied the correlations between other aspects of respondents' demographics.
In Figure~\ref{fig:part2:Q1vQ4Q8}, the correlations between the respondents' current position, age and country of employment are investigated.
Regarding country, there might be in principle a correlation with position reflecting the differences in the numbers of positions available, employment conditions, available career paths and time taken to reach a permanent position. However, due to the finite sample size, trends seen in Figure~\ref{fig:part2:Q1vQ4Q8} are more likely to reflect the sample of people who chose to respond to the survey.
Predictably, there is a strong positive correlation between age and position.
However, above age 30, positions are slightly more distributed amongst ages, with a large fraction in the PostDoc or research fellow category, even for respondents above age 45.

\begin{figure}[ht!]
    \centering
        \subfloat[]{\label{fig:Q3vQ1}\includegraphics[width=0.49\textwidth]{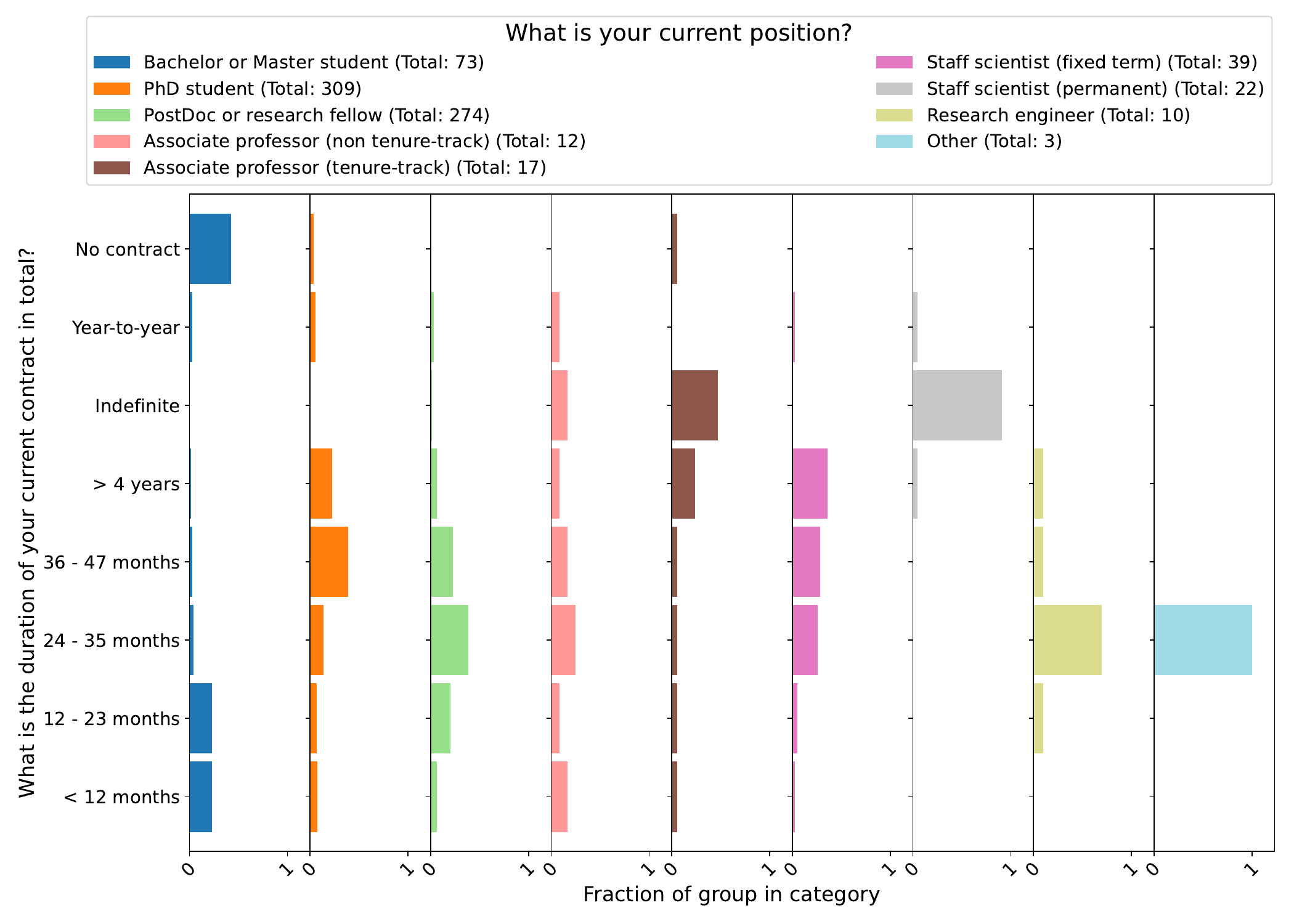}}
        \subfloat[]{\label{fig:Q3vQ2}\includegraphics[width=0.49\textwidth]{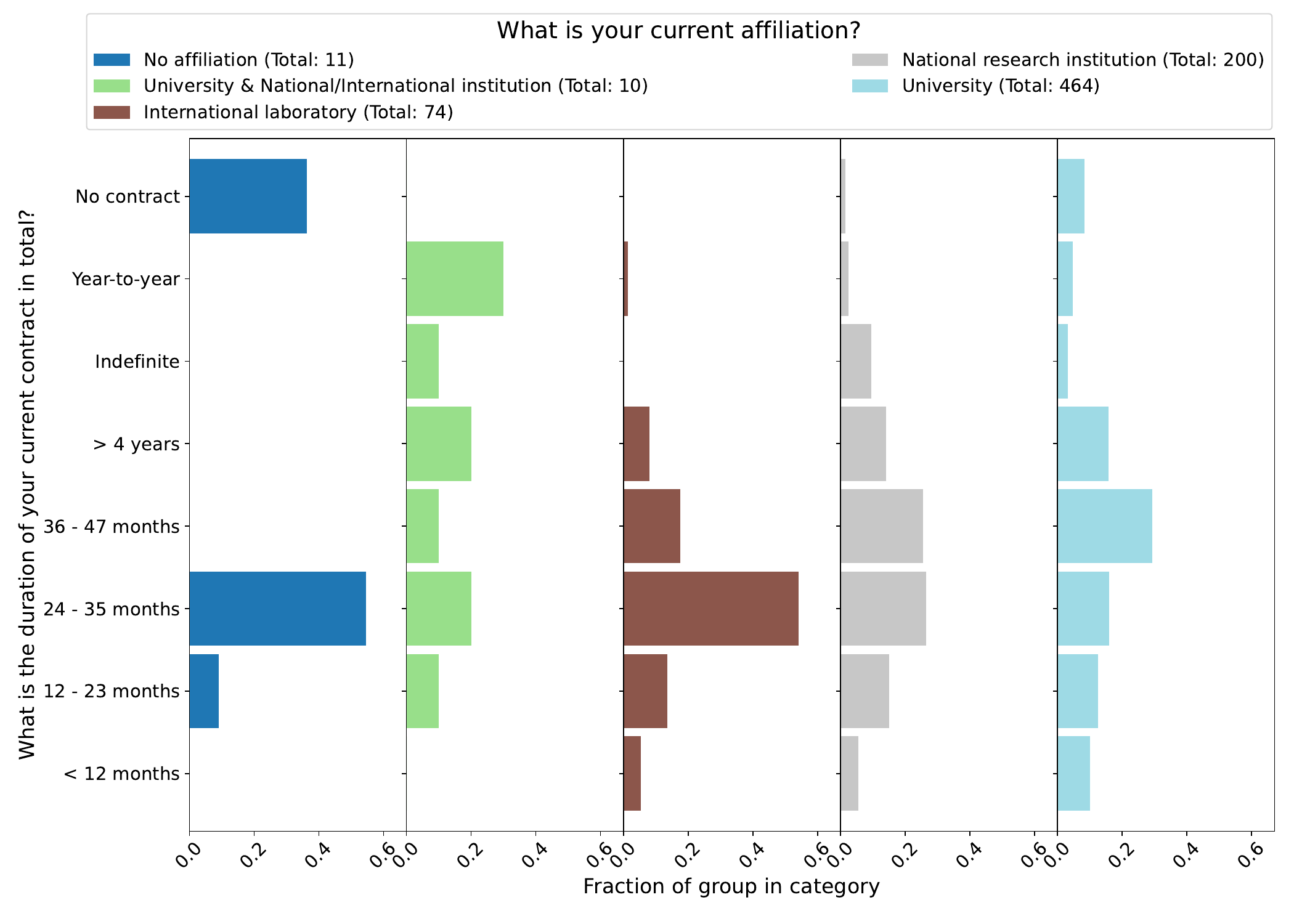}}
    \caption{(Q3 v Q1--2) Correlations between duration of respondents' contracts and their current position or affiliation. Fractions are given out of all respondents.}
    \label{fig:part2:Q3vQ1Q2}
\end{figure}

In Figure~\ref{fig:part2:Q3vQ1Q2}, the correlations between the respondents' current position, affiliation and duration of contract are displayed.
Focus on respondents who do not have a contract, in addition to undergraduate students, we found that a small fraction of PhD students and even associate professors (tenure track) do not have a contract.
Respondents who do not have a contract primarily also do not have an affiliation.

\begin{figure}[ht!]
    \centering
        \subfloat[]{\label{fig:part2:Q9vQ7}\includegraphics[width=0.49\textwidth]{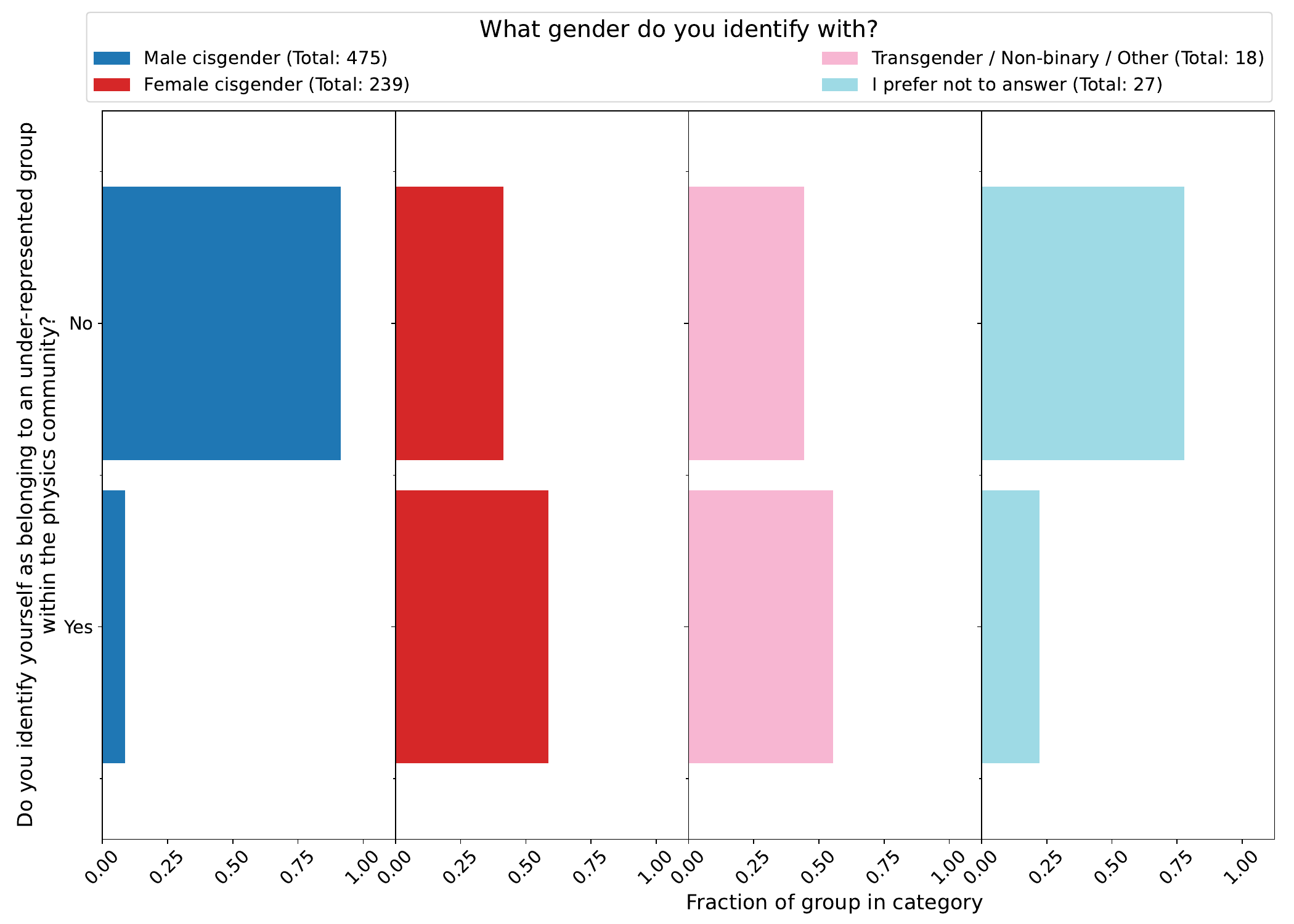}}
        \subfloat[]{\label{fig:part2:Q10vQ7}\includegraphics[width=0.49\textwidth]{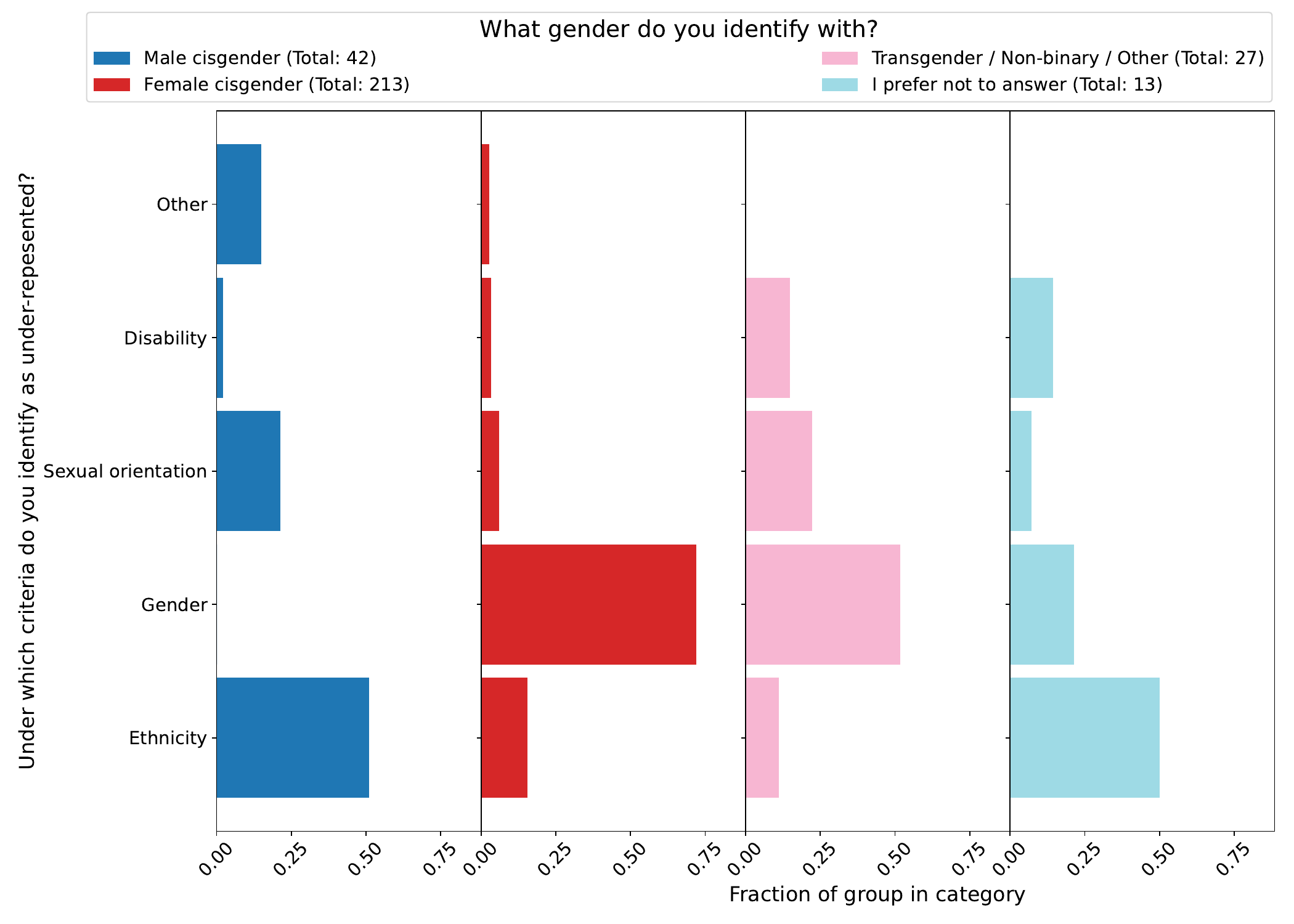}}
    \caption{(Q9--10 v Q7) Correlations between respondents' gender and (a) belonging to an (b) category of (with multiple answers allowed per respondent), under-represented group. Fractions are given out of all respondents, or for all respondents who answer the question in (b).}
    \label{fig:part2:Q9Q10vQ7}
\end{figure}

In Figure~\ref{fig:part2:Q9vQ7}, the correlations between respondents' gender and whether they belonging to an under-represented group are shown.
Most of the respondents who identify themselves as male cisgender feel they don't belong to an under-represented group.
On the other hand, just over half of the respondents who identify themselves as female cisgender, transgender, non binary or other answered that they do.

To understand more the differences between the gender categories and how they feel under-represented, the correlations between the respondents' gender and category of under representation are examined in Figure~\ref{fig:part2:Q10vQ7}.
For the few cis-gendered male respondents who consider themselves as under-represented, this is dominantly due to ethnicity and never due to gender.
On the other hand gender is the dominant choice for other respondents.

\begin{figure}[ht!]
    \centering
        \subfloat[]{\label{fig:part2:Q7vQ6}\includegraphics[width=0.49\textwidth]{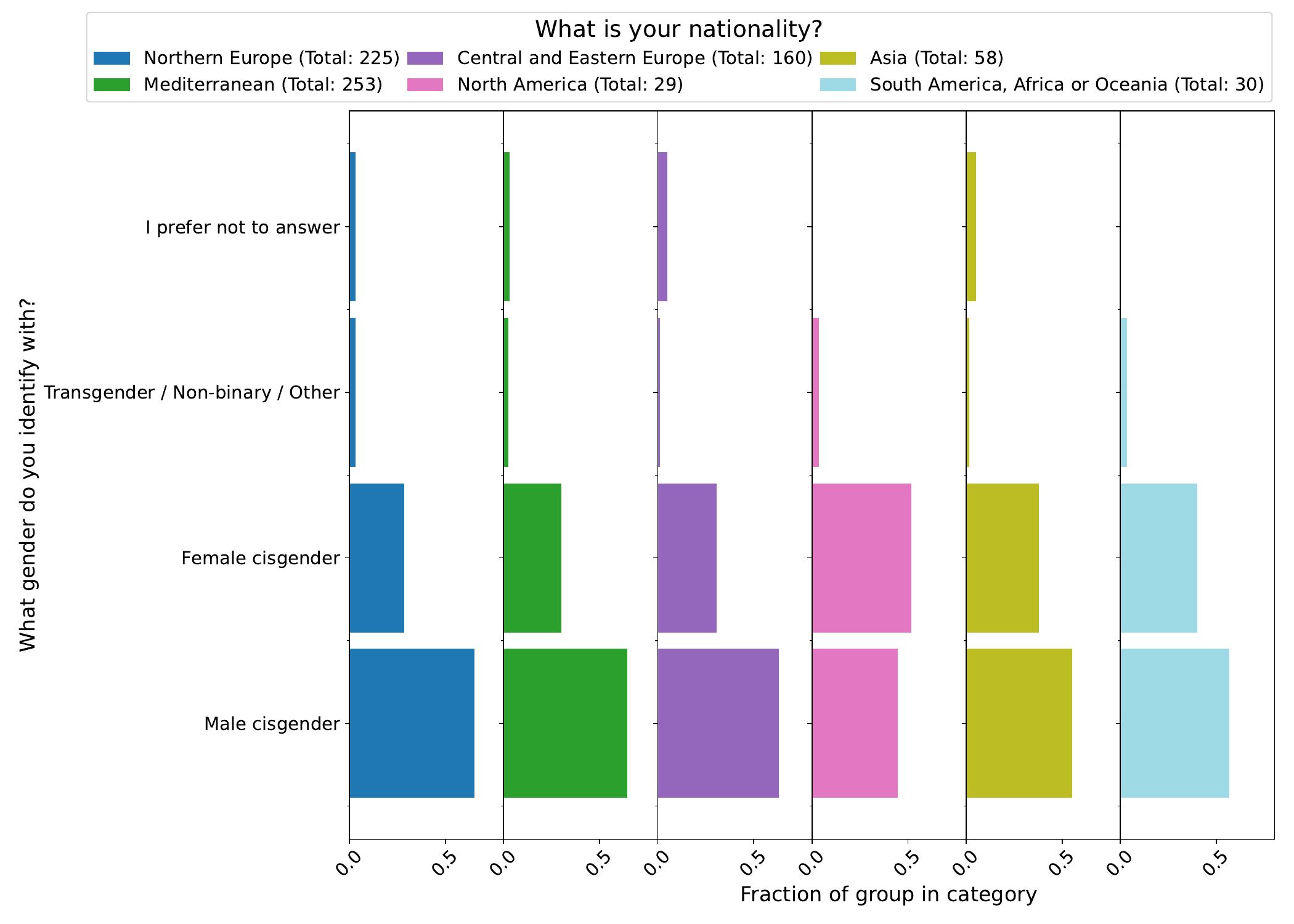}}
        \subfloat[]{\label{fig:part2:Q10vQ6}\includegraphics[width=0.49\textwidth]{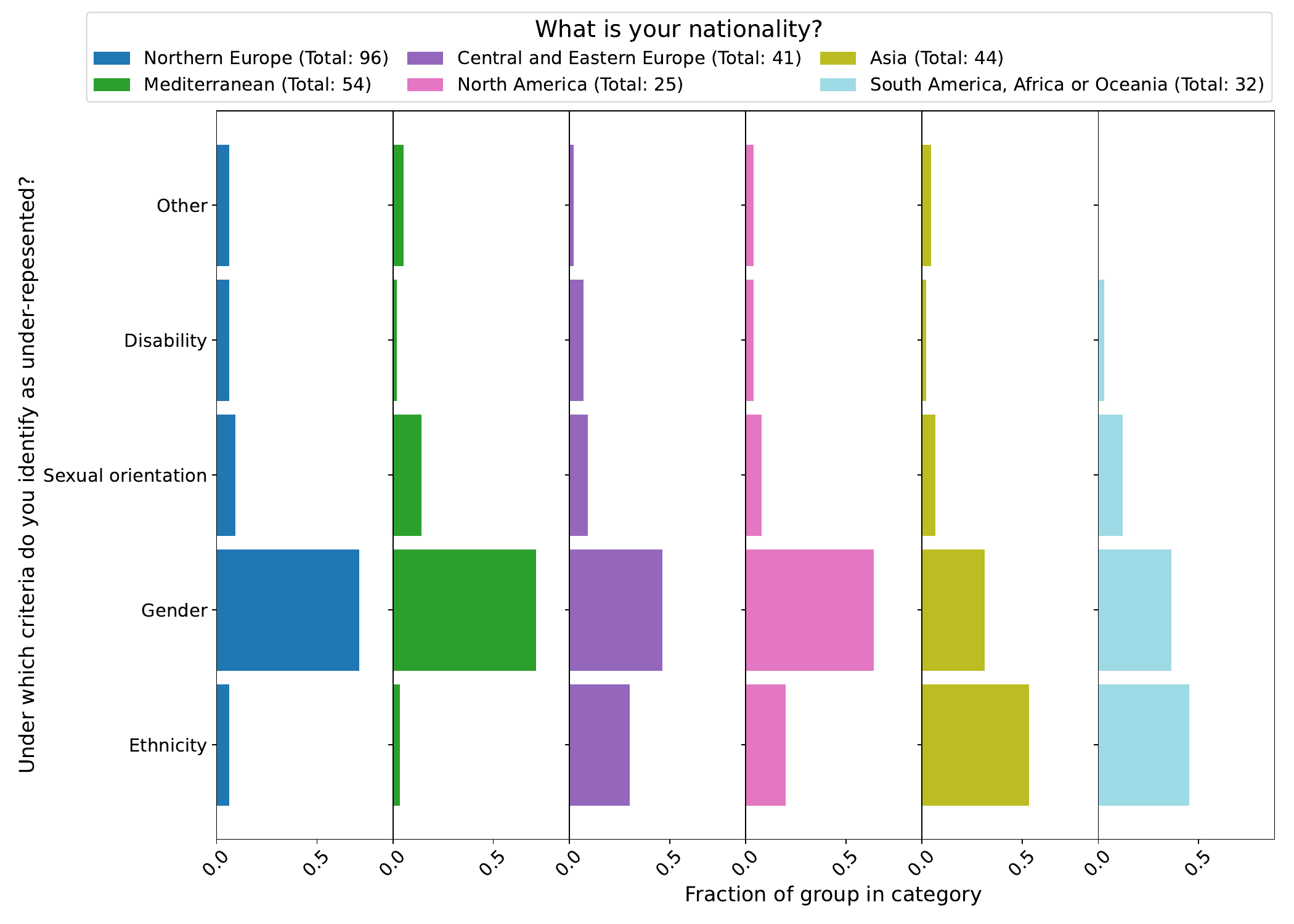}}
    \caption{(Q7,10 v Q6) Correlations between respondents' nationality and (a) gender and (b) belonging to a category of under-represented group (multiple answers allowed per respondent). Fractions are given out of all respondents, or for all respondents who answer the question in (b)}
    \label{fig:part2:Q7Q10vQ6}
\end{figure}

This topic was investigated further by considering correlations between the respondents' nationality and category of under representation, shown in Figure~\ref{fig:part2:Q7Q10vQ6}.
Gender is the most common criteria given for European and North American respondents (we note that the proportion of cis-gender female respondents is also higher for these nationalities).
On the other hand, ethnicity is the most common criteria for the respondents who are Asian, South American, African or Oceanian.

\begin{figure}[ht!]
    \centering
        \subfloat[]{\label{fig:part2:Q11vQ1}\includegraphics[width=0.49\textwidth]{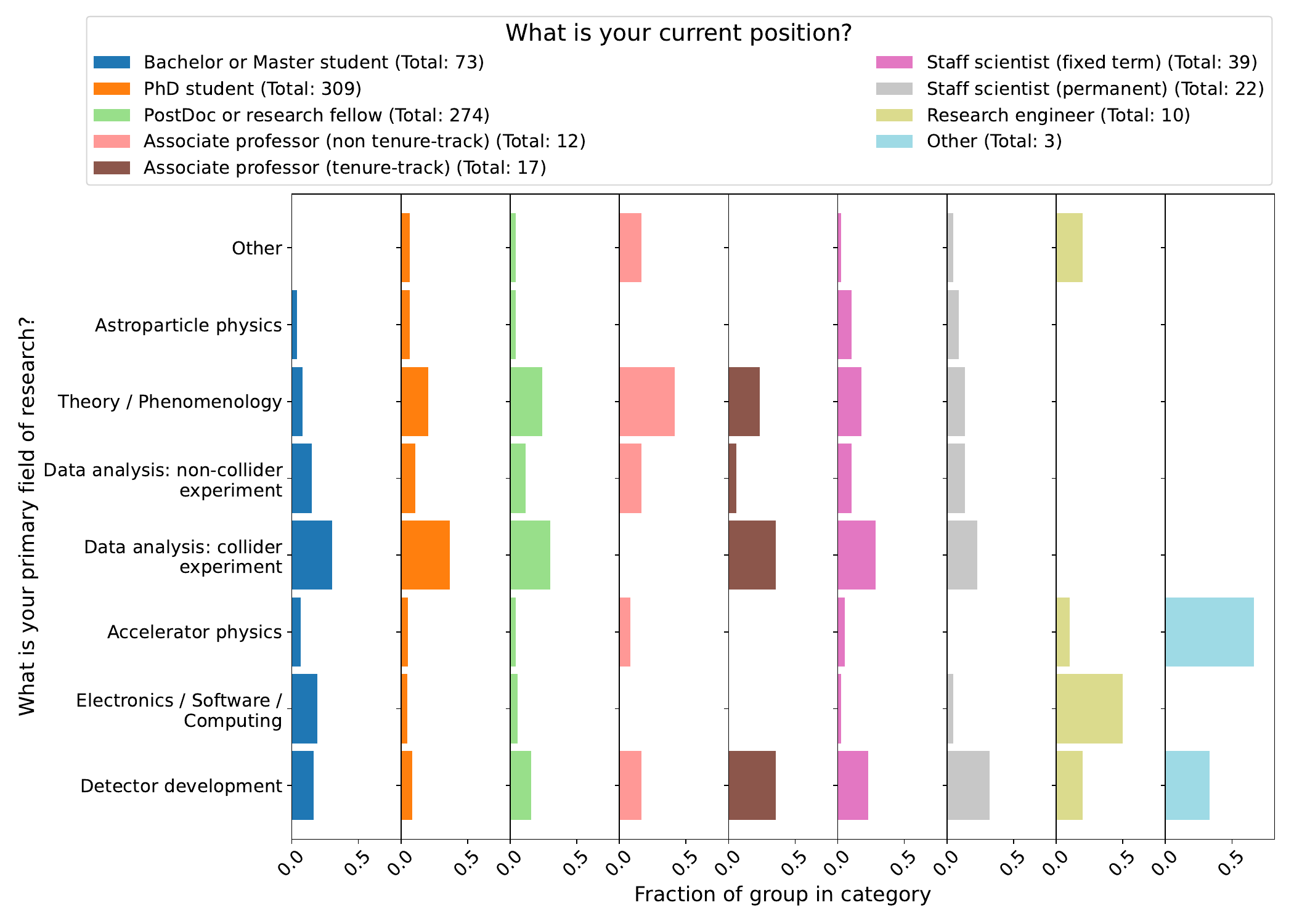}}
        \subfloat[]{\label{fig:part2:Q11vQ4}\includegraphics[width=0.49\textwidth]{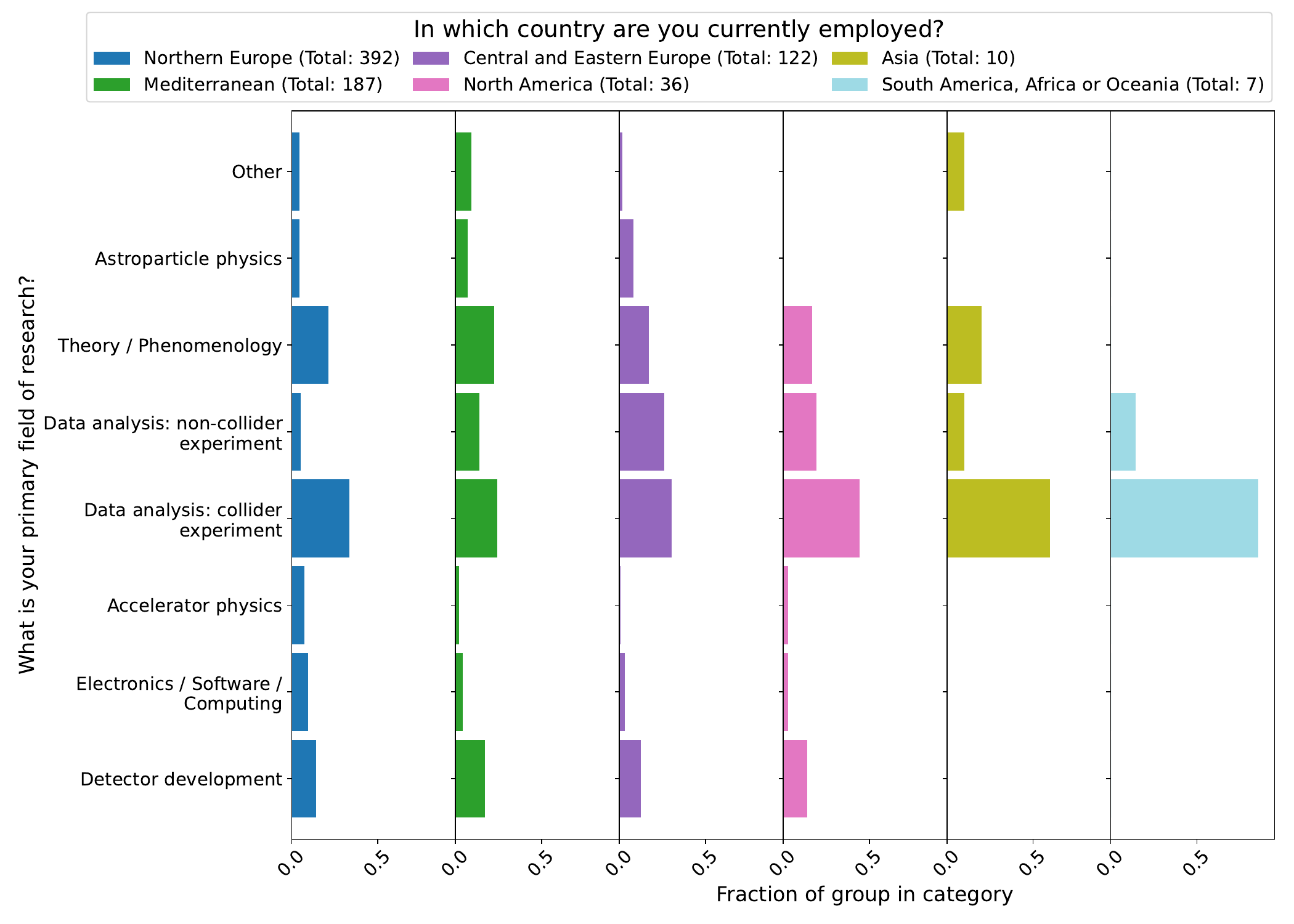}}\\
        \subfloat[]{\label{fig:part2:Q11vQ7}\includegraphics[width=0.49\textwidth]{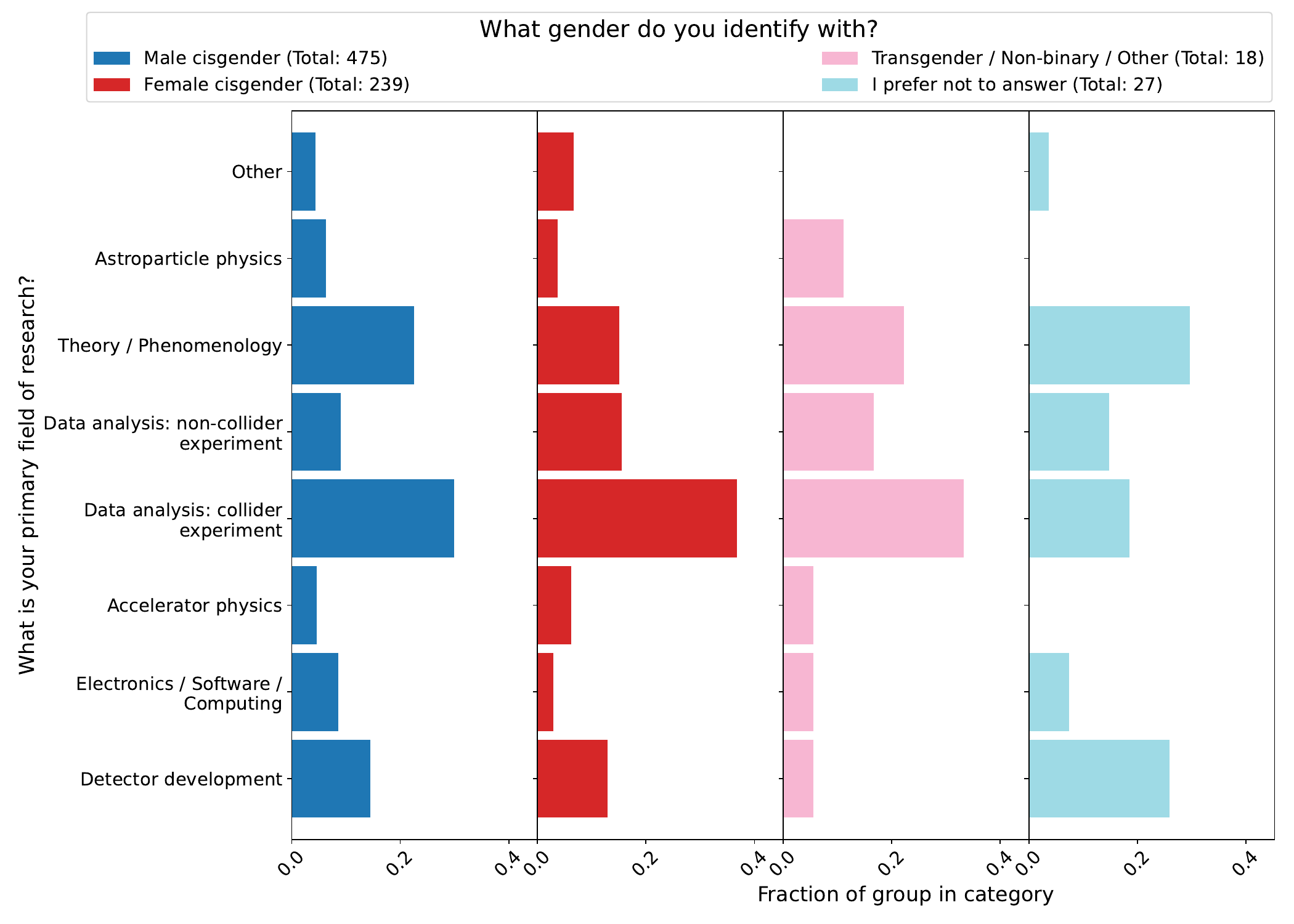}}
        \subfloat[]{\label{fig:part2:Q12vQ11}\includegraphics[width=0.49\textwidth]{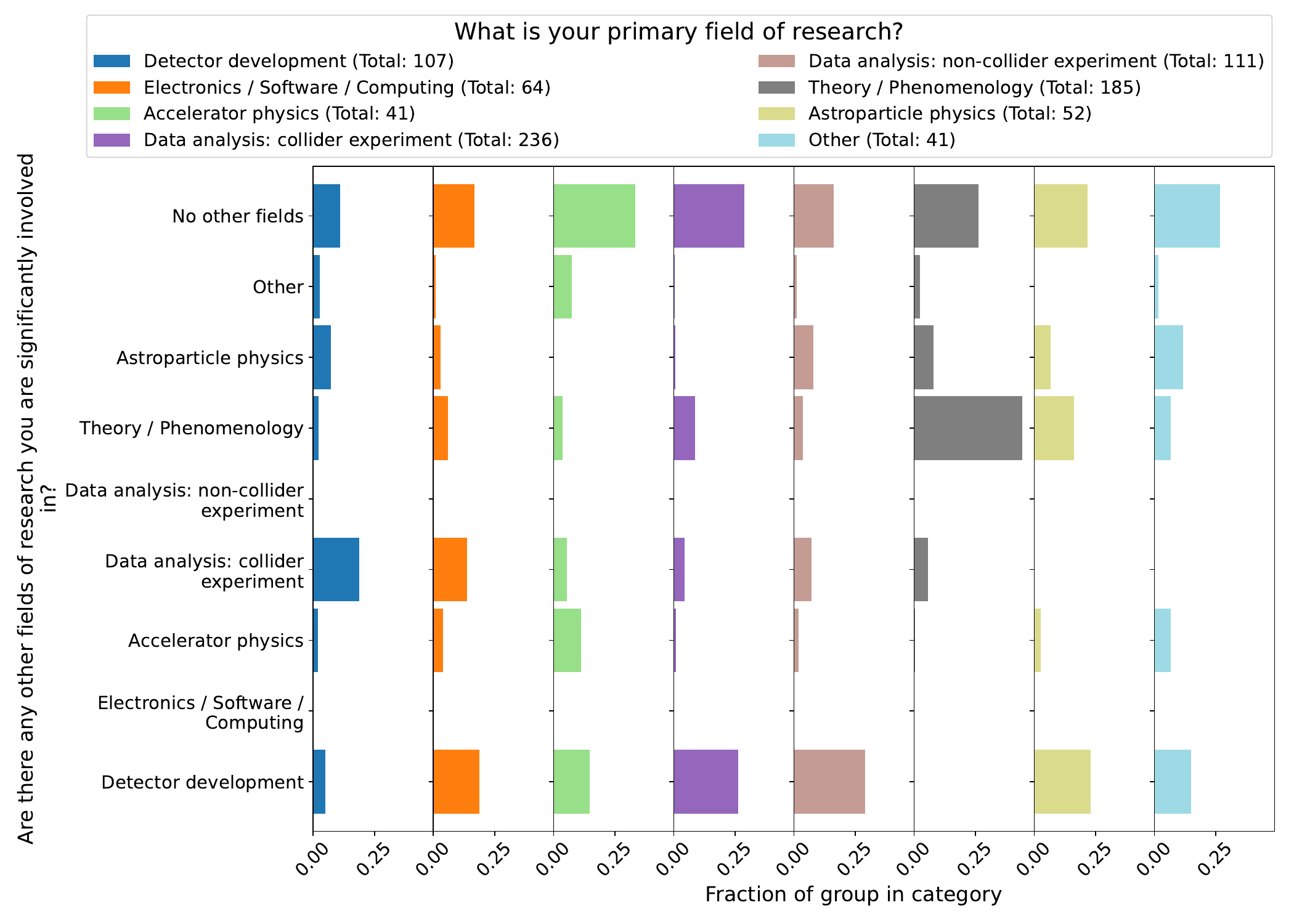}}
    \caption{(Q11 v Q1,4,7; Q12 v Q11) Further examples of correlations between respondent demographics. When considering secondary fields of research, multiple answers were allowed per respondent. Fractions are given out of all respondents, or for all respondents who answer the question in (b).}
    \label{fig:part2:Q11vQ1Q4Q7_Q12vQ11}
\end{figure}

Finally, to better understand the sample of people who responded to the survey and provide extra context for the results we show some more example correlations between respondent demographics in Figure~\ref{fig:part2:Q11vQ1Q4Q7_Q12vQ11}.

\clearpage

%%%%===========================================================================================
\subsection{Work within a research group or collaboration}

%----------------------------------------------------------------------------------------
\subsubsection{Research groups}

\begin{figure}[b!]
    \centering
        \subfloat[]{\label{fig:part2:Q80vQ15}\includegraphics[width=0.49\textwidth]{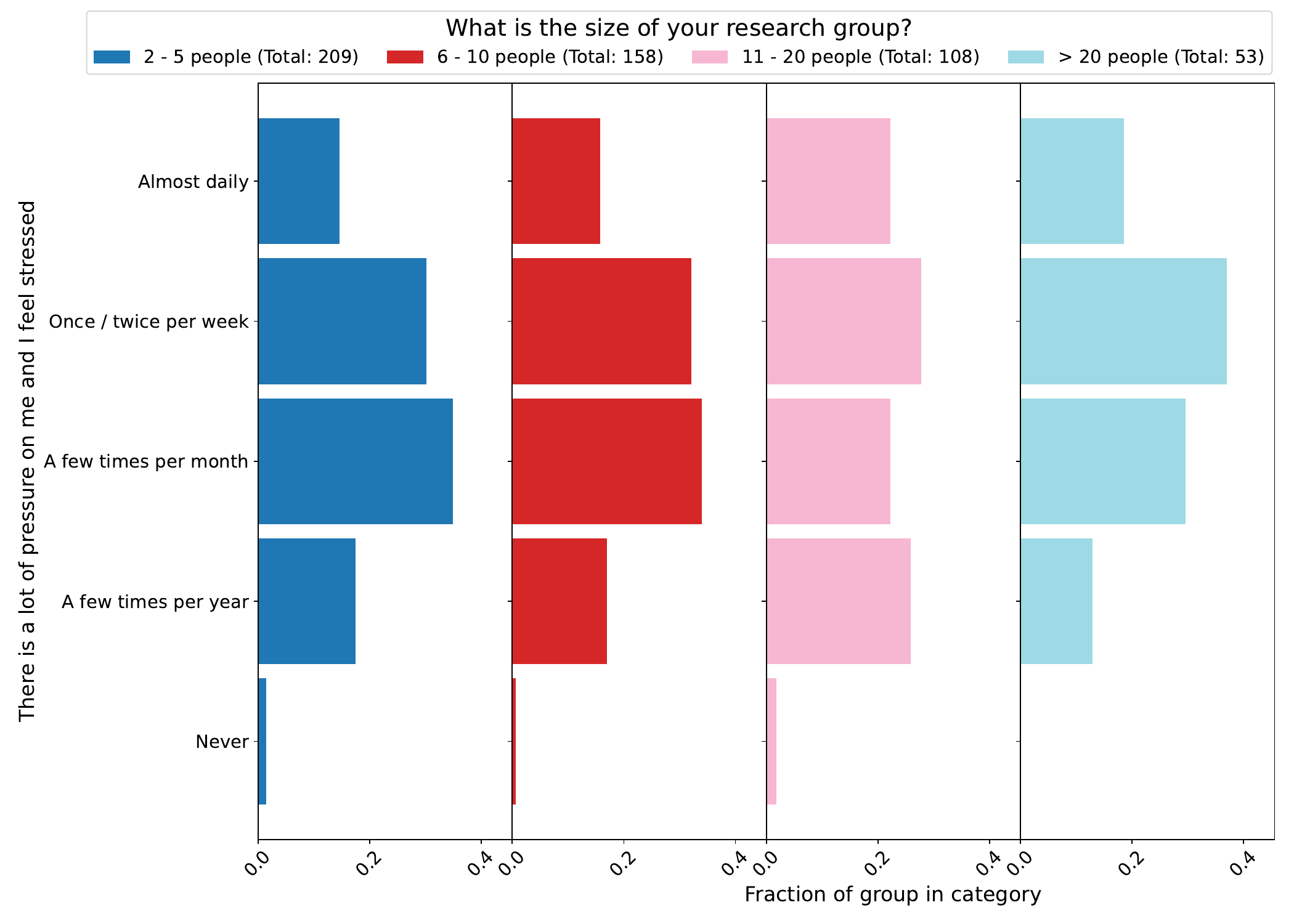}}
        \subfloat[]{\label{fig:part2:Q94vQ15}\includegraphics[width=0.49\textwidth]{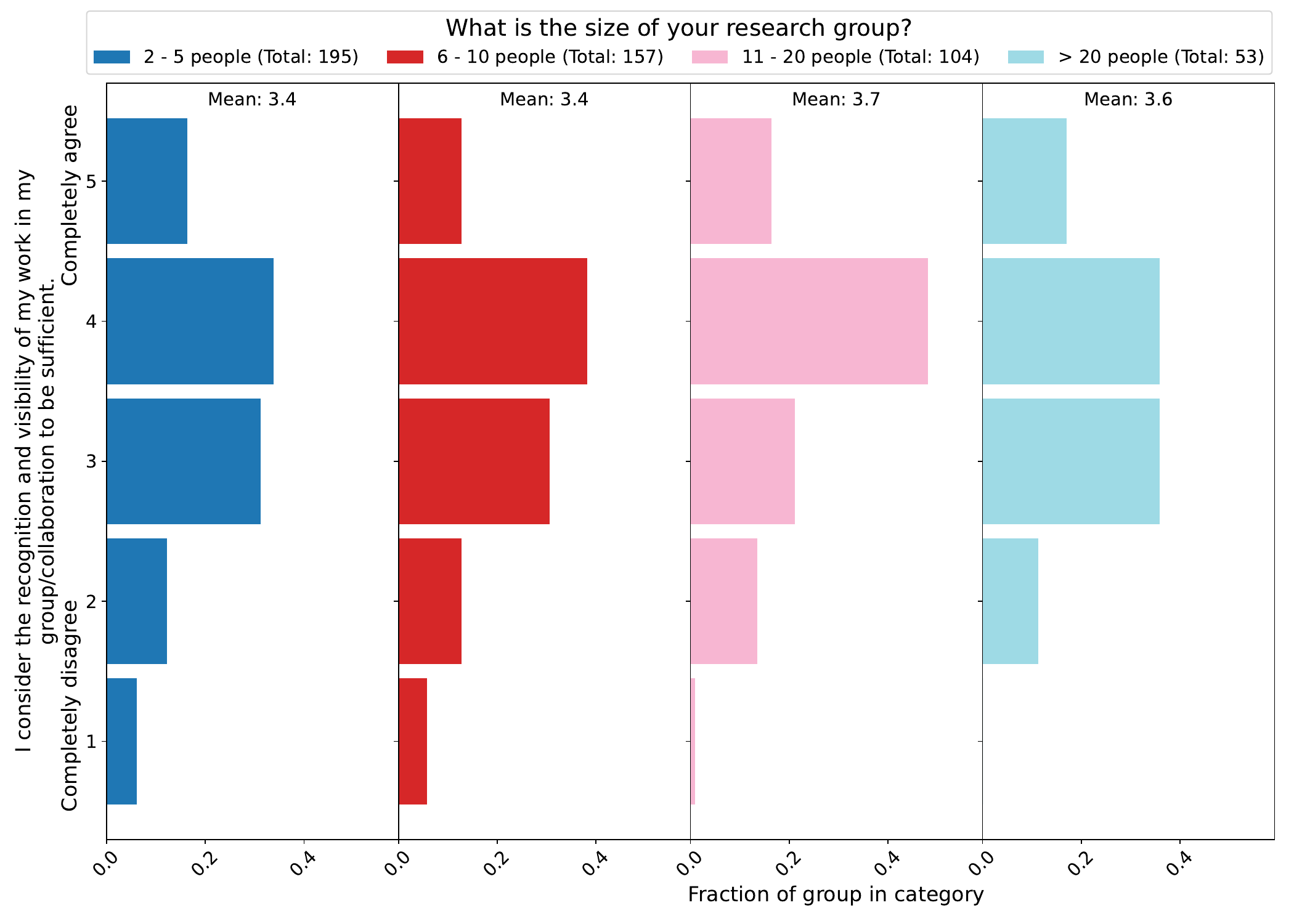}}\\
        \subfloat[]{\label{fig:part2:Q29vQ15}\includegraphics[width=0.49\textwidth]{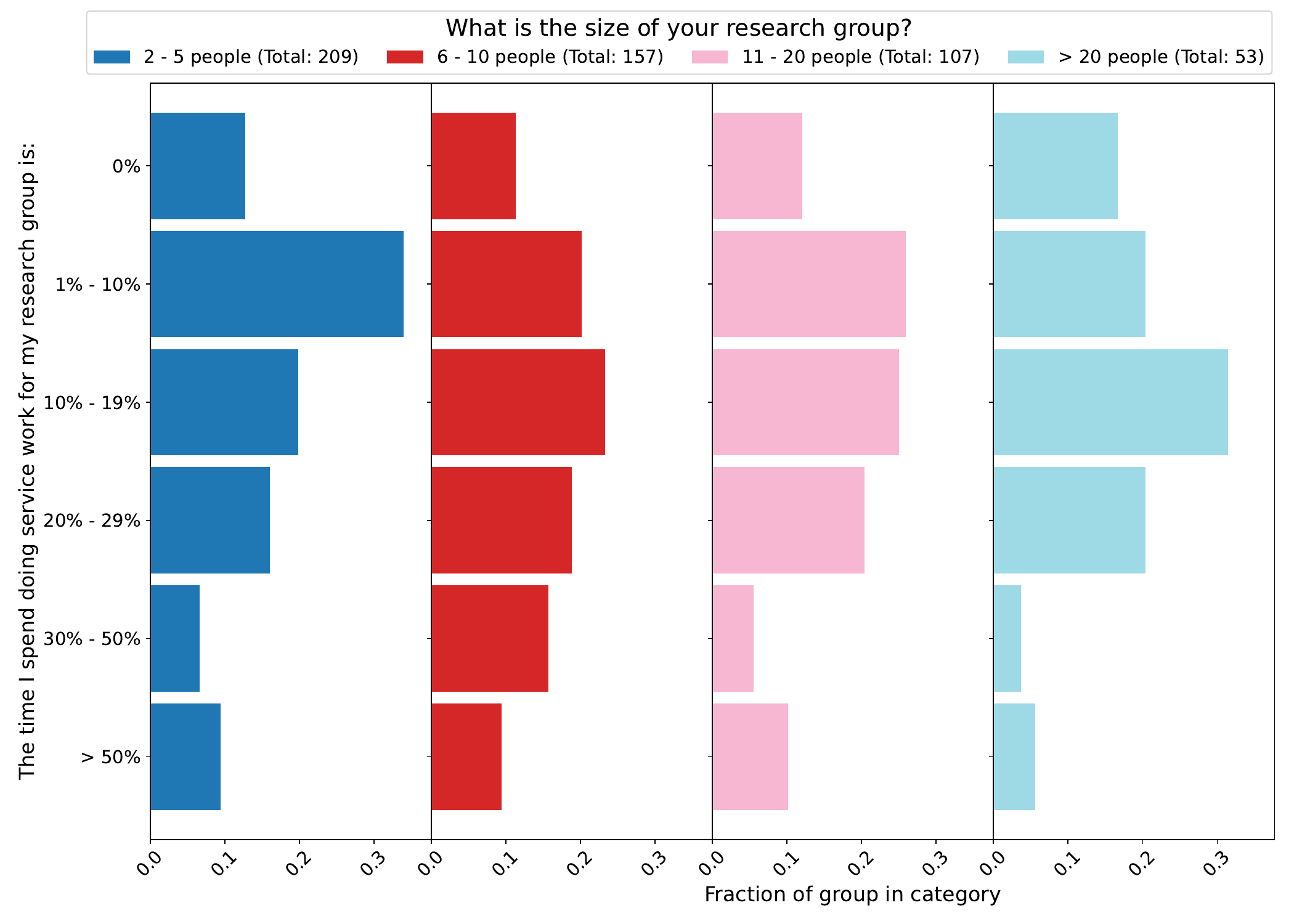}}
    \caption{(Q80,94,29 v Q15) Correlations with respondents' research group size. Fractions are given out of all respondents who are part of a research group and answered the questions.}
    \label{fig:part2:Q80Q94Q29vQ15}
\end{figure}

Next, the correlations between questions related to work in a research group, and others, are discussed.
The first study concerns the influence of research group size.
From the results, several conclusions can be drawn.
Firstly, shown in Figure~\ref{fig:part2:Q80Q94Q29vQ15}, respondents working in larger research groups feel stressed and under pressure more frequently, however they also agree slightly more often that the recognition and visibility of their work within their group is sufficient.
Secondly, we see that the mode time the respondents spend doing service work for their research group increases with the size of their research group.
Finally, we did not find any correlation between group size and how well-informed respondents feel about various aspects related to careers and training opportunities or how much they agree that there is room for them to express their original/new ideas within the group (not plotted here).

\begin{figure}[b!]
    \centering
        \subfloat[]{\label{fig:part2:Q17vQ1}\includegraphics[width=0.49\textwidth]{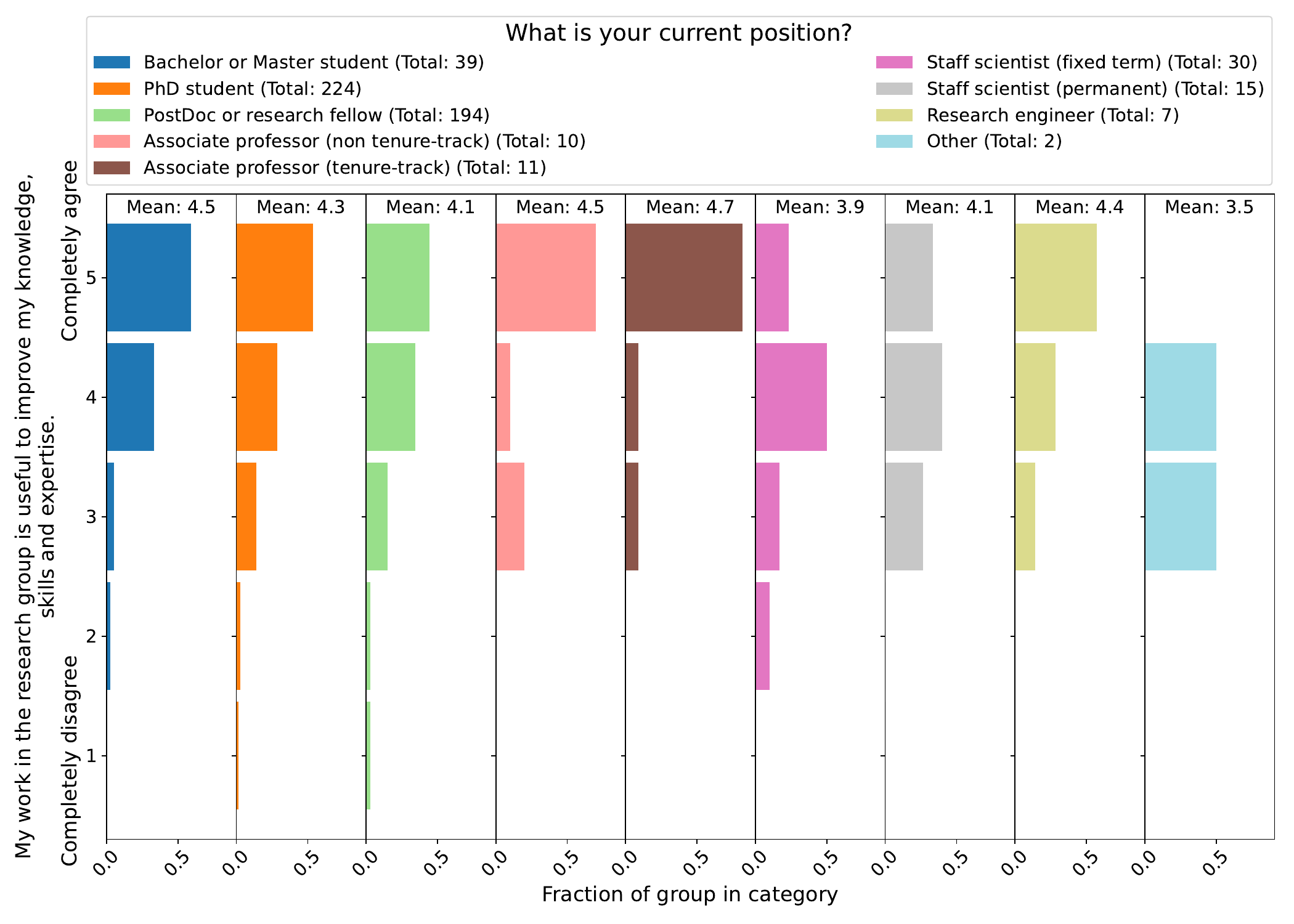}}
        \subfloat[]{\label{fig:part2:Q17vQ8}\includegraphics[width=0.49\textwidth]{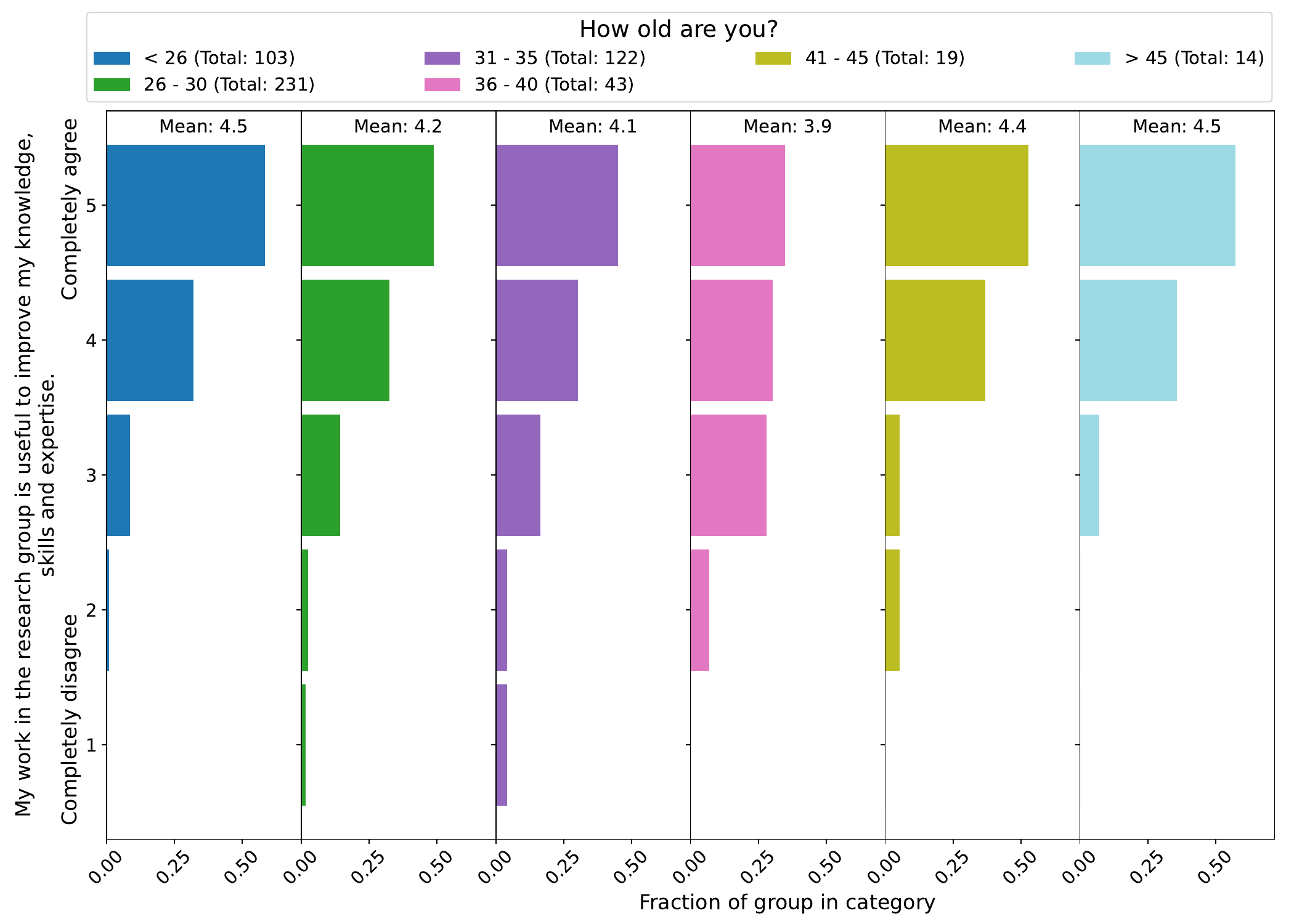}}
    \caption{(Q17 v Q1,8) Correlations between respondents' perceptions of how useful their work in their group is and their position or age. Fractions are given out of all respondents who are part of a research group and answered the questions.}
    \label{fig:part2:Q17vQ1Q8}
\end{figure}

In Figure~\ref{fig:part2:Q17vQ1Q8}, we show correlations with respondents' perceptions of how useful work in their research group is to improving their knowledge, skills and expertise.
Weak correlations with position and age were seen, with fixed-term staff scientists and respondents aged 36-40 less positive about this question than others.
We also studied potential correlations with gender and geography and saw no significant trends.

\begin{figure}[ht!]
    \centering
        \subfloat[]{\label{fig:part2:Q18vQ1}\includegraphics[width=0.49\textwidth]{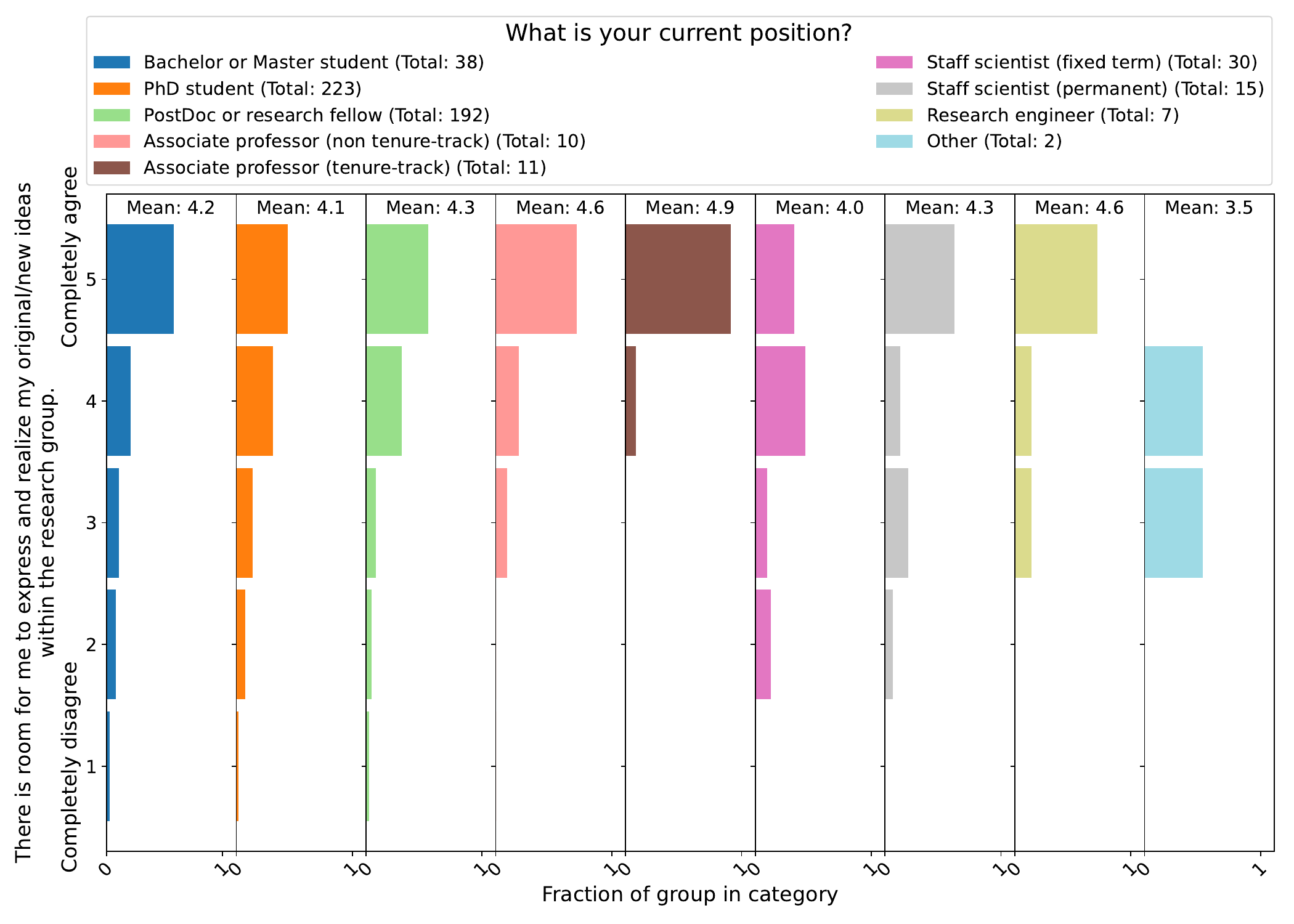}}
        \subfloat[]{\label{fig:part2:Q18vQ4}\includegraphics[width=0.49\textwidth]{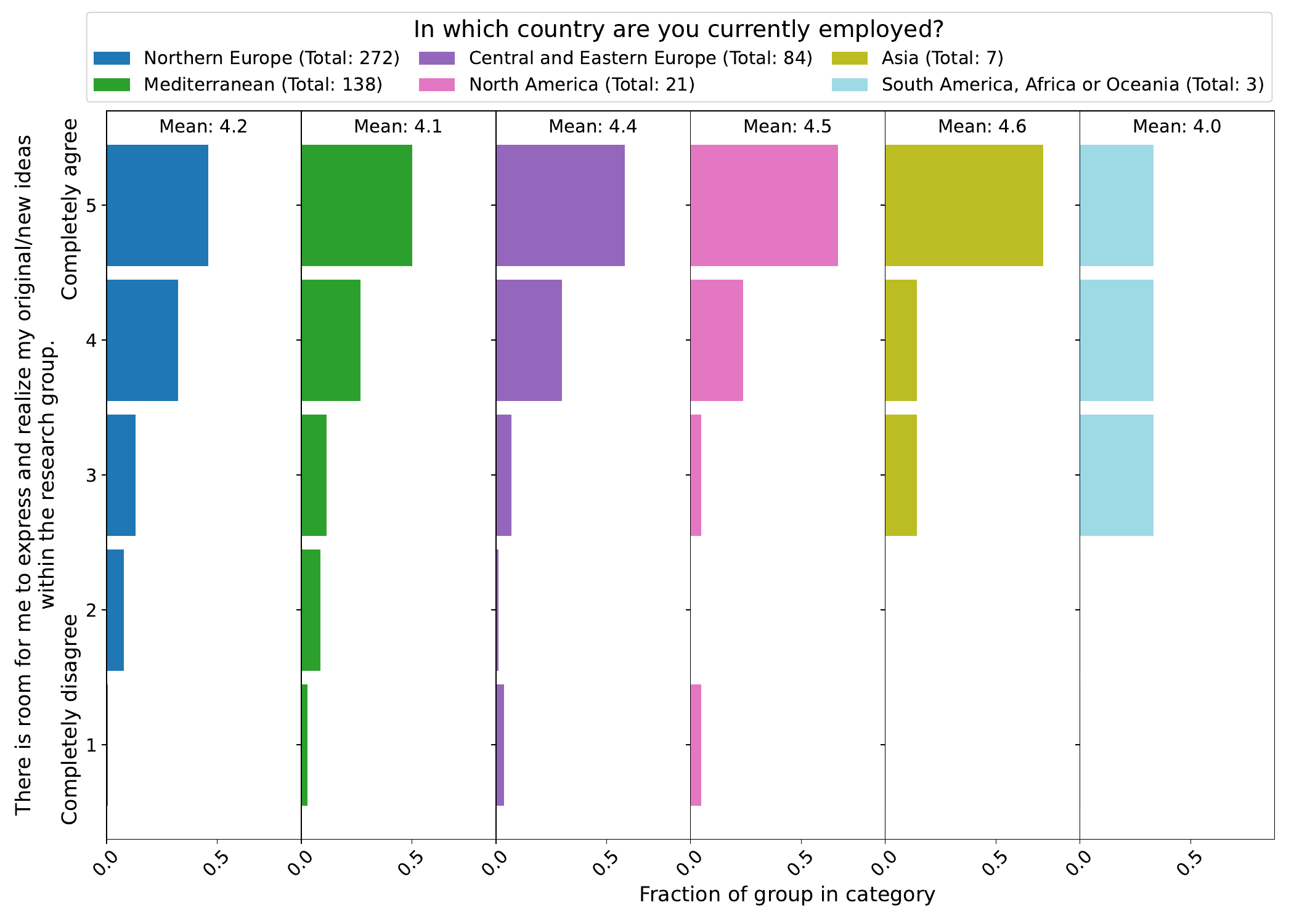}}\\
        \subfloat[]{\label{fig:part2:Q18vQ8}\includegraphics[width=0.49\textwidth]{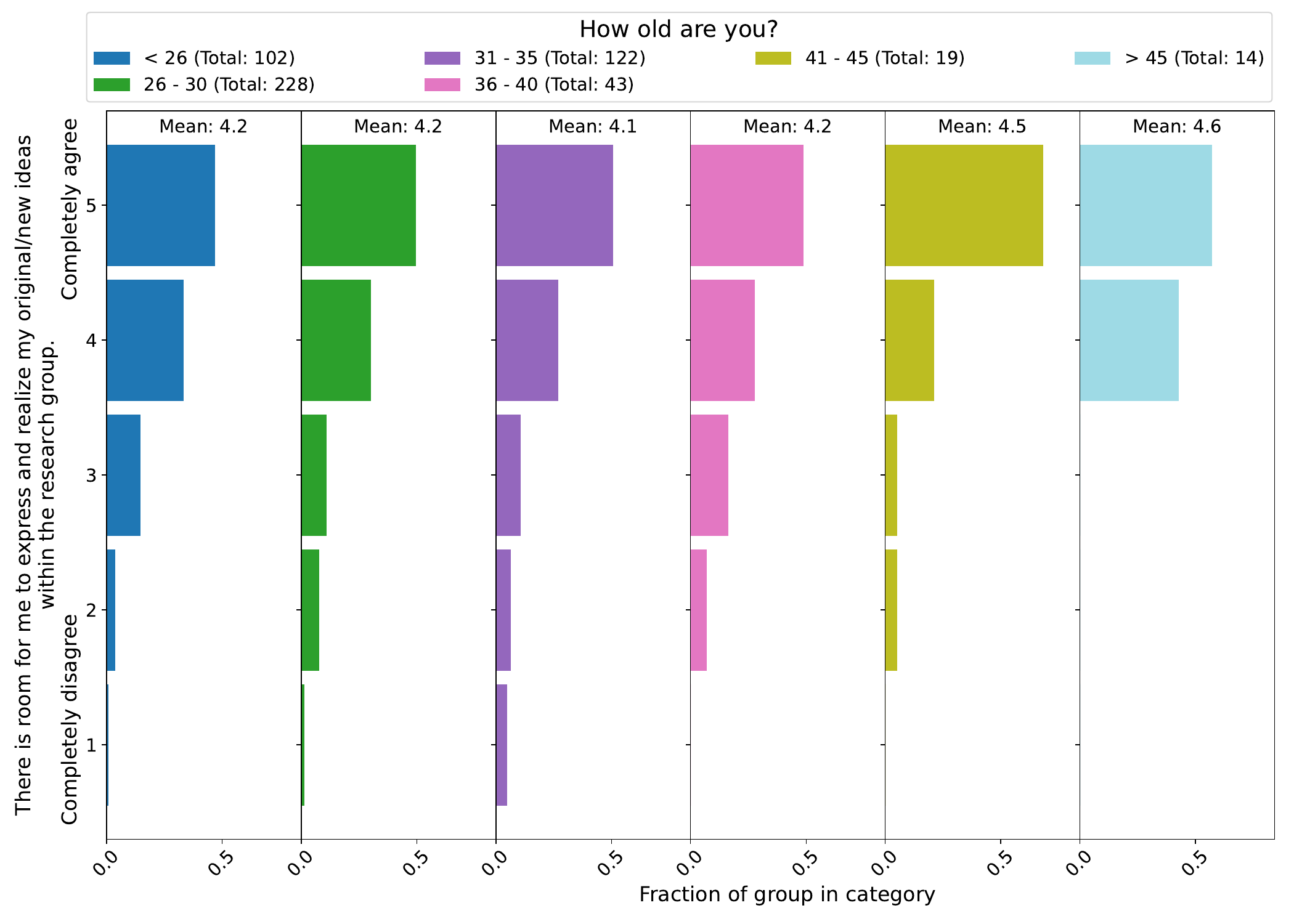}}
    \caption{(Q18 v Q1,4,8) Correlations between respondents' opportunities to express new ideas in their research group, and position, country of employment and age. Fractions are given out of all respondents who are part of a research group and answered the questions.}
    \label{fig:part2:Q18vQ1Q4Q8}
\end{figure}

\begin{figure}[ht!]
    \centering
        \subfloat[]{\label{fig:part2:Q21vQ8}\includegraphics[width=0.49\textwidth]{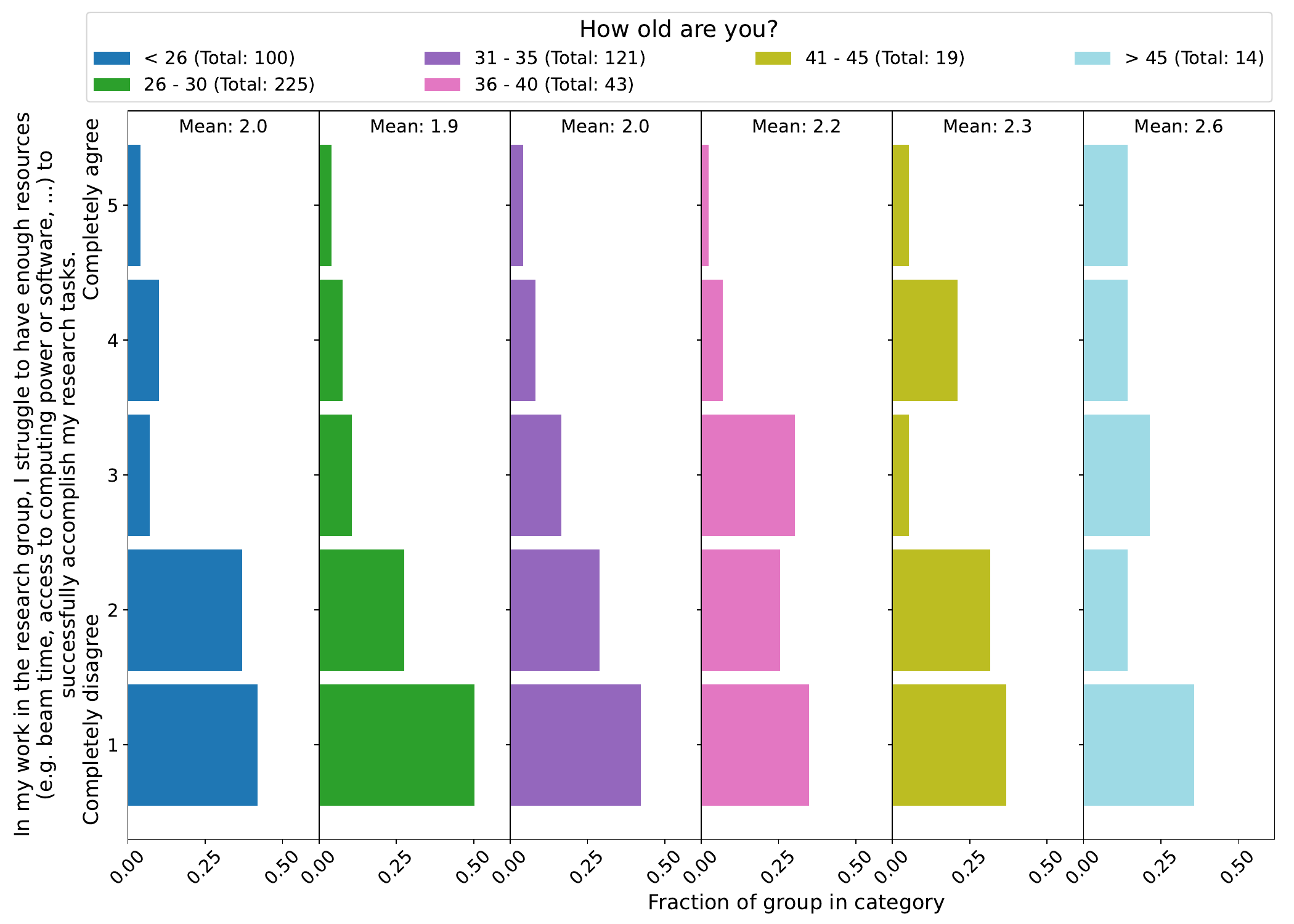}}
        \subfloat[]{\label{fig:part2:Q21vQ11}\includegraphics[width=0.49\textwidth]{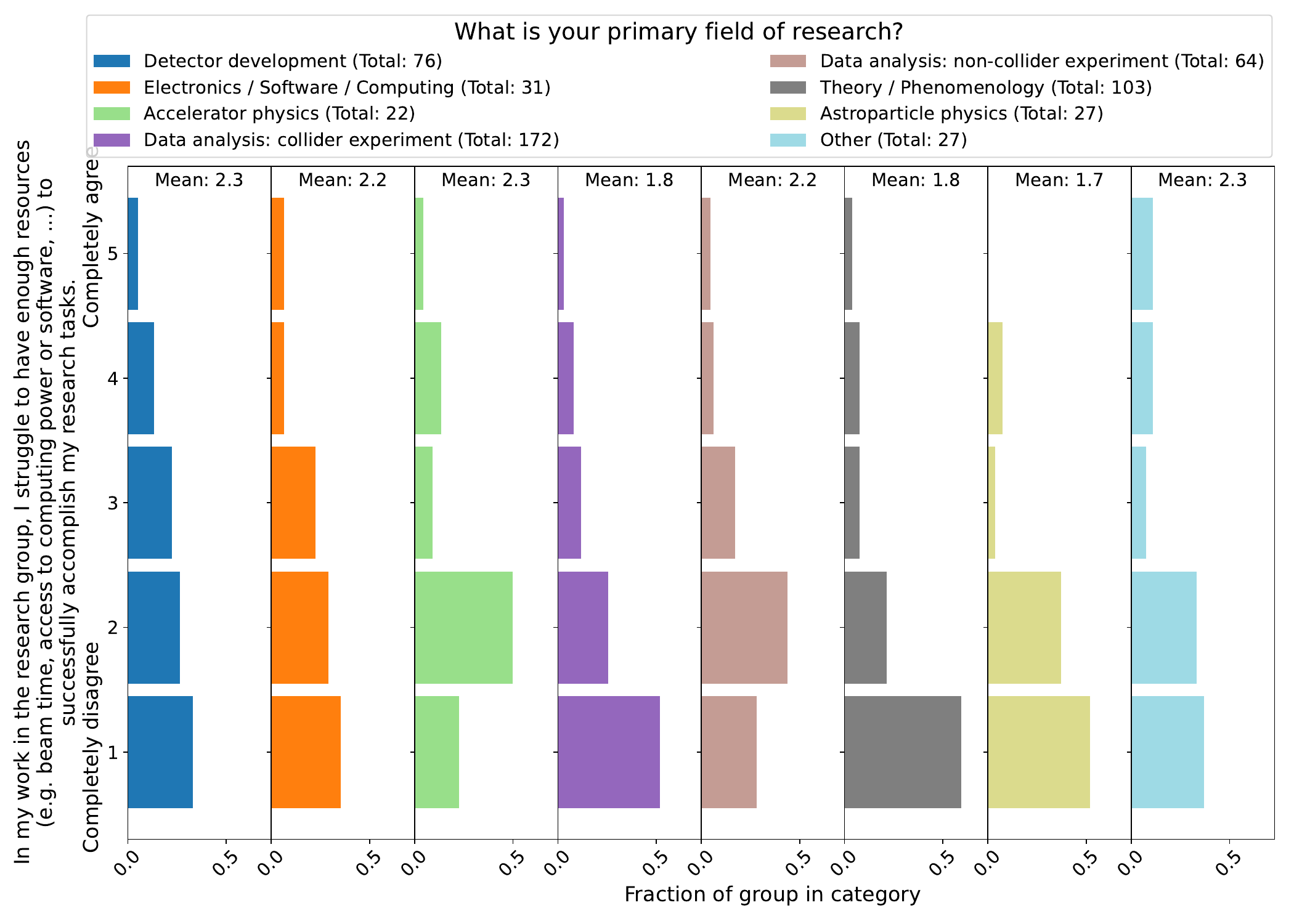}}
    \caption{(Q21 v Q8,11) Correlations between respondents' resource constraints in research group work and age or field of research. Fractions are given out of all respondents who are part of a research group and answered the questions.}
    \label{fig:part2:Q21vQ8Q11}
\end{figure}

\begin{figure}[ht!]
    \centering
        \subfloat[]{\label{fig:part2:Q22vQ1}\includegraphics[width=0.49\textwidth]{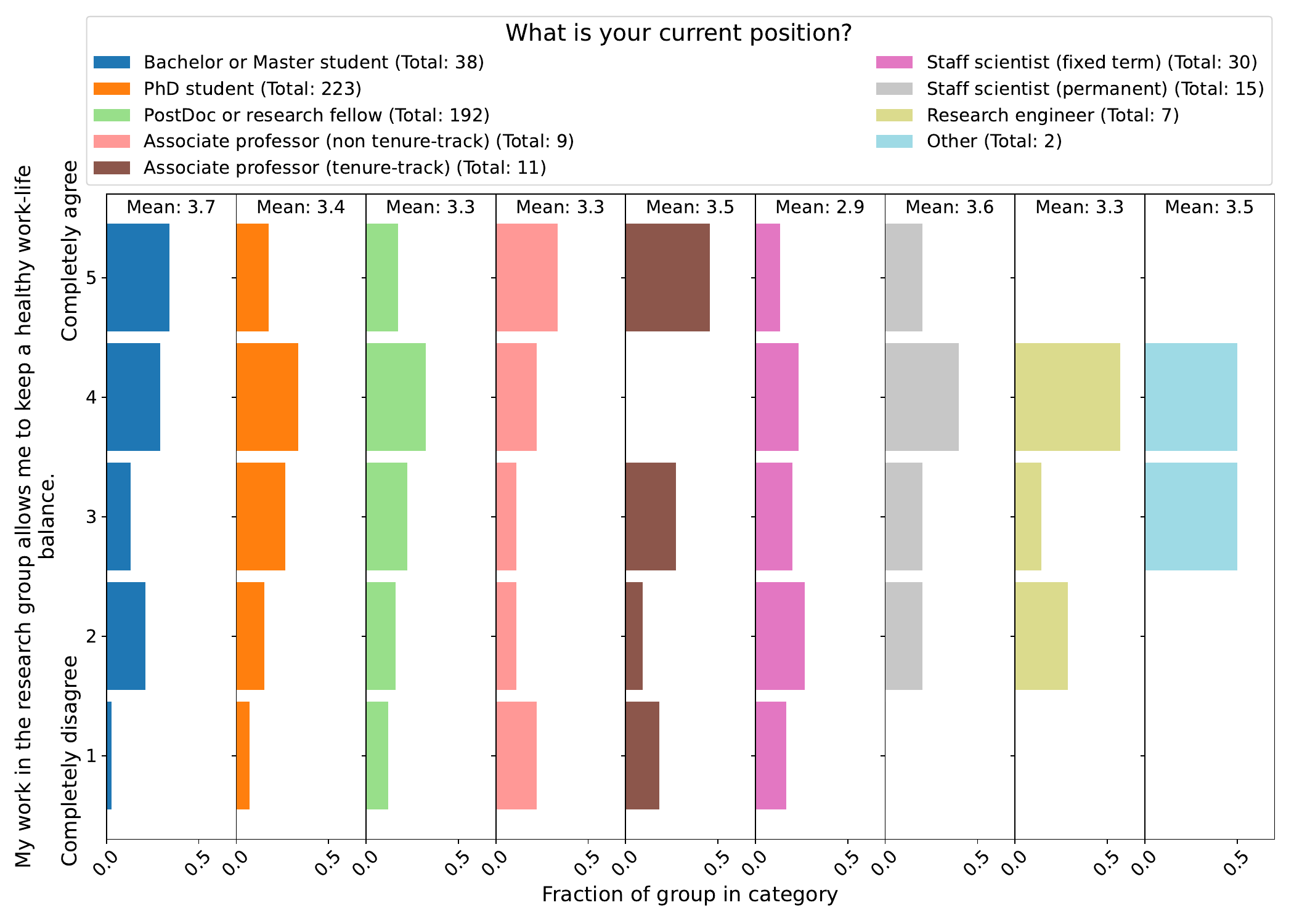}}
        \subfloat[]{\label{fig:part2:Q22vQ4}\includegraphics[width=0.49\textwidth]{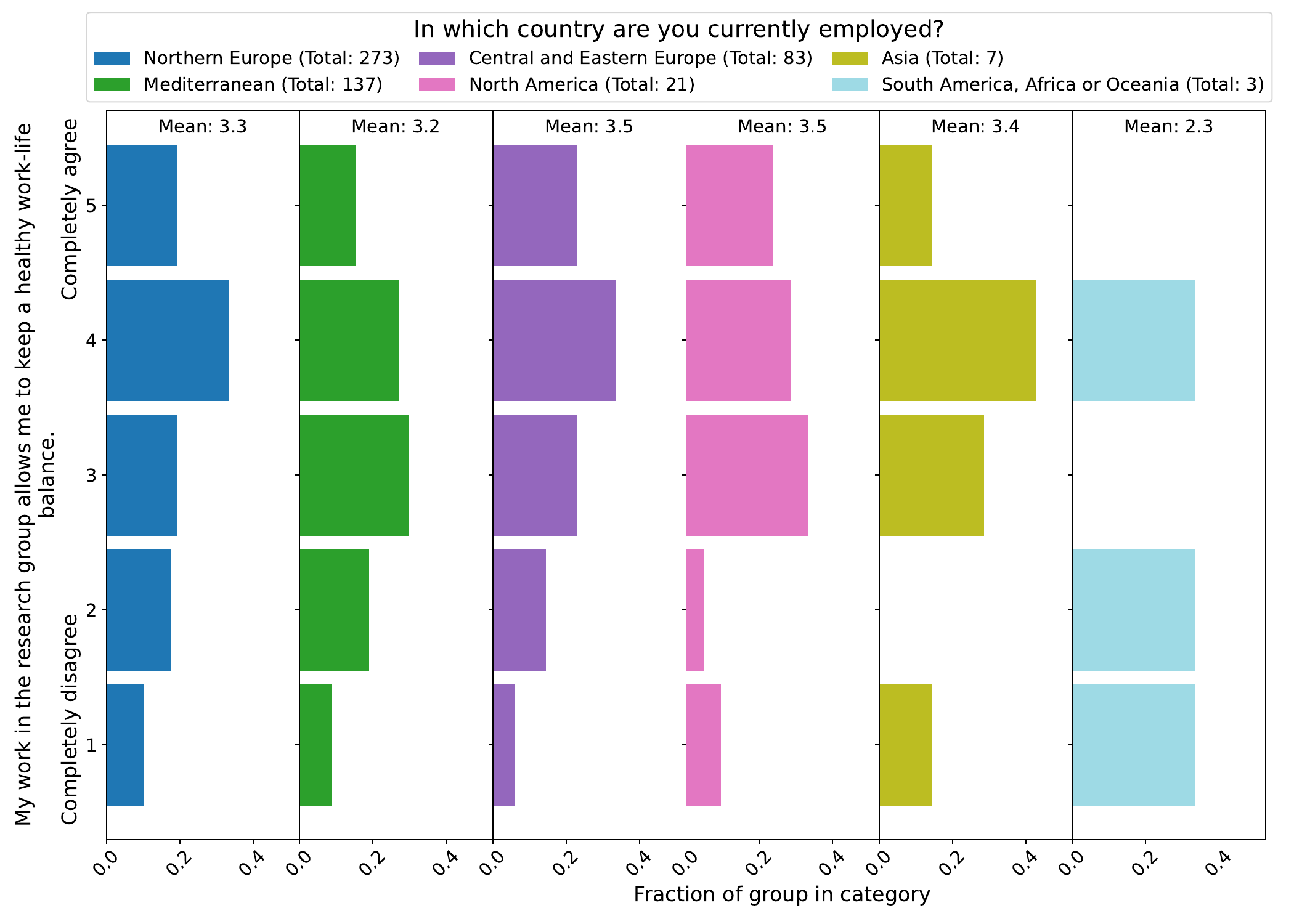}}\\
        \subfloat[]{\label{fig:part2:Q22vQ7}\includegraphics[width=0.49\textwidth]{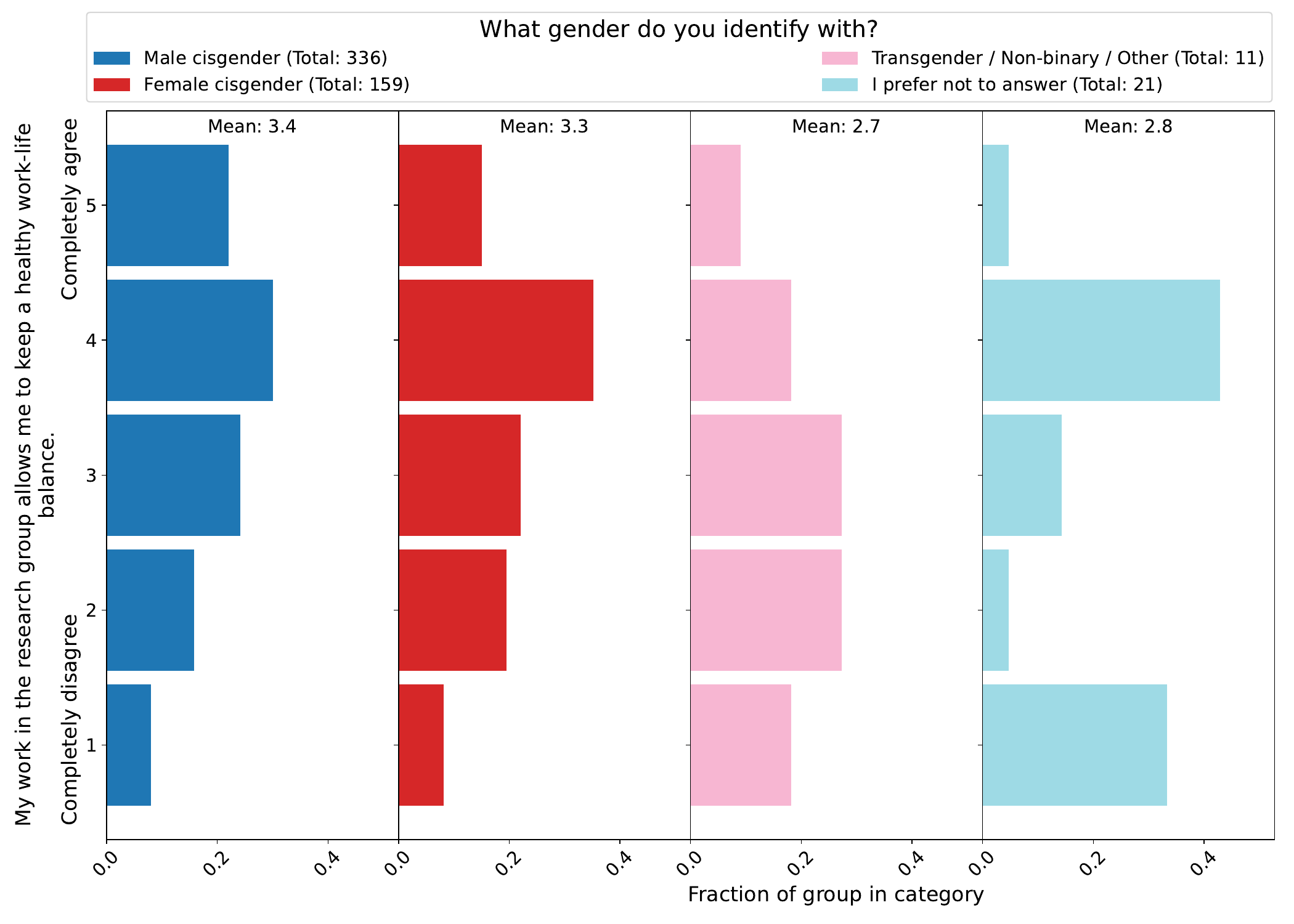}}
        \subfloat[]{\label{fig:part2:Q22vQ8}\includegraphics[width=0.49\textwidth]{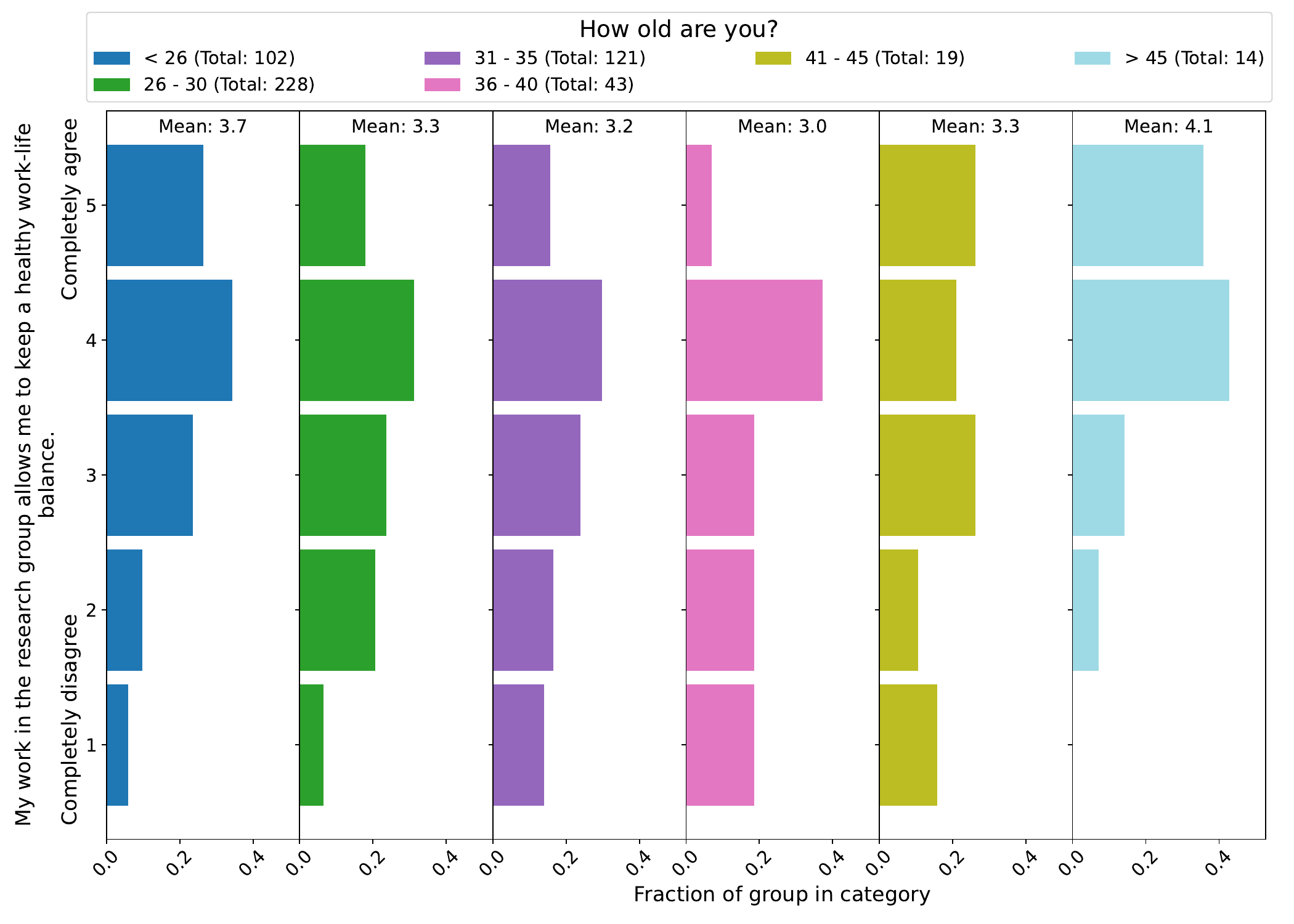}}
    \caption{(Q22 v Q1,4,7,8) Correlations between respondents' work-life balance in research group work and position, country of employment, gender and age. Fractions are given out of all respondents who are part of a research group and answered the questions.}
    \label{fig:part2:Q22vQ1Q4Q7Q8}
\end{figure}

\begin{figure}[ht!]
    \centering
        \subfloat[]{\label{fig:part2:Q27vQ4}\includegraphics[width=0.49\textwidth]{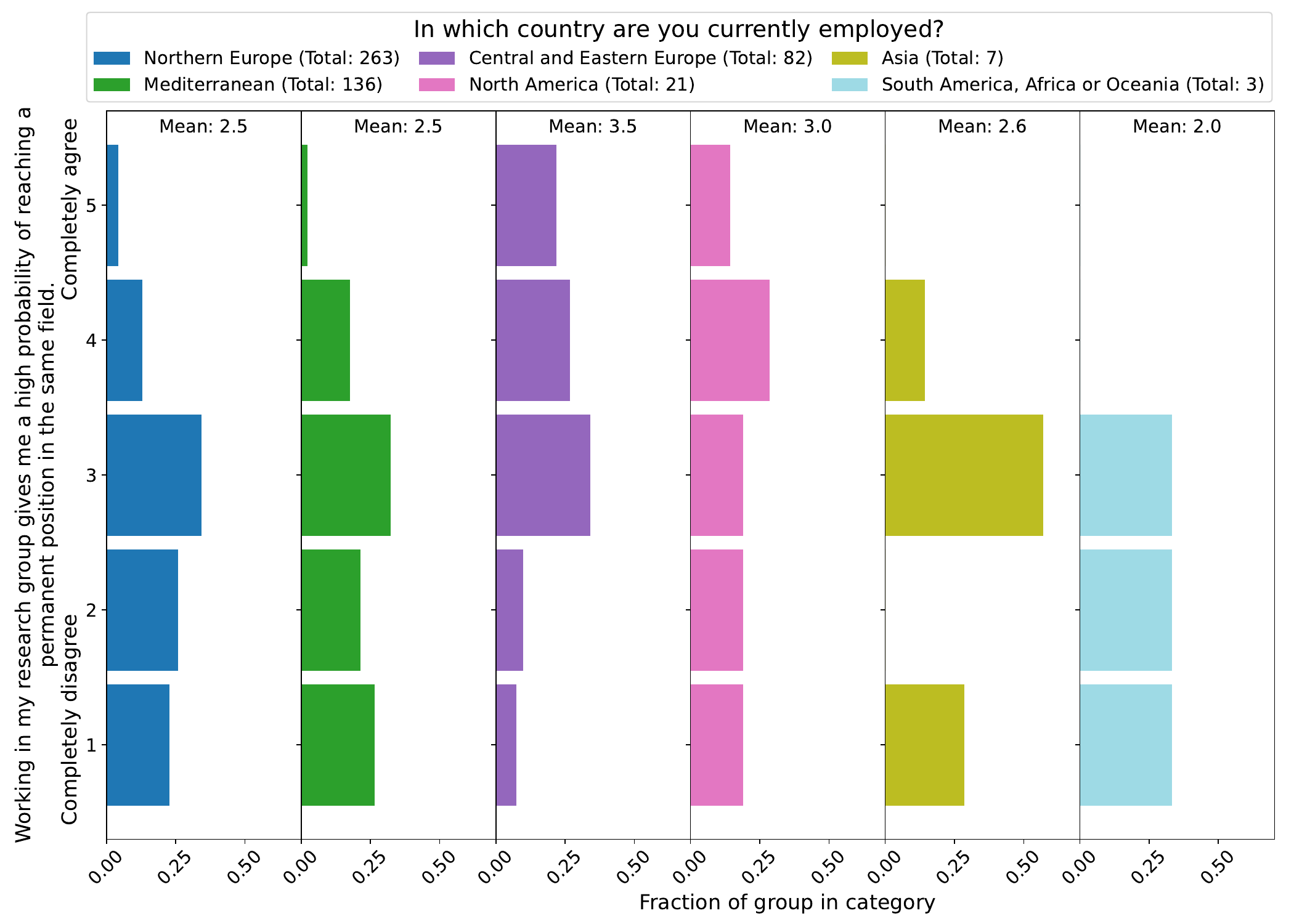}}
        \subfloat[]{\label{fig:part2:Q27vQ8}\includegraphics[width=0.49\textwidth]{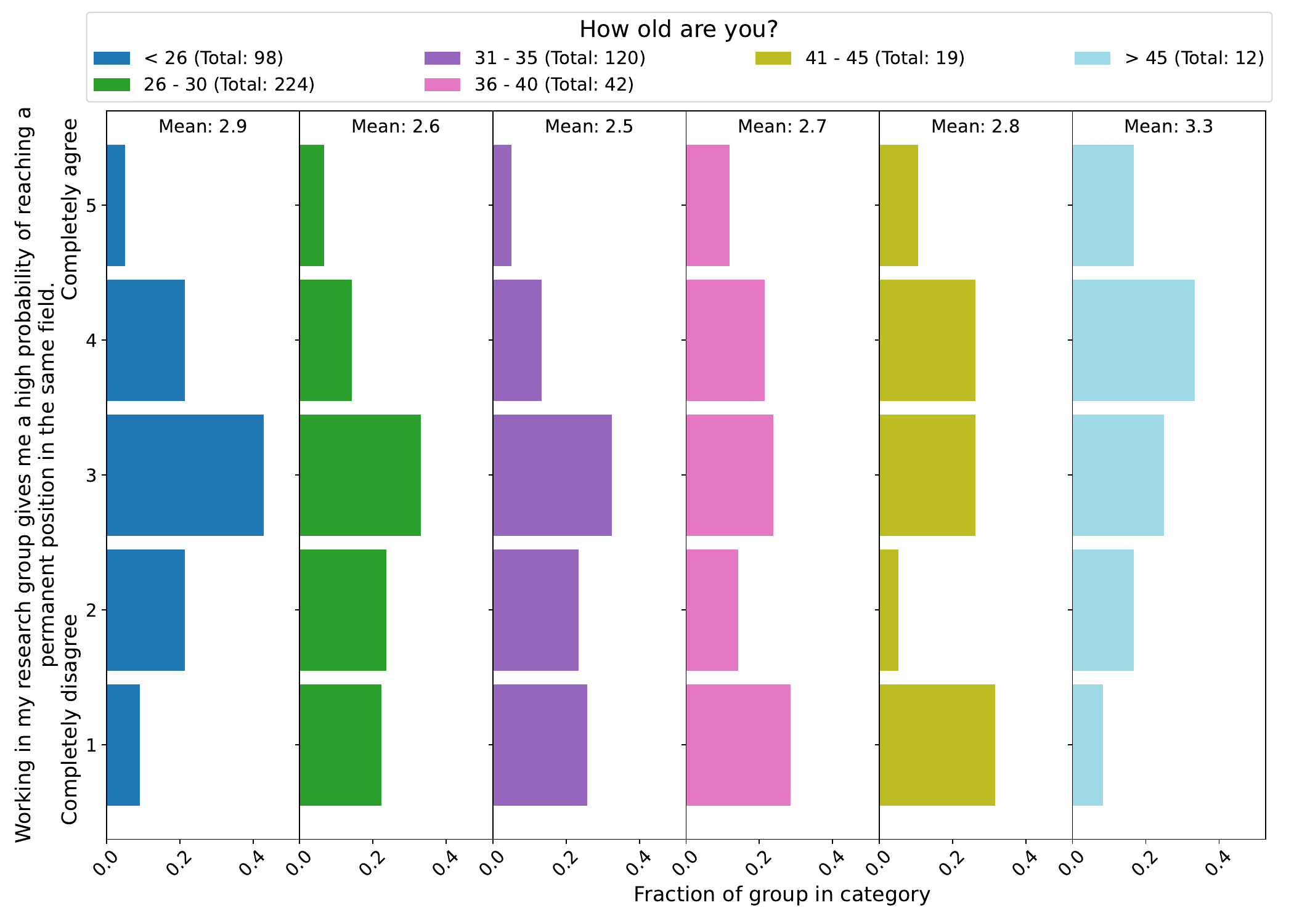}}
    \caption{(Q27 v Q4,8) Correlations between respondents' perceived career advancement opportunities inn relation to their research group and current position, country of employment, nationality, gender and age. Fractions are given out of all respondents who are part of a research group and answered the questions.}
    \label{fig:part2:Q27vQ4Q8}
\end{figure}

\begin{figure}[ht!]
    \centering
    \subfloat[]{\label{fig:part2:Q23vQ92}\includegraphics[width=0.49\textwidth]{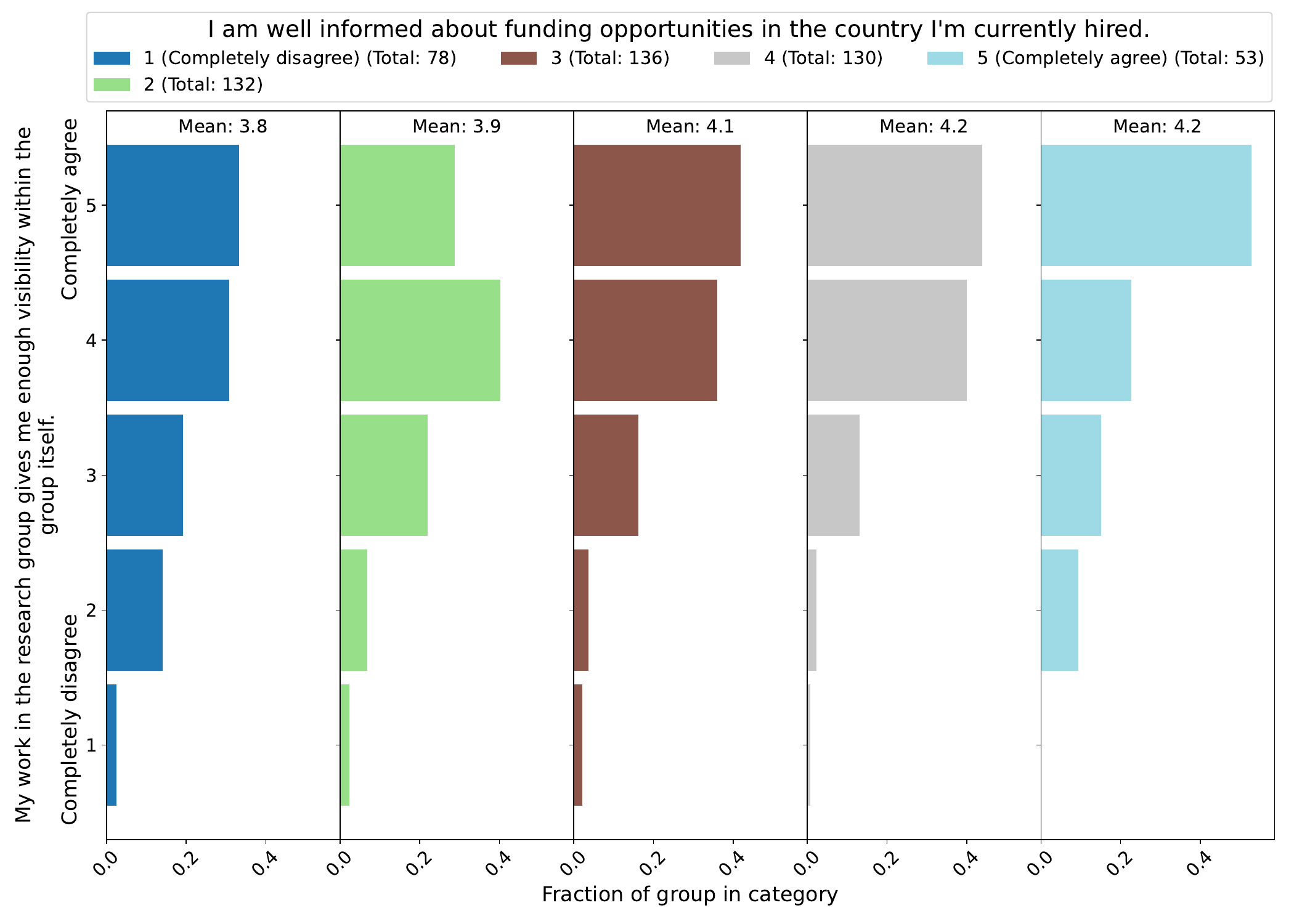}}
    \subfloat[]{\label{fig:part2:Q23vQ97}\includegraphics[width=0.49\textwidth]{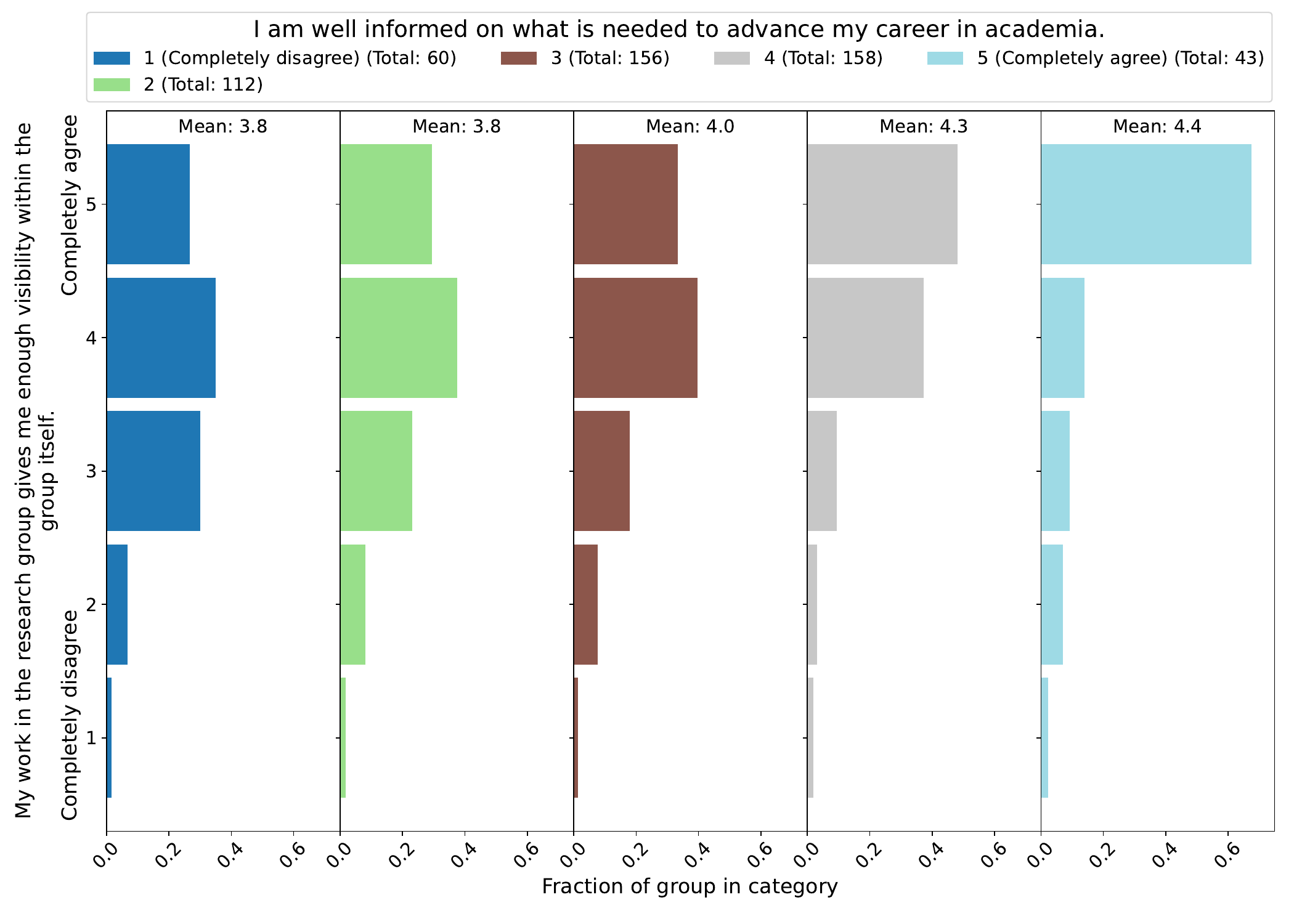}}\\
    \subfloat[]{\label{fig:part2:Q25vQ92}\includegraphics[width=0.49\textwidth]{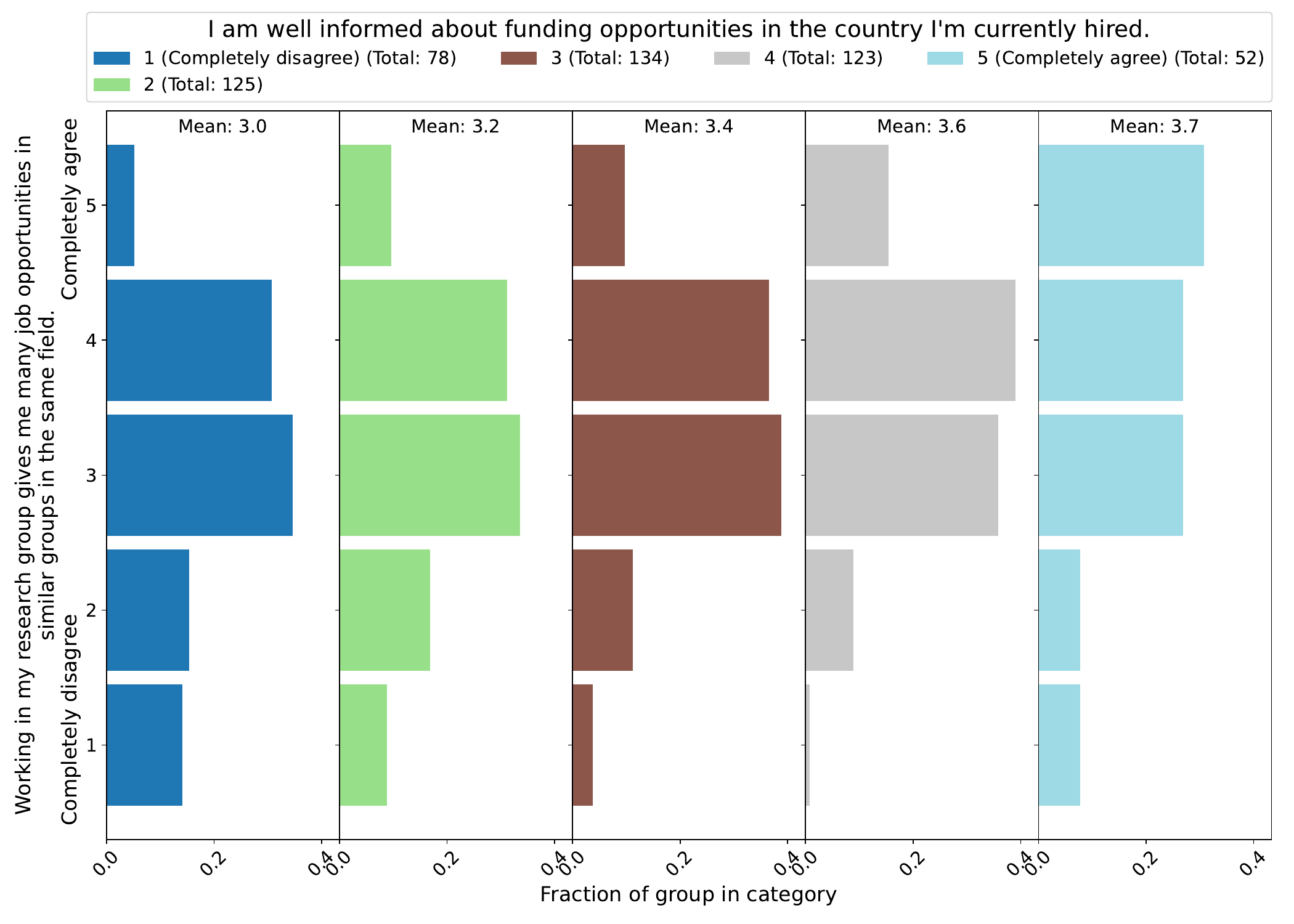}}
    \subfloat[]{\label{fig:part2:Q25vQ97}\includegraphics[width=0.49\textwidth]{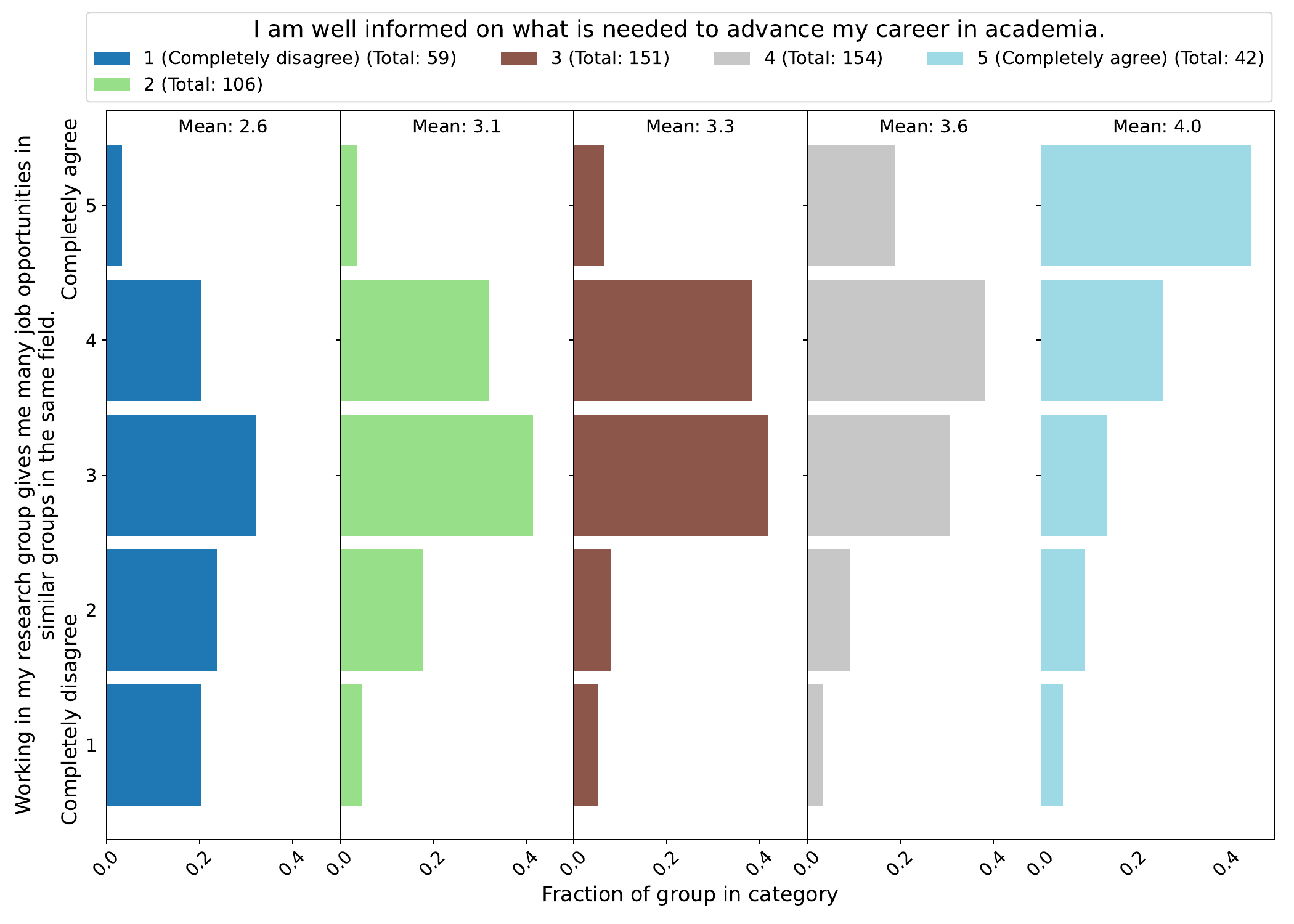}}\\
    \subfloat[]{\label{fig:part2:Q27vQ92}\includegraphics[width=0.49\textwidth]{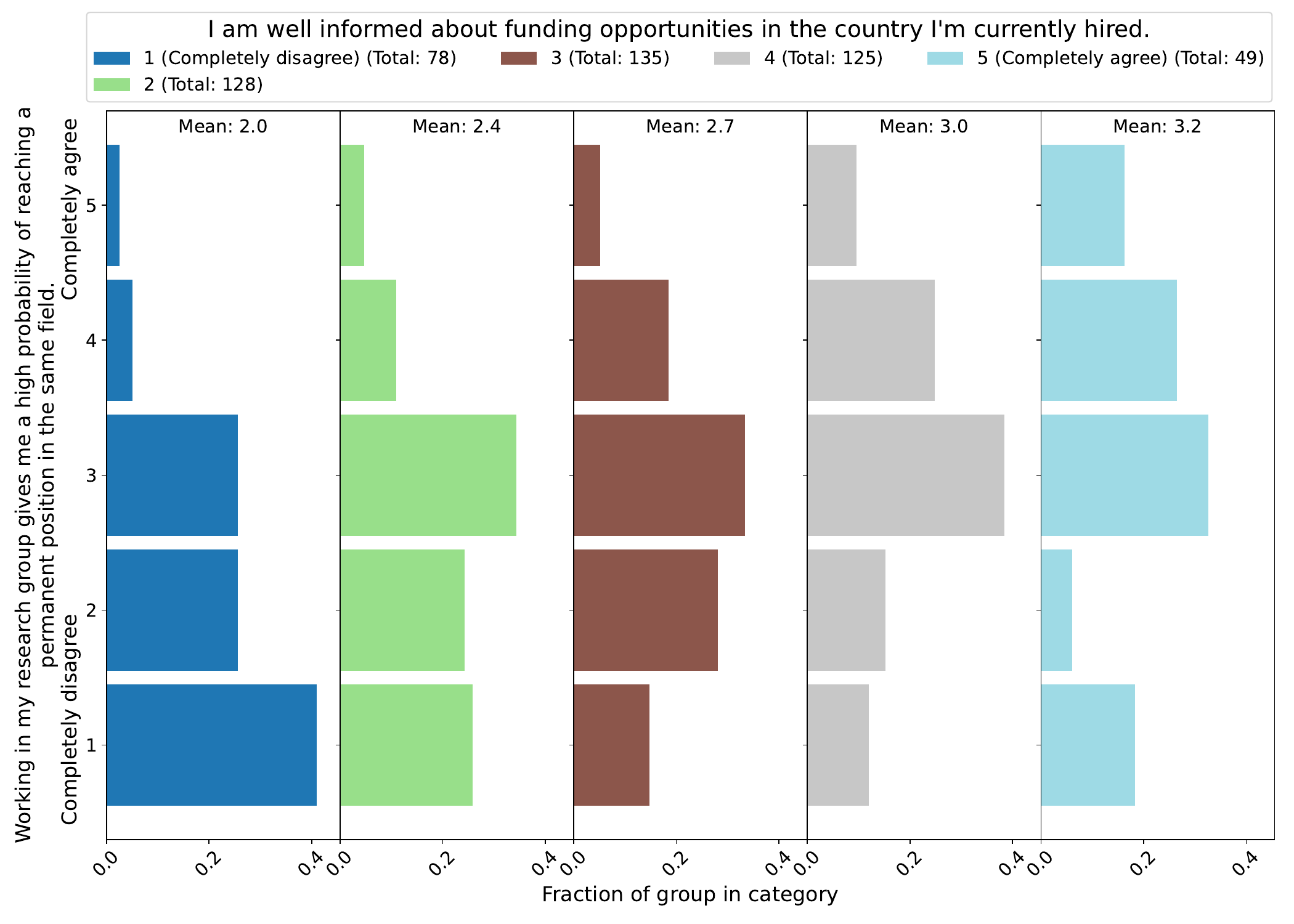}}
    \subfloat[]{\label{fig:part2:Q27vQ97}\includegraphics[width=0.49\textwidth]{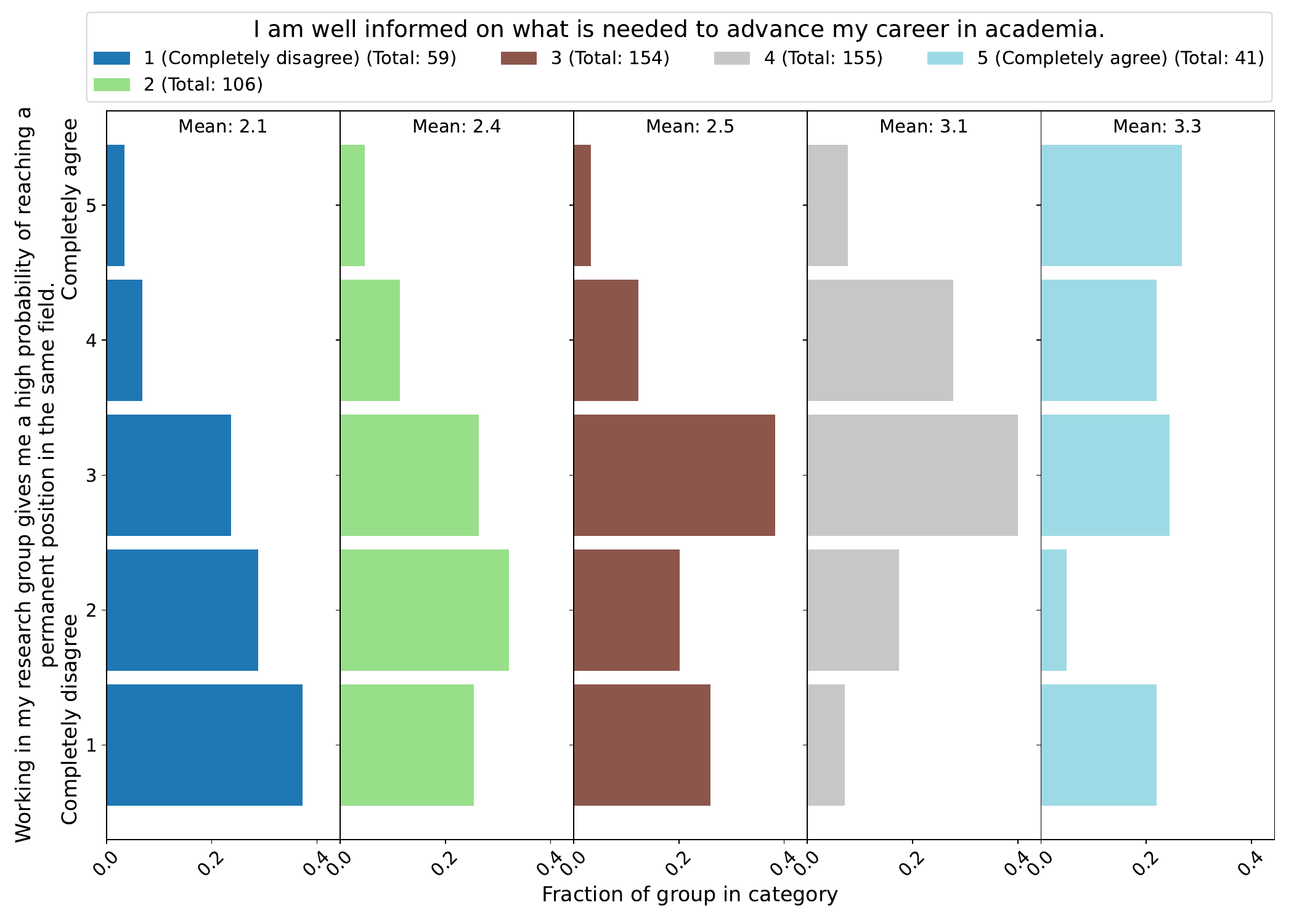}}
    \caption{(Q55, 60 v Q23,25,27) Correlations between the respondents' awareness on the funding opportunities in the country they are currently hired and questions related to visibility and opportunities in the same field resulting from their research group work. Fractions are given out of all respondents who are part of a research group and answered the questions.}
    \label{fig:part2:Q23Q25Q27vQ92Q97}
\end{figure}

Building on the examination of respondents' perceptions of the value of their work, Figure~\ref{fig:part2:Q18vQ1Q4Q8} shows correlations with respondents' ability to express and implement original ideas within the research group.
We observed a positive correlation between room to express ideas and seniority of position.
Similarly, respondents older than 40 feel the most positive about this.
Respondents employed in Central and Eastern Europe, Northern America and Asia feel more positive about this topic.
We saw no other strong correlations between this topic and respondent demographics.

\FloatBarrier
\pagebreak
We next studied respondents' challenges in finding resources for fulfilling their research tasks, and the correlations found are shown in Figure~\ref{fig:part2:Q21vQ8Q11}.
We see that age correlates positively with an increased likelihood of struggling.
Additionally, we find that respondents working in fields that are more theoretical, or which rely more on centralised resources and large experiments, struggle less for resources in their research group.

The respondents' work-life balance was next studied, with correlations found shown in Figure~\ref{fig:part2:Q22vQ1Q4Q7Q8}.
Some groups of respondents exhibited a more negative, or at least wider ranging, response to the healthiness of their work-life balance.
These groups, to name a few, are fixed term staff scientists and non-tenure-track associate professors; respondents employed by a country in the Mediterranean or South America, Africa or Oceania; respondents who identify as transgender, non binary or other-gendered; or respondents aged 31 to 40.
We note, however, that these categories generally have a low sample size.

Another question concerns respondents' career advancement opportunities.
Interesting correlations found between this and respondents' demographics are displayed in Figure~\ref{fig:part2:Q27vQ4Q8}.
We observe that respondents employed in Central and Eastern Europe or North America feel substantially more positive than others about work in their research group giving them a high probability of reaching a permanent position in their field.
We also see a correlation with age, where respondents become more negative as they age, until they reach 36 years and above, where they become more positive.

\begin{figure}[ht!]
    \centering
    \subfloat[]{\label{fig:Q24vQ61}\includegraphics[width=0.49\textwidth]{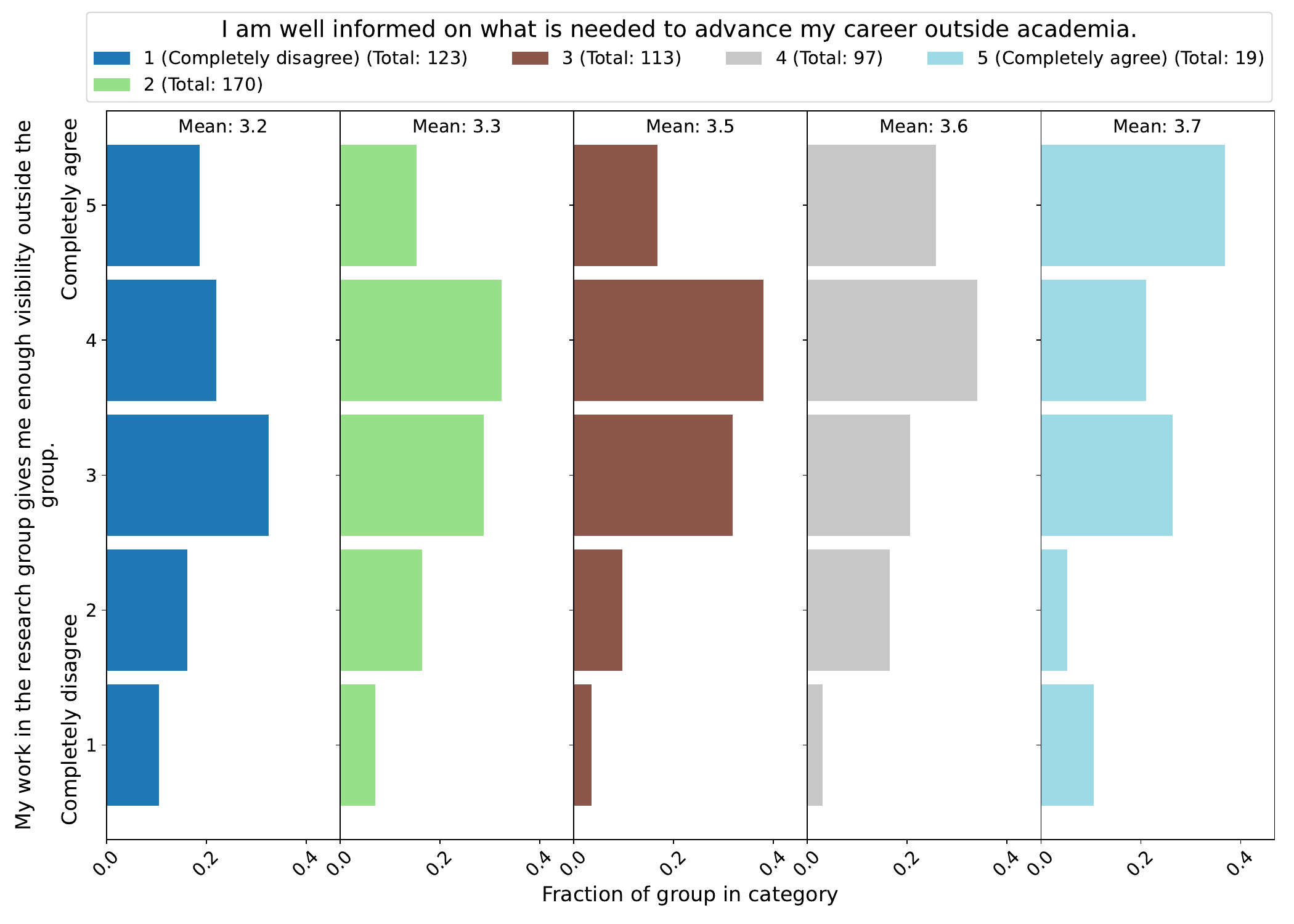}}
    \subfloat[]{\label{fig:Q28vQ61}\includegraphics[width=0.49\textwidth]{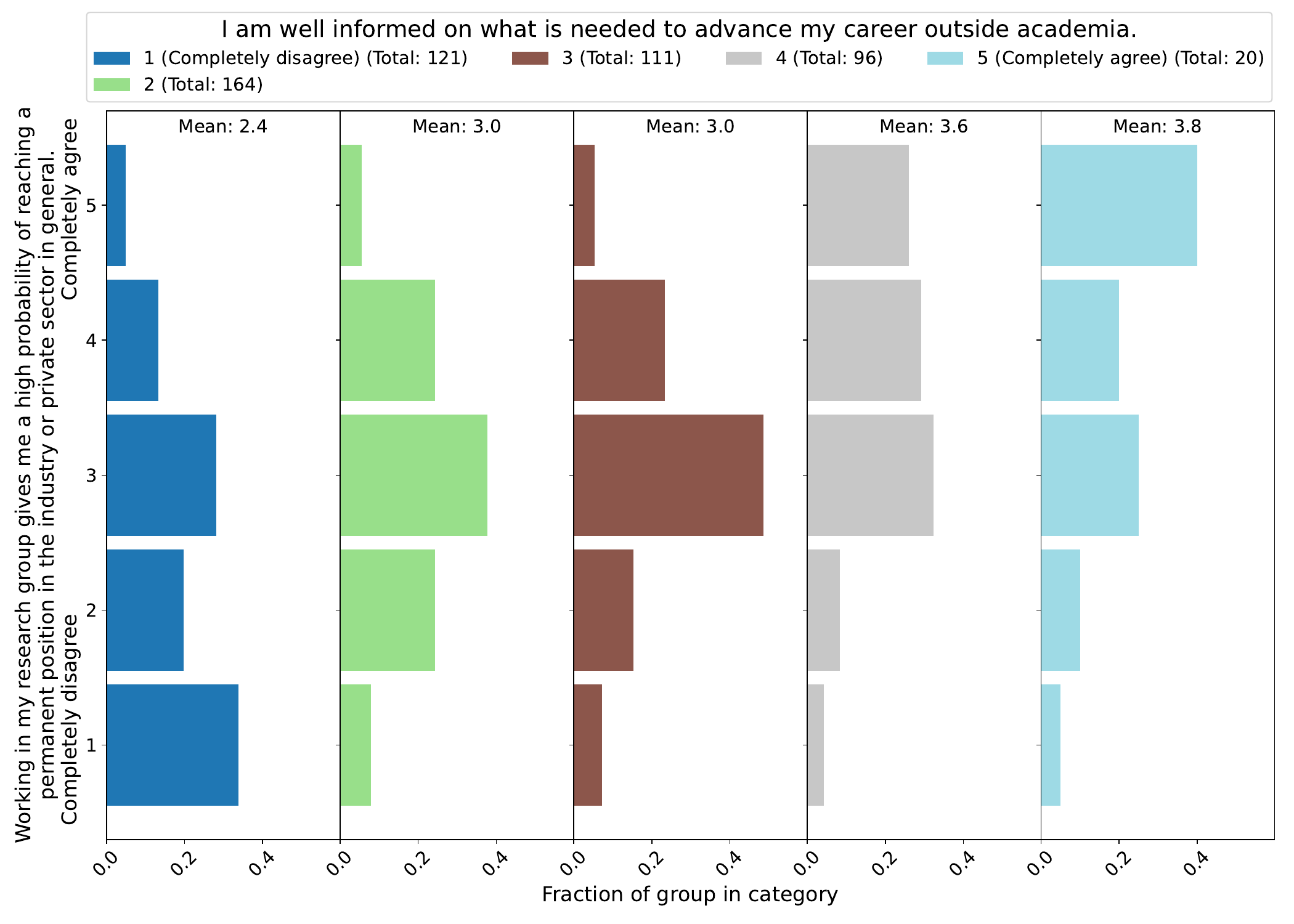}}
    \caption{(Q61 v Q24,28) Correlations between the respondents' perception of what is needed to advance their career outside academia and questions related to visibility and opportunities in other research groups resulting from their research group work. Fractions are given out of all respondents who are part of a research group and answered the questions.}
    \label{fig:part2:Q61vQ24Q26Q28}
\end{figure}

Concluding our analysis of respondents' perspectives in research groups, we consider respondents' visibility and job prospects within their group correlated to how well informed they feel about funding and career opportunities in Figures~\ref{fig:part2:Q23Q25Q27vQ92Q97}--\ref{fig:part2:Q61vQ24Q26Q28}.
We see strong positive correlations between respondent's feelings of adequate visibility within their group and being well informed about funding opportunities where they are hired and how to advance their career in academia in general.
Similarly strong positive correlations are seen between respondents' feelings about job opportunities or permanent positions in their field relating to their work within their research group, and feeling informed.
Finally we close this section by observing positive correlations between how respondents feel about visibility outside of their research group, or the probability of reaching a permanent position outside of academia, and feeling informed about advancing their careers outside of academia.

%%%%%%%%======================================================================================
\FloatBarrier
\subsubsection{Collaborations}

\begin{figure}[ht!]
    \centering
        \subfloat[]{\label{fig:part2:Q39vQ1}\includegraphics[width=0.49\textwidth]{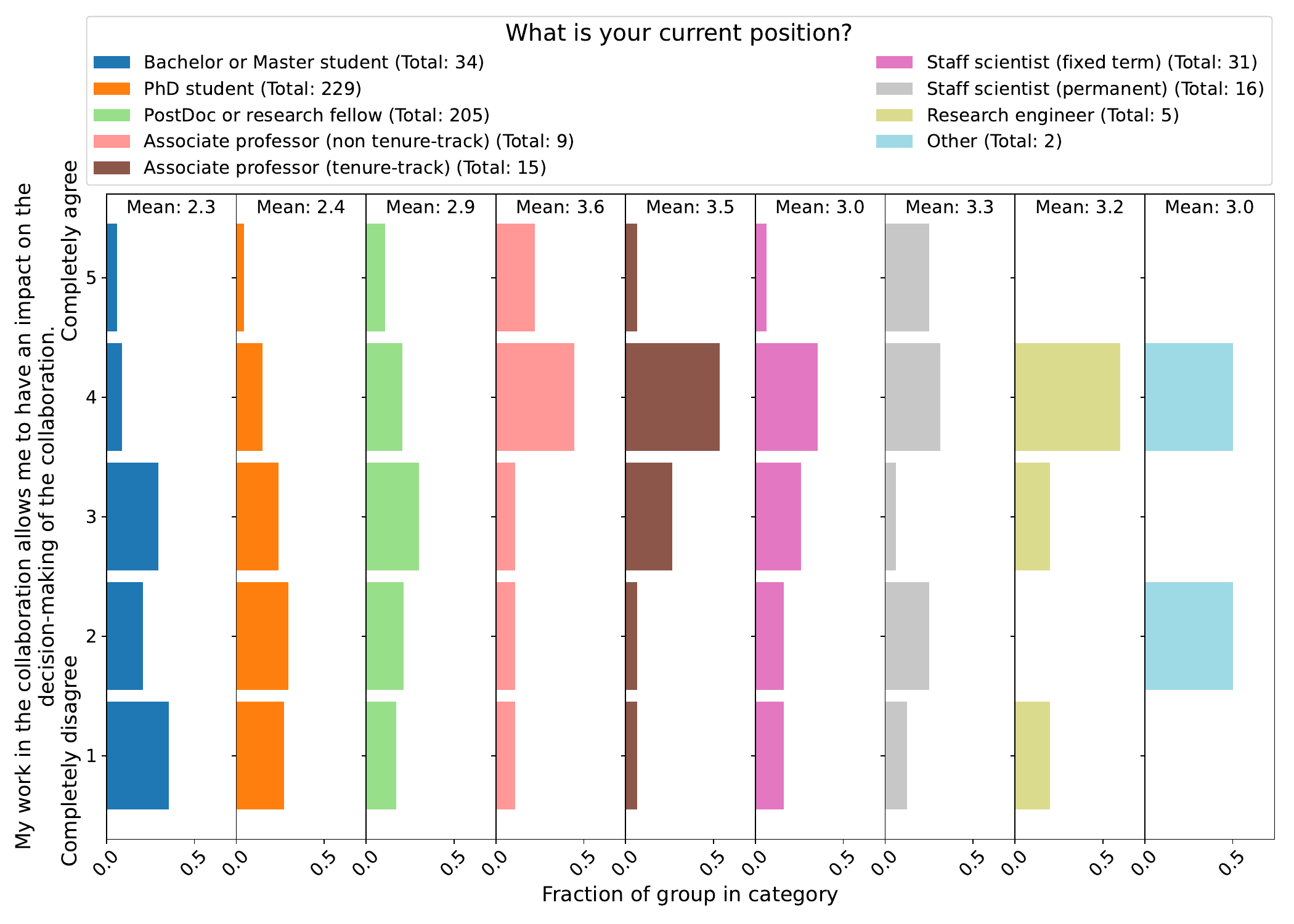}}
        \subfloat[]{\label{fig:part2:Q39vQ8}\includegraphics[width=0.49\textwidth]{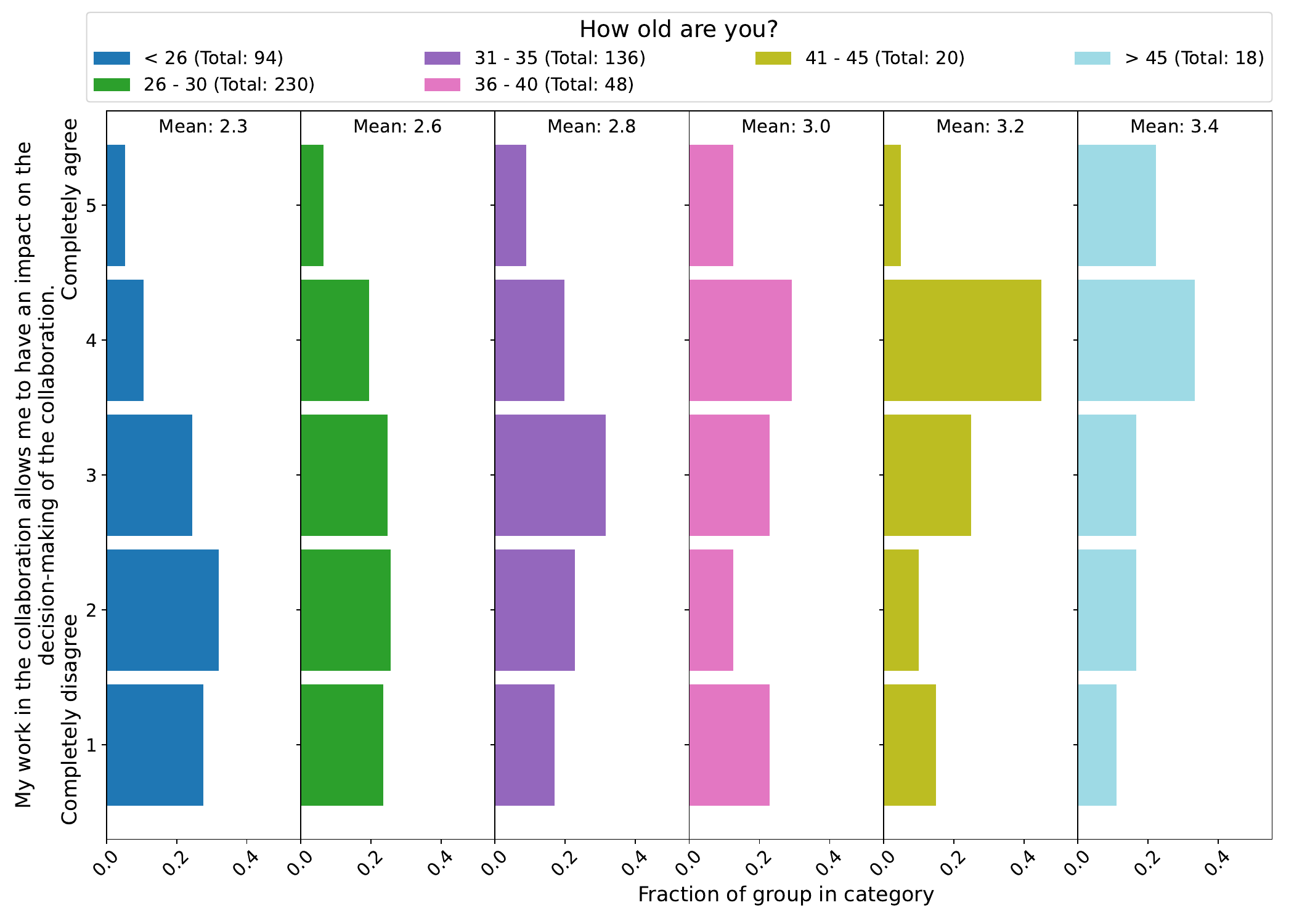}}
    \caption{(Q39 v Q1,8) Correlation between respondent's agreement that their work in a collaboration allows them to have an impact on the decision-making of the collaboration, and current position or age. Fractions are given out of all respondents who are part of a collaboration and answered the question.}
    \label{fig:part2:Q39vQ1Q8}
\end{figure}

In this section, the correlations between questions related to work in a scientific collaboration, and others, are discussed.
Figure~\ref{fig:part2:Q39vQ1Q8} compares the demographics of current position and age with how much respondents agree that they have an impact on the decision-making of the collaboration they work in.
Considering current position, it is evident that those in more junior positions feel they have less of an impact on decision-making compared to more senior members. The same correlation is present when comparing with age; younger collaboration members feel they have less of an impact than older members.

\begin{figure}[ht!]
    \centering
        \subfloat[]{\label{fig:part2:Q42vQ1}\includegraphics[width=0.49\textwidth]{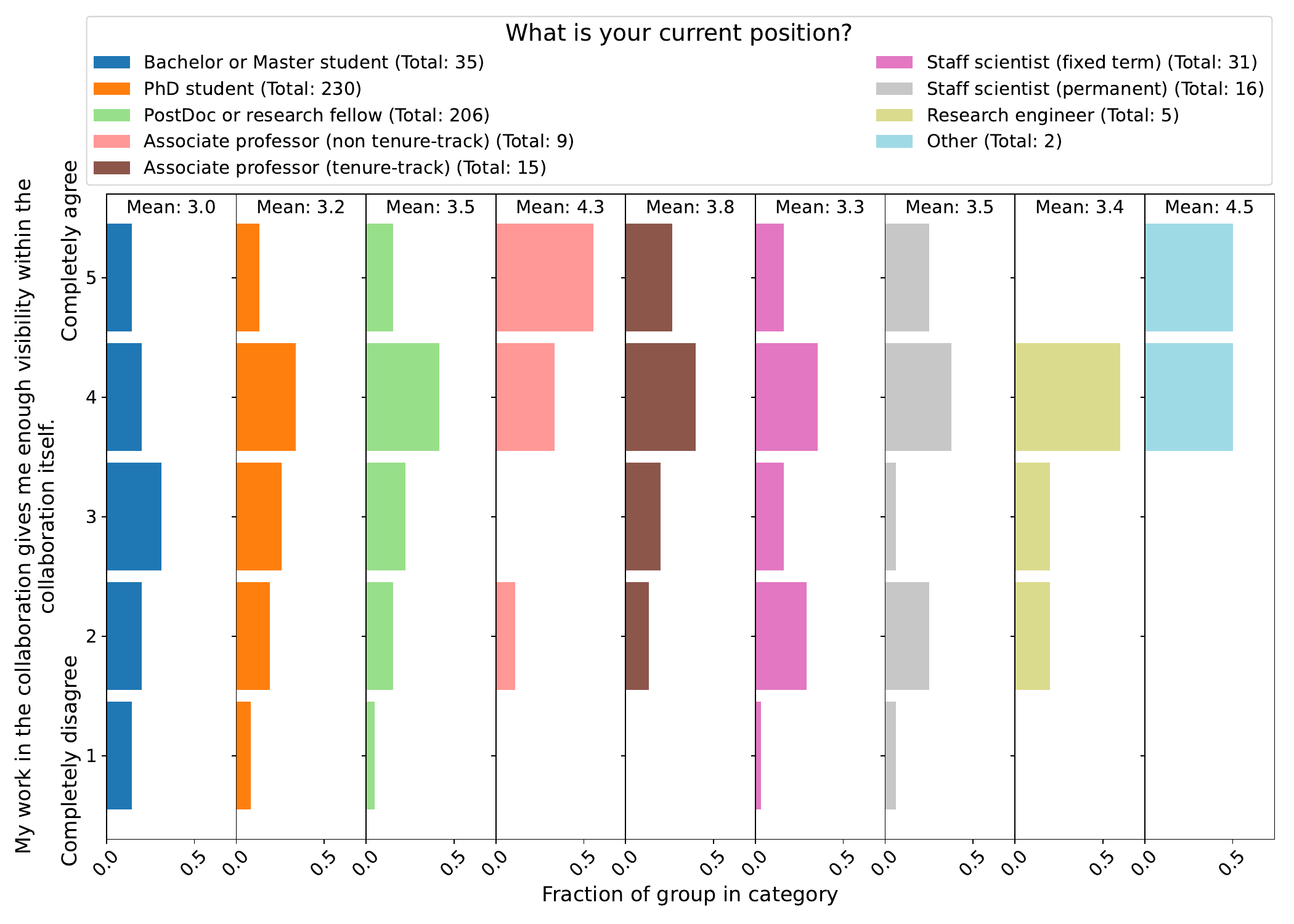}}
        \subfloat[]{\label{fig:part2:Q42vQ8}\includegraphics[width=0.49\textwidth]{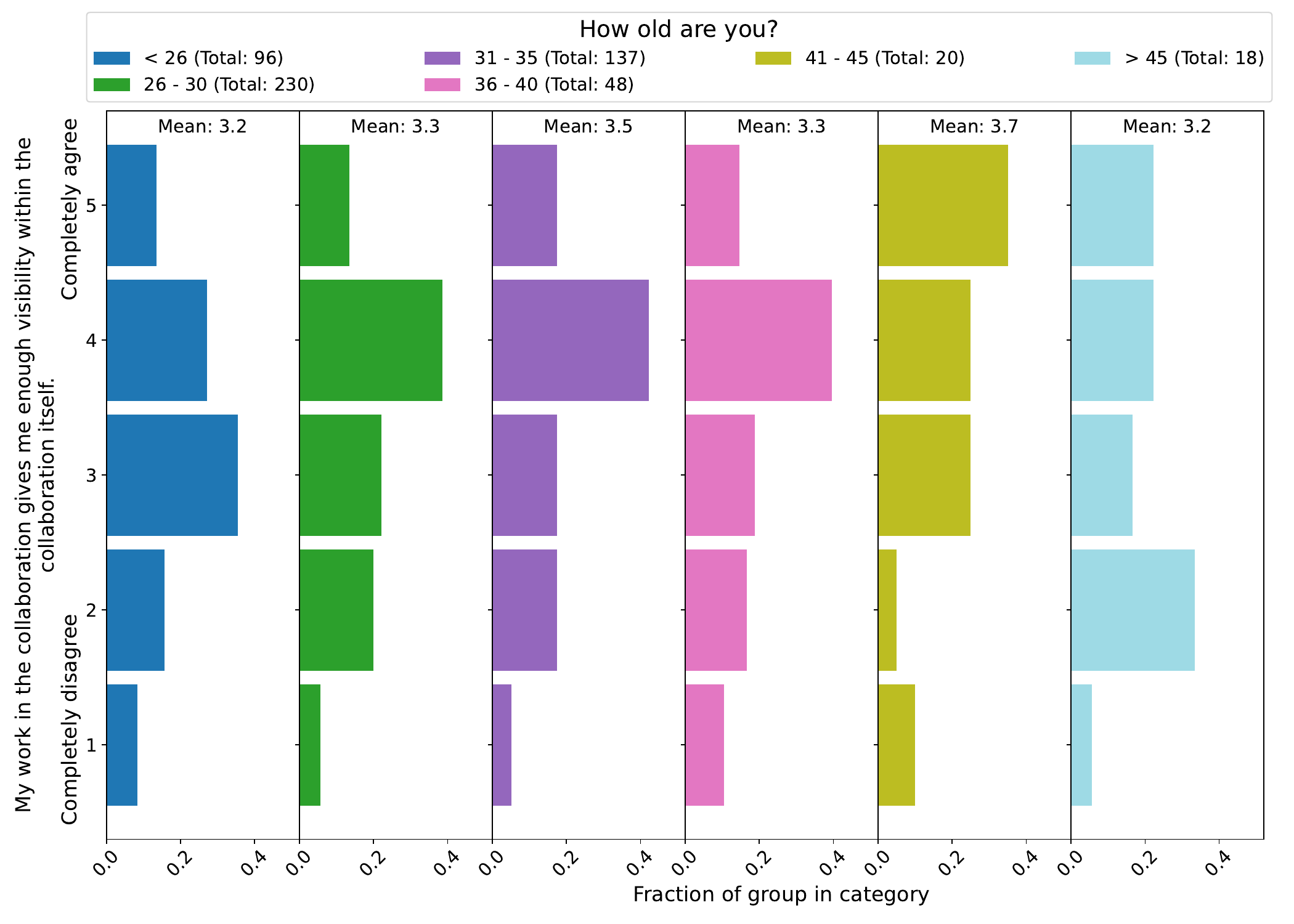}}\\
        \subfloat[]{\label{fig:part2:Q43vQ1}\includegraphics[width=0.49\textwidth]{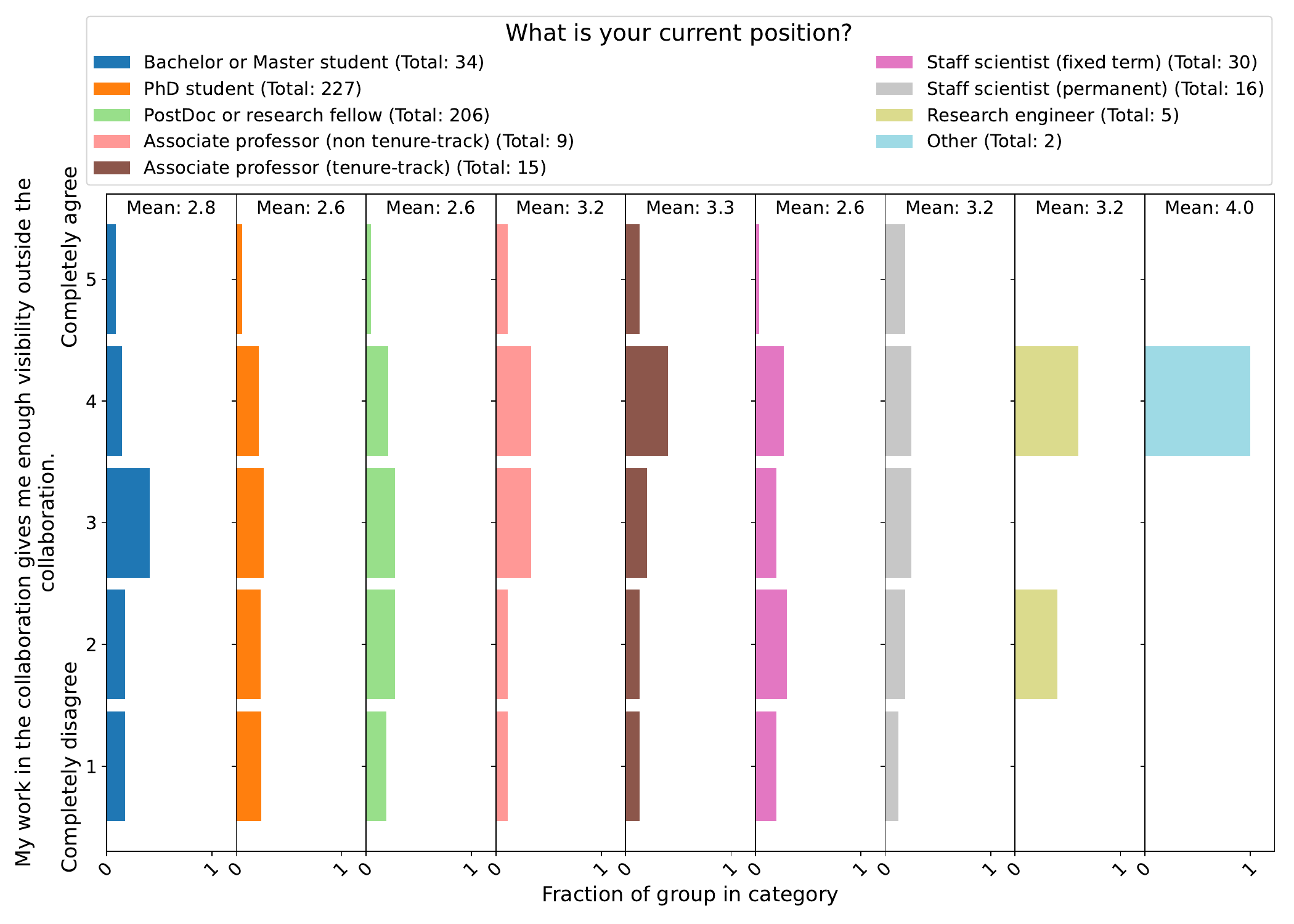}}
        \subfloat[]{\label{fig:part2:Q43vQ8}\includegraphics[width=0.49\textwidth]{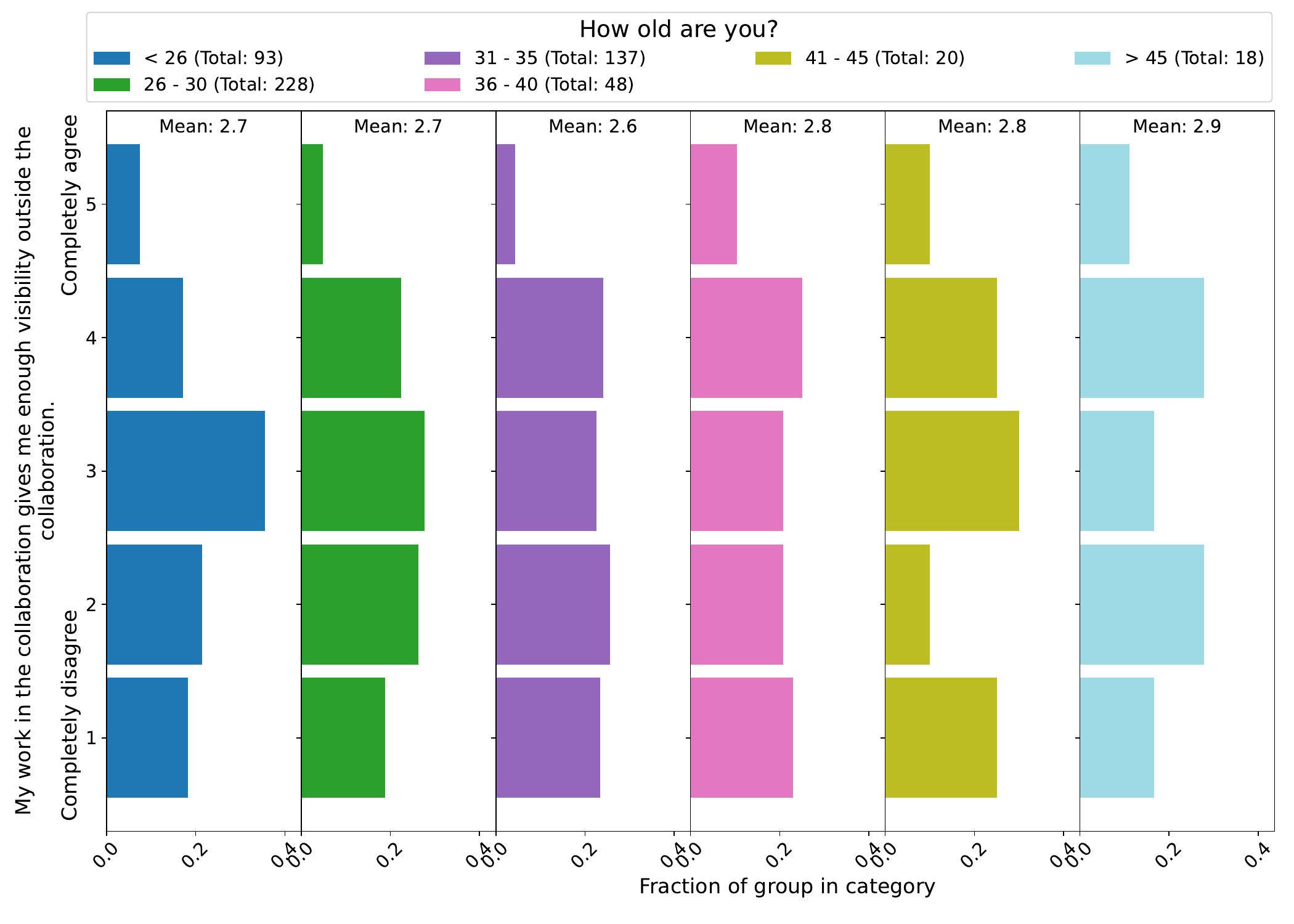}}
    \caption{(Q42--43 vs Q1,8) Correlation between whether respondents feel that their work within a collaboration gives them enough visibility within/outside of the collaboration, and their current position or age. Fractions are given out of all respondents who are part of a collaboration and answered the question.}
    \label{fig:part2:Q42Q43vQ1Q8}
\end{figure}

Figure \ref{fig:part2:Q42Q43vQ1Q8} compares whether respondents feel that they have enough visibility within their collaboration and their
current position, and age.
We observe a positive correlation between visibility and seniority of position, but note that associate professors who are not tenure-track feel more positive about their visibility than those who are, perhaps relating to tenure-track respondents being more likely to have more departmental or teaching responsibilities to compete with their collaboration work, or non tenure-track respondents actively trying harder to achieve visibility.
A similar trend is seen with age, with a drop in visibility for the oldest respondents.
Figure~\ref{fig:part2:Q42Q43vQ1Q8} also shows the same correlations between current position or age and collaboration work providing enough visibility outside the collaboration.
Here, we see similar, but slightly weaker, correlations.
No strong correlations were seen with other respondent demographics.

\begin{figure}[ht!]
    \centering
        \subfloat[]{\label{fig:part2:Q37vQ33}\includegraphics[width=0.49\textwidth]{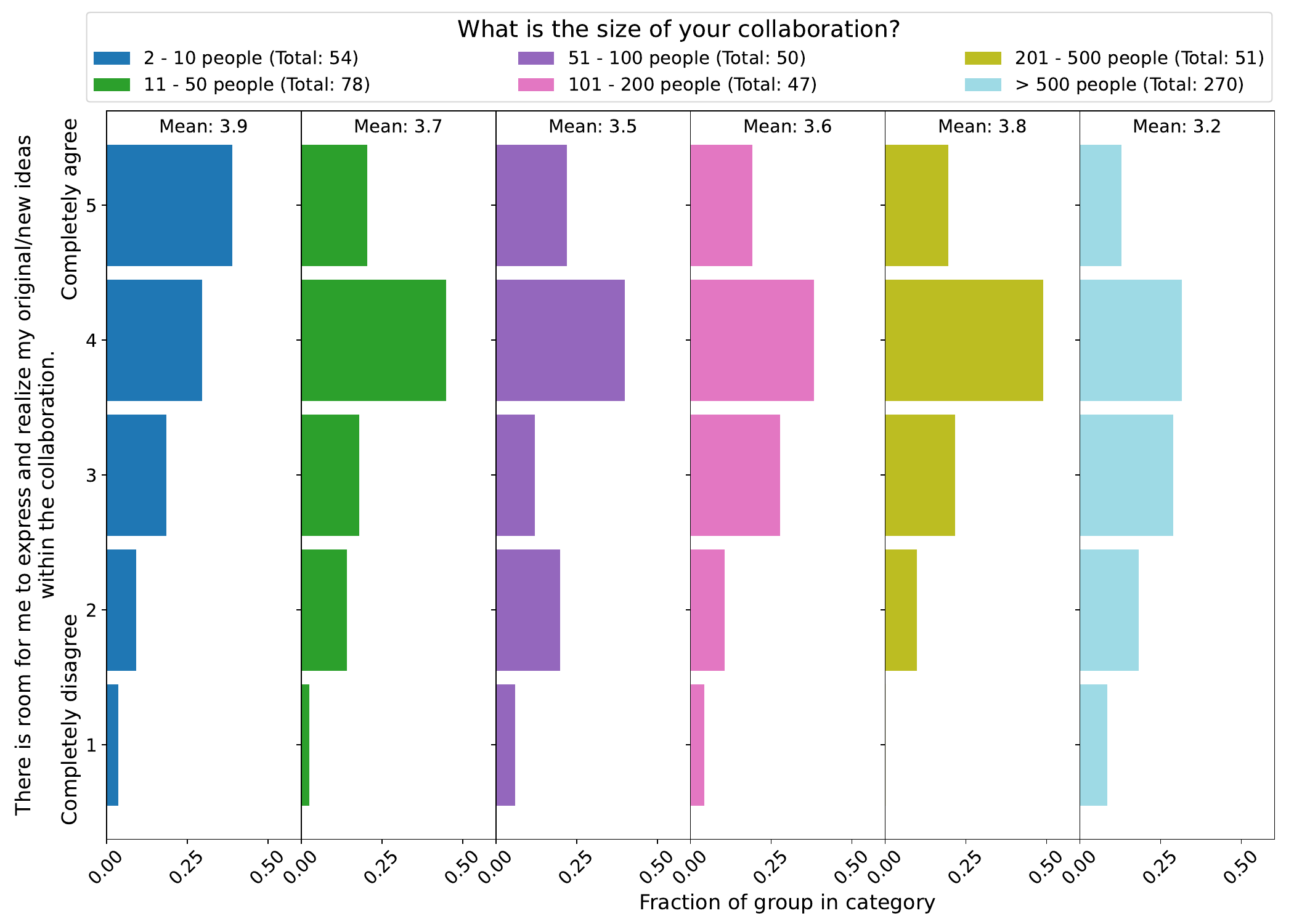}}
        \subfloat[]{\label{fig:part2:Q39vQ33}\includegraphics[width=0.49\textwidth]{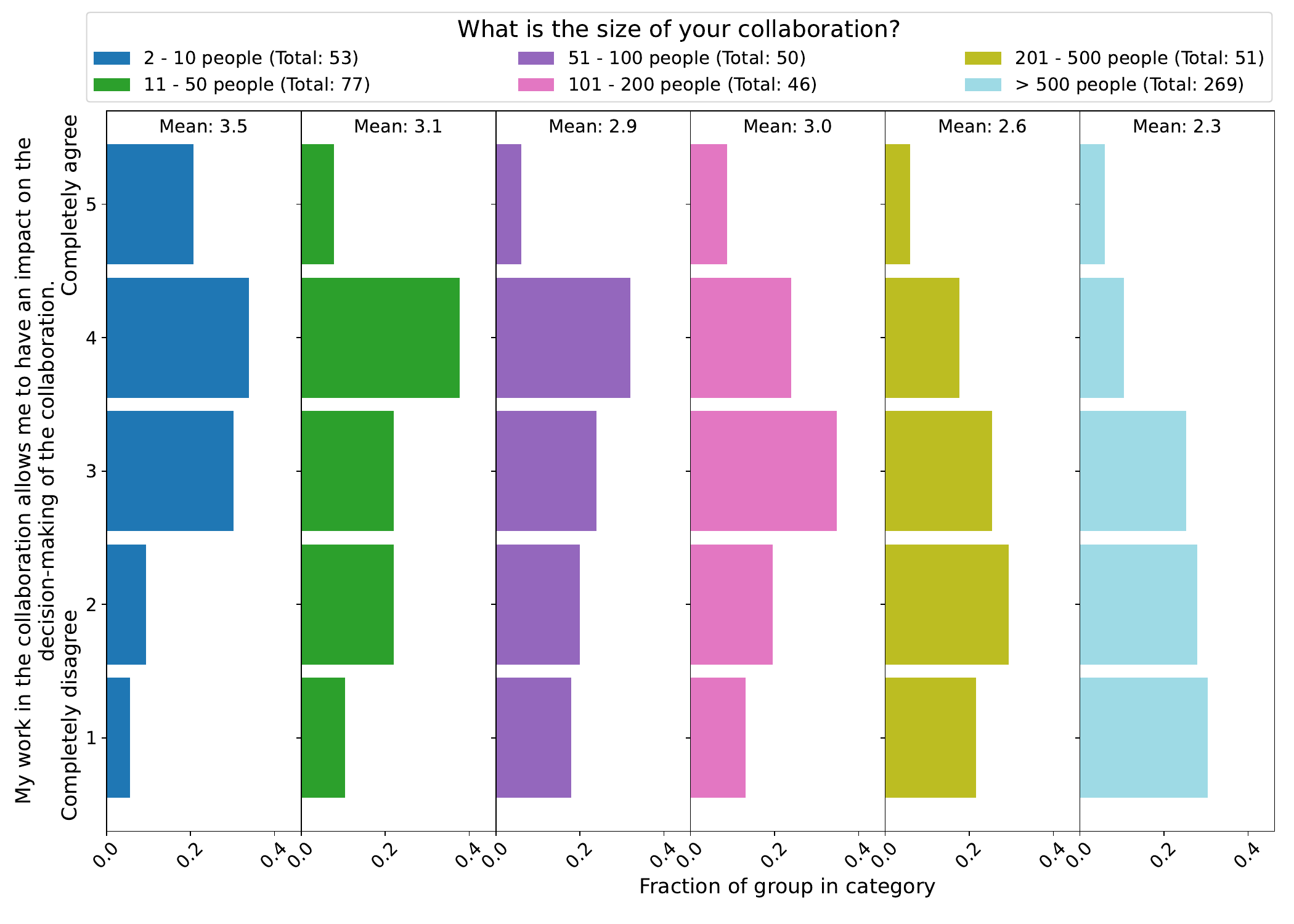}}\\
        \subfloat[]{\label{fig:part2:Q41vQ33}\includegraphics[width=0.49\textwidth]{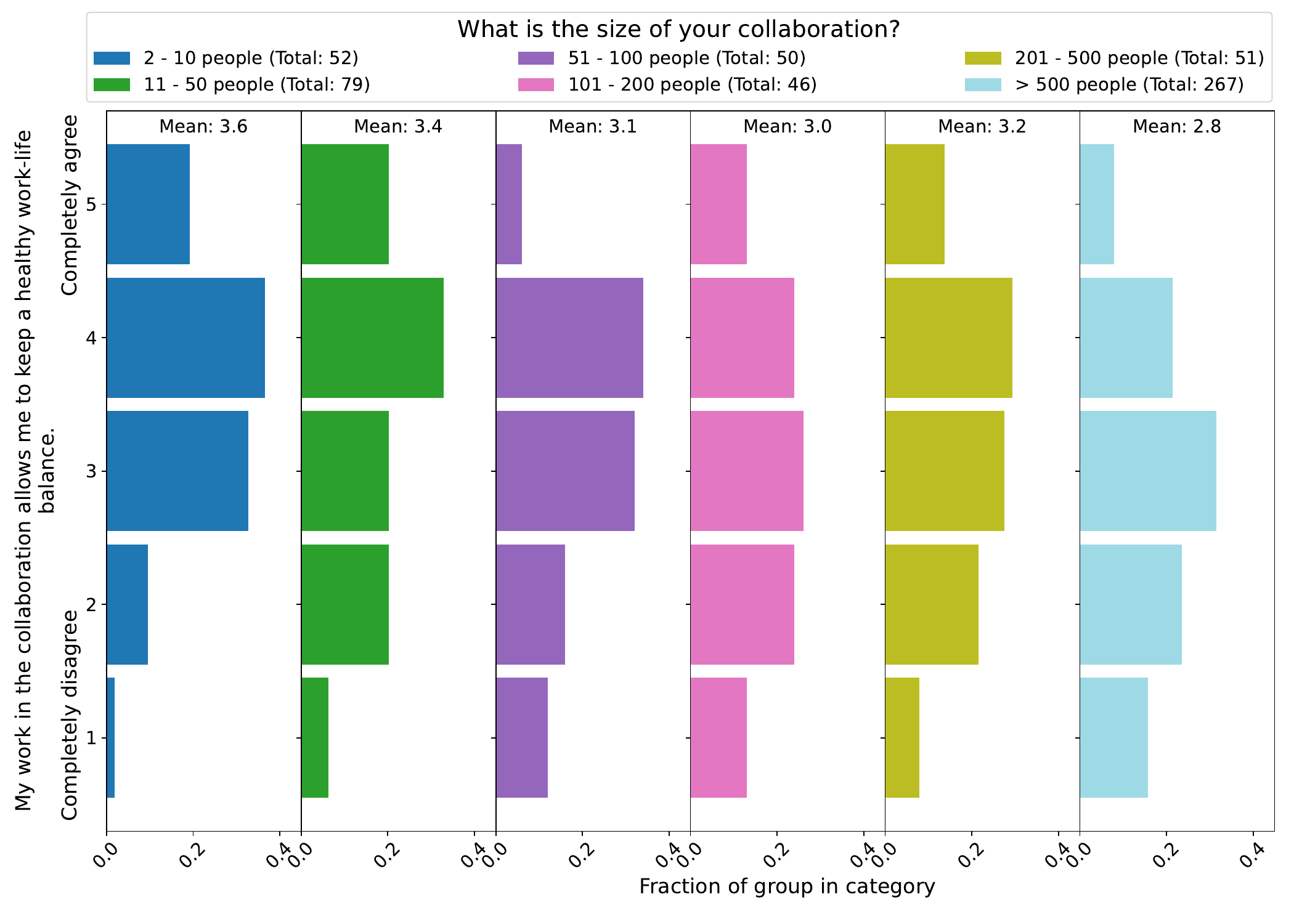}}
        \subfloat[]{\label{fig:part2:Q42vQ33}\includegraphics[width=0.49\textwidth]{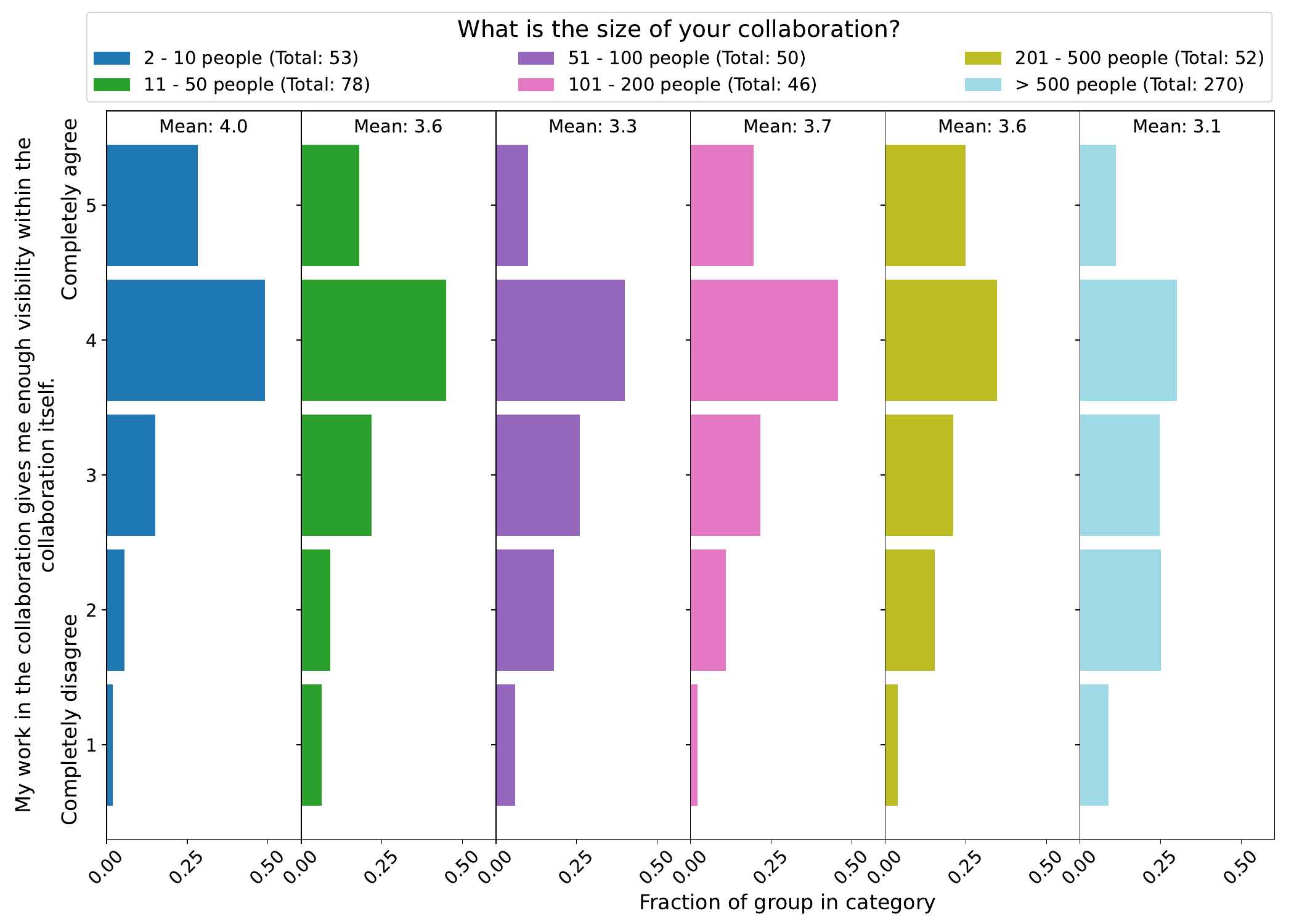}}
    \caption{(Q33 v Q37,39,41,42) Correlations between size of a collaboration and possibility to express respondent’s ideas, impact on decision making, work-life balance, and visibility within the collaboration. Fractions are given out of all respondents who are part of a collaboration and answered the question.}
    \label{fig:part2:Q33vQ37Q39Q41Q42}
\end{figure}

Figure~\ref{fig:part2:Q33vQ37Q39Q41Q42} presents correlations found with respondents views on collaboration work relating to impact, visibility and work-life balance, and collaboration size.
Firstly, we see a clear negative correlation between having an impact on decision-making with collaboration size; respondents in larger collaborations more commonly do not agree that they have an impact on decision-making.
We also see a more fluctuating correlation between visibility and collaboration size.
Finally, we see negative correlations between collaboration size and the collaboration work allowing respondents to achieve a good work-life balance, or realise their original ideas.
We found very little correlation between collaboration size and how well-informed respondents feel about funding, training and job opportunities.
We also saw no strong correlations with collaboration size and other questions in the survey, in particular no correlation was found between collaboration size and frequency of overtime work or feeling stressed.

\begin{figure}[ht!]
    \centering
         \subfloat[]{\label{fig:part2:Q41vQ34}\includegraphics[width=0.49\textwidth]{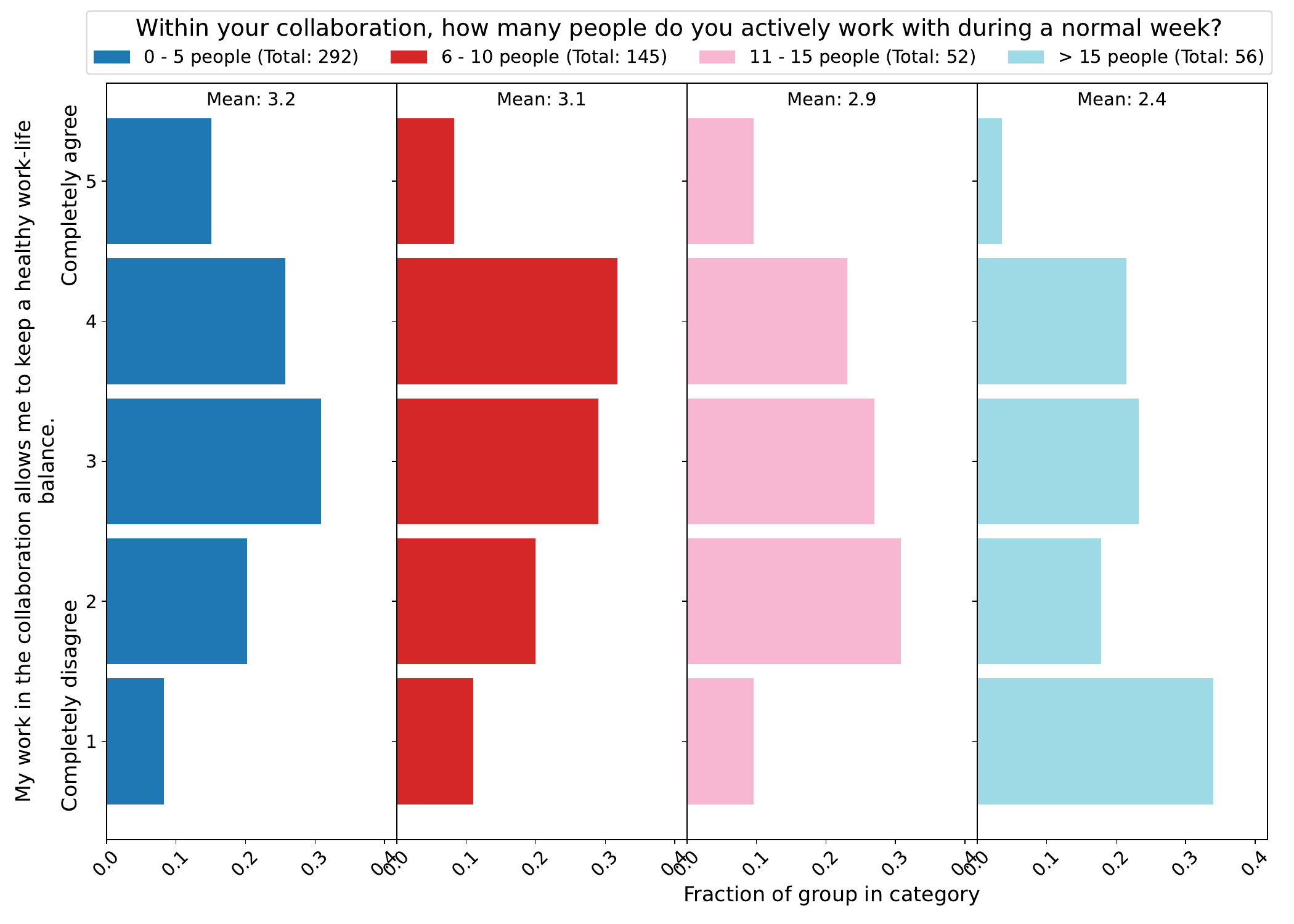}}
         \subfloat[]{\label{fig:part2:Q41vQ36}\includegraphics[width=0.49\textwidth]{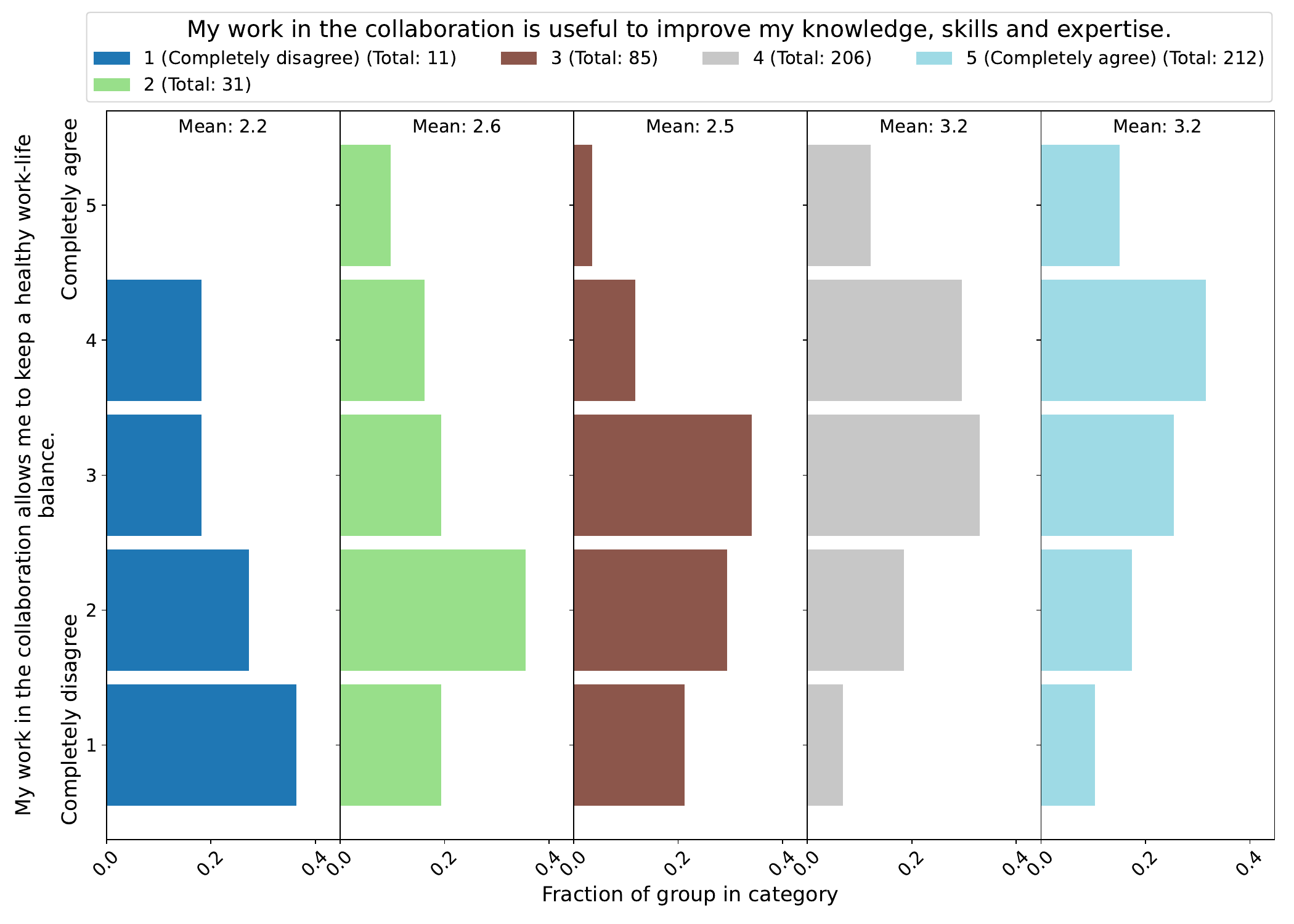}}\\
         \subfloat[]{\label{fig:part2:Q41vQ42}\includegraphics[width=0.49\textwidth]{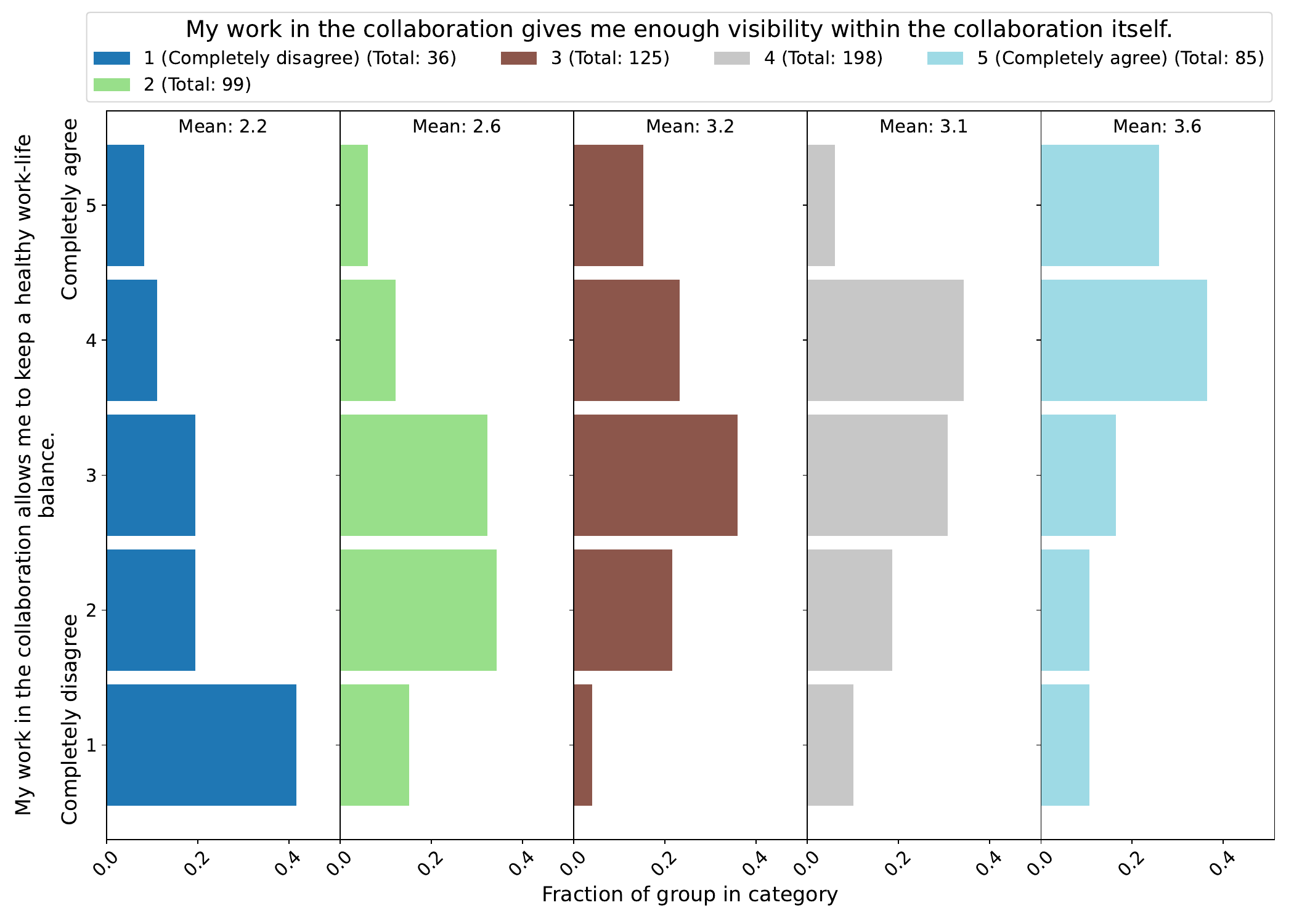}}
         \subfloat[]{\label{fig:part2:Q41vQ49}\includegraphics[width=0.49\textwidth]{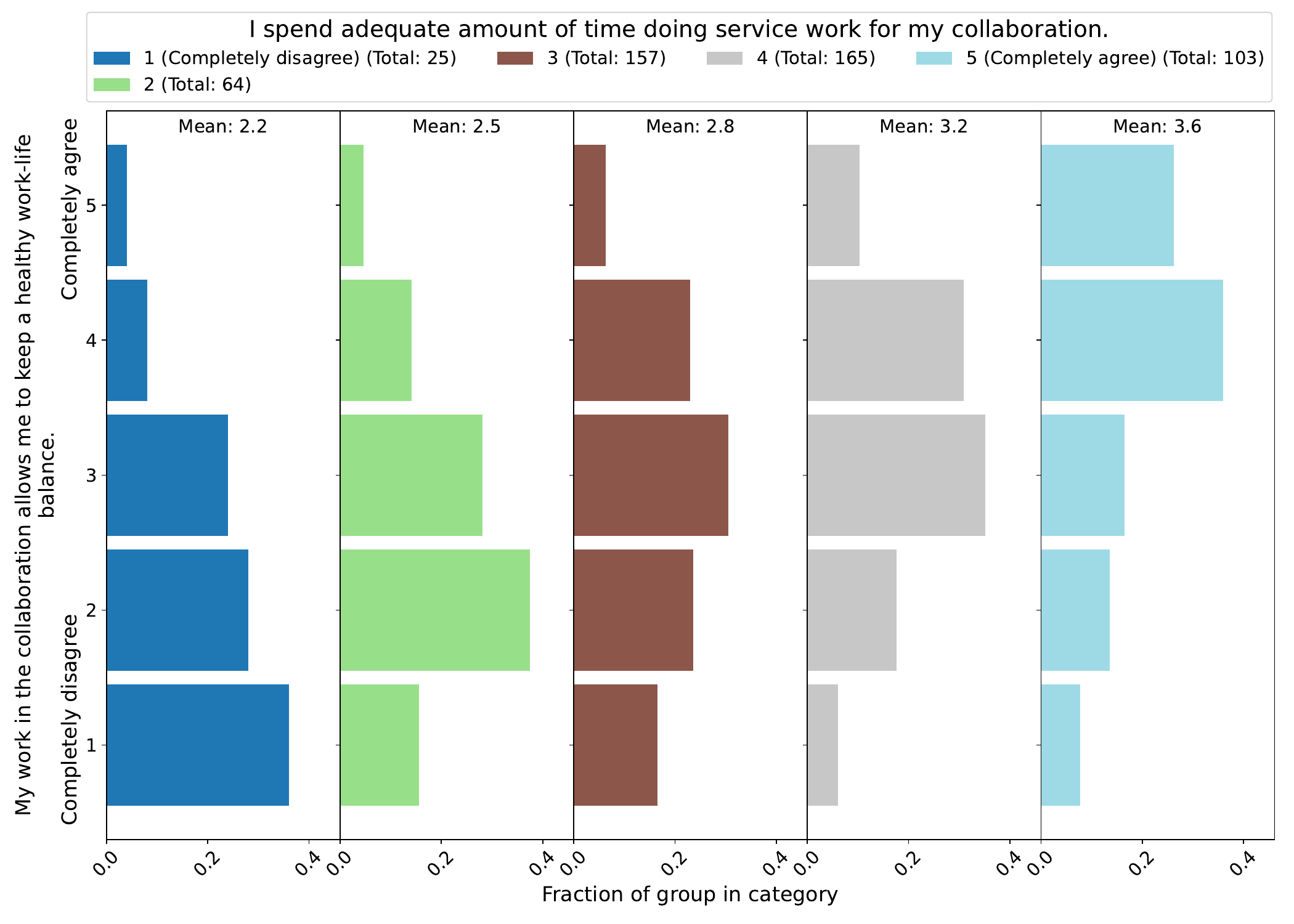}}\\
         \subfloat[]{\label{fig:part2:Q41vQ50}\includegraphics[width=0.49\textwidth]{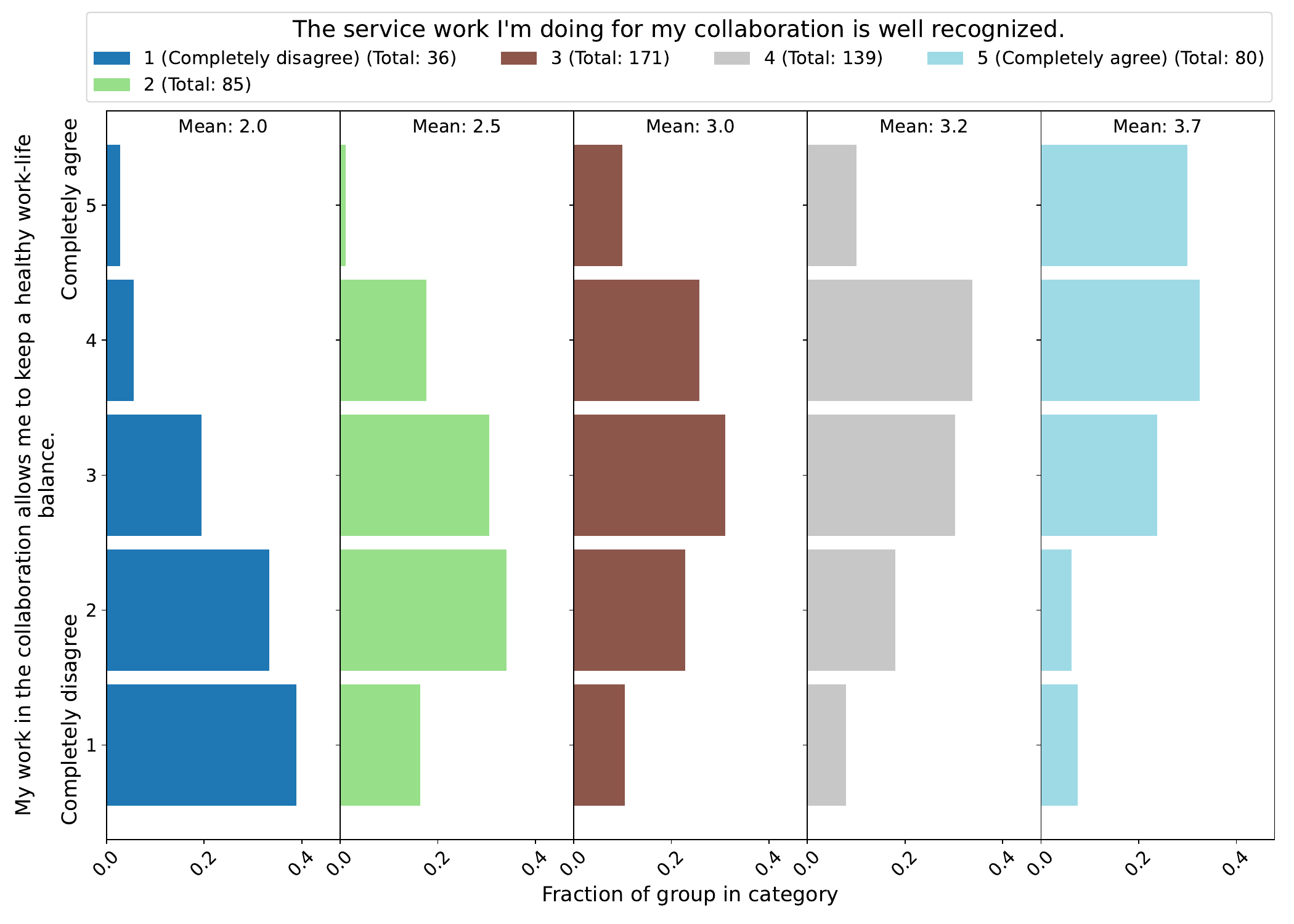}}
         \subfloat[]{\label{fig:part2:Q41vQ80}\includegraphics[width=0.49\textwidth]{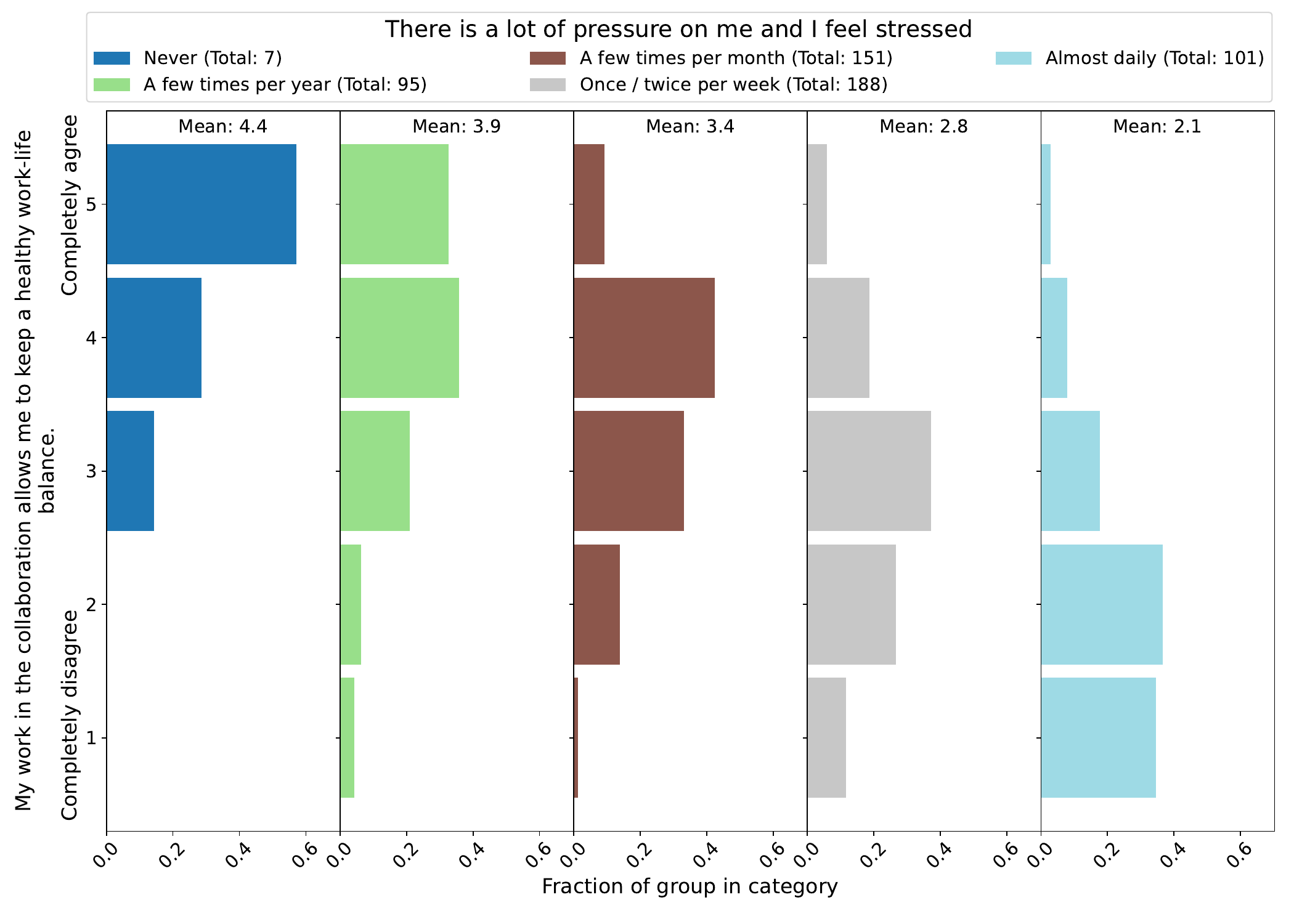}}
    \caption{(Q41 v Q34,36,42,49,50,80) Correlations between work-life balance within a collaboration with other factors such as number of people respondents actively work with, the usefulness of collaboration, visibility; service work, and feelings of stress. Fractions are given out of all respondents who are part of a collaboration and answered the question.}
    \label{fig:part2:Q41vQ34Q36Q42Q49Q50Q80}
\end{figure}

Figure~\ref{fig:part2:Q41vQ34Q36Q42Q49Q50Q80} shows correlations between work-life balance within a collaboration and other questions.
Firstly, we see that respondents who work with more people in a week in their collaboration, achieve a worse work-life balance.
Considering whether collaboration work is useful for improving knowledge, skills and expertise, we see a strong correlation where respondents who are more positive about the usefulness of their collaboration work are more positive about their work-life balance too.
The same positive correlation is seen when considering visibility within the collaboration, adequate time spent doing service work, and recognition for service work.
Finally, not surprisingly, there is a strong negative correlation between frequency of feeling stressed or under pressure and obtaining a healthy work-life balance. 

%%%%%%%========================================================================================
\FloatBarrier
\subsection{The diversity of physics programs}

\begin{figure}[ht!]
    \centering
        \subfloat[]{\label{fig:Q52vQ1}\includegraphics[width=0.49\textwidth]{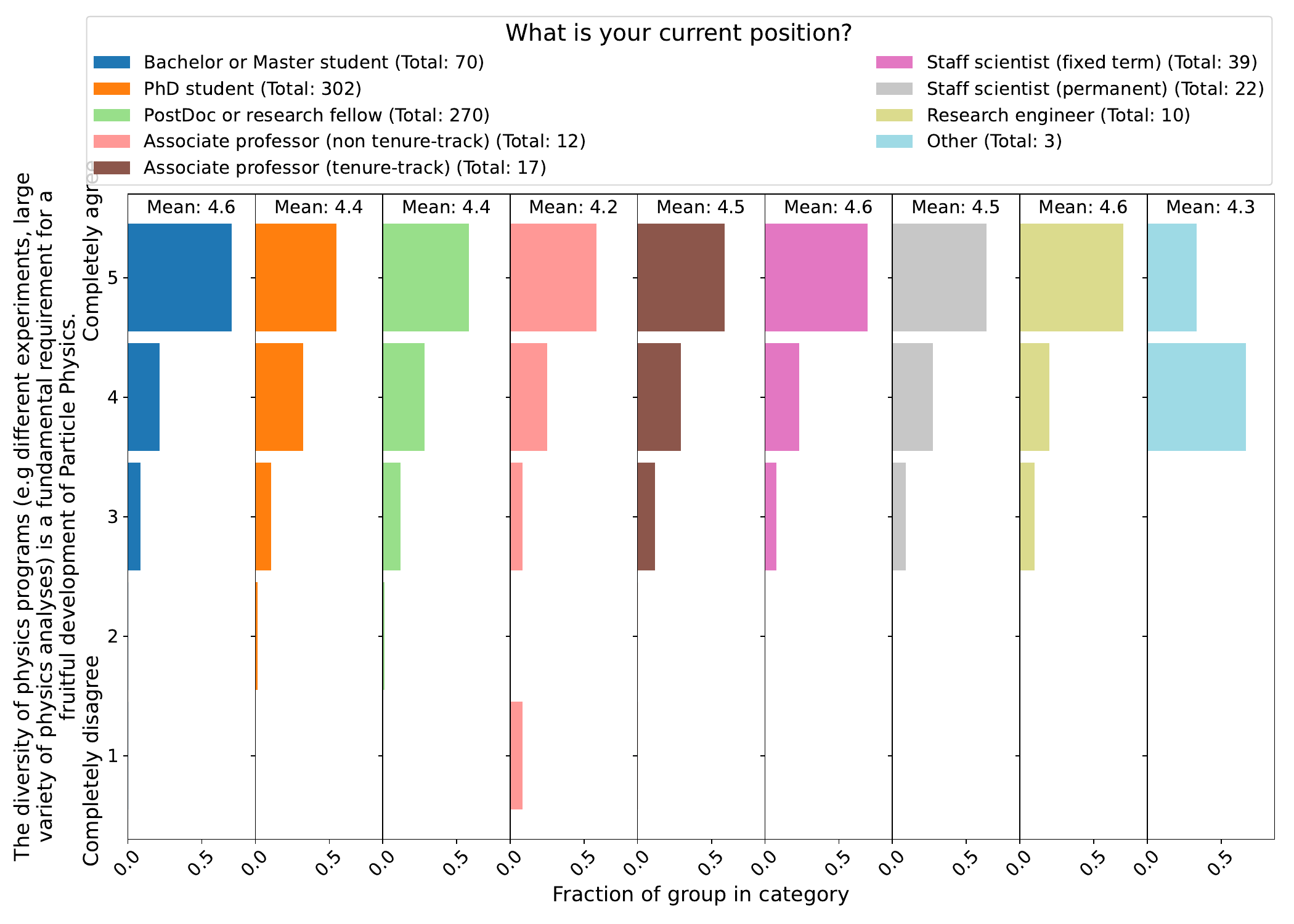}}
        \subfloat[]{\label{fig:Q52vQ4}\includegraphics[width=0.49\textwidth]{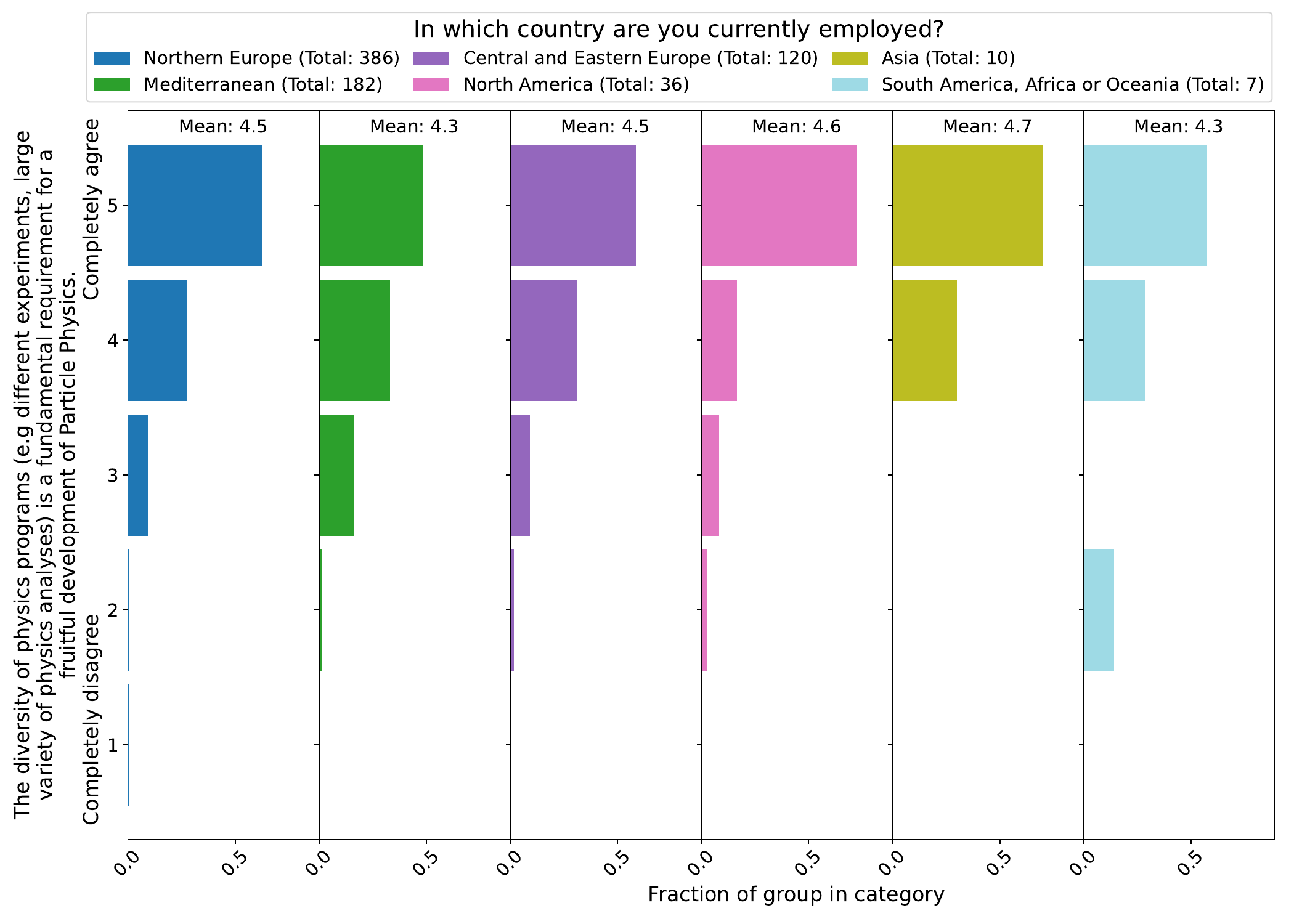}}
    \caption{(Q52 v Q1,4) Correlation between demographics and agreement with the statement that having a diverse physics programs is fundamental for a fruitful development of particle physics. Fractions are given out of those who answered the questions.}
    \label{fig:part2:Q52vQ1Q4}
\end{figure}

Figure~\ref{fig:part2:Q52vQ1Q4} displays the correlation between the respondents' agreement that the diversity of physics programs is a fundamental requirement for a fruitful development of particle physics, and respondent demographics.
The results show very little correlation between the two aspects, indicating that all demographics of respondents seem to largely agree on the statement. 

\begin{figure}[ht!]
    \centering
    \includegraphics[width=0.7\textwidth]{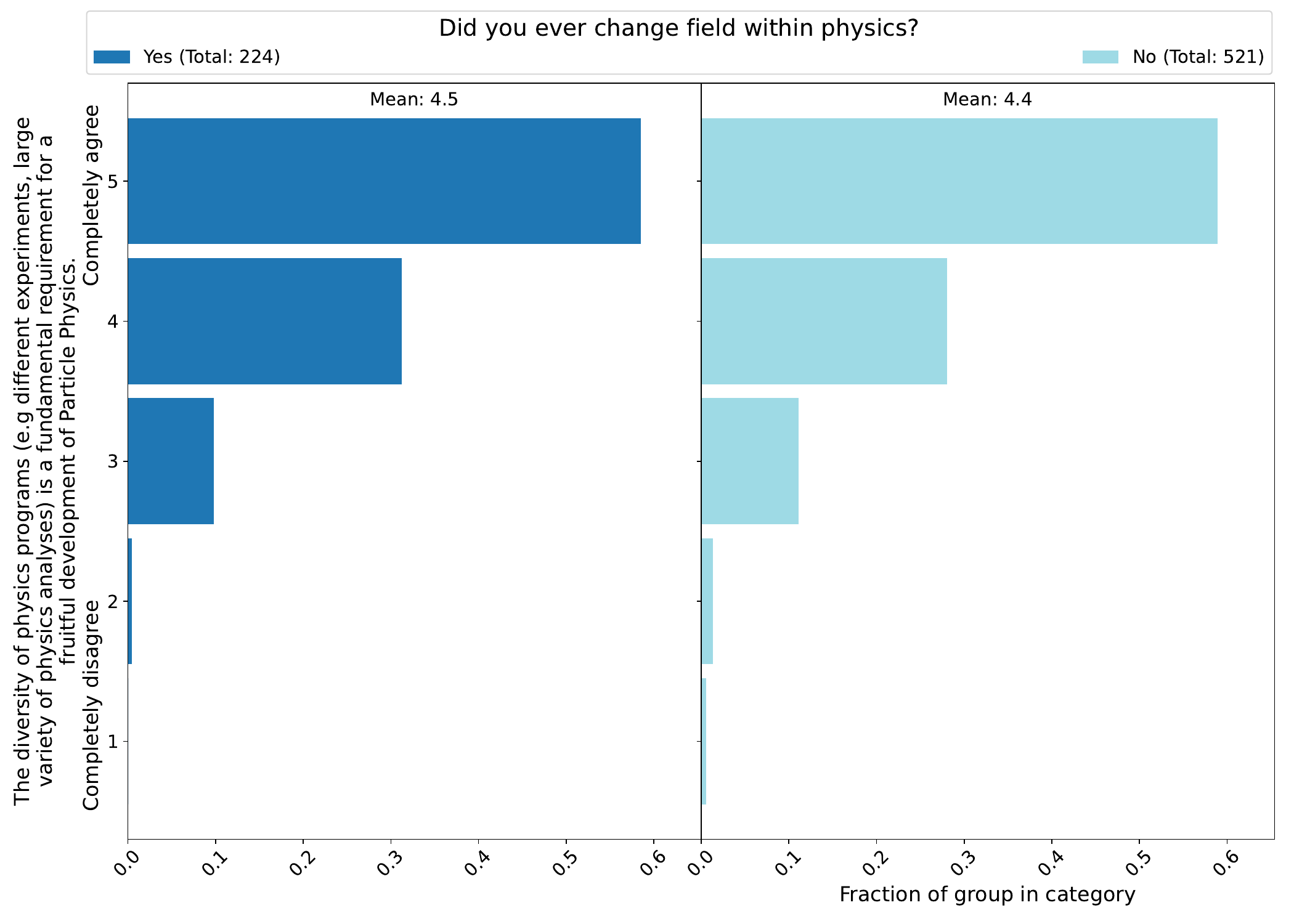}
    \caption{(Q52 v Q84) Correlation between changing field within physics and agreement with the statement that having a diverse physics programs is fundamental for a fruitful development of particle physics. Fractions are given out of those who answered the questions.}
    \label{fig:part2:Q52vQ84}
\end{figure}

Figure~\ref{fig:part2:Q52vQ84} displays the correlation between the agreement with the same statement and whether respondents have already changed field within physics.
The results show very little correlation between the two aspects.
In our studies, we observed no other strong correlations for this topic.

%%%%%%%========================================================================================
\FloatBarrier
\subsection{Career perspectives and information sharing}

%----------------------------------------------------------------------------------------
\subsubsection{Information on career prospects}

\begin{figure}[ht!]
    \centering
        \subfloat[]{\label{fig:part2:Q56vQ1}\includegraphics[width=0.49\textwidth]{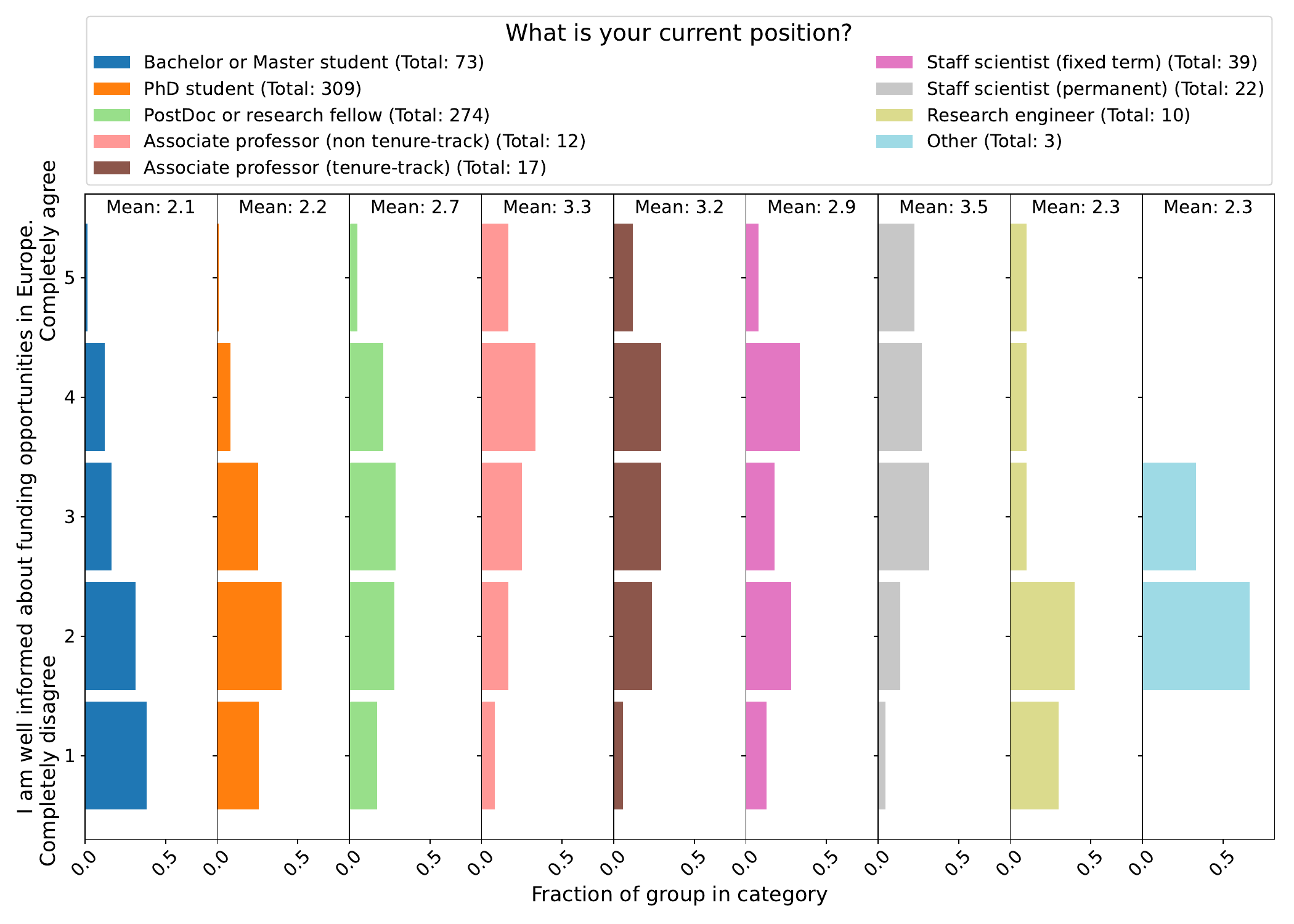}}
        \subfloat[]{\label{fig:part2:Q56vQ4}\includegraphics[width=0.49\textwidth]{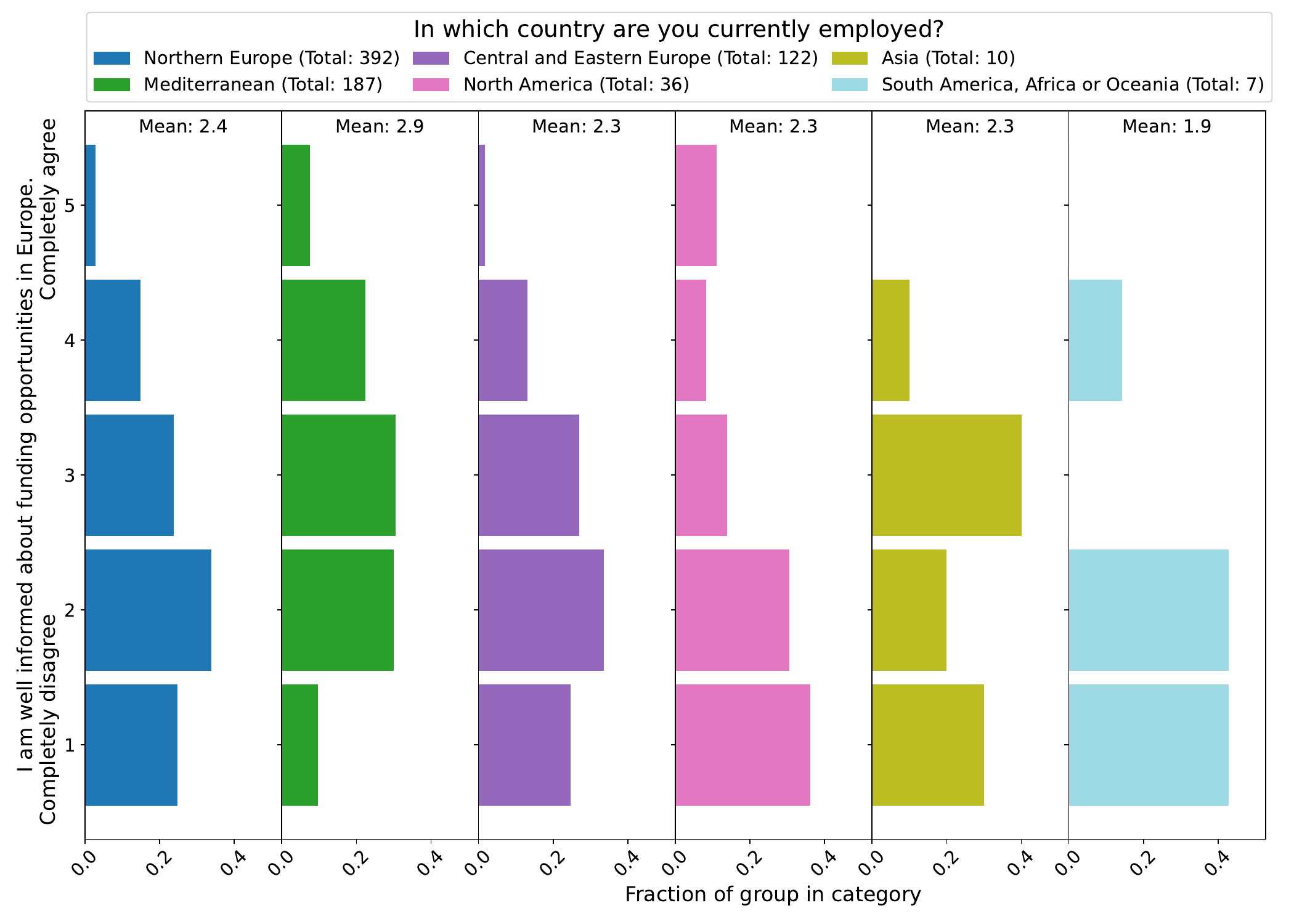}}\\
        \subfloat[]{\label{fig:part2:Q57vQ4}\includegraphics[width=0.49\textwidth]{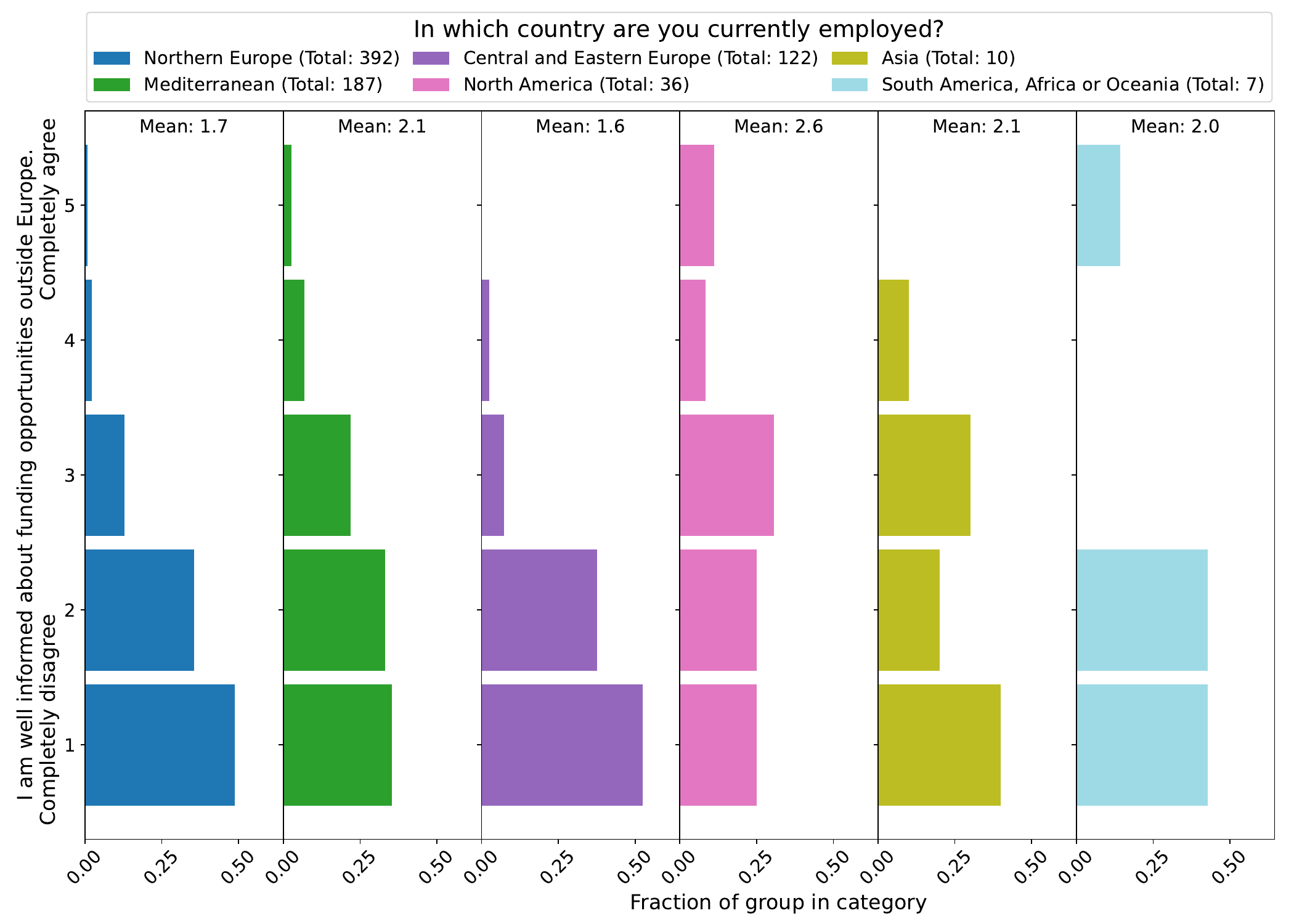}}
        \subfloat[]{\label{fig:part2:Q58vQ1}\includegraphics[width=0.49\textwidth]{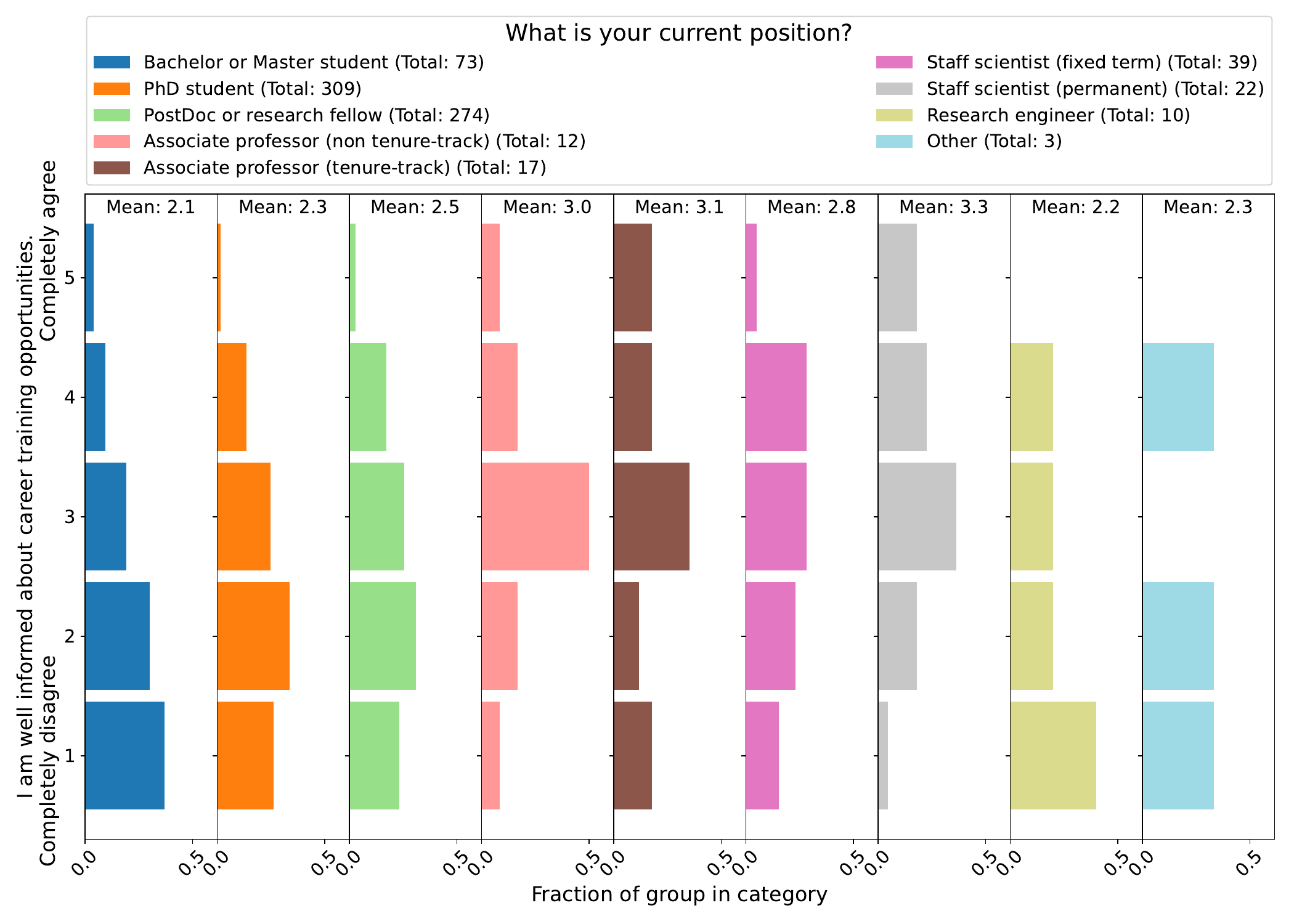}}
    \caption{(Q56 v Q1,4; Q57 v Q4; Q58 v Q1) Correlations between how well-informed respondents feel and selected demographics. Fractions are given out of all respondents who answered the questions.}
    \label{fig:part2:Q56vQ1Q4_Q57vQ4_Q58vQ1}
\end{figure}

In this section we show correlations found between how well-informed respondents feel about different aspects of career progression, and other questions.
When asked how well-informed respondents feel about funding opportunities in the country they are hired, answers are not significantly correlated with most of the respondents' demographics, in particular the country where they are hired.
However, we see a strong positive correlation with position, age and contract length; for example, respondents with more senior positions are more well-informed.
This correlation is also observed when considering how well-informed respondents feel about funding opportunities inside or outside of Europe, shown in Figure~\ref{fig:part2:Q56vQ1Q4_Q57vQ4_Q58vQ1}.
We also find that when comparing respondents employed within Europe to those employed outside of it, those within Europe are only slightly more well-informed about European funding opportunities, whilst respondents employed in Europe are less well-informed about non-European funding opportunities.

When asked how well-informed respondents feel about career training and job application training opportunities, the correlations are similar.
No strong correlations are seen with demographics such as country of employment, field of research or type of institution.
Conversely, we see that respondents in more senior positions become more aware of training opportunities, particularly those in tenure-track or permanent positions as shown in Figure~\ref{fig:part2:Q56vQ1Q4_Q57vQ4_Q58vQ1}.
Similar correlations are seen when asking how well-informed respondents feel about what is needed to advance their careers in academia.
Finally, when asked how well-informed respondents feel about what is needed to advance their careers outside of academia, or where to find advice and guidance about career progression, correlations are similar but weaker.

\begin{figure}[ht!]
    \centering
    \includegraphics[width=0.65\textwidth]{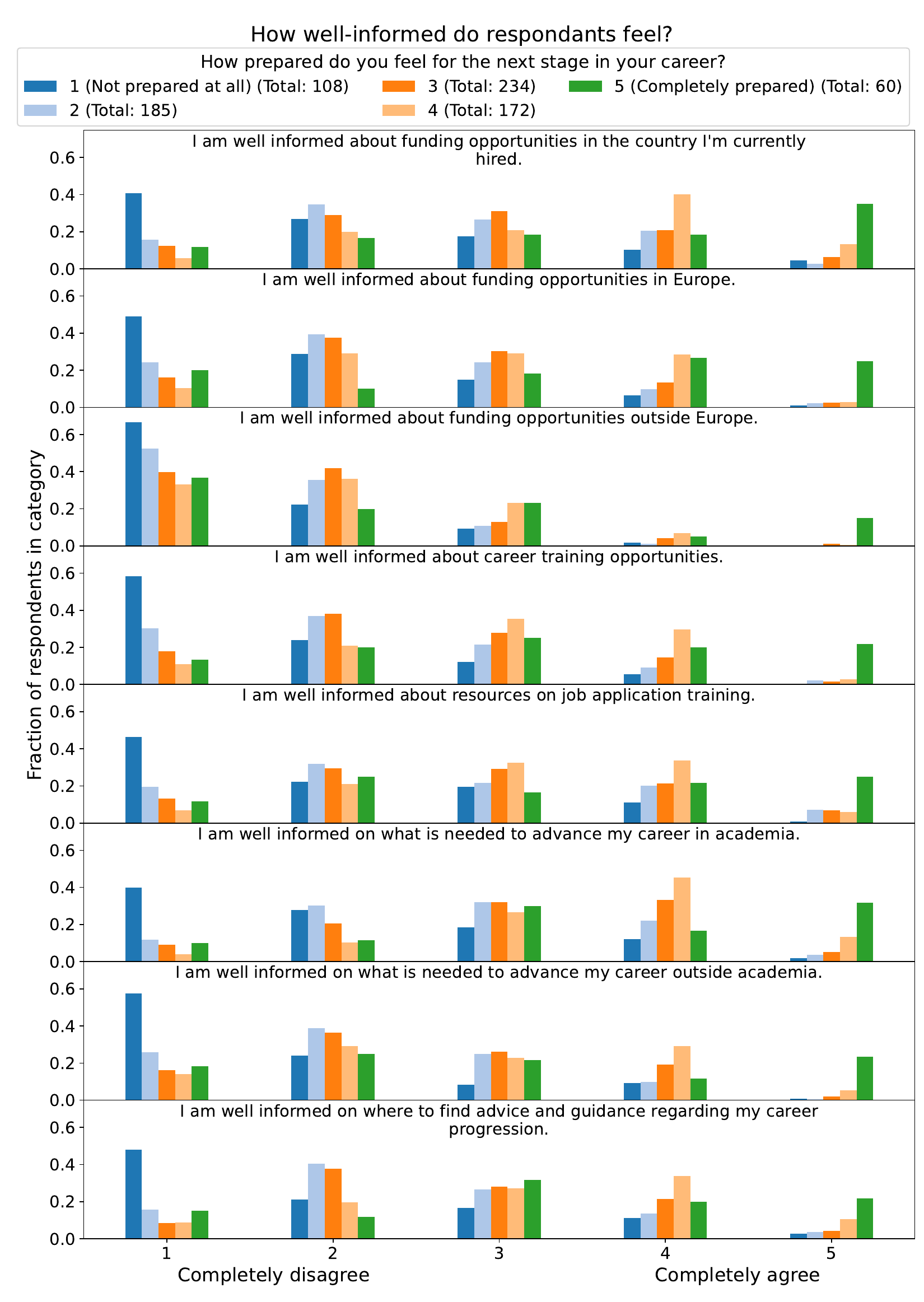}
    \caption{(Q55--Q62 v 64) Correlations between how well-informed respondents' feel about funding and training opportunities, and how prepared they feel for the next stage in their career. Fractions are given out of all respondents.}
    \label{fig:part2:FeelInformedvQ64}
\end{figure}

\begin{figure}[ht!]
    \centering
    \includegraphics[width=0.7\textwidth]{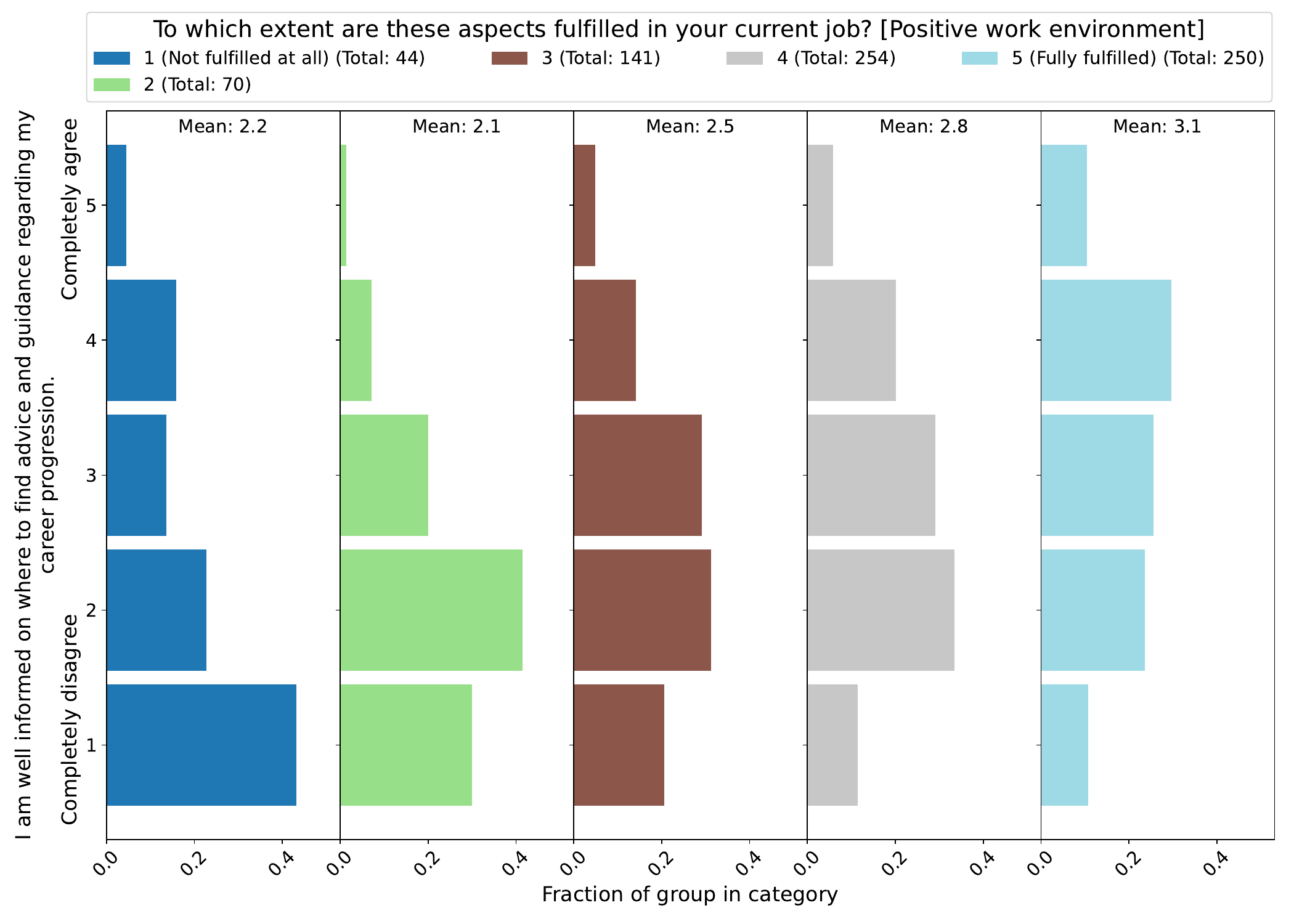}
    \caption{(Q62 v Q74) Correlations between how well-informed respondents' feel about where to find advice and guidance regarding career progression, and how much they agree they have a positive work environment. Fractions are given out of all respondents.}
    \label{fig:part2:Q62vQ74}
\end{figure}

\begin{figure}[ht!]
    \centering
    \includegraphics[width=0.7\textwidth]{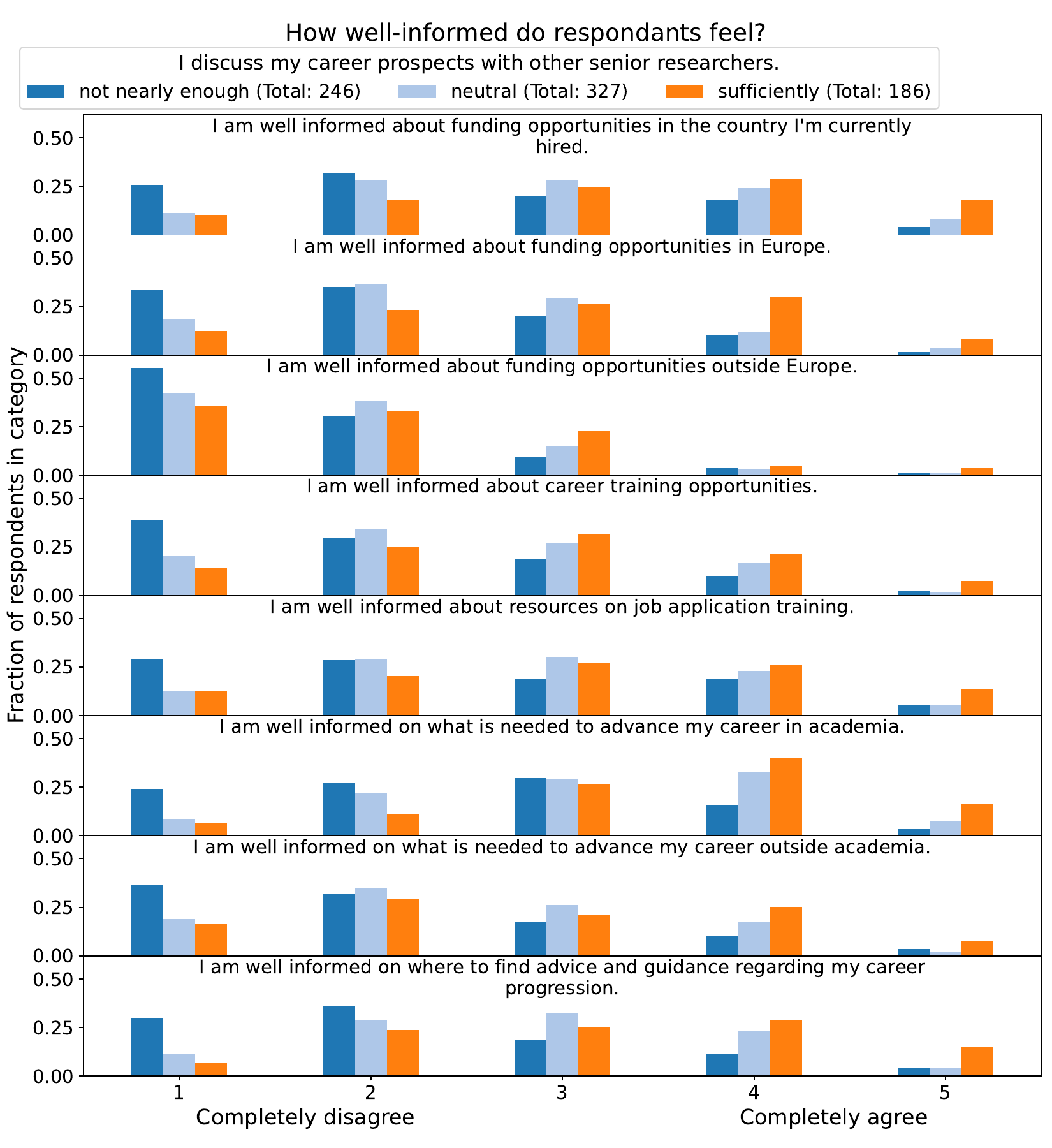}
    \caption{(Q55--62 v Q66) Correlations between how well-informed respondents' feel about funding and training opportunities, and how satisfied they feel about their level of career-prospect discussions with senior researchers. Fractions are given out of all respondents.}
    \label{fig:part2:FeelInformedvQ66}
\end{figure}

Following this, we studied how well-informed respondents feel about funding and training opportunities, correlated with other questions unrelated to demographics.
We saw a very strong correlation with how prepared respondents feel for the next stage in their careers, as shown in Figure~\ref{fig:part2:FeelInformedvQ64}.
Respondents also feel more well-informed about these issues when they have a positive work environment, shown for example in Figure~\ref{fig:part2:Q62vQ74}.
How well-informed respondents are increases substantially when they are also satisfied with the level of discussion about career prospects they have with their senior researchers, as shown in Figure~\ref{fig:part2:FeelInformedvQ66}.
The same correlation is also present considering discussion with respondents' supervisors.

\begin{figure}[ht!]
    \centering
        \subfloat[]{\label{fig:part2:Q64vQ7}\includegraphics[width=0.49\textwidth]{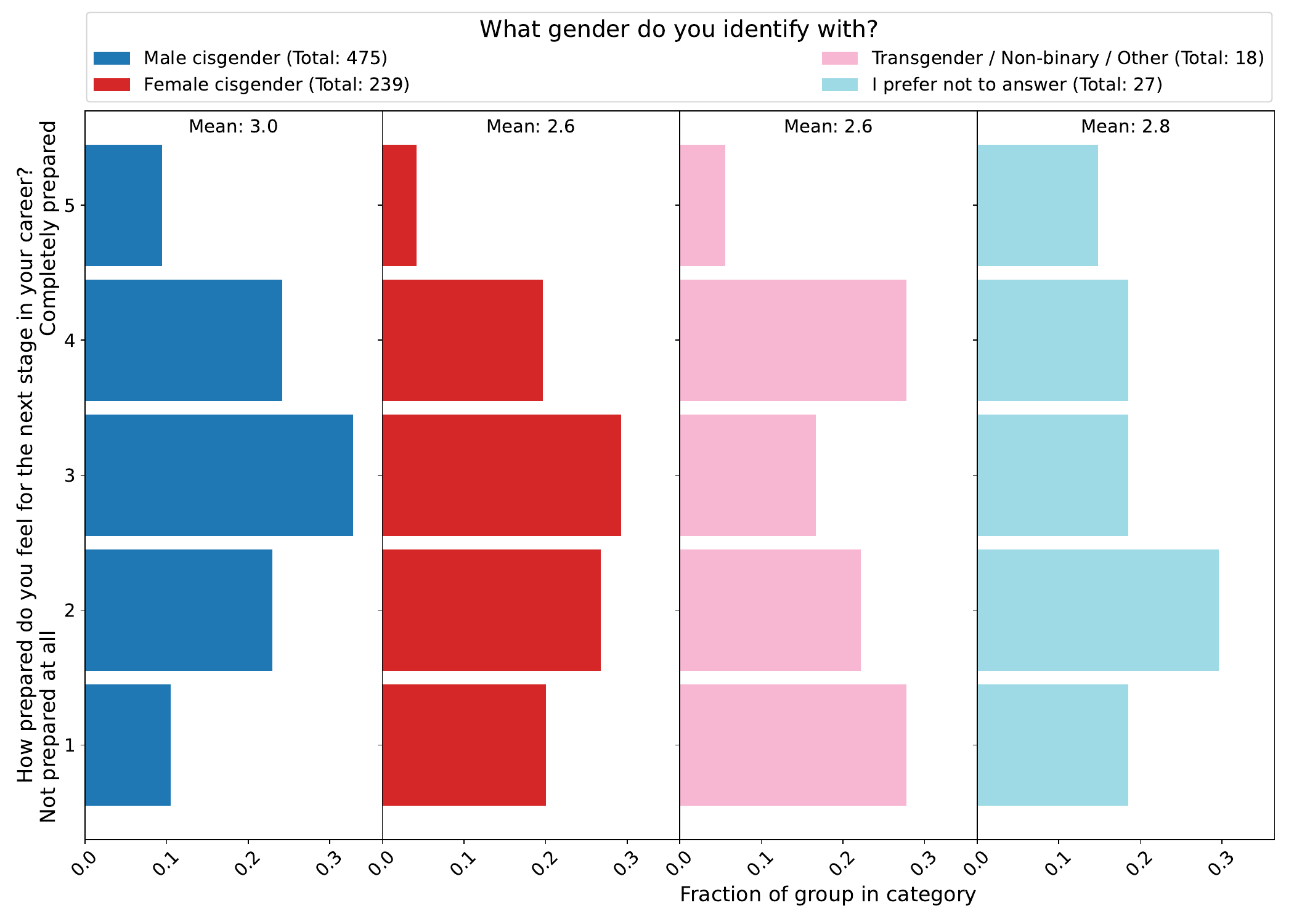}}
        \subfloat[]{\label{fig:part2:Q64vQ1}\includegraphics[width=0.49\textwidth]{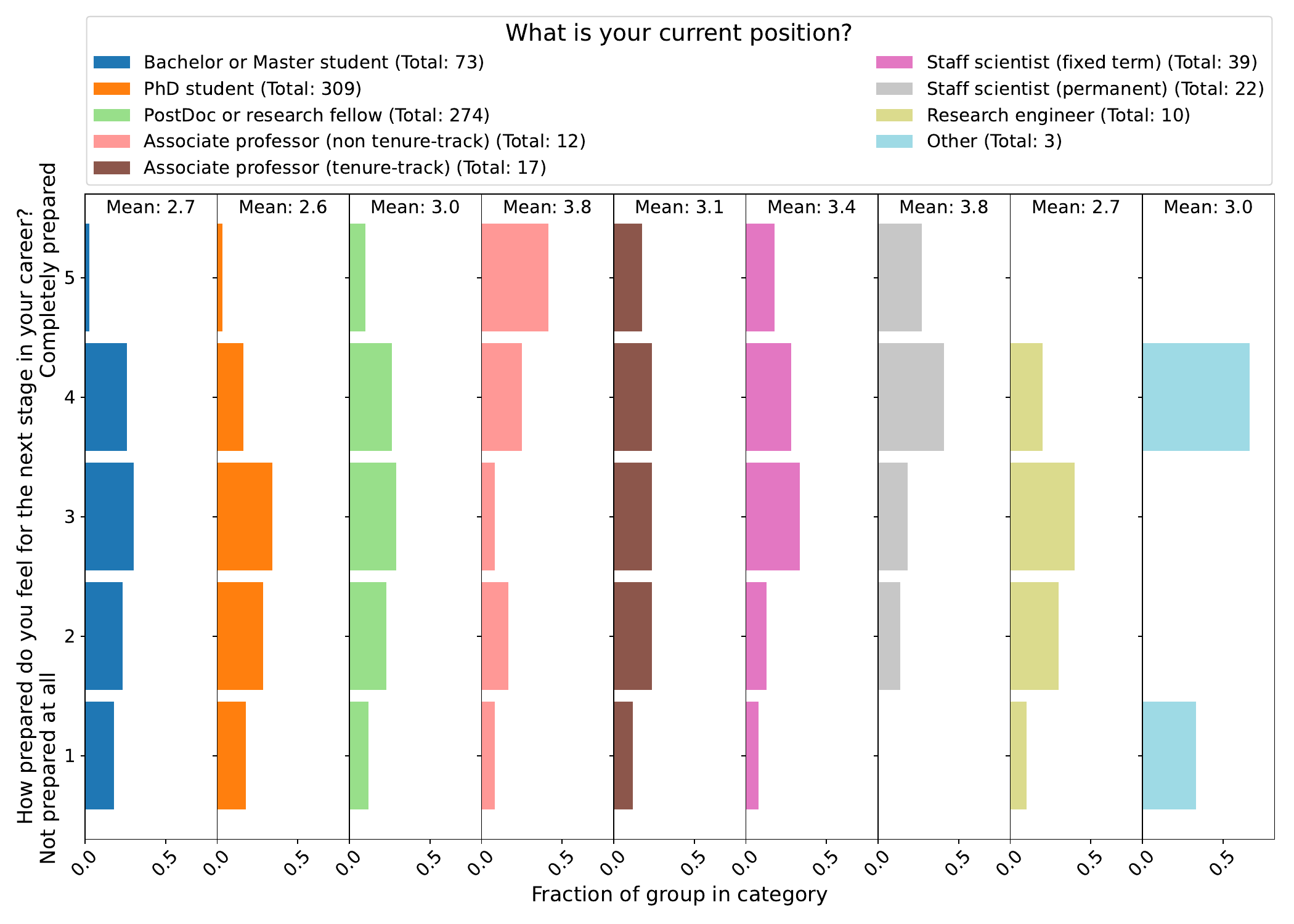}}
    \caption{(Q64 v Q1,7) Correlations between how prepared respondents feel for the next stage in their careers, and selected demographics. Fractions are given out of all respondents.}
    \label{fig:part2:Q64vQ1Q7}
\end{figure}

Considering next how prepared respondents feel for the next stage in their careers, selected correlations are shown in Figure~\ref{fig:part2:Q64vQ1Q7}.
We see that cisgender males feel somewhat more prepared than respondents of other genders.
We also see a strong correlation with respondents' position, where there is a striking jump between postdocs and non tenure-track associate professors, and a difference between associate professors and staff scientists.

\begin{figure}[ht!]
    \centering
        \subfloat[]{\label{fig:part2:Q66vQ5}\includegraphics[width=0.49\textwidth]{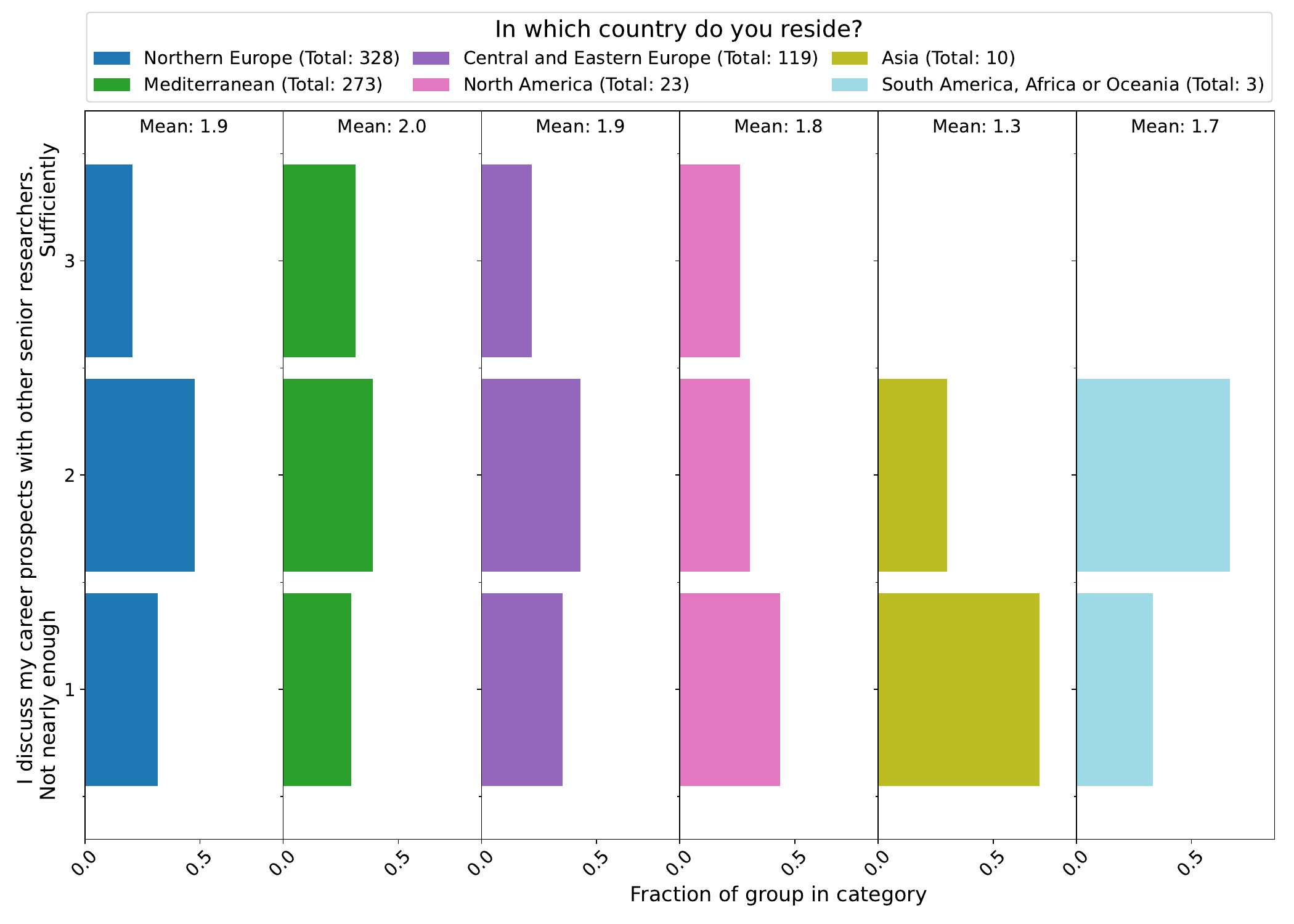}}
        \subfloat[]{\label{fig:part2:Q67vQ5}\includegraphics[width=0.49\textwidth]{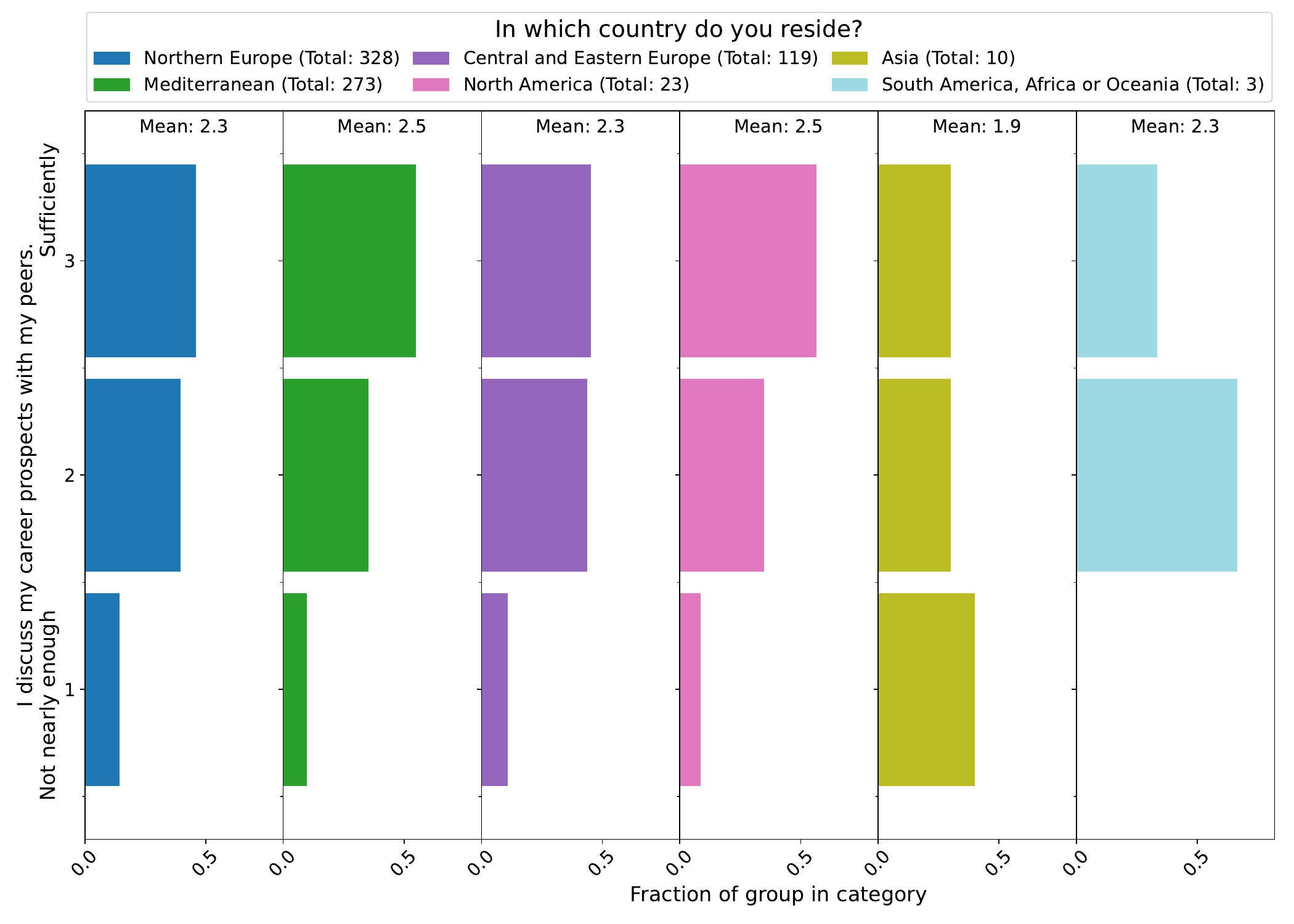}}
    \caption{(Q66--67 v Q5) Correlations between how sufficient respondents' discussion of career prospects with country of residence. Fractions are given out of all respondents.}
    \label{fig:part2:Q66Q67vQ5}
\end{figure}

\begin{figure}[ht!]
    \centering
    \includegraphics[width=0.7\textwidth]{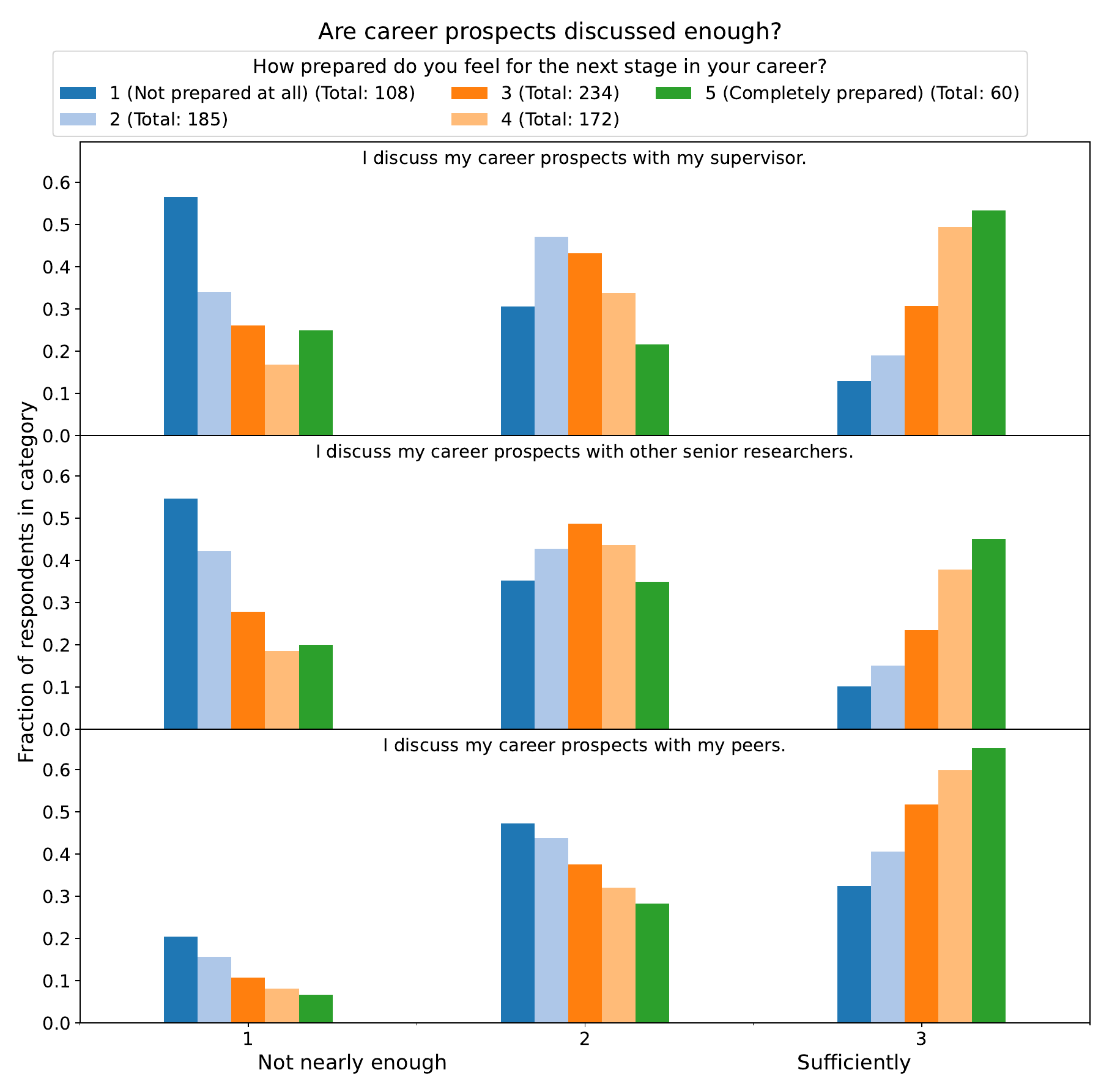}
    \caption{(Q65--67 v Q64) Correlations between satisfied respondents feel about their level of career-prospect discussions with colleagues, and how prepared they feel for the next stage in their career. Fractions are given out of all respondents.}
    \label{fig:part2:DiscussionvQ64}
\end{figure}

Moving to whether respondents feel that they discuss their career prospects with others enough, there are no strong correlations between demographics and discussions with respondents' supervisors.
When considering discussion with other senior researchers or their peers, we see in Figure~\ref{fig:part2:Q66Q67vQ5} that respondents residing in Asia are less satisfied with the amount of discussion (although there is a low sample size in this category).
For completion, it is also very clear from Figure~\ref{fig:part2:DiscussionvQ64} that feeling prepared for the next career stage is highly correlated with sufficient discussion with supervisors, other senior researchers and respondents' peers.

%----------------------------------------------------------------------------------------
\FloatBarrier
\pagebreak
\subsubsection{Valuing research skills}

\begin{figure}[ht!]
    \centering
        \includegraphics[width=0.8\textwidth]{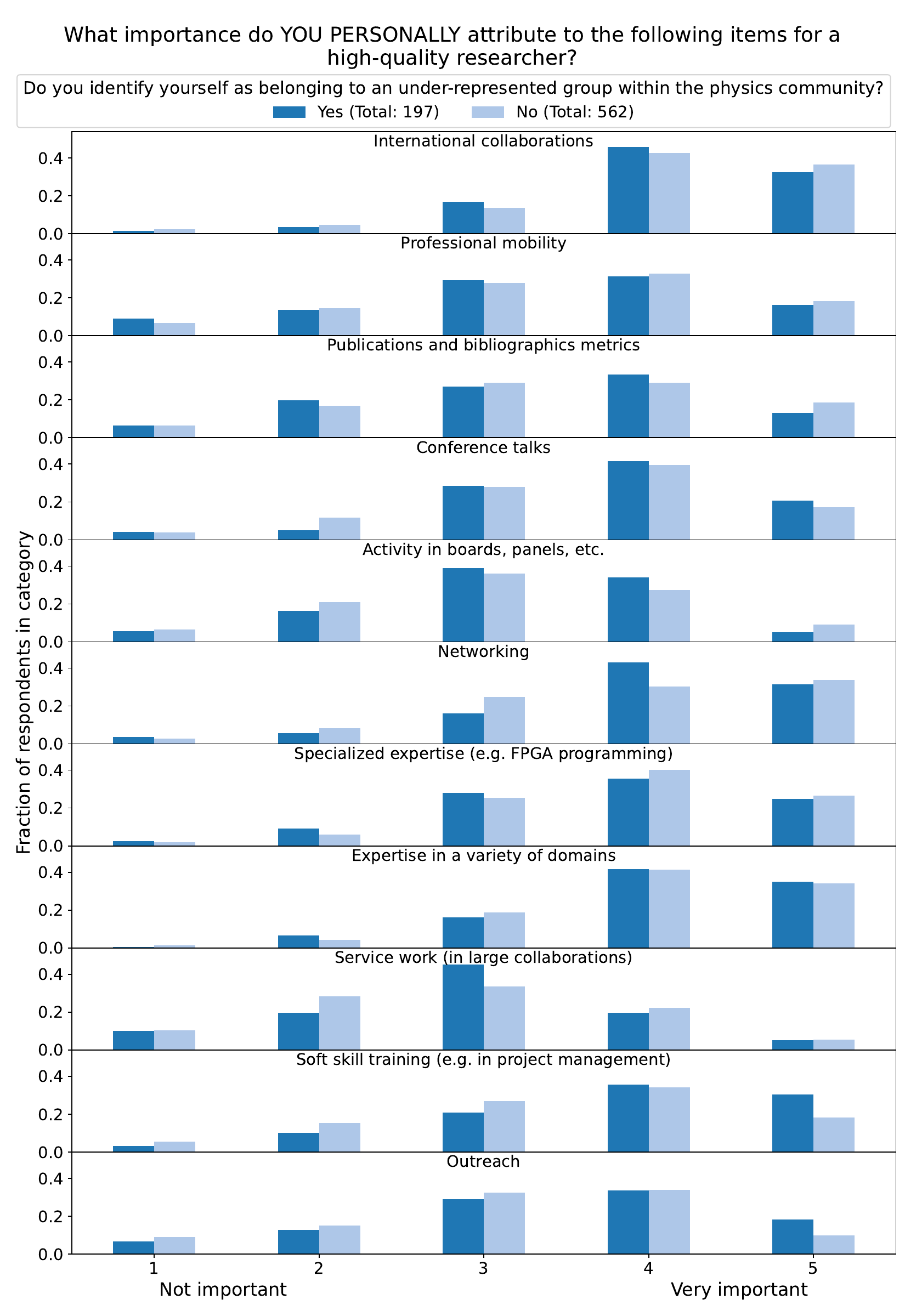}
    \caption{(Q68 v Q9) Correlations between how important respondents feel various items related to a successful academic career are, personally, and whether they belong to an under-represented community. Fractions are given out of all respondents.}
    \label{fig:part2:Q68vQ9}
\end{figure}

\begin{figure}[ht!]
    \centering
        \includegraphics[width=0.8\textwidth]{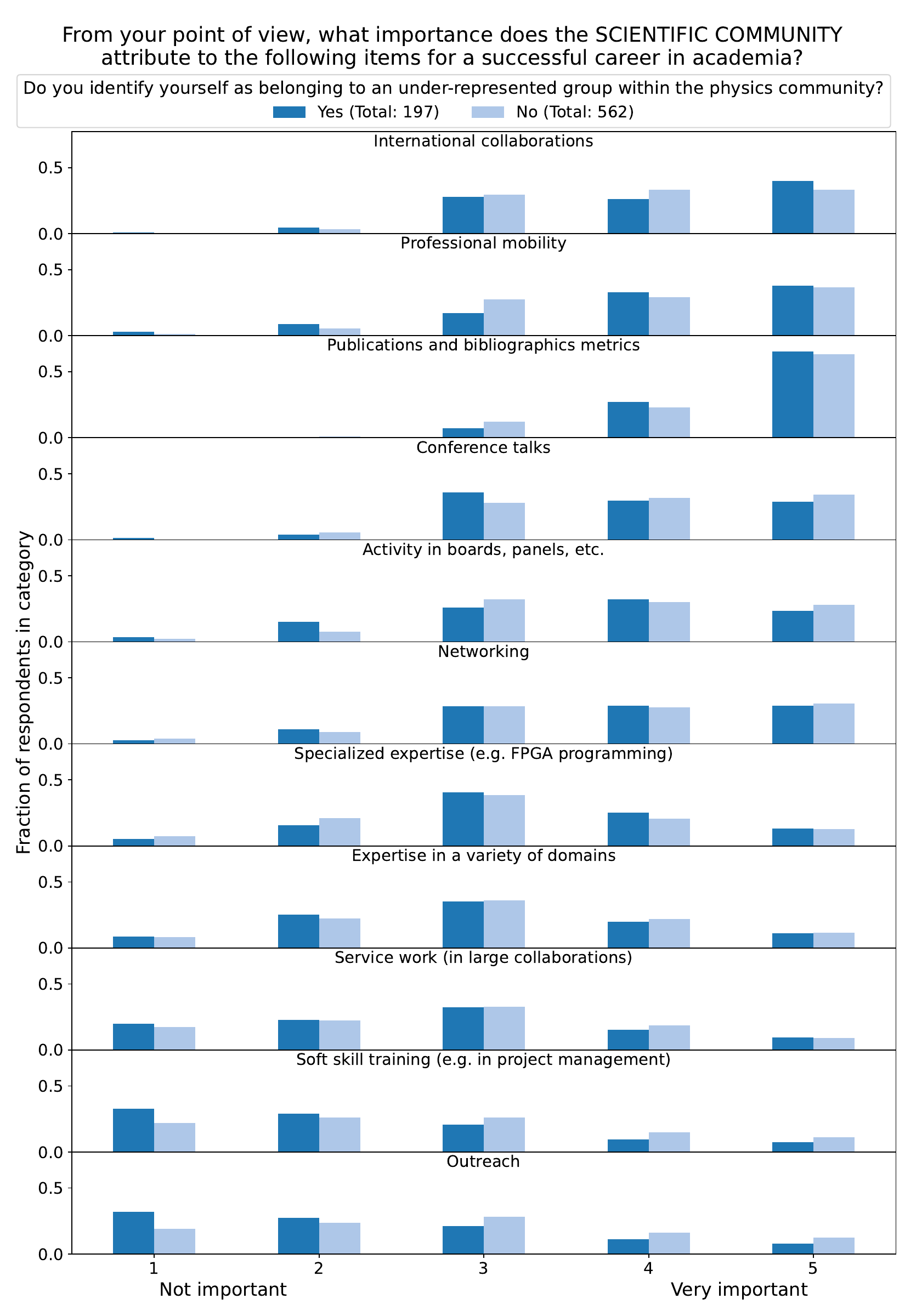}
    \caption{(Q69 v Q9) Correlations between how important respondents feel various items related to a successful academic career are, to the scientific community, and whether they belong to an under-represented community. Fractions are given out of all respondents.}
    \label{fig:part2:Q69vQ9}
\end{figure}

\begin{figure}[ht!]
    \centering
        \includegraphics[width=0.8\textwidth]{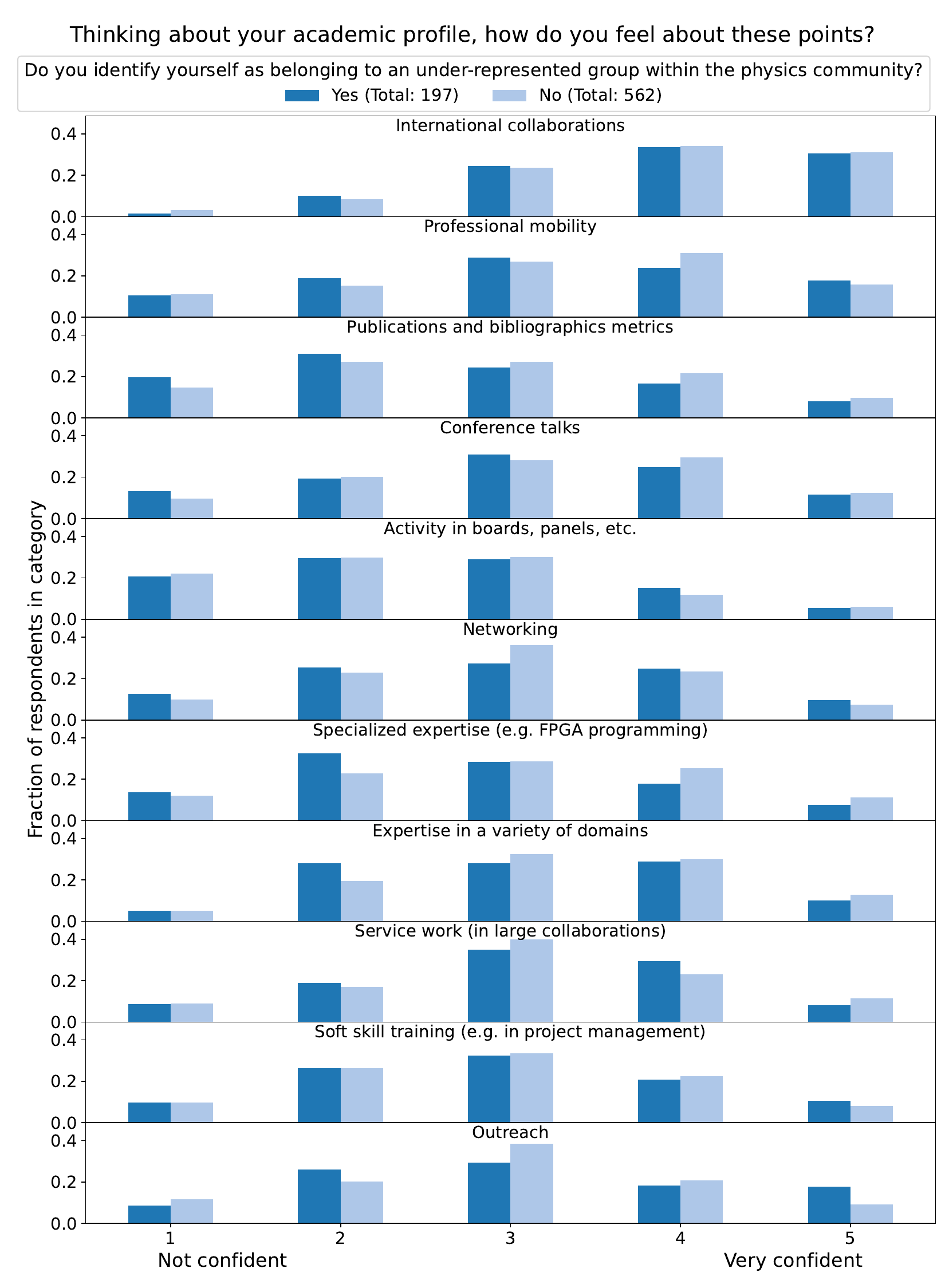}
    \caption{(Q70 v Q9) Correlations between how confident respondents feel about various items related to a successful academic career and whether they belong to an under-represented community. Fractions are given out of all respondents.}
    \label{fig:part2:Q70vQ9}
\end{figure}

Next, we investigate how important respondents feel various items (listed in Appendix~\ref{app:questions}) related to a successful academic career are personally or to the scientific community, along with how confident they feel about these items.
First we studied correlations with respondent demographics.
We consider in Figures~\ref{fig:part2:Q68vQ9},~\ref{fig:part2:Q69vQ9} and~\ref{fig:part2:Q70vQ9} how these metrics correlate to whether respondents belong to an under-represented group.
We see, for example, that these respondents value more soft skill training and outreach, but believe they are less important to the scientific community.

\begin{figure}[ht!]
    \centering
        \subfloat[]{\label{fig:part2:Q68avQ6}\includegraphics[width=0.49\textwidth]{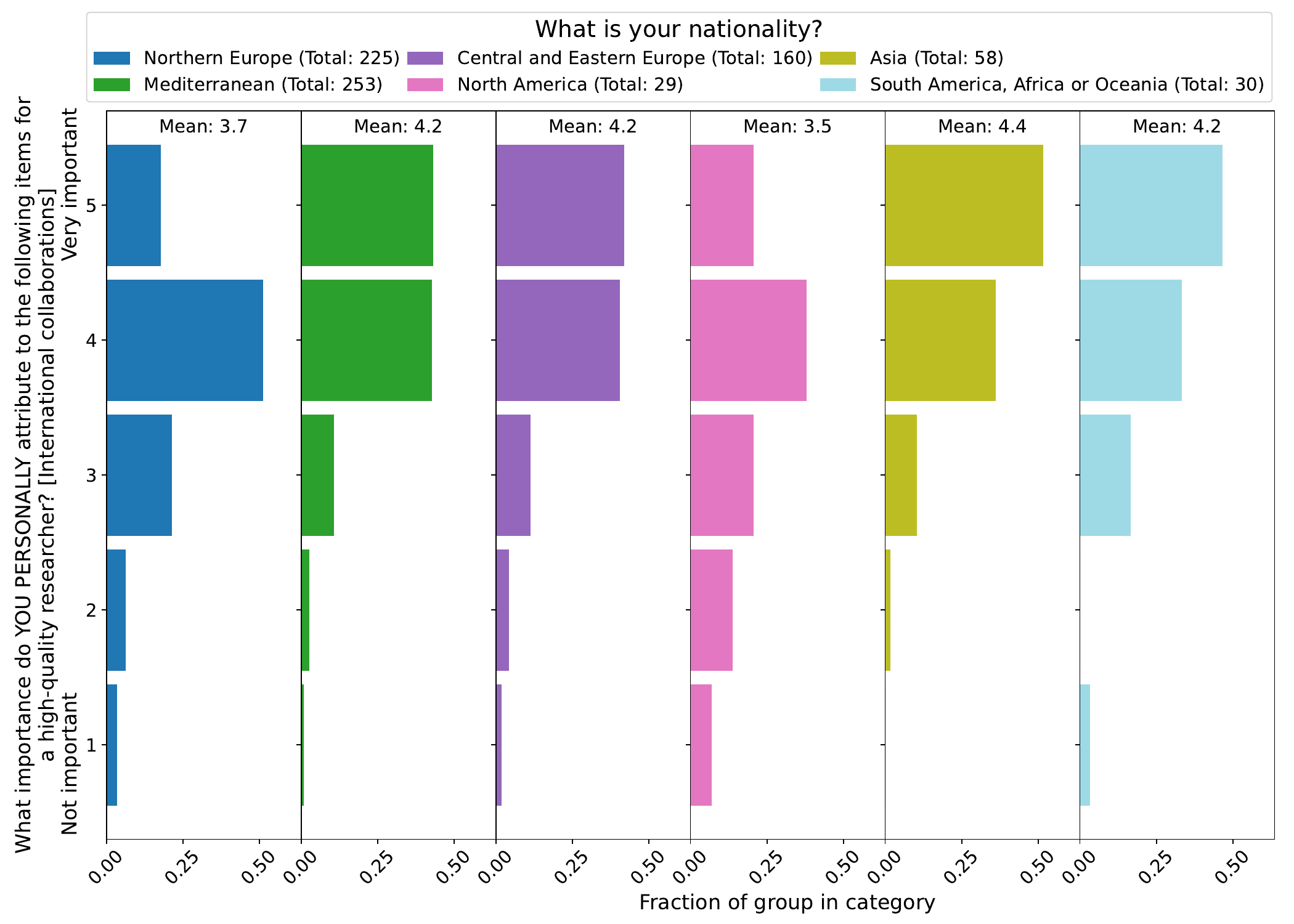}}
        \subfloat[]{\label{fig:part2:Q69avQ6}\includegraphics[width=0.49\textwidth]{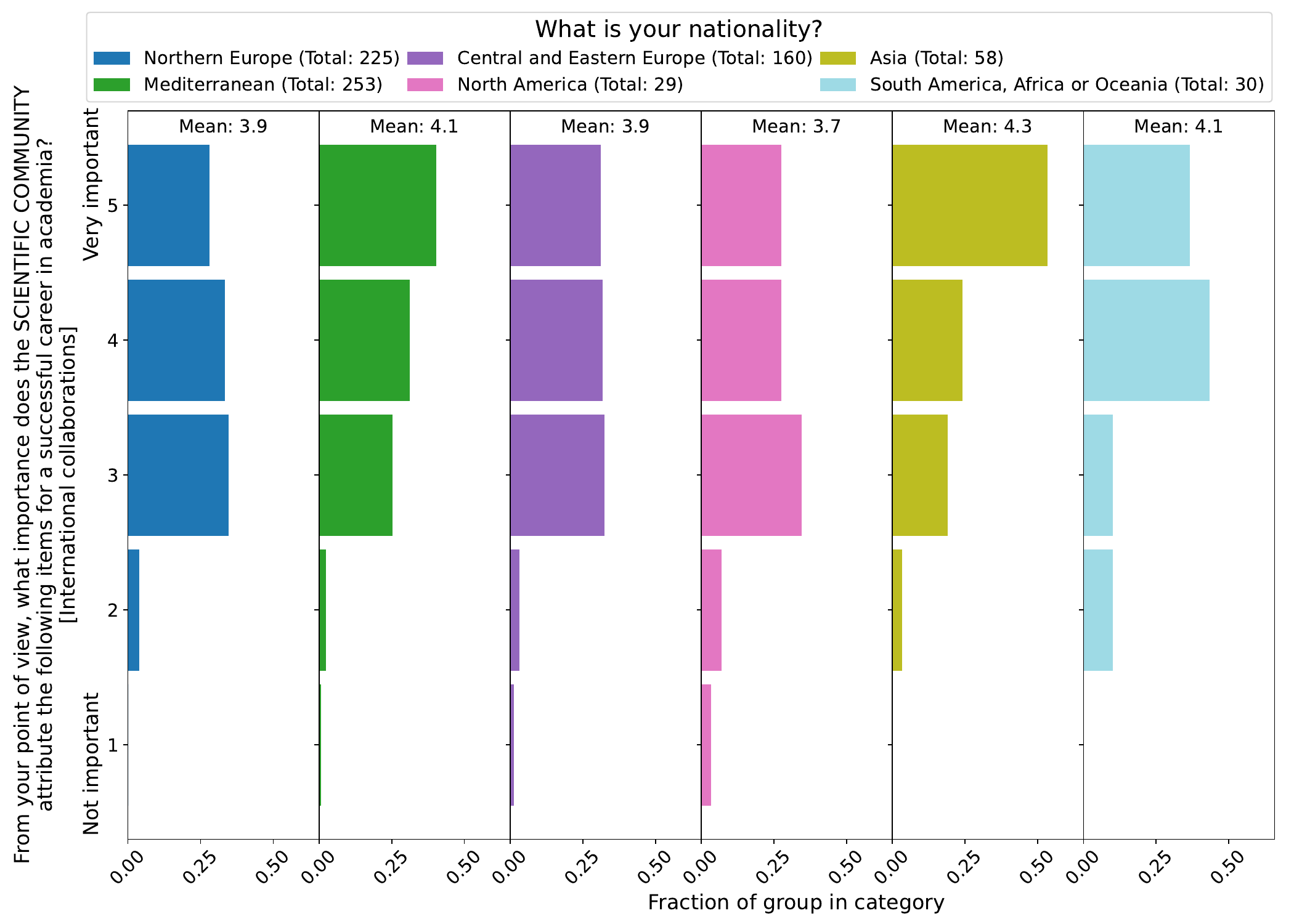}}\\
        \subfloat[]{\label{fig:part2:Q70avQ6}\includegraphics[width=0.49\textwidth]{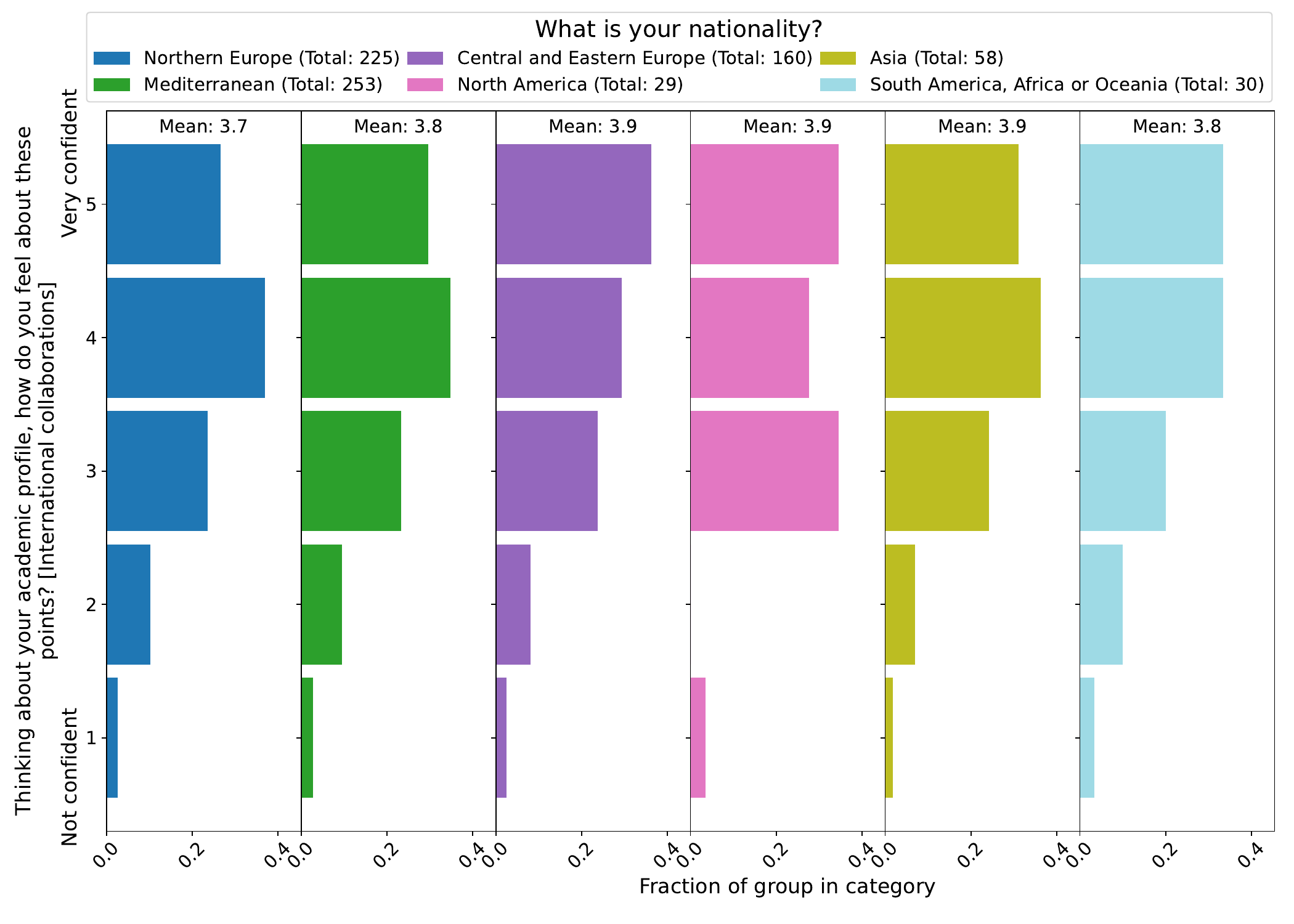}}
        \subfloat[]{\label{fig:part2:Q68avQ14}\includegraphics[width=0.49\textwidth]{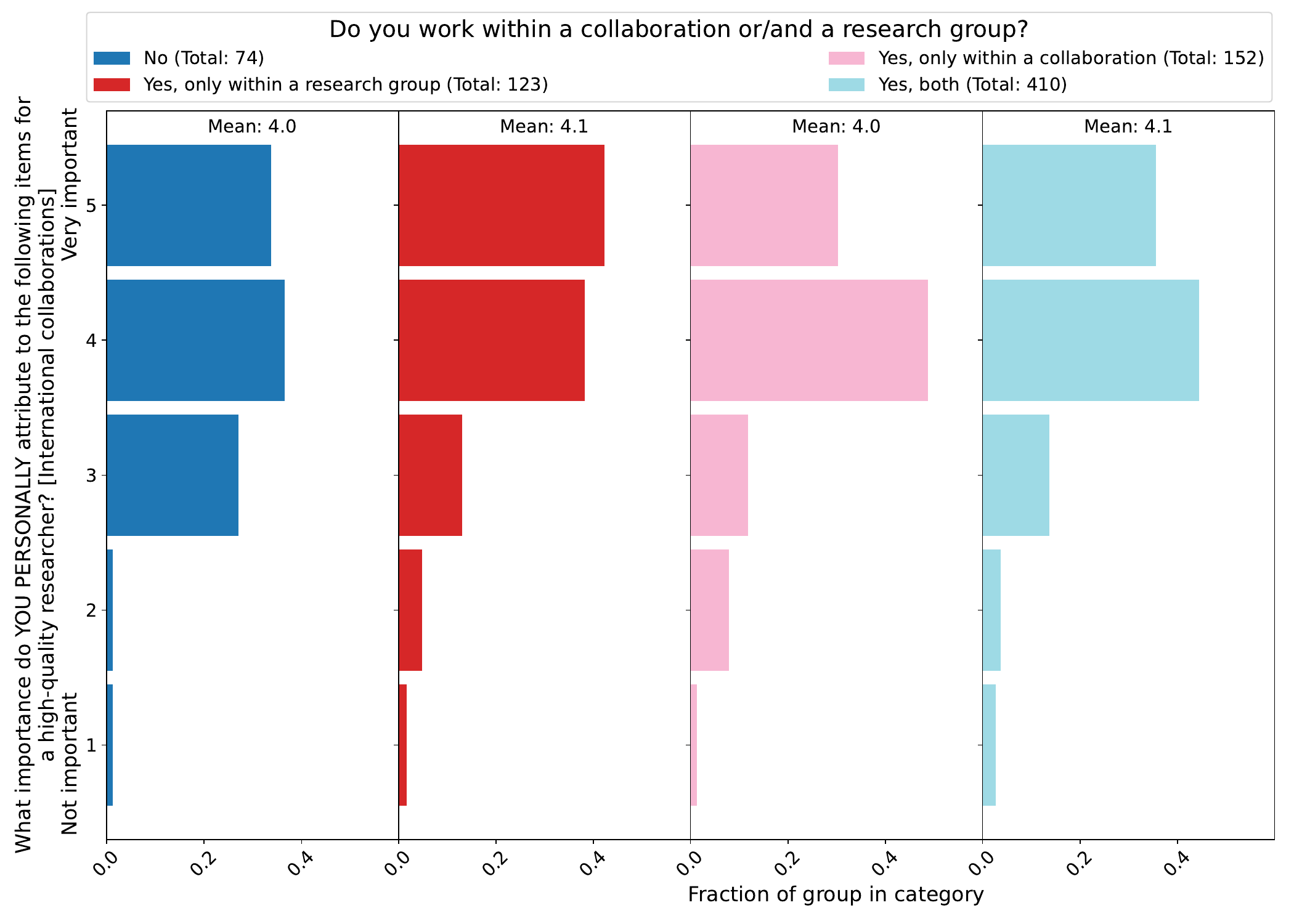}}\\
        \subfloat[]{\label{fig:part2:Q69avQ14}\includegraphics[width=0.49\textwidth]{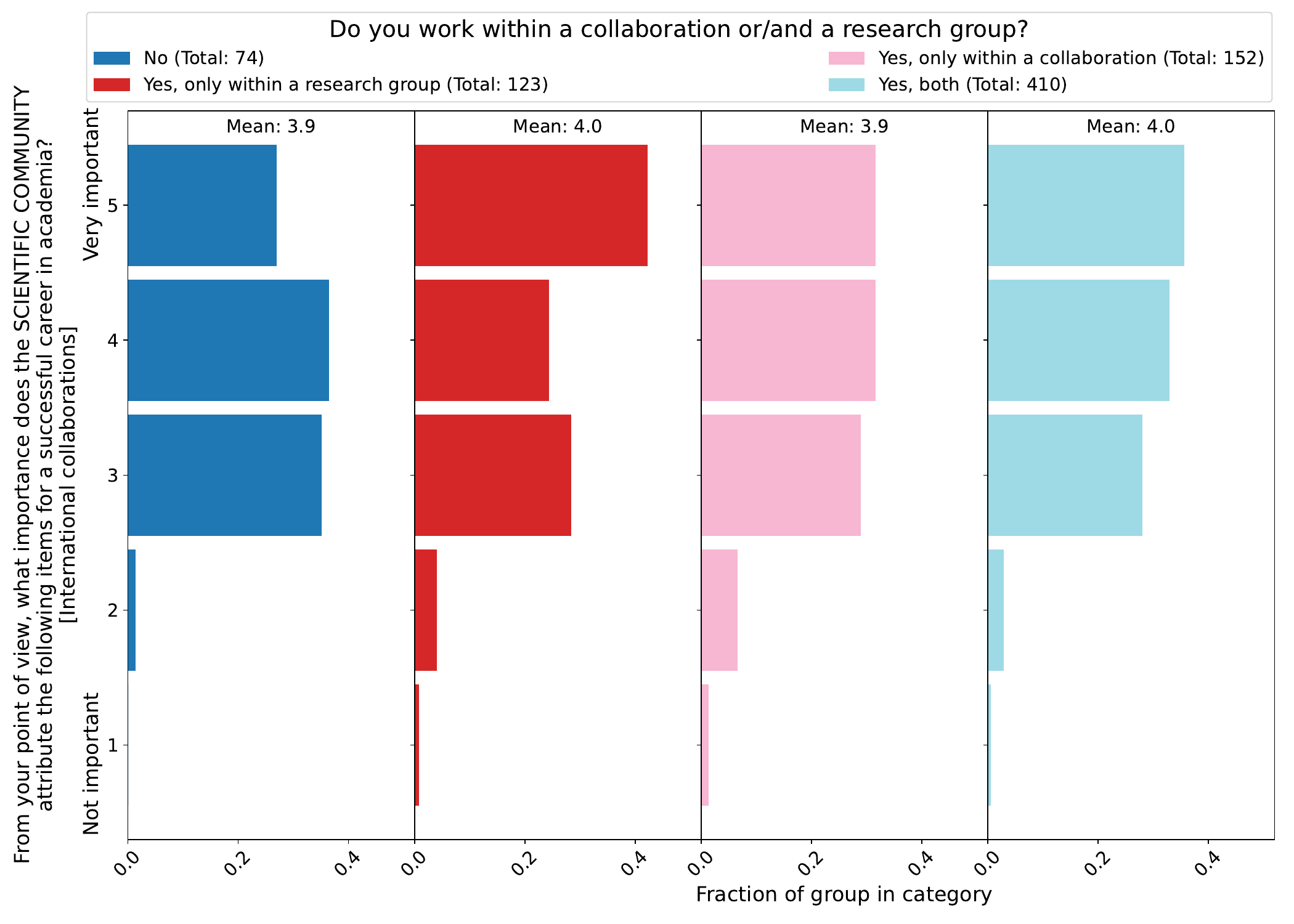}}
    \caption{(Q68a--70a v Q6,14) Correlations between how important international collaborations are considered to be for a successful academic career and selected demographics. Fractions are given out of all respondents who answered the questions.}
    \label{fig:part2:Q68-70avQ6Q14}
\end{figure}

\FloatBarrier
\pagebreak
The importance of international collaborations (shown in Figure~\ref{fig:part2:Q68-70avQ6Q14}) to respondents is weakly correlated to nationality, with those from Northern Europe and North America viewing this as less important.
However, when considering importance to the scientific community and personal profile this correlation disappears.
There is also no clear correlation observed when considering if whether respondents are a member of the collaboration and/or research group.

\begin{figure}[ht!]
    \centering
        \subfloat[]{\label{fig:part2:Q68bvQ1}\includegraphics[width=0.49\textwidth]{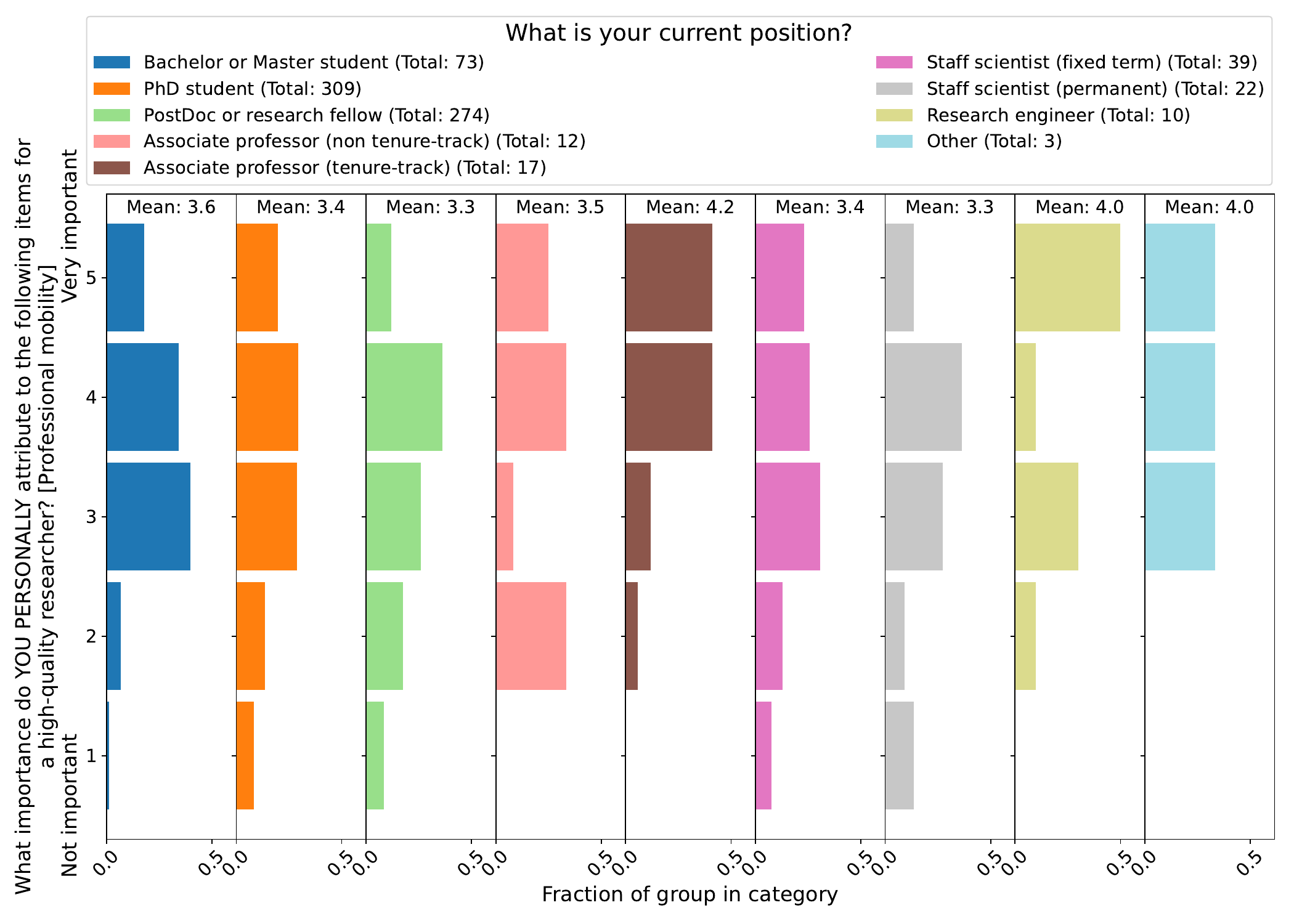}}
        \subfloat[]{\label{fig:part2:Q68bvQ6}\includegraphics[width=0.49\textwidth]{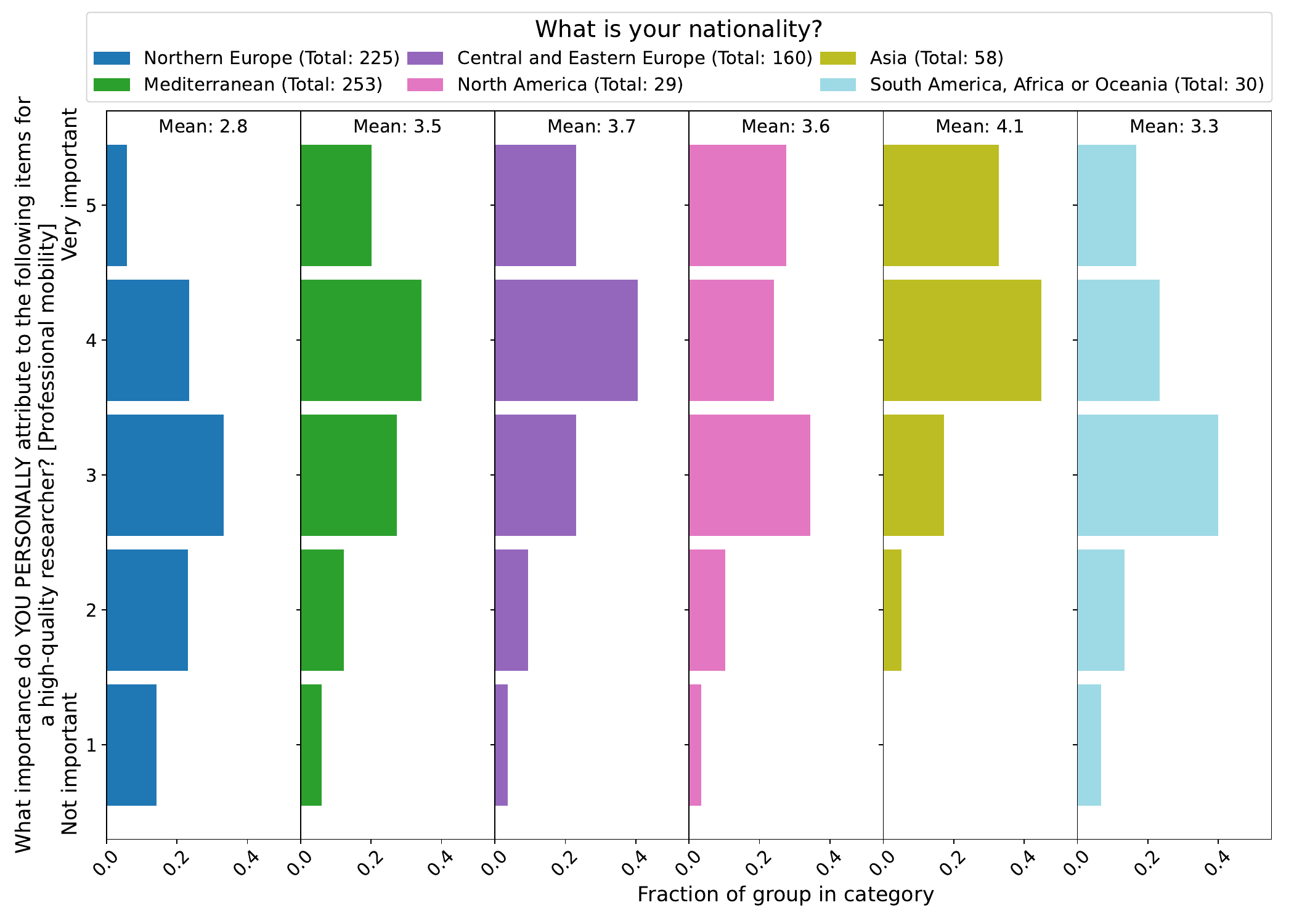}}\\
        \subfloat[]{\label{fig:part2:Q69bvQ1}\includegraphics[width=0.49\textwidth]{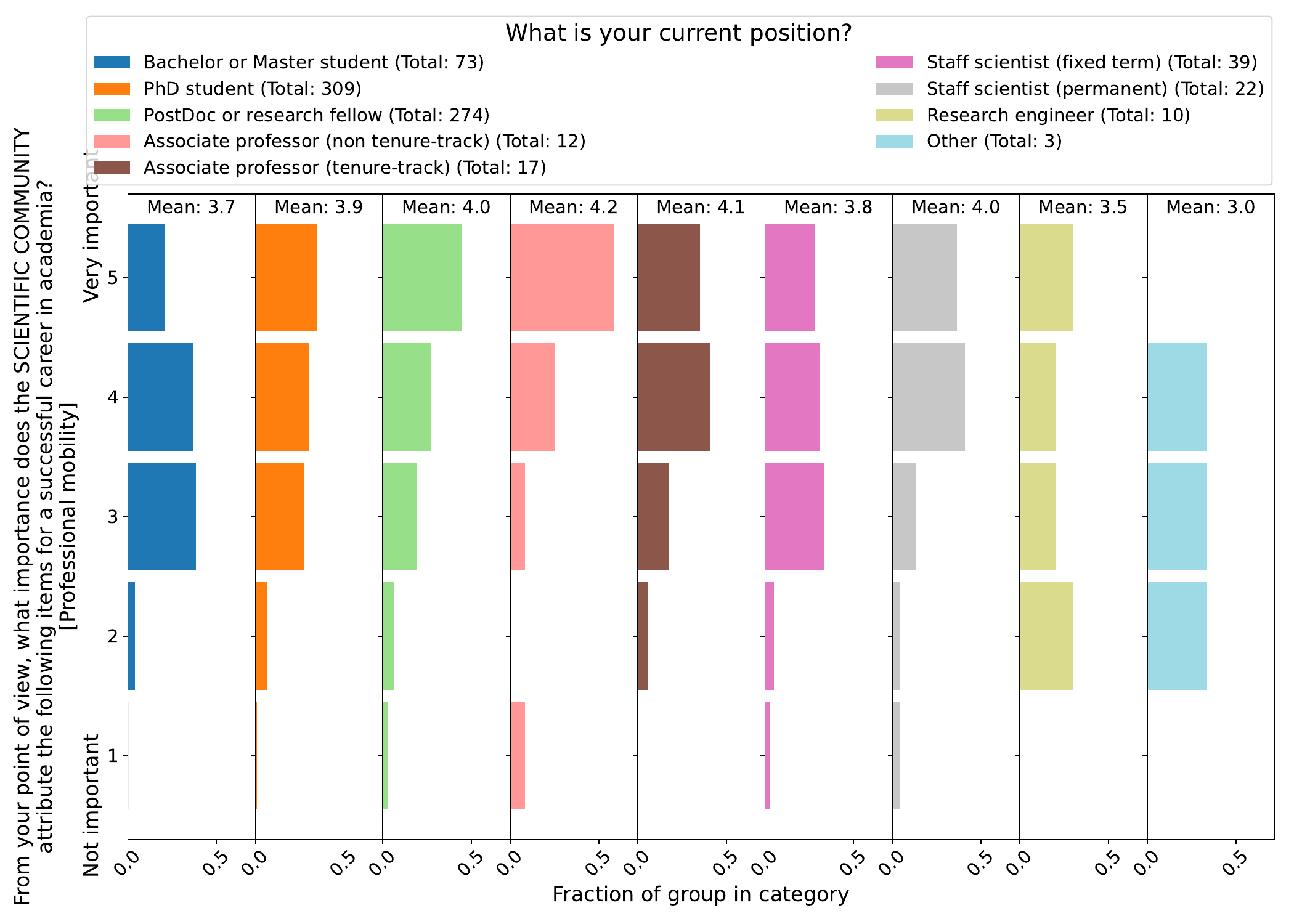}}
        \subfloat[]{\label{fig:part2:Q69bvQ6}\includegraphics[width=0.49\textwidth]{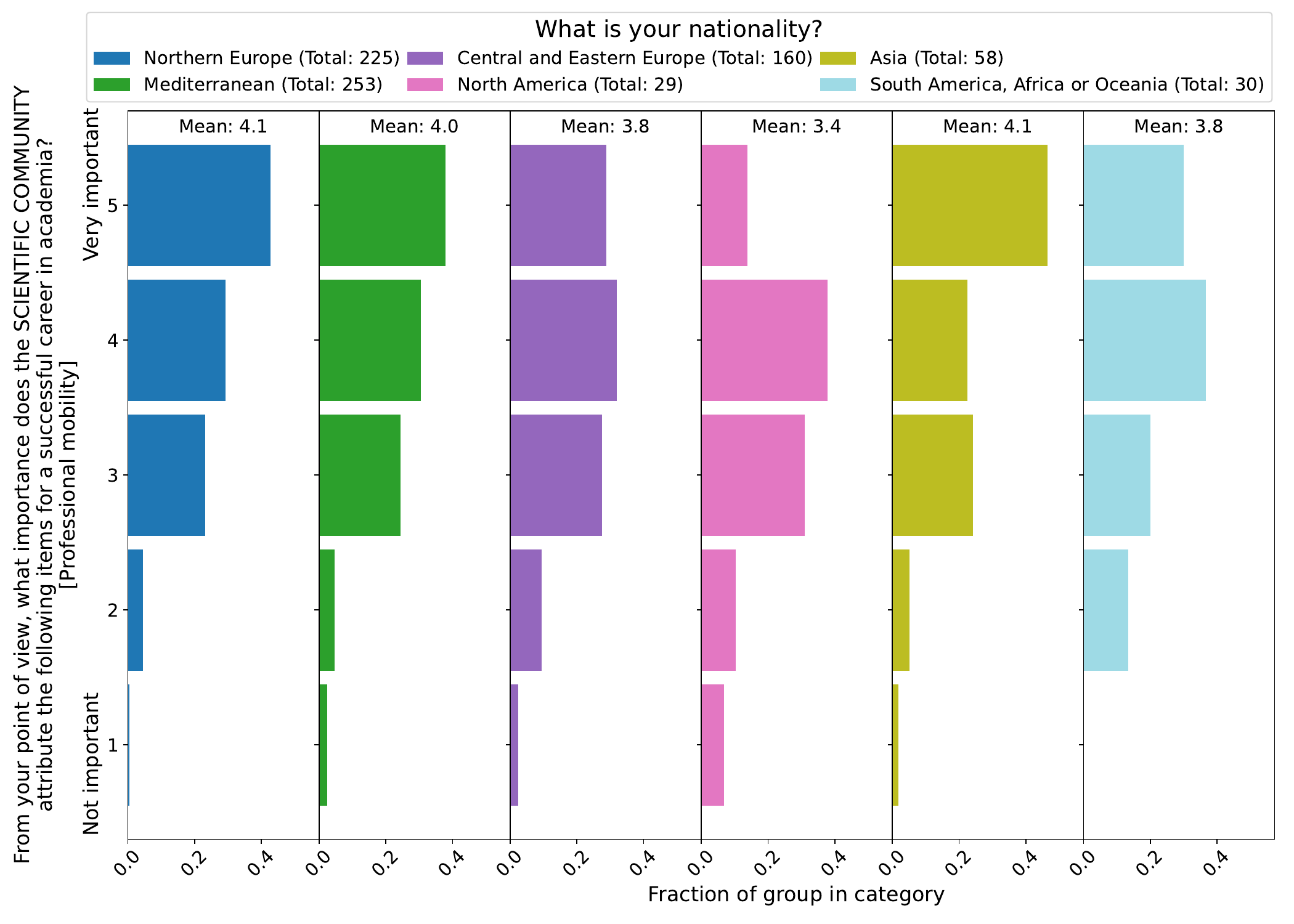}}\\
        \subfloat[]{\label{fig:part2:Q70bvQ1}\includegraphics[width=0.49\textwidth]{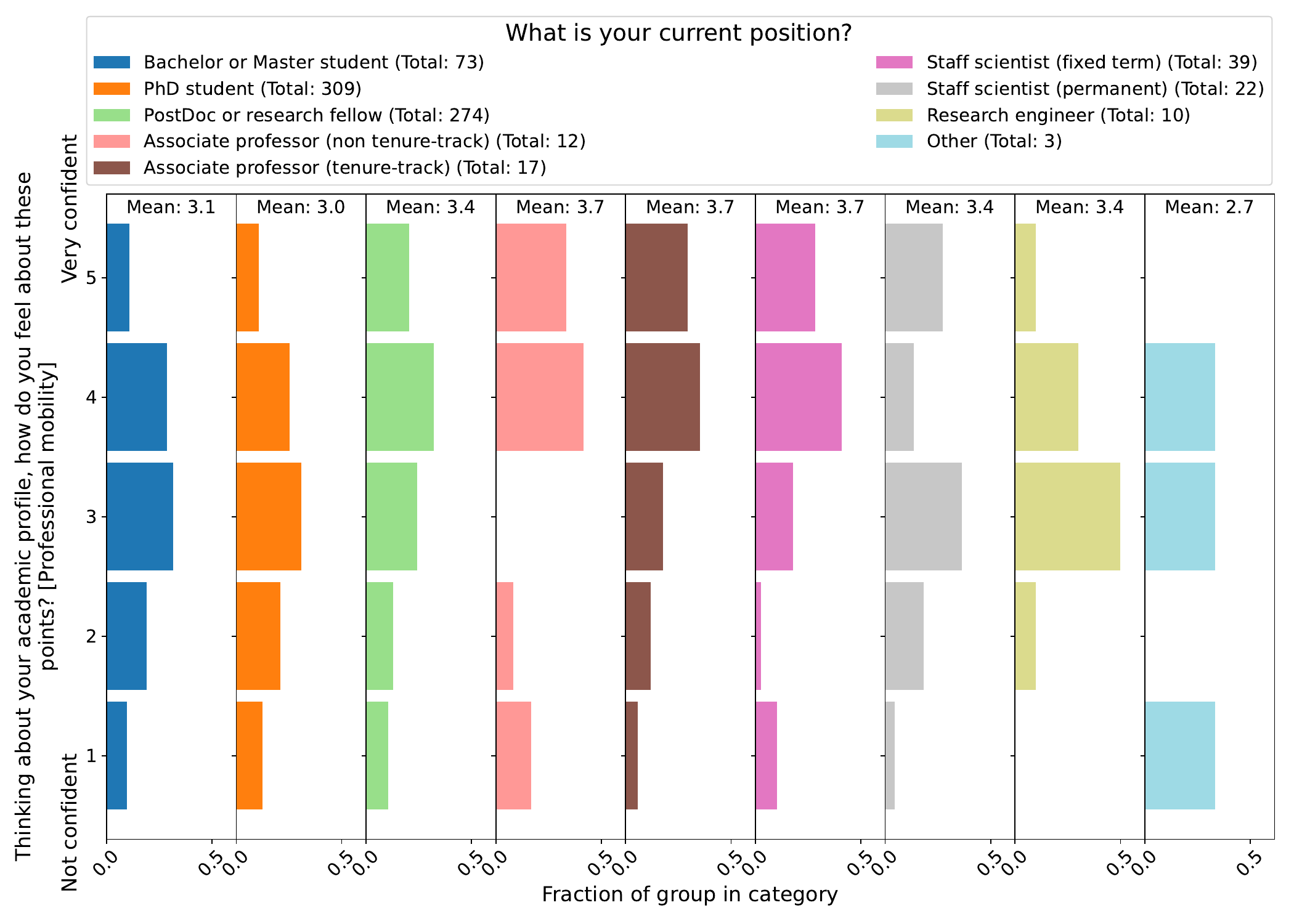}}
        \subfloat[]{\label{fig:part2:Q70bvQ6}\includegraphics[width=0.49\textwidth]{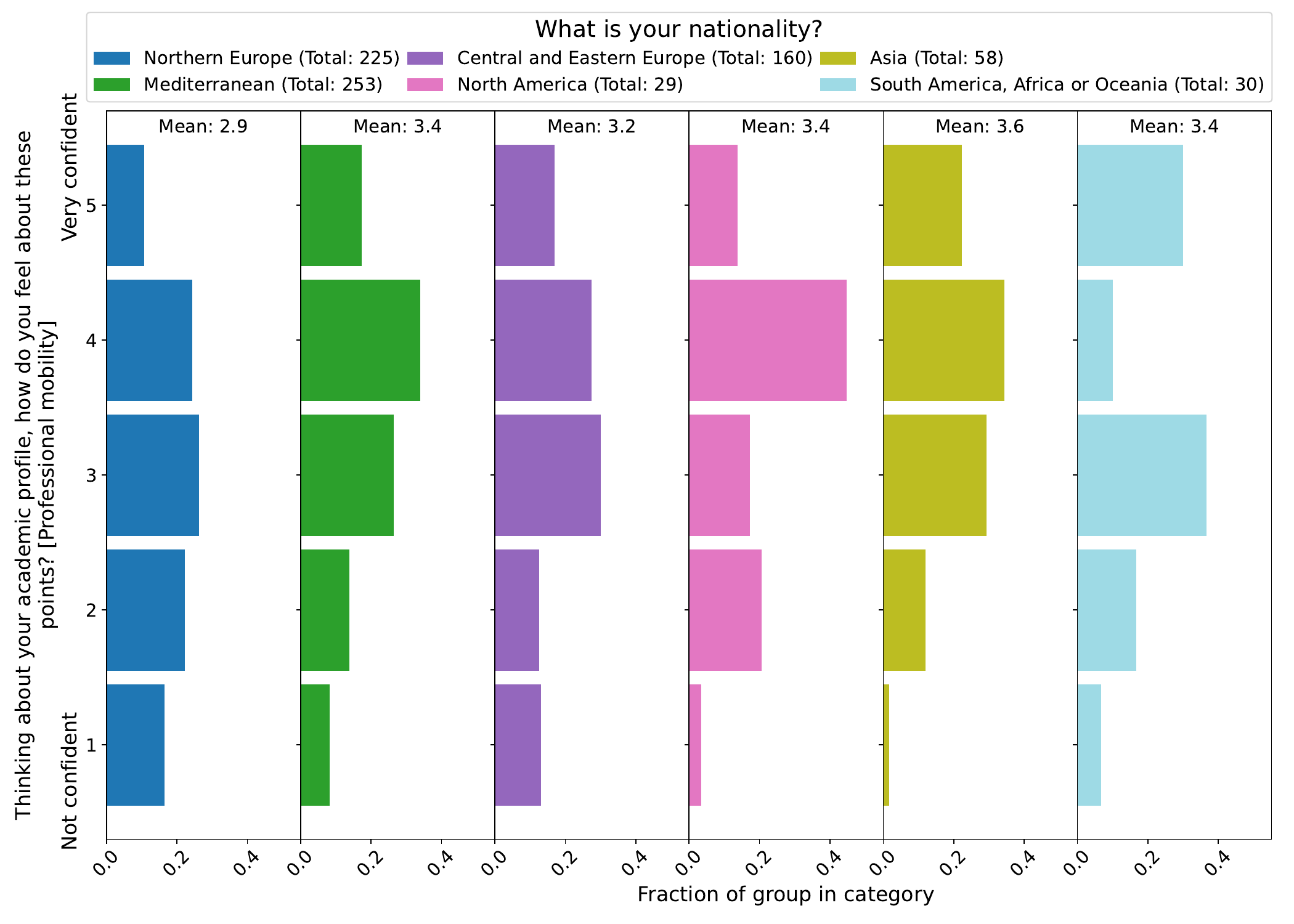}}
    \caption{(Q68b--70b v Q1,6) Correlations between how important professional mobility is considered to be for a successful academic career and selected demographics. Fractions are given out of all respondents who answered the questions.}
    \label{fig:part2:Q68-70bvQ1Q6}
\end{figure}

Considering the importance of professional mobility in Figure~\ref{fig:part2:Q68-70bvQ1Q6}, we see a correlation with nationality.
Respondents from Asia view this as very important, and are confident about it, whilst Europeans view it as more important to the community than personally (especially in the case of Northern Europeans), and are less confident about mobility in relation to their profile.
Only North American respondents believe that professional mobility is less important to the scientific community than to themselves.
Respondents who are tenure-track associate professors viewed professional mobility as most important to themselves. However, tenure and non-tenure track associate professors view this as most important to the scientific community, and also more fulfilled for their profile, alongside fixed term staff scientists.

\begin{figure}[ht!]
    \centering
        \subfloat[]{\label{fig:part2:Q68cvQ1}\includegraphics[width=0.49\textwidth]{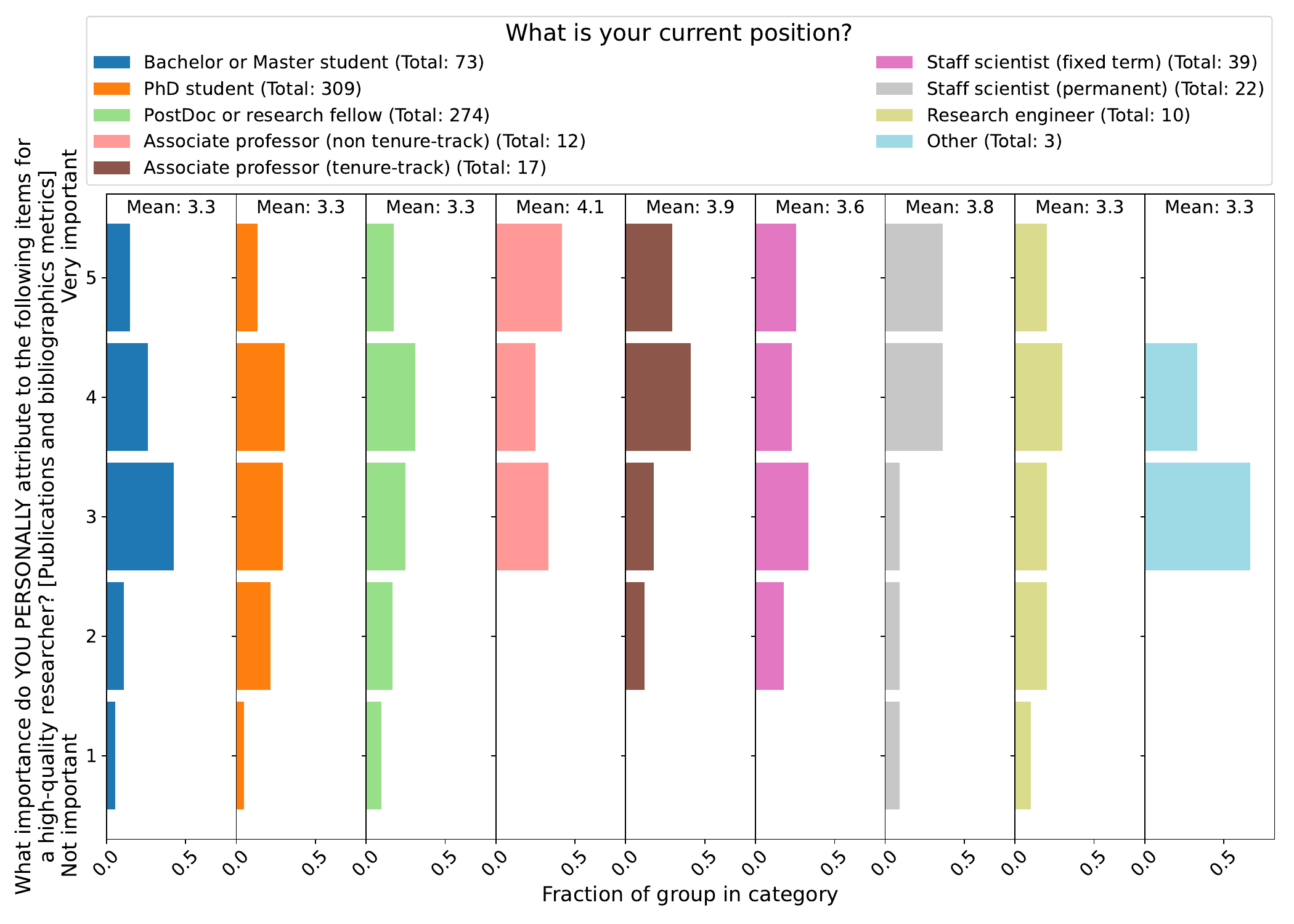}}
        \subfloat[]{\label{fig:part2:Q68cvQ6}\includegraphics[width=0.49\textwidth]{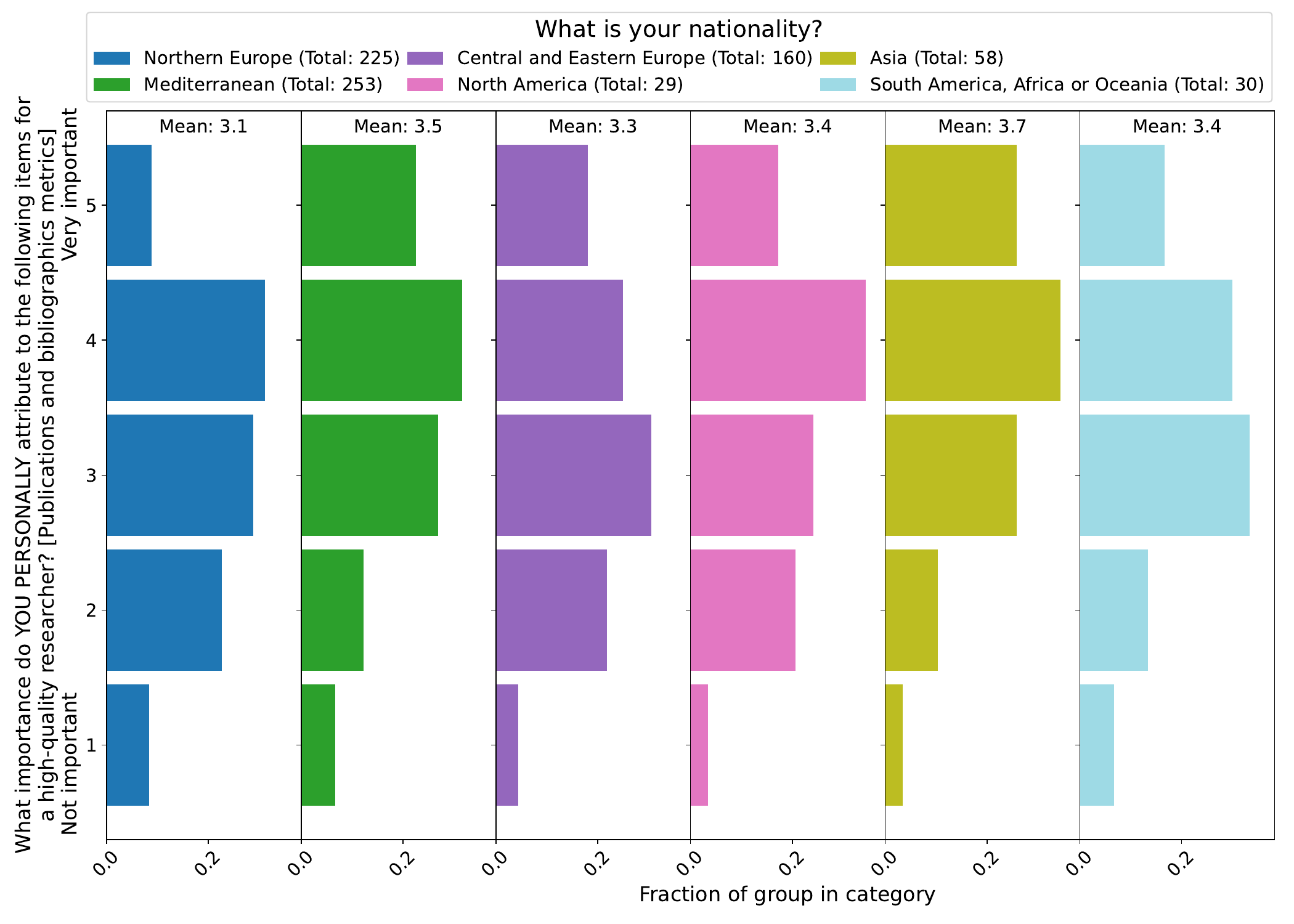}}\\
        \subfloat[]{\label{fig:part2:Q69cvQ1}\includegraphics[width=0.49\textwidth]{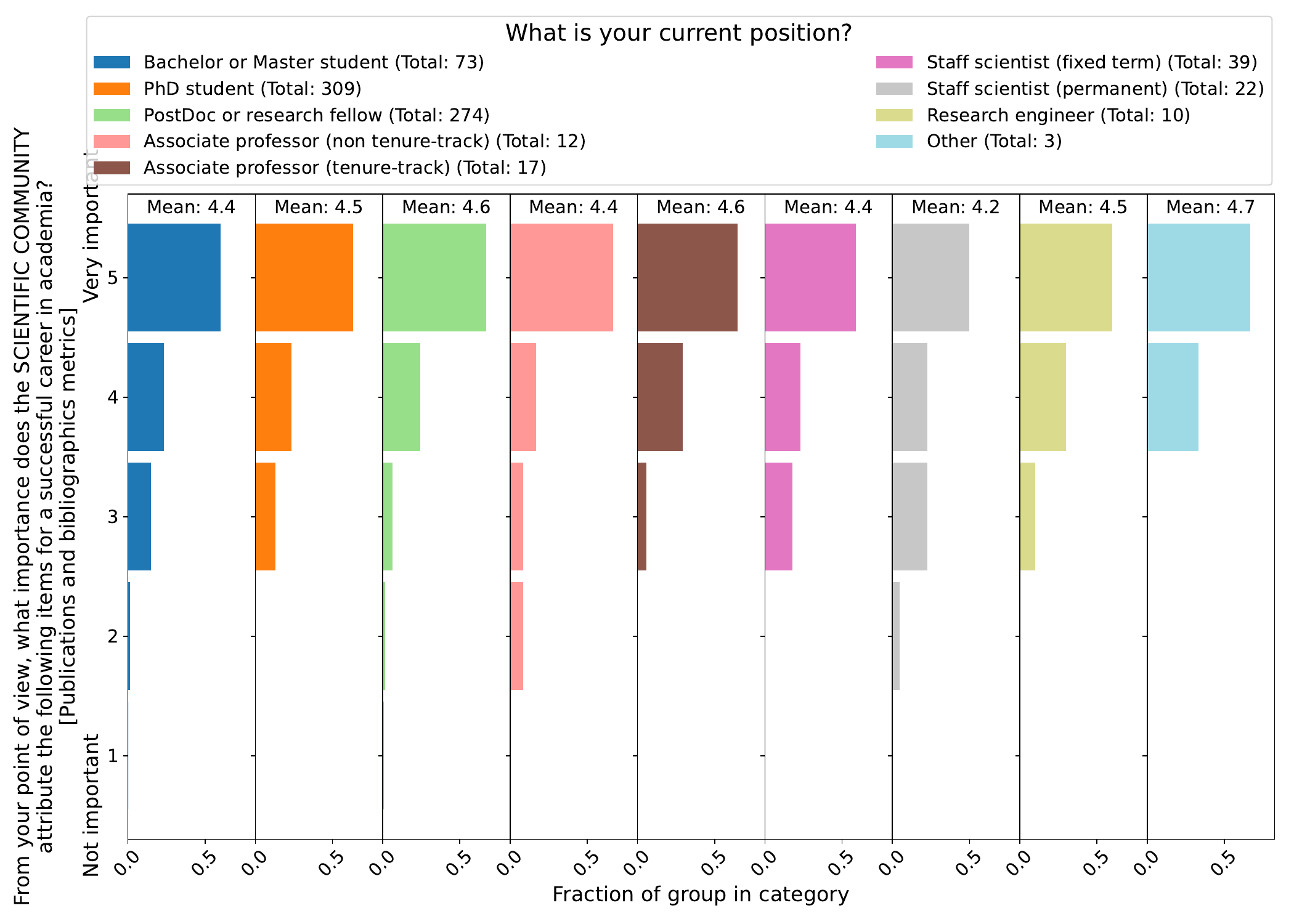}}
        \subfloat[]{\label{fig:part2:Q69cvQ6}\includegraphics[width=0.49\textwidth]{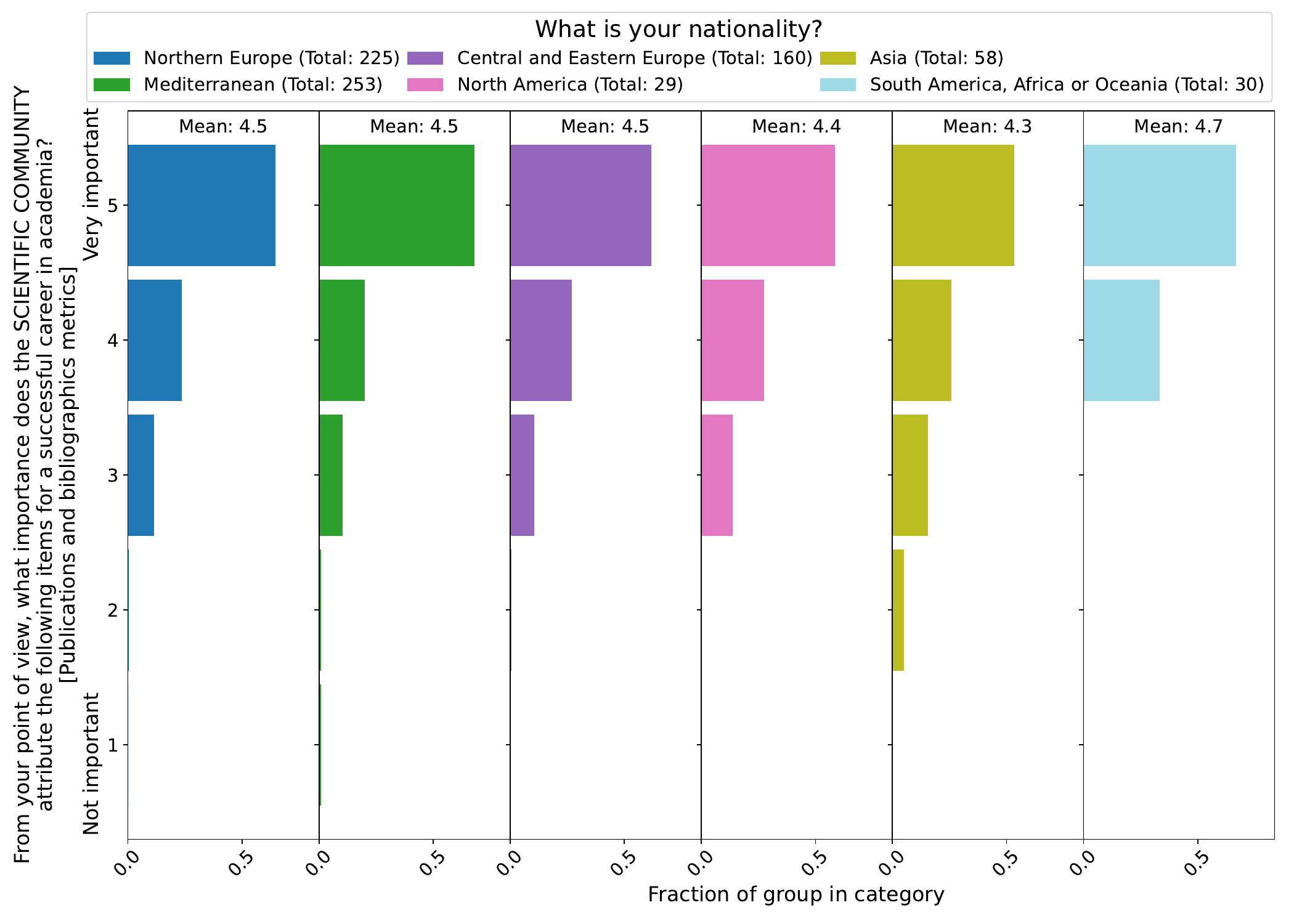}}\\
        \subfloat[]{\label{fig:part2:Q70cvQ1}\includegraphics[width=0.49\textwidth]{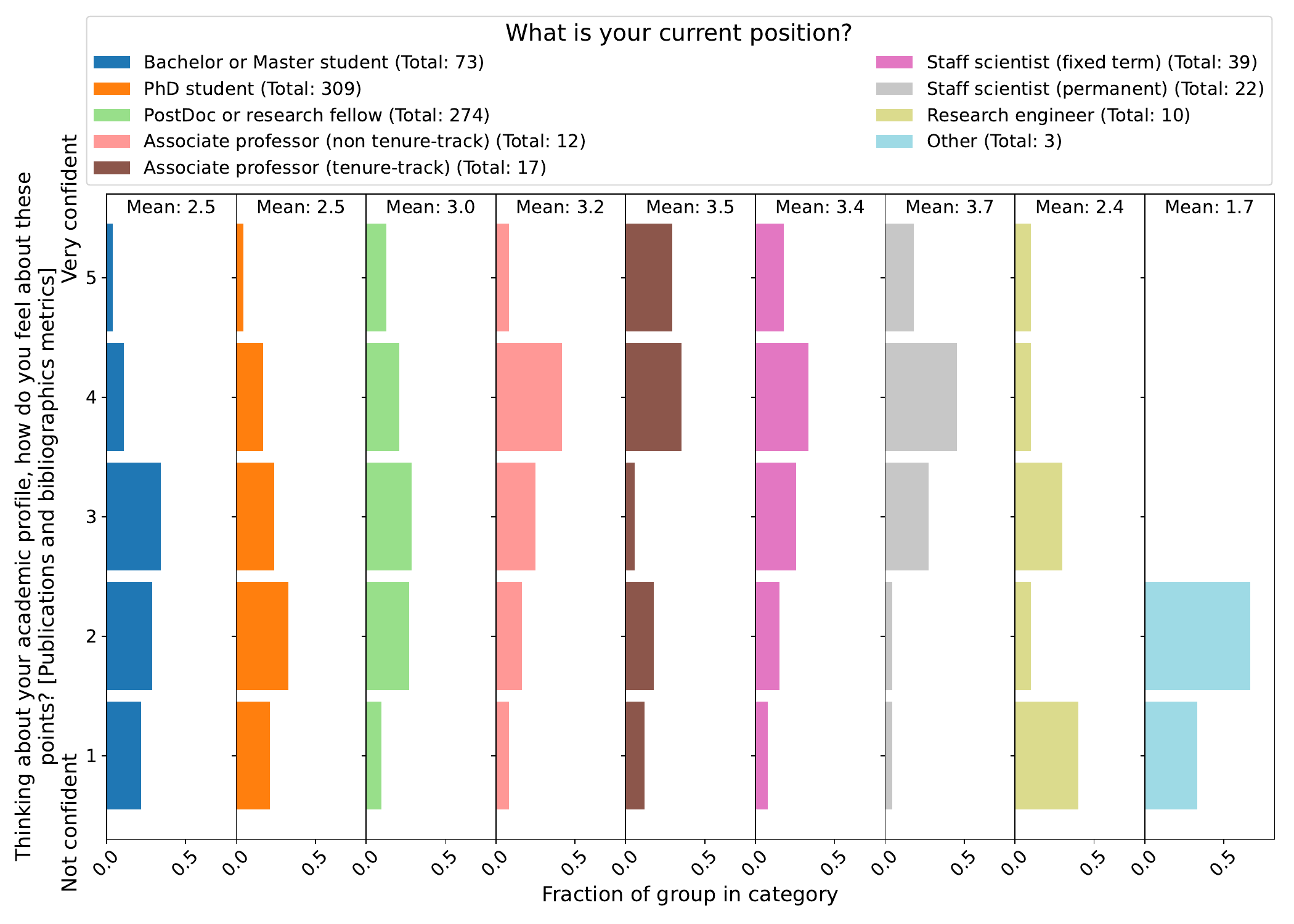}}
        \subfloat[]{\label{fig:part2:Q68cvQ11}\includegraphics[width=0.49\textwidth]{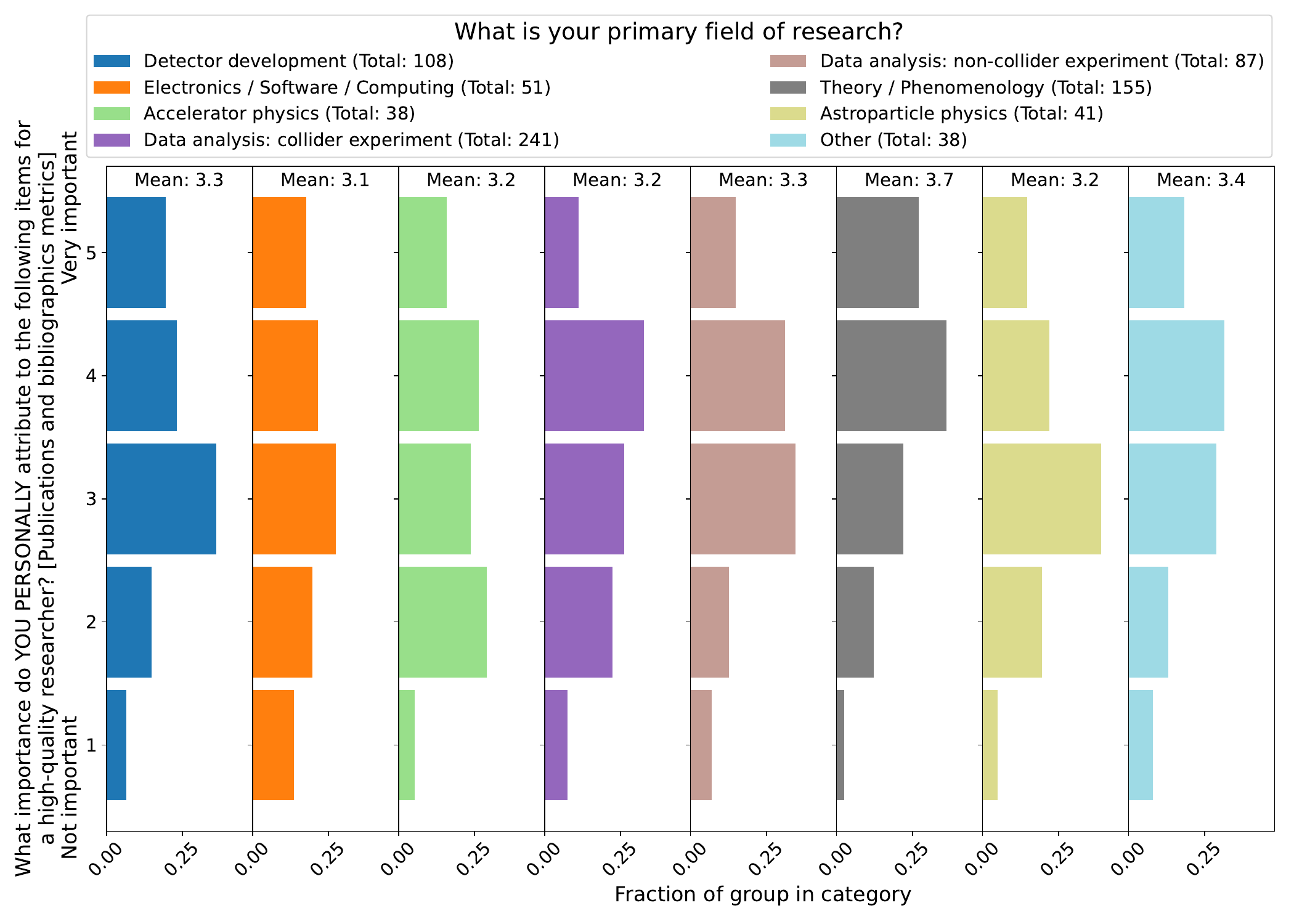}}
    \caption{(Q68c--70c v Q1,6,11) Correlations between how important publications and bibliographic metrics are considered to be for a successful academic career and selected demographics. Fractions are given out of all respondents who answered the questions.}
    \label{fig:part2:Q68-70cvQ1Q6Q11}
\end{figure}

Considering publications and bibliographic metrics in Figure~\ref{fig:part2:Q68-70cvQ1Q6Q11}, whilst respondents from all nationalities and positions view this as equally important to the scientific community, Asian respondents view this as substantially more important personally than Northern Europeans.
Furthermore, respondents in more senior positions both consider this to be more important personally, and are more confident about it.
We also observed these trends considering the metric of conference talks.
Respondents working in theory/phenomenology consider publications and bibliographic metrics to be more personally important than those working in other fields.

\begin{figure}[ht!]
    \centering
        \subfloat[]{\label{fig:part2:Q68gvQ11}\includegraphics[width=0.49\textwidth]{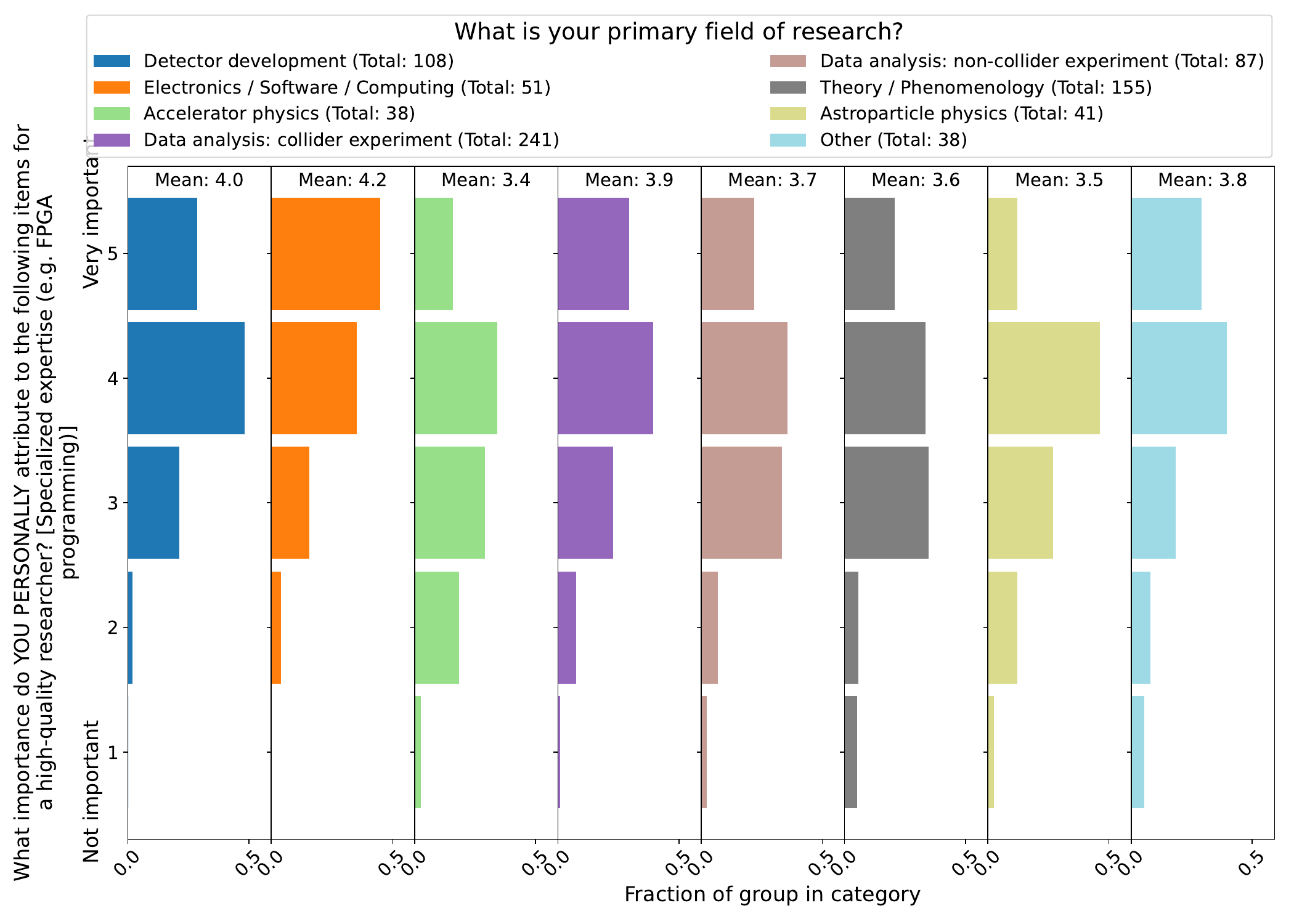}}
        \subfloat[]{\label{fig:part2:Q70gvQ11}\includegraphics[width=0.49\textwidth]{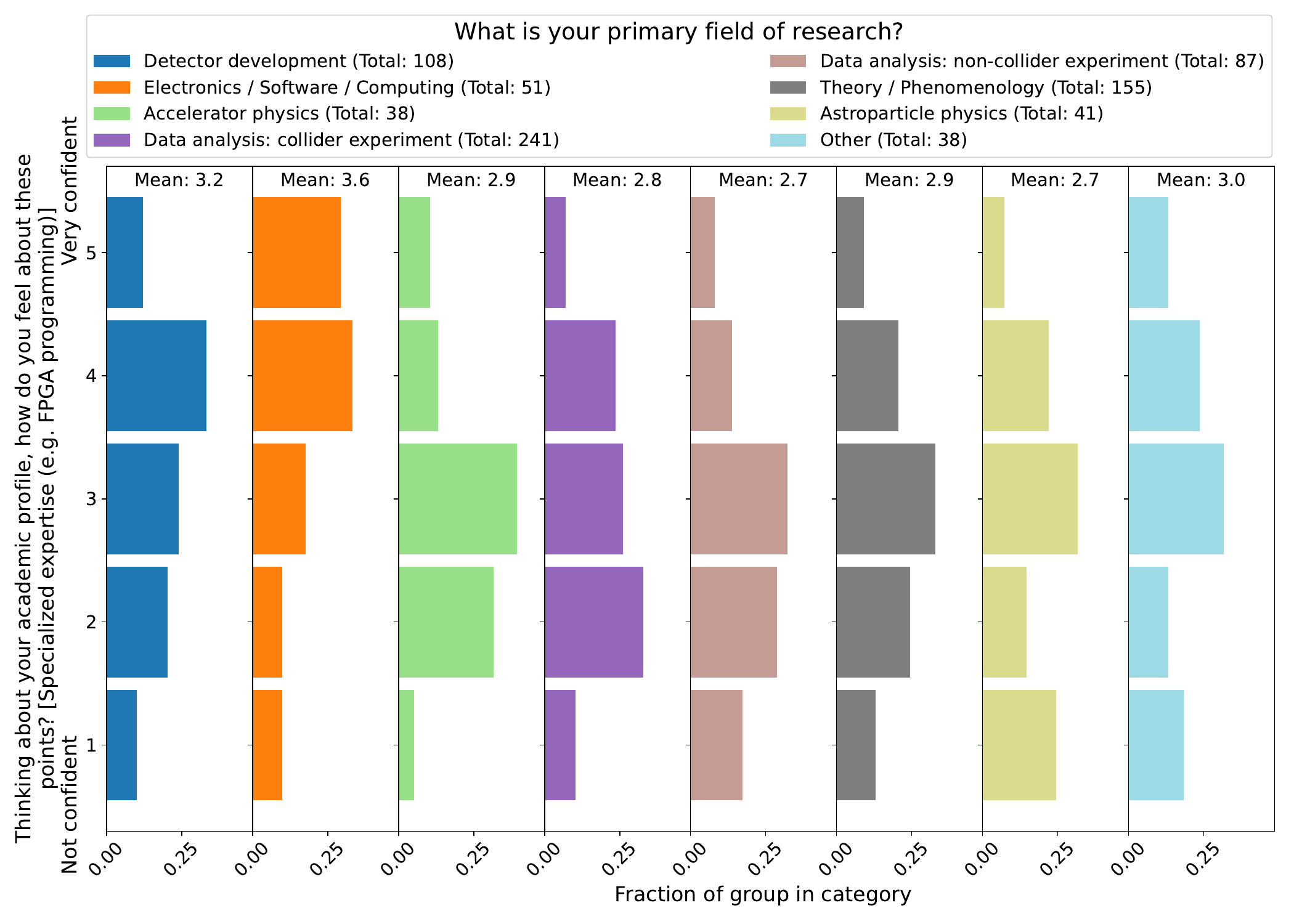}}
    \caption{(Q68g--70g v Q11,14) Correlations between how important specialised expertise is considered to be for a successful academic career and selected demographics. Fractions are given out of all respondents.}
    \label{fig:part2:Q68-70gvQ11Q14}
\end{figure}

Figure~\ref{fig:part2:Q68-70gvQ11Q14} shows correlations between the perceived importance of specialised expertise and field of research.
While no correlation was seen for how important the scientific community views this, we see that respondents working in electronics/software/computing view this as most important (and are by far the most confident about it) and accelerator physicists the least.

\begin{figure}[ht!]
    \centering
        \subfloat[]{\label{fig:part2:Q69hvQ1}\includegraphics[width=0.49\textwidth]{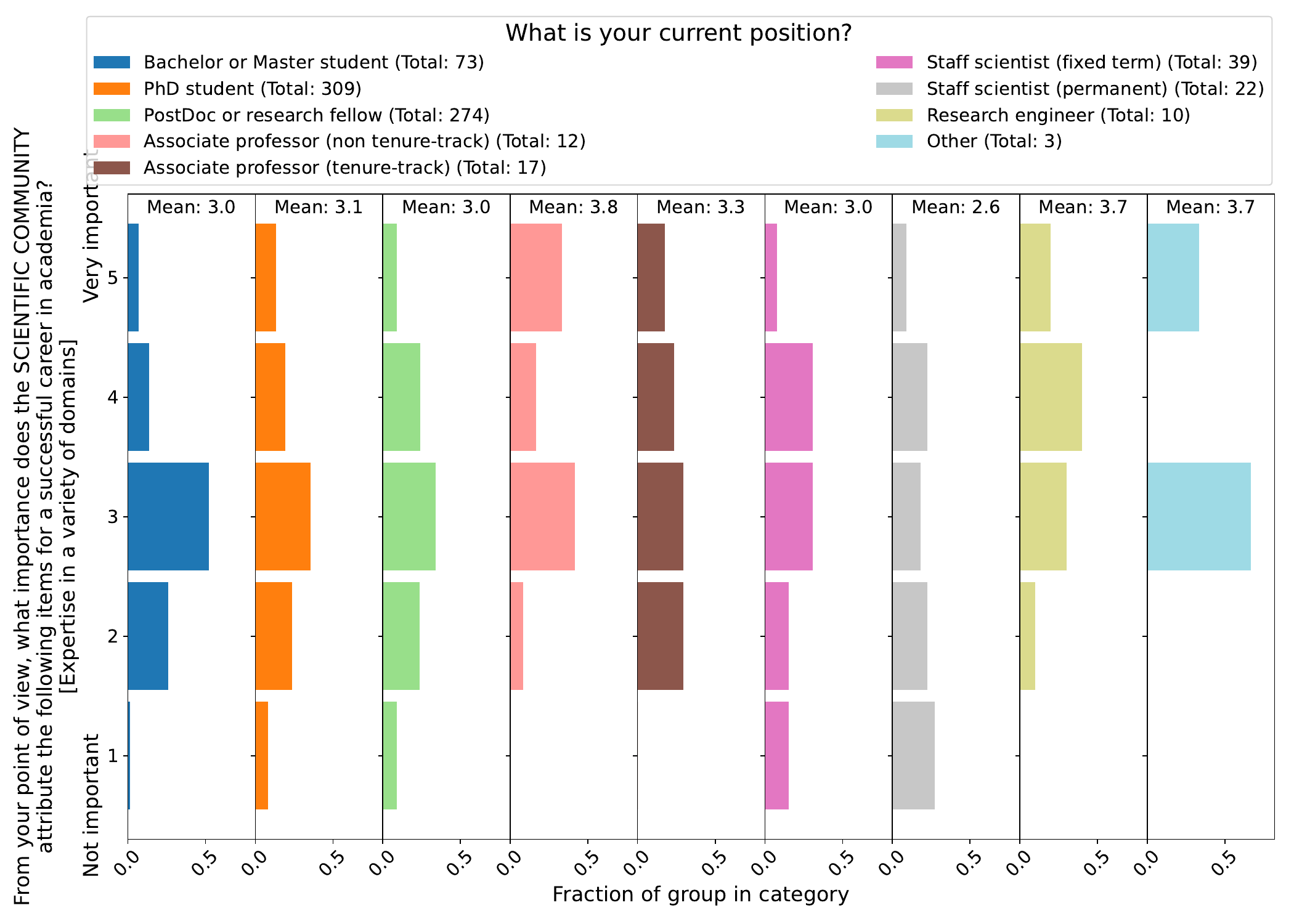}}
        \subfloat[]{\label{fig:part2:Q69hvQ6}\includegraphics[width=0.49\textwidth]{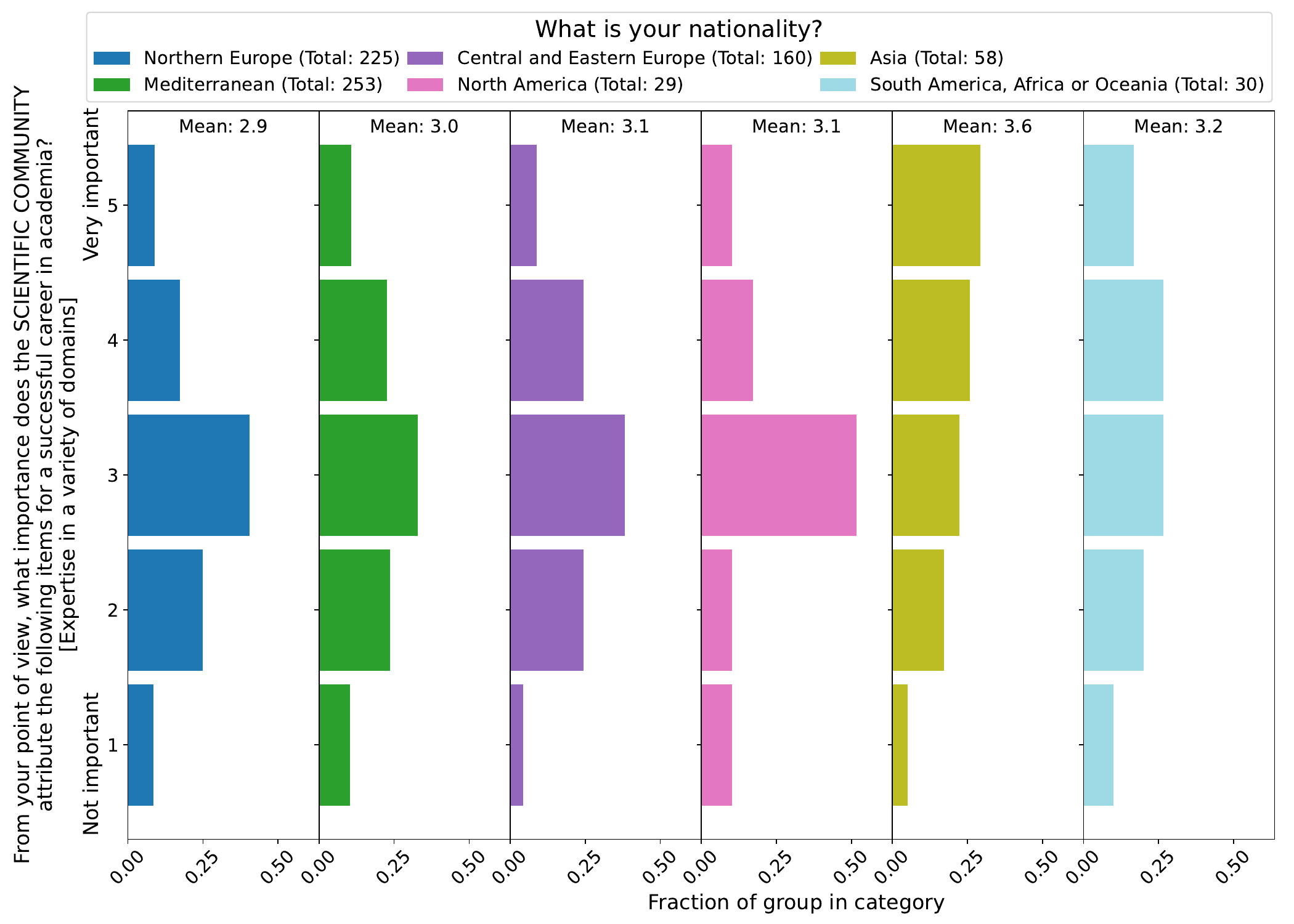}}\\
        \subfloat[]{\label{fig:part2:Q70ivQ1}\includegraphics[width=0.49\textwidth]{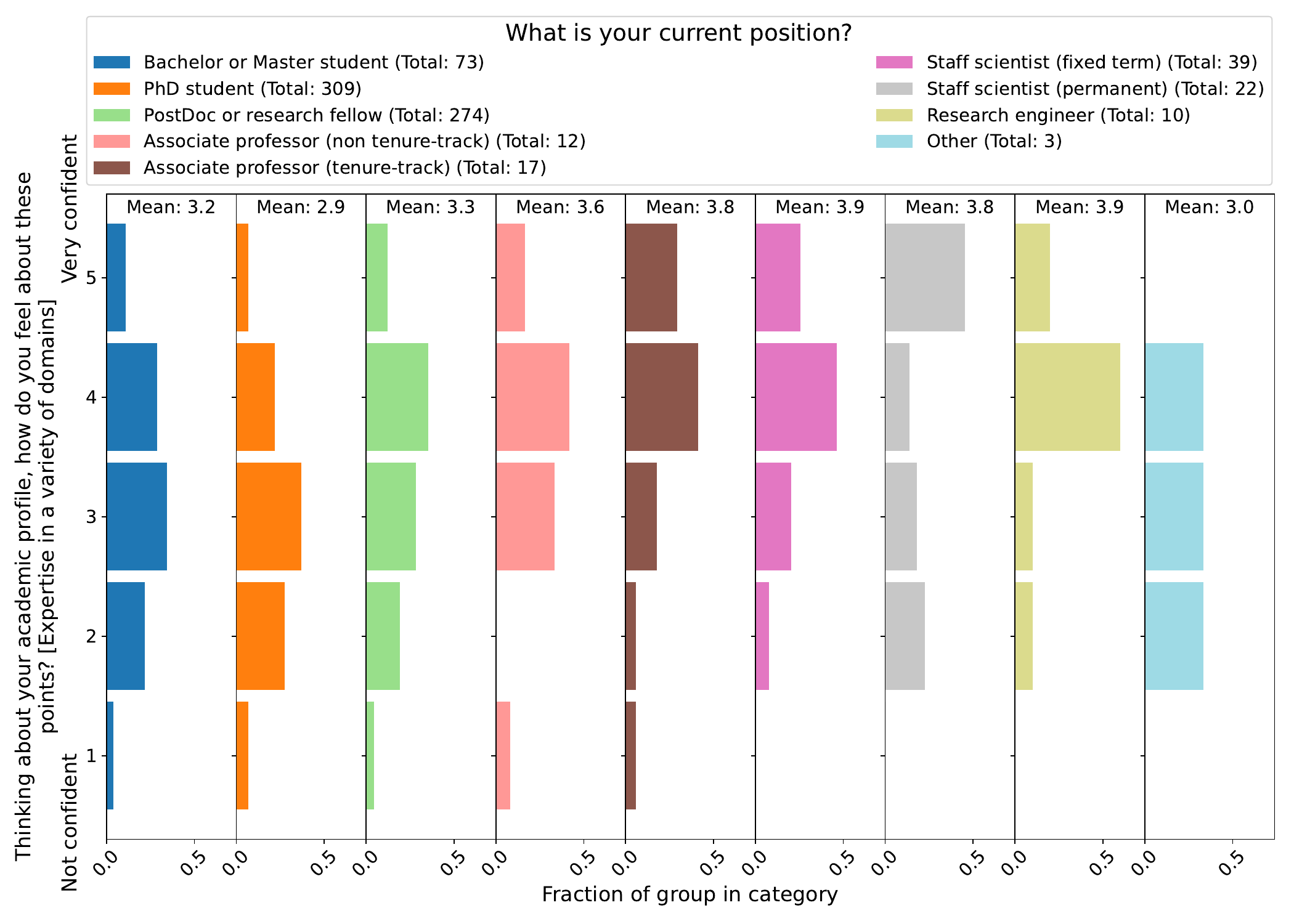}}
        \subfloat[]{\label{fig:part2:Q70ivQ11}\includegraphics[width=0.49\textwidth]{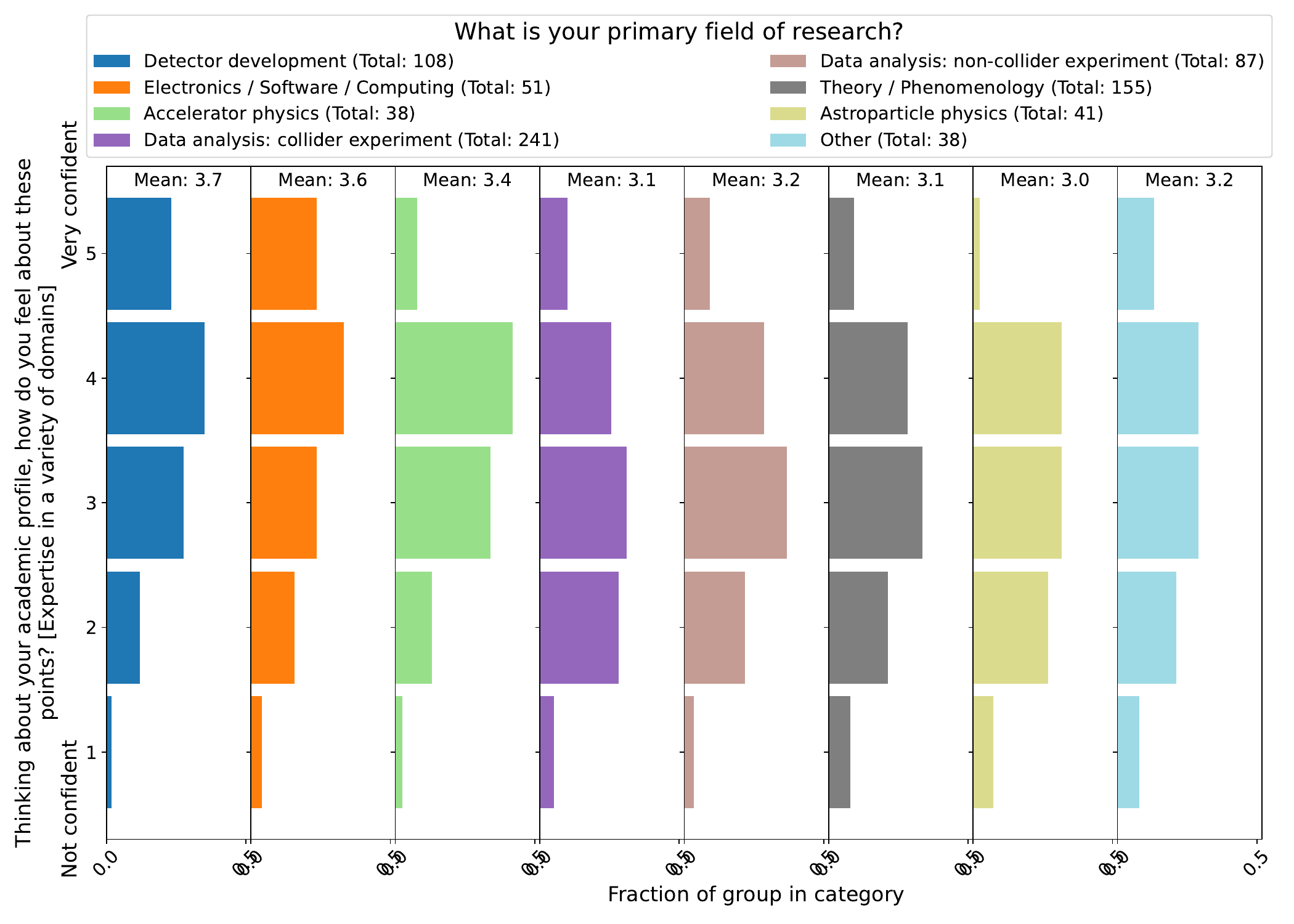}}\\
        \subfloat[]{\label{fig:part2:Q68hvQ1}\includegraphics[width=0.49\textwidth]{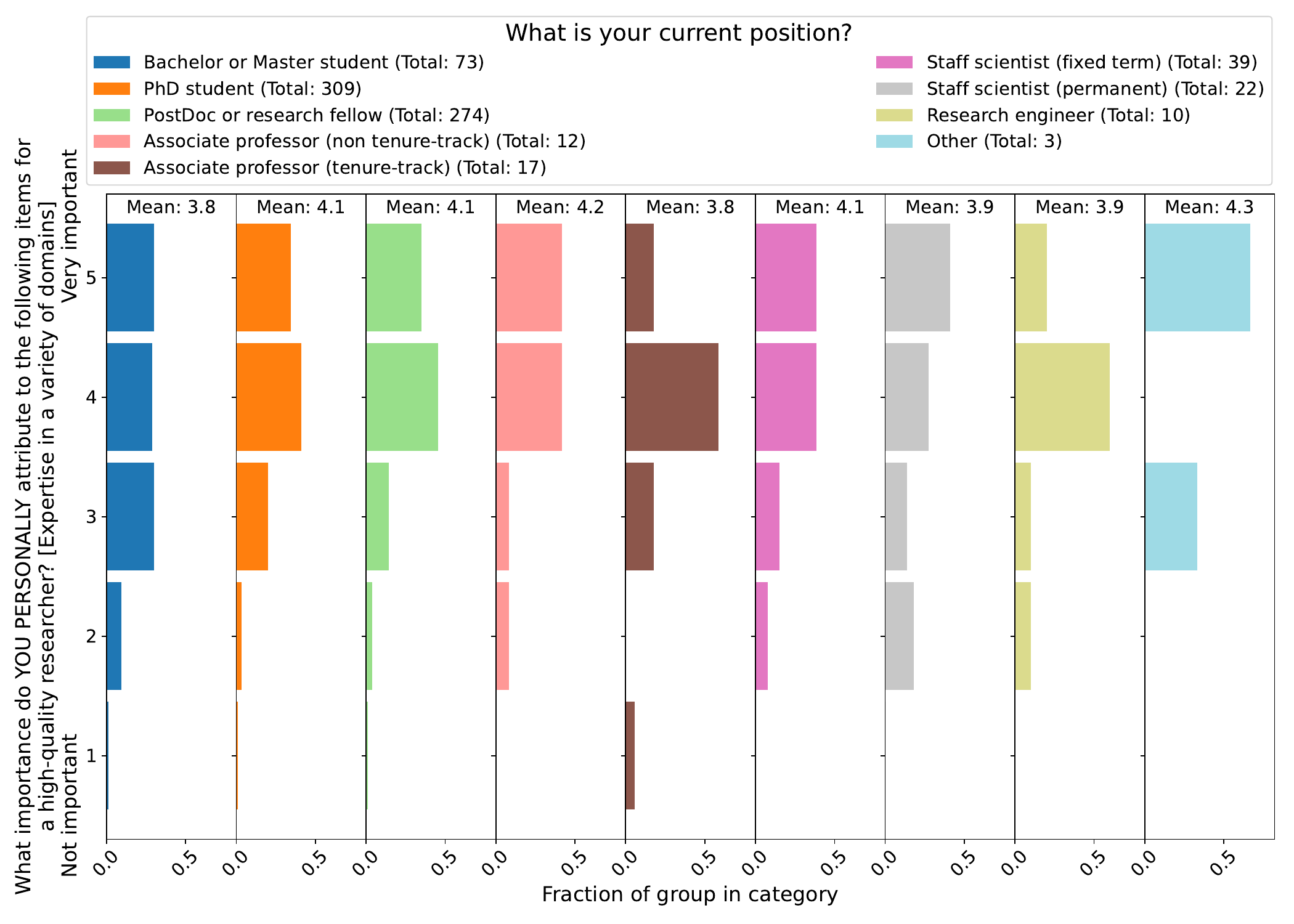}}
    \caption{(Q68h--70i v Q1,6,11) Correlations between how important having expertise in a variety of domains is considered to be for a successful academic career and selected demographics. Fractions are given out of all respondents who answered the questions.}
    \label{fig:part2:Q68-70hvQ1Q6Q11}
\end{figure}

In contrast, when considering the importance of research breadth, we saw no correlation in perceived importance between fields, while respondents working in detector development or electronics/software/computing are the most confident about it, as Figure~\ref{fig:part2:Q68-70hvQ1Q6Q11} shows.
Similar to other metrics, Asian respondents view research breadth as more important to the scientific community than Northern Europeans.
We also see that respondents with more senior positions feel this is less important to them personally, despite generally being more confident about it.
Research breadth is considered to be most important to the scientific community by non tenure-track associate professors and least important by permanent staff scientists.

\begin{figure}[ht!]
    \centering
    \includegraphics[width=0.8\textwidth]{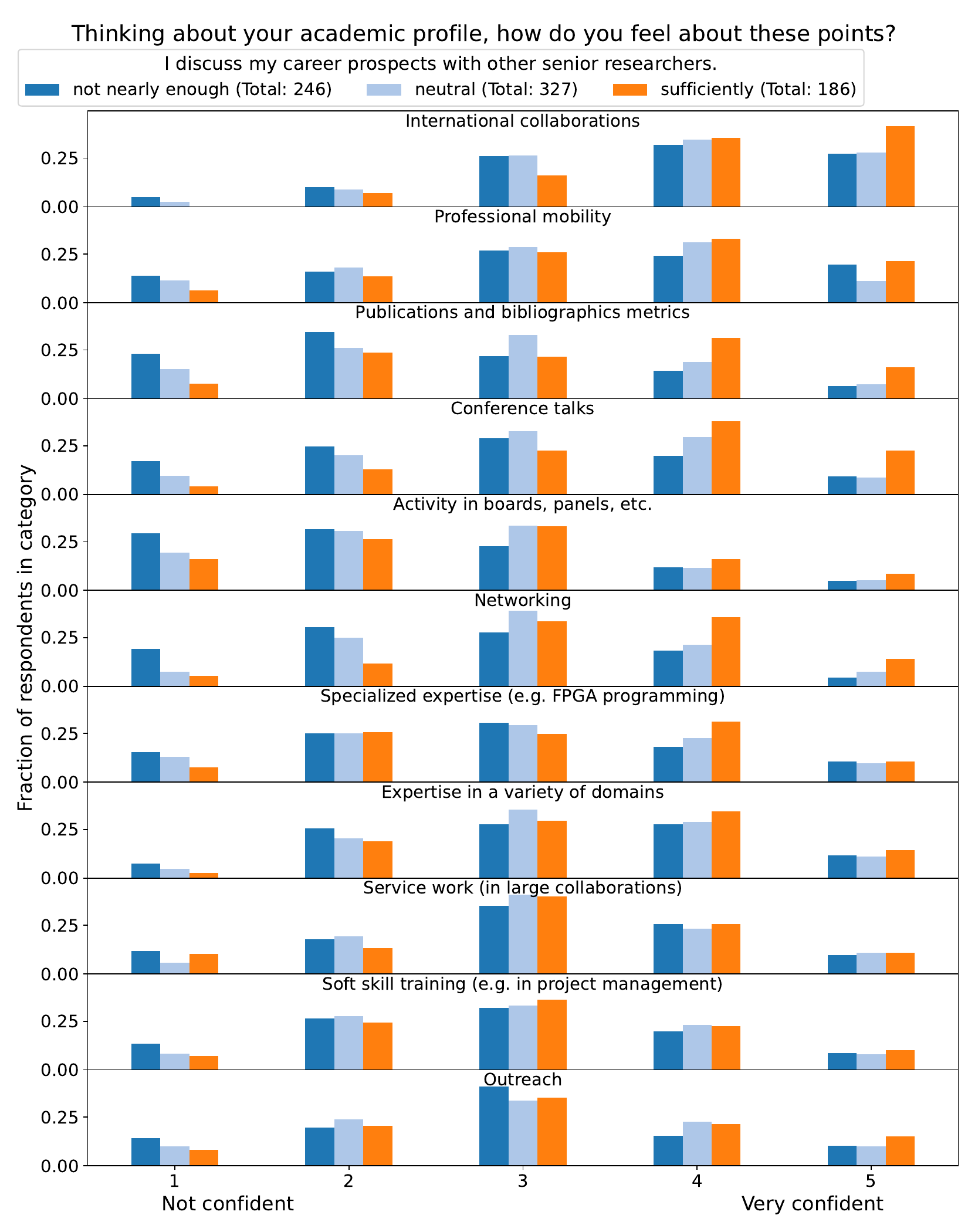}
    \caption{(Q70 v Q66) Correlations between how confident respondents feel about various items related to a successful academic career and how sufficient their discussion about career prospects is with senior researchers. Fractions are given out of all respondents.}
    \label{fig:part2:Q70vQ66}
\end{figure}

\begin{figure}[ht!]
    \centering
    \includegraphics[width=0.8\textwidth]{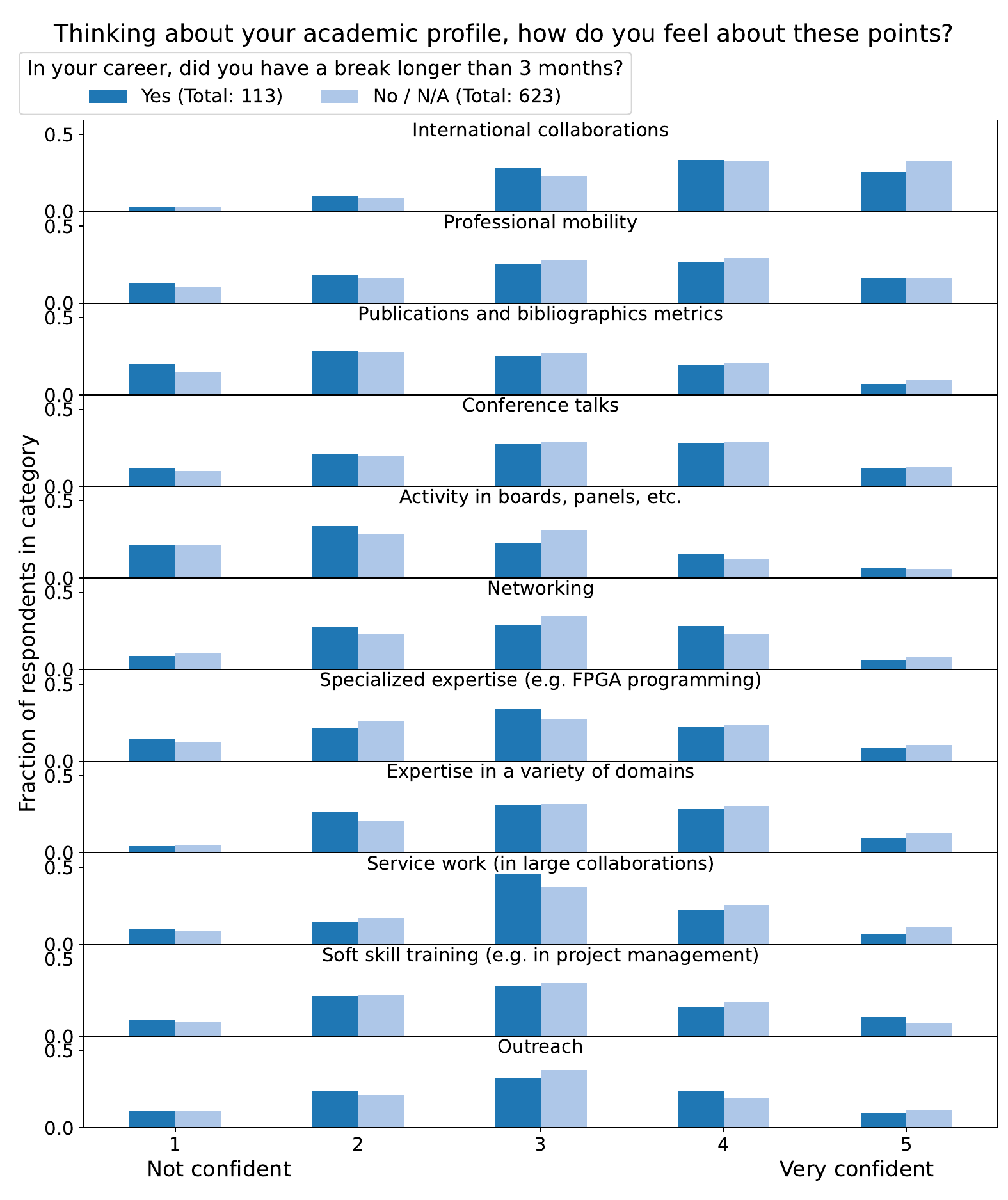}
    \caption{(Q70 v Q83) Correlations between how confident respondents feel about various items related to a successful academic career and whether they have had a career break longer than 3 months. Fractions are given out of all respondents.}
    \label{fig:part2:Q70vQ83}
\end{figure}

Next, we investigate correlations between how important respondents feel the items attributed to a successful career in academia are, and how confident they feel about them, with questions unrelated to demographics.
We find that these views are generally not correlated with questions outside of respondent demographics.
However, we found that those who discuss career prospects sufficiently with senior researchers and those who haven't taken a career break longer than 3 months feel more confident about the items, as shown in Figures~\ref{fig:part2:Q70vQ66}--\ref{fig:part2:Q70vQ83}.

%%%%%%%%%%%%%===================================================================================================
\FloatBarrier
\subsection{Work-life balance, career planning and mobility}

%----------------------------------------------------------------------------------------
\subsubsection{Work-life balance}

\begin{figure}[ht!]
    \centering
        \subfloat[]{\label{fig:part2:Q71vQ1}\includegraphics[width=0.49\textwidth]{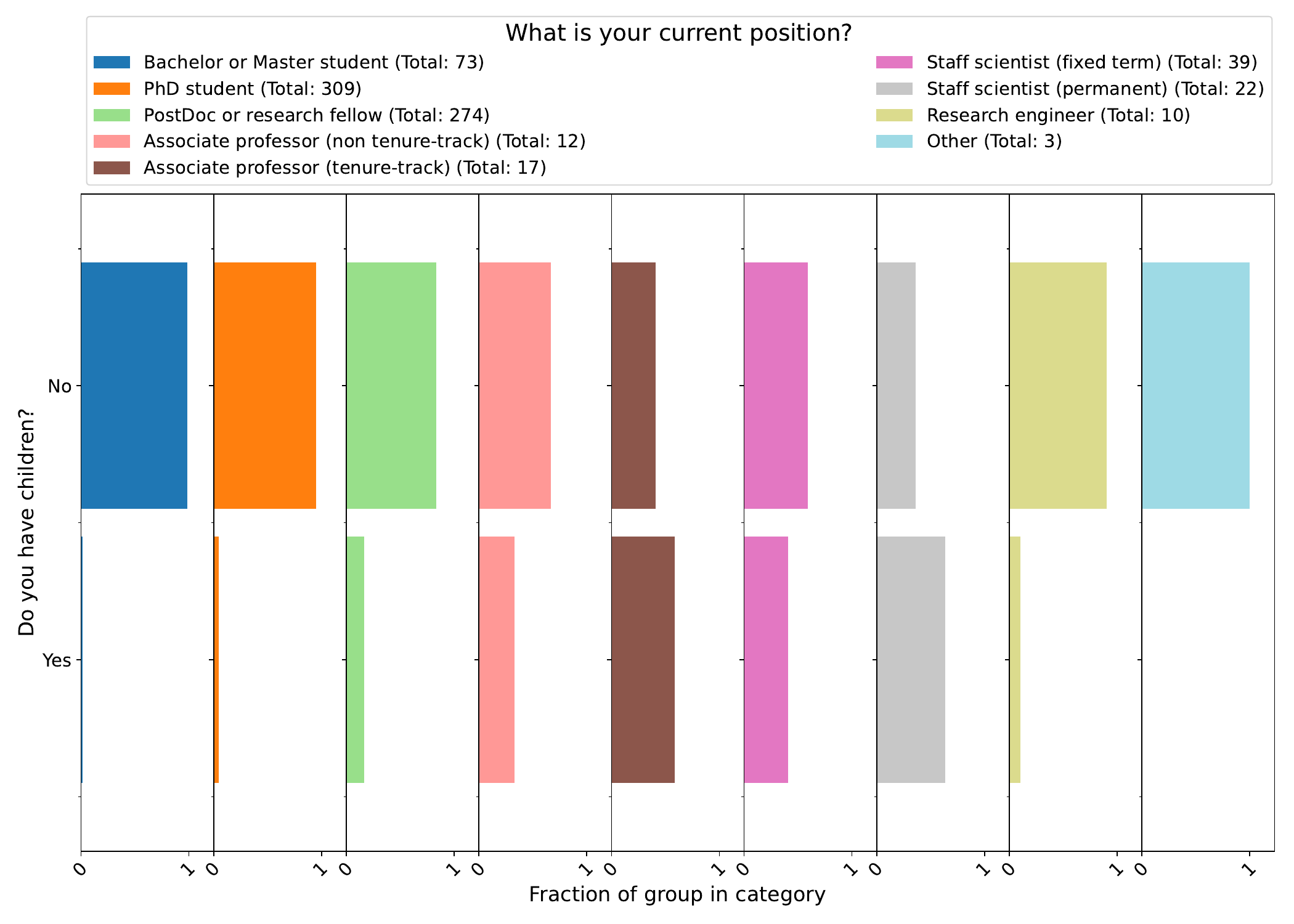}}
        \subfloat[]{\label{fig:part2:Q71vQ6}\includegraphics[width=0.49\textwidth]{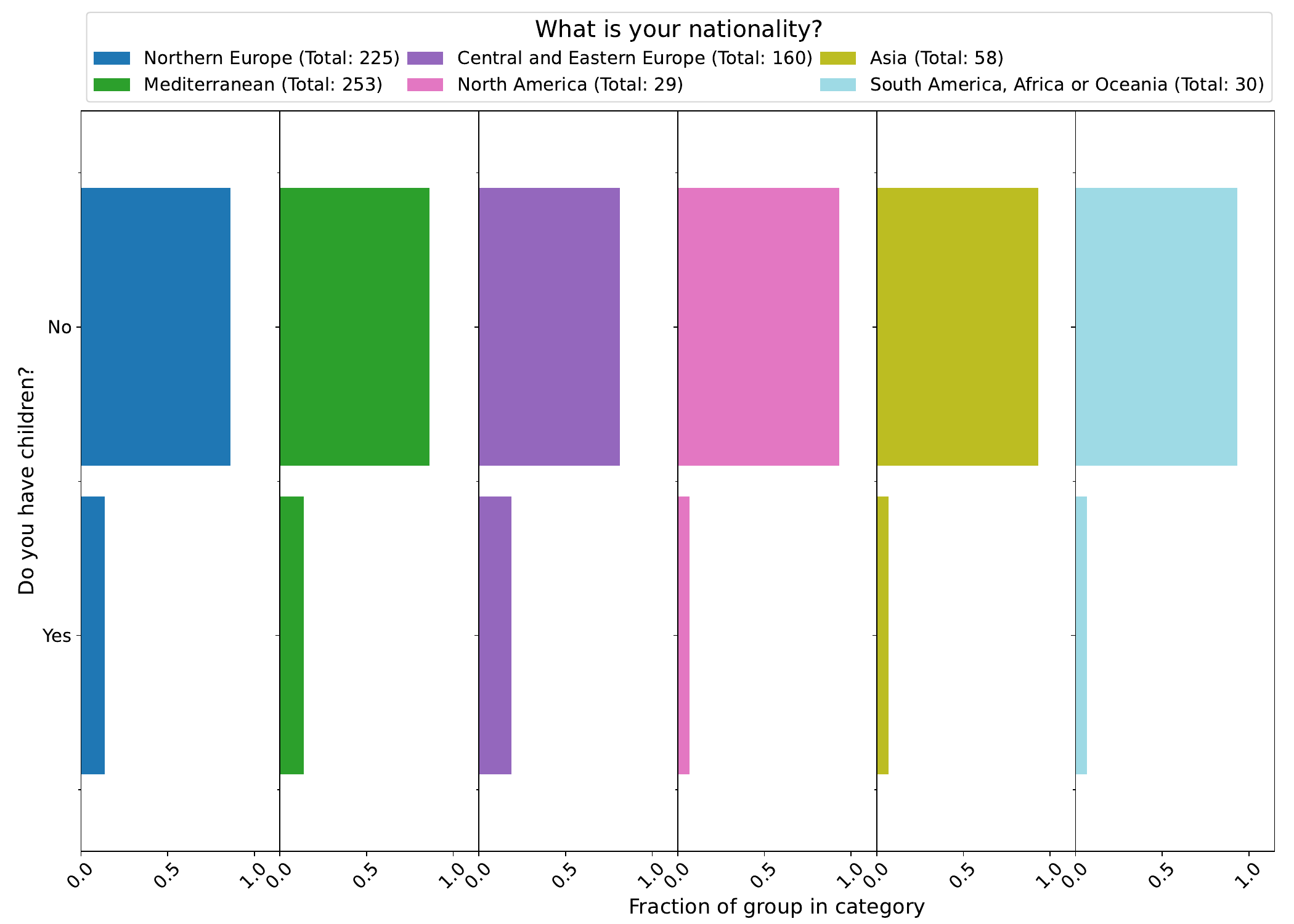}}\\
        \subfloat[]{\label{fig:part2:Q71vQ8}\includegraphics[width=0.49\textwidth]{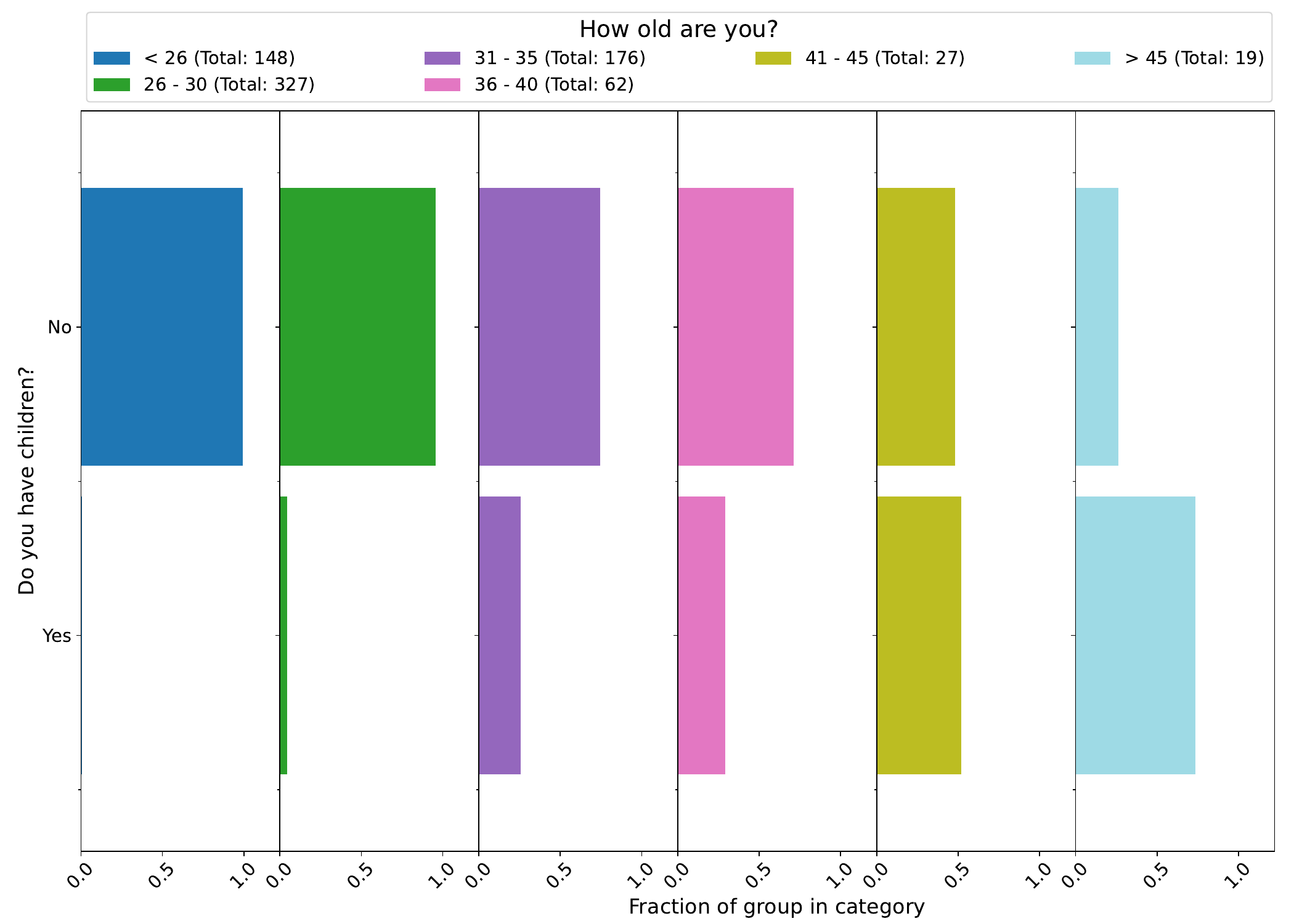}}
        \subfloat[]{\label{fig:part2:Q72vQ6}\includegraphics[width=0.49\textwidth]{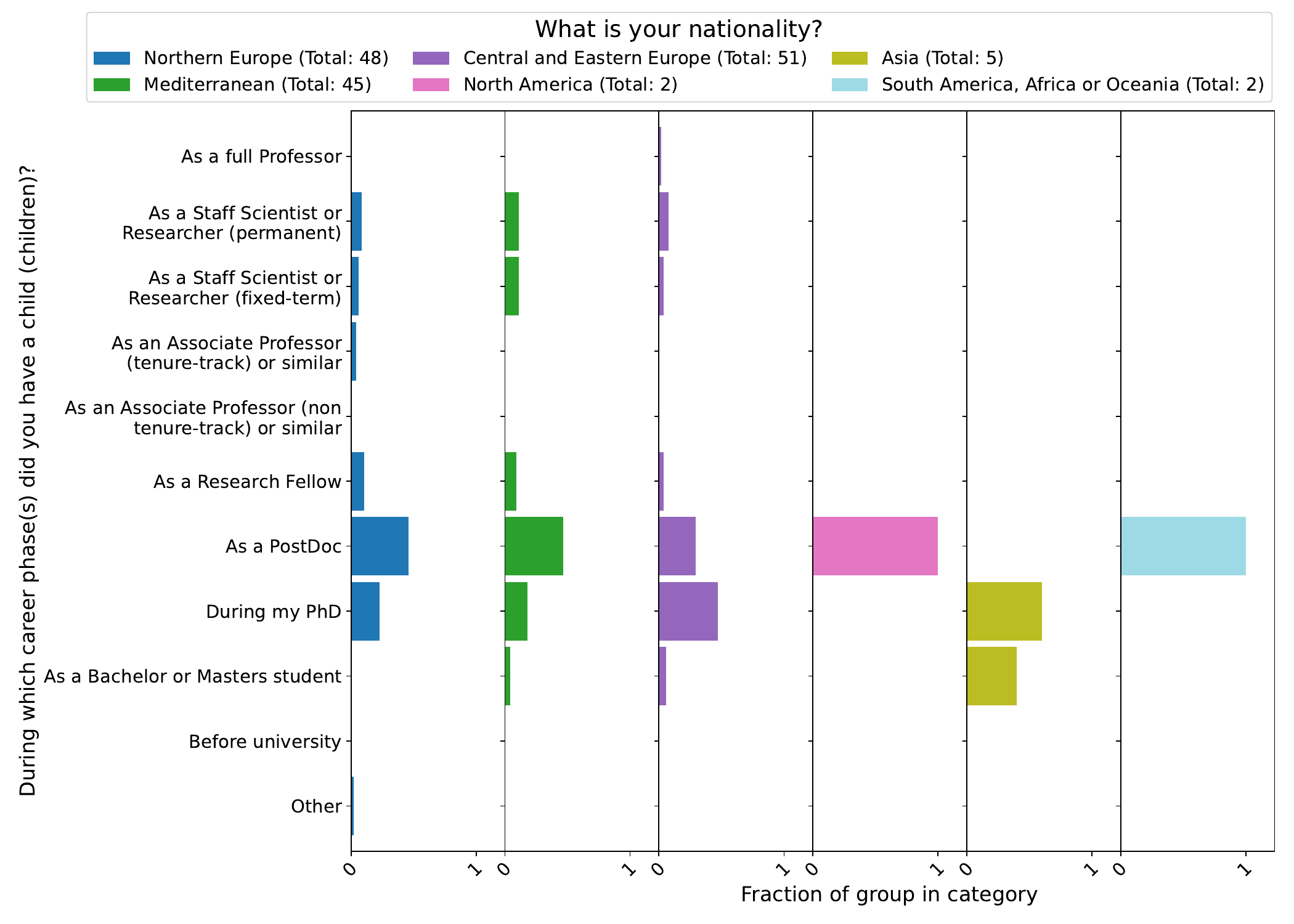}}
    \caption{(Q71 v Q1,6,8; Q72 v Q6) Correlations between whether respondents have children and at what stage in their career, and selected demographics. Fractions are given out of all respondents who answered the questions.}
    \label{fig:part2:Q71vQ1Q3Q5Q6Q8_Q72vQ6}
\end{figure}

In Figure~\ref{fig:part2:Q71vQ1Q3Q5Q6Q8_Q72vQ6}, selected correlations between whether respondents have children, and their demographics, are summarised.
As expected, the older respondents are, and the more senior their academic position, the more likely they are to have children.
We also see some geographic correlation, with more European respondents indicating they have children than those with a different nationality.
Within Europe, Central and Eastern Europeans responded that they have children slightly more frequently than those from other regions.
This correlation is similar but slightly weaker when considering country of residence (not shown).
In Figure~\ref{fig:part2:Q72vQ6}, the career phase(s) within which respondents had children are correlated with nationality.
Here, we can see a much higher portion of Central and Eastern Europeans had a child during their PhD than other respondents.

\begin{figure}[ht!]
    \centering
        \subfloat[]{\label{fig:part2:Q73bvQ8}\includegraphics[width=0.49\textwidth]{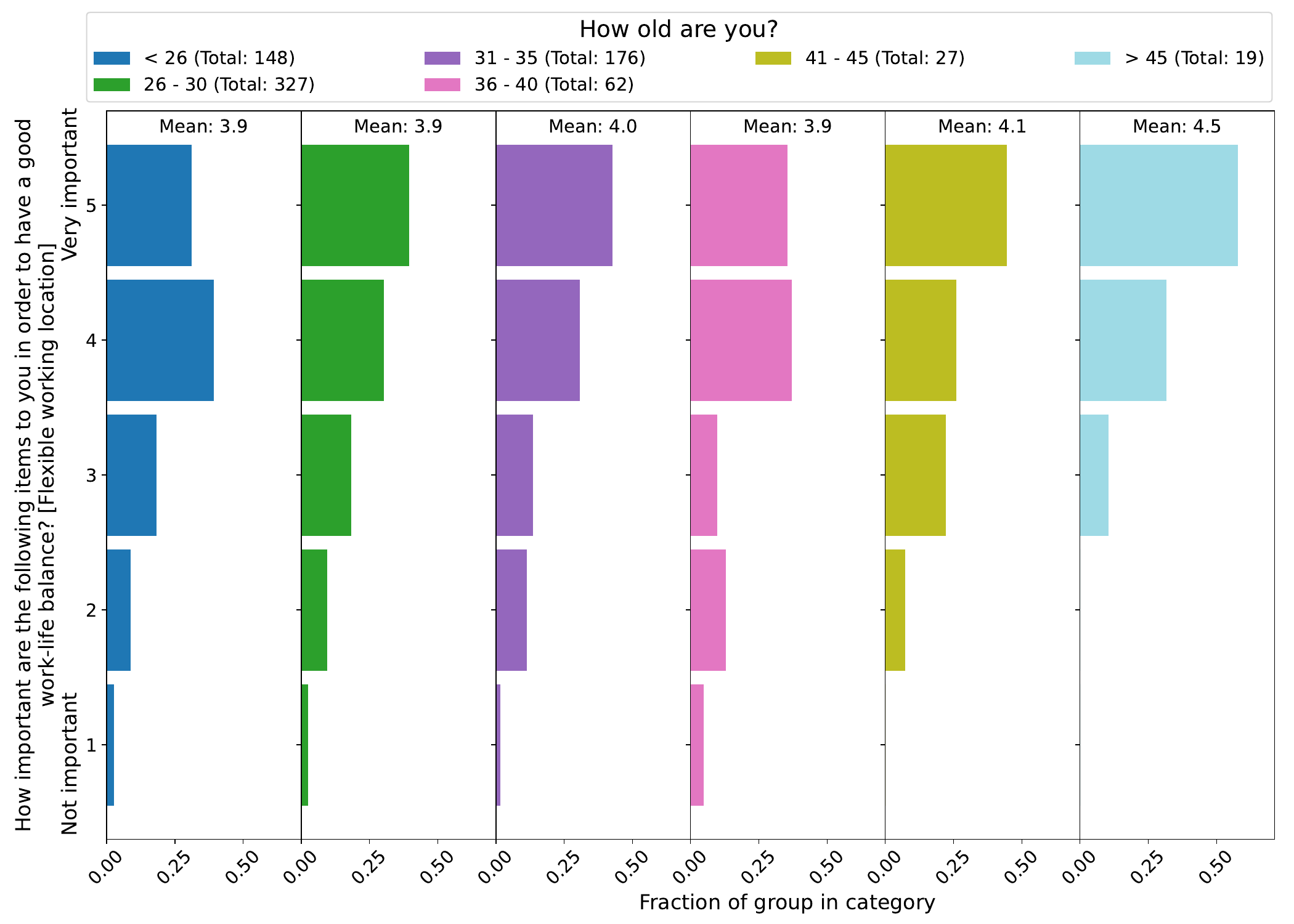}}
        \subfloat[]{\label{fig:part2:Q73cvQ4}\includegraphics[width=0.49\textwidth]{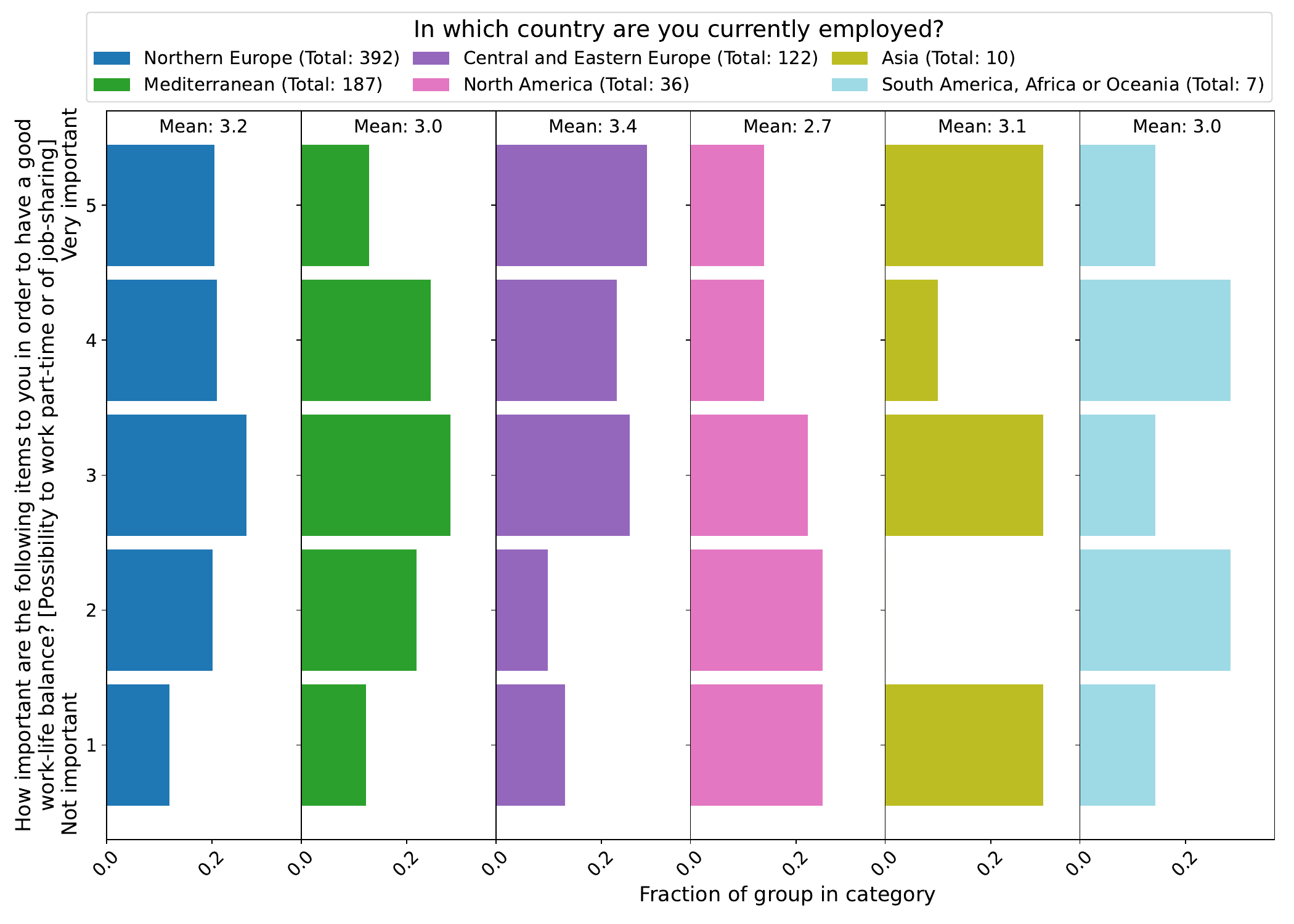}}\\
        \subfloat[]{\label{fig:part2:Q73cvQ7}\includegraphics[width=0.49\textwidth]{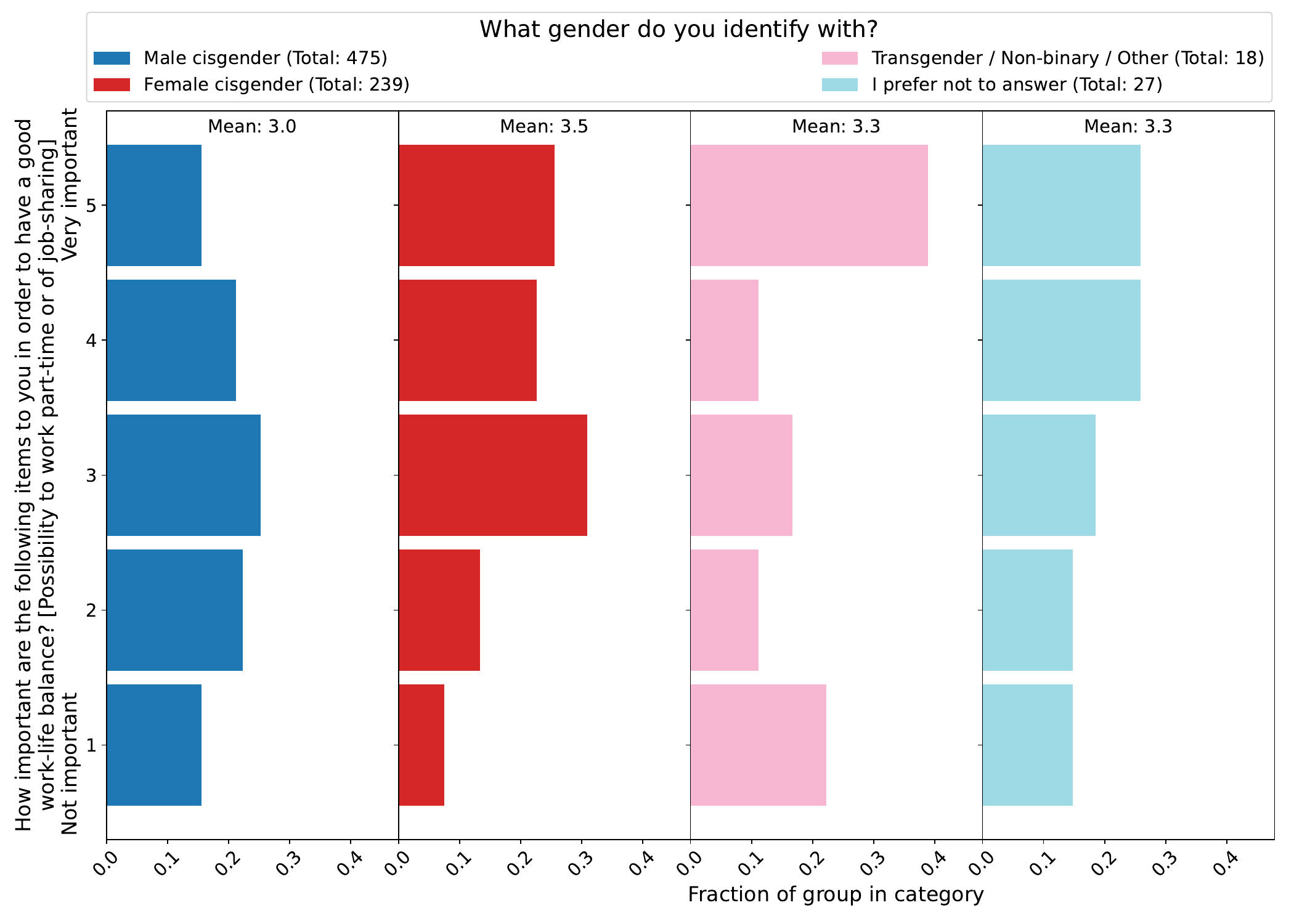}}
        \subfloat[]{\label{fig:part2:Q73dvQ1}\includegraphics[width=0.49\textwidth]{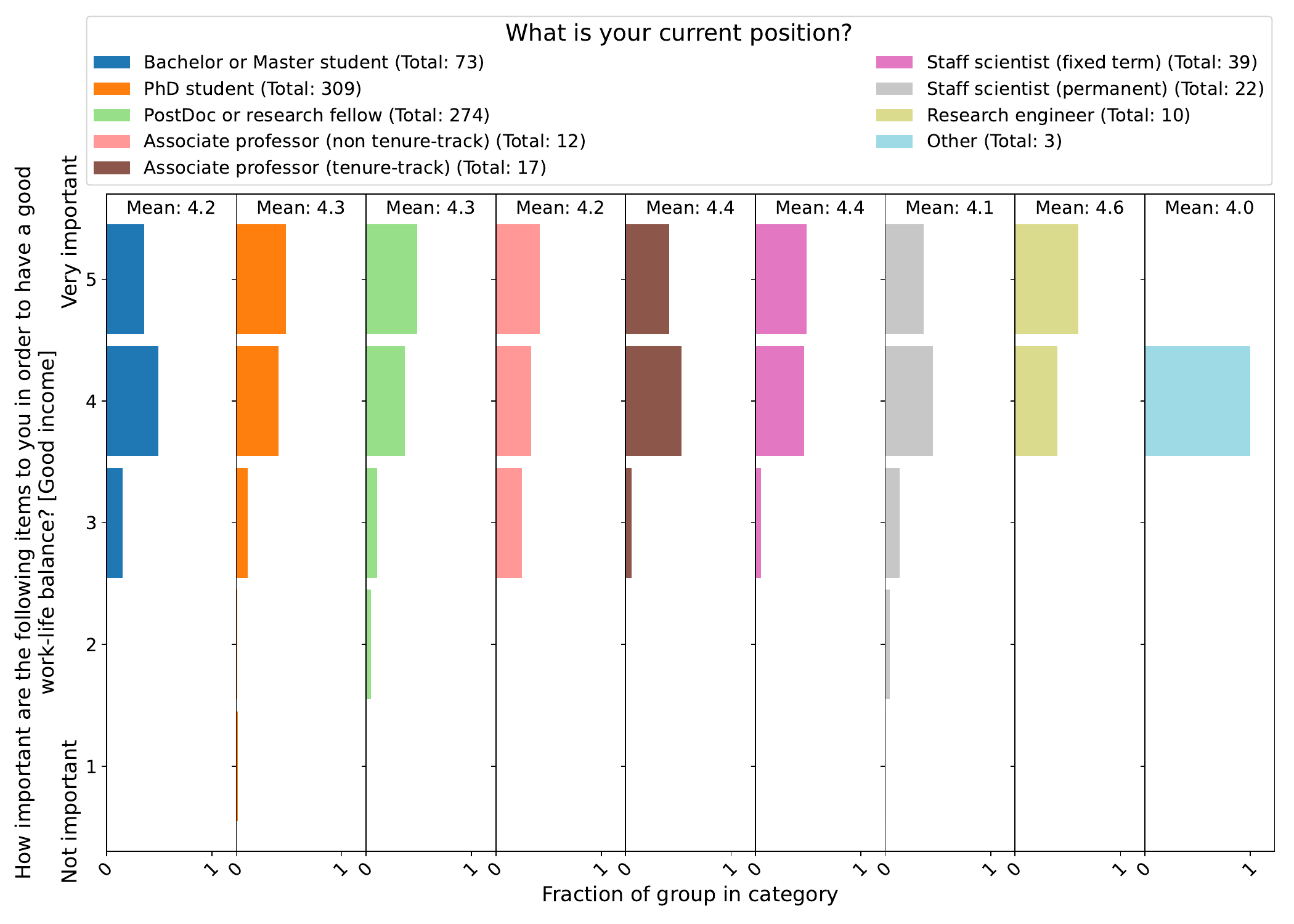}}\\
        \subfloat[]{\label{fig:part2:Q73dvQ4}\includegraphics[width=0.49\textwidth]{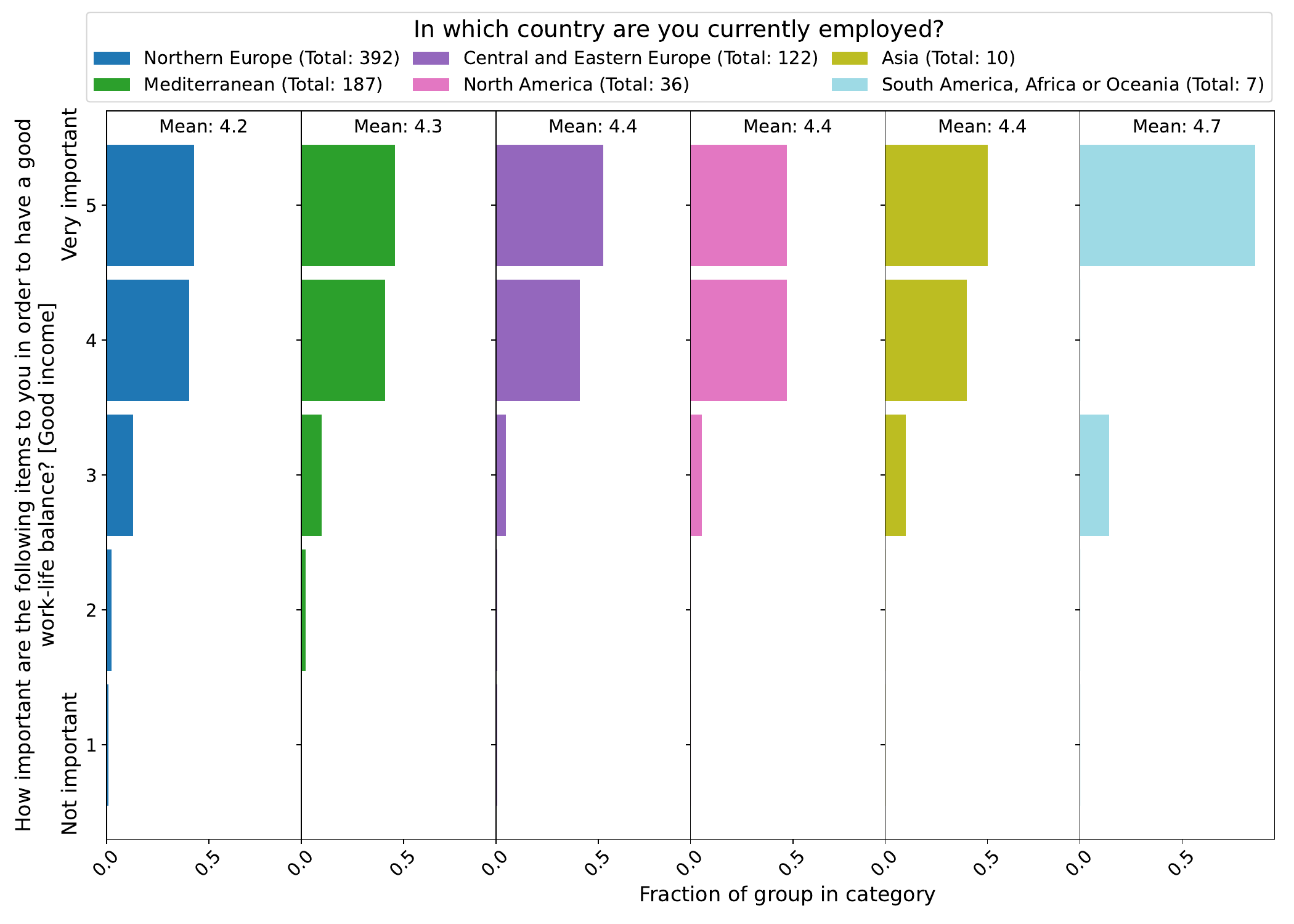}}
        \subfloat[]{\label{fig:part2:Q73evQ8}\includegraphics[width=0.49\textwidth]{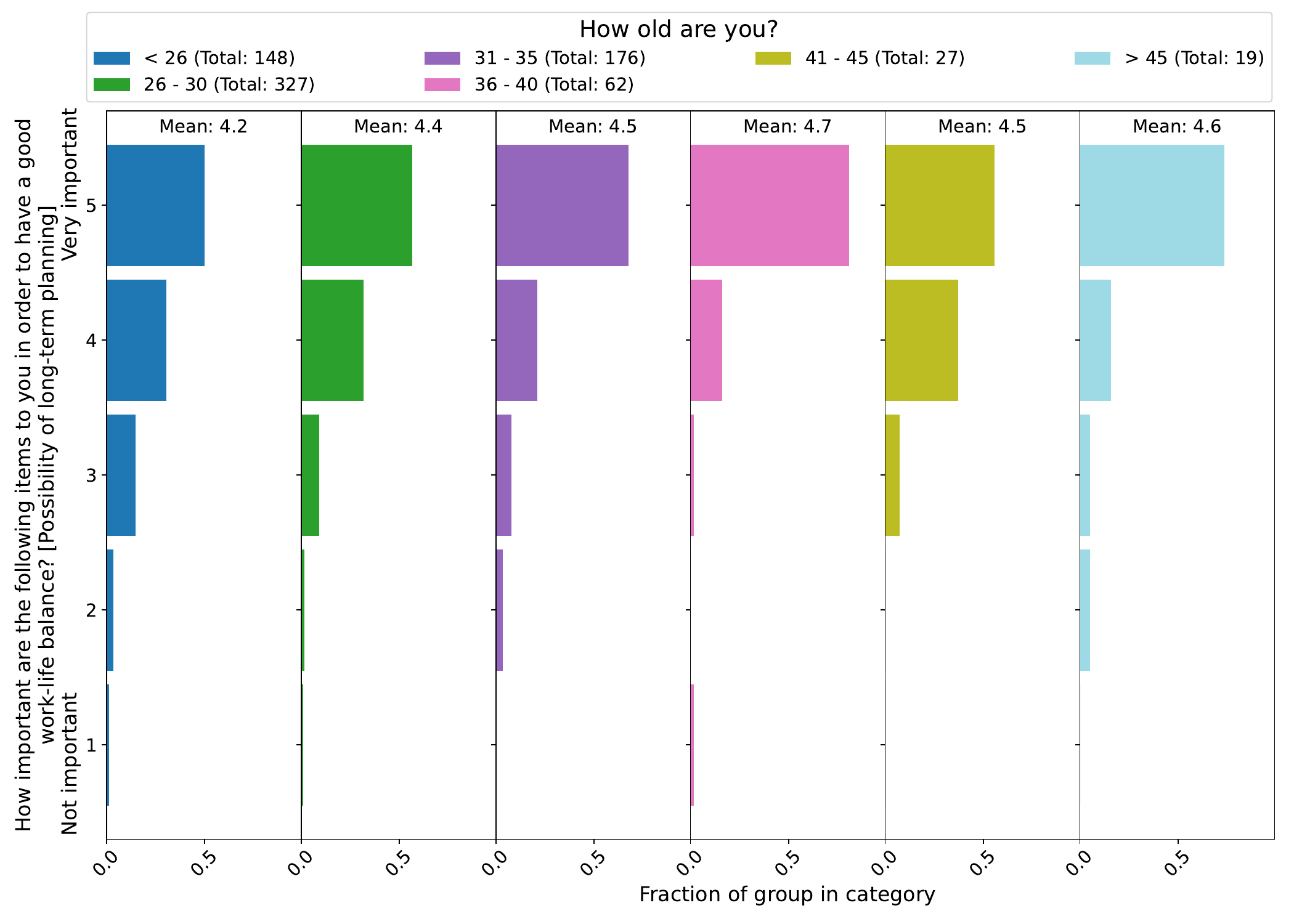}}
    \caption{(Q73b v Q8; Q73c v Q4,7; Q73d v Q1,4; Q73e v Q8) How important various items are to respondents in order to have a good work-life balance, correlated with selected demographics Fractions are given out of all respondents who answered the questions..}
    \label{fig:part2:Q73vQ1Q6Q7Q8}
\end{figure}

\begin{figure}[ht!]
    \centering
        \subfloat[]{\label{fig:part2:Q74bvQ11}\includegraphics[width=0.49\textwidth]{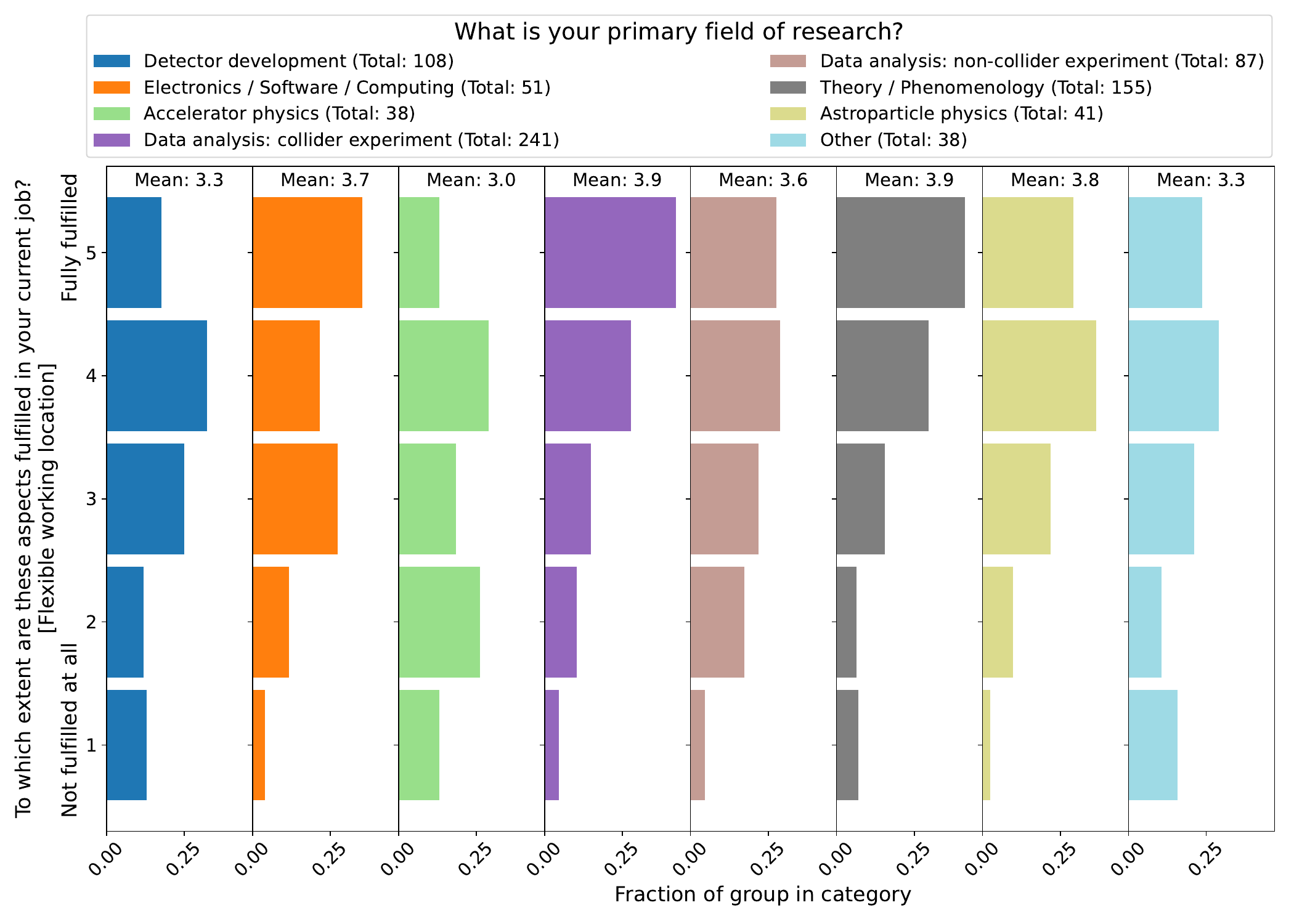}}
        \subfloat[]{\label{fig:part2:Q74cvQ4}\includegraphics[width=0.49\textwidth]{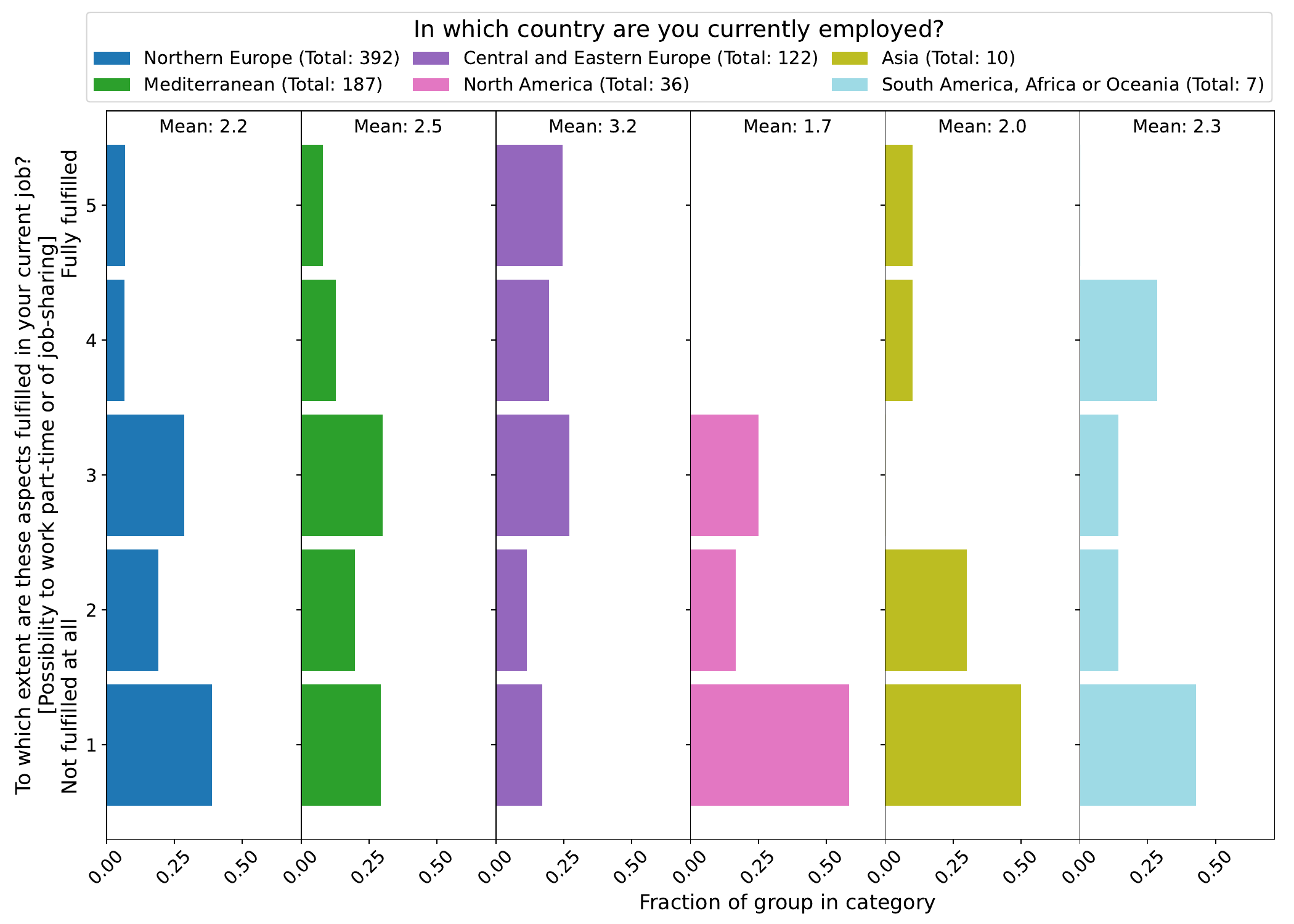}}\\
        \subfloat[]{\label{fig:part2:Q74cvQ7}\includegraphics[width=0.49\textwidth]{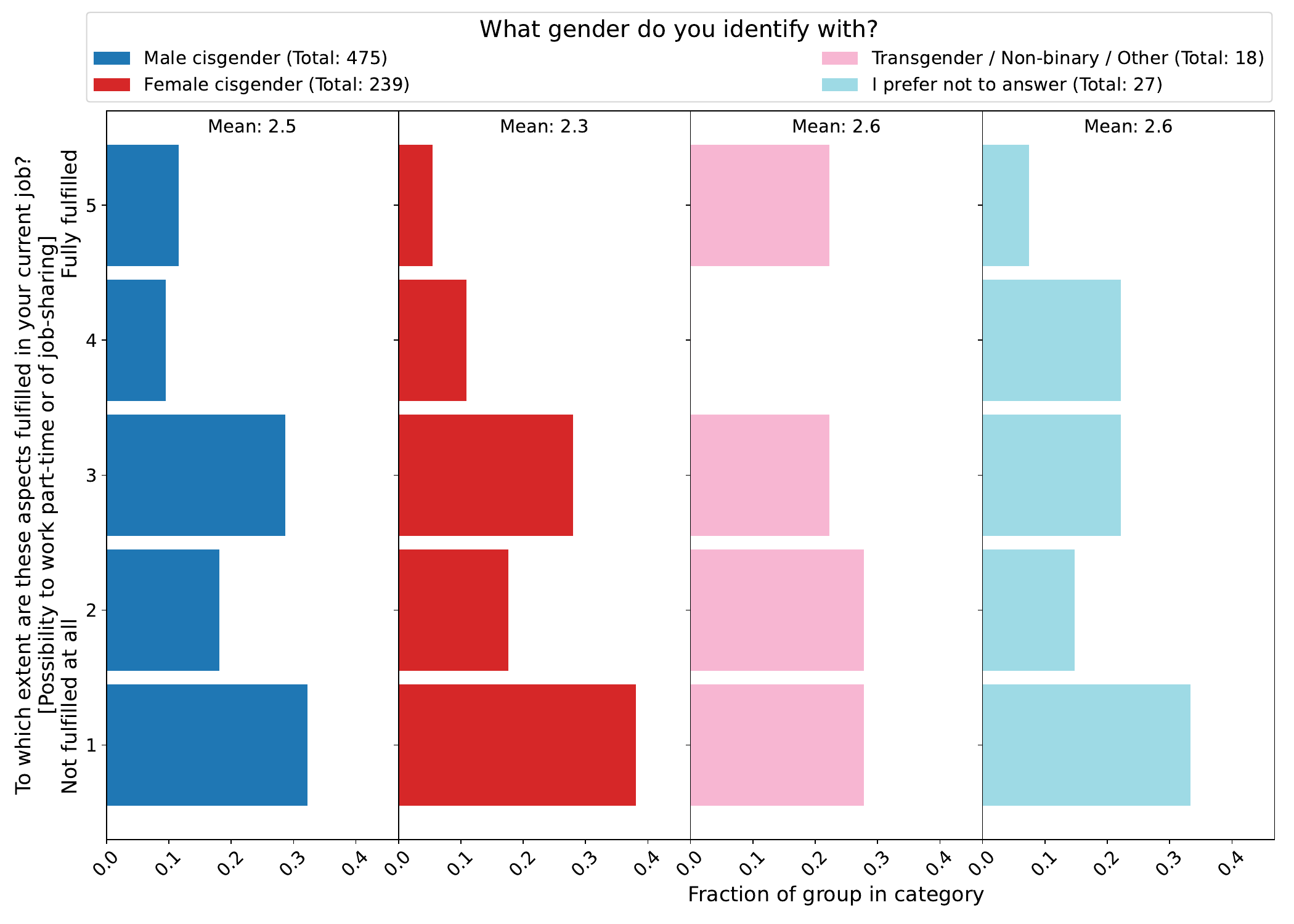}}
        \subfloat[]{\label{fig:part2:Q74dvQ1}\includegraphics[width=0.49\textwidth]{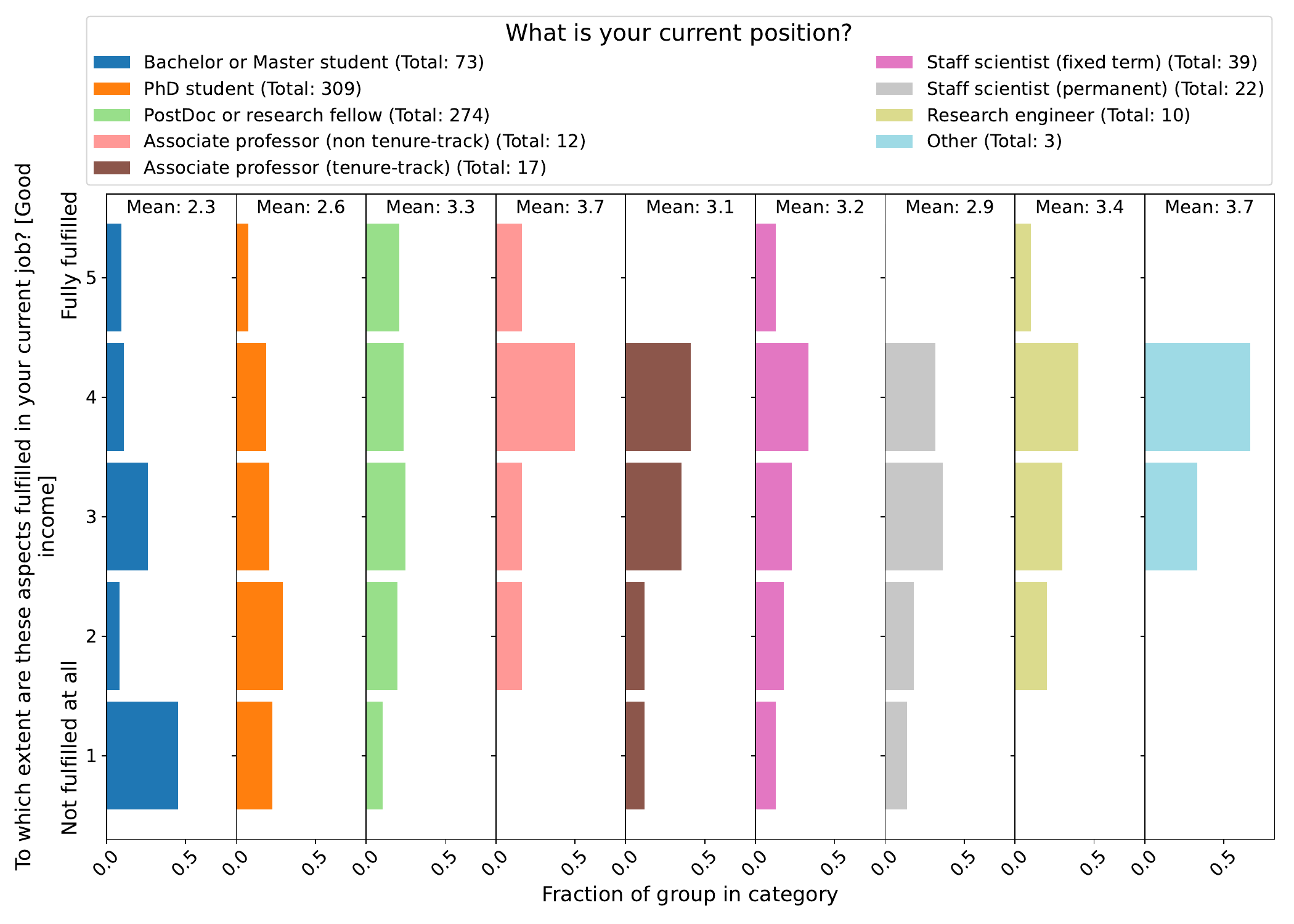}}\\
        \subfloat[]{\label{fig:part2:Q74dvQ2}\includegraphics[width=0.49\textwidth]{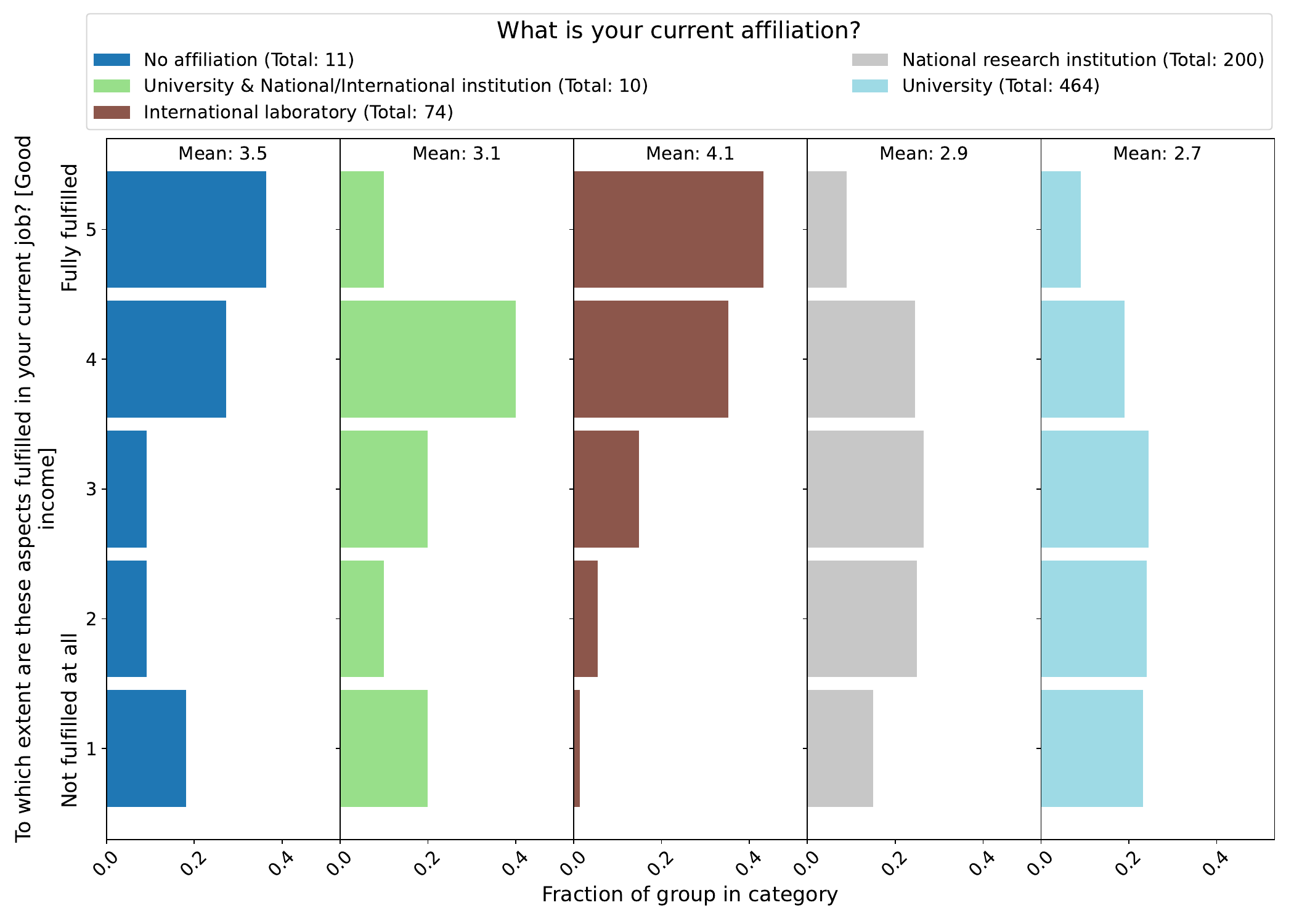}}
        \subfloat[]{\label{fig:part2:Q74dvQ4}\includegraphics[width=0.49\textwidth]{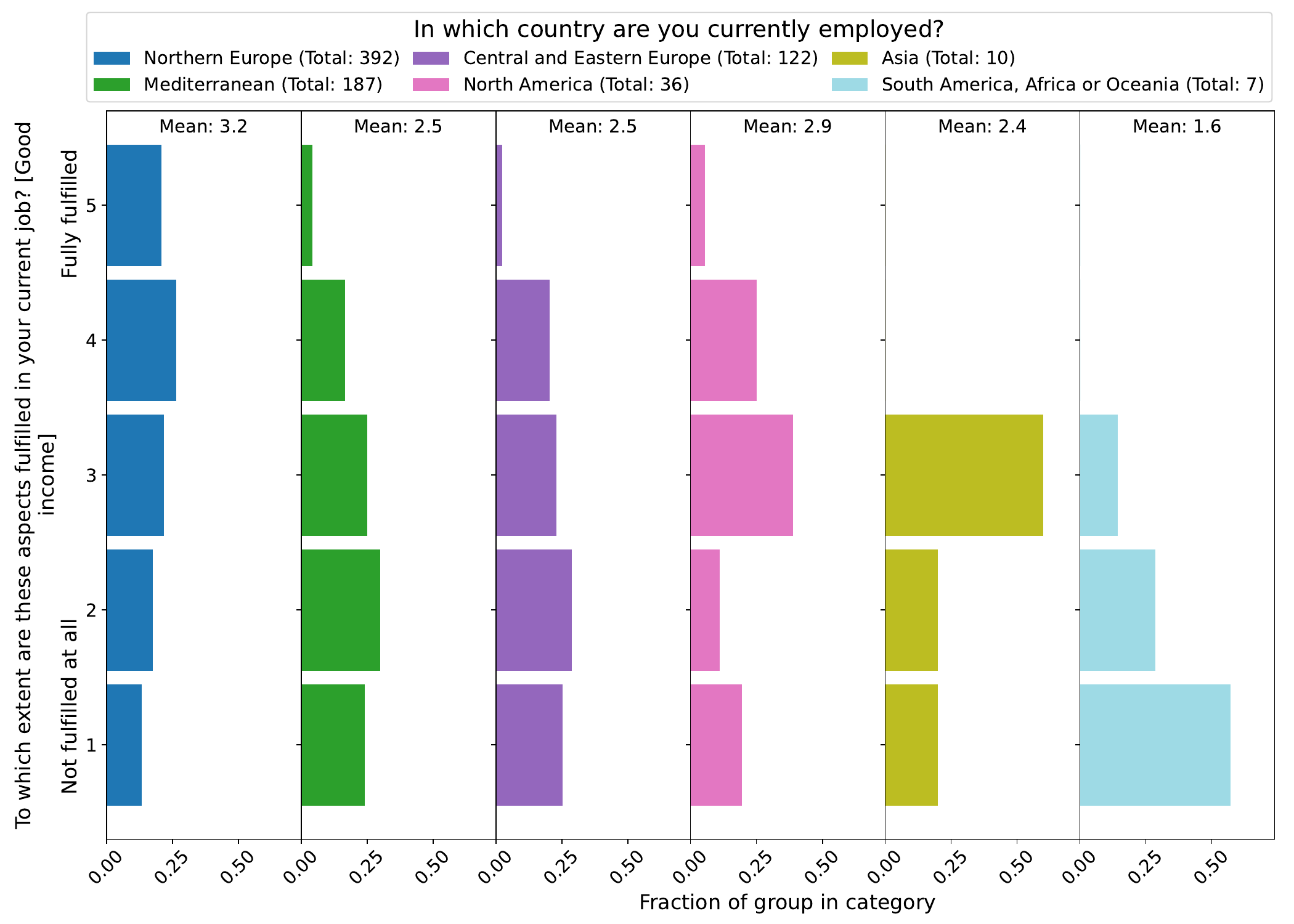}}
    \caption{(Q74b v Q11; Q74c v Q4,7; Q74d v Q1,2,4) How respondents view various items related to a good work-life balance being fulfilled in their current job, correlated with selected demographics. Fractions are given out of all respondents who answered the questions.}
    \label{fig:part2:Q74vQ1Q2Q4Q7Q11}
\end{figure}

\begin{figure}[ht!]
    \centering
        \subfloat[]{\label{fig:part2:Q74dvQ11}\includegraphics[width=0.49\textwidth]{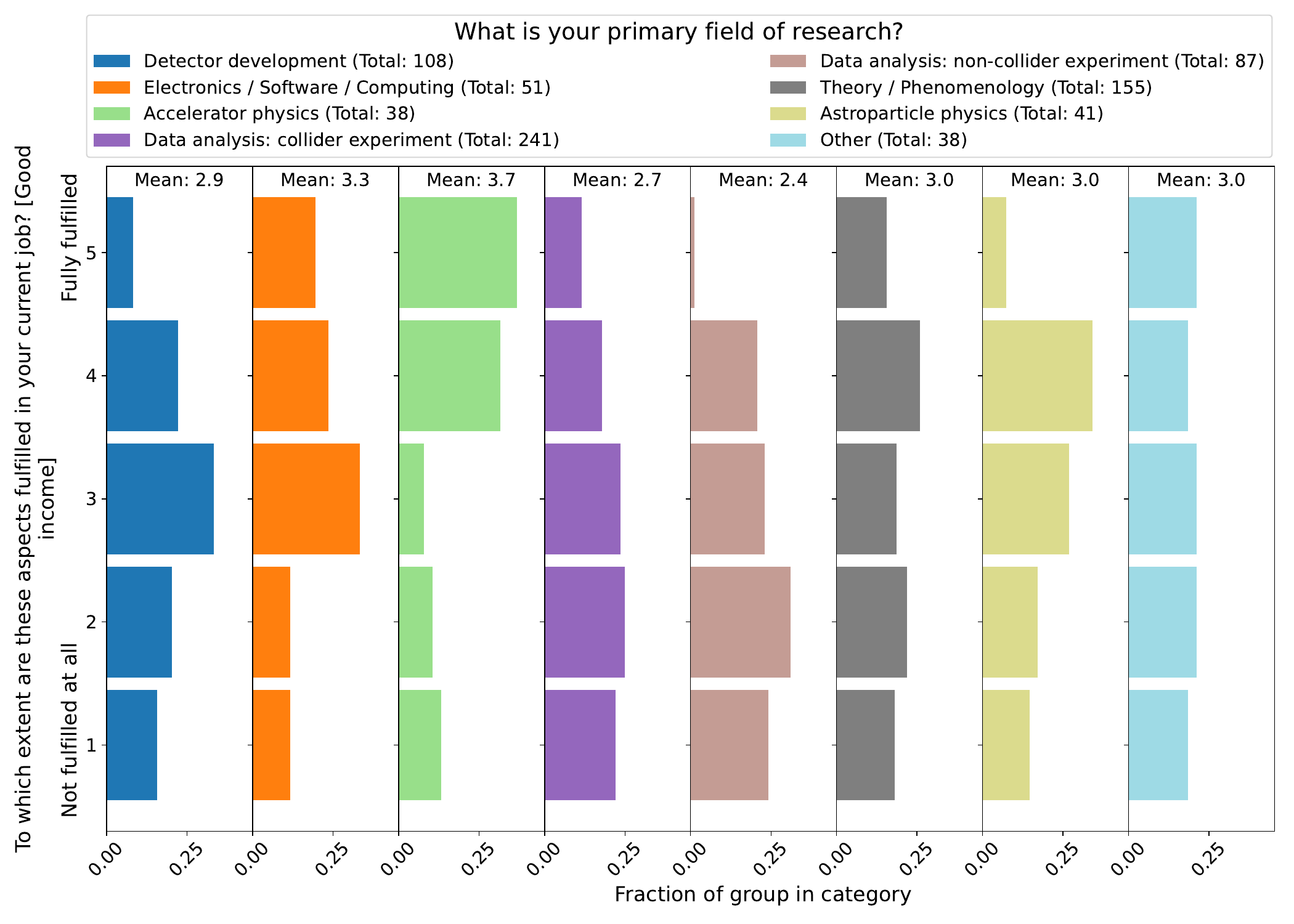}}
        \subfloat[]{\label{fig:part2:Q74evQ1}\includegraphics[width=0.49\textwidth]{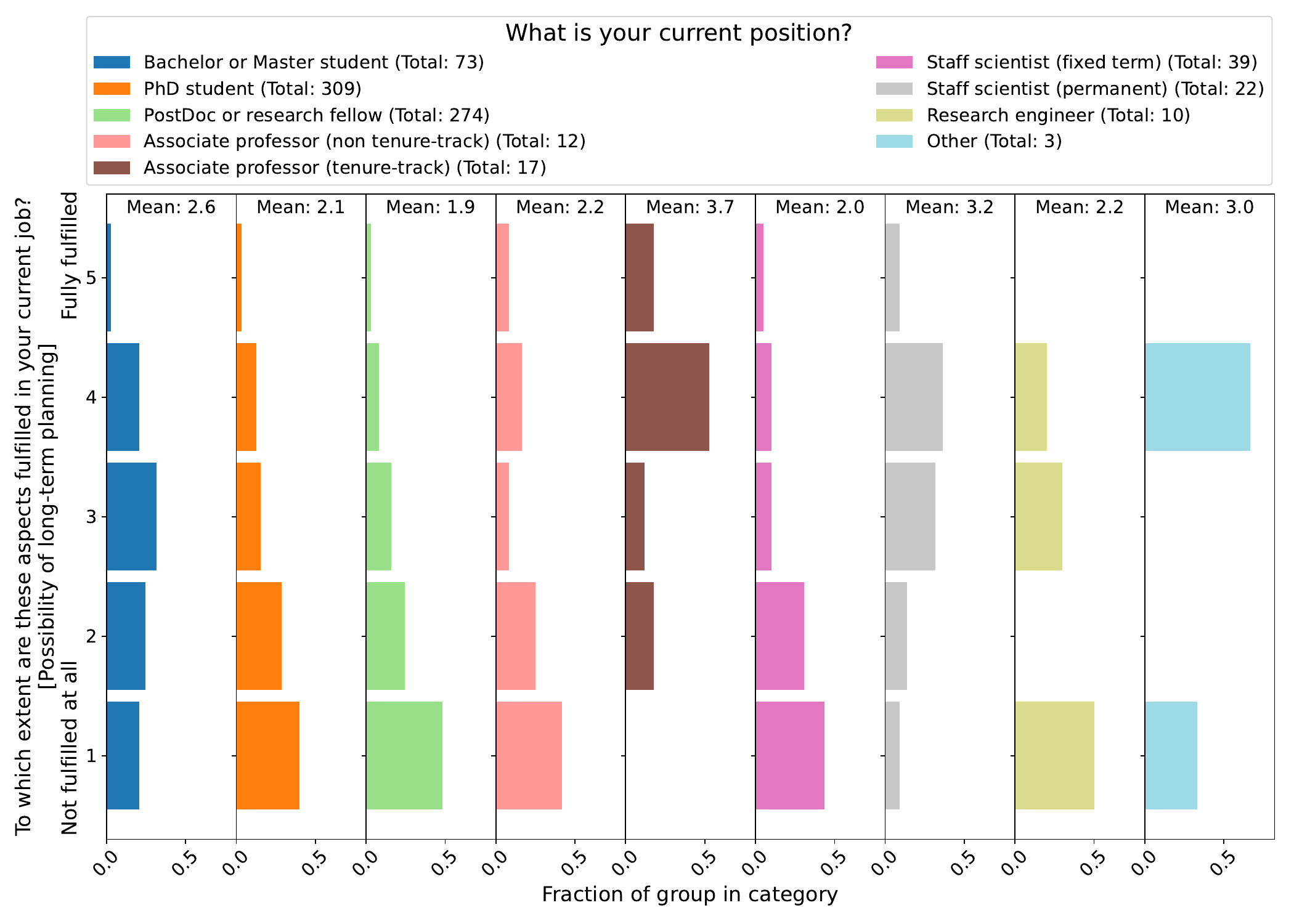}}\\
        \subfloat[]{\label{fig:part2:Q74evQ4}\includegraphics[width=0.49\textwidth]{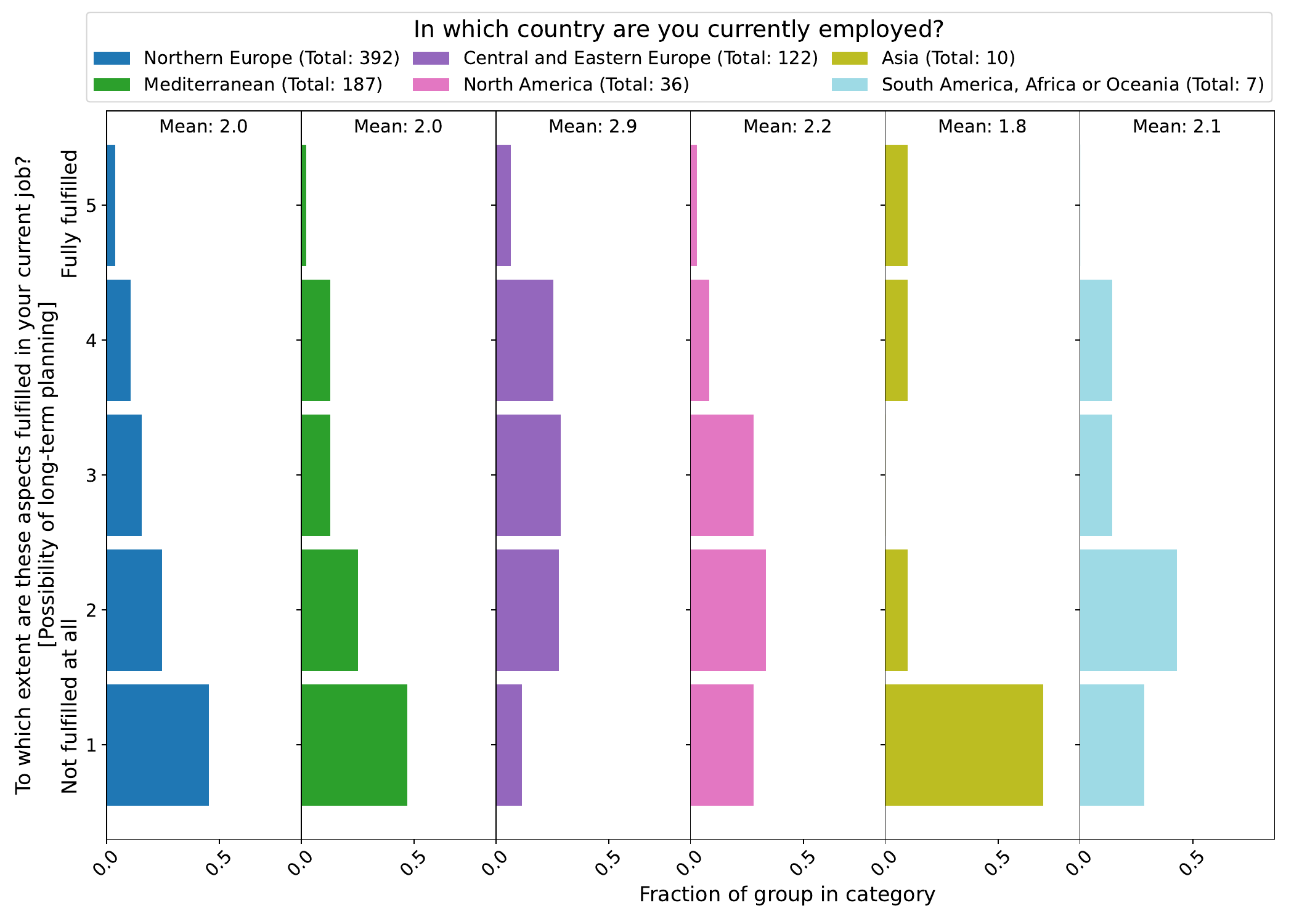}}
        \subfloat[]{\label{fig:part2:Q74evQ11}\includegraphics[width=0.49\textwidth]{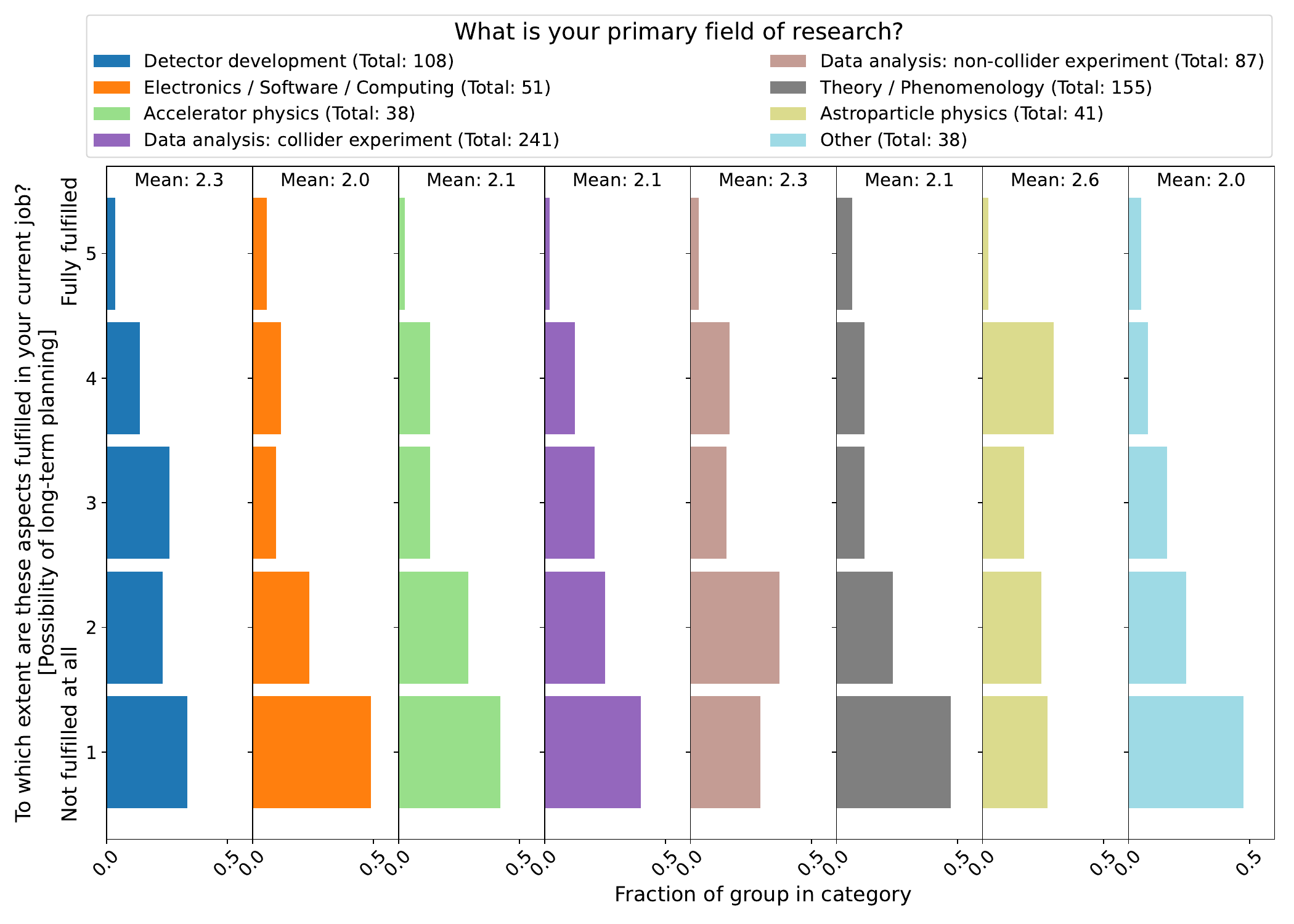}}
    \caption{(Q74d v Q11; Q74e v Q1,4,11) How respondents view various items related to a good work-life balance being fulfilled in their current job, correlated with selected demographics. Fractions are given out of all respondents who answered the questions.}
    \label{fig:part2:Q74vQ1Q4Q11}
\end{figure}

Correlations were next studied between views regarding how important various items (listed in Appendix~\ref{app:questions}) are to respondents in order to have a good work-life balance, and demographics.
Several interesting correlations are shown in Figure~\ref{fig:part2:Q73vQ1Q6Q7Q8}.
Additionally, we investigated the correlations between to what extent respondents feel the items relating to work-life balance are fulfilled in their current job, and demographics.
Selected correlations in this case are shown in Figures~\ref{fig:part2:Q74vQ1Q2Q4Q7Q11}--\ref{fig:part2:Q74vQ1Q4Q11}.

The importance and fulfilment of flexible working hours and a positive work environment weren't found to be strongly correlated to respondent demographics.
The importance of flexible working location is similarly distributed across most demographics, and it is shown in Figure~\ref{fig:part2:Q73bvQ8} weakly positively correlated with respondents' ages.
Fields that require work with hardware offer less flexible work locations to respondents than more theoretical or data analysis related fields, as seen in Figure~\ref{fig:part2:Q74bvQ11}.

The possibility of working part-time or of job-sharing is shown in Figure~\ref{fig:part2:Q73cvQ4} to be less important for respondents employed in North America than elsewhere.
However, Central and Eastern Europe appears to offer this the most and North America the least, according to Figure~\ref{fig:part2:Q74cvQ4}.
We see in Figure~\ref{fig:part2:Q73cvQ7}, it is also most important for cisgender female respondents, despite them feeling it is the least fulfilled in their current job, as shown in Figure~\ref{fig:part2:Q74cvQ7}.

A good income is more important for research engineer respondents than those with other positions, according to Figure~\ref{fig:part2:Q73dvQ1}.
Furthermore, respondents working in accelerator physics and electronics are more satisfied with their salaries than those in other fields, as shown in Figure~\ref{fig:part2:Q74dvQ11}.
It is also weakly correlated with country of employment, being slightly less important to those employed in Northern Europe in Figure~\ref{fig:part2:Q73dvQ4}.
This is interesting given that we see in Figure~\ref{fig:part2:Q74dvQ4} that respondents employed in Northern Europe are the most satisfied with their income.
Considering their current jobs in Figure~\ref{fig:part2:Q74dvQ1}, student respondents are less satisfied with their pay than older colleagues, though it is similarly important to all.
This is not unexpected since they are likely to be paid the least.
However, one can see that also respondents with tenure-track or permanent positions are less satisfied with their income. 
Finally, we see in Figure~\ref{fig:part2:Q74dvQ2} that respondents working in international laboratories are more satisfied with their salaries than respondents working at universities or national research institutions.

We see in Figure~\ref{fig:part2:Q74evQ1} that respondents with tenure or permanent positions have a much better possibility of long-term planning than respondents in other positions.
In Figure~\ref{fig:part2:Q74evQ4} we see that respondents employed in Central and Eastern Europe are more positive about the possibility of long-term planning in their current job relative to elsewhere.
Furthermore, respondents' jobs in astroparticle physics offer a better possibility of long-term planning than other fields, but the level of fulfilment seen in Figure~\ref{fig:part2:Q74evQ11} is still very low.

\begin{figure}[ht!]
    \centering
        \subfloat[]{\label{fig:part2:Q74avQ75a}\includegraphics[width=0.49\textwidth]{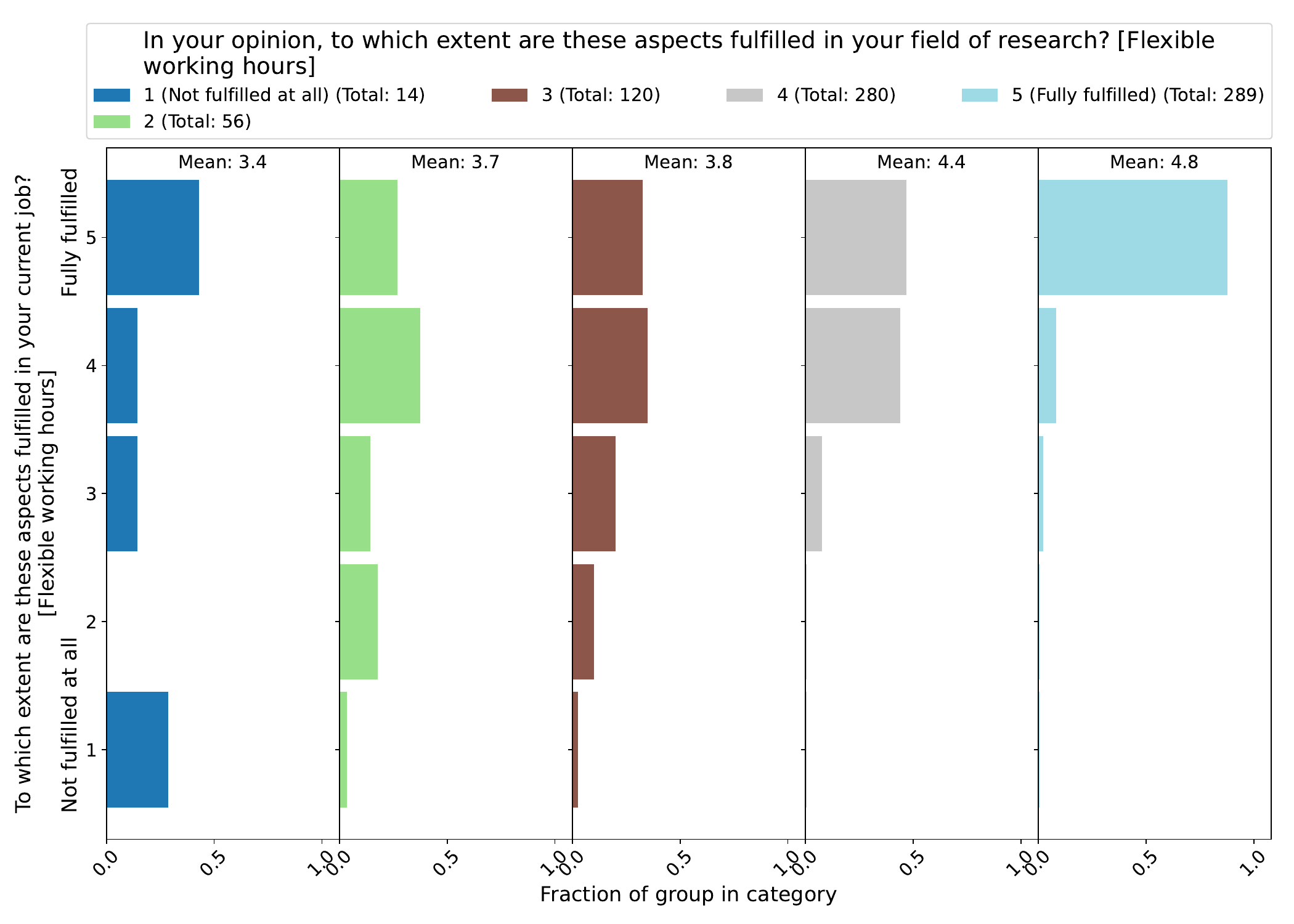}}
        \subfloat[]{\label{fig:part2:Q74bvQ75b}\includegraphics[width=0.49\textwidth]{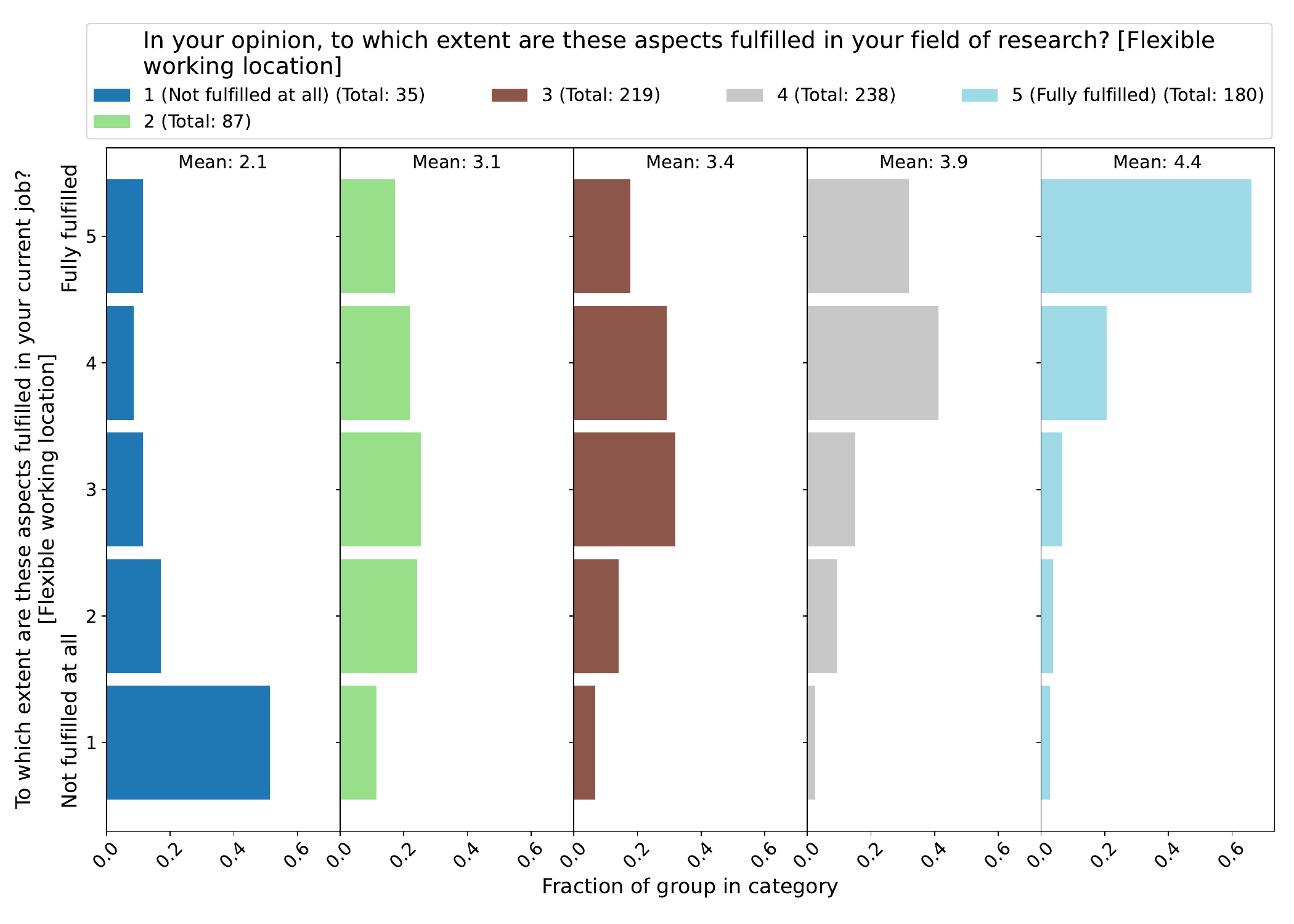}}\\
        \subfloat[]{\label{fig:part2:Q74cvQ75c}\includegraphics[width=0.49\textwidth]{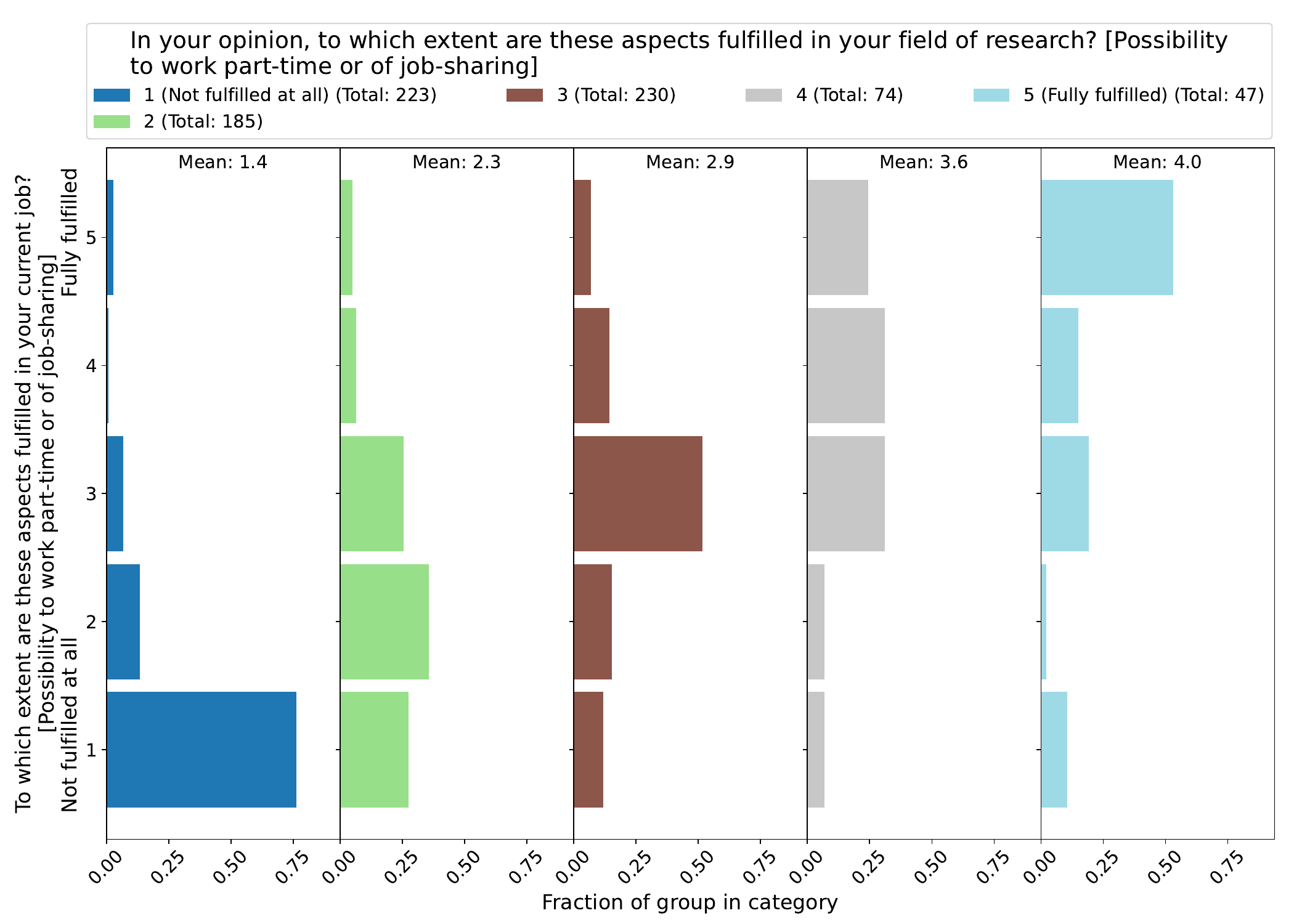}}
        \subfloat[]{\label{fig:part2:Q74dvQ75d}\includegraphics[width=0.49\textwidth]{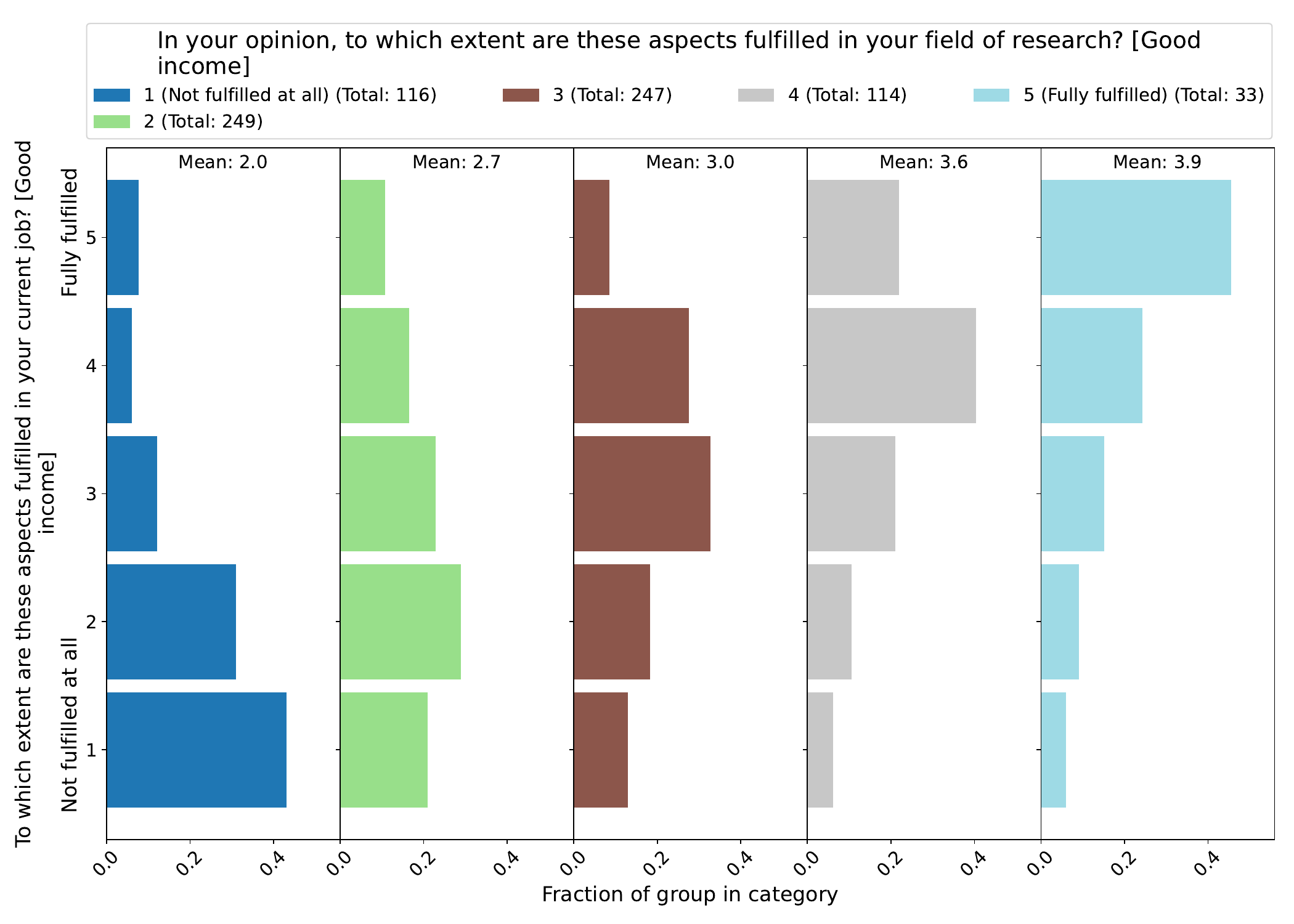}}\\
        \subfloat[]{\label{fig:part2:Q74evQ75e}\includegraphics[width=0.49\textwidth]{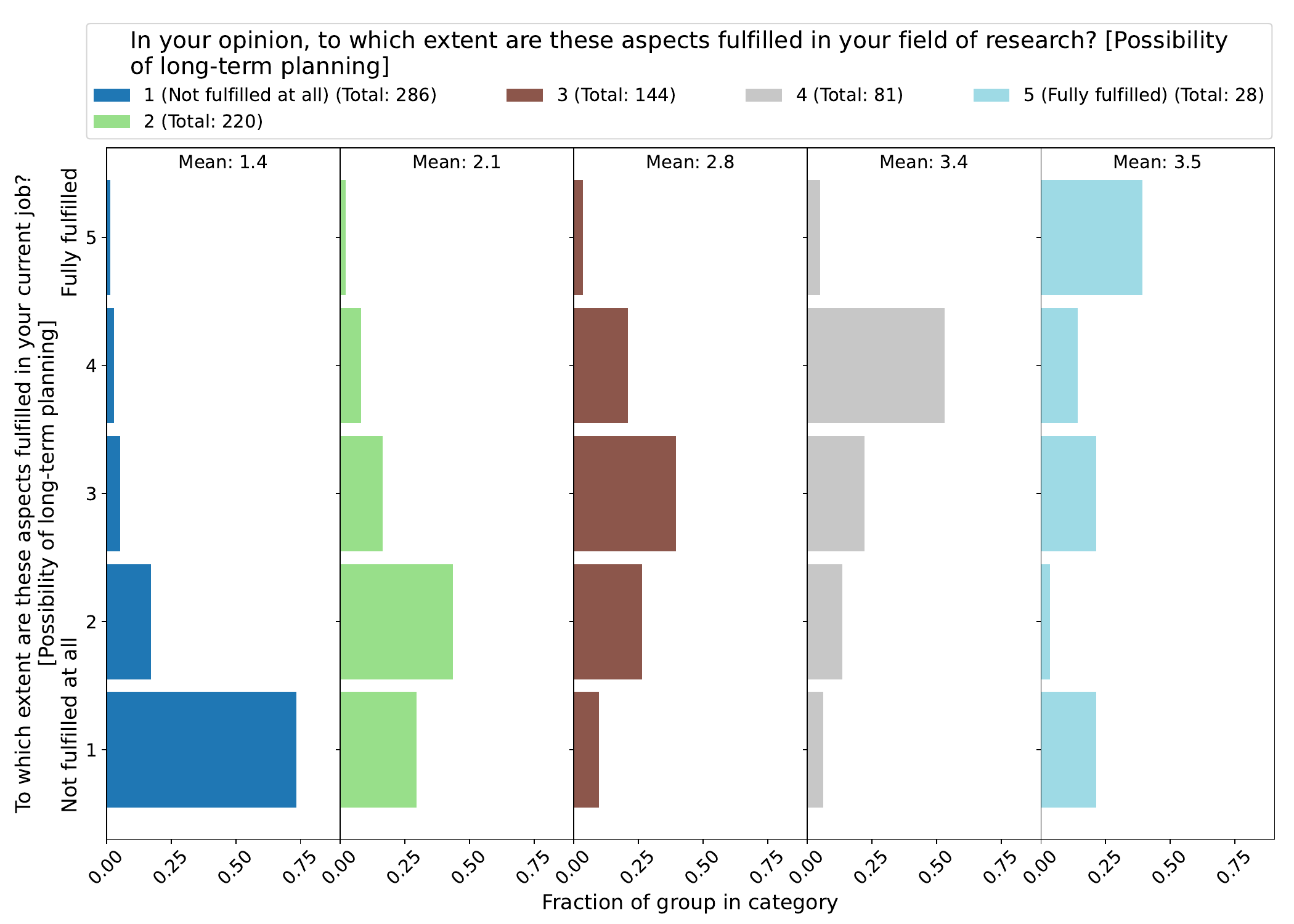}}
        \subfloat[]{\label{fig:part2:Q74fvQ75f}\includegraphics[width=0.49\textwidth]{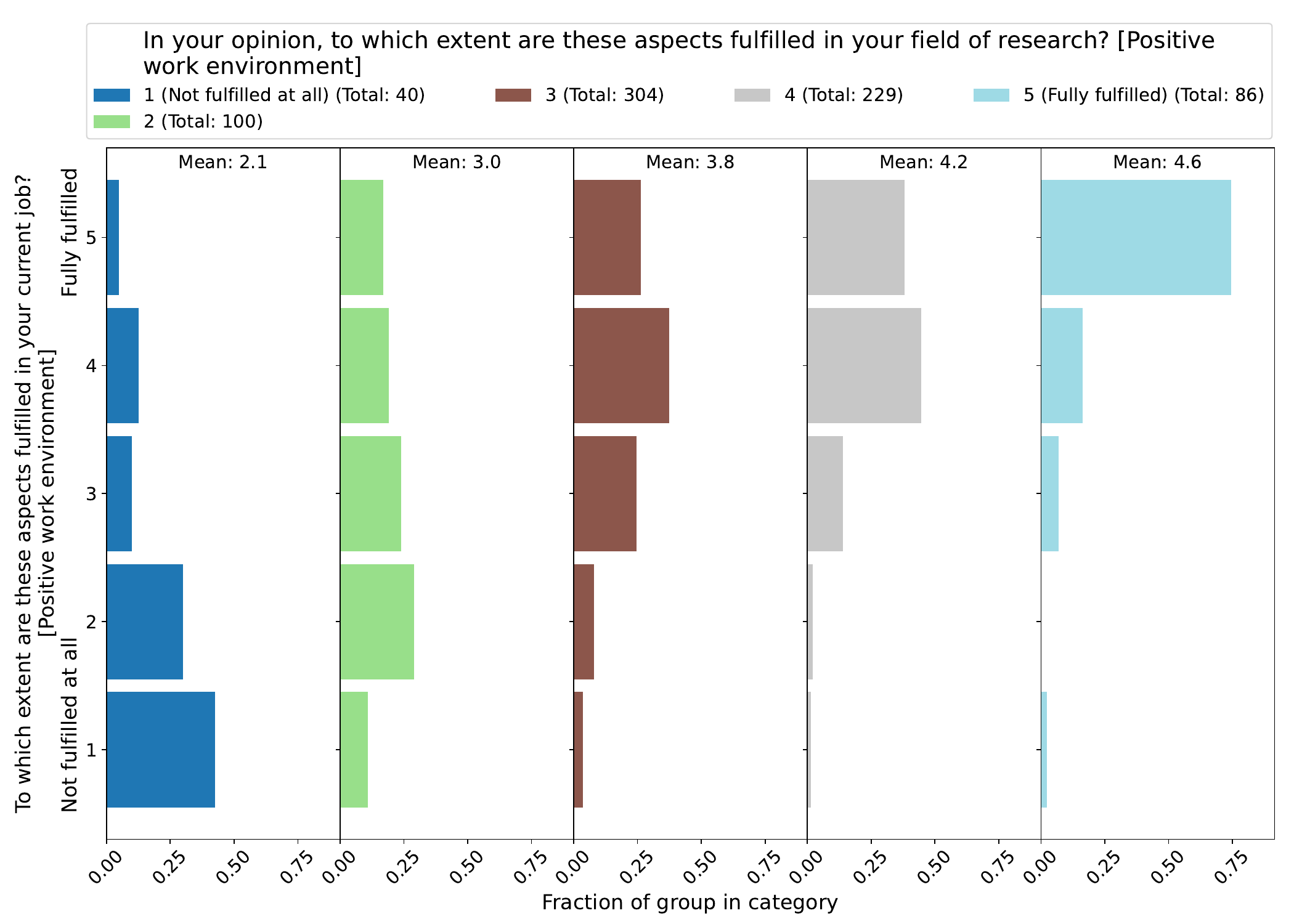}}
    \caption{(Q74 v Q75) How respondents view various items related to a good work-life balance being fulfilled in their current job, compared to their field in general. Fractions are given out of all respondents.}
    \label{fig:part2:Q74vQ75}
\end{figure}

The next study looked for correlations between how well respondents think the same items as above related to work-life balance are fulfilled in their research field, and other questions.
We found that the correlations were all very similar to those presented in Figure~\ref{fig:part2:Q73vQ1Q6Q7Q8}.
However, in general respondents are more negative about how the items are fulfilled in their field, in comparison to their current job.
Some examples of this trend are shown in Figure~\ref{fig:part2:Q74vQ75}.

\begin{figure}[ht!]
    \centering
        \subfloat[]{\label{fig:part2:Q76avQ4}\includegraphics[width=0.49\textwidth]{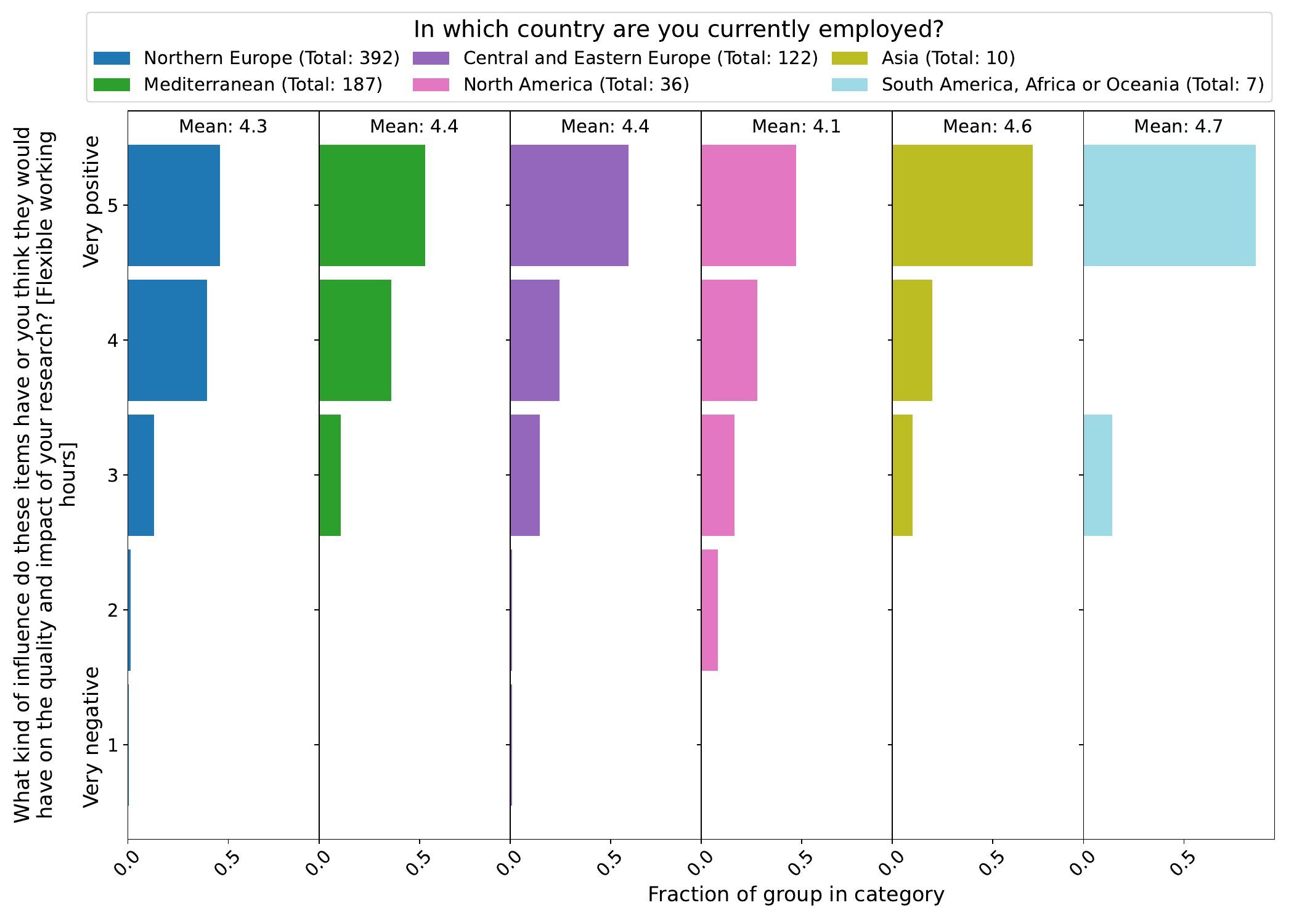}}
        \subfloat[]{\label{fig:part2:Q76bvQ4}\includegraphics[width=0.49\textwidth]{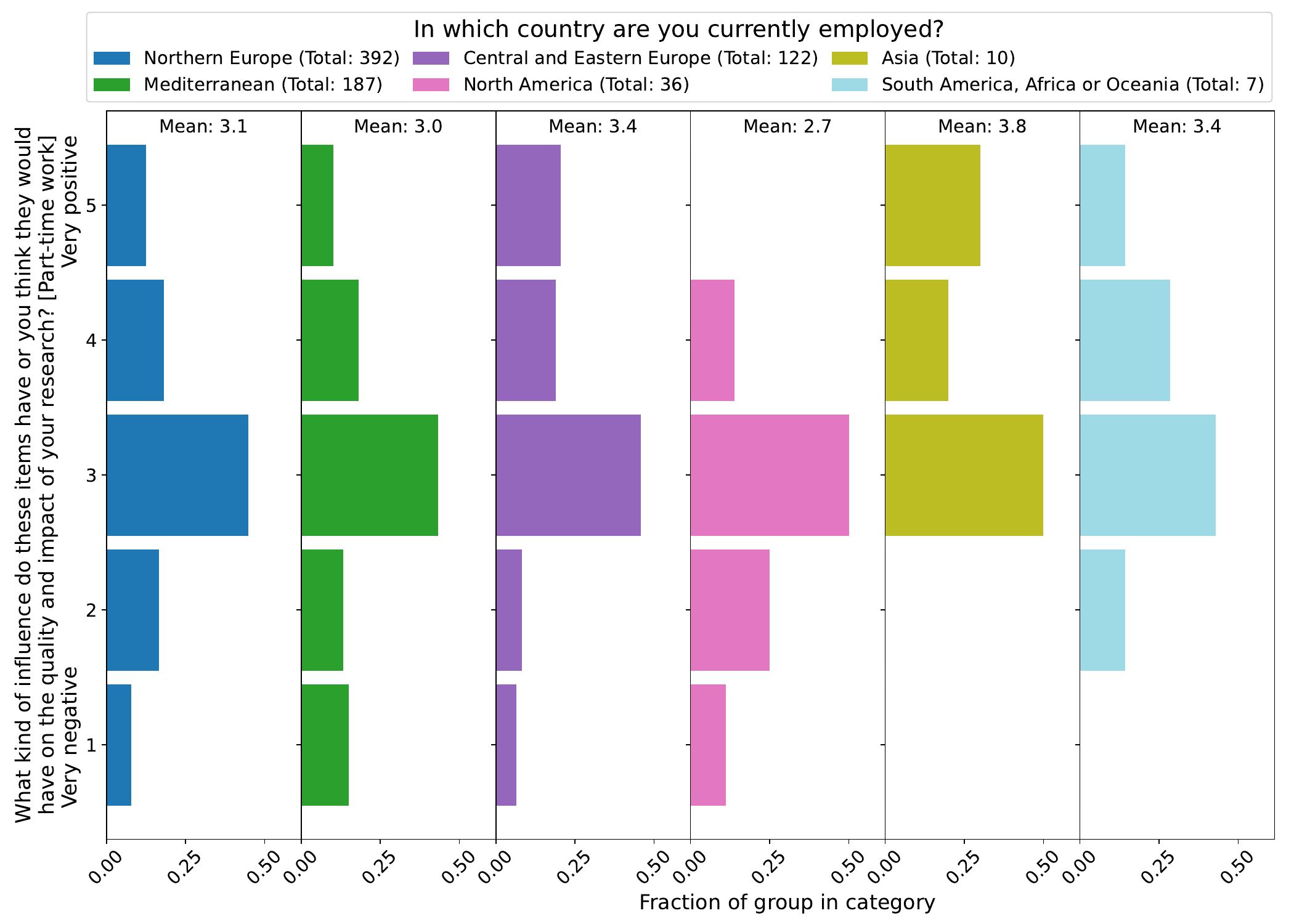}}\\
        \subfloat[]{\label{fig:part2:Q76cvQ5}\includegraphics[width=0.49\textwidth]{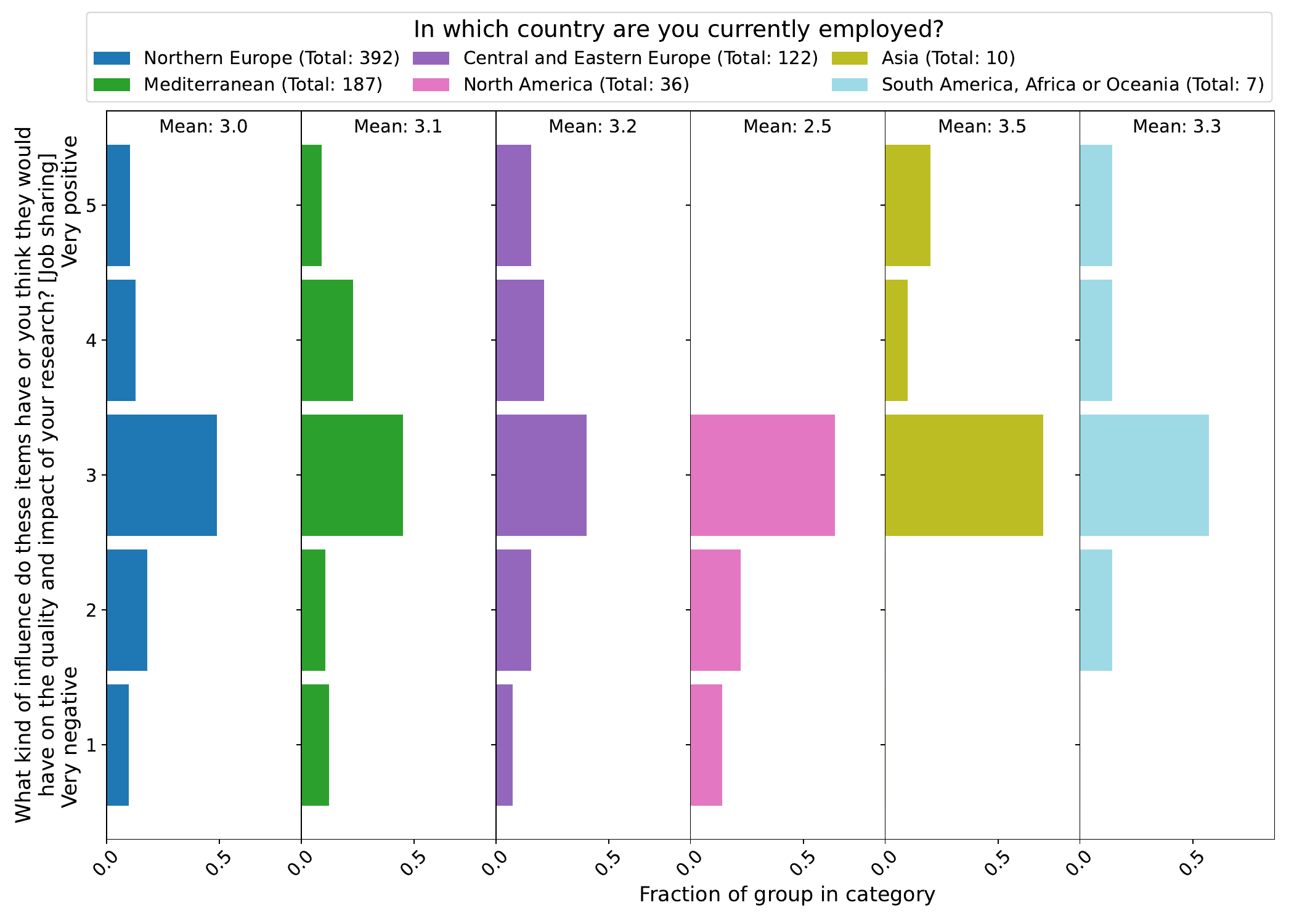}}
        \subfloat[]{\label{fig:part2:Q76dvQ5}\includegraphics[width=0.49\textwidth]{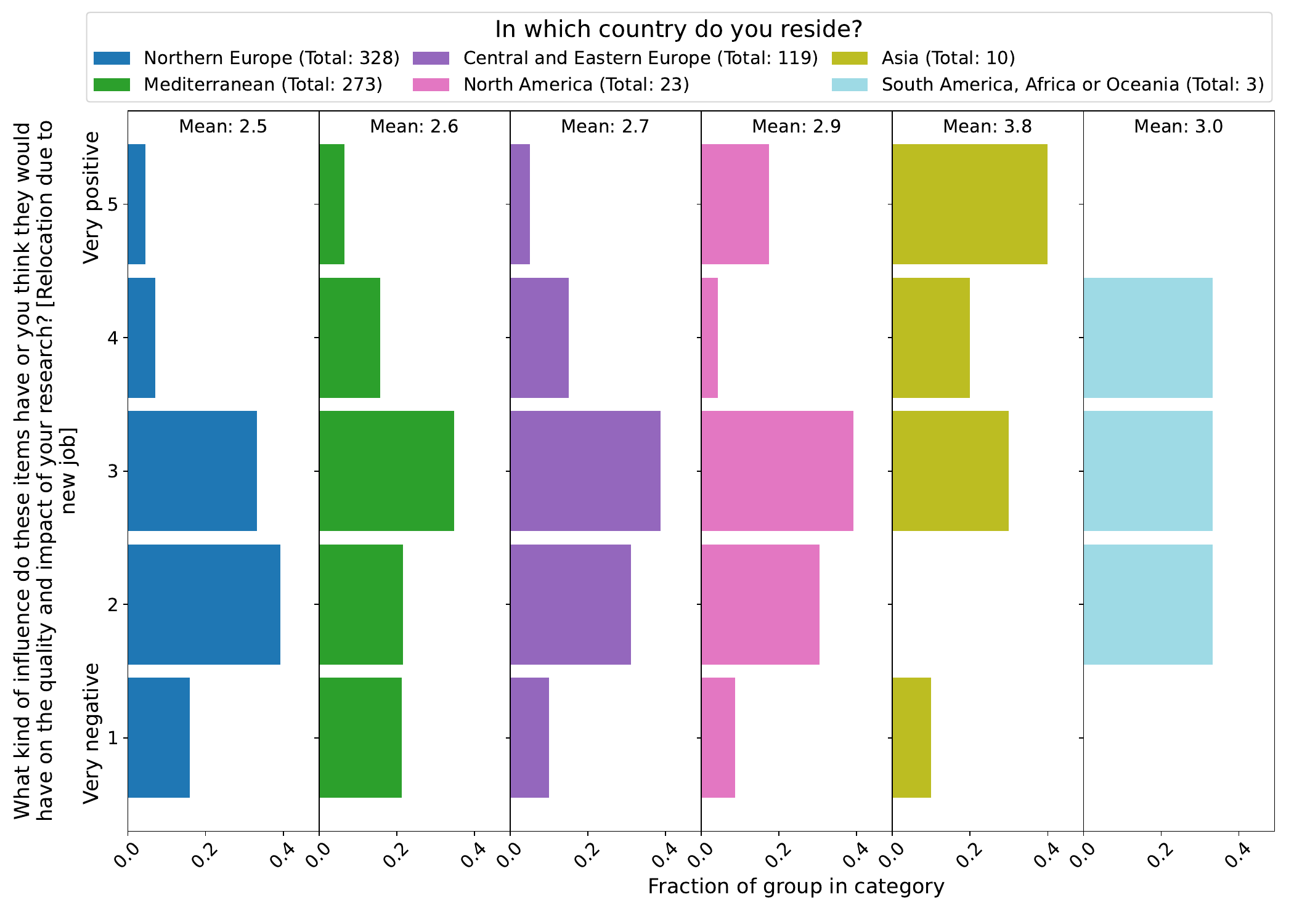}}\\
        \subfloat[]{\label{fig:part2:Q76dvQ6}\includegraphics[width=0.49\textwidth]{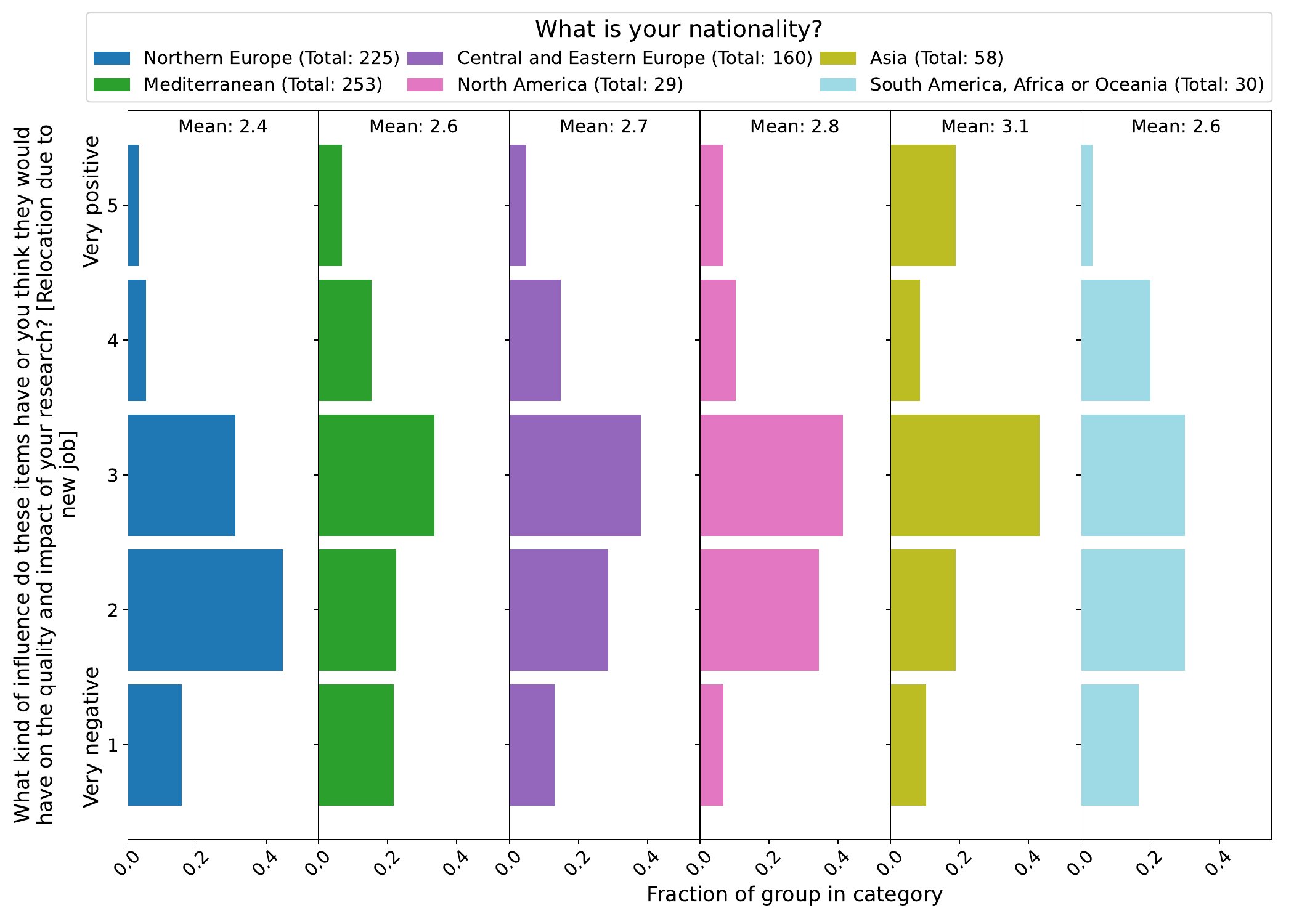}}
    \caption{(Q76a v Q4; Q76b v Q4;, Q76c v Q4;, Q76d v Q5,6) Correlations between the possible influence of given items on the quality and impact of respondents' research, and selected demographics. Fractions are given out of all respondents who answered the questions.}
    \label{fig:part2:Q76vQ4Q5}
\end{figure}

The correlations between respondents views on the influence certain items (listed in Appendix~\ref{app:questions}) could/would have on the quality and impact of their research, and respondent demographics, were next studied.
Selected correlations are shown in Figure~\ref{fig:part2:Q76vQ4Q5}.
We see that respondents employed in North America are more negative about the (possible) impact of flexible hours, part-time work, and job sharing on their research than those employed elsewhere.
This is consistent with our earlier observations that these items were less important to, and less fulfilled for, the respondents employed in North America.
We see that respondents who reside outside of Europe view relocating for a new job as having a more positive impact on their research than others.
This is particularly true for those residing in Asia, given a low sample size.
We observed the same trend when considering country of employment, and a weaker correlation when considering nationality.

\begin{figure}[ht!]
    \centering
    \includegraphics[width=0.6\textwidth]{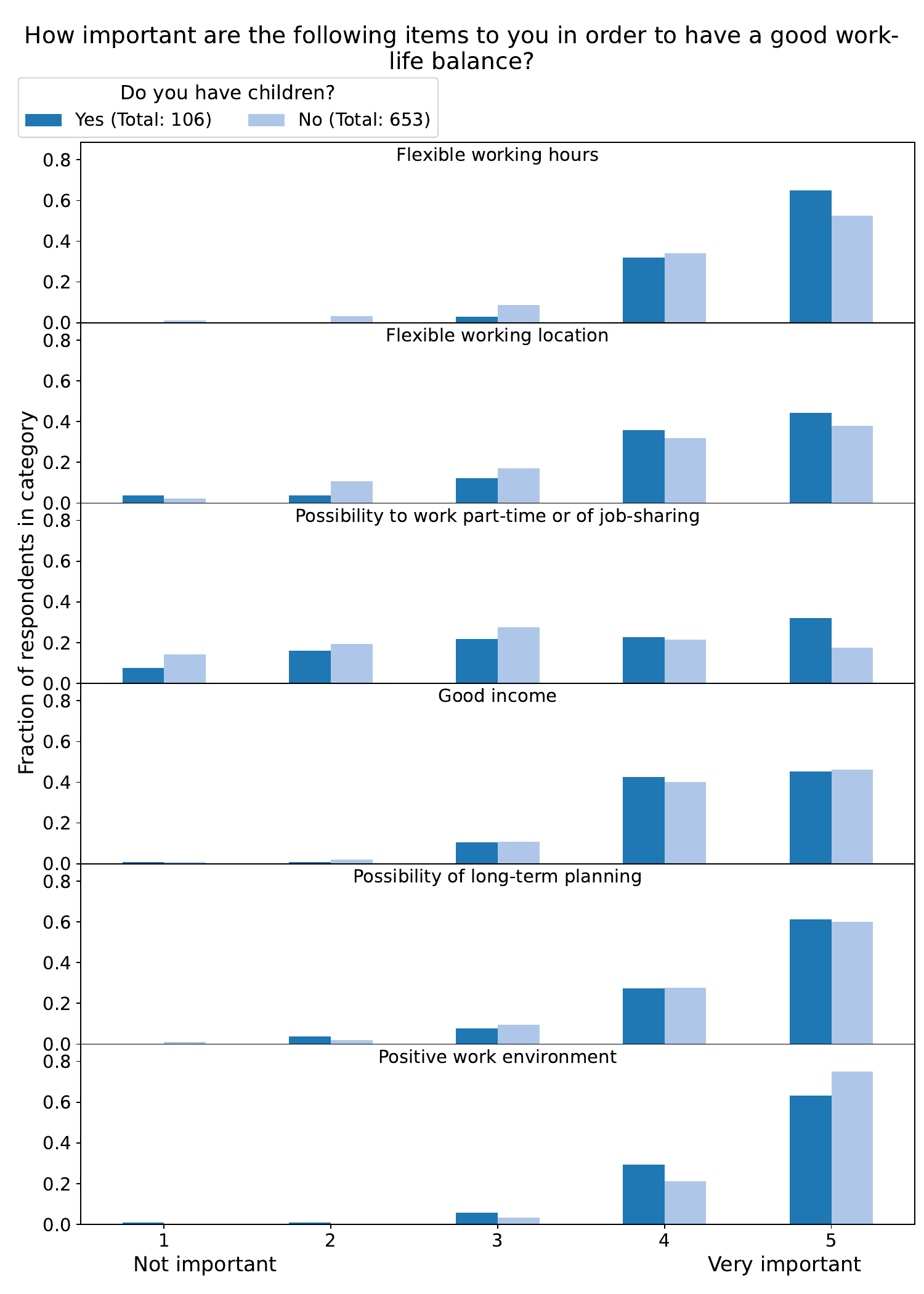}
    \caption{(Q73 v Q71) How important various items are to respondents in order to have a good work-life balance, correlated with whether they have children. Fractions are given out of all respondents.}
    \label{fig:part2:Q73vQ71}
\end{figure}

We now consider correlations between the importance and fulfilment respondents feel about items related to work-life balance and questions unrelated to demographics.
Considering whether respondents have children, we find in Figure~\ref{fig:part2:Q73vQ71} that unsurprisingly flexible working hours/location and the possibility of part-time working are more important to those with children.

\clearpage

\begin{figure}[h!]
    \centering
    \includegraphics[width=0.7\textwidth]{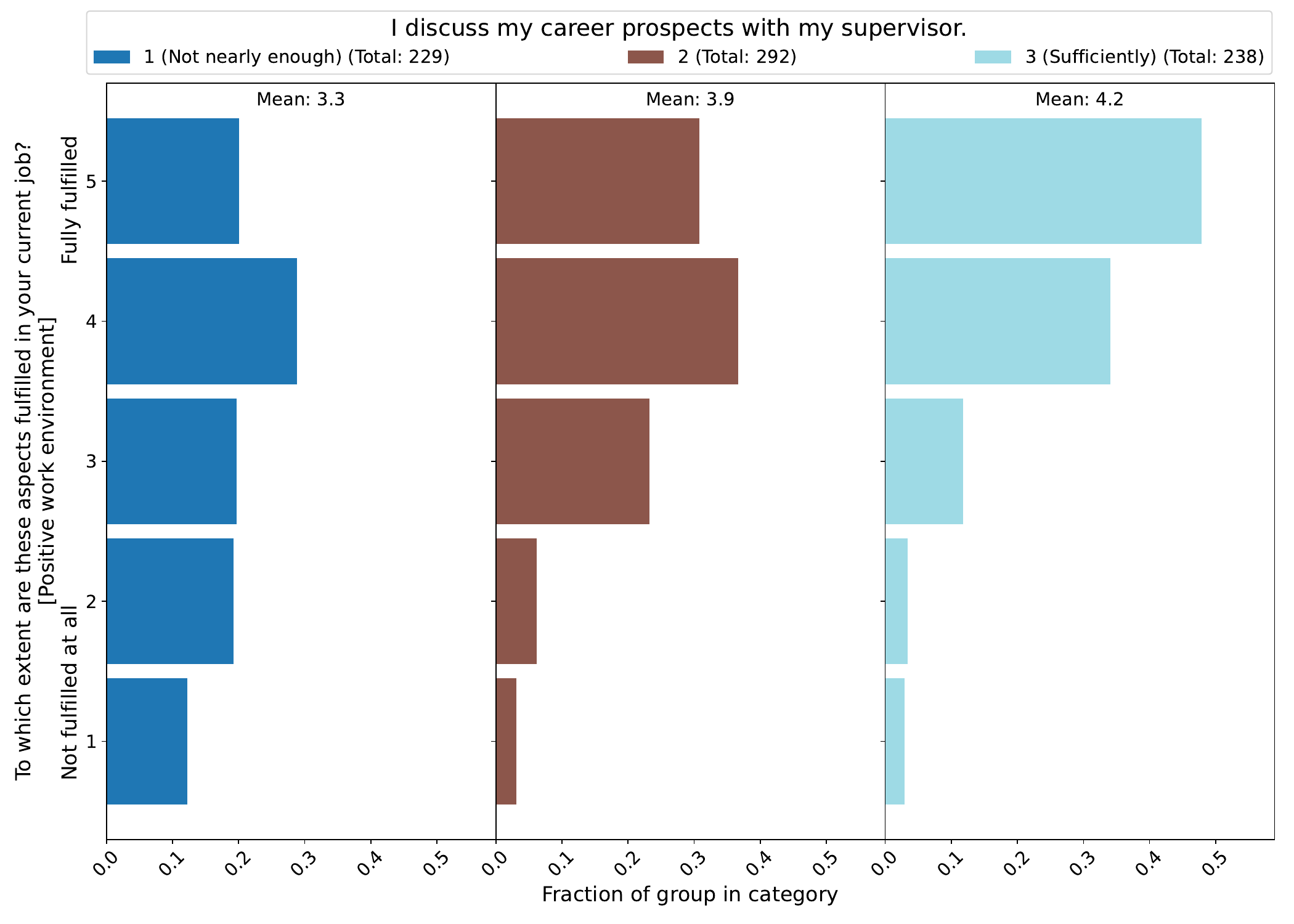}
    \caption{(Q74f v Q65) Correlations between how fulfilled respondents feel about a positive work environment in their current job and how much they discuss career prospects with their supervisors. Fractions are given out of all respondents.}
     \label{fig:part2:Q89vQ65}
\end{figure}

In Figure~\ref{fig:part2:Q89vQ65} we see a strong positive correlation between those who discuss their career prospects sufficiently with their supervisor and those who feel that they have a positive work environment in their current job.

\begin{figure}[ht!]
    \centering
        \subfloat[]{\label{fig:part2:Q79vQ1}\includegraphics[width=0.49\textwidth]{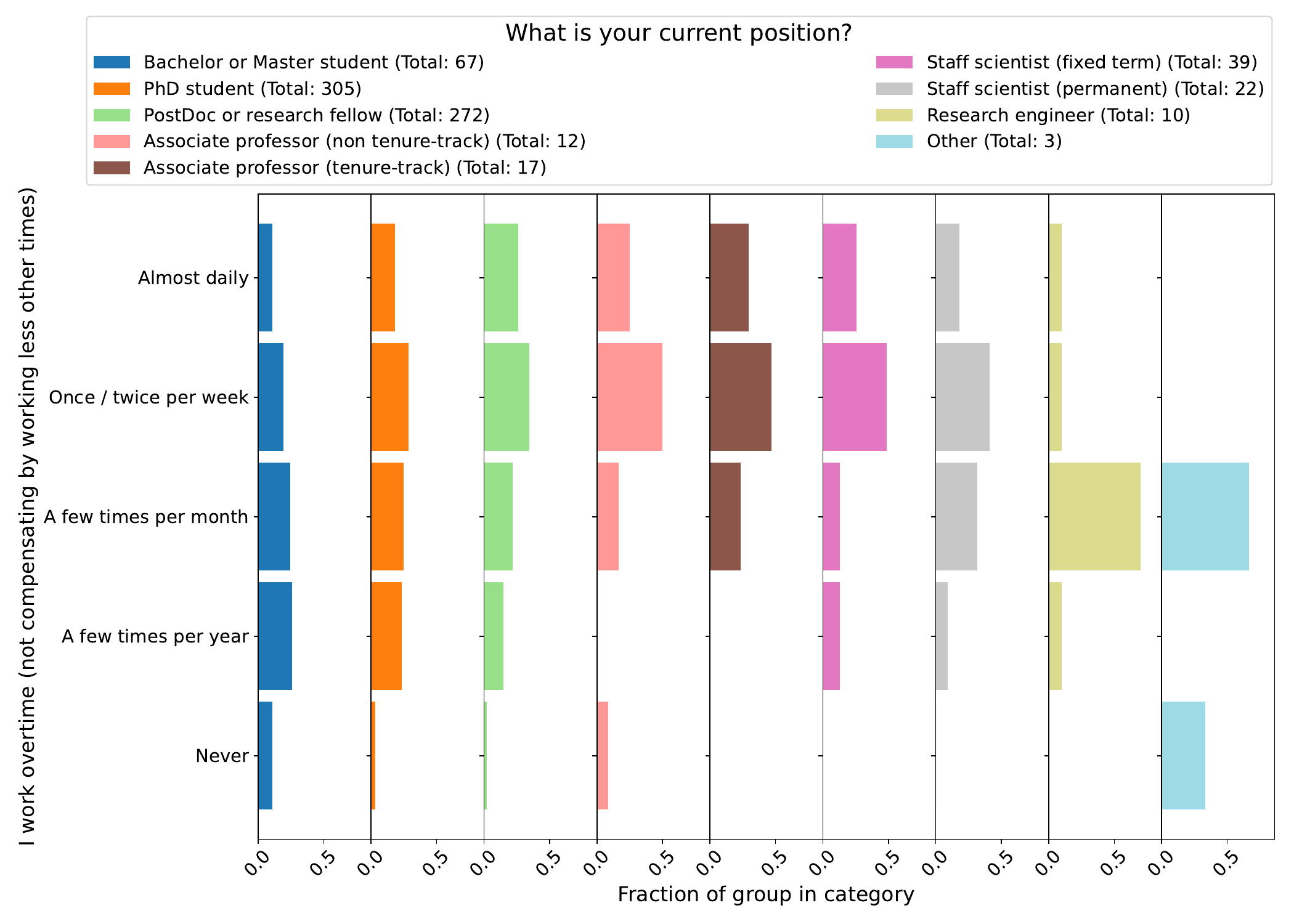}}
        \subfloat[]{\label{fig:part2:Q80vQ1}\includegraphics[width=0.49\textwidth]{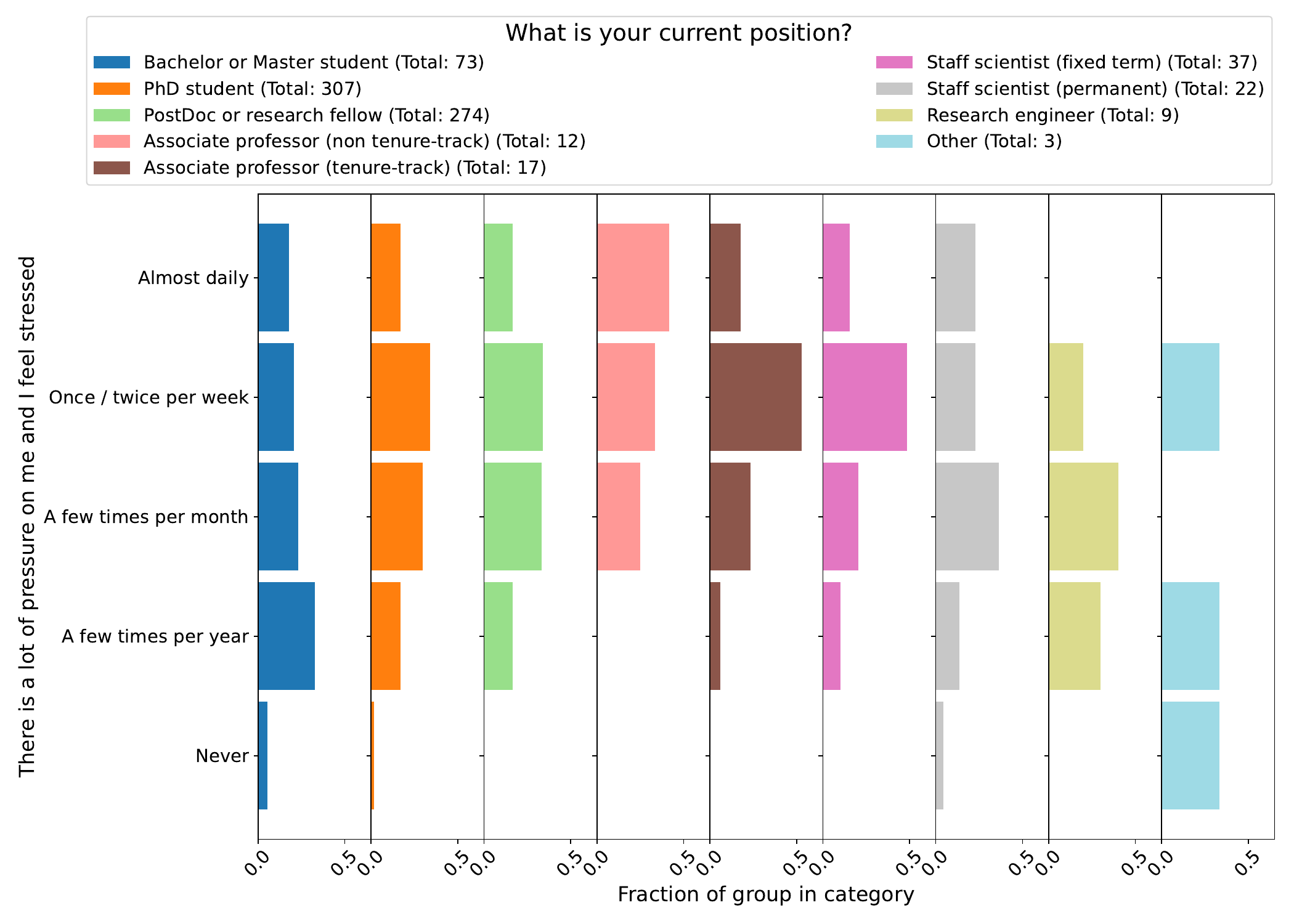}}\\
        \subfloat[]{\label{fig:part2:Q79vQ6}\includegraphics[width=0.49\textwidth]{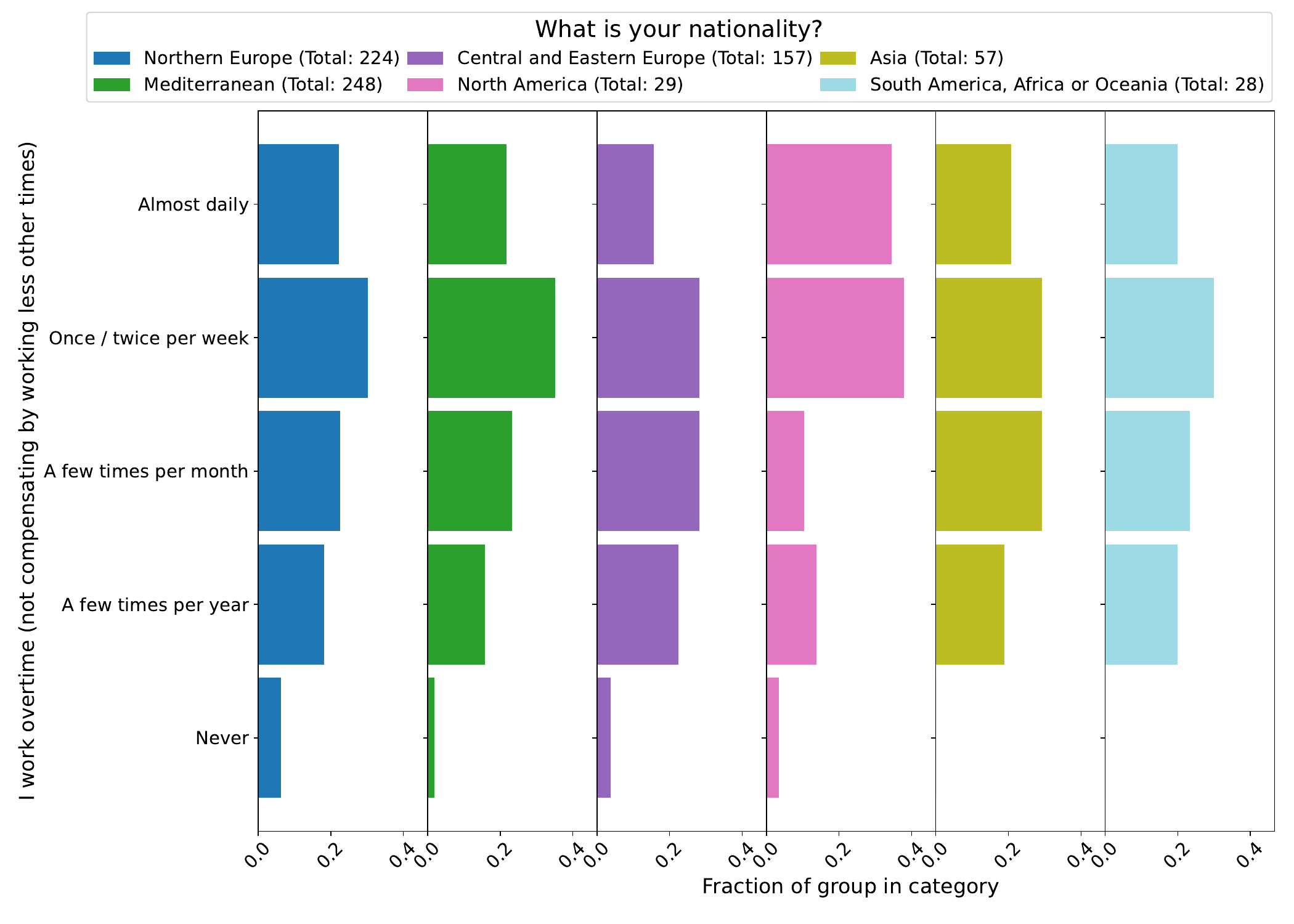}}
        \subfloat[]{\label{fig:part2:Q80vQ6}\includegraphics[width=0.49\textwidth]{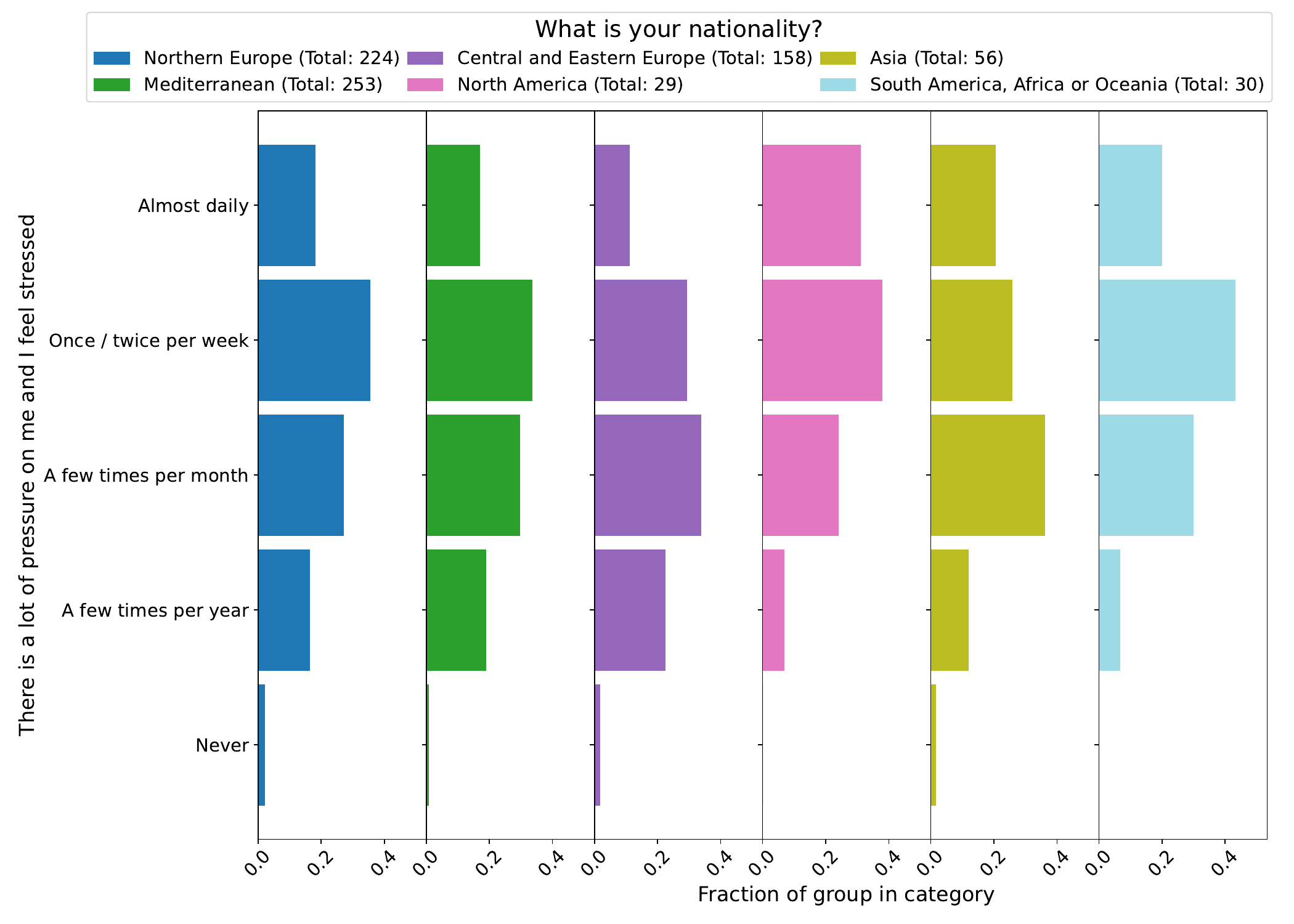}}\\
        \subfloat[]{\label{fig:part2:Q79vQ10}\includegraphics[width=0.49\textwidth]{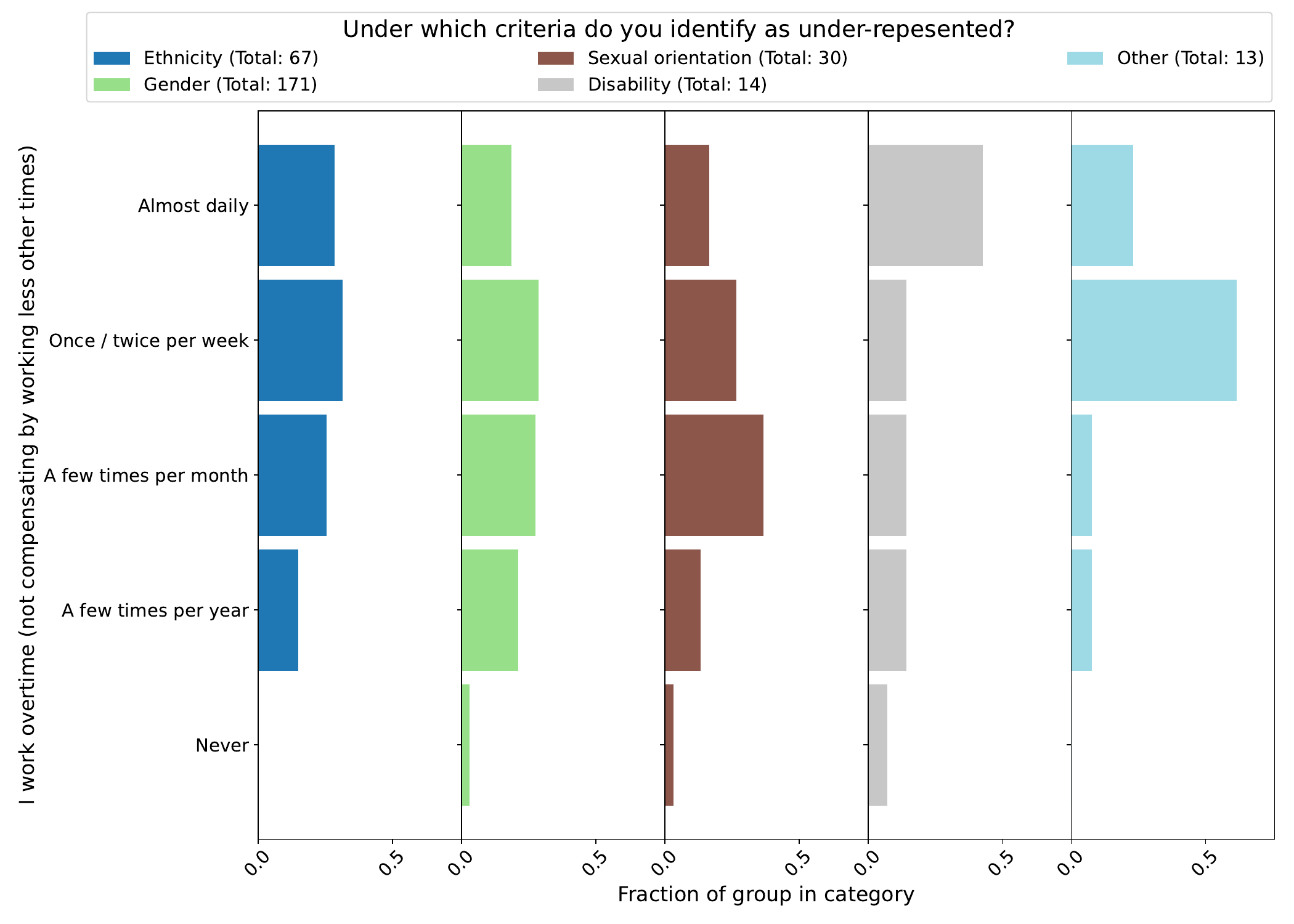}}
    \caption{(Q79--80 v Q1,6,10) Correlations between how often respondents work overtime (not compensating by working less other times) and selected demographics. Fractions are given out of all respondents who answered the questions.}
    \label{fig:part2:Q79Q80vQ1Q6}
\end{figure}

\begin{figure}[ht!]
    \centering
     \includegraphics[width=0.6\textwidth]{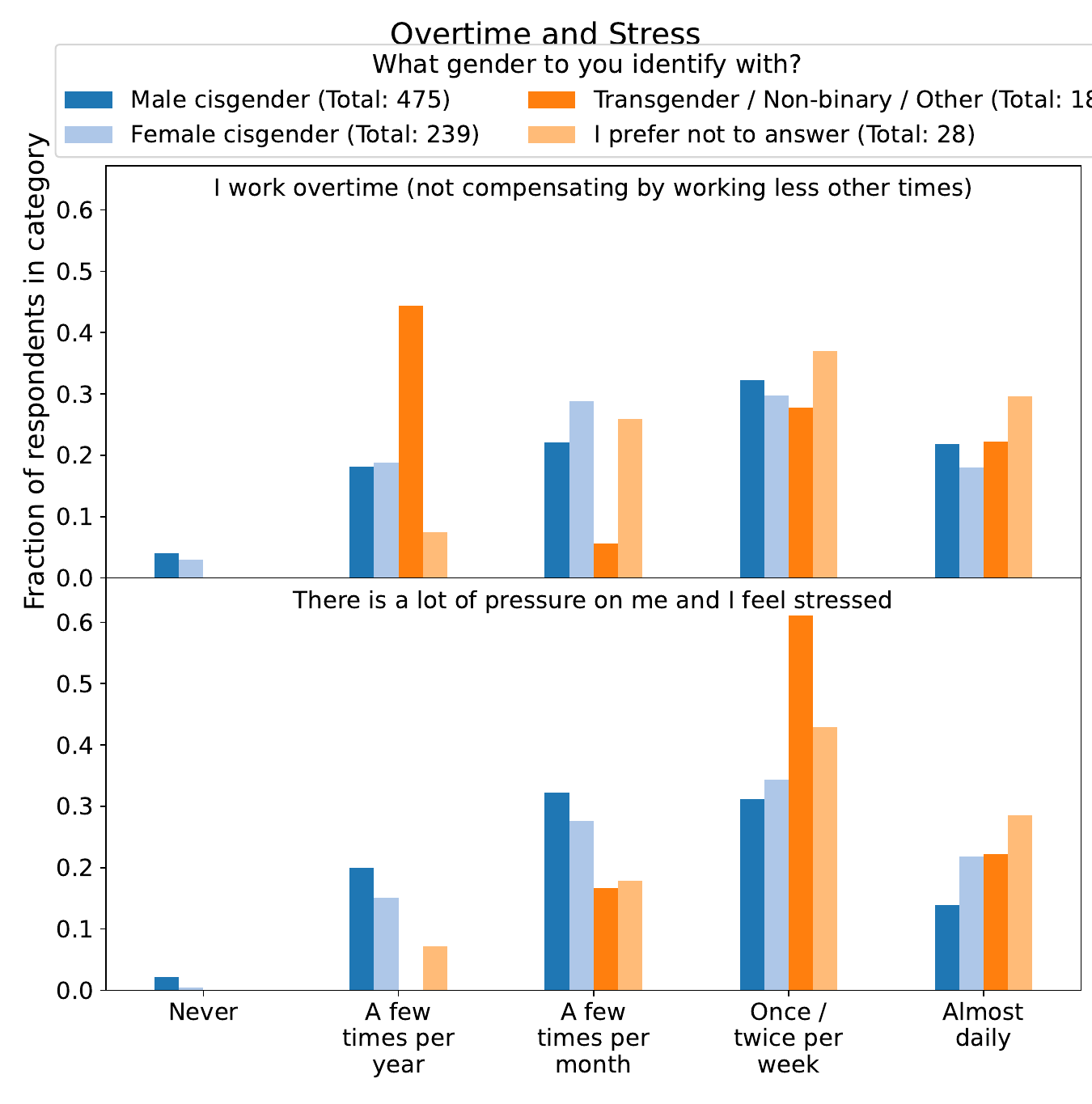}
    \caption{(Q79--80 v Q7) Correlations between how much overtime respondents work or how stressed they feel, and their gender. Fractions are given out of all respondents who answered the questions.}
    \label{fig:part2:OvertimeStressvQ7}
\end{figure}

\begin{figure}[ht!]
    \centering
     \includegraphics[width=0.6\textwidth]{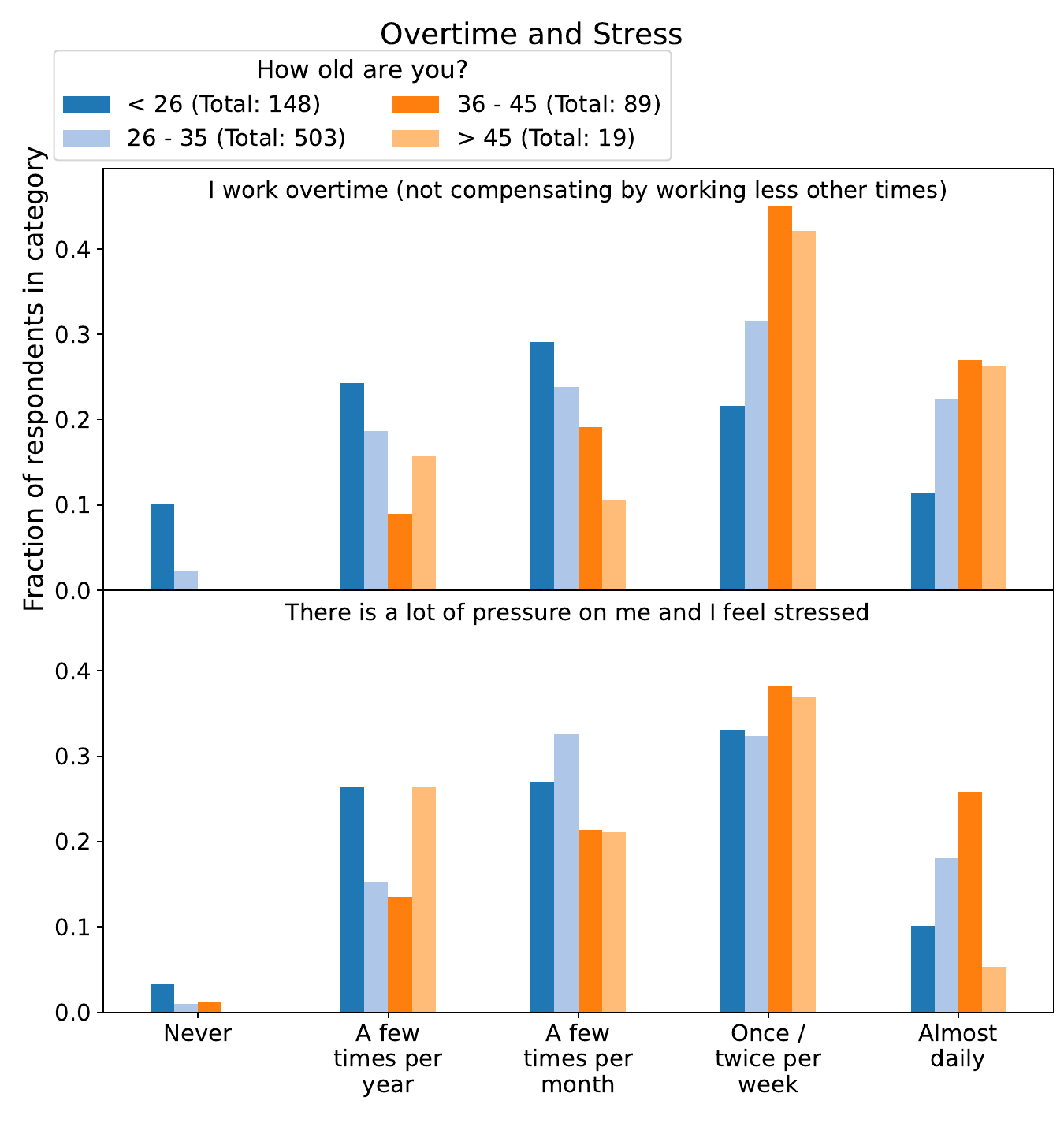}
    \caption{(Q79--80 v Q8) Correlations between how much overtime respondents work or how stressed they feel, and their age-group. Fractions are given out of all respondents who answered the question.}
    \label{fig:part2:OvertimeStressvQ8}
\end{figure}

\begin{figure}[ht!]
    \centering
     \includegraphics[width=0.6\textwidth]{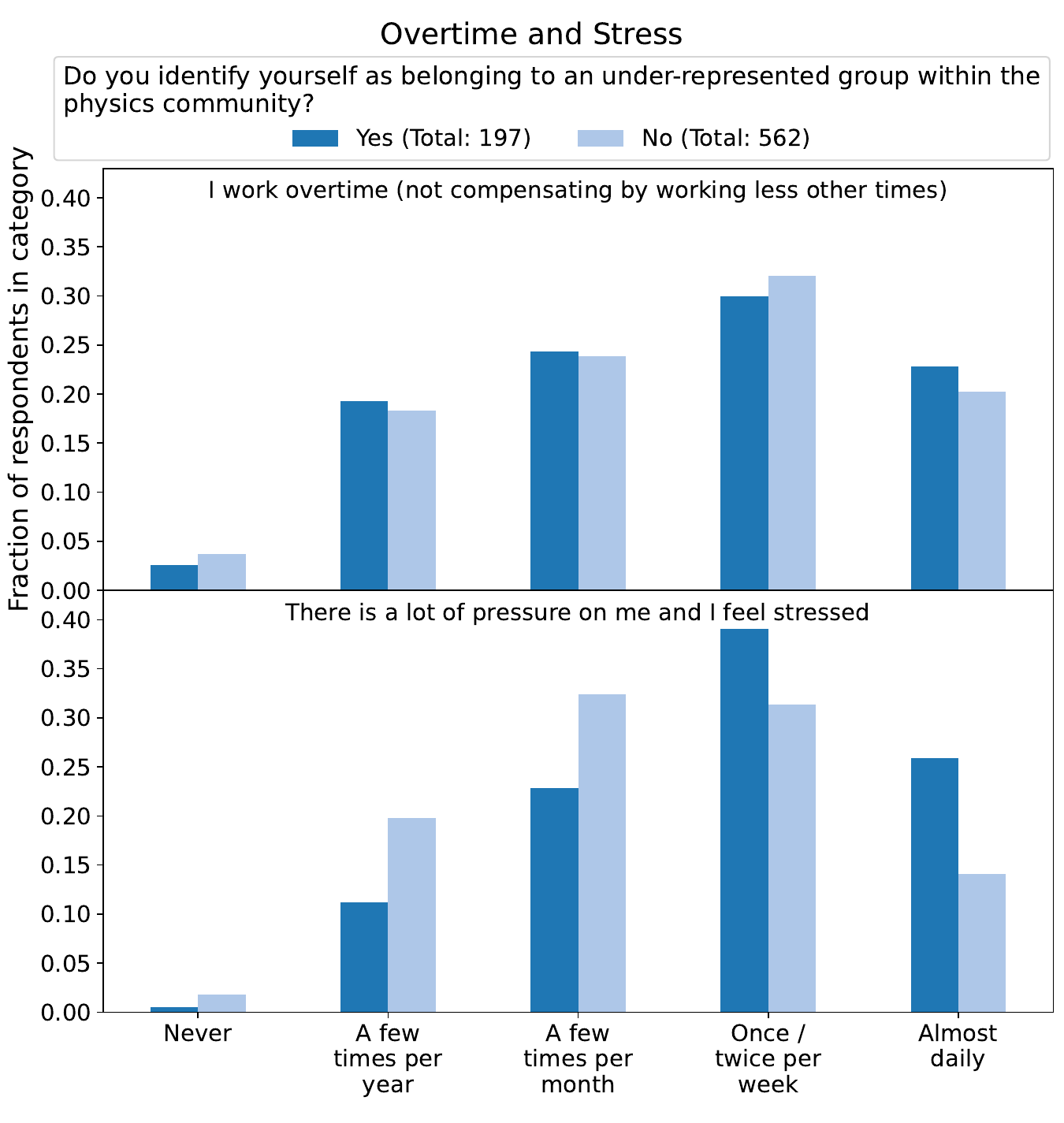}
    \caption{(Q79--80 v Q9) Correlations between how much overtime respondents work or how stressed they feel, and whether they identify as part of an under-represented group within physics. Fractions are given out of all respondents who answered the questions.}
    \label{fig:part2:OvertimeStressvQ9}
\end{figure}

\begin{figure}[ht!]
    \centering
     \includegraphics[width=0.6\textwidth]{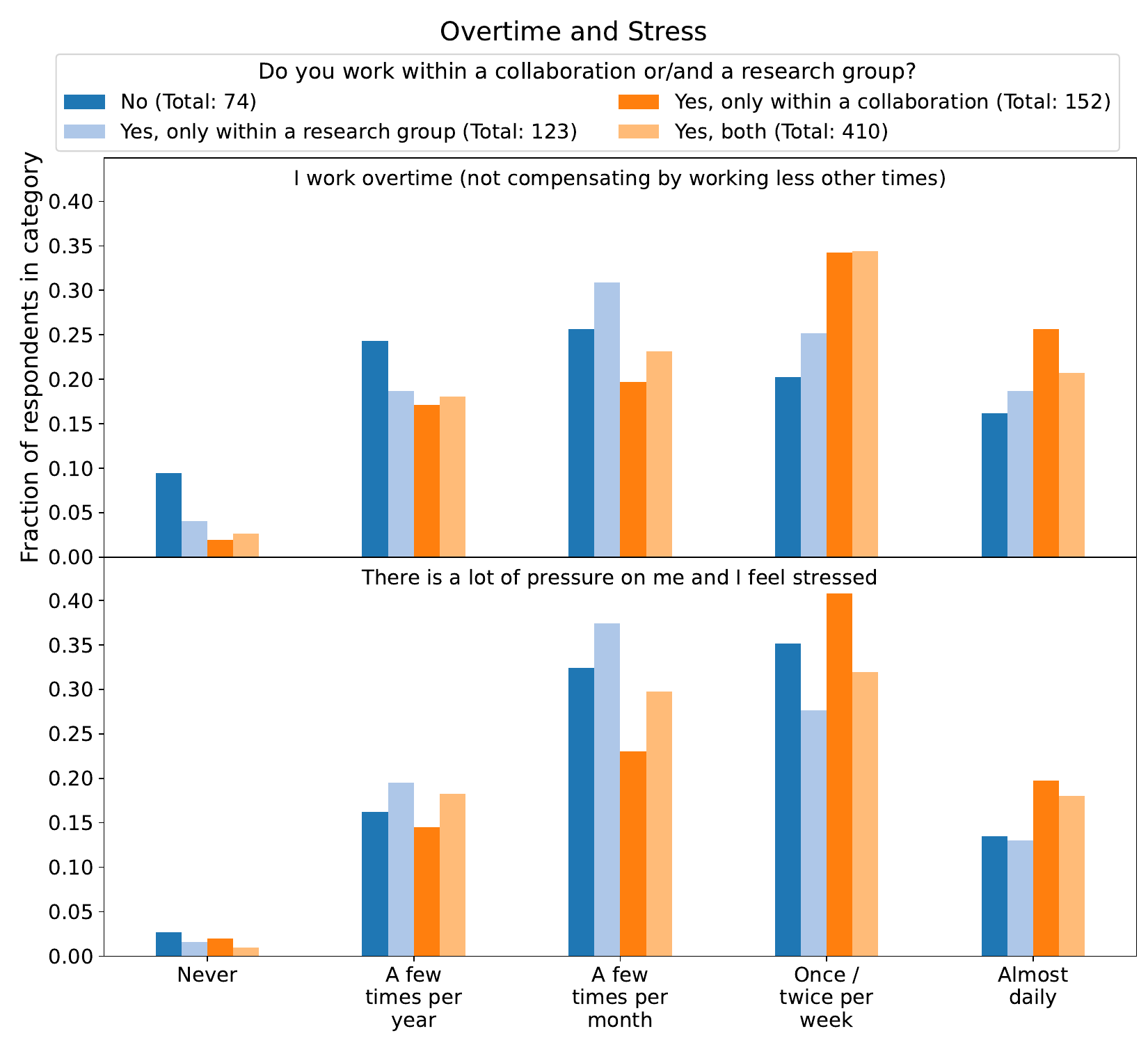}
    \caption{(Q79--80 v Q14) Correlations between how much overtime respondents work or how stressed they feel, and whether they work in a research group and/or collaboration. Fractions are given out of all respondents who answered the questions.}
    \label{fig:part2:OvertimeStressvQ14}
\end{figure}

We now move to studying the stress levels and amount of overtime work respondents experience.
In Figures~\ref{fig:part2:Q79Q80vQ1Q6}--\ref{fig:part2:OvertimeStressvQ14} these are correlated with selected demographics.
We observe that older and more senior respondents work overtime more frequently.
Additionally, the more senior positions respondents have, the more frequently they feel stressed and under pressure, until they obtain permanent/tenure-track positions, when this frequency drops.
We also confirmed that the same trend appears for contract length: stress levels rise until an indefinite contract is obtained.
North American respondents work overtime more often than those from elsewhere, with Central and Eastern European respondents working overtime least often.
This trend was also observed regarding country of employment and residence, not shown here for brevity.
Similar behaviour is also seen when considering frequency of feeling stressed and under pressure.
Comparing genders, given the limited sample sizes, cisgender male respondents work overtime slightly more than cisgender females, but feel stressed and under pressure less frequently than other respondents.
We see that respondents who identify as belonging to an under-represented group work overtime slightly more frequently than those who don't, and feel stressed and under pressure far more frequently.
Among under-represented respondents, those who have disability, work overtime most often (but the sample size is very limited).
However, no strong correlation was seen here with regards to reported stress levels.
Working in a collaboration is more strongly correlated with respondents working overtime or feeling stressed than working in a research group.

\begin{figure}[ht!]
    \centering
    \includegraphics[width=0.6\textwidth]{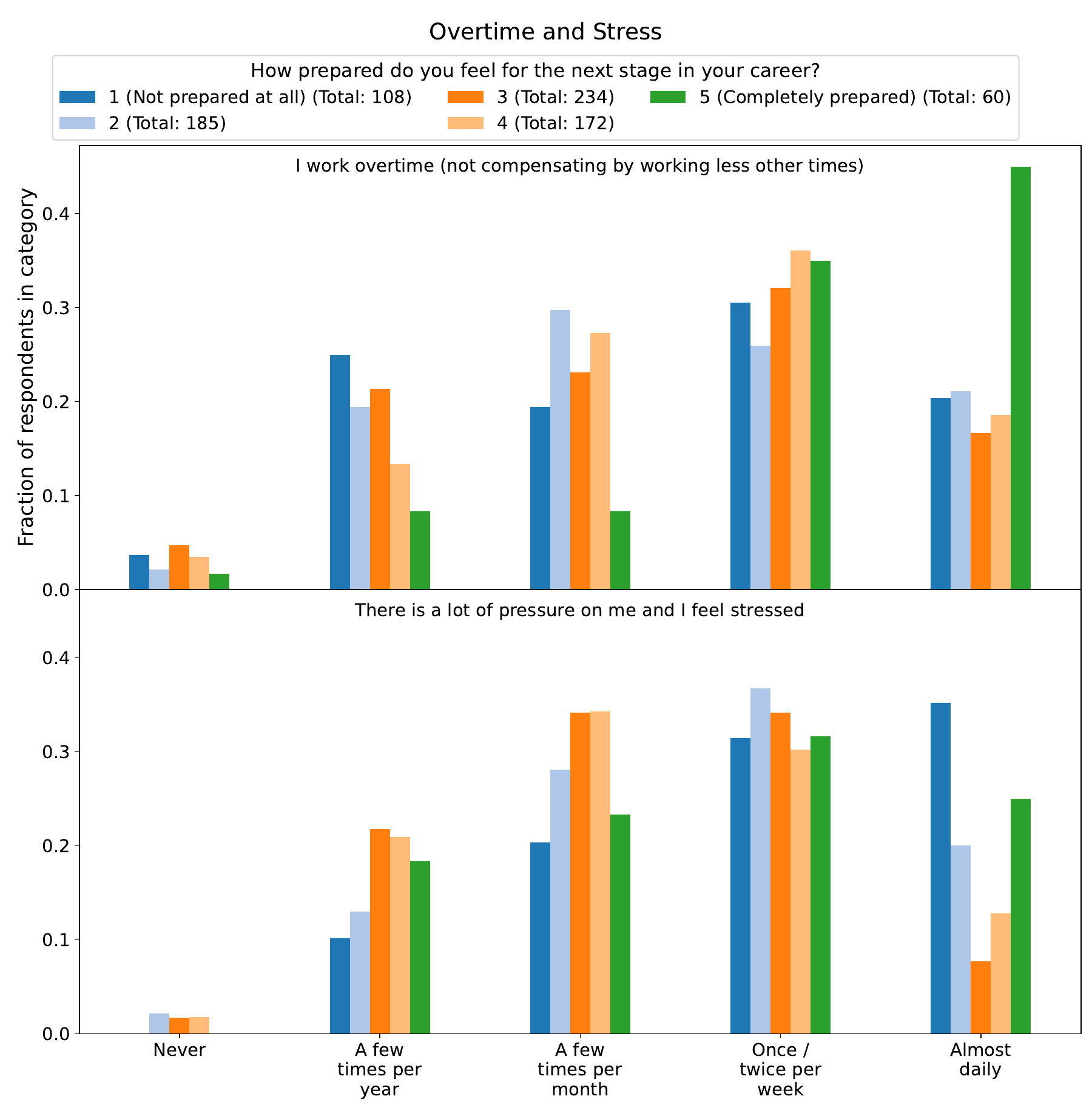}
    \caption{(Q79--80 v Q64) Correlations between how much overtime respondents do or how frequently they feel stressed and how prepared they feel for the next stage of their careers. Answers are given on a scale of 1--5 where 1 indicates completely y unprepared, and 5 indicates completely prepared. Fractions are given out of all respondents who answered the questions.}
    \label{fig:part2:OvertimeStressvQ64}
\end{figure}

Moving to correlations on this topic unrelated to demographics, a strong trend is seen in Figure~\ref{fig:part2:OvertimeStressvQ64} where respondents who feel the most prepared for the next stage in their careers are also working overtime the most frequently.
In the same figure we also see an interesting trend where the respondents who feel stressed and under pressure the most frequently are bimodal between those who are the most and least prepared for the next stage in their careers.

\begin{figure}[ht!]
    \centering
    \includegraphics[width=0.6\textwidth]{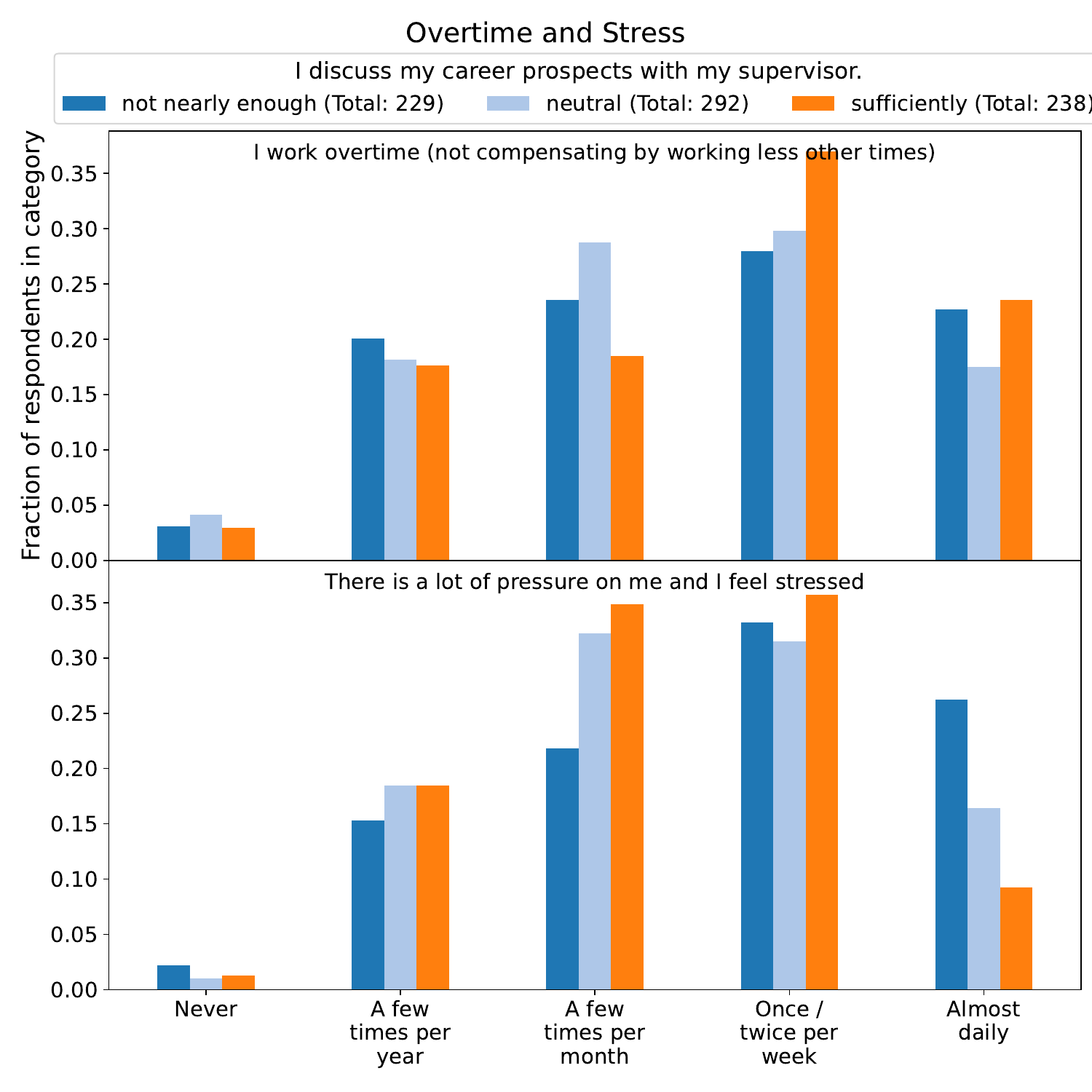}
    \caption{(Q79--80 v Q65) Correlations between how much overtime respondents do or how frequently they feel stressed and how sufficiently they discuss their career prospects with supervisors. Fractions are given out of all respondents who answered the questions.}
    \label{fig:part2:OvertimeStressvQ65}
\end{figure}

\begin{figure}[ht!]
    \centering
    \includegraphics[width=0.55\textwidth]{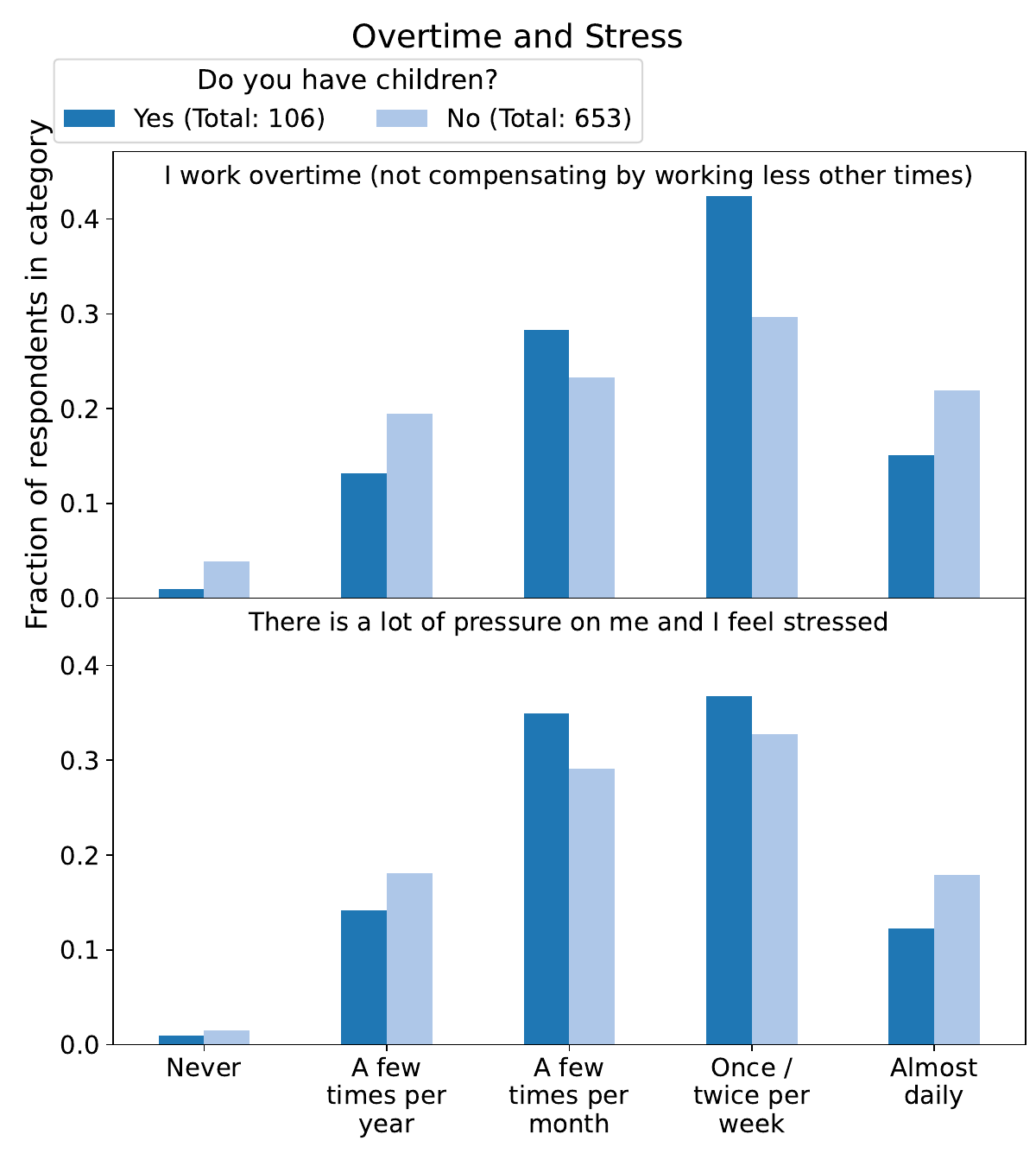}
    \caption{(Q79--80 v Q71) Correlations between how much overtime respondents do or how frequently they feel stressed and whether they have children.  Fractions are given out of all respondents who answered the questions.}
    \label{fig:part2:OvertimeStressvQ71}
\end{figure}

\begin{figure}[ht!]
    \centering
    \includegraphics[width=0.55\textwidth]{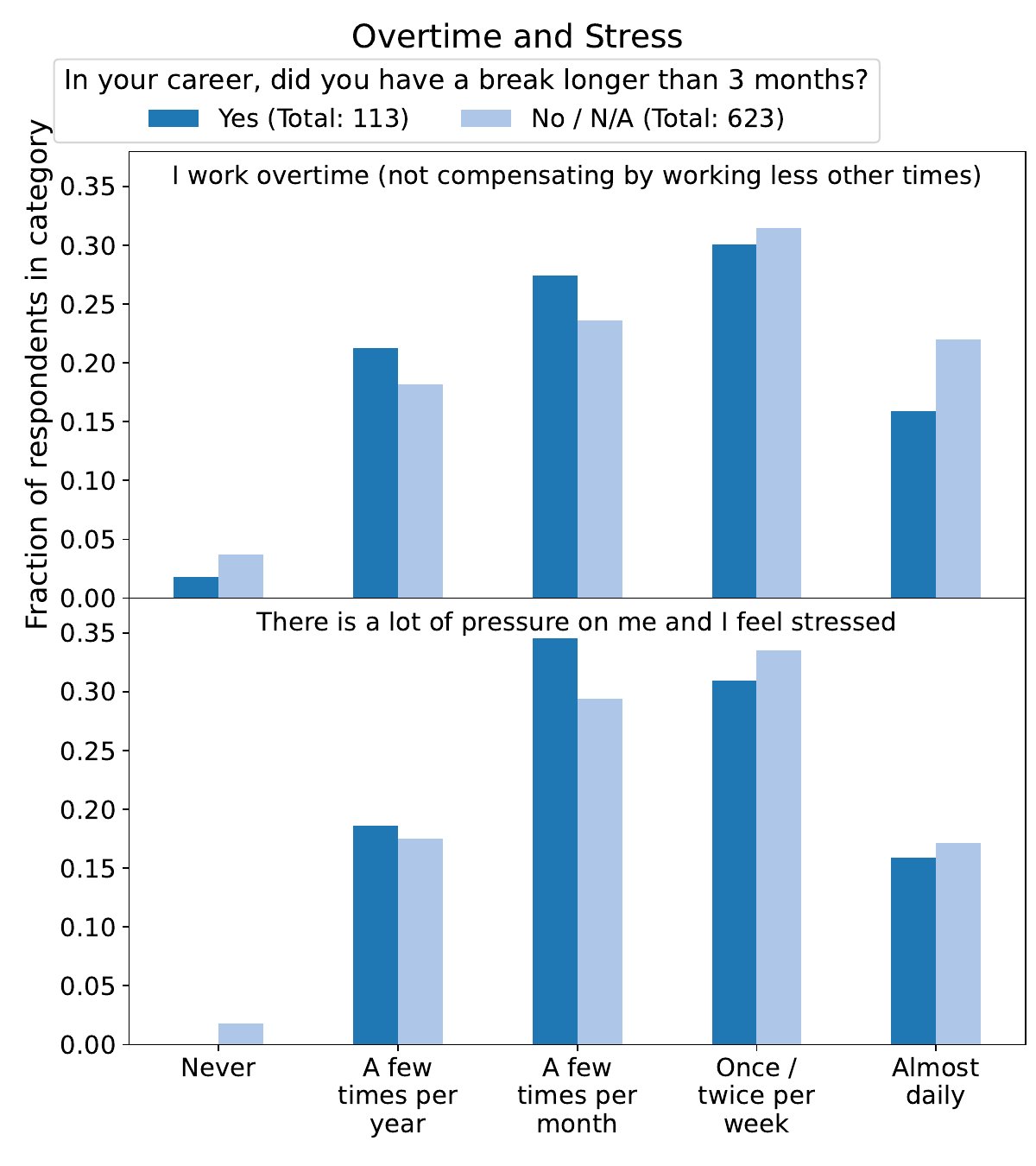}
    \caption{(Q79--80 v Q83) Correlations between how much overtime respondents do or how frequently they feel stressed and whether they've had a career break. Fractions are given out of all respondents who answered the questions.}
    \label{fig:part2:OvertimeStressvQ83}
\end{figure}

Considering discussion with their supervisors, we see in Figure~\ref{fig:part2:OvertimeStressvQ65} that those who feel this is sufficient are working overtime very frequently, though they are least likely to feel stressed or under pressure almost daily.
Respondents with children manage to work overtime ``almost daily'' less often than those without, but work overtime ``once/twice per week'' more often, as shown in Figure~\ref{fig:part2:OvertimeStressvQ71}.
Similarly, Figure~\ref{fig:part2:OvertimeStressvQ83} shows that respondents who have had a career break work overtime and feel stressed slightly less often than those who haven't.

%----------------------------------------------------------------------------------------
\clearpage
\subsubsection{Career mobility and leaving academia}

\begin{figure}[ht!]
    \centering
        \subfloat[]{\label{fig:part2:Q81vQ5}\includegraphics[width=0.49\textwidth]{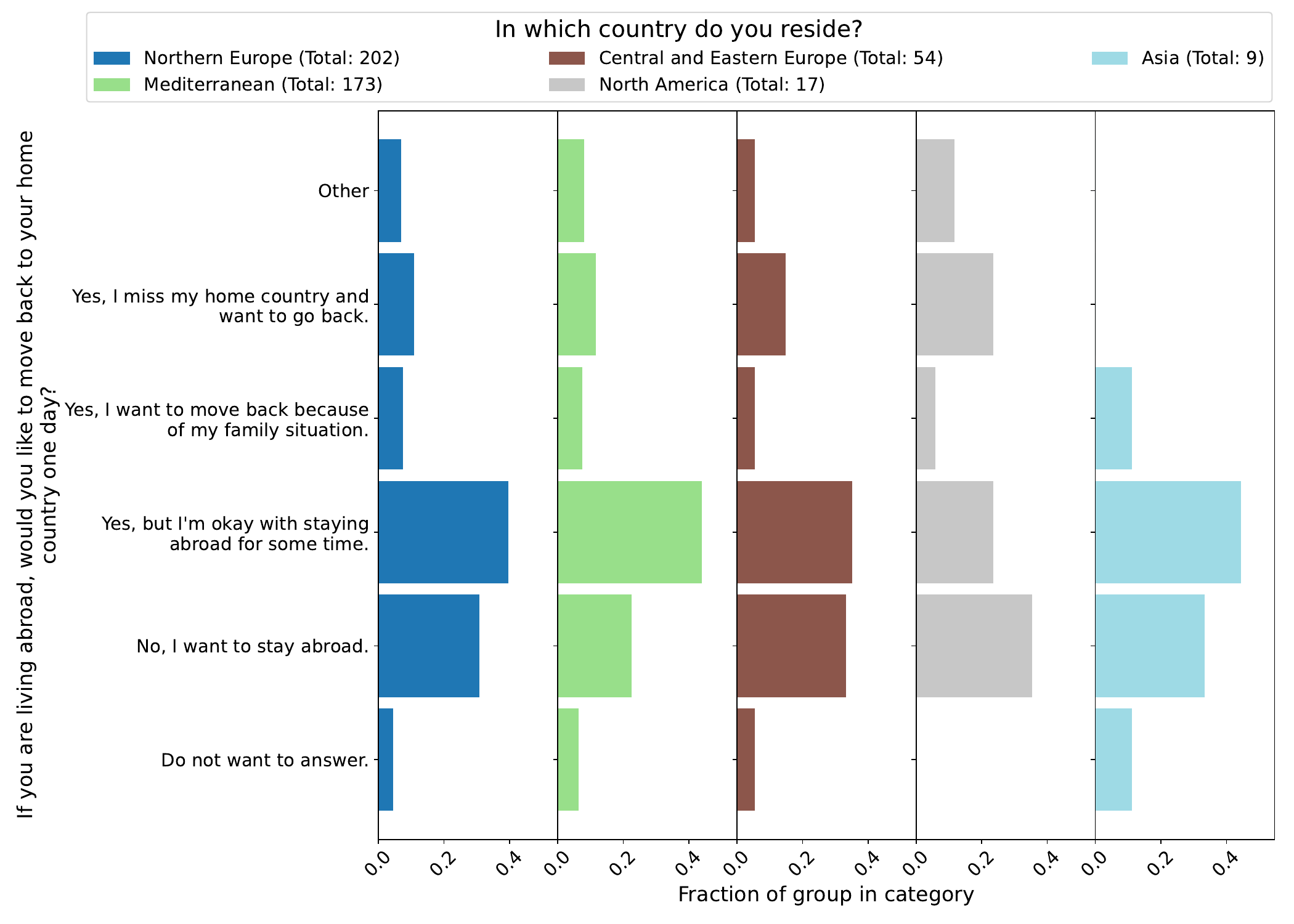}}
        \subfloat[]{\label{fig:part2:Q81vQ6}\includegraphics[width=0.49\textwidth]{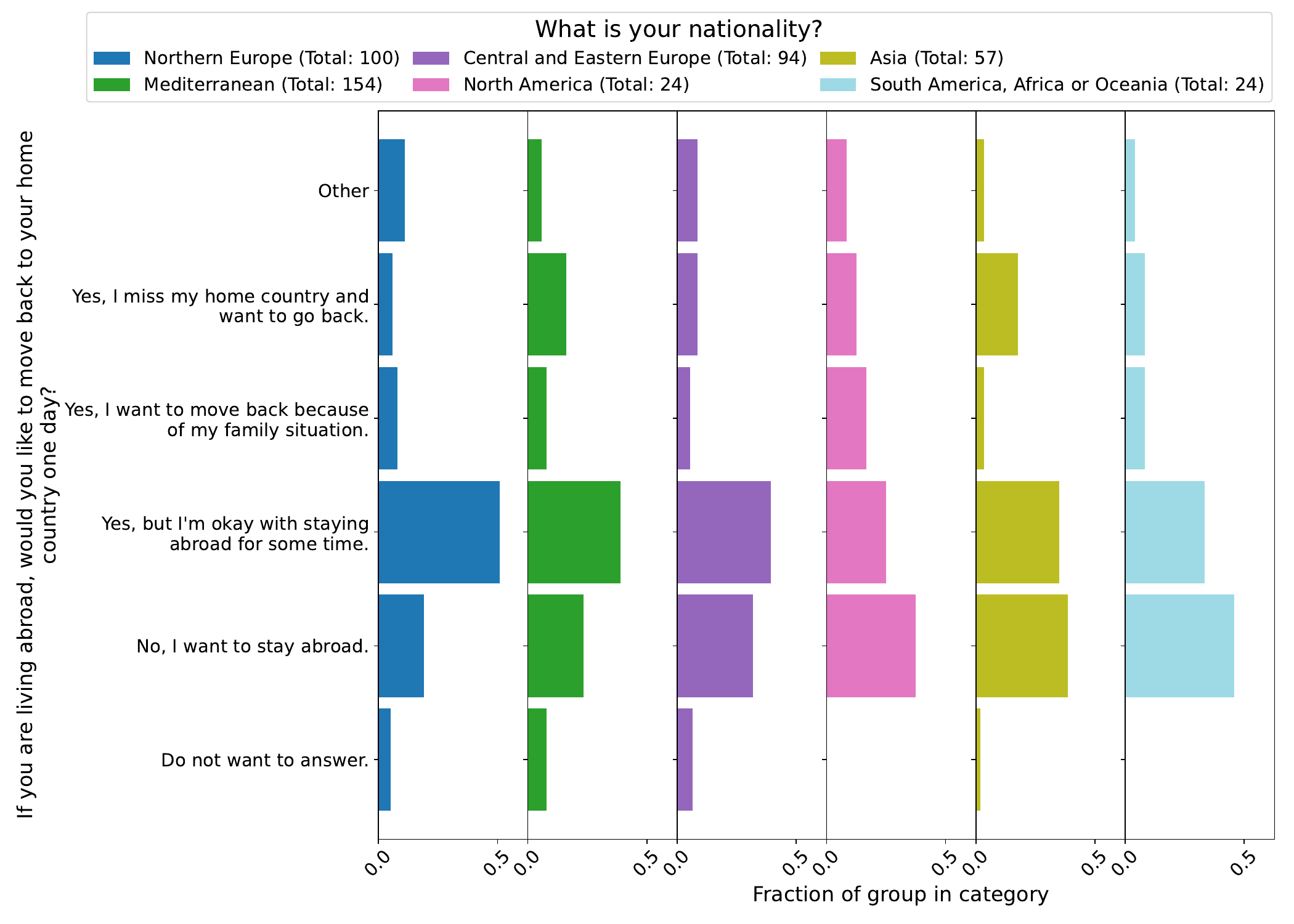}}\\
        \subfloat[]{\label{fig:part2:Q81vQ7}\includegraphics[width=0.49\textwidth]{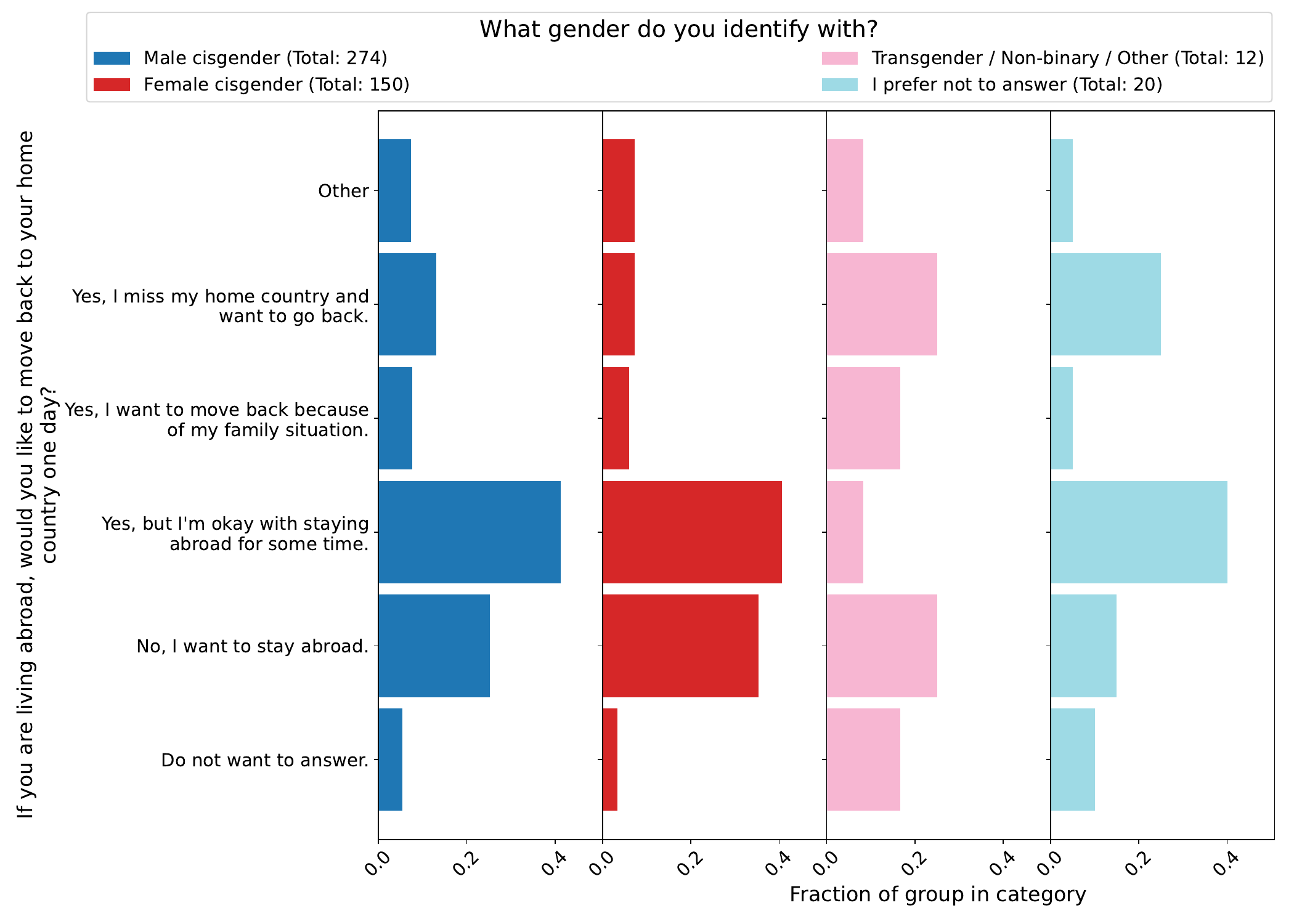}}
        \subfloat[]{\label{fig:part2:Q81vQ8}\includegraphics[width=0.49\textwidth]{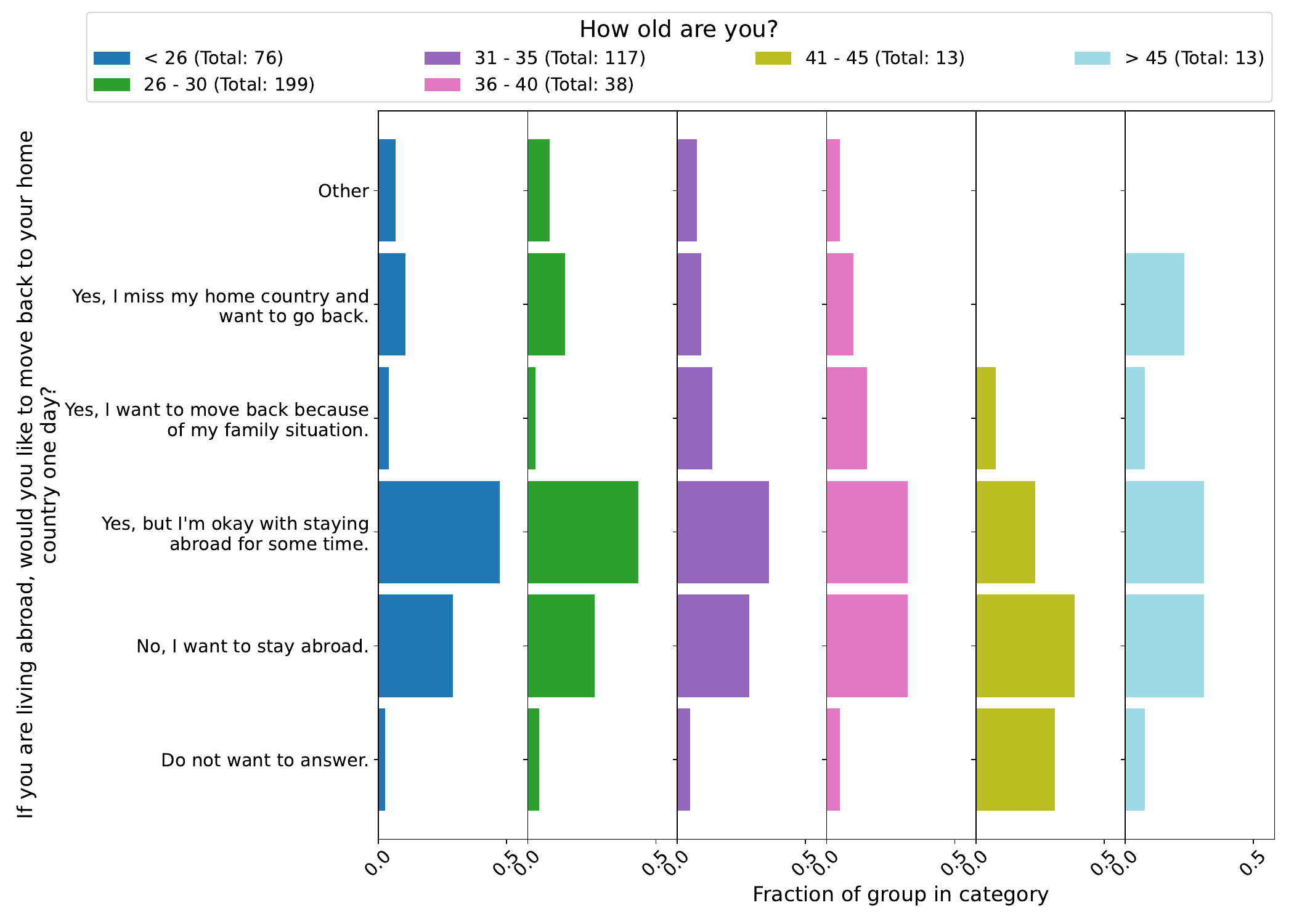}}\\
    \caption{(Q81 v Q5--8) Answers to the question on if respondents living abroad would like to come back to their country one day correlated with some aspects of participant profile and demographics. Fractions are given out of all respondents who answered the questions.}
    \label{fig:part2:Q81vsQ5Q6Q7Q8}
\end{figure}

We now move to questions regarding mobility.
In Figure~\ref{fig:part2:Q81vsQ5Q6Q7Q8}, answers to whether respondents who are living abroad would like to come back to their country one day are correlated with selected demographics.
We see that respondents residing abroad in North America are more likely to want to return home soon than respondents residing elsewhere, and conversely that respondents residing abroad in Asia are more likely to wish to stay abroad either for a while or permanently.
On the other hand, North American or Asian respondents are more likely to want to move back home soon than those with other nationalities.
Cisgender female respondents are the happiest to remain abroad.
Older respondents living abroad are generally happier to remain abroad.
Of those that want to return home, unsurprisingly this is more often due to a family situation as respondents age, until above 40 where missing their home country becomes more dominant again.

\begin{figure}[ht!]
    \centering
        \subfloat[]{\label{fig:part2:Q82vQ4}\includegraphics[width=0.49\textwidth]{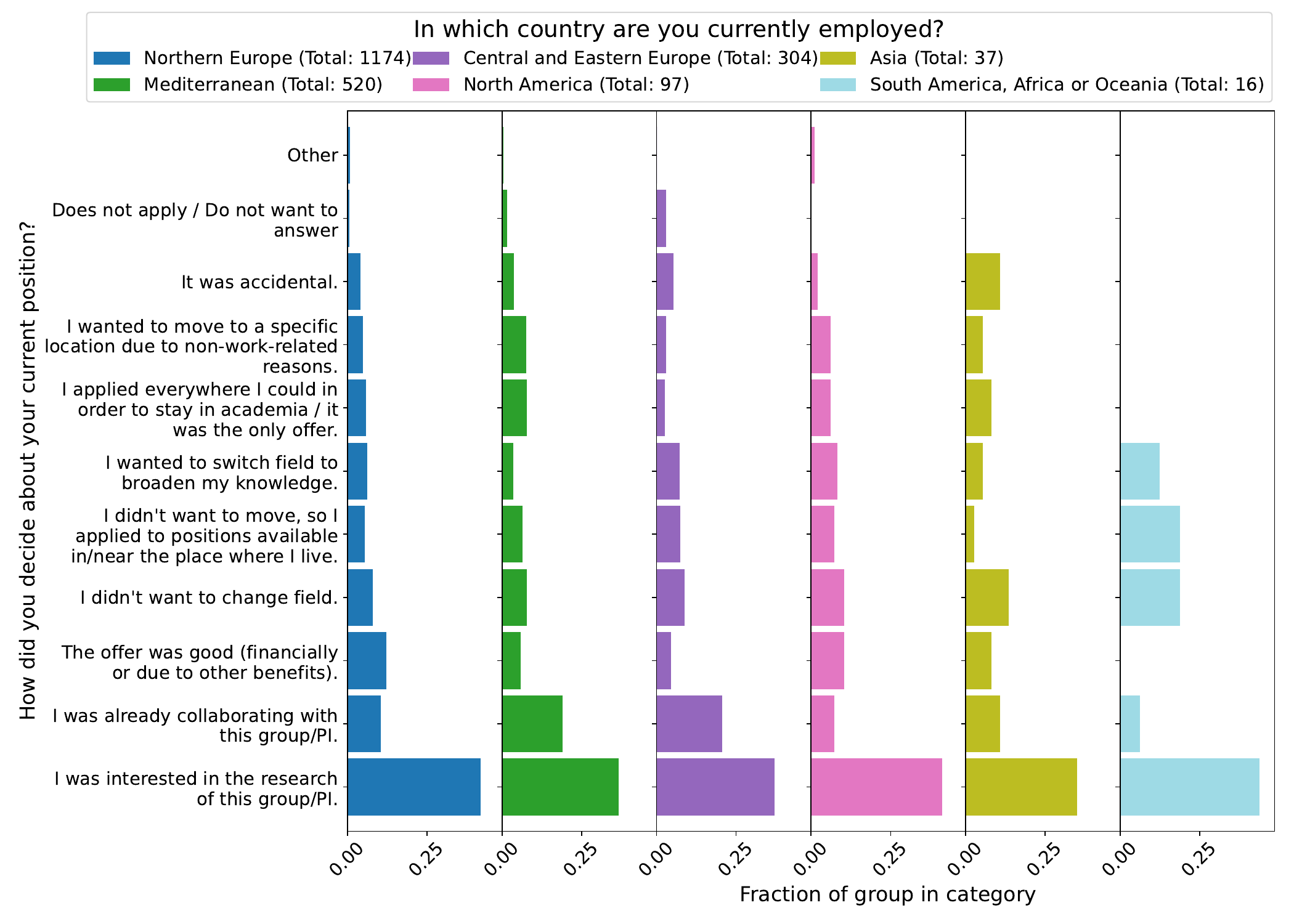}}
        \subfloat[]{\label{fig:part2:Q82vQ6}\includegraphics[width=0.49\textwidth]{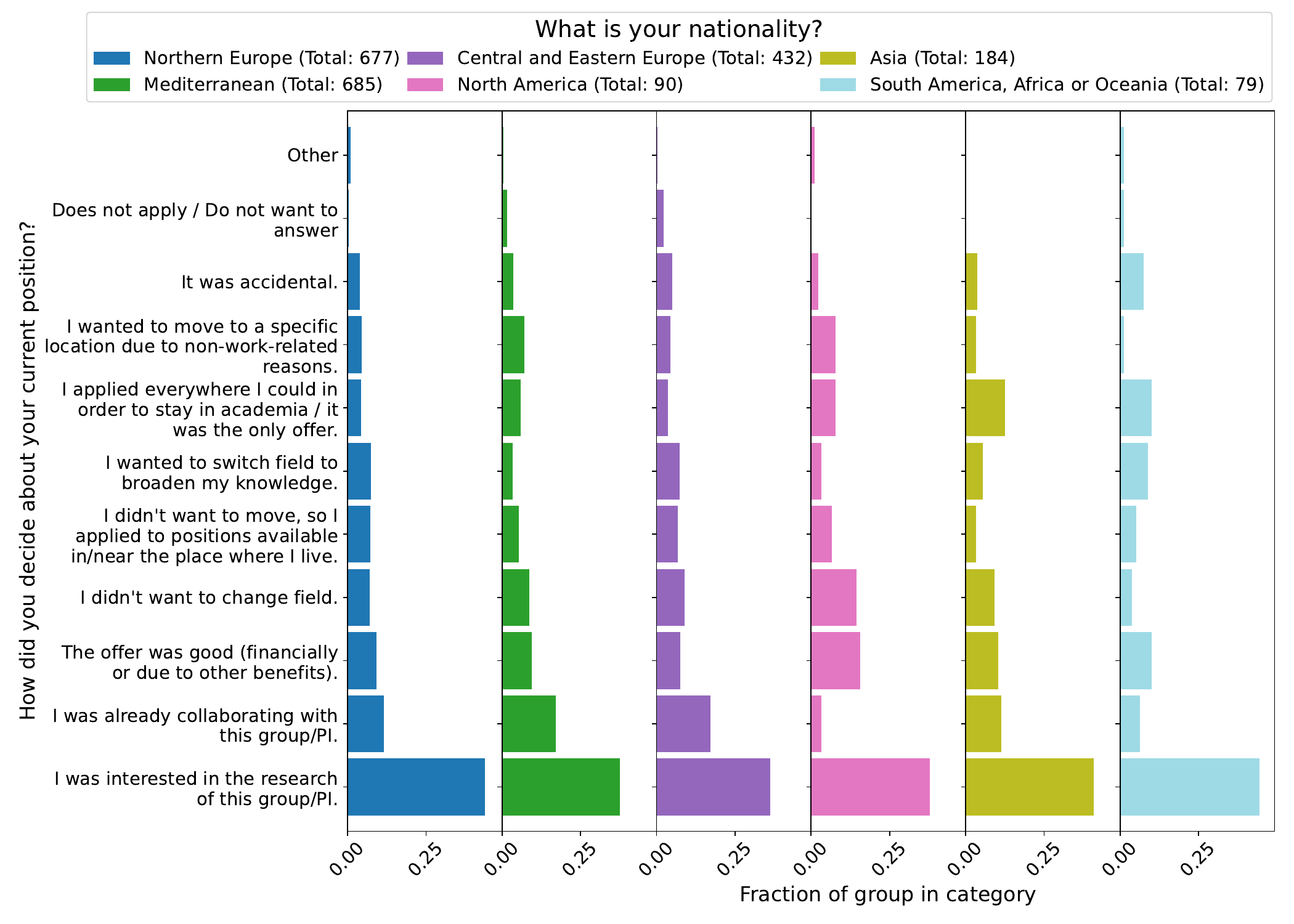}}
    \caption{(Q82 v Q4,6) Correlations between factors used by respondents to decide their current position and demographics. Multiple answers were allowed per respondent. Fractions are given out of all respondents who answered the questions.}
    \label{fig:part2:Q82vQ4Q6}
\end{figure}

In Figure~\ref{fig:part2:Q82vQ4Q6} we show correlations that were spotted between the question of how respondents decided on their current position and geographical demographics.
For example, respondents employed in Asia claim the decision was accidental more often than those employed elsewhere.
Respondents employed in the Mediterranean or Central and Eastern Europe point more often to existing collaboration with their group as a factor than respondents employed elsewhere.
North American respondents point more to the offer being good, whilst Asian respondents point more to taking the only offer after applying broadly.

\begin{figure}[ht!]
    \centering
    \includegraphics[width=0.7\textwidth]{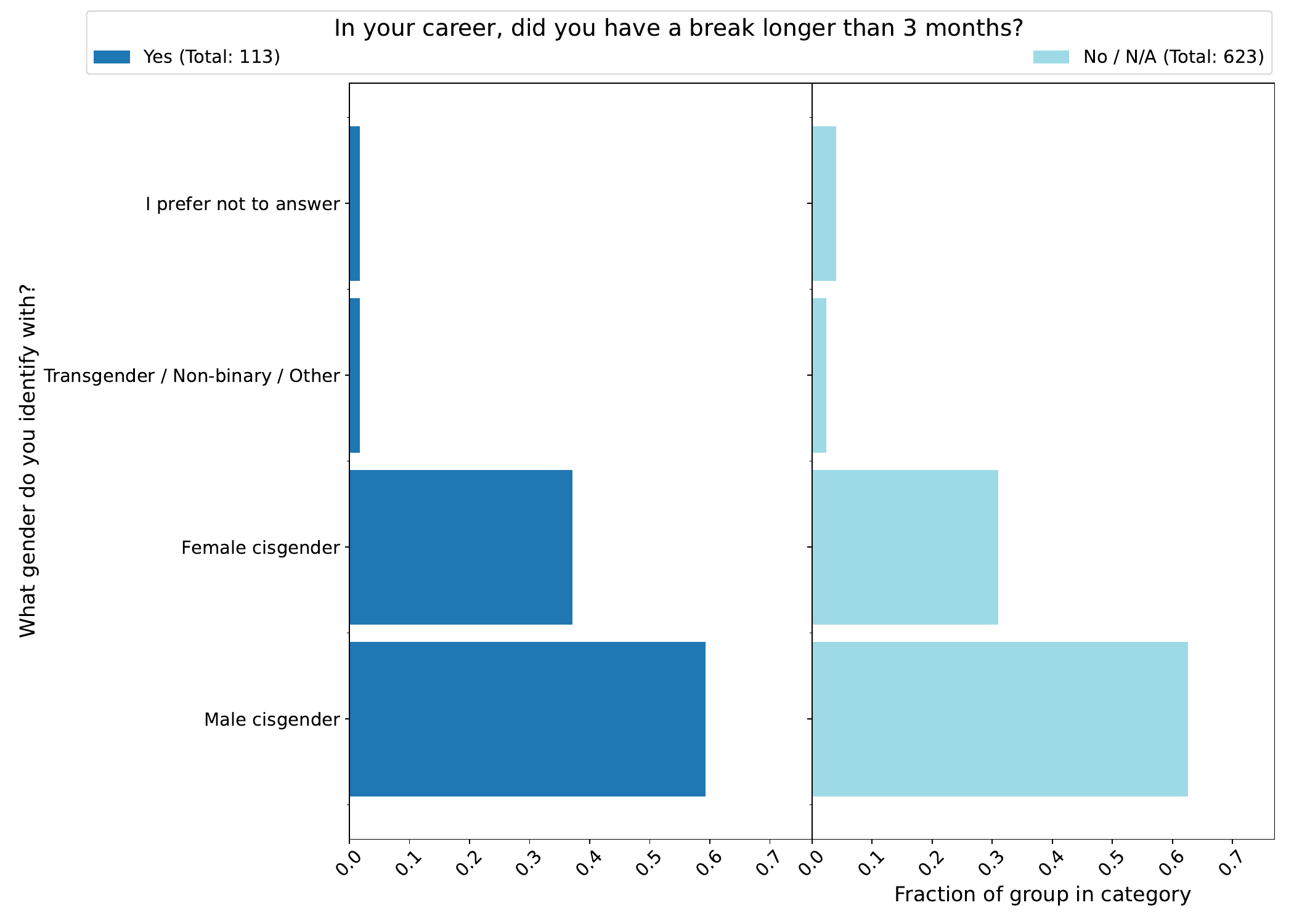}
    \caption{(Q83 v Q7) Correlations between whether respondents have had a career break longer than 3 months and their gender. Fractions are given out of all respondents.}
    \label{fig:part2:Q83vQ7}
\end{figure}

Figure~\ref{fig:part2:Q83vQ7} shows whether respondents have had a career break longer than 3 months, correlated with gender, where we see that respondents who haven't taken a career break are more likely to be cisgender males than those who have.
The correlation seen here is not as strong as one might suspect given an assumption of women taking more time off for maternity leave or child-care than men, this could be --- for example --- due to improvement in paternity leave opportunities/willingness, or that the survey doesn't consider those who didn't resume their career after having children.
Other than the obvious correlation that a career break is more likely in older respondents, no other interesting correlations were observed.

\begin{figure}[ht!]
    \centering
    \includegraphics[width=0.7\textwidth]{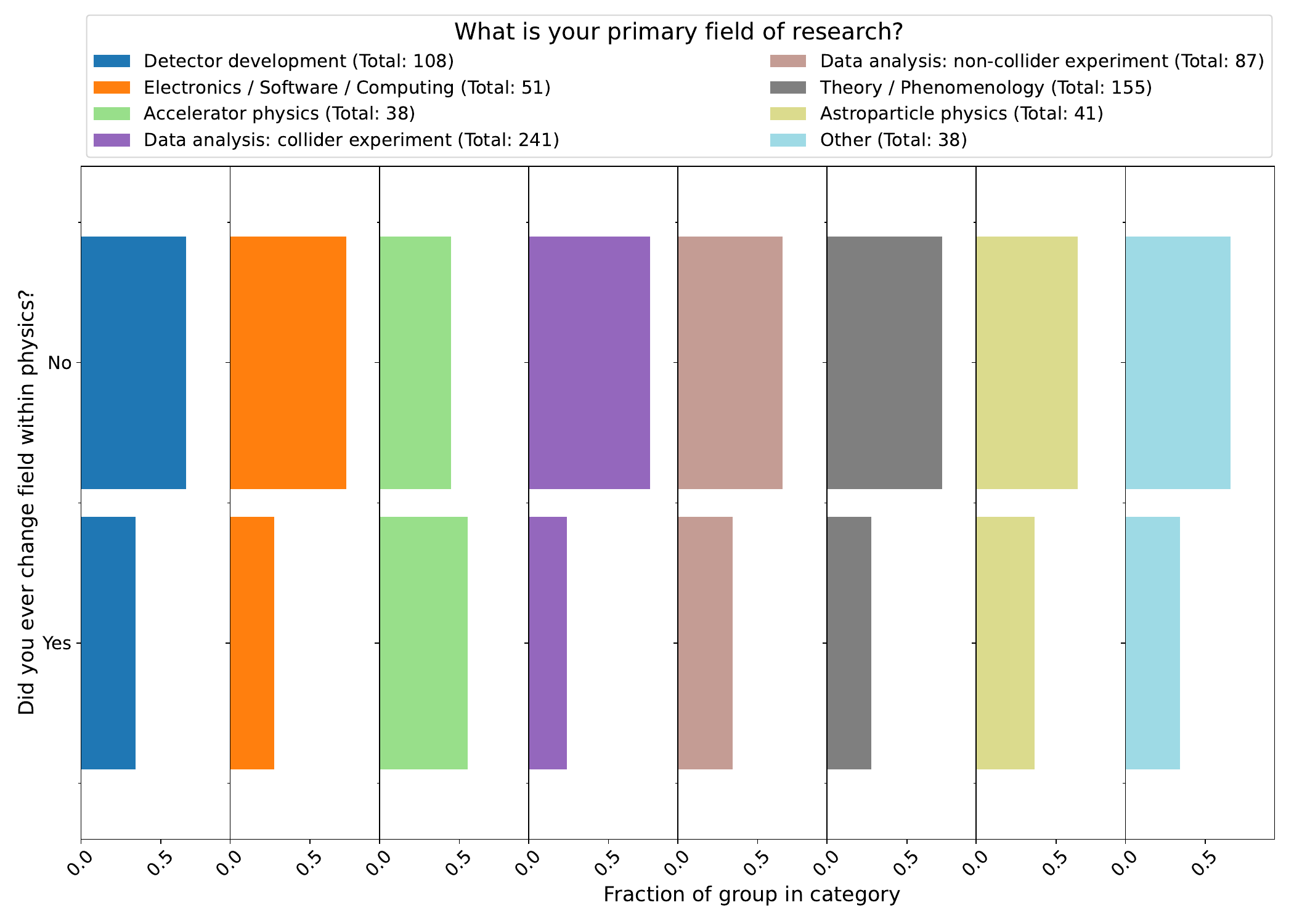}
    \caption{(Q84 v Q11) Correlations between whether respondents have changed their field of research and their current field of research. Fractions are given out of all respondents.}
    \label{fig:part2:Q84vQ11}
\end{figure}

In Figure~\ref{fig:part2:Q84vQ11} we correlate whether respondents have ever changed their research field with their current field; no other interesting correlations were seen.
Respondents working currently in accelerator physics area have changed their field of research the most often, and those in data analysis the least.

\begin{figure}[ht!]
    \centering
        \subfloat[]{\label{fig:part2:Q85vQ4}\includegraphics[width=0.49\textwidth]{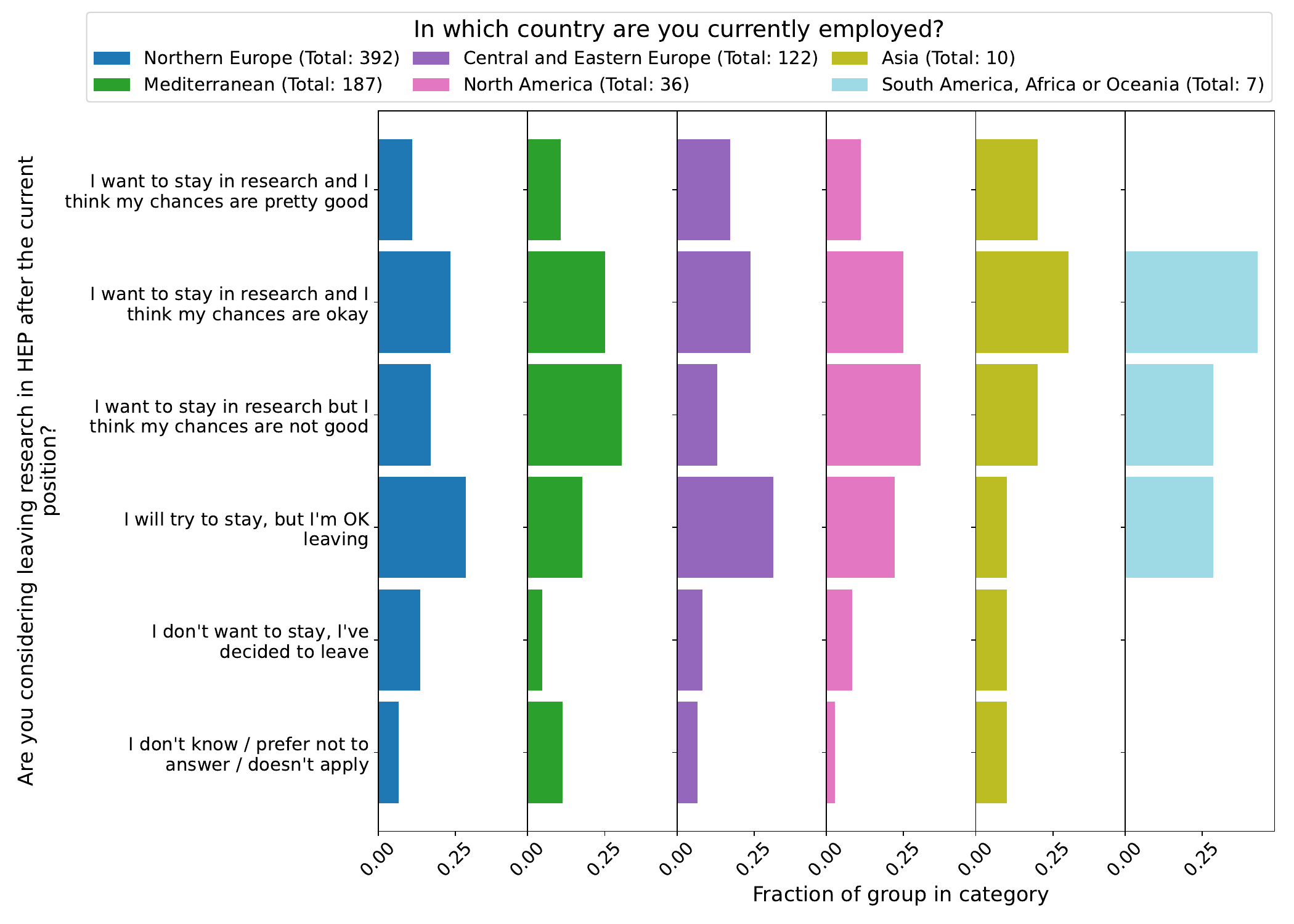}}
        \subfloat[]{\label{fig:part2:Q86vQ4}\includegraphics[width=0.49\textwidth]{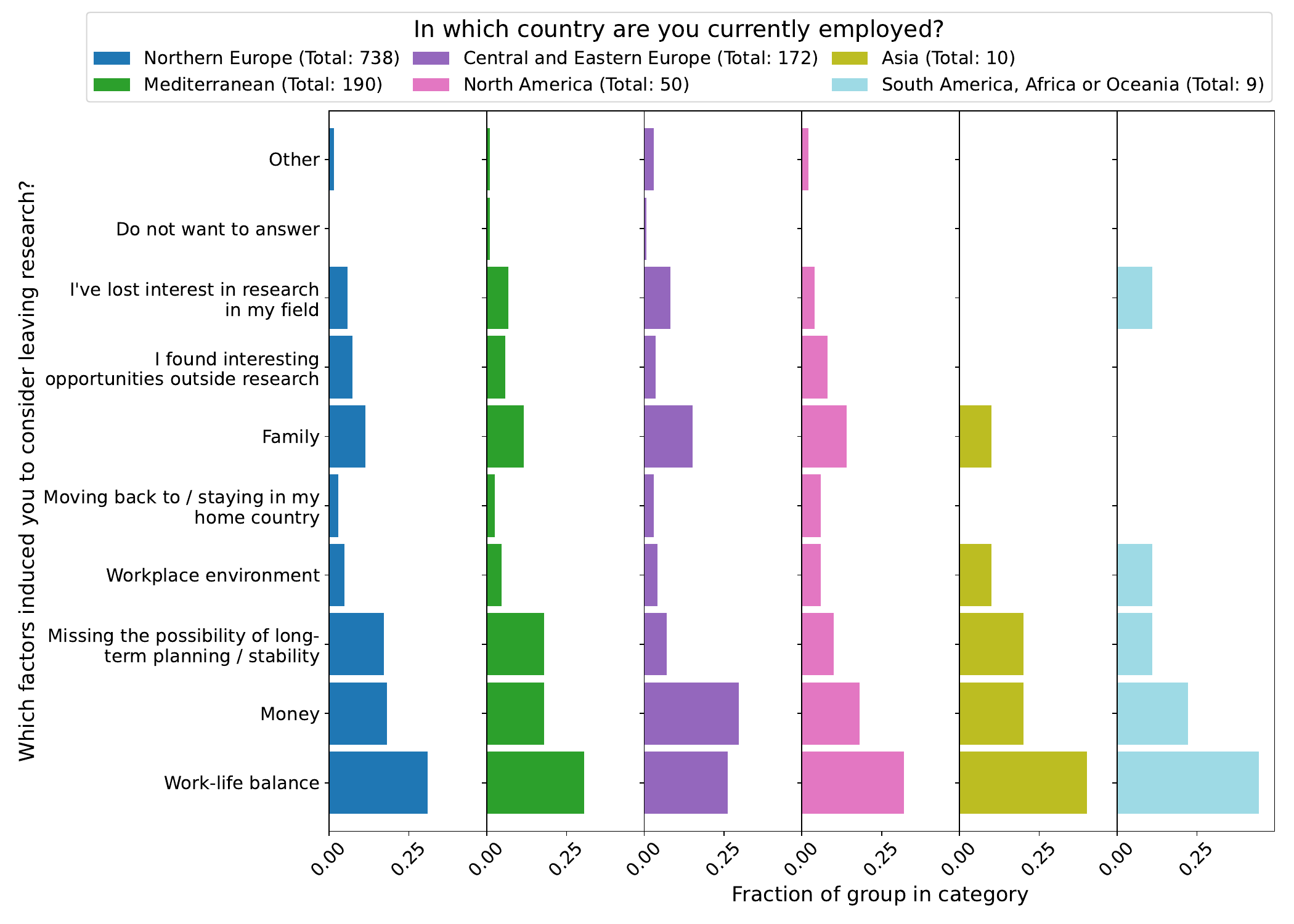}}\\
        \subfloat[]{\label{fig:part2:Q85vQ9}\includegraphics[width=0.49\textwidth]{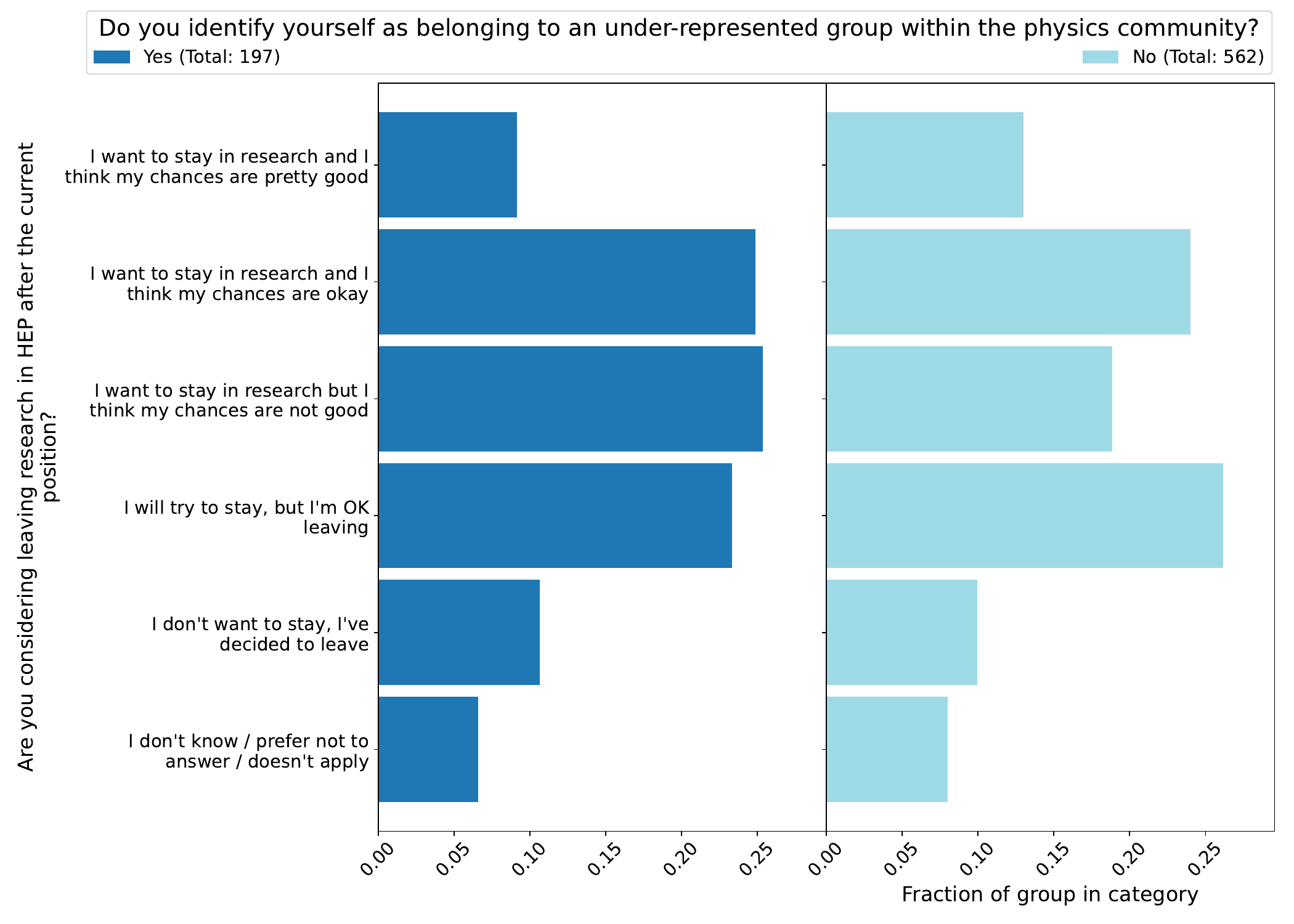}}
        \subfloat[]{\label{fig:part2:Q86vQ9}\includegraphics[width=0.49\textwidth]{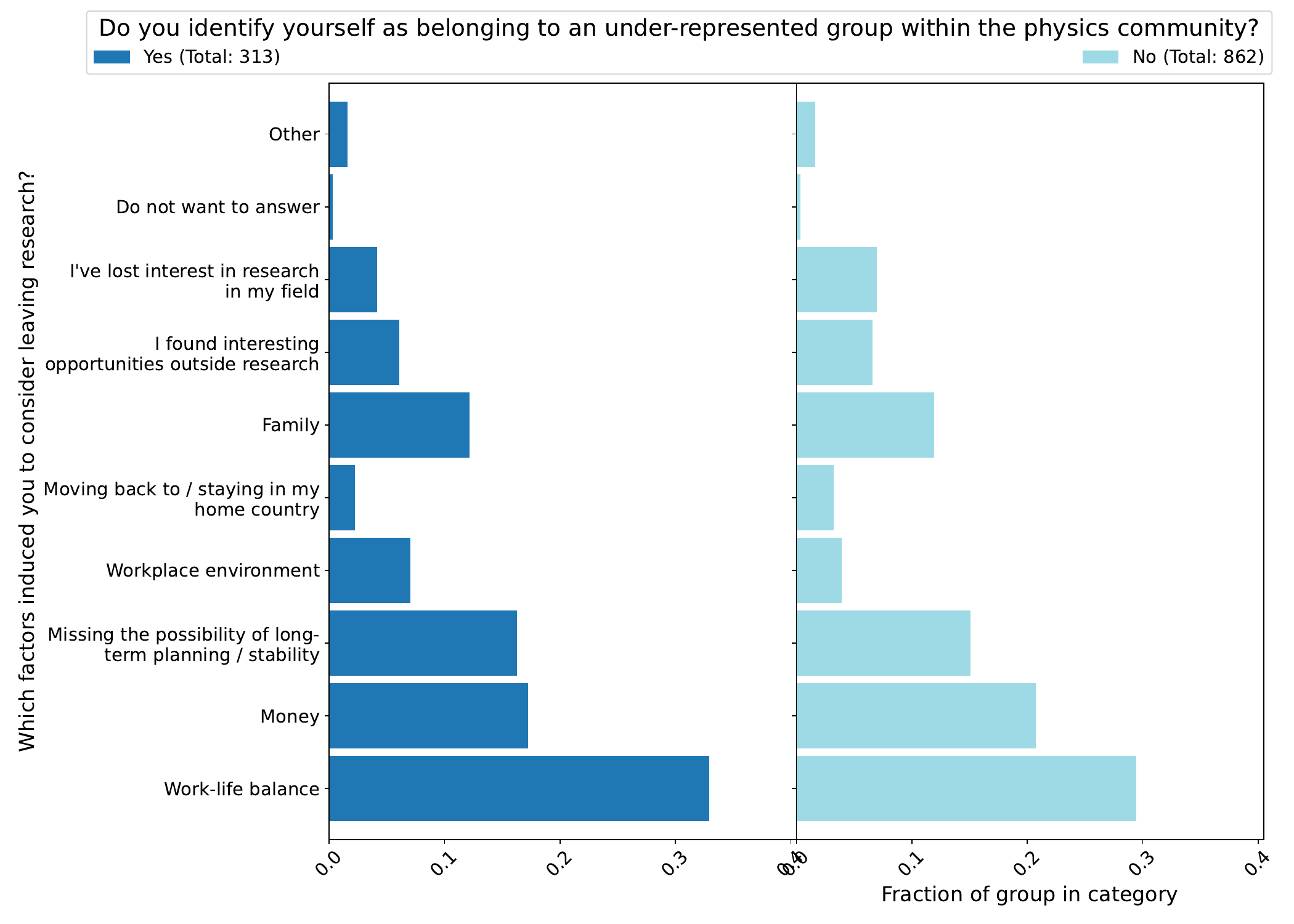}}\\
        \subfloat[]{\label{fig:part2:Q85vQ8}\includegraphics[width=0.49\textwidth]{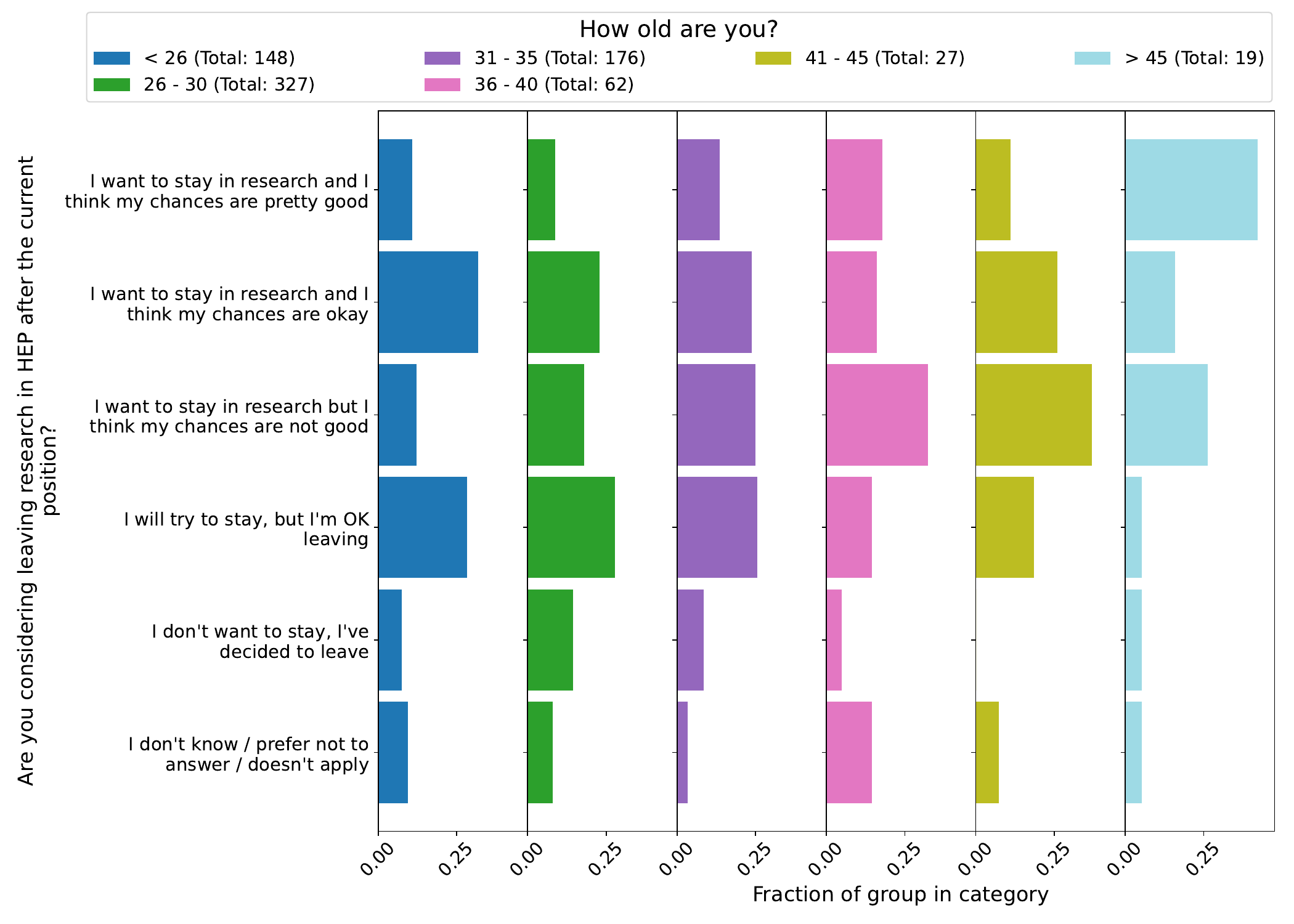}}
        \subfloat[]{\label{fig:part2:Q85vQ10}\includegraphics[width=0.49\textwidth]{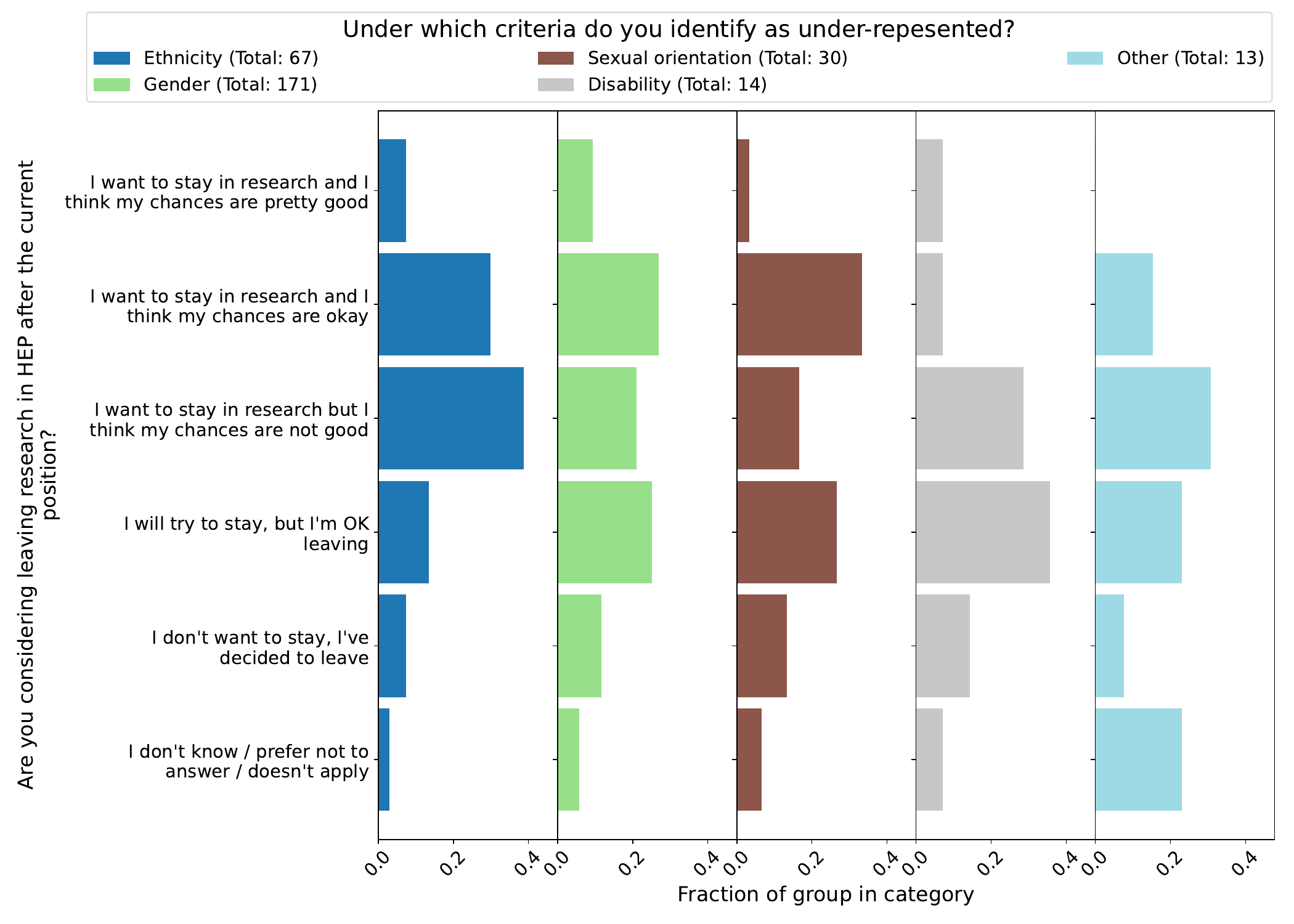}}
    \caption{(Q85 v Q4,8,9,10; Q86 v Q4,9) Correlations between whether respondents are considering leaving research after their current position and selected demographics. Fractions are given out of all respondents who answered the questions.}
    \label{fig:part2:Q85vQ4Q8Q9Q10_Q86vQ4Q9}
\end{figure}

Finally we turn to whether respondents are considering leaving research after their current job, and why.
Interesting correlations with demographics are shown in Figure~\ref{fig:part2:Q85vQ4Q8Q9Q10_Q86vQ4Q9}.
Considering country of employment, of the respondents who want to stay in research, those employed in Central and Eastern Europe are the most confident about their chances (though they are also most likely to be OK leaving) and those in North America the least confident.
Respondents who are employed in Northern Europe are the most likely to want to leave research.
The older respondents get, the less positive they are about their chance to stay in research, until they reach an age of 45.
Respondents who identify as under-represented and want to stay in research are less confident about their chances of staying in research than those who don't, though they are also more likely to be OK leaving.
Here, respondents with disabilities see their chances as especially bad.
We note that cisgender male respondents compared to others show the same pattern as comparing non-under-represented respondents to under-represented ones (not shown here). 

Addressing reasons for wanting to leave research, we see that respondents employed in Central and Eastern Europe are the most likely to indicate money or family, whilst elsewhere work-life balance and missing the possibility of long-term planning or stability are more dominant.
We also see that respondents in under-represented groups chose work-life balance and work-place environment more often.

\begin{figure}[ht!]
    \centering
        \subfloat[]{\label{fig:part2:Q85vQ62}\includegraphics[width=0.49\textwidth]{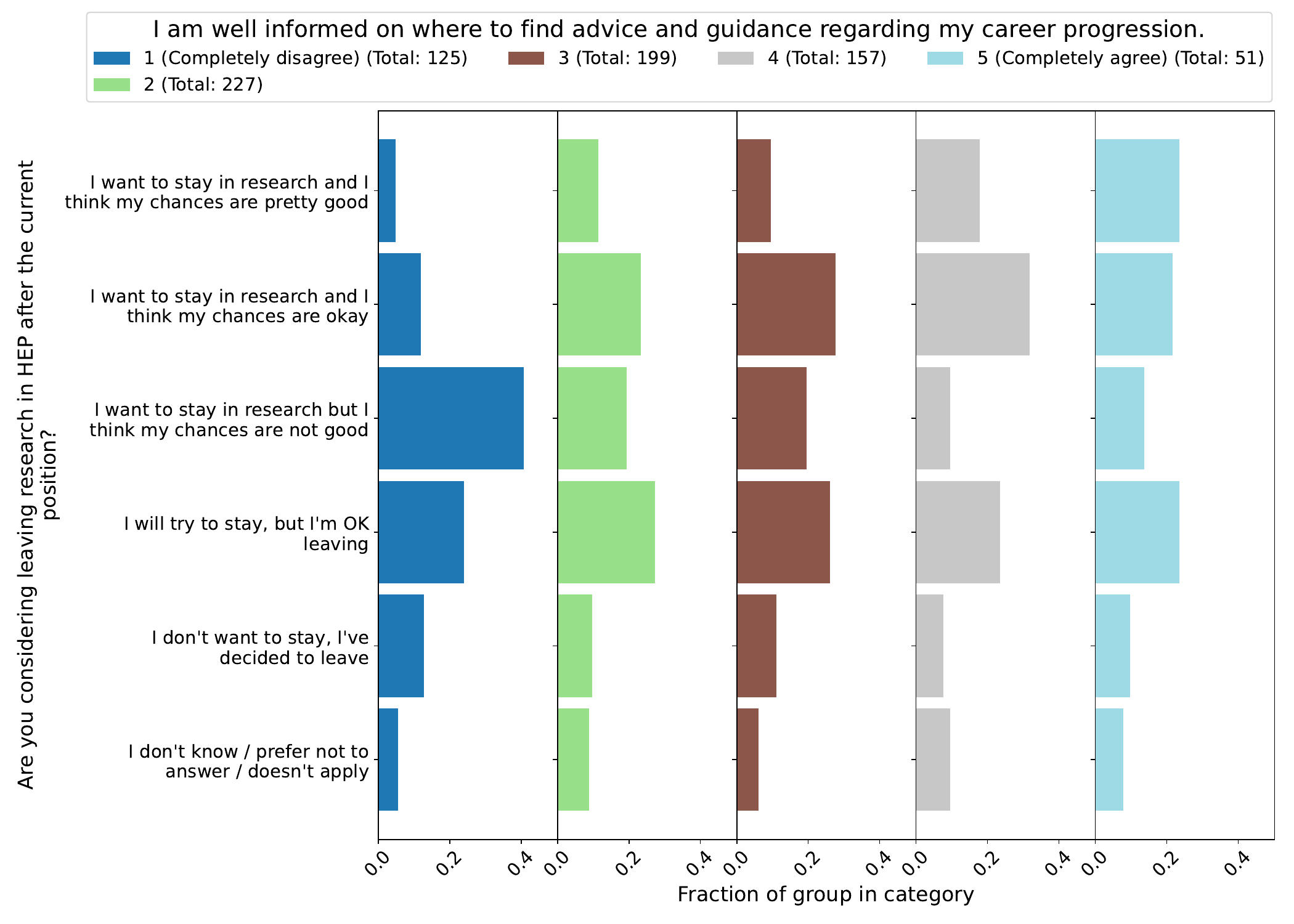}}
        \subfloat[]{\label{fig:part2:Q85vQ75f}\includegraphics[width=0.49\textwidth]{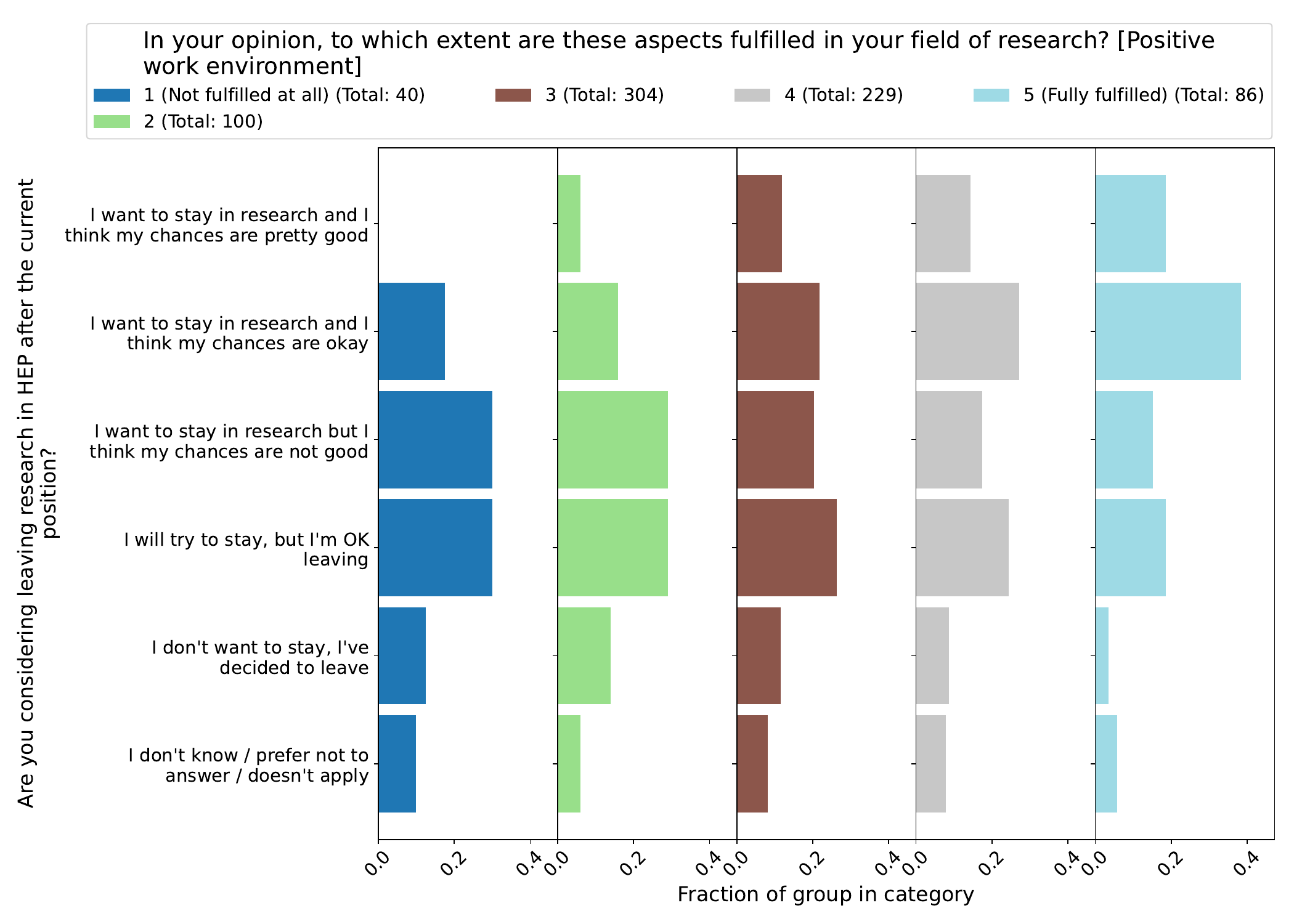}}\\
        \subfloat[]{\label{fig:part2:Q85vQ80}\includegraphics[width=0.49\textwidth]{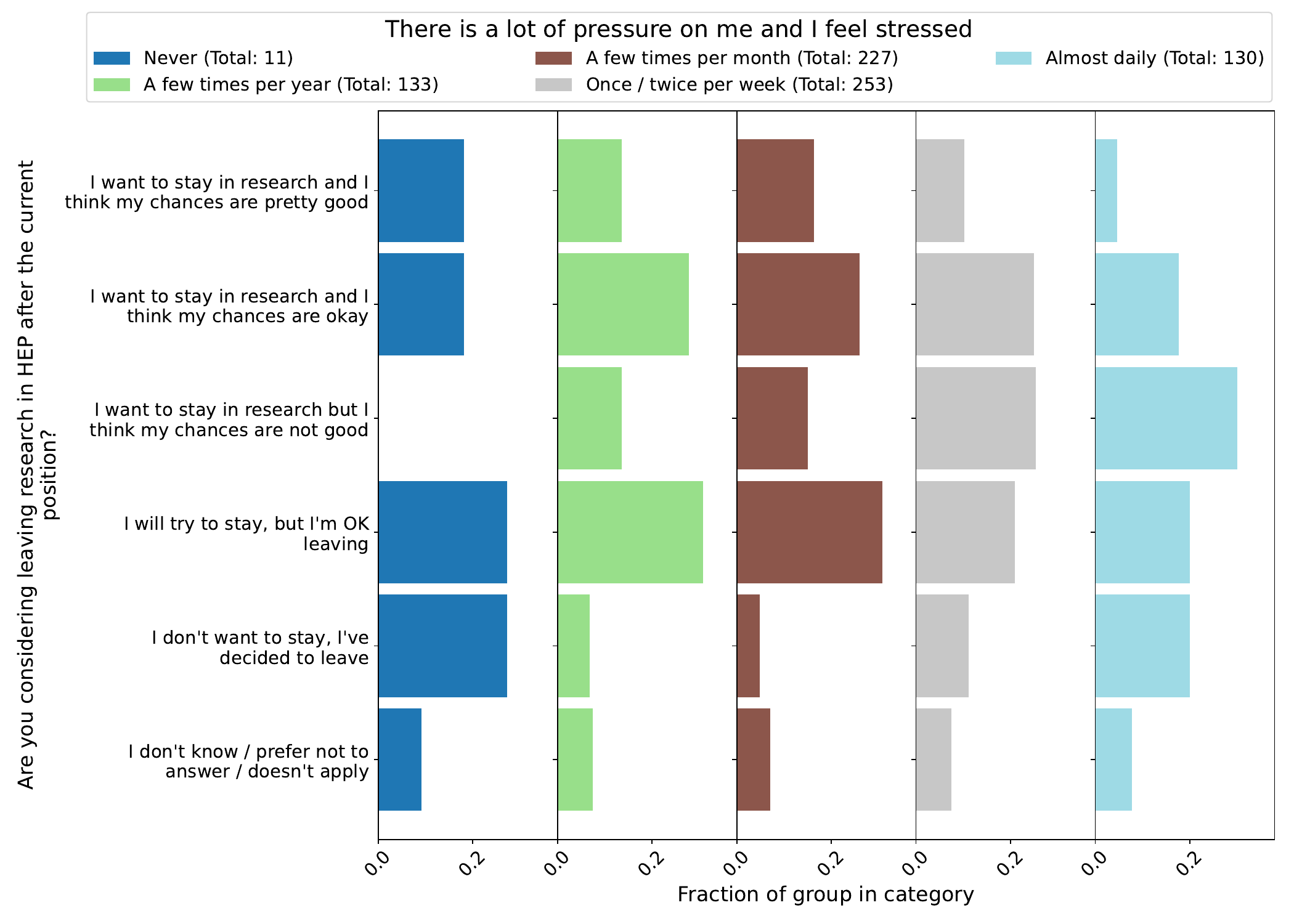}}
        \subfloat[]{\label{fig:part2:Q85vQ83}\includegraphics[width=0.49\textwidth]{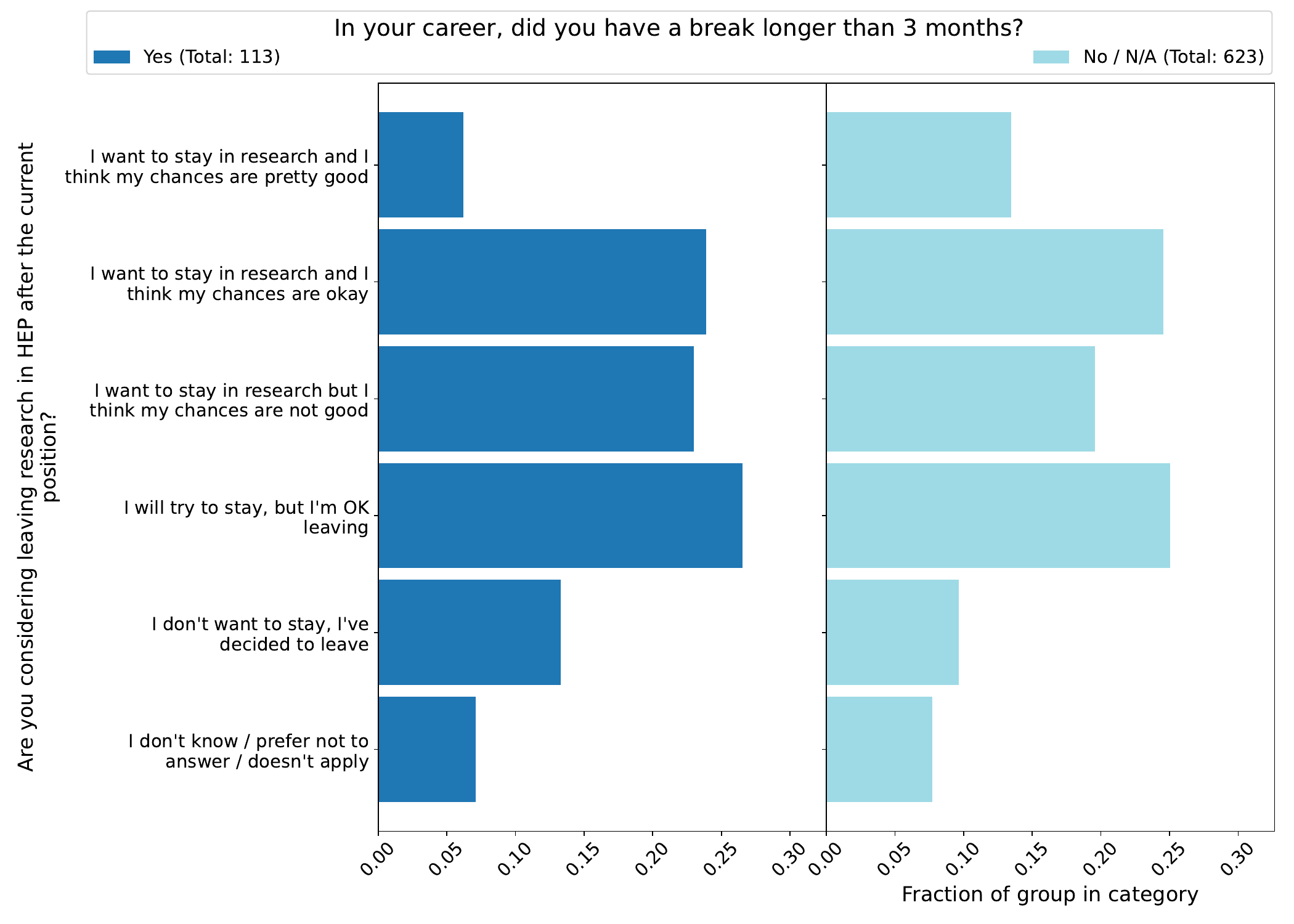}}\\
        \subfloat[]{\label{fig:part2:Q85vQ94}\includegraphics[width=0.49\textwidth]{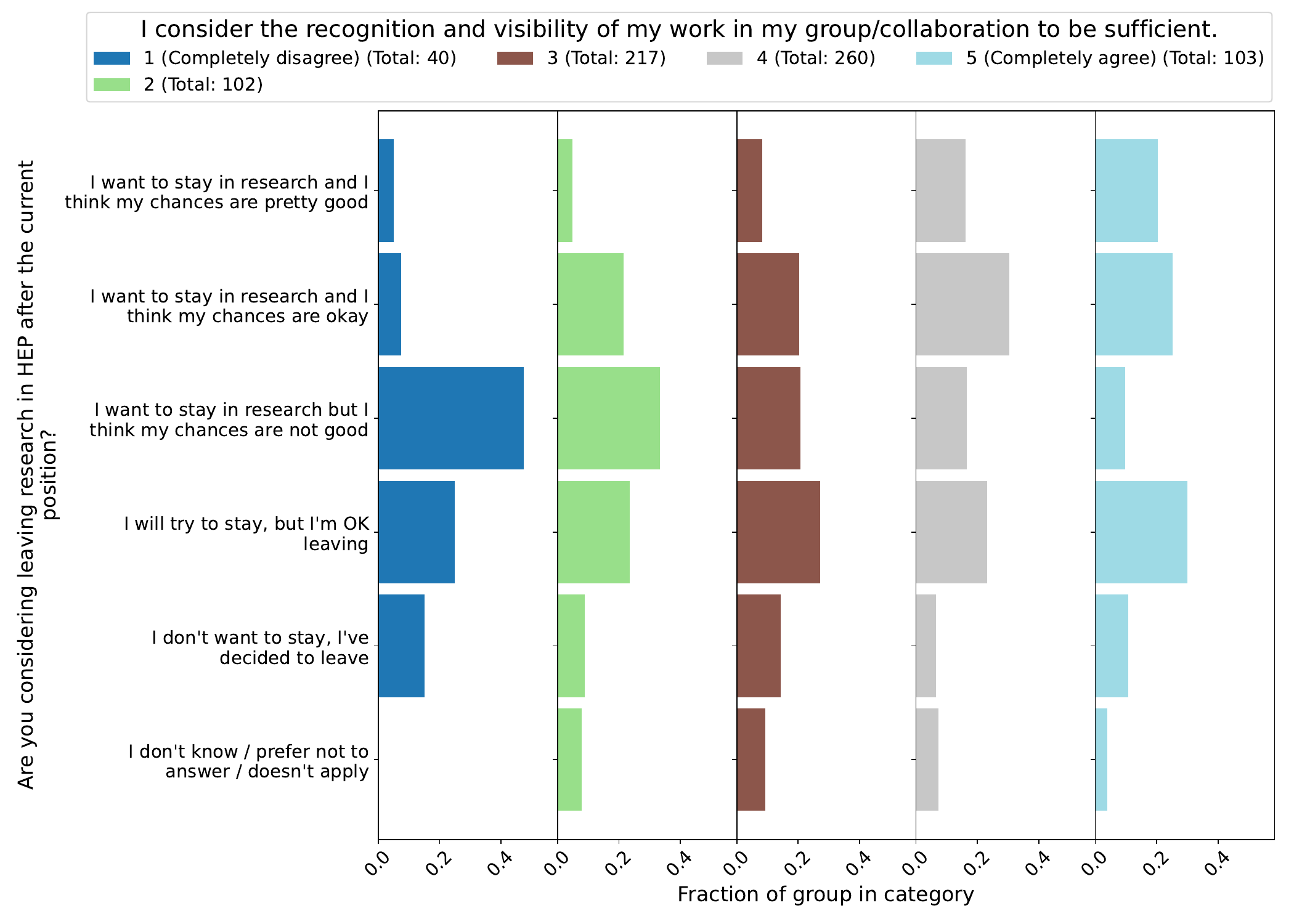}}
    \caption{(Q85 v Q62,75f,80,83,94) Correlations between whether respondents are considering leaving research after their current position and selected questions. Fractions are given out of all respondents who answered the questions.}
    \label{fig:part2:Q85vQ62Q75Q80Q83Q94}
\end{figure}

We now study whether respondents are considering leaving research in HEP after their current position, correlated with questions unrelated to demographics.
For respondents who want to stay in research, we see in Figure~\ref{fig:part2:Q85vQ62Q75Q80Q83Q94} positive correlations between how good they think their chances are and several factors: how well-informed respondents are about where to find advice and guidance regarding career progressions, how fulfilled respondents think a positive work environment is in their field, and how sufficient respondents feel the recognition and visibility of their work is.
These respondents are also less likely to have decided to leave research.
On the other hand we see a negative correlation in Figure~\ref{fig:part2:Q85vQ83} with whether respondents have had a career break, with those that have either feeling less confident about their chances of staying in research or being more likely to want to leave it.

\begin{figure}[ht!]
    \centering
        \subfloat[]{\label{fig:part2:Q86vQ80}\includegraphics[width=0.49\textwidth]{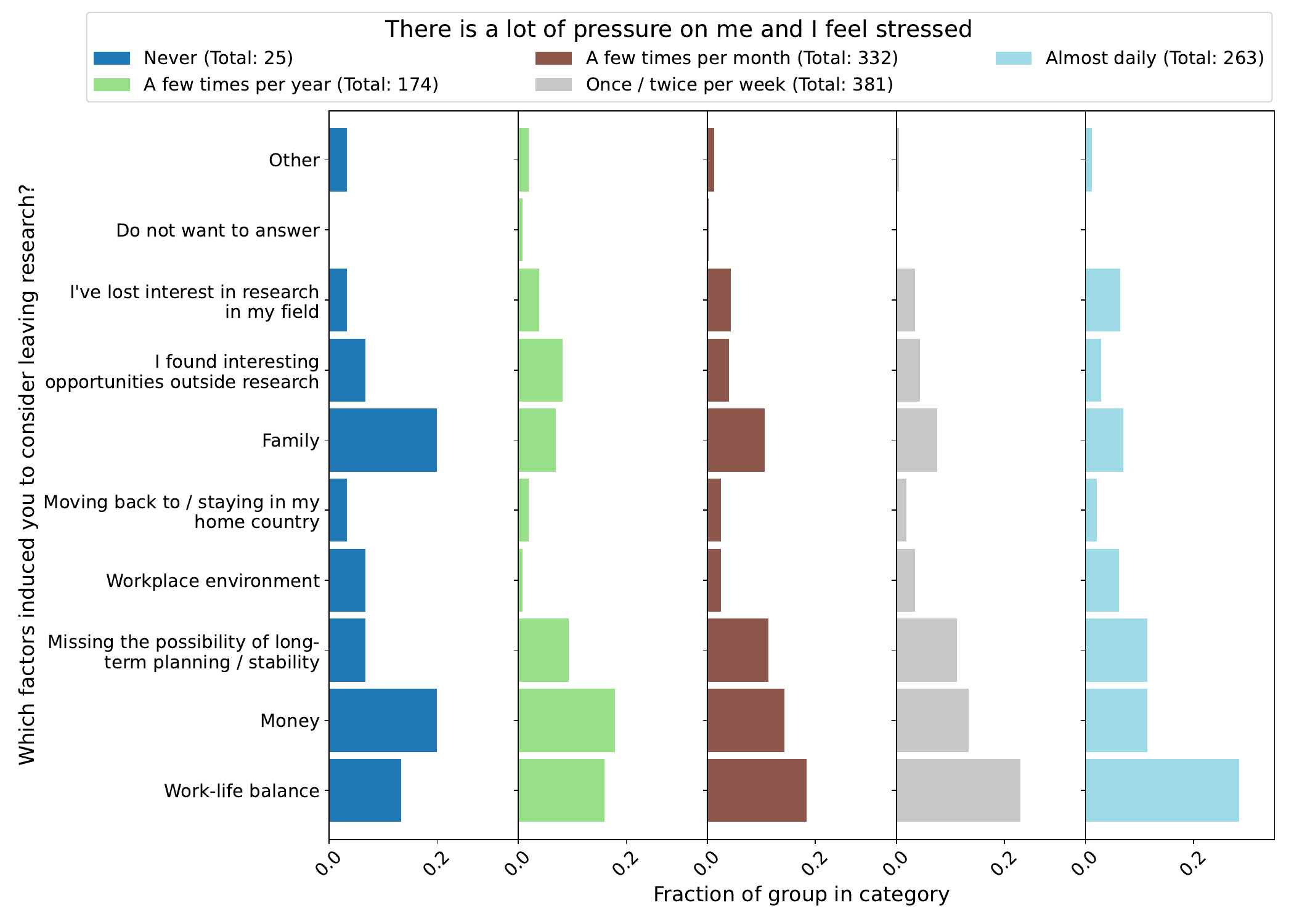}}
        \subfloat[]{\label{fig:part2:Q86vQ83}\includegraphics[width=0.49\textwidth]{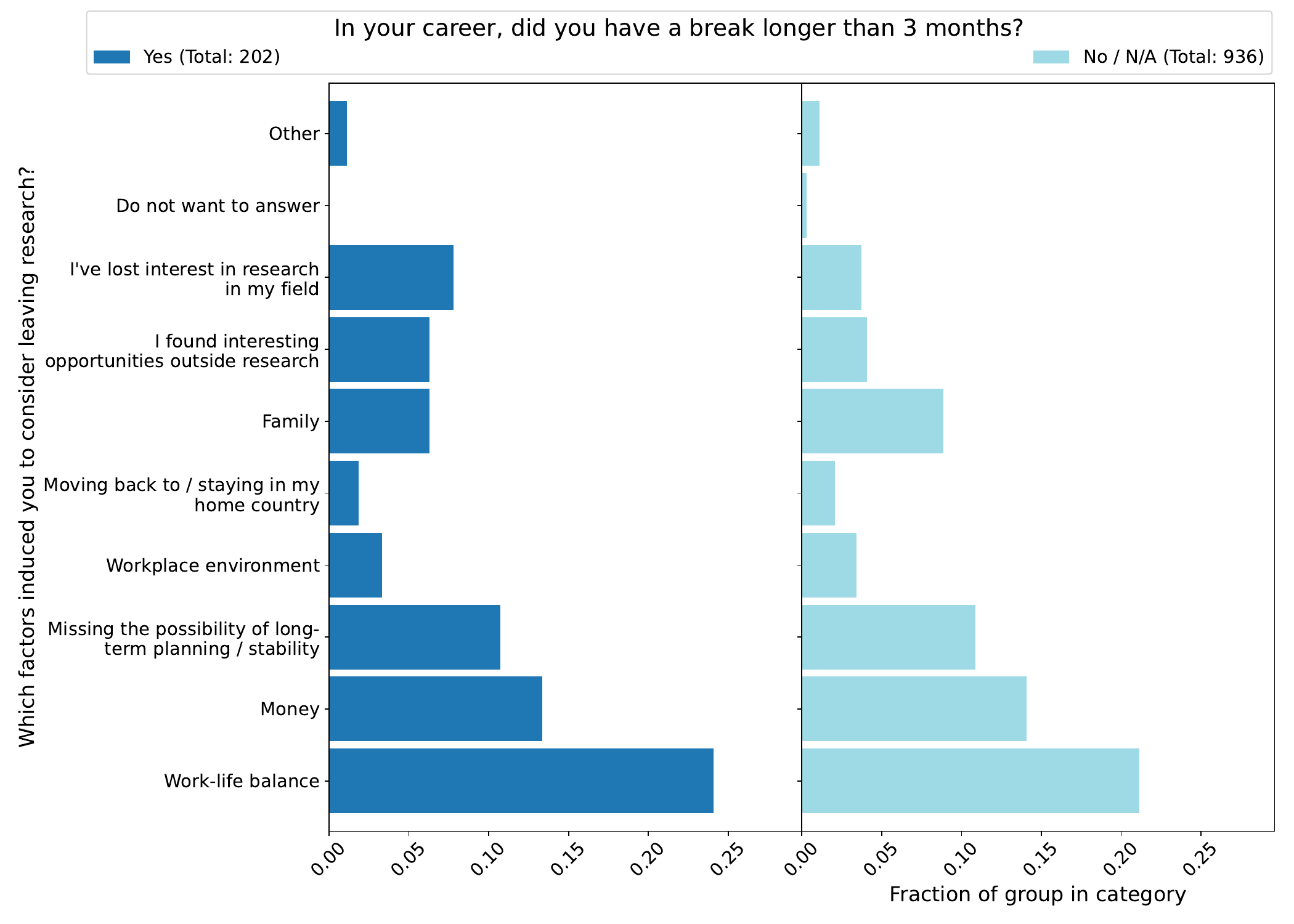}}\\
    \caption{(Q86 v Q80,83) Correlations between what factors are leading respondents to consider leaving research after their current position and selected questions. Fractions are given out of all respondents.}
    \label{fig:part2:Q86vQ80Q83}
\end{figure}

For those respondents who are considering leaving, some examples of correlations with the reasons for leaving are shown in Figure~\ref{fig:part2:Q86vQ80Q83}.
As expected, for those feeling stressed and under pressure more frequently, work-life balance is a more likely factor.
For those who have taken a career break, work-life balance and losing interest in research are more important, whilst family is a less cited reason.

\clearpage

%%%%%%%%%%%%%===================================================================================================

\subsection{Discriminatory or abusive treatment}

\begin{figure}[ht!]
    \centering
    \subfloat[]{\label{fig:part2:Q87vQ4}\includegraphics[width=0.49\textwidth]{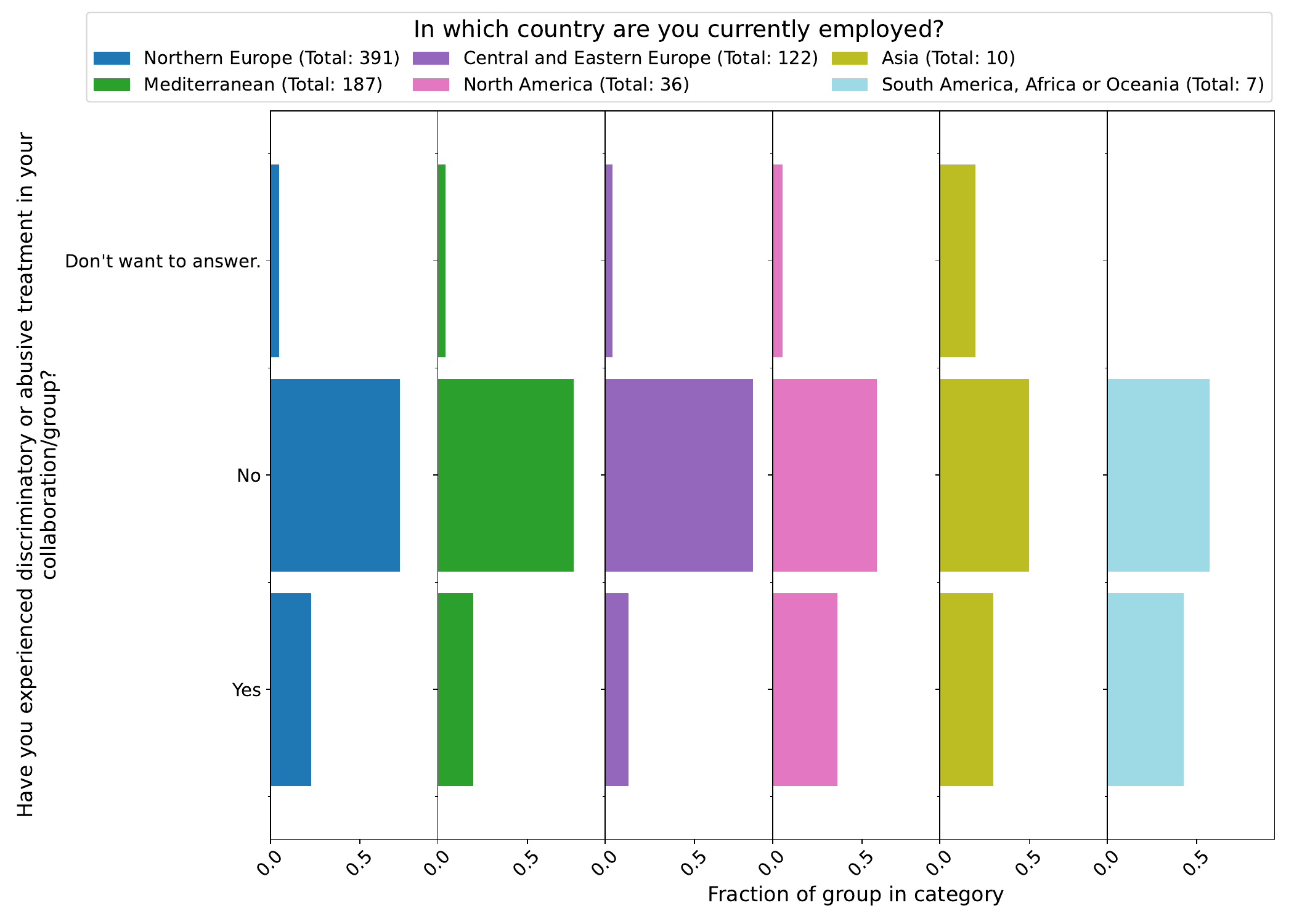}}
    \subfloat[]{\label{fig:part2:Q87vQ7}\includegraphics[width=0.49\textwidth]{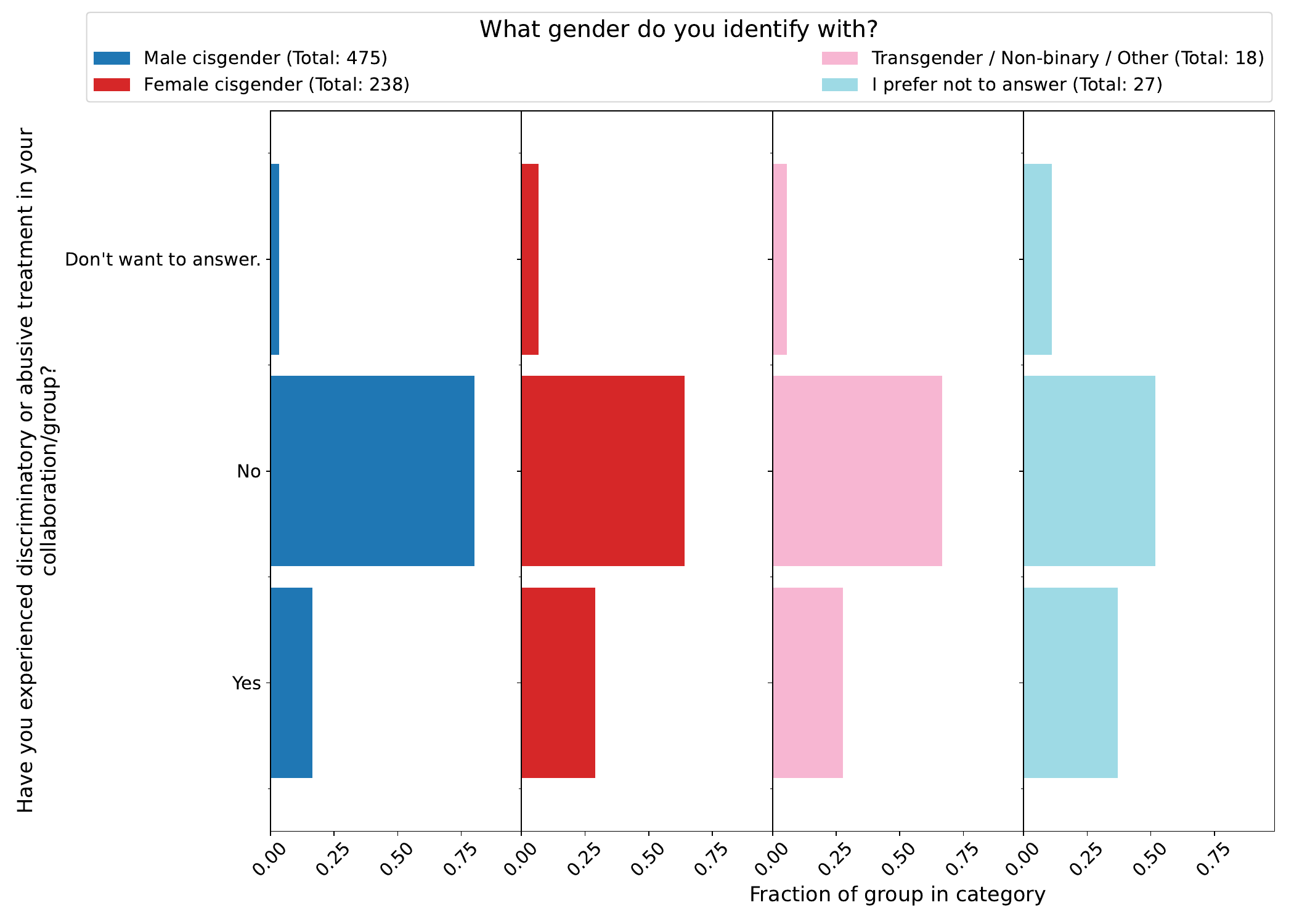}}\\
    \subfloat[]{\label{fig:part2:Q87vQ8}\includegraphics[width=0.49\textwidth]{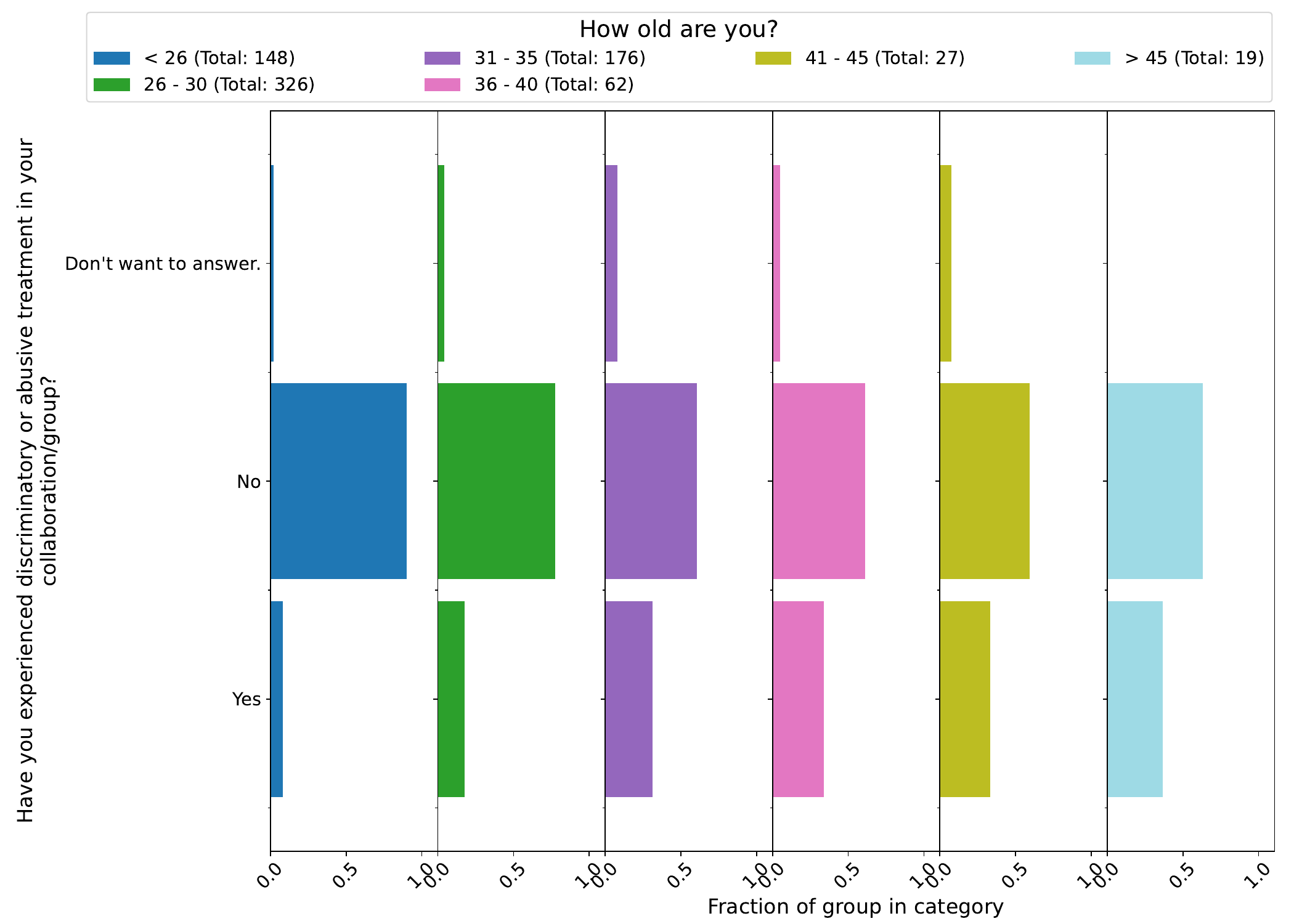}}
    \subfloat[]{\label{fig:part2:Q87vQ9}\includegraphics[width=0.49\textwidth]{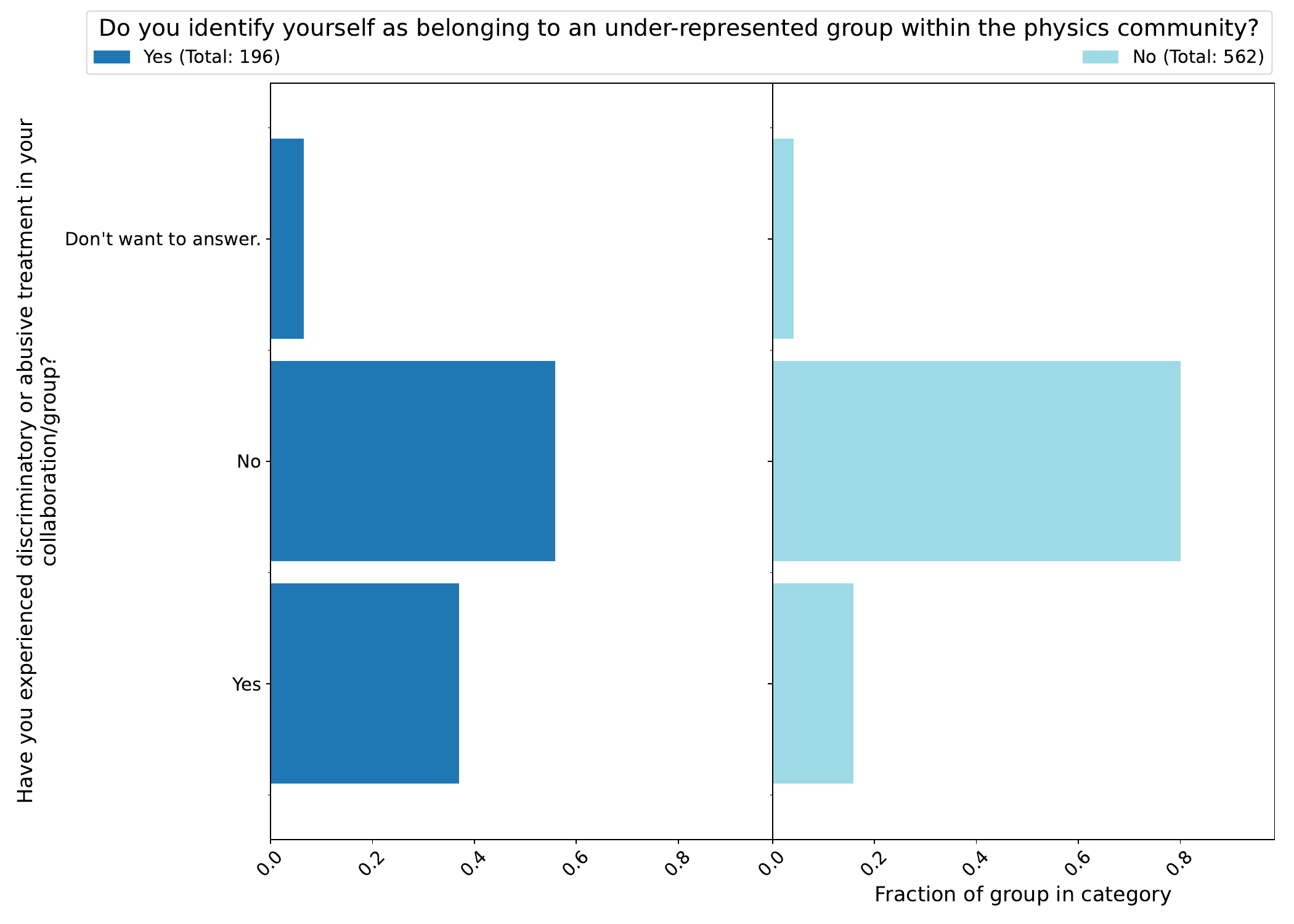}}\\
    \subfloat[]{\label{fig:part2:Q87vQ10}\includegraphics[width=0.49\textwidth]{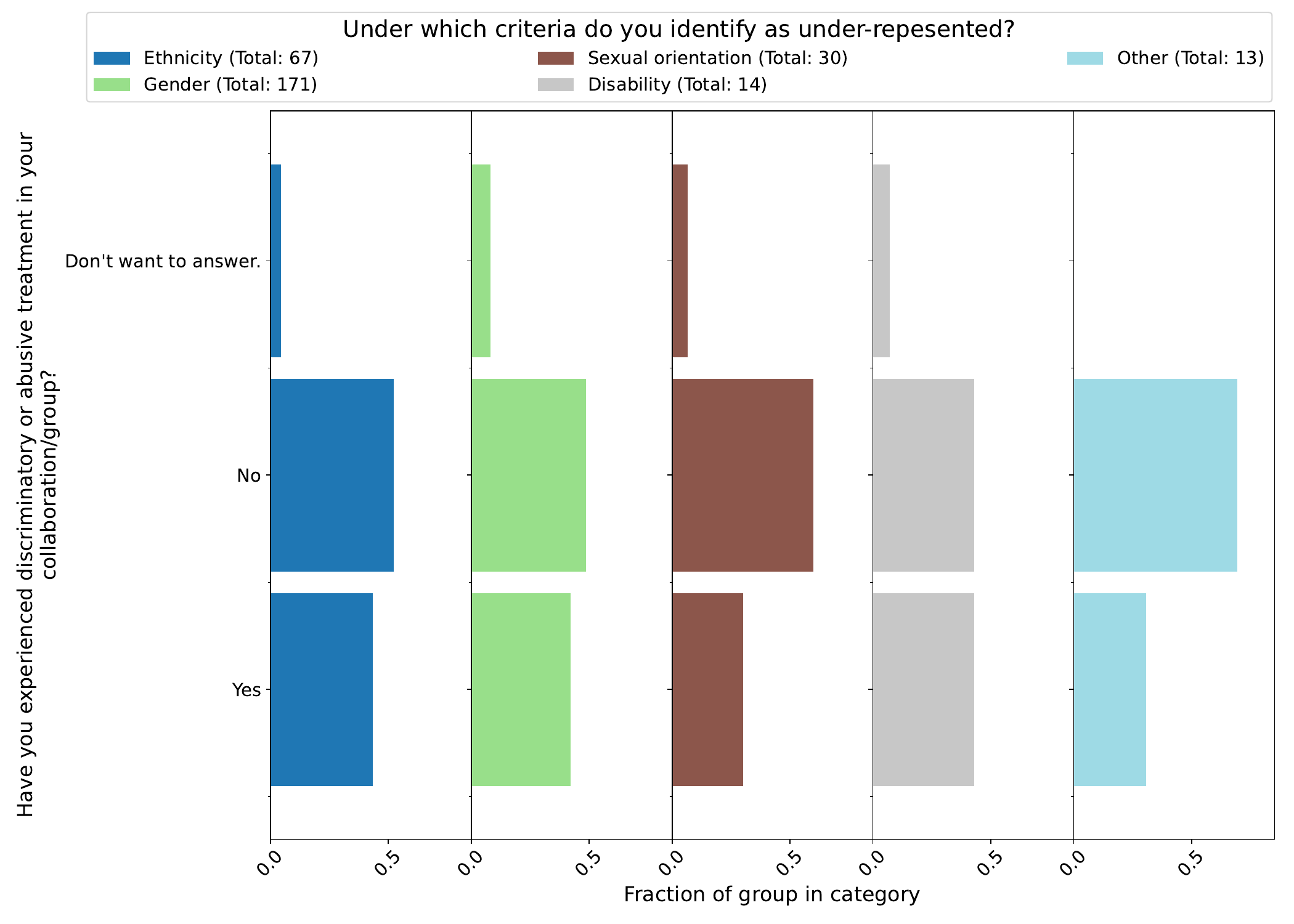}}
    \caption{(Q87 v Q4,7--10) Respondents' experience of discriminatory or abusive treatment in their collaboration/group correlated with demographics. Fractions are given out of all respondents who answered the questions.}
    \label{fig:part2:Q87vQ4Q7Q8Q9Q10}
\end{figure}

In Figure~\ref{fig:part2:Q87vQ4Q7Q8Q9Q10} we consider correlations between whether respondents have ever experienced discriminatory or abusive treatment in their collaboration/group, and demographics.
The responses are somewhat dependent on country of employment/nationality.
For example, we see that respondents employed in Central and Eastern Europe claim they have experienced less discriminatory/abusive treatment than respondents from the rest of the world.
Cisgender males, and younger respondents, are less likely to respond that have experienced discriminatory or abusive treatment.
Respondents who consider themselves as belonging to an under-represented group have experienced discriminatory or abusive treatment significantly more often than respondents who do not.
Within this category, gender, ethnicity and disability are more correlated with experiencing discriminatory treatment than sexual orientation.

\begin{figure}[ht!]
    \centering
        \subfloat[]{\label{fig:part2:Q88vQ3}\includegraphics[width=0.49\textwidth]{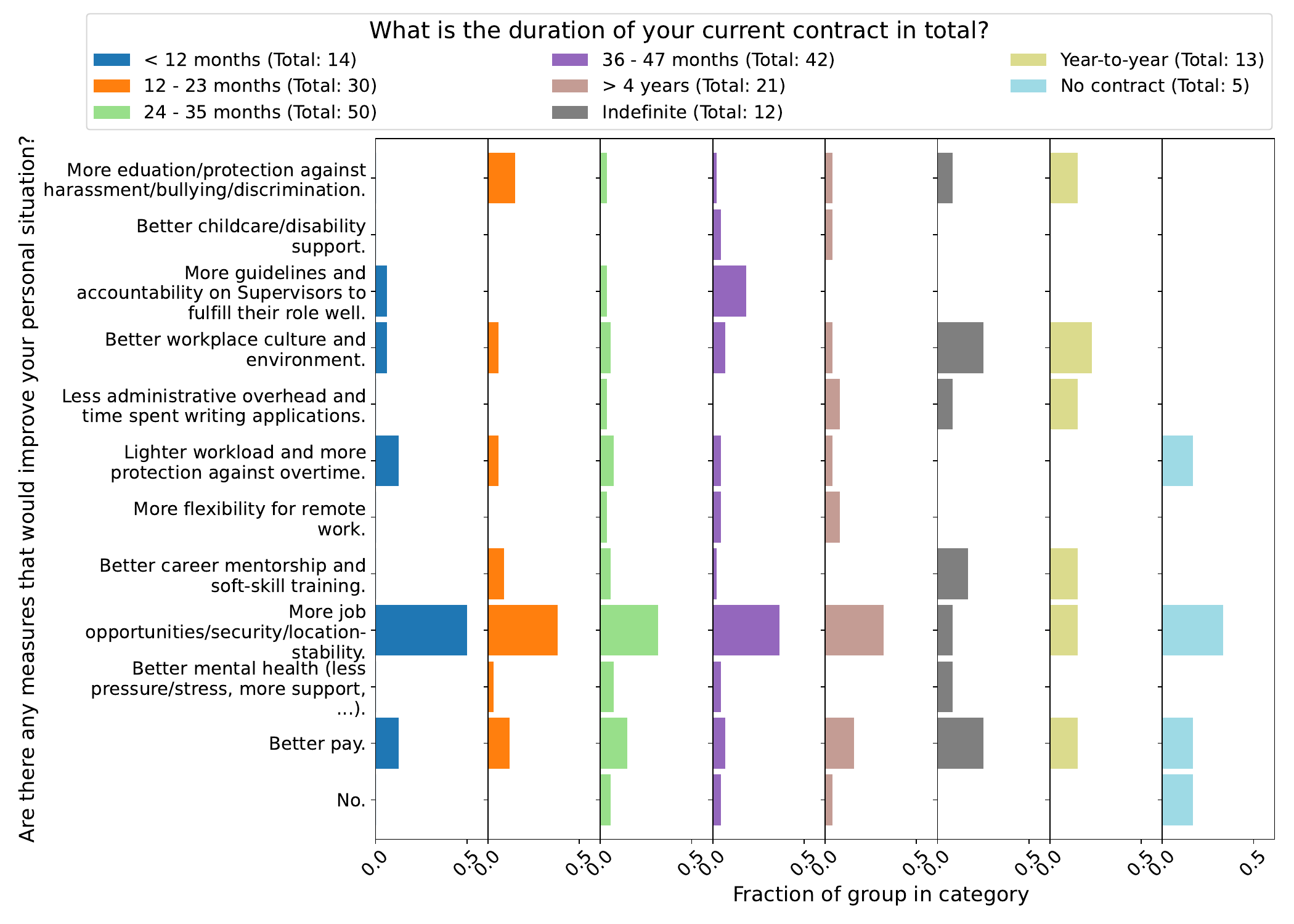}}
        \subfloat[]{\label{fig:part2:Q88vQ4}\includegraphics[width=0.49\textwidth]{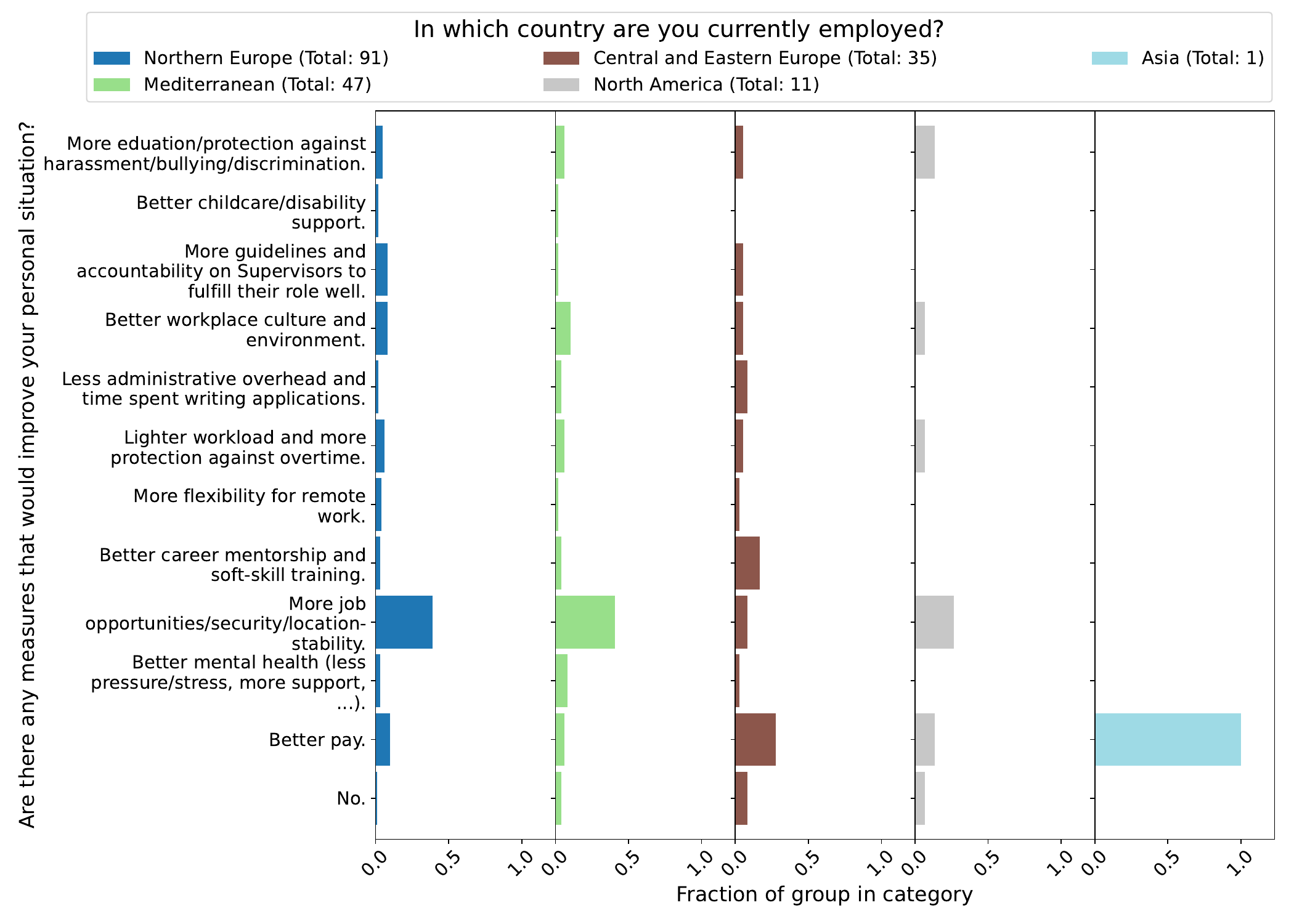}}\\
        \subfloat[]{\label{fig:part2:Q88vQ7}\includegraphics[width=0.49\textwidth]{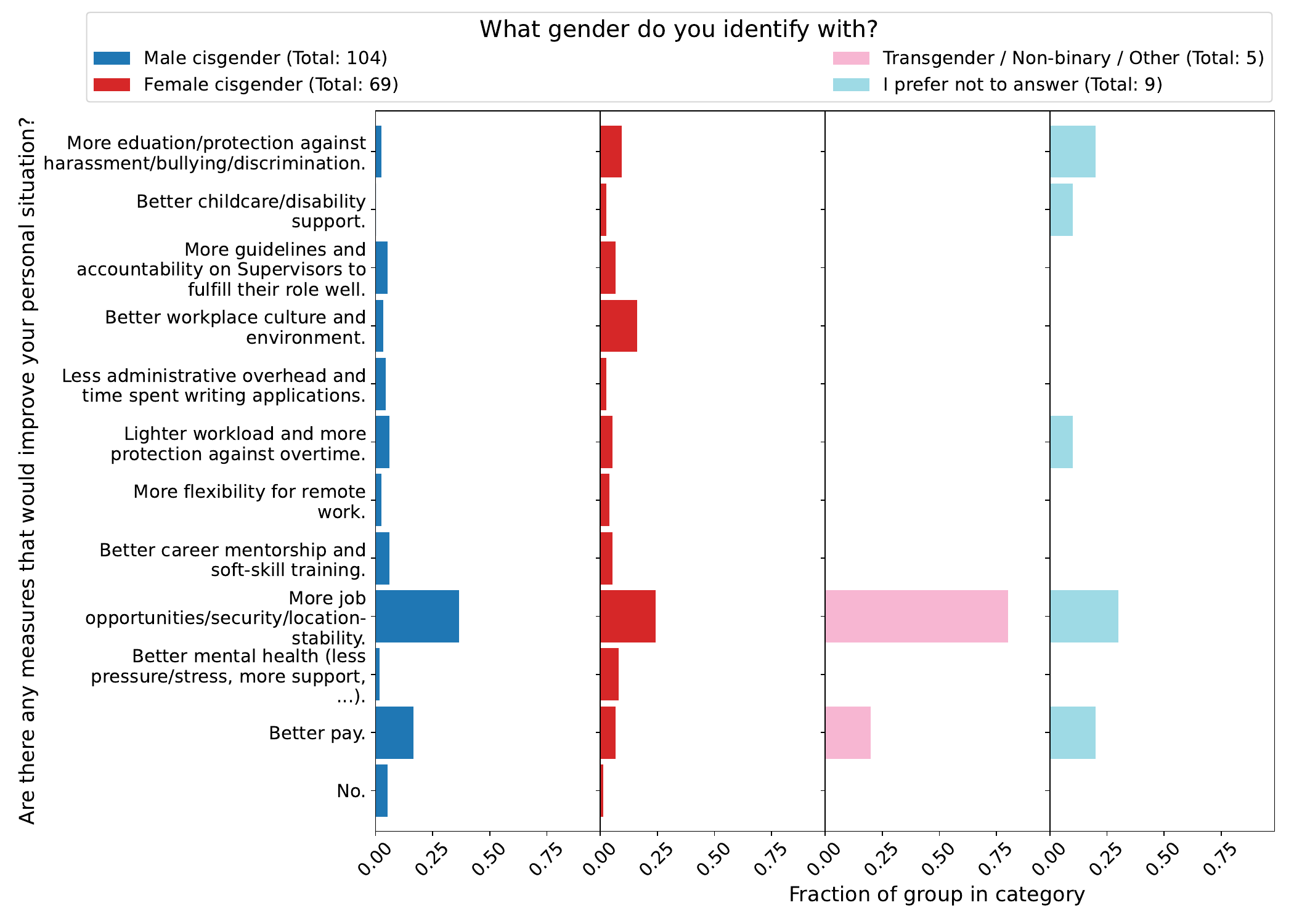}}
        \subfloat[]{\label{fig:part2:Q88vQ9}\includegraphics[width=0.49\textwidth]{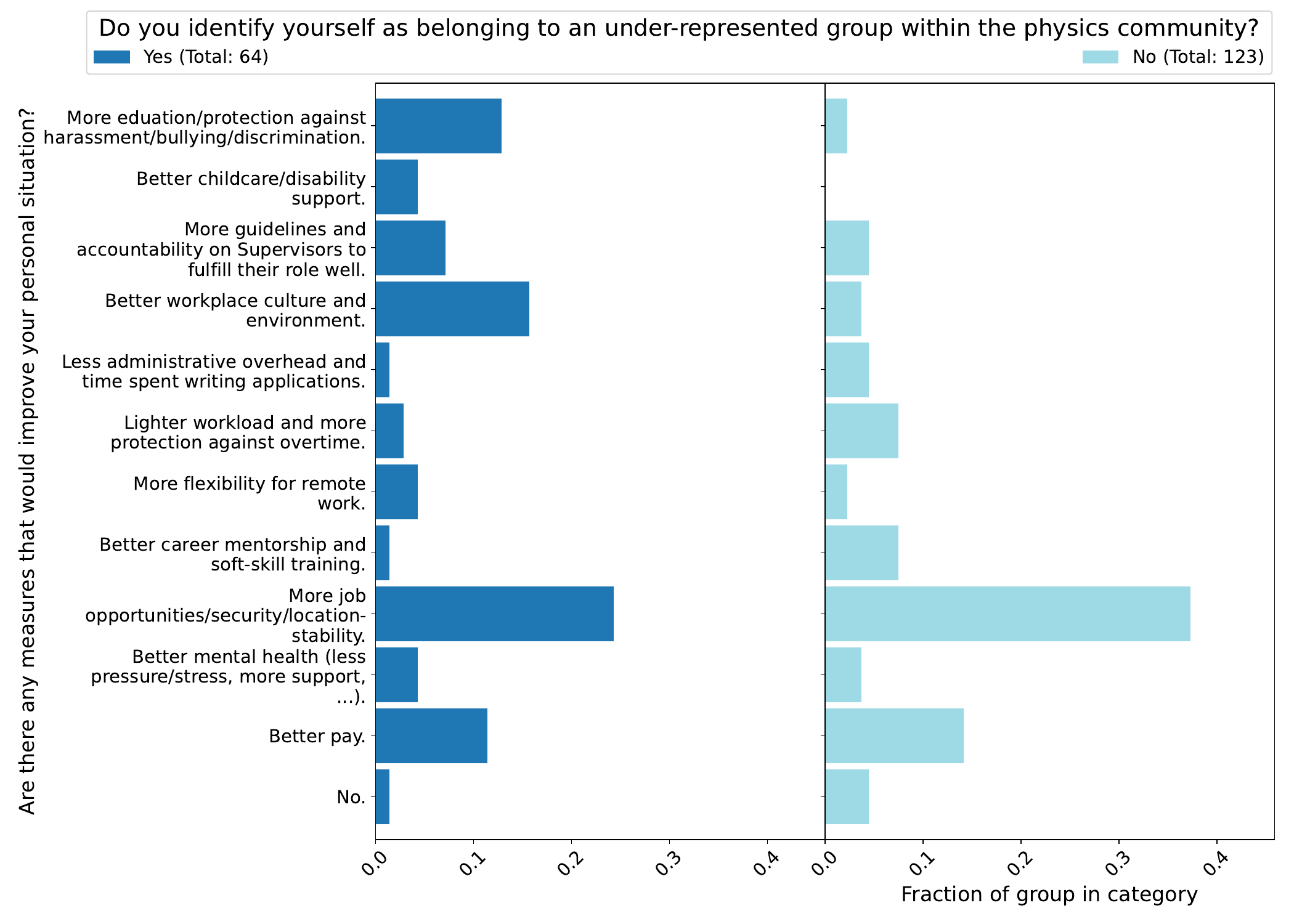}}
\caption{(Q88 v Q3,4,7,9) Answers to the (open) question on whether any measures would improve the personal situation of the participant correlated with some aspects of the participant profile and demographics. The free format answers of the participants were grouped in the categories after collecting the answers. Multiple answers were allowed per respondent. Fractions are given out of all respondents who answered the questions.}
    \label{fig:part2:Q88vsQ3Q4Q7Q9}
\end{figure}

Respondents were next asked in an open question if there are any measures that would improve their situation.
The answers were grouped into categories and correlated with respondent demographics in Figure~\ref{fig:part2:Q88vsQ3Q4Q7Q9}.
We see that for respondents with no permanent contract, more job opportunities, job security and job location-stability are more important than for respondents with indefinite contracts and are the leading factors that would improve their situation in their opinion.
For respondents with indefinite contracts better pay and better workplace culture become the most important factors.
We also see geographic correlations.
For example, better pay is the most important factor for respondents employed in Central and Eastern Europe or Asia, whereas more job opportunities, job security and job location-stability are the most important factors elsewhere.
The plots also show that more education/protection against harassment/bullying/discrimination and better workplace culture and environment are more important to cisgender female respondents than cisgender males, and more important to respondents belonging to an under-represented group than to respondents who don't.

\begin{figure}[ht!]
    \centering
    \includegraphics[width=0.6\textwidth]{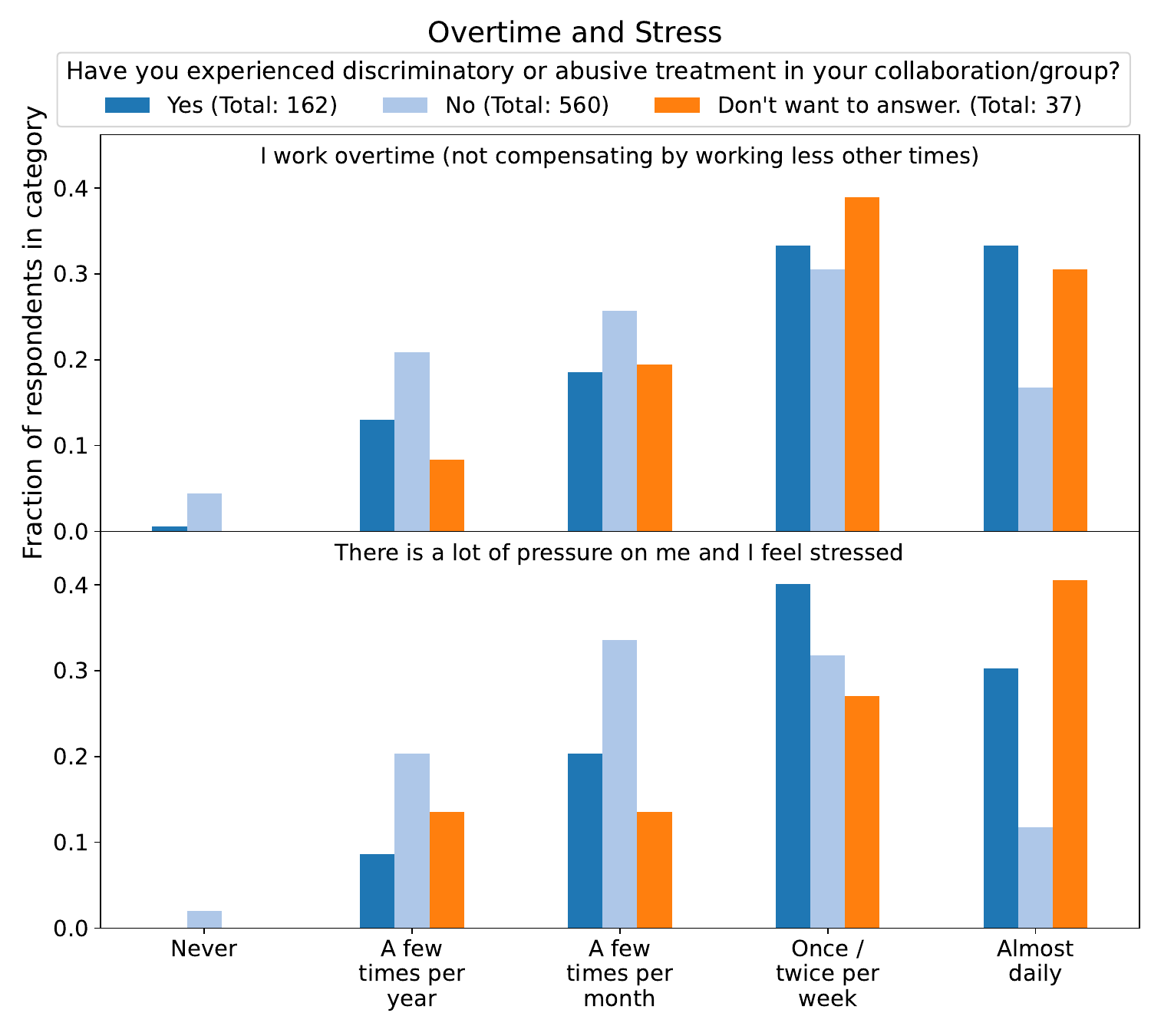}
    \caption{(Q79--80 v Q87) Correlations between how much overtime respondents do or how frequently they feel stressed and whether they've experienced abusive or discriminatory treatment at work.}
    \label{fig:part2:OvertimeStressvQ87}
\end{figure}

\begin{figure}[ht!]
    \centering
        \subfloat[]{\label{fig:part2:Q74fvQ87}\includegraphics[width=0.49\textwidth]{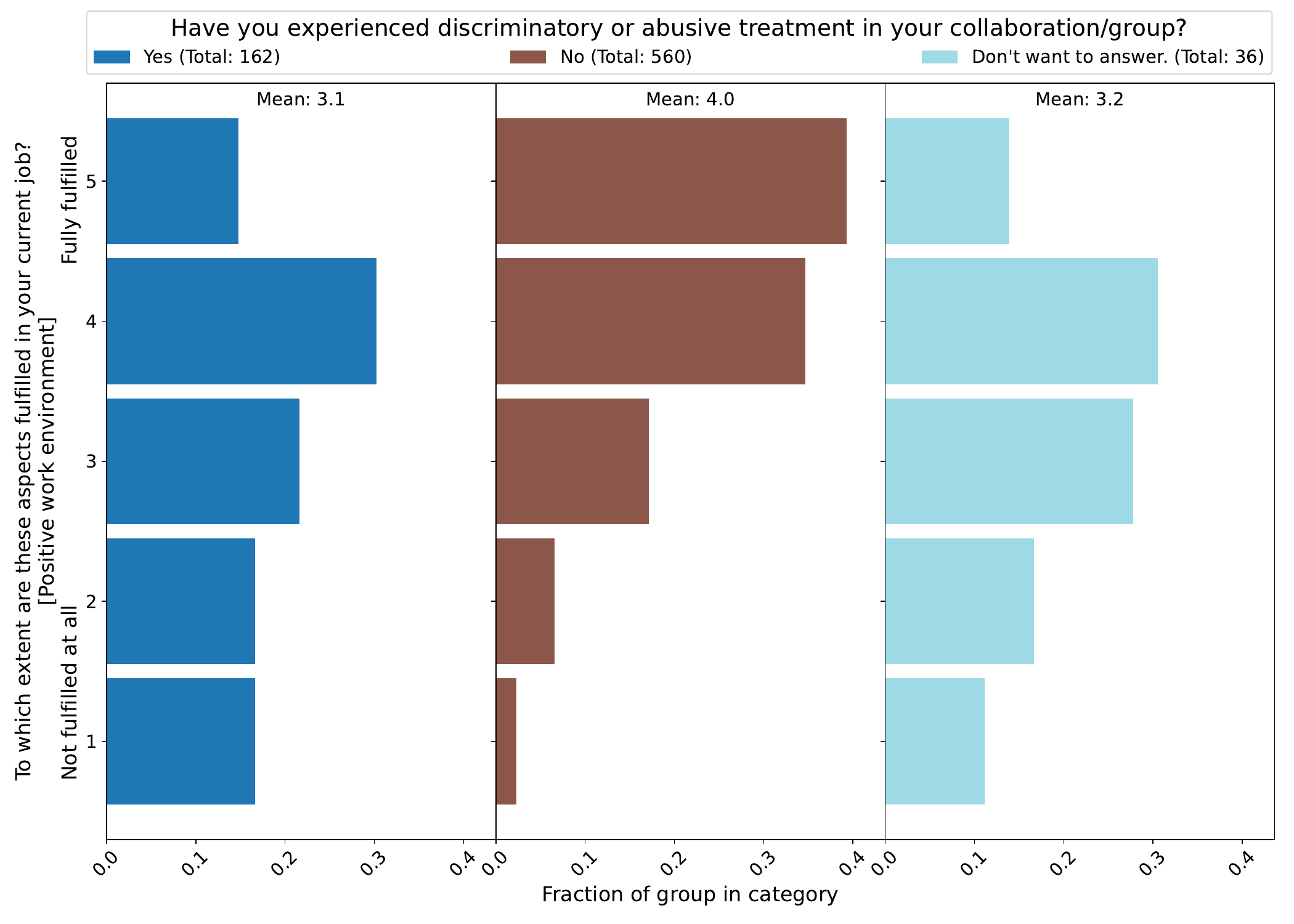}}
        \subfloat[]{\label{fig:part2:Q85vQ87}\includegraphics[width=0.49\textwidth]{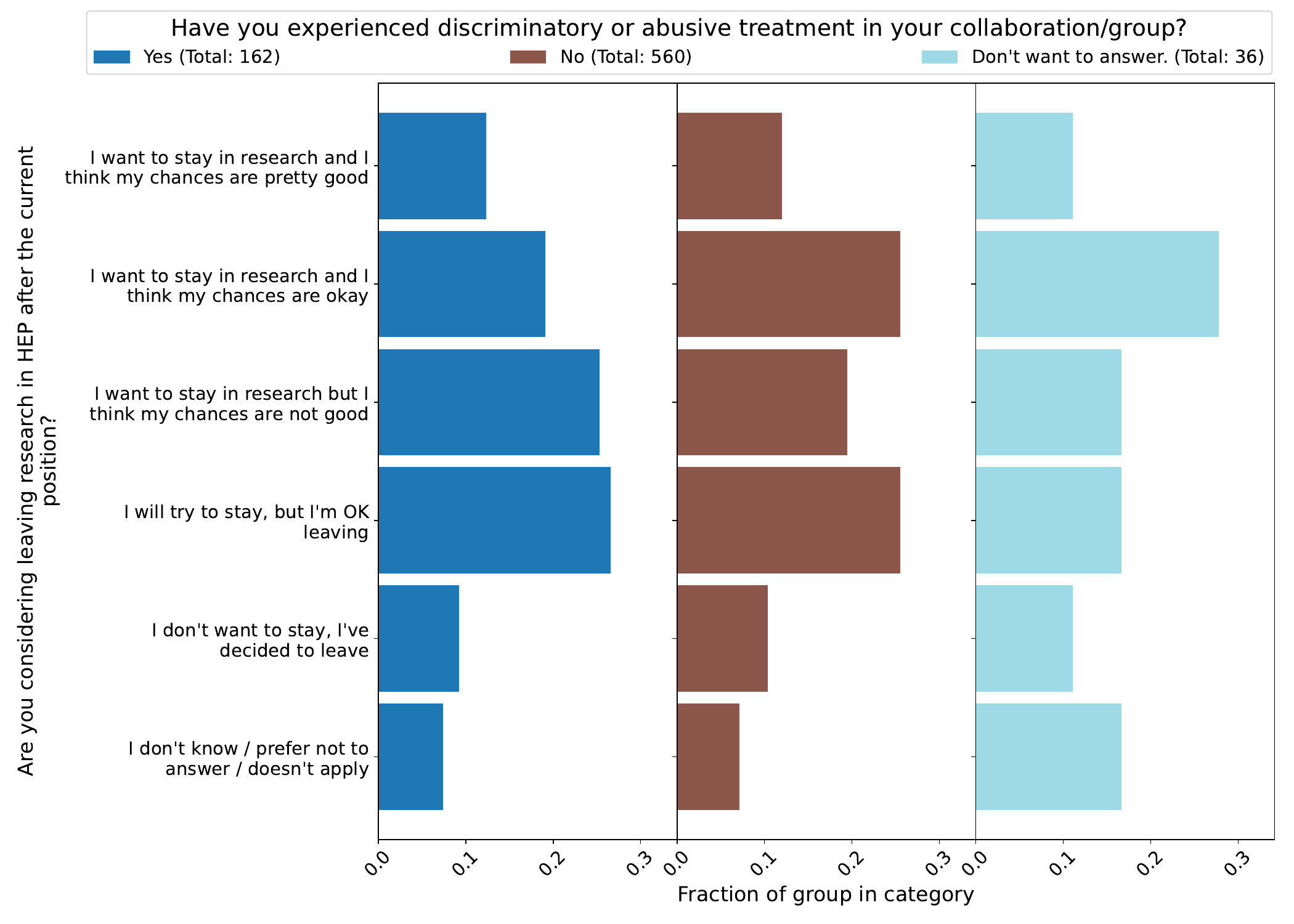}}\\
        \subfloat[]{\label{fig:part2:Q86vQ87}\includegraphics[width=0.49\textwidth]{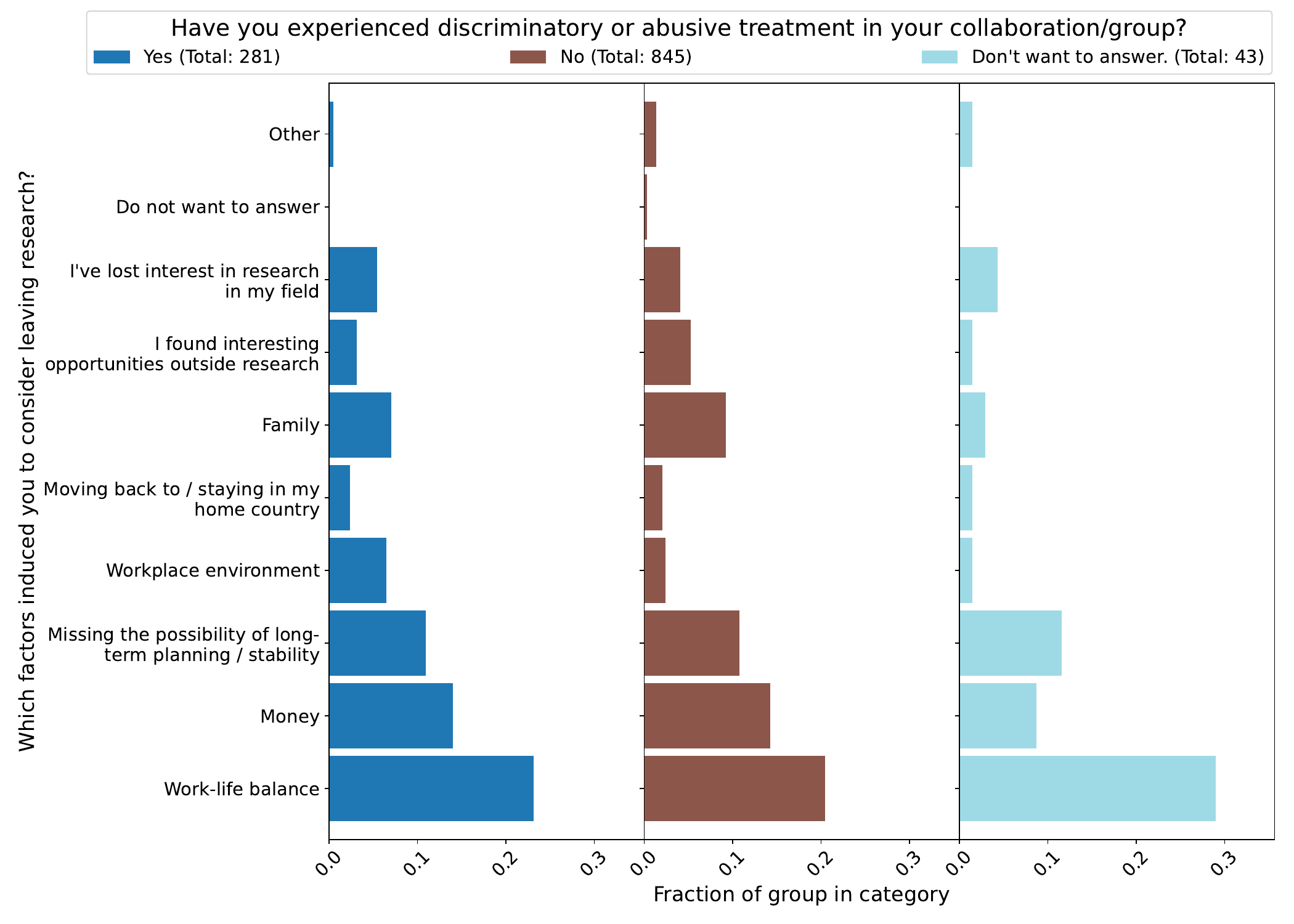}}
        \subfloat[]{\label{fig:part2:Q94vQ87}\includegraphics[width=0.49\textwidth]{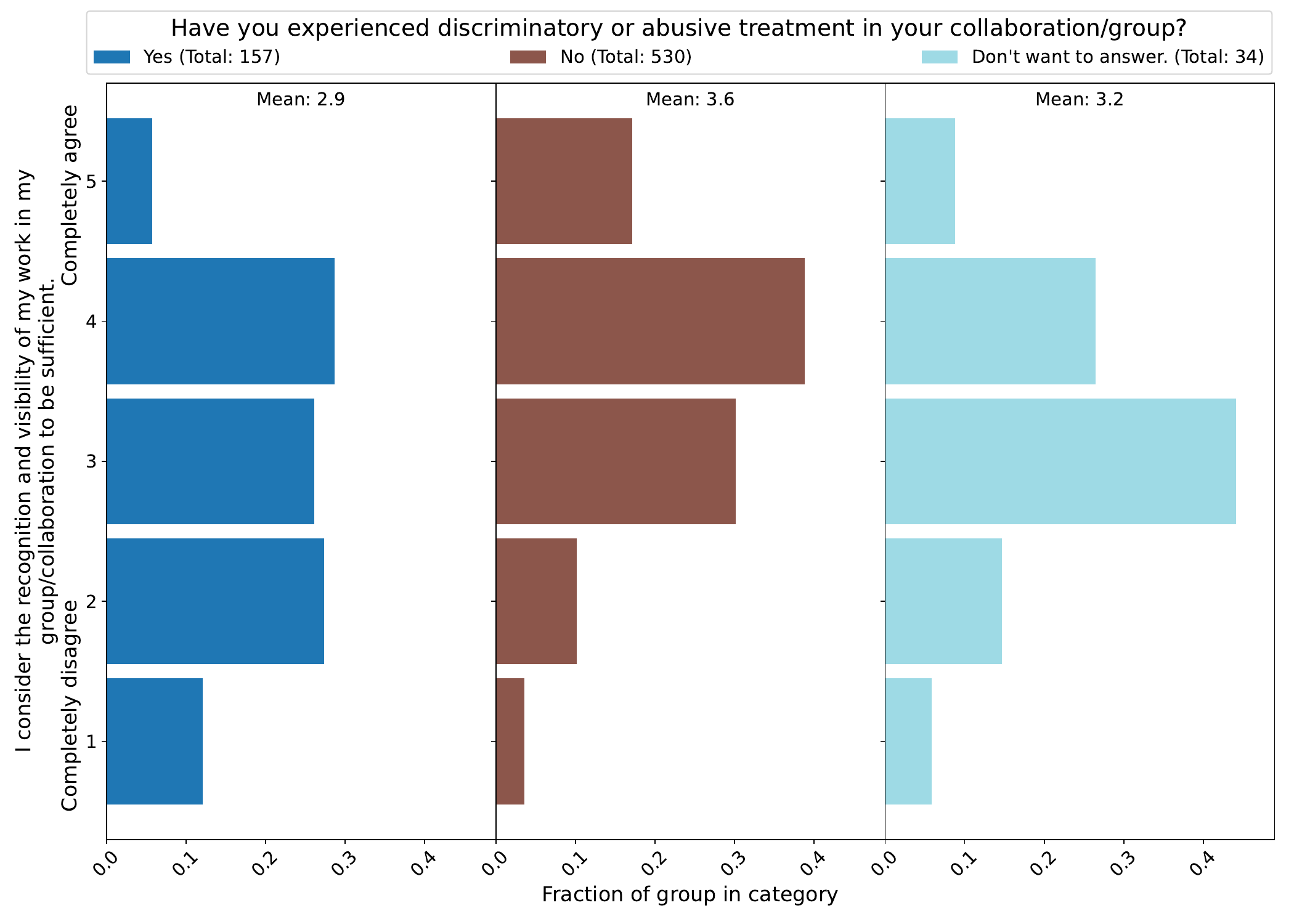}}
    \caption{(Q74f,85,86,94 v Q87) Correlations between respondents' answers to various questions and whether they've experienced abusive or discriminatory treatment at work. Fractions are given out of all respondents who answered the question.}
    \label{fig:part2:Q74Q85Q86Q94vQ87}
\end{figure}

We now move to correlations between this topic and questions unrelated to demographics.
In Figure~\ref{fig:part2:OvertimeStressvQ87} it is clear that for respondents who have received this treatment, the frequency of stress and working overtime is much higher than for other respondents.
Some more correlations found are shown in Figure~\ref{fig:part2:Q74Q85Q86Q94vQ87}.
We find that respondents who haven't received discriminatory or abusive treatment feel that their current job has a more positive work environment, and that the recognition and visibility of their work is sufficient.
Additionally, for respondents who want to stay in research after their current position, those who have received this treatment feel less confident about their chances.
However, we don't see a larger fraction of respondents who want to leave research.
This is consistent with the correlations seen with factors inducing respondents to leave research.
For those who have experienced discriminatory or abusive treatment, workplace environment and losing interest in the research are somewhat more important but ultimately work-life balance, money and lack of stability remain dominant.

%%%%%%%%%%%%%===================================================================================================
\FloatBarrier
\subsection{Recognition and visibility}

\begin{figure}[ht!]
    \centering
        \subfloat[]{\label{fig:part2:Q89vQ1}\includegraphics[width=0.49\textwidth]{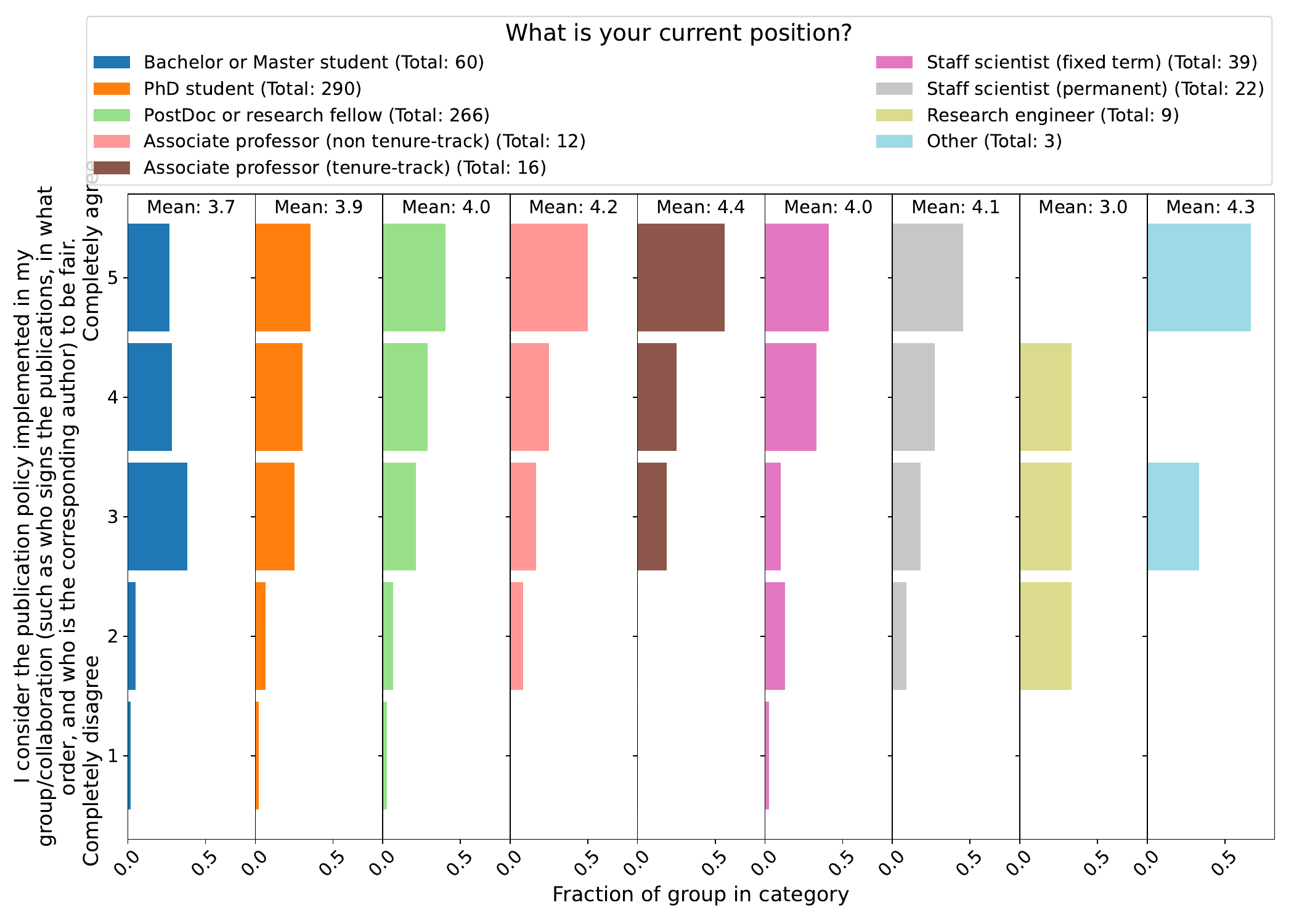}}
        \subfloat[]{\label{fig:part2:Q92vQ1}\includegraphics[width=0.49\textwidth]{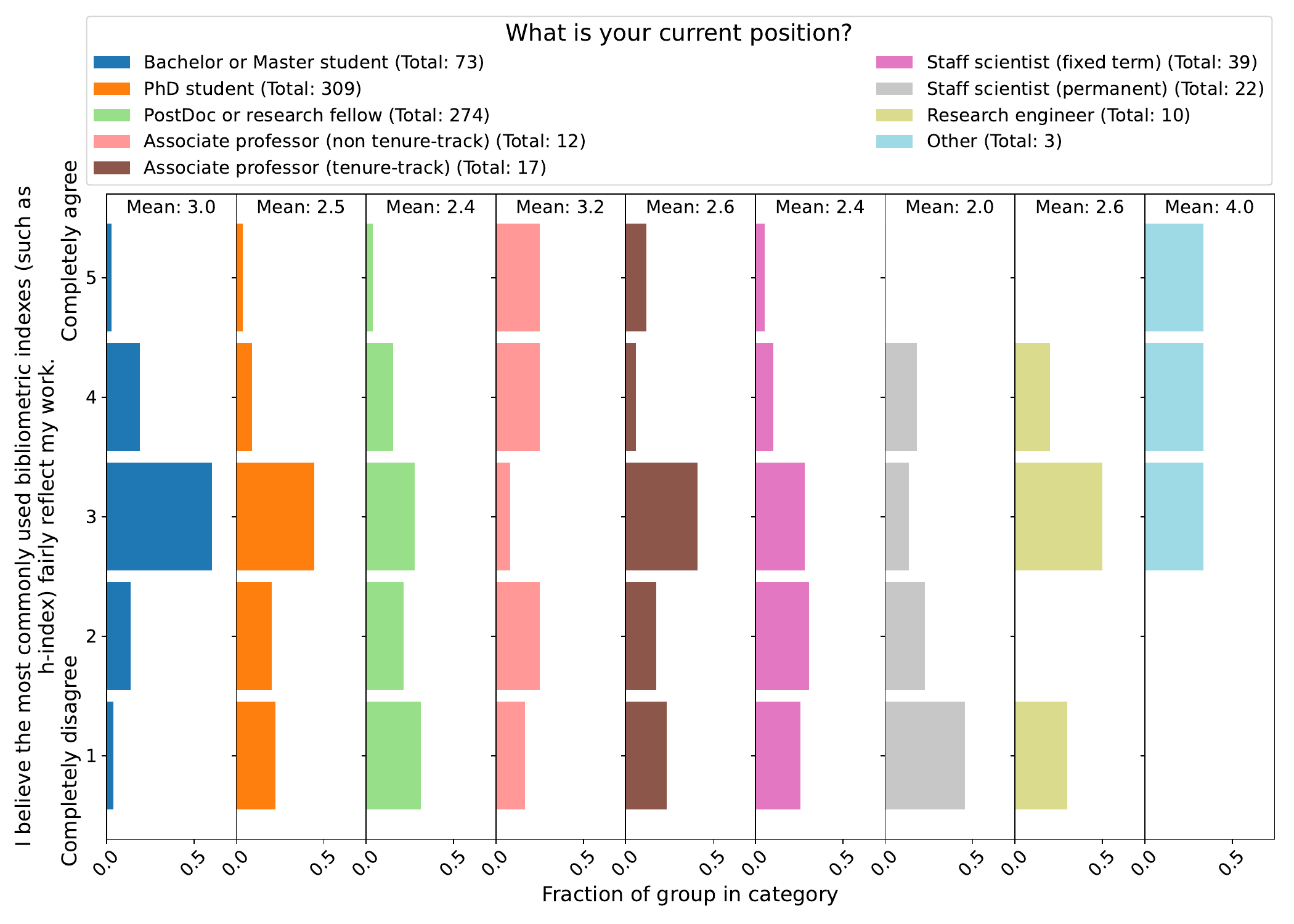}}\\
        \subfloat[]{\label{fig:part2:Q92vQ8}\includegraphics[width=0.49\textwidth]{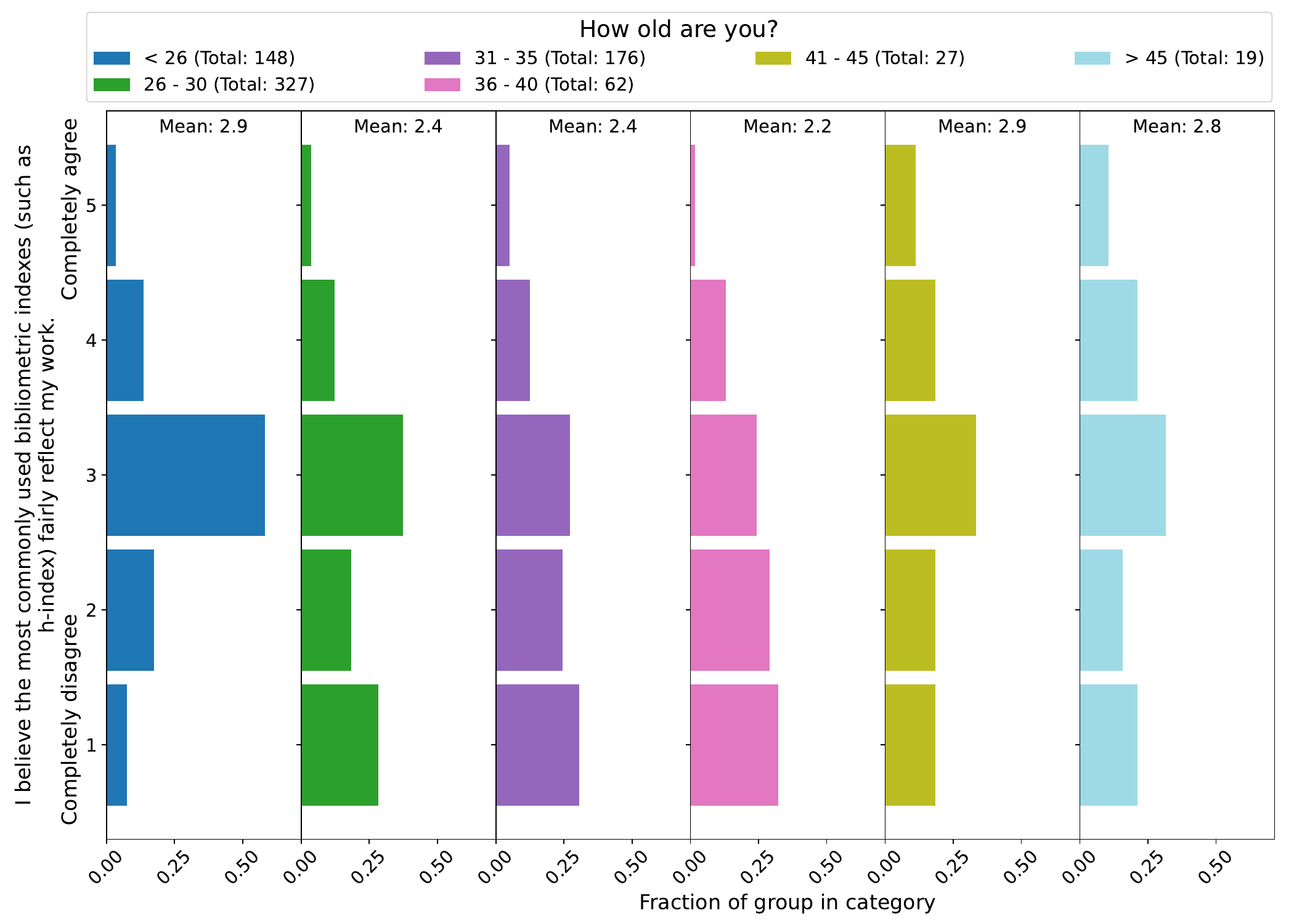}}
        \subfloat[]{\label{fig:part2:Q92vQ11}\includegraphics[width=0.49\textwidth]{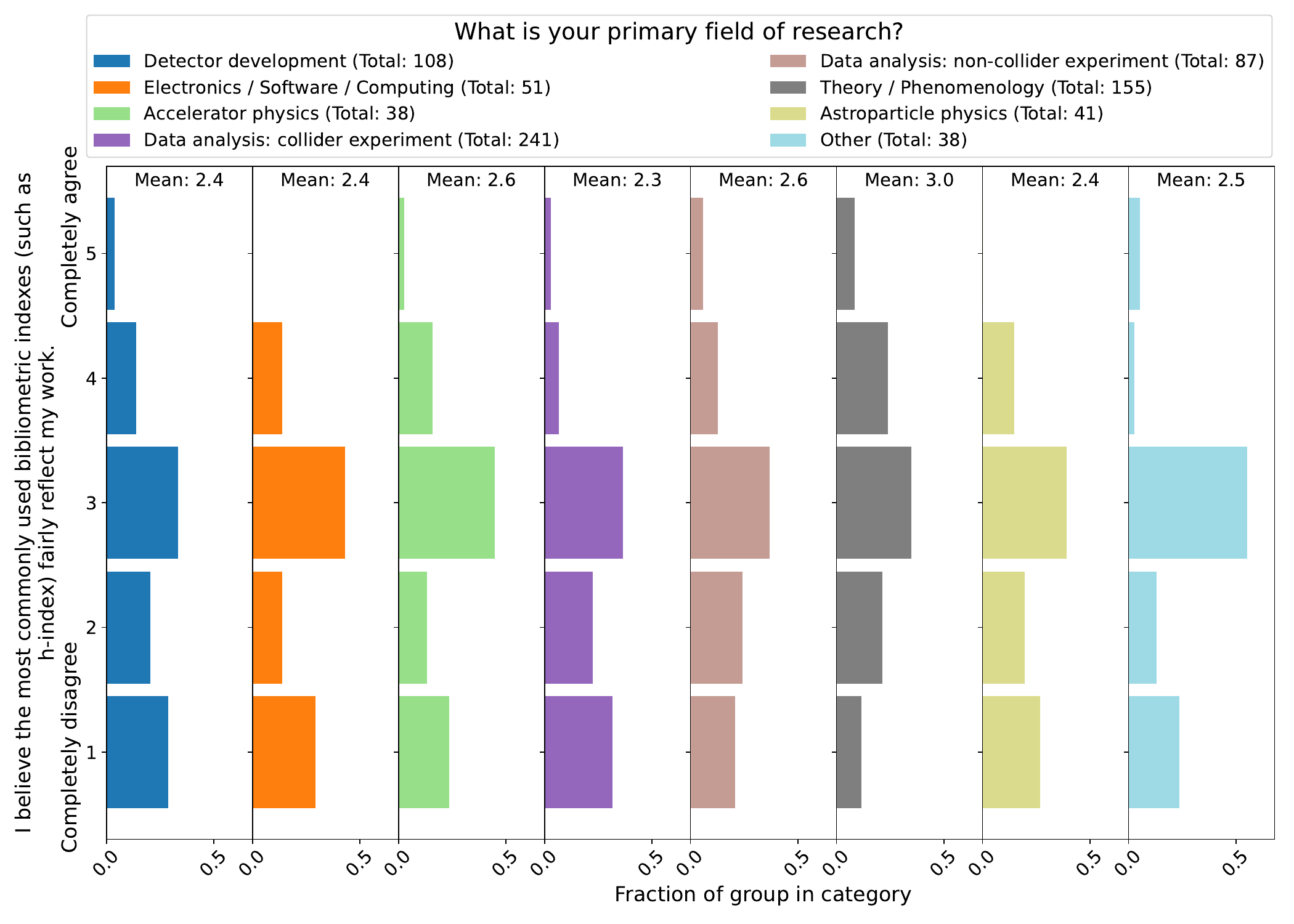}}\\
        \subfloat[]{\label{fig:part2:Q94vQ11}\includegraphics[width=0.49\textwidth]{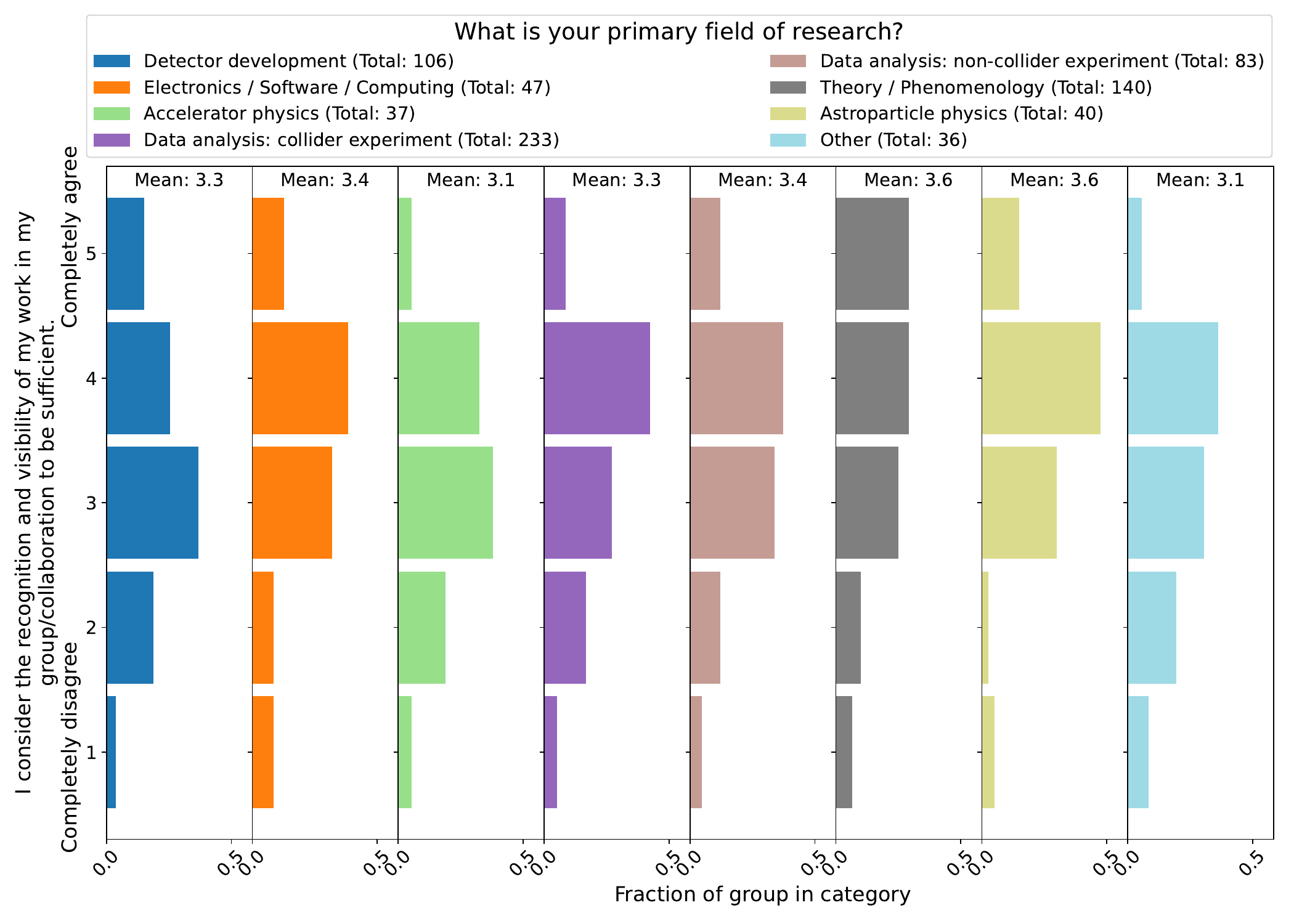}}
        \subfloat[]{\label{fig:part2:Q95vQ11}\includegraphics[width=0.49\textwidth]{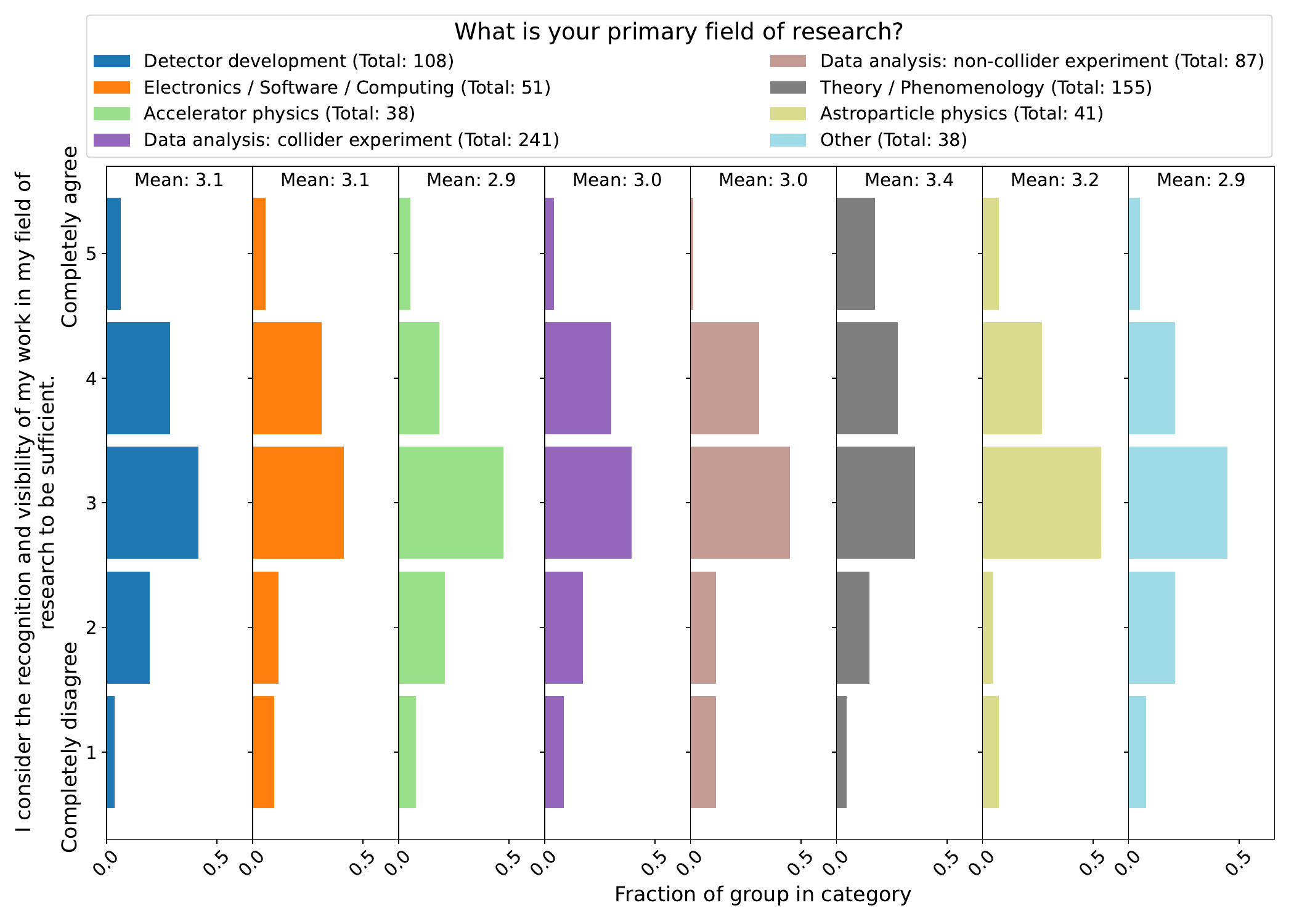}}
    \caption{(Q89 v Q1; Q92 v Q1,8,11; Q94,95 v Q11) Correlations between respondents views on fairness and recognition, and selected demographics. Fractions are given out of those who answered the questions.}
    \label{fig:part2:Q89vQ1_Q92vQ1Q8Q11_Q94Q95vQ11}
\end{figure}

\begin{figure}[ht!]
    \centering
    \includegraphics[width=0.7\textwidth]{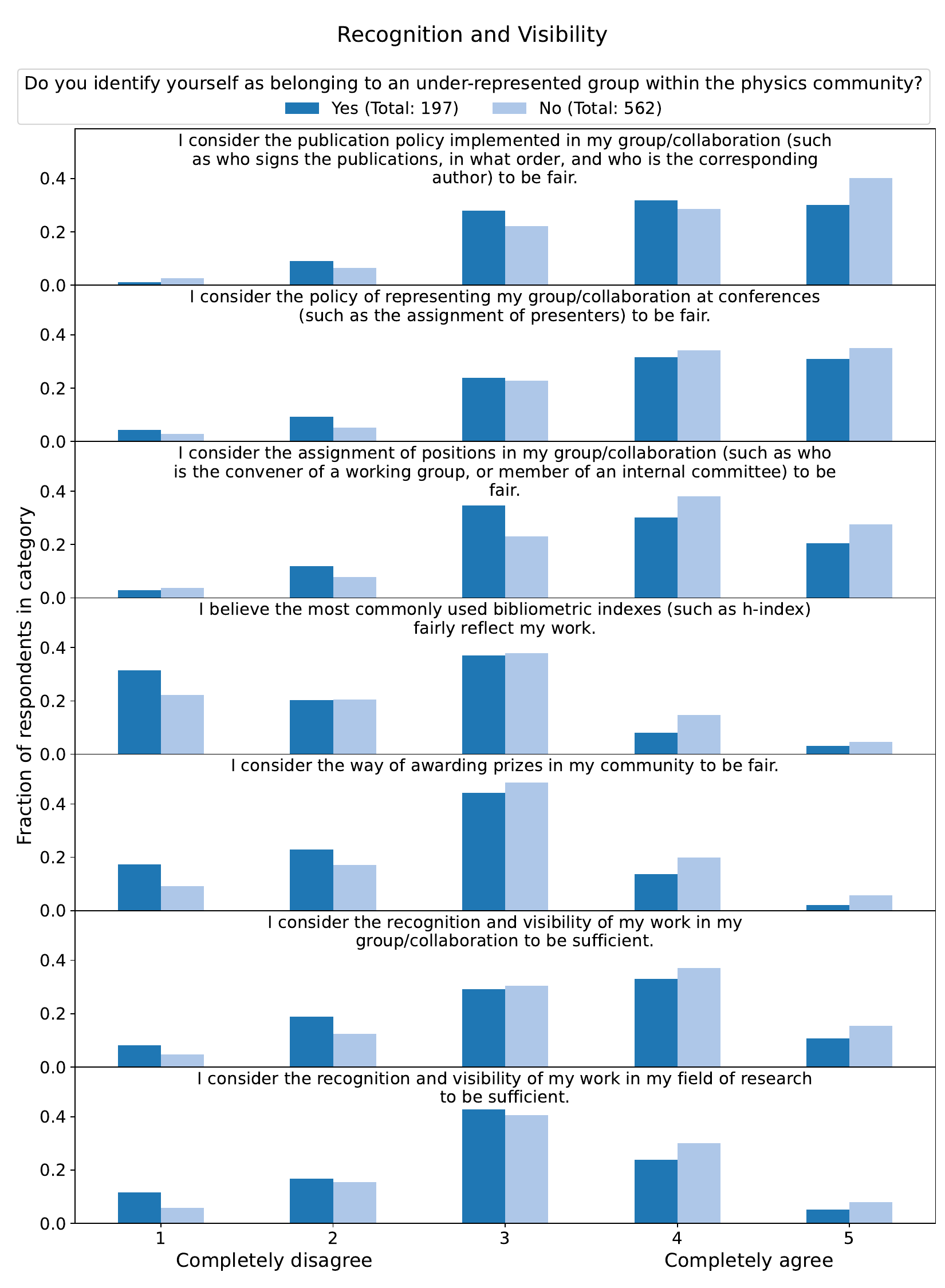}
    \caption{(Q89--95 v Q9) Correlations between respondents views on fairness and recognition, and whether they identify as belonging to an under-represented group. Fractions are given out of those who answered the questions.}
    \label{fig:part2:FairnessRecognitionvQ9}
\end{figure}

Moving to respondents views on their recognition and visibility, interesting correlations are shown in Figures~\ref{fig:part2:Q89vQ1_Q92vQ1Q8Q11_Q94Q95vQ11}--\ref{fig:part2:FairnessRecognitionvQ9}.
Correlations with respondent demographics are similar when considering views on the fairness of group/collaboration policy on publications and conference talks.
Respondents are more positive about these if they have more senior positions, and longer contracts (not shown).
Furthermore, views on these questions and others in the same category are consistently more negative for respondents who identify as belonging to an under-represented group within the physics community, as shown in Figure ~\ref{fig:part2:FairnessRecognitionvQ9}.
There is little correlation between views on fairness and recognition, and other properties of respondent demographics.

Considering instead respondent's view on the fairness of common bibliometric indices to reflect their work, we see strong demographic dependence is again in relation to age and position.
Students are neutral about these indices, and as the position becomes more senior the indices are generally viewed as less fair.
Opinion also becomes more negative with age, up until the 40s where it becomes more neutral.
In addition, respondents working in theory or phenomenology are less negative about the fairness of these indices.
Respondent's views on the fairness of how prizes are awarded in their community correlates very similarly to views on bibliometric indices (not plotted).

When asked how strongly respondents agree that their recognition and visibility within their group or collaboration is sufficient, responses were generally not strongly correlated with demographics.
Two exceptions to this are through field of research, where astroparticle physicists and theorist/phenomenologists agree more strongly than respondents working in other fields.
In contrast, when considering recognition within their whole field, astroparticle physicists are more neutral.

\begin{figure}[ht!]
    \centering
    \includegraphics[width=0.7\textwidth]{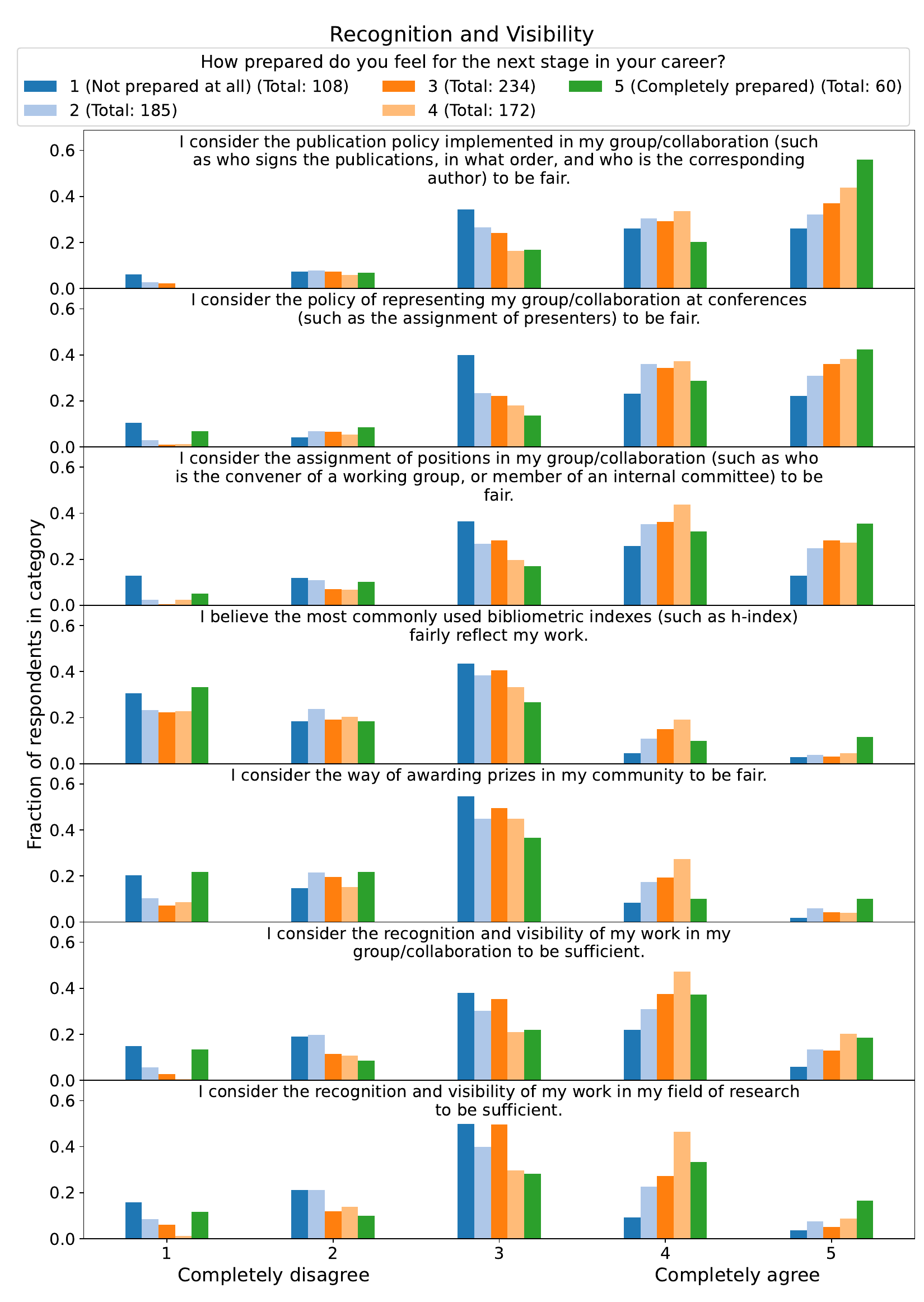}
    \caption{(Q89--95 v Q64) Correlations between respondents views on fairness and recognition, and how prepared they feel for their next career stage. Fractions are given out of those who answered the questions.}
    \label{fig:part2:FairnessRecognitionvQ64}
\end{figure}

\begin{figure}[ht!]
    \centering
    \includegraphics[width=0.7\textwidth]{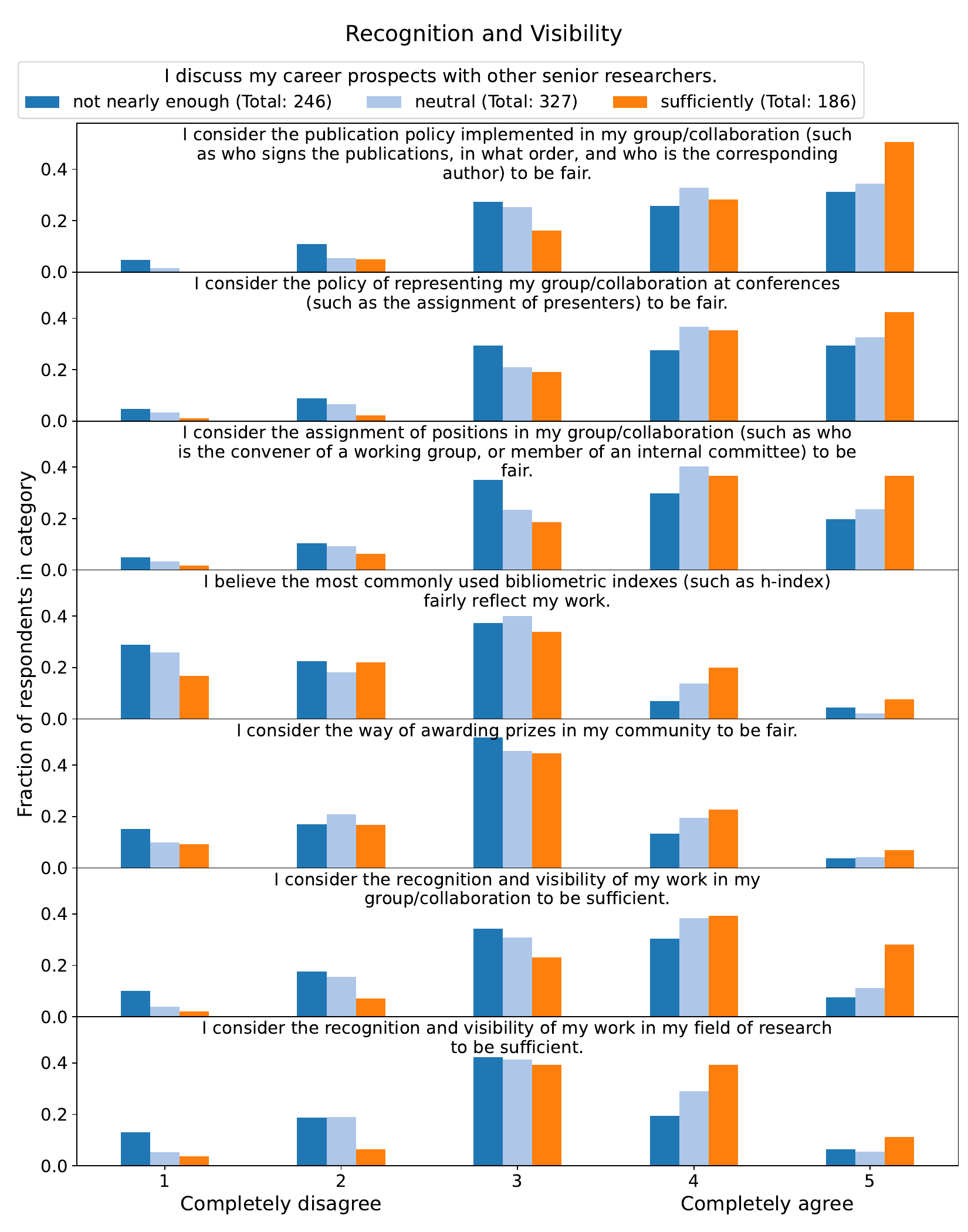}
    \caption{(Q89--95 v Q66) 
     Correlations between respondents views on fairness and recognition, and how sufficient their discussion with senior researchers is. Fractions are given out of those who answered the questions.}
    \label{fig:part2:FairnessRecognitionvQ66}
\end{figure}

\begin{figure}[ht!]
    \centering
    \includegraphics[width=0.7\textwidth]{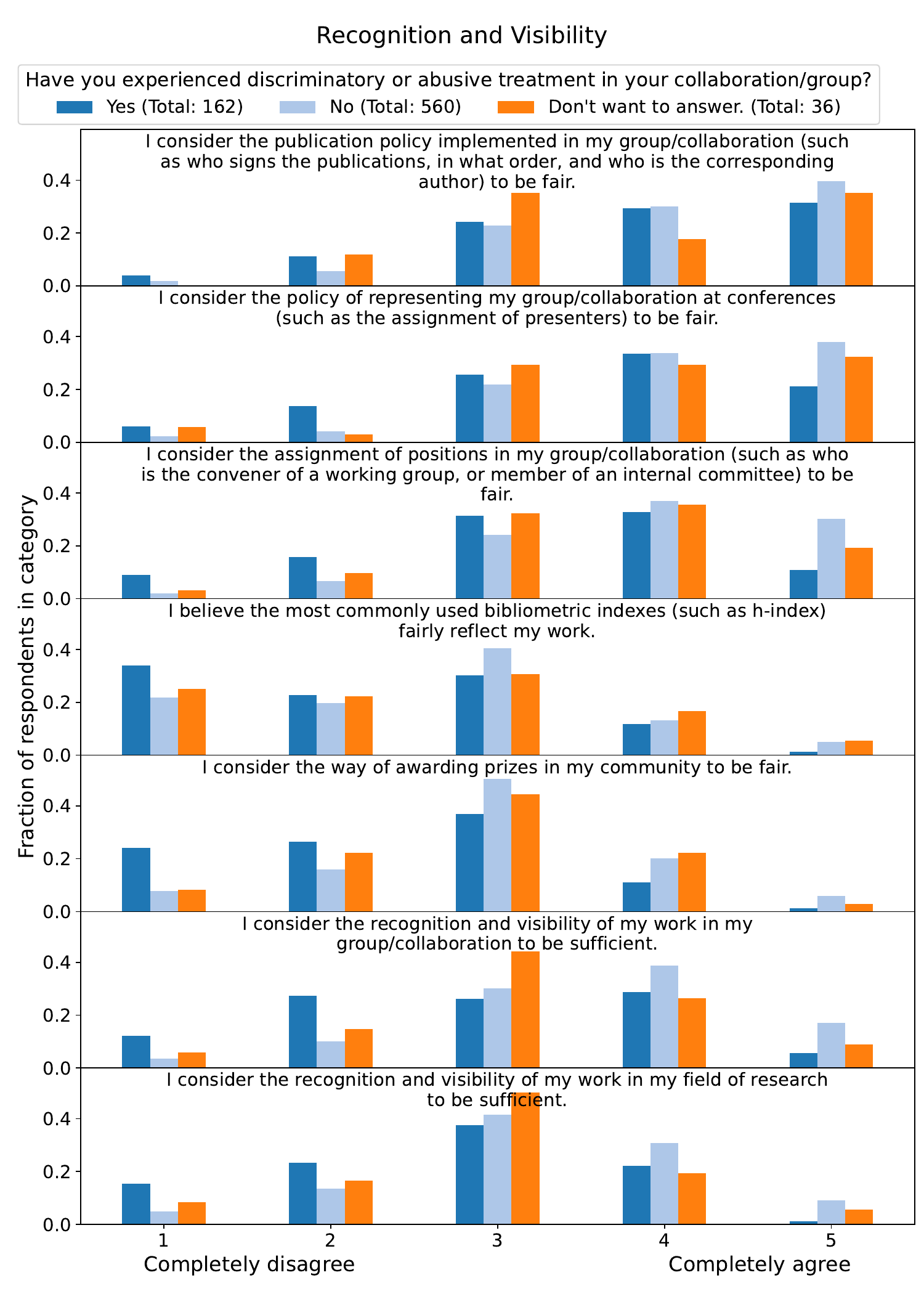}
    \caption{(Q89--95 v Q87) Correlations between respondents views on fairness and recognition, and whether they have experienced discriminatory or abusive treatment in their group/collaboration. Fractions are given out of those who answered the questions.}
    \label{fig:part2:FairnessRecognitionvQ87}
\end{figure}

We now consider correlations between respondents' views on recognition and visibility, and other questions not related to demographics.
In Figure \ref{fig:part2:FairnessRecognitionvQ64} we find a strong positive correlation between how fair respondents consider various policies and metrics to be and how sufficient their work's recognition is, and how prepared they feel for the next stage in their careers.
From Figure~\ref{fig:part2:FairnessRecognitionvQ66}, we see a strong positive correlation between how much respondents discuss their career prospects with senior researchers, and how positive they feel about their recognition and visibility, representation policies, and bibliometric indices.
This correlation is seen, though more weakly, when considering instead discussion with respondents' supervisor or peers (not shown).
In Figure~\ref{fig:part2:FairnessRecognitionvQ87} it is shown that respondents who have experienced discriminatory or abusive treatment in their group/collaboration instead feel much less positive about these statements.
In contrast, there is very little difference in feelings about recognition between respondent's who have or haven't changed field, or who have or haven't taken a career break of more than three months (not shown).

%%%%%%%%%%%%%%%%%%%%%%%%%%%%%%%%%%%%%%%%%%%%%%%%%%%%%%%%%%%%%%%%%%%%%%%%%%%%%%%%%%%%%%%%%%%%%%%%
\clearpage
\section{Conclusions and community recommendations}
\label{sec:conclusion}

The responses to two final open-box questions in the survey, in conjunction with the rest of the analysis, were used to produce a set of recommendations for the ECFA ECR Panel and the particle physics community.
One open question, which collected over 130 responses, asked for respondents' ideas for any event that the ECFA Early-Career Researchers Panel could organise to help them develop their careers.
The other open question asked for feedback on the survey.

We recommend that the ECFA ECR panel considers the following activities in their future plans:
\begin{itemize}
    \item Existing resources that list job openings and technical/soft-skill training opportunities should be better documented and publicised amongst the HEP community. Discussions should be had on how to better encourage supervisors/PIs to support ECRs pursuing training and career-development outside their direct line of work.
    \item Identify what kinds of soft-skill and career-development training opportunities are not already readily widely available, and help to organise events to provide them.
    \item Publish statistics on the profiles of people who have successfully obtained permanent positions in HEP academia recently, to ensure that ECR perceptions of what skills and experiences are useful to achieve this goal are accurate.
    \item Host an event focused on jobs in industry, inviting people who were previously involved in HEP and have since found success in other careers to discuss their work and the skills needed to transition.
\end{itemize}

More broadly, we recommend that members of the HEP community all consider the following:
\begin{itemize}
    \item Foster a culture of open discussion of career prospects and challenges with colleagues at all levels.
    \item Try to provide longer job contracts wherever possible.
    \item Try to avoid fostering a culture where professional mobility is a prerequisite for a successful career.
    \item Support flexible working hours and locations where feasible.
    \item Provide more assistance (administrative, housing, childcare support etc.) for people moving to another city or country for a new position. 
    \item Avoid a work culture where unpaid overtime work is required or encouraged.
    \item As a supervisor/PI, provide career mentorship and a supportive work environment, beyond technical aspects of the research project.
    \item Provide training and support where needed to allow supervisors and line-managers to act as better mentors and provide a supportive work environment.
    \item Increase awareness, and provide training, to support a workplace that encourages good mental health, and is free from discriminatory treatment or harassment at all levels.
    \item Support research assessment beyond traditional bibliometric indices, such as the `Declaration of Research Assessment' proposes~\cite{DORA}. For example, using narrative CVs for job applications, and improving collaboration/group policies to improve recognition of service work.
    \item Organise national events where ECRs and non-ECRs can discuss issues facing ECRs, such as academic funding issues and working culture. An example of a similar `Town Hall' event hosted in the UK can be found in Ref.~\cite{UKFCTownhall}.
\end{itemize}

For anyone aiming to produce a similar survey in the future, we have the following suggestions for developments beyond this iteration.
\begin{itemize}
    \item Survey length should be minimised for a given goal, and a predicted time needed to fill the survey should be provided to prospective respondents.
    \item Consider sectioning the survey such that relevant parts could be skipped by a respondent for whom it isn't applicable (for example different sections targeting PhD student or PostDoc respondents).
    \item To compliment this survey, more focus could be placed in the future on topics related to:
        \begin{itemize}
            \item ECR diversity;
            \item experiences of ECRs in under-represented communities and specific challenges faced with regards to work-life balance and career development;
            \item ECRs' mental health;
            \item the impact of socio-economic background on the ECRs' opportunities;
            \item ECR views on career prospects and the future within the context of future experiments and research programmes within HEP;
            \item the kinds of discrimination, bullying and harassment witnessed by ECRs as well as experienced personally by them;
            \item differences between ECR and non-ECR views.
        \end{itemize} 
\end{itemize}

We conclude from the survey analysis that ECRs in HEP research face challenges in many areas, alongside positive experiences and outlooks in others.
It is clear that the lack of long-term planning and stability caused by fixed-term jobs and frequent relocation is overall the biggest concern ECRs face, which is not unexpected.
We hope that this report quantifies, and raises awareness of, ECR concerns throughout the HEP community and to our more senior members in particular, such that we can start to improve academic culture over time.
The results obtained in this survey related to work-life balance could be compared to analysis performed in Ref~\cite{Janssens:2023ufo}.
Finally, we would like to sincerely thank everyone who took the time to fill out the survey, and made this report possible.

%%%%%%%%%%%%%%%%%%%%%%%%%%%%%%%%%%%%%%%%%%%%%%%%%%%%%%%%%%%%%%%%%%%%%%%%%%%%%%%%%%%%%%%%%%%%%%%%
\appendix
\FloatBarrier

\section{List of questions}
\label{app:questions}

This appendix contains the list of questions (or statements) presented in the survey. The questions were grouped into topics which reflect their subject matter and the structure of this document. This is indicated here in bold.
Most of the questions were obligatory, with one answer allowed per respondent. Deviations from this set-up are indicated below.\\
\newcounter{Counter-ListOfQuestions}
\vspace{0.5cm}

\noindent \textbf{Demographics of respondents}
\begin{enumerate}
\item What is your current position? (Figure~\ref{fig:part1:Q1})
\item What is your current affiliation? (Figure~\ref{fig:part1:Q2})
\item What is the duration of your current contract in total? (Figure~\ref{fig:part1:Q3})
\item In which country are you currently employed? (Figure~\ref{fig:part1:Q4})
\item In which country do you reside? (Figure~\ref{fig:part1:Q5})
\item What is your nationality? (Figure~\ref{fig:part1:Q6})
\item What gender to you identify with? (Figure~\ref{fig:part1:Q7})
\item How old are you? (Figure~\ref{fig:part1:Q8})
\item Do you identify yourself as belonging to an underrepresented group within the physics community?
\item Under which criterion do you identify as under-represented? (Figure~\ref{fig:part1:Q10}) (voluntary, multiple answers per respondent allowed)
\setcounter{Counter-ListOfQuestions}{\value{enumi}}
\end{enumerate}

\noindent \textbf{Field of work}
\begin{enumerate}
\setcounter{enumi}{\value{Counter-ListOfQuestions}}
\item What is your primary field of research? (Figure~\ref{fig:part1:Q11})
\item Are there any other fields of research you are significantly involved in? (Figure~\ref{fig:part1:Q12}) (multiple answers per respondent allowed)
\item In which stage is the experiment you are working in? (Figure~\ref{fig:part1:Q13}) (voluntary)
\item Do you work within a collaboration or/and a research group? (Figure~\ref{fig:part1:Q14})
\setcounter{Counter-ListOfQuestions}{\value{enumi}}
\end{enumerate}

\noindent \textbf{Work within a research Group}\\
Only respondents who are a member of a research group were invited to answer questions from this part of the survey.

\begin{enumerate}
\setcounter{enumi}{\value{Counter-ListOfQuestions}}
\item What is the size of your research group? If you are part of multiple research groups, please consider the one you mostly work for. (Figure~\ref{fig:part1:Q15})
\item Within your research group, how many people do you actively work with during a normal week? (Figure~\ref{fig:part1:rgroup_you_work_with})
\item My work in the research group is useful to improve my knowledge, skills and expertise. (Figure~\ref{fig:part1:rgroup_aspects_of_work}) (voluntary)
\item There is room for me to express and realise my original/new ideas within the research group. (Figure~\ref{fig:part1:rgroup_aspects_of_work}) (voluntary)
\item My work in the research group is too focused on my own research so because of that, I feel isolated from other research aspects of the whole project. (Figure~\ref{fig:part1:rgroup_aspects_of_work}) (voluntary)
\item My work in the research group allows me to have an impact on the decision-making of the project. (Figure~\ref{fig:part1:rgroup_aspects_of_work}) (voluntary)
\item In my work in the research group, I struggle to have enough resources (e.g. beam time, access to computing power or software, ...) to successfully accomplish my research tasks. (Figure~\ref{fig:part1:rgroup_aspects_of_work}) (voluntary)
\item My work in the research group allows me to keep a healthy work-life balance. (Figure~\ref{fig:part1:rgroup_aspects_of_work}) (voluntary)
\item My work in the research group gives me enough visibility within the group itself. (Figure~\ref{fig:part1:rgroup_visibility}) (voluntary)
\item My work in the research group gives me enough visibility outside the group. (Figure~\ref{fig:part1:rgroup_visibility}) (voluntary)
\item Working in my research group gives me many job opportunities in similar groups in the same field.(Figure~\ref{fig:part1:rgroup_job}) (voluntary)
\item Working in my research group gives me many job opportunities in other research groups.(Figure~\ref{fig:part1:rgroup_job}) (voluntary)
\item Working in my research group gives me a high probability of reaching a permanent position in the same field.(Figure~\ref{fig:part1:rgroup_job}) (voluntary)
\item Working in my research group gives me a high probability of reaching a permanent position in the industry or private sector in general. (Figure~\ref{fig:part1:rgroup_job}) (voluntary)
\item The time I spend doing service work for my research group is... (Figure~\ref{fig:part1:rgroup_amount_of_service_work}) (voluntary)
\item The time I spend doing service work for my research group is adequate. (Figure~\ref{fig:part1:rgroup_service_work}) (voluntary)
\item The service work I'm doing for my research group is well recognized. (Figure~\ref{fig:part1:rgroup_service_work}) (voluntary)
\item The service work I'm doing for my research group is useful for my career. (Figure~\ref{fig:part1:rgroup_service_work}) (voluntary)
\setcounter{Counter-ListOfQuestions}{\value{enumi}}
\end{enumerate}

\noindent \textbf{Work within a collaboration}\\
Only respondents who are a member of a collaboration were invited to answer questions from this part of the survey.

\begin{enumerate}
\setcounter{enumi}{\value{Counter-ListOfQuestions}}
\item What is the size of your collaboration? If you are part of multiple collaborations, please consider the one you mostly work for. (Figure~\ref{fig:part1:Q33})
\item Within your collaboration, how many people do you actively work with during a normal week? (Figure~\ref{fig:part1:Q34})
\item How do you consider the size of your collaboration to be? (Figure~\ref{fig:part1:consider-collab-size})
\item My work in the collaboration is useful to improve my knowledge, skills and expertise. (Figure~\ref{fig:part1:aspects_of_work_in_collab}) (voluntary)
\item There is room for me to express and realise my original/new ideas within the collaboration. (Figure~\ref{fig:part1:aspects_of_work_in_collab}) (voluntary)
\item My work in the collaboration is too focused on my own research so because of that, I feel isolated from other research aspects of the whole project. (Figure~\ref{fig:part1:aspects_of_work_in_collab}) (voluntary)
\item My work in the collaboration allows me to have an impact on the decision-making of the collaboration. (Figure~\ref{fig:part1:aspects_of_work_in_collab}) (voluntary)
\item In my work in the collaboration, I struggle to have enough resources (e.g. beam time, access to computing power or software, ...) to successfully accomplish my research tasks. (Figure~\ref{fig:part1:aspects_of_work_in_collab}) (voluntary)
\item My work in the collaboration allows me to keep a healthy work-life balance. (Figure~\ref{fig:part1:aspects_of_work_in_collab}) (voluntary)
\item My work in the collaboration gives me enough visibility within the collaboration itself. (Figure~\ref{fig:part1:visibility_collaboration}) (voluntary)
\item My work in the collaboration gives me enough visibility outside the collaboration. (Figure~\ref{fig:part1:visibility_collaboration}) (voluntary)
\item Working in my collaboration gives me many job opportunities in similar groups of the same collaboration. (Figure~\ref{fig:part1:job_collaboration}) (voluntary)
\item Working in my collaboration gives me many job opportunities in other collaborations. (Figure~\ref{fig:part1:job_collaboration}) (voluntary)
\item Working in my collaboration gives me a high probability of reaching a permanent position in the same field. (Figure~\ref{fig:part1:job_collaboration}) (voluntary)
\item Working in my collaboration gives me a high probability of reaching a permanent position in the industry or private sector in general. (Figure~\ref{fig:part1:job_collaboration}) (voluntary)
\item The time I spend doing service work for my collaboration is... (Figure~\ref{fig:part1:amount-of-service-work}) (voluntary)
\item I spend adequate amount of time doing service work for my collaboration. 
 (Figure~\ref{fig:part1:service_work_collaboration}) (voluntary)
\item The service work I'm doing for my collaboration is well recognized. (Figure~\ref{fig:part1:service_work_collaboration}) (voluntary)
\item The service work I'm doing for my collaboration is useful for my career. (Figure~\ref{fig:part1:service_work_collaboration}) (voluntary)
\setcounter{Counter-ListOfQuestions}{\value{enumi}}
\end{enumerate}

\noindent \textbf{Diversity of Physics programs}
\begin{enumerate}
\setcounter{enumi}{\value{Counter-ListOfQuestions}}
\item The diversity of physics programs (e.g different experiments, large variety of physics analyses) is a fundamental requirement for a fruitful development of Particle Physics. (Figure~\ref{fig:part1:Q52}) (voluntary)
\item Working in experiments that are under construction or in planning is ... for early-career researchers. (Figure~\ref{fig:part1:Q53}) (voluntary)
\item Working in experiments that are under construction or in planning offers ... career prospects for early-career researchers. (Figure~\ref{fig:part1:Q54}) (voluntary)
\setcounter{Counter-ListOfQuestions}{\value{enumi}}
\end{enumerate}

\noindent \textbf{Career perspective and planning}
\begin{enumerate}
\setcounter{enumi}{\value{Counter-ListOfQuestions}}
\item I am well informed about funding opportunities in the country I'm currently hired. (Figure~\ref{fig:part1:FeelInformed})
\item I am well informed about funding opportunities in Europe. (Figure~\ref{fig:part1:FeelInformed})
\item I am well informed about funding opportunities outside Europe. (Figure~\ref{fig:part1:FeelInformed})
\item I am well informed about career training opportunities. (Figure~\ref{fig:part1:FeelInformed})
\item I am well informed about resources on job application training. (Figure~\ref{fig:part1:FeelInformed})
\item I am well informed on what is needed to advance my career in academia. (Figure~\ref{fig:part1:FeelInformed})
\item I am well informed on what is needed to advance my career outside academia. (Figure~\ref{fig:part1:FeelInformed})
\item I am well informed on where to find advice and guidance regarding my career progression. (Figure~\ref{fig:part1:FeelInformed})
\item I get informed about funding or job opportunities on. (Figure~\ref{fig:part1:Q63}) (multiple answers per respondent allowed)
\item How prepared do you feel for the next stage in your career? (Figure~\ref{fig:part1:Q64})
\item I discuss my career prospects with my supervisor. (Figure~\ref{fig:part1:Discussion})
\item I discuss my career prospects with other senior researchers. (Figure~\ref{fig:part1:Discussion})
\item I discuss my career prospects with my peers. (Figure~\ref{fig:part1:Discussion})
\item What importance do YOU PERSONALLY attribute to the following items for a high-quality researcher? (a)-(k) (Figure~\ref{fig:part1:Q68})
\item From your point of view, what importance does the SCIENTIFIC COMMUNITY attribute the following items for a successful career in academia? (a)-(k) (Figure~\ref{fig:part1:Q69})
\item Thinking about your academic profile, how do you feel about these points? (a)-(k) (Figure~\ref{fig:part1:Q70})
\begin{enumerate}
\item International collaborations
\item Professional mobility
\item Publications and bibliographic metrics
\item Conference talks
\item Activity in boards, panels, etc.
\item Networking
\item Specialised expertise (e.g. FPGA programming)
\item Expertise in a variety of domains
\item Service work (in large collaborations)
\item Soft skill training (e.g. in project management)
\item Outreach
\end{enumerate}
\setcounter{Counter-ListOfQuestions}{\value{enumi}}
\end{enumerate}

\noindent \textbf{Work-life balance}
\begin{enumerate}
\setcounter{enumi}{\value{Counter-ListOfQuestions}}
\item Do you have children? (Figure~\ref{fig:part1:Q71})
\item During which career phase(s) did you have a child (children)? (Figure~\ref{fig:part1:Q72})
\item How important are the following items to you in order to have a good work-life balance? (a)-(f) (Figure~\ref{fig:part1:Q73})
\item To which extent are these aspects fulfilled in your current job? (a)-(f) (Figure~\ref{fig:part1:Q74})
\item In your opinion, to which extent are these aspects fulfilled in your field of research? (a)-(f) (Figure~\ref{fig:part1:Q75})
\begin{enumerate}
\item Flexible working hours
\item Flexible working location
\item Possibility to work part-time or of job-sharing
\item Good income
\item Possibility of long-term planning
\item Positive work environment
\end{enumerate}
\item What kind of influence do these items have or you think they would have on the quality and impact of your research? (Figure~\ref{fig:part1:Q76})
\begin{enumerate}
\item Flexible working hours
\item Part-time work
\item Job sharing
\item Relocation due to new job
\end{enumerate}
\item To undertake a new position in HEP... (Figure~\ref{fig:part1:Q77}) (multiple answers per respondent allowed)
\item Which problems or difficulties did you or your family members encounter when moving abroad for your new position? (Figure~\ref{fig:part1:Q78}) (voluntary, multiple answers per respondent allowed)
\item I work overtime (not compensating by working less other times). (Figure~\ref{fig:part1:overtimeStress})
\item There is a lot of pressure on me and I feel stressed. (Figure~\ref{fig:part1:overtimeStress})
\item If you are living abroad, would you like to move back to your home country one day? (Figure~\ref{fig:part1:Q81})
\item How did you decide about your current position? (Figure~\ref{fig:part1:Q82}) (multiple answers per respondent allowed)
\item In your career, did you have a break longer than 3 months? (Figure~\ref{fig:part1:Q83})
\item Did you ever change field within physics?
\item Are you considering leaving research in HEP after the current position? (Figure~\ref{fig:part1:Q85})
\item Which factors induced you to consider leaving research? (Figure~\ref{fig:part1:Q86}) (multiple answers per respondent allowed)
\setcounter{Counter-ListOfQuestions}{\value{enumi}}
\end{enumerate}

\noindent \textbf{Discriminatory or abusive treatment}
\begin{enumerate}
\setcounter{enumi}{\value{Counter-ListOfQuestions}}
\item Have you experienced discriminatory or abusive treatment in your collaboration/group? (Figure~\ref{fig:part1:Q87})
\item Are there any measures that would improve your personal situation? (Figure~\ref{fig:part1:Q88})
\setcounter{Counter-ListOfQuestions}{\value{enumi}}
\end{enumerate}

\noindent \textbf{Recognition and visibility}
\begin{enumerate}
\setcounter{enumi}{\value{Counter-ListOfQuestions}}
\item I consider the publication policy implemented in my group/collaboration (such as who signs the publications, in what order, and who is the corresponding author) to be fair. (Figure~\ref{fig:part1:FairnessRecognition}) (voluntary)
\item I consider the policy of representing my group/collaboration at conferences (such as the assignment of presenters) to be fair. (Figure~\ref{fig:part1:FairnessRecognition}) (voluntary)
\item I consider the assignment of positions in my group/collaboration (such as who is the convener of a working group, or member of an internal committee) to be fair. (Figure~\ref{fig:part1:FairnessRecognition}) (voluntary)
\item I believe the most commonly used bibliometric indexes (such as h-index) fairly reflect my work. (Figure~\ref{fig:part1:FairnessRecognition})
\item I consider the way of awarding prizes in my community to be fair. (Figure~\ref{fig:part1:FairnessRecognition})
\item I consider the recognition and visibility of my work in my group/collaboration to be sufficient. (Figure~\ref{fig:part1:FairnessRecognition}) (voluntary)
\item I consider the recognition and visibility of my work in my field of research to be sufficient. (Figure~\ref{fig:part1:FairnessRecognition})
\setcounter{Counter-ListOfQuestions}{\value{enumi}}
\end{enumerate}

\noindent \textbf{Final questions, feedback and remarks}
\begin{enumerate}
\setcounter{enumi}{\value{Counter-ListOfQuestions}}
\item In your opinion, is there any event the ECFA Early-Career Researchers Panel could organize to help you develop your career?
\item What are your most pressing questions in terms of career development? (Figure~\ref{fig:part1:Q97})
\item Do you have any other feedback about this survey?
\setcounter{Counter-ListOfQuestions}{\value{enumi}}
\end{enumerate}

%-----------------------------------------------------------------------------------------------
\section{Nationality groups}
\label{app:AppNationalityGroups}
For answers relating to countries, the following groupings are used to define regions with a larger sample size:
\begin{itemize}
    \item Northern Europe: Finland, Sweden, Norway, Netherlands, United Kingdom, Ireland, Belgium, Switzerland, Germany, Austria and Denmark;
    \item Mediterranean: France, Italy, Spain, Portugal, Greece;
    \item Central and Eastern Europe: Poland, Czech Republic, Slovakia, Slovenia, Bulgaria, Romania, Ukraine, Hungary, Belarus, Lithuania, Croatia, Georgia, Bosnia and Herzegovina, Moldova, Serbia and Russia;
    \item North America: United States of America, Canada and Mexico;
    \item South America: Peru, Chile, Brazil, Argentina, Colombia, Venezuela and Guatemala;
    \item West / Central Asia: Turkey, Kazakhstan, Israel and Iran;
    \item South Asia: Pakistan, India and Bangladesh;
    \item East Asia: China, Japan, Korea and Taiwan;
    \item Southeast Asia: Vietnam, Thailand and Malaysia;
    \item Africa: Zambia, Morocco, Egypt, Sudan and Algeria;
    \item Oceania: Australia and New Zealand.
\end{itemize}
Countries not included here had no responses.
%-----------------------------------------------------------------------------------------------
\section{Direct quotations}
\label{app:quotes}

In this appendix, direct (and adequately anonymous) quotations from the survey are taken, to provide a more detailed and honest report on the views of some of the ECRs. These do not necessarily reflect the views of the ECFA ECR panel members.

\vspace{0.5cm}
\noindent Direct quotations of answers given in response to question 78:``Which problems or difficulties did you or your family members encounter when moving abroad for your new position?'' in the open `Other' box:
\begin{itemize}
    \item ``I declined jobs not to leave my partner'';
    \item ``I destroyed any chances I ever had in finding companionship'';
    \item ``I never moved (all positions at the same university), to avoid a long-range relationship'';
    \item ``I searched for a position where I did not need to move, because my relationship is more important to me'';
    \item ``I would not like to move because of family and relationship'';
    \item ``leaving academia due to the contemporary mobility requirements'';
    \item ``my partner moved with me, and now we refuse to be separated'';
    \item ``difficulty in maintaining long distance relationships'';
    \item ``I cannot ask my partner to quit his/her permanent job to follow me'';
    \item ``I have managed to avoid moving since finding a partner and having children, but this has come at the expense of my career'';
    \item ``long distance relationship with partner'';
    \item ``missing intimacy with my partner'';
    \item ``partner was/is herself completing a PhD in another country'';
    \item ``relationship problems due to long distance'';
    \item ``relationships ultimately ended due to relocation'';
    \item ``long distance relationship over a decade'';
    \item ``having to relocate every couple of years, difficulties in calling a place home'';
    \item ``not enough documentation on the paper-work to be done after moving to a different country/continent'';
    \item ``missing support from family for childcare'';
    \item ``huge amount of time for the logistics of each move'';
    \item ``administration: registration, insurance, bank account, ...'';
    \item ``difficulties with bureaucracy being a foreign resident in the country'';
    \item ``Too difficult to find housing, zero support is given to PhD students'';
    \item ``residency/visa applications'';
    \item ``moving alone mid-pandemic was not the greatest experience''.
\end{itemize}

\vspace{0.5cm}
\noindent Direct quotations of answers given in response to question 86 ``Which factors induced you to consider leaving research?'' in the open `Other' box:
\begin{itemize}
    \item ``competitiveness, constant pressure and stress, lack of acknowledgement, lack of guidance in projects'';
    \item ``competition for PhD places'';
    \item ``difficult career path forward'';
    \item ``do not want to move abroad'';
    \item ``double rent and flights are very expensive on an average salary. Children living apart is very difficult'';
    \item ``geographical restriction of postdoc and professor market'';
    \item ``having to take a break (illness, child birth,... ) can be a serious problem (e.g. short - term contract does not get extended)'';
    \item ``I don't want to relocate to an other country'';
    \item ``I have been applying for permanent faculty positions over the last two years. The market is very rough. I personally felt that my work is not valued and, in several hiring decisions, it is connections and power dynamics which play a major role. At some point, I felt I had enough of this broken system and it is time to explore new directions'';
    \item ``I want the person who pays my salary/hires new people to also be the person who sets my work priorities. Them being separate really sucks'';
    \item ``I want to decide where to live based on the place, not based on where I get a job'';
    \item ``lack of motivation to pursue research in the field'';
    \item ``limited job opportunities'';
    \item ``long term HEP prospects are not very good because of political and economical crises'';
    \item ``moving to a place with gay-friendly social life'';
    \item ``my research does not seem to have direct impact to the immediate issues of society'';
    \item ``need to follow next another passion of mine and want to spend more time with other people'';
    \item ``no chance for a growth because of useless supervisors'';
    \item ``perspective'';
    \item ``preexisting psychological issues largely amplified by the pandemic'';
    \item ``staying in my home country'';
    \item ``the prospect of moving to a different country every two or three years to maybe then get a permanent position when I'm forty. Considering the amount of doctoral candidates and the amount of permanent positions, it is quite likely that I won't get a permanent position anyways, so I might just leave straight away'';
    \item ``trying and learning something new'';
    \item ``war'';
    \item ``academia is a terrible field with terrible prospects'';
    \item ``change research subject to different than HEP'';
    \item ``lack of self-realization'';
    \item ``lots of bureaucracy, not plausible situation with funding of basic research, small chance to get a grant'';
    \item ``not enough available tenure-track positions''.
\end{itemize}

\vspace{0.5cm}
\noindent 

Direct quotations of answers given in the open `Other' box in response to question 88 ``Are there any measures that would improve your personal situation?'' and question 96 ``In your opinion, is there any event the ECFA Early-Career Researcher Panel could organise to help you develop your career?'' which concern serious issues HEP is suffering from from the point of view of ECRs.
We have grouped these questions together since the answers were often broader than specifically suggesting events, and the topics are highly related.

\begin{itemize}
    \item ``Yes. My personal situation would improve immeasurably if permanent, tenure-track positions at universities, in the field of high-energy physics as a whole, especially LHC experimentation, were filled based upon individual candidate achievement and impact upon physics as a whole rather than a disgusting, malevolent, and destructive combination of whim, cronyism, and random chance. Hiring biases in HEP-EX are real and are getting worse. I have explicitly built an objectively impressive, above-and-beyond CV that stretches outside the bounds of the major LHC experiments -- but is firmly grounded in impressive work within them -- and I have received zero offers in many years as a post-doc. I'm well-known in the field -- and outside -- and I have been on some short lists but I have been passed over for all jobs in favour of less experienced, less impactful, less dynamic people (a situation that has nothing to do with gender, by the way; I am only comparing myself to other cis-men and I support, 100\%, all universities who hire my physicist colleagues who are women) -- and some of these people who do get the jobs I've applied for I've never heard of. I have solicited advice for how to improve my application over the years, hoping to see trends -- and there are no trends. The advice is random and contradictory. But the senior people who give this advice *all* speak with conviction about, "This is how to get a job in our field." When I point out how their advice explicitly contradicts advice I've received elsewhere, I receive excuses, dissembling, and noise. HEP-EX is dying because those making the tenure-track hiring decisions have outmoded, atavistic biases built into them, they select candidates who reinforce those biases, and there is absolutely no way to disrupt this wheel. I have given years and years of my life, ideas, inventiveness, enthusiasm, and knowledge to this field because I love it, and I will soon be forced to leave it because ... well, in fact, no one can tell me why. Permanent hiring in HEP-EX is based on whim, cronyism, and chance -- but we pretend that it is based upon objective metrics of achievement.''
    \item ``Yes, I was alone with children under 5 years of age which obviously affected my overall efficiency. Due to this reason now I am unable to apply for further position as my supervisor suggested that it would be very difficult for me to work further. However I would very much like to stay here...''
    \item ``The academia in general has taken up false values including hyper-mobility, career uncertainty, poor compensation and maximisation of the number of publications. Some of these issues interfere making things even worse. The issues have gotten worse as "business logic" has been brought to academia - without proper compensation or definition of working hours, not to speak of uncertain career prospects. A complete reform is in place, but not in sight.''
    \item ``Academia turns more and more into a rat-race. Even if one doesn't fall for it and manages oneself properly, the task of dealing with other people who will walk over heads, lie, steal for some promise of a nice position in the future, is time-consuming and unpleasant. I don't see this improving any time soon, instead I see more and more postdoc and PhD grants, which makes me wonder if it's worth staying in academia even if I am able to and get a nice position somewhere.''
    \item ``People should focus on science without making strong comments on issues not related to science like political and religious issues assuming that everyone agrees with them.''
    \item ``Combat short-term funding. Ideally every position should have a clear local career path associated with it (which might involve mobility: only you know you can come back afterwards).''
    \item ``Current post-doc challenges forces people to live super uncertain lives for uncertain duration. This needs to be addressed for our field to continue as all works rely on post-docs.''
    \item ``Discussions about mobility: research should be possible without the need of constant relocation, loosing actual (in person) social networks, and the trouble to find a match with partner position (or the rest of the family).''
    \item ``ECFA Early-Career Researchers Panel could organise something for the senior researchers to help them understand how the situation for the ECR community is different (both what is better now and what is worse) from when they were ECR.''
    \item ``I don't really think so as our field and science in general suffers from problems that are too big to be solved by ECFA alone.''
    \item ``it is important to involve more researchers all over the world and make them aware of the issues facing academia today (...). It is unclear to me whether organising any event will bring about reforms but, at the very least, it can bring some level of awareness.''
    \item ``Maybe one where PhD students can talk about our situations how do we feel and how people face the difficult situations and how they solve it. It's good to know that you are not the only one feeling stupid, depressed or burnout.''
    \item ``No - I think there needs to be a general shift in the way senior scientists think about recognition. Recognition is currently based on how well a student/postdoc networks with senior scientists to get ``high-quality'' reference letters. I have seen too many times that postdocs who are poorly seen by their peers get advanced.''
    \item ``No, because a major change in research politics would be necessary first.''
    \item ``No. The issues in careers in HEP can't be solved by focusing on how to ``develop'' the candidate, but by changing the expectations placed upon researchers. If one was really interested in a more inclusive workspace, with better work-life balance, with better long-term prospects and less mental health issues, the field itself would have to be changed. Non-compensated overtime should be banned, mobility should not be taken into account, average first permanent position age should be significantly lowered, PhD to permanent position ratio should be altered significantly, and nonsensical arbitrary metrics should be dropped. None of these issues can be solved by this panel, and while I truly appreciate the intentions, efforts and sincerity of this group, pushing young researchers to best-adapt in order to advance in a fundamentally broken field risks to further entrenching and reinforcing these issues.''
    \item ``No. The whole funding model needs to be reformed, this is very unlikely to happen.''
    \item ``No. There is no lack of information nor support structures. What is missing are real long-term perspectives and adequately paid permanent positions.''
    \item ``There are a lot of hidden biases in many important decisions, women less likely to get selected, non-white people less favoured, etc. I believe education can go a long way into minimising these things.''
\end{itemize}

\addcontentsline{toc}{section}{Bibliography}
\printbibliography[title=References]

@misc{ECFAECRPanel,
    url={https://ecfa.web.cern.ch/ecfa-early-career-researchers-panel},
    title={The ECFA Early-Career Researchers Panel (2021-2022)},
      note = {Accessed: 2022-12-01}
}

@misc{UKFCTownhall,
  title = {UK Future Collider Town-Hall},
  url = {https://conference.ippp.dur.ac.uk/event/1201/},
  note = {Accessed: 02/2024}
}

@misc{DORA,
    title = {Declaration of Research Assessment},
    url = {https://sfdora.org/read/},
    note = {Accessed: 02/2024}
}

@techreport{report2021-2022,
      author        = "{ECFA Early-Career Researcher Panel}",
      collaboration = "ECFA Early-Career Researcher Panel",
      title         = "{The ECFA Early Career Researcher's Panel: composition, structure, and activities, 2021 -- 2022}",
      archivePrefix = "arXiv",
      eprint        = "2212.11238",
      year          = "2022",
      url           = "https://arxiv.org/abs/2212.11238",
      note          = "Editors: Jan-Hendrik Arling, Emanuele Bagnaschi, Xabier Cid Vidal, Katherine Dunne, Viktoria Hinger, Armin Ilg, Henning Kirschenmann, Steven Schramm, Paweł Sznajder, Sarah Williams, Valentina Zaccolo"
}

@misc{instSchools,
    title = "Collection of Schools for Early Career Researchers in Particle Physics Instrumentation",
    url = "https://early-career-instrumentation.web.cern.ch/",
      note = {Accessed: 2022-12-01}
}

@misc{inspireHEP,
    title = "InspireHEP",
    url = "https://inspirehep.net/jobs",
    note = {Accessed: 02/2024}
}

@misc{AcademicJobsOnline,
    title = "AcademicJobsOnline",
    url = "https://academicjobsonline.org/",
    note = {Accessed: 02/2024}
}

@misc{LinkedIn,
    url = "https://www.linkedin.com/",
    title = "LinkedIn",
    note = {Accessed: 02/2024}
}

@misc{Researchgate,
    url = "https://www.researchgate.net/",
    title = "Researchgate",
    note = {Accessed: 02/2024}
}

@misc{Indeed,
    url = "https://www.indeed.com/",
    title = "Indeed",
    note = {Accessed: 02/2024}
}

@misc{EURAXESS,
    url = "https://euraxess.ec.europa.eu/jobs",
    title = "EURAXESS",
    note = {Accessed: 02/2024}
}

@misc{Stepstone,
    url = "https://www.stepstone.de/en/",
    title = "Stepstone",
    note = {Accessed: 02/2024}
}

@misc{Xing,
    url = "https://www.xing.com/en",
    title = "Xing",
    note = {Accessed: 02/2024}
}

@misc{JobsAcUK,
    url = "https://www.jobs.ac.uk/",
    title = "jobs.ac.uk",
    note = {Accessed: 02/2024}
}

@misc{AcademicsDE,
    url = "https://www.academics.de/",
    title = "academics.de",
    note = {Accessed: 02/2024}
}

@misc{PolytechnicPositions,
    url = "https://polytechnicpositions.com/",
    title = "PolytechnicPositions",
    note = {Accessed: 02/2024}
}

@misc{FindaPHD,
    url = "https://www.findaphd.com/",
    title = "findaphd",
    note = {Accessed: 02/2024}
}

@misc{CERNAlumni,
    url = "https://alumni.cern/page/career",
    title = "CERN Alumni",
    note = {Accessed: 02/2024}
}

@misc{AASJobRegister,
    url = "https://jobregister.aas.org/",
    title = "AAS Job Register",
    note = {Accessed: 02/2024}
}

@misc{EuroScienceJobs,
    url = "https://www.eurosciencejobs.com/",
    title = "EuroScienceJobs",
    note = {Accessed: 02/2024}
}

@article{Janssens:2023ufo,
    author = "Janssens, Kamiel and Ueda, Michiko",
    title = "{Probing depressive symptoms and the desire to leave academia among scientists in large, international collaborations in STEM}",
    eprint = "2308.05107",
    archivePrefix = "arXiv",
    primaryClass = "physics.soc-ph",
    month = "7",
    year = "2023"
}

\end{document}